\documentclass[letterpaper,12pt,titlepage,oneside,final]{book}
 
\newcommand{\href}[1]{#1} 

\usepackage{ifthen}
\newboolean{PrintVersion}
\setboolean{PrintVersion}{false}

\usepackage{amsmath,amssymb,amstext} 
\usepackage[pdftex]{graphicx} 

\usepackage[pdftex,pagebackref=false]{hyperref} 
\hypersetup{
    plainpages=false,       
    unicode=false,          
    pdftoolbar=true,        
    pdfmenubar=true,        
    pdffitwindow=false,     
    pdfstartview={FitH},    
    pdfnewwindow=true,      
    colorlinks=true,        
    linkcolor=blue,         
    citecolor=magenta,        
    filecolor=magenta,      
    urlcolor=cyan           
}
\ifthenelse{\boolean{PrintVersion}}{   
\hypersetup{	
    citecolor=black,%
    filecolor=black,%
    linkcolor=black,%
    urlcolor=black}
}{} 

\let\origdoublepage\cleardoublepage
\newcommand{\clearemptydoublepage}{%
  \clearpage{\pagestyle{empty}\origdoublepage}}
\let\cleardoublepage\clearemptydoublepage

\usepackage{xcolor}

\usepackage[justification=justified,
   format=plain]{caption} 
\usepackage{subcaption}
\usepackage{ragged2e}
\usepackage{todonotes}
\usepackage{bm} 
\usepackage{amsmath,amssymb,mathtools}
\usepackage{epigraph}
\usepackage[utf8]{inputenc}
\usepackage[greek.polutoniko,english]{babel}
\usepackage[T1]{fontenc}
\usepackage{threeparttable}
\usepackage{tcolorbox} 

\newcommand{\cc}[1]{\textsf{\textbf{#1}}}

\newcommand{\ket}[1]{| #1 \rangle} 		
\newcommand{\bra}[1]{\langle #1 |} 		
\newcommand{\braket}[2]{\langle #1 | #2 \rangle}

\newcommand{\RR}{\mathbb{R}} 
\newcommand{\CC}{\mathbb{C}} 
\newcommand{\ZZ}{\mathbb{Z}} 
\newcommand{\QQ}{\mathbb{Q}} 
\newcommand{\Sph}{\mathbb{S}} 
\newcommand{\EE}{\mathbb{E}} 
\newcommand{\HH}{\mathbb{H}} 
\newcommand{\MM}{\mathbb{M}} 
\newcommand{\dS}{\mathrm{dS}} 
\newcommand{\TT}{\mathbb{T}} 

\newcommand{\ABsquare}[2]{%
\mathord{\tikz[baseline=-0.5ex,scale=1.5,rotate=#2]
\fill[#1] (-0.8ex,-0.8ex) rectangle (0.8ex,0.8ex);}
}

\newcommand{\ABrhombus}[2]{%
\mathord{\tikz[baseline=0.2ex,scale=3,rotate around={#2:(0.603553ex, 0.25ex)}]
\fill[#1]
(0,0) --
(0.7071ex,0) --
(1.2071ex,0.5ex) --
(0.5ex,0.5ex) --
cycle;}
}

\newcommand{\CNP}{\texttt{CNP}}
\newcommand{\sCNP}{\texttt{sCNP}}
\newcommand{\cCG}{\mathrm{I}_{2}^{\infty}(\lambda_0^+)} 

\newcommand{\dd}{\mathrm{d}}

\newcommand{\fixedblock}[2]{%
  \fcolorbox{#1}{#1!20}{%
    \parbox[c][3em][c]{3em}{%
      \centering
      \scalebox{0.85}{$\begin{matrix}#2\end{matrix}$}%
    }%
  }%
}

\newcommand{\inflblock}[2]{%
  \fcolorbox{#1}{#1!20}{%
    \parbox[c][5em][c]{5em}{%
      \centering
      \scalebox{0.85}{$\begin{matrix}#2\end{matrix}$}%
    }%
  }%
}

\newcommand{\GramBlock}[2]{%
  \fcolorbox{#1}{#1!20}{%
    \parbox[c][3em][c]{6.5em}{%
      \centering
      \scalebox{0.75}{$\begin{matrix}#2\end{matrix}$}%
    }%
  }%
}

\newcommand{\BigGramBlock}[2]{%
  \fcolorbox{#1}{#1!20}{%
    \parbox[c][3em][c]{11em}{%
      \centering
      \scalebox{0.75}{$\begin{matrix}#2\end{matrix}$}%
    }%
  }%
}

\newcommand{\Gal}{\mathrm{Gal}}
\newcommand{\Tr}{\mathrm{Tr}}

\newcommand{\Cint}[2]{\mathrm{ \bf I} [ #1 , #2 ]  }
\usepackage{oplotsymbl}
\usepackage{scalerel} 
\usepackage{relsize} 
\definecolor{ppBlueAccentL40}{HTML}{9CC2E5}
\definecolor{ppBlueAccentL60}{HTML}{B4C6E7}
\newcommand{\cdiamond}{{
\textcolor{ppBlueAccentL60}{\mathlarger{\blacklozenge}} 
}}

\usepackage{amsthm}
\usepackage{thmtools}
\usepackage{thm-restate}
\newtheorem{theorem}{Theorem}

\newtheorem{lemma}{Lemma}
\newtheorem*{thesis}{Physical Extended Church--Turing Thesis}
\newtheorem*{notile}{No-Tile Theorem}

\newcounter{namedstatement}

\newenvironment{statement}[1]
{%
  \refstepcounter{namedstatement}%
  \begin{center}
  \begin{minipage}{\linewidth}
  \begin{quote}
  \textbf{#1:}
}
{%
  \end{quote}
  \end{minipage}
  \end{center}
}

\newcounter{namedconjecture}

\newenvironment{conjecture}[1]
{%
  \refstepcounter{namedconjecture}%
  \begin{center}
  \begin{minipage}{\linewidth}
  \begin{quote}
  \textbf{#1:}
}
{%
  \end{quote}
  \end{minipage}
  \end{center}
}

\begin{document}

\pagestyle{empty}
\pagenumbering{roman}

\begin{titlepage}
        \begin{center}
        \vspace*{0cm}

        \Huge
        {\bf Spacetime Quasicrystals} \\ {\Large and} \\ {\bf Computational Complexity Theory} \\ {\Large in the} \\ {\bf Road Towards Quantum Gravity}

        \vspace*{0.5cm}

        \normalsize
        by \\

        \vspace*{0.5cm}

        \Large
        Sotirios Mygdalas \\ \large{(\textgreek{Σωτήριος Μυγδαλάς})}\\

        \vspace*{3cm}

        \normalsize
        A thesis \\[0.5ex] 
        presented to the University Of Waterloo\\[0.5ex] 
        in fulfillment of the \\[0.5ex]
        thesis requirement for the degree of \\[0.5ex]
        Doctor of Philosophy\\[0.5ex] in\\[0.5ex] Physics\\

        \vspace*{1.5cm}

        Waterloo, Ontario, Canada, 2026 \\

        \vspace*{0.5cm}

        \copyright\ Sotirios Mygdalas 2026 \\
        \end{center}
\end{titlepage}

\pagestyle{plain}
\setcounter{page}{2}

\cleardoublepage 

\newpage
\phantomsection    
 

\addcontentsline{toc}{chapter}{Examining Committee}
\begin{center}\textbf{Examining Committee Membership}\end{center}
  \noindent
The following served on the Examining Committee for this thesis. The decision of the Examining Committee is by majority vote.
  \bigskip
  
  \noindent
\begin{tabbing}
Internal-External Member: \=  \kill 
External Examiner: \>  Felix Flicker \\ 
\> Senior Lecturer, School of Physics, University of Bristol \\
\end{tabbing} 
  \bigskip
  
  \noindent
\begin{tabbing}
Internal-External Member: \=  \kill 
Supervisor(s): \> Latham Boyle \\
\>  Reader, Higgs Centre for Theoretical Physics, University of Edinburgh \\\\
\> Niayesh Afshordi \\
\> Professor, Dept. of Physics and Astronomy, University of Waterloo \\
\end{tabbing}
  \bigskip
  
  \noindent
  \begin{tabbing}
Internal-External Member: \=  \kill 
Internal Member: \> Avery Broderick \\
\> Professor, Dept. of Physics and Astronomy, University of Waterloo \\
\end{tabbing}
  \bigskip
  
  \noindent
\begin{tabbing}
Internal-External Member: \=  \kill 
Internal-External Member: \> Achim Kempf \\
\> Professor, Dept. of Applied Mathematics, University of Waterloo \\
\end{tabbing}
  \bigskip
  
  \noindent
\begin{tabbing}
Internal-External Member: \=  \kill 
Other Member(s): \> Lucien Hardy \\
\> Research Faculty, Perimeter Institute for Theoretical Physics \\
\end{tabbing}

\cleardoublepage
\phantomsection    

 \addcontentsline{toc}{chapter}{Author's Declaration}
 \begin{center}\textbf{Author's Declaration}\end{center}

 \noindent  
This thesis consists of material all of which I authored or co-authored: see Statement of Contributions included in the thesis. This is a true copy of the thesis, including any required final revisions, as accepted by my examiners.
I understand that my thesis may be made electronically available to the public.

\cleardoublepage
\phantomsection    

 \addcontentsline{toc}{chapter}{Statement of Contributions}
 \begin{center}\textbf{Statement of Contributions}\end{center}

The thesis contains two parts (see Outline \ref{sec:ThesisOutline}).

Part I is based on a short paper \cite{fox2026semiclassicalgravityefficientlysolves} that has been written in collaboration with Matthew Fox (University of Colorado, Boulder) and Chaitanya Karamchedu (University of Maryland). This project was born several years ago with Matthew, during our Master's year at the Perimeter Scholar's International (PSI) program in 2021-2022. Matthew attended a talk at Perimeter on the Schr\"odinger-Newton equation and its foundations, while I was working on a winter school project on collapse models based on that equation (with Latham Boyle). We soon realized that Matthew's interest in computational complexity theory and mine in gravitational physics could lead to an interesting project, so we started working together. It was not until 2024-2025, though, that we were able to convert our original ideas into something more concrete. Teamed up with our friend Chai, we wrote an essay for the 2026 Gravity Research Foundation Essay Competition. This evolved into a paper, which is currently under review at \textit{Physical Review Letters} (PRL) (submitted June 24, 2026). All authors contributed equally to the writing of the paper and the conceptualization of the ideas. The most important difference in Part I from the paper \cite{fox2026semiclassicalgravityefficientlysolves} is that the introduction (Chapter \ref{ch:complexity}) is slightly more expanded here, and we added three figures.


Part II is based on a considerably longer paper \cite{mygdalas2026spacetimequasicrystals} that I co-authored with my PhD supervisor Latham Boyle (Higgs Centre for Theoretical Physics, University of Edinburgh and Perimeter Institute). The project started in 2022 in my PSI Master's Essay, where we mentioned the first speculations about how a Lorentzian quasicrystal -- \textit{if} it existed -- could be relevant for string theory compactifications. Building these quasicrystals mathematically and understanding their properties has been the main part of my PhD work since then. While Part II contains the paper \cite{mygdalas2026spacetimequasicrystals}, it is a substantial extension of that work in the following sense:  
\begin{itemize}
    \item The introduction in Chapter \ref{ch:MatBackgroundPartII} has been expanded from the paper, especially the part referring to Coxeter's theory; there are more figures here and more thorough explanations.
    \item Chapter \ref{ch:QuasicrystalsMinkowski} has been revised as well, especially the parts about the non-crystallographic symmetries in Sections \ref{sec:MathematicalBackgroundLorentzian} and \ref{sec:SpacetimeCNP}. I added more figures throughout and another example of a spacetime quasicrystal in Section \ref{sec:ExamplesSpacetimeQuasicrystals}.
    \item Chapters \ref{ch:GeneralizedSpacetimeCNP} and \ref{ch:sQC_CausalSets} are entirely new and are based on ongoing unpublished work that I have been doing on my own since April 2026. This thesis is the first time that these ideas are reported, apart from Lemma \ref{lemma:EigenvalueSharing} in Appendix \ref{appendix:FoldingUnfolding} and the related discussion in Section \ref{sec:RevisitingNonCrystallographicSymmetry} that was worked out by Alex Thorne \cite{Thorne2026} in a Master's thesis at the University of Edinburgh, for which I served as the co-supervisor (with Latham Boyle). I acknowledge useful discussions with Latham Boyle and Yasaman Yazdi (Dublin Institute for Advanced Studies) for the content of these chapters, and especially with Alex Thorne (see Acknowledgements) 
    \item All the appendices, apart from Section \ref{app:inflation} and Appendix \ref{app:SymmetriesLorentzianLattices} that were copied from our paper \cite{mygdalas2026spacetimequasicrystals} have been written from scratch for this thesis. Appendix \ref{appendix:FoldingUnfolding}, in particular, might be of independent mathematical interest.
\end{itemize}
Other minor revisions have been carried out throughout the thesis. At the time of writing, the paper \cite{mygdalas2026spacetimequasicrystals} is peer review at \textit{Physical Review Research} (PRR) (submitted February 24, 2026), and we are expecting the first referee report. More publications are to follow from the new results of this thesis.

Unless otherwise stated, all figures in this thesis have been produced by me, either in \texttt{Mathematica} or by drawing them by hand on my reMarkable tablet. 

Finally, I have been heavily benefited from interactions with Anthropic's \texttt{Claude Code}, who helped me drastically revise my own code and build more efficient new functions, especially for the new results of Chapter \ref{ch:sQC_CausalSets}. I have also used Google's \texttt{Gemini} quite frequently for browsing through the literature, and as a research assistant (using the Pro Extended version), to bounce ideas off of. 

\noindent  
\cleardoublepage
\phantomsection    

\addcontentsline{toc}{chapter}{Abstract}
\begin{center}\textbf{Abstract}\end{center}

\textbf{Part I: Semiclassical Gravity Efficiently Solves \cc{NP}-Complete Problems}

Assuming the gravitational field is classical and that it couples to quantum fields via the semiclassical Einstein field equations, we show that the weak-field dynamics of a massive and non-relativistic qubit can in principle be used to solve an $\cc{NP}$-complete problem in polynomial time. We attribute this vast computational power to the non-linear qubit dynamics afforded by the semiclassical Einstein field equations. Consequently, the above two assumptions entail a violation of the Physical Extended Church--Turing Thesis, which we regard as evidence for the quantization of gravity.

\textbf{Part II: Spacetime Quasicrystals} 

Self-similar quasicrystals (like the famous Penrose and Ammann-Beenker tilings) are exceptional geometric structures in which long-range order, quasi-periodicity, non~crystallographic orientational symmetry, and discrete scale invariance are tightly interwoven in a beautiful way. Here, we show how such structures may be generalized from Euclidean space to Minkowski spacetime. We construct the first examples of such {\it Lorentzian quasicrystals} (the spacetime analogues of the Penrose or Ammann-Beenker tilings), and point out key novel features of these structures (compared to their Euclidean cousins). Then, we present some ideas about how such spacetime quasicrystals might relate to reality. 

First, we mention a speculative yet intriguing scenario in which our infinite $(3+1)$D universe is embedded (like one of our spacetime quasicrystal examples) in a particularly symmetric $(9+1)$D torus $\TT^{9,1}$ (which was previously found to yield the most symmetric toroidal compactification of the superstring). We suggest how this picture might help explain the mysterious seesaw relationship $M_{\rm Pl}M_{\rm vac}\approx M_{\rm EW}^{2}$ between the Planck, vacuum energy, and electroweak scales ($M_{\rm Pl}$, $M_{\rm vac}$, $M_{\rm EW}$).

Second, we explore causal set theory by replacing the traditionally random causal sets with our ordered and highly symmetric quasi-crystalline causal sets. We provide suggestive evidence that our spacetime quasicrystals do beat Poisson random sets in the number-volume correspondence in both ($1+1$)D and ($3+1$)D, with the latter case conjectured to be impossible. Finally, we attempt to understand these results by extending the concept of hyperuniformity from space to spacetime, and we show that our $(1+1)$D quasicrystal is the first structure shown to be indeed spacetime hyperuniform (we leave testing the $(3+1)$D quasicrystal for a future work).

\cleardoublepage
\phantomsection    

\addcontentsline{toc}{chapter}{Acknowledgements}
\begin{center}\textbf{Acknowledgements}\end{center}

Defending this thesis marks not only the end of my PhD, but also my stay in Waterloo and Perimeter (at least for now!). As a result, this is the perfect opportunity to reflect and be grateful for all the support that I received throughout this journey. 

First of all, let me take some time to properly thank my advisor, Latham Boyle, for... everything! Literally! I do not even know where to begin, so let me start from the very beginning. Back in 2022, Latham agreed to take me as his PhD student, but he knew he would not stay at Perimeter for much longer (and indeed, in July 2023 he left for the Higgs Centre at the University of Edinburgh). Somehow he managed to strike the best deal for my interest: on the one hand, he secured funds for my PhD at Perimeter till the end (I am grateful to Neil Turok as well for that), and on the other hand, he negotiated some more funds on the Edinburgh side for me to visit for several months per year.  

Let me pause for a second and thank the School of Physics and Astronomy and the Higgs Center for Theoretical Physics at the University of Edinburgh for their hospitality in the following occasions (Fall 2023, July 2024, Fall 2024, July 2025, Fall 2025 and July 2026), without which finishing this work would have been impossible. Specific thanks to Fiona Mitchell for helping me arrange all of my visits.

Back to Latham now. Latham has such a zest for physics and mathematics, and such broad interests and knowledge that interacting with him has been an absolute pleasure all these years. I have learned how to make -- or search for -- connections between seemingly unrelated things and I will always cherish our moments on the blackboard trying to figure out something together. Finishing a PhD is difficult when your advisor is along the corridor, let-alone when they are on a different continent! But we did it (almost, with the permission of this committee), and I could not be any more grateful for knowing that Latham was there for me when I wanted him. He introduced me to the fascinating world of aperiodic order and together we achieved to construct the first spacetime quasicrystals. 


Even though we did not end up working together, I would like to thank Niayesh Afshordi for being the voice-of-reason in all my PhD advisory committee meetings, for always emphasizing that I should write my ideas\footnote{Actually, in my most recent PhD committee meeting, I got a report back saying: ``Defence planned; now, write, write, write!''. Now, I am not sure who wrote it, but whoever did, I blame them for this thesis being 300+ pages!} and publish papers, which pushed me to work harder and eventually do so.

Many thanks to my examining committee, in particular to my external examiner, Felix Flicker, for pointing out several corrections that could me made to the thesis to improve clarity and readability. I am also grateful for the support of my advisory committee throughout this PhD. 

Next, I would like to thank my collaborators and the people that helped me with my research all these years:

Let's start from Part I. Working with my friends and fellow PhD students Matthew Fox and Chaitanya (Chai) Karamchedu on this project that we affectionately call \cc{GRAVIgate} has been a phenomenal experience for me! Not only did I learn so much from them (I know all of my complexity theory from Matthew), but also writing up our paper \cite{fox2026semiclassicalgravityefficientlysolves} together was really a process I adored and definitely taught me one or two things about how to write. I will say one thing: we had been going back and forth about what is the perfect word to use in almost each sentence that conveys our thoughts more accurately, which is why it took us two more months to put our work on the arXiv from the moment we had the first essay ready for the Gravity Research Foundation competition. I am very happy with the final product and am looking forward to what else we can work all together.

Not all projects make it to one's PhD thesis or to a publishable result in the strict timeline of a PhD. This has happened to me with another project I have been working for almost two years now with my friend and fellow PhD student Cole Coughlin (Perimeter Institute). Cole has an infectious curiosity and enthusiasm about theoretical physics; you can hear him talking about shapes and his voice would echo in the full Perimeter building. We figured out so many things together when we started on a mission to reformulate the full standard model of particle physics in terms of rotations, projections and whatnot of polytopes in higher dimensions. I learned so many things from him already and I cannot wait to go back to our project after this thesis is successfully submitted. A highlight in our collaboration so far is the fact that we once made a rhombic dodecahedron from Lee Smolin's hangers -- sorry Lee! 

On that note, I would like to thank Kirill Krasnov for teaching me so many things about spinors, the standard model and unification, and for hosting me at the University of Nottingham. Kirill, thank you for listening to my shenanigans and for helping me think more mathematically in order to communicate my ideas to you. 

Next in the list is my former MSc-student, now PhD-brother, Alex Thorne (University of Edinburgh). Alex was looking into a Master's project at the intersection of quantum error-correction, quantum gravity and quasicrystals when we first met at the TJ Walls cafe in Edinburgh last summer. After talking to Latham, we decided to give him the project of tiling our spacetime quasicrystals. This turned out to be a wild goose hunt, so we quickly changed the project to studying the symmetries of spacetime quasicrystals. Alex took on the challenge so well: not only did he come back with insightful comments on the things we discussed after each meeting, but he constantly raised important points on what to try next. His Lemma \ref{lemma:EigenvalueSharing} was the first step towards unlocking more of the non-crystallographic symmetries of spacetime quasicrystals. Talking to Alex helped me piece together several of my slightly disconnected thoughts and come up with the framework of foldings and unfoldings discussed in this thesis. I am looking forward to discussing more with him now that he is in his PhD and write our paper together!

I would like to thank Yasaman Yazdi (Dublin Institute for Theoretical Physics) for all of our interactions ever since we met this December in Dublin, where she invited me to give a talk. I have learned a lot about causal sets by talking to Yasaman and without her I would not even know where to start from the results of Chapter \ref{ch:sQC_CausalSets}. Also, a special thanks to Paul Steinhardt for introducing me to the concept of hyperuniformity during a short visit to Princeton in November 2025, and to Fabian Wagner (University of Heidelberg) for brainstorming with me how to generate more efficiently spacetime quasicrystals (and for taking interest in my work and giving us the first citation of our paper \cite{mygdalas2026spacetimequasicrystals}).

My PhD mentor, Selim Hotinli, has been instrumental in helping me finish this thesis. Not only did he teach me how to use AI tools more efficiently, but also paid for half a year of subscription to Anthropic's \texttt{Claude}, which definitely sped up the coding process. 

I am indebted to Padelis Papadopoulos -- whom I affectionately call ``the Captain'' -- for organizing stimulating physics discussions among peers and for cultivating part of my thinking about physical theories. 

Perimeter is a very special place and perhaps the sole institute for theoretical physics whose character is not measured by how old the building is (it's new and beautiful!) but by the amazing people that are inside it everyday. From the people at the security desk that will make your day with a good morning (thanks Anik), to the cleaning stuff that will keep you company even late at night (thanks Randy, for making it `another day in paradise' for us), everyone is amazing. Special thanks to Dan Lynch from the Black Hole Bistro and his staff; you guys make the bistro such a beautiful environment to be. Thanks to the people in Academic Programs, in particular to Angela Hovdestad, Maïté Dupuis and Sunny Tsang, for always being helpful with grad affairs, and to the grad students that served with me for three and a half years in Perimeter's Grad Committee; I want to believe that we brought some change all together. Thanks to Outreach too, for giving me the chance to practice my science communication skills. 

I could literally spend hours thanking people at Perimeter, but let me close this part of the acknowledgements by expressing my utmost gratitude for the help I received from the IT department at Perimeter on August 11, when I accidentally dropped coffee all over my laptop during a defence presentation practice and I almost fried my laptop! I told them that the only copy of my presentation was inside that laptop and they did everything in their power to rescue my SSD and save it! They did, and I am really thankful for all their help during these very stressful two days before my defence on August 14, 2026!

I would also like to thank the Savvas Chamberlain Family Foundation and the Hellenic Heritage Foundation for their continuous support through the Anaximandros Fellowship -- and personally, Savvas and Christine Chamberlain, Maria Harper, Tula Alexopoulos, Michael Lagopoulos and Tony Lee. I consider all of you amazing friends that played a very important part of my stay in Canada. Thank you for your generosity and for choosing to support me through thick and thin. I would also want to thank the AHEPA Educational Foundation for the 2024/25 Mavroyiannis Scholarship. 

A big thank you to Perimeter Institute and the Stavros Niarchos Foundation (SNF) in Athens for inviting us to participate in their SNF Nostos 2026 conference in June 2026. Their hospitality was amazing and I thank them for keeping me caffeinated and energized during the last moments of writing of this PhD thesis. A special thank you to Olga Michalopoulos for being a constant source of insipiration, and for making sure I was not distracted by the conference events and was making progress with my thesis every day. 

I met several people at Perimeter that I got fortunate enough to call my friends during my PhD. I thank Conner Dailey, Suvendu Giri, Fransisco Borjes, Dalila Pirvu, Hassan Khalvati, Jury Radkovski, José de Jesús Padua Argüelles, Robin Oberfrank, Kiana Salehi, Alice Chen, and my officemates, Yìlè Yīng and Marina Maciel Ansanelli.  

As my PhD-brother Vatsalya Vaibhav has been calling me, I was fortunate enough to be a \textit{quasi-local} in Edinburgh.  I have met lots of people and made lots of friends there. I thank Vatsalya, Sam Bateman and Kostas Tzanavaris for stimulating discussions about physics and life. Thanks also to Maria Polackova, Asaad Elkhidir and Maegan Anderson for making my stays much more enjoyable.

Finally, let me thank the most important person in my life, without whom completing (or even starting) this PhD might have been impossible: my partner, Jordan Krywonos. Jordan has been my support for four years now and believed in me in moments that I did not even believe in myself. She has been always there for me and I am so proud of everything we achieved together in this PhD. I could not have had a better partner to rely on than her, and I am very grateful for the fact that coming to Canada and Perimeter was the reason I met her and my life is happier ever since.

\newpage 
\noindent\textbf{General Acknowledgments}

\textbf{Part I: Semiclassical Gravity Efficiently Solves \cc{NP}-Complete Problems}

The authors are grateful to Eivind J{\o}rstad, Charlie Cummings, Mathew Bub, Thomas Galley, Flaminia Giacomini, and Andr{\'e} Gro{\ss}ardt for discussions related to this project. We are particularly grateful to Ning Bao for explaining the details of the Bao--Bouland--Jordan algorithm in \cite{Bao_2016}.

\textbf{Part II: Spacetime Quasicrystals}

We thank Nikolay Bogachev, 
Mikhail Belolipetsky,
Richard Borcherds,
Gerald H\"ohn, Tobias Hartnick, 
Henna Koivusalo,  
Jury Radkovski, 
Ben Webster,
and Kendrick Smith for helpful discussions.

Research at Perimeter Institute is supported in part by the Government of Canada through the Department of Innovation, Science and Economic Development Canada and by the Province of Ontario through the Ministry of Colleges and Universities.

\cleardoublepage
\phantomsection    

\addcontentsline{toc}{chapter}{Dedication}
\begin{center}\textbf{Dedication}\end{center}

To my parents, Asterios (\textgreek{Αστέριος}) and Athanasia (\textgreek{Αθανασία}), who have wholeheartedly supported me throughout this academic journey and always encouraged me to pursue my dreams and my aspirations. The unconditional love I have been getting throughout the years from my family, including my sister Stella (\textgreek{Στέλλα}) and my two grandparents that are still with us, Ioanna (\textgreek{Ιωάννα}) and Athanasios (\textgreek{Αθανάσιος}), is remarkable. Making them proud is definitely one of the driving forces that made me finish this PhD degree.

\cleardoublepage
\phantomsection    

\renewcommand\contentsname{Table of Contents}
\tableofcontents
\cleardoublepage
\phantomsection    

\addcontentsline{toc}{chapter}{List of Figures}
\listoffigures
\cleardoublepage
\phantomsection		

\addcontentsline{toc}{chapter}{List of Tables}
\listoftables
\cleardoublepage
\phantomsection		



\newpage
\cleardoublepage
\setlength{\epigraphwidth}{0.6\textwidth}
\epigraph{``\it \textgreek{>Ασπούδαστος περί γεωμετρίας μηδείς ε>ισίτω...}'' \hspace{+1ex}\footnotemark \\[1ex] (Let no-one uninterested in geometry enter)}{\textgreek{>Ακαδημία Πλάτωνoς, >Αρχαία >Αθήνα, $\sim$ 387 π.Χ.}\\[1ex](Plato's Academy, Ancient Athens, $\sim$ 387 BC}
\footnotetext{
A tribute to the Perimeter Institute’s iconic saying, once gloriously decorating its entrance, welcoming its many visitors. Now, gracefully retired, yet immortalized on the wall beside the entrance with its famous typo at last set right. Funny how some charming imperfections can somehow survive even in an institute devoted to fundamental truths. For twenty five years since its conception, the misspelled saying was mercifully unnoticed by those stampeding each morning towards the coffee station. Even the few among us who understand Greek, though potentially bewildered, did not say anything; but this stops now. The author is grateful to Brian Lasher and Jen Giller for their help and eagerness to correct the saying (from \textgreek{>Ασπο\underline{τ}ύδαστος} to \textgreek{>Ασπούδαστος}).}

\newpage
\pagenumbering{arabic}

\addcontentsline{toc}{chapter}{Preface}
 \begin{center}\textbf{Preface}\end{center}

\noindent 
The notation used in this thesis reflects my own personal taste. An example is the choice of $\texttt{CNP}$ -- instead of the C\&P used in our paper \cite{mygdalas2026spacetimequasicrystals} -- to denote the cut-and-project algorithm.\footnote{We hope that no-one feels grammatically offended by this choice. If someone does, though, I am wondering whether having some Mac `n' Cheese or Fish `n' Chips, or listening to Rock `n' Roll or Guns `n' Roses, could help. If not, a visit to an In-N-Out-Burger would definitely help! Inspired perhaps by the latter, the middle `n' is capitalized and carries no apostrophes here. Also, \texttt{CNP} looks better to us as an acronym than \texttt{CnP} or -- the perhaps cursed -- \texttt{C`n'P}!} Throughout the thesis we use three different fonts:

\textsf{This font is used to describe {\bf complexity classes}, i.e.~like \cc{NP}.}

\texttt{This font is used to describe the cut-and-project algorithm \CNP{}/\sCNP{} and other coding-related things, like Mathematica or C.}

This font is used for the rest of the text.

\cleardoublepage
\phantomsection    



\chapter{Motivation}

\section{Is Quantum Gravity Necessary?}

The road to quantum gravity is not easy. Stuck in our armchairs with little to no experimental guidance, it is prudent to ponder why we think gravity is quantum-mechanical at all. Could it be that the contrary is true, that gravity is not fundamentally quantum but classical? Of course, this question is hardly new. It has a long and rich history, and traces back to the seminal works of M{\o}ller \cite{Lichnerowicz:1962gnh} and Rosenfeld \cite{ROSENFELD1963353} in the 1960s.

Treating gravity classically but matter quantum-mechanically embodies \emph{semiclassical gravity}, which naturally divides into two parts. The first part is the effect of curved spacetimes on quantum fields. This part is well-understood (full books like \cite{Wald:1995yp,Birrell_Davies_1982,Mukhanov:2007zz,Parker:2009uva} have been written on it) and it has proven to be invaluable because of Hawking's discovery that gravitational fields can create particles \cite{hawking1975particle}. The second part is the effect of quantum fields on curved spacetimes. This part is not nearly as understood as the first, largely because it is not clear what it means to superpose spacetimes, which appears to be an inevitability \cite{Wal84}.

Within the quantum gravity community, semiclassical gravity is generally regarded as an approximation to an as yet undiscovered theory of quantum gravity \cite{Wald:1995yp,Birrell_Davies_1982,Mukhanov:2007zz,Parker:2009uva}. However, despite many ongoing experiments (e.g., \cite{MarlettoVidal_GravityMediatedEntanglement_2017, Berglund_2023, Namdar_TripleInterference, tobar2024detecting}), there is no definitive experimental evidence that gravity is quantized. As a result, quantum gravity is supported primarily by theoretical arguments, such as those by Eppley and Hannah \cite{eppley1977necessity} and Page and Geilker \cite{Page_1981}. However, many of these arguments are considered inconclusive \cite{mattingly2005quantum,mattingly2006eppley,Carlip_2008, Grossardt_2022, Bahrami_2014, kent2018AgainstEH}, highlighting the need for a new case for the quantization of gravity.

In Part I, we provide such an argument by exploring the computational consequences of a fundamentally semiclassical world. In particular, we show that if semiclassical gravity (as described by the semiclassical Einstein field equations) is correct, then it is possible to solve $\cc{NP}$-complete problems in polynomial time. As this challenges commonly held beliefs about what is and what is not efficiently computable by physical devices (beliefs which are encapsulated in a strong and physical version of the Church--Turing Thesis), our results suggest that semiclassical gravity cannot be a correct description of physical reality.

Our main result hinges on an algorithm due to Bao, Bouland, and Jordan \cite{Bao_2016}, which itself is based on an earlier algorithm by Abrams and Lloyd \cite{Abrams_1998}. Essentially, the Abrams--Lloyd algorithm showed how to use Weinberg's non-linear theory of quantum mechanics \cite{WEINBERG1989336ha} to efficiently solve hard computational problems (e.g., $\cc{NP}$-complete problems). Several years later, the Bao--Bouland--Jordan algorithm extended this result and showed that it is, in fact, a more general phenomenon: basically \emph{any} non-linear theory of quantum mechanics can be used to efficiently solve hard computational problems. Our conclusions follow from combining the Bao--Bouland--Jordan algorithm with arguments due to Bahrami et al.~\cite{Bahrami_2014} and Giulini et al.~\cite{Giulini_2023} that the weak-field evolution of a massive qubit in semiclassical gravity is necessarily non-linear. Incidentally, our work is consistent with related work by Kent et al. \cite{Kent_2005,Kent_2021,Fedida_2026}, which we discuss in Section~\ref{sec:GRAVIgateDiscussion}.

\section{Roaming Over the Landscape of Quantum Gravity Theories}
\label{sec:QGTheories}

Having another reason to believe that gravity must be quantized, we now turn to what do we mean by ``quantum gravity''. 

Over a century ago, Einstein (in fact, Minkowski) taught us that {space and time are inseparably united} into a single entity: \text{spacetime}, the stage on which materials live and evolve according to physical laws. General relativity took a bolder step further, revealing spacetime as a dynamical stage that affects and gets affected by the materials, and whose {geometry manifests as gravity}. At the same time, {quantum theory} showed that quantities long assumed continuous can be {fundamentally discrete}. Macroscopically, materials and spacetime appear classical, smooth and continuous. Yet, through a microscope one sees the granularity of a slab of matter, with its atoms obeying the laws of quantum theory at these tiny scales. The question is: ``if gravity is geometry and `quantumness' is discreteness, \textbf{would one see similar {`geometric atoms'} if we could peak into spacetime's microscopic structure?}''

There are several approaches to quantum gravity (QG) in the literature and any attempt to provide a comprehensive review of each one of them here would be futile. For this reason, we contain ourselves to simply mentioning some of them. 

\subsection{Discretum vs Continuum}

We do so by placing them into two very broad categories: discrete and continuous theories. Here, we focus on whether or not they describe spacetime in terms of a continuous manifold (on which perhaps quantum fields or strings live) or replace this smooth structure with some discrete notion of spacetime. In the discrete category, we have
\begin{itemize}
    \item \textbf{Loop Quantum Gravity (LQG)} \cite{Ashtekar_2021,rovelli2015covariant,Engle_2023}: background-independent quantization of Einstein's general relativity rewritten in terms of connection variables. It has two formulations: first, its Hamiltonian version, based on canonical quantization and an inherent $(3+1)$ splitting of spacetime. Quantizing the variables on a 3D spatial slice yields the quantum states of spatial geometry, known as the \textit{spin networks}. Second, its covariant, or Lagrangian, formulation, based on the path integral. This leads to a sum over all possible spacetime histories (geometries) in 4D, known as \textit{spinfoams}.    
    \item \textbf{Causal Dynamical Triangulations (CDTs)} \cite{ambjorn2024causaldynamicaltriangulationsgateway,ambjornCDTs,Loll_2019}: a non-perturbative lattice path-integral approach that sums over all possible ways to glue flat simplicial geometric building blocks (4-simplices, the analogues of triangles in higher dimensions) together to build a triangulated spacetime.
    \item \textbf{Causal Set Theory (CST)} \cite{Surya_2019}: replaces the continuous spacetime entirely with a discrete, partially ordered set of fundamental, indivisible spacetime elements (``events''). The partial order explicitly represents the causal structure, meaning geometry is fundamentally discrete and causality is the foundational organizing principle.
\end{itemize}
Perhaps the most discrete theory out of all discrete theories is CST; this is because in the other approaches one usually takes the limit that the discrete building blocks vanish, for example by taking the edges of a triangulation to be zero in order to recover the continuum. 

In the continuous category, we have
\begin{itemize}
    \item \textbf{String Theory (ST)} \cite{Polchinski:1998rq,Polchinski:1998rr}: replaces 0D point-like particles with 1D vibrating strings moving through a higher-dimensional spacetime, and higher-dimensional objects known as branes. Gravity emerges naturally as the lowest vibrational mode of a closed string (the graviton), and the framework naturally incorporates gauge interactions and supersymmetry.
    \item \textbf{Asymptotic Safety (AS)} \cite{Niedermaier:2006wt,percacci2008asymptoticsafety,eichhorn2019asymptoticallysafeguidequantum,eichhorn2026asymptoticallysafequantumgravity}: postulates that gravity remains a conventional quantum field theory of metrics up to arbitrarily high energies, avoiding standard perturbative non-renormalizability via a non-trivial, non-Gaussian ultraviolet (UV) fixed point. Under the renormalization group flow, the gravitational couplings are driven to finite, stable values at the Planck scale.
    \item \textbf{Holographic Ideas \cite{takayanagi2025emergentholographicspacetimequantum} and $AdS/CFT$ Correspondence \cite{Hubeny_2015}:} postulates that the entire informational content of a volume of space can be encoded on a lower-dimensional boundary of that region, much like a 2D hologram contains 3D information. Its most famous mathematical realization is the \textit{$AdS/CFT$ correspondence}, which demonstrates a duality between a theory of quantum gravity in a bulk volume (Anti-de Sitter space) and a non-gravitational quantum field theory defined on its boundary (Conformal Field Theory).
\end{itemize}
A good place to look for a broad overview of all of these approaches is \cite{loll2022quantumgravity30questions} and \cite{Buoninfante_2025}. To this list we shall add ideas that attempt to unify the standard model (SM) of particle physics with gravity \cite{Krasnov_2018}, theories based on effective field theory techniques, higher-order gravities and more.

\subsection{Hints of Fractality Coming from Multiple Directions}

Hints of a fractal nature of spacetime pervade modern theoretical physics. Our most successful physical theories - like the Standard Model, for example - are Quantum Field Theories (QFTs), and any sensible QFT ‘flows’ to a conformal fixed point under the renormalization group either at high (UV) or low (IR) energies, where scaling symmetry becomes important \cite{francesco2012conformal}. Close to their fixed points, various very distinct condensed-matter systems share \textit{universal properties}: a scale-invariant power-spectrum, characterized by a common set of exponents. When gravity is brought within the QFT-framework, this idea is
realized in Asymptotic Safety \cite{percacci2008asymptoticsafety,eichhorn2026asymptoticallysafequantumgravity}, where spacetime is continuous and this fractal scaling symmetry is promoted to a UV-principle. Even outside of the QFT-paradigm, though, in CDTs for instance, where spacetime is
now described through a triangulation, simulations show a dynamical dimensional reduction from 4 to 2 dimensions
happening in the UV, which can only occur for a fractal (see \S 7.1 in \cite{Loll_2019}).\footnote{Here, `dimension' refers to the \textit{`spectral dimension'} $d_s \in \RR$ (measuring the spread of a process into space over time; it depends on the topology of the underlying space and the diffusion rate), as opposed to the usual \textit{`topological dimension'} $d_t \in \ZZ$ (calculated by considering coverings of a topological space by open sets; it is 0 for a point, 1 for a line, 2 for a surface etc). While smooth manifolds have $d_t = d_s$ in general, a fractal can have $d_t \neq d_s$. There is also the `fractal (Hausdorff) dimension' $d_f \in \RR$: this is a statistical index showing how completely a self-similar shape fills space as you change the scale of measurement.} In the landscape of String Theory, there is
the vastly unexplored corner of quasi-crystalline orbifold compactifications, an old, \cite{Harvey:1987da} but recently revisited \cite{Baykara2024Quasicrystalline} idea that spacetime can look like an
orbifold with ‘forbidden’ symmetries. Even cosmological observations hint at an early universe characterized by near-scale invariant power-spectrum \cite{Planck_2020}, suggesting that such symmetry may be woven into the fabric of spacetime itself.

In Part II, we take \textit{discreteness} (i.e.~``the quantum'') and \textit{fractality} (i.e.~``scaling symmetry'') seriously, constructing the first examples of spacetime quasicrystals with these properties and examining their applications in different fields. These structures combine discreteness, fractality, symmetry, and order all at once, and are a novel example of discrete Lorentzian geometries with high degree of symmetry.

\section{Quasicrystals and Where to Find Them}
\label{sec:QuasicrystalsIntro}

\setlength{\epigraphwidth}{0.6\textwidth}
\epigraph{\justifying{
``... [The Penrose tiles] are a marvellous example of how a mathematical discovery, made with no inkling of its application to reality, may turn out to have been anticipated by Mother Nature!''}}{Martin Gardner (in his memoirs \cite{gardner2013undiluted})}

\subsection{A Journey from Space to Spacetime}
\label{sec:QC_Journey_Intro}

\begin{figure}[!t]
    \begin{minipage}{0.55\textwidth}
        \includegraphics[width=.95\linewidth]{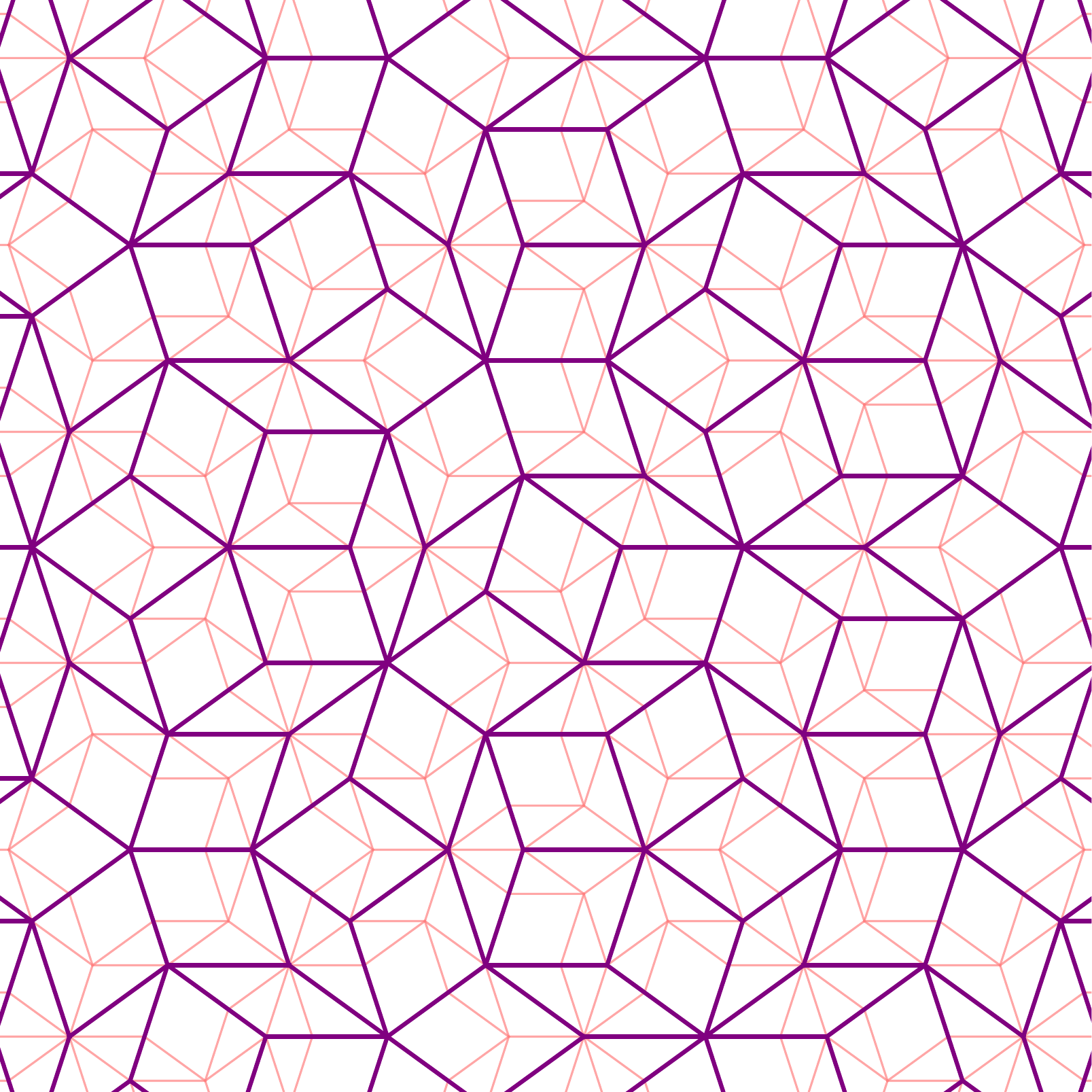}
	\end{minipage}
    \begin{minipage}{0.35\textwidth}
        \includegraphics[width=.95\linewidth]{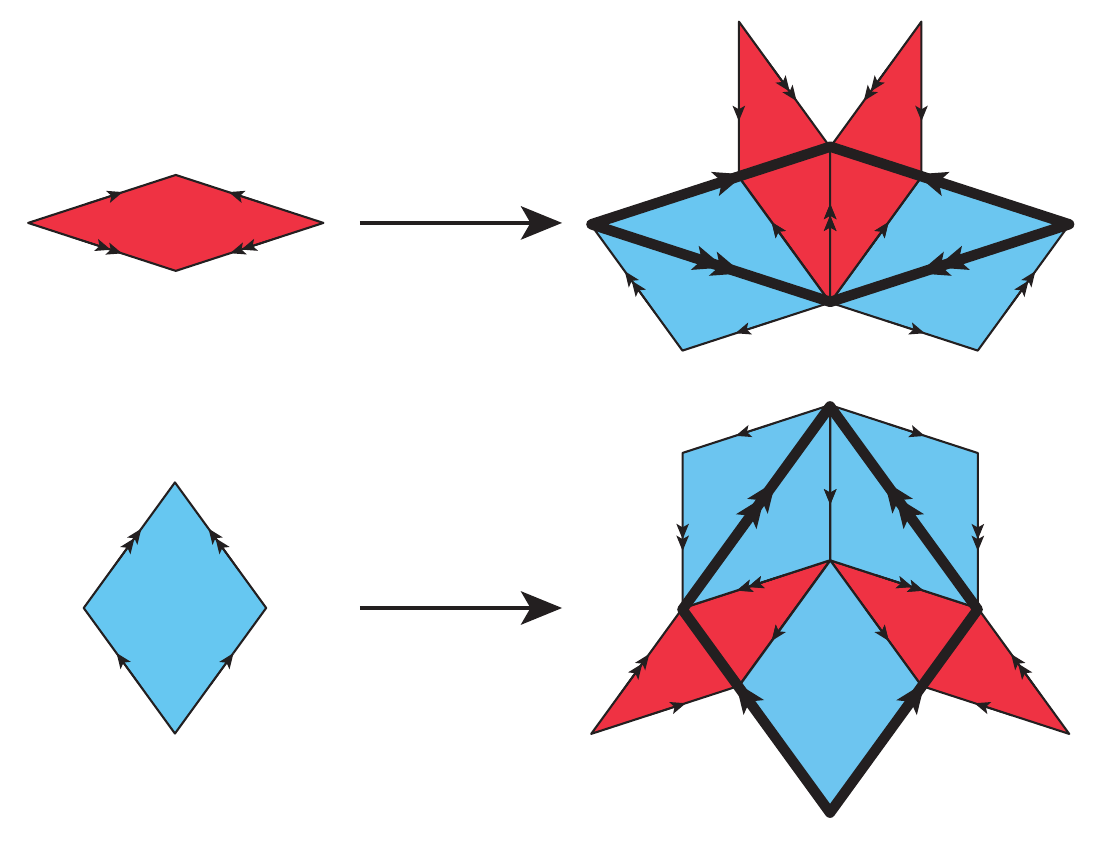}
        \includegraphics[width=.95\linewidth]{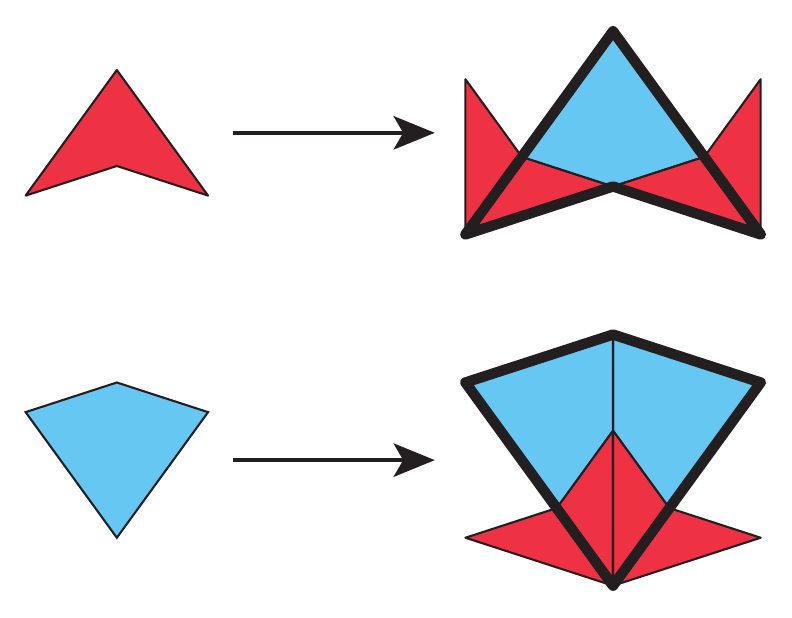}
	\end{minipage}
	\caption[A patch of the Penrose Tiling with the substitution rules.]{Left: a patch of the Penrose tiling \cite{penrose1974role} given by rhombs in purple, underlaid by its ``inflation'' in pink. Right top: the thin (red) and thick (blue) Penrose rhombs, along with their edge-matching rules and inflation rules. Right bottom: Conway's version of the tiles: ``darts'' (red) and ``kites'' (blue) are given with their inflation rules. The two versions are equivalent since tilings built from kites and darts are mutually locally derivable from tilings by rhombs. For an introduction to the Penrose tiling, see  Ref.~\cite{gardner1977extraordinary}. Figure created by Latham Boyle and reproduced from \cite{Boyle:2024qzn}.}
	\label{fig:penroseTiling}
\end{figure}

Self-similar quasicrystals (like the Penrose tiling \cite{penrose1974role, gardner1977extraordinary} in Fig.~\ref{fig:penroseTiling}) are mathematical structures whose beautiful and unexpected properties have fascinated mathematicians since their discovery in the 1970s \cite{gardner1977extraordinary, grunbaum1987tilings, baake1999guidemathematicalquasicrystals, baake2013aperiodic,STEINHARDT1990112}. Among many other fascinating features, they:~(i)~are intrinsically non-periodic, yet perfectly long-range ordered (with a pure-point diffraction spectrum, \textit{like} a crystal);~(ii) have a ``non-crystallographic'' orientational symmetry -- too large to be compatible with any periodic pattern; for this reason, it is a ``forbidden'' (for crystals) symmetry; and (iii) embody a kind of discrete scale-invariance or self-similarity that is also impossible in a periodic pattern. Moreover, these three features are tightly intertwined -- a manifestation of the deep connections between the geometry of lattices \cite{conway2013sphere}, Coxeter's theory of reflection symmetry \cite{coxeterSimplices, coxeterDiscreteGroups, coxeterCompleteEnumeration, humphreys_1990}, and algebraic number theory \cite{serre2012arithmetic, lang1994algebraic, cassels1967algebraic}.  

Such patterns were initially thought to be purely mathematical curiosities, but beautiful mathematical structures often have a way of finding a place in the real world! In the 1980s, these patterns turned out to be the blueprints for a new class of materials (quasicrystals), first discovered in the lab \cite{Schechtman1984}, and later found to also occur naturally \cite{bindi2009natural} (forming {\it e.g.}\ in the birth of the Solar System \cite{hollister2014impact}, in lightning strikes \cite{bindi2023electrical}, and in the first atomic bomb test \cite{bindi2021accidental}). A very vivid narrative behind these discoveries is given in \cite{steinhardt2019second}. Since then, quasicrystals have almost entirely been studied in that context -- {\it i.e.}\ within condensed-matter physics, where their nature is spatial, like the materials they describe. More recently, they have also appeared in other, completely different situations: as the blueprints for quantum error correcting codes \cite{li2024penrosetilingquantumerrorcorrecting}, and in the context of discretizing scale-invariant/conformally-invariant systems and ``discrete holography'' \cite{Boyle:2018uiv, Boyle:2024qzn}, all of which suggests a possible connection to quantum gravity.

All these are examples of applications of \textit{purely spatial} quasicrystals, either in Euclidean or in hyperbolic space. Until now, nobody had succeeded in constructing analogous structures in \textit{spacetime}, where causality and Lorentz symmetry are fundamental. This gap has implicitly limited the role that aperiodic order could play in relativistic spacetime contexts and quantum gravity.

Motivated both by these physical connections and also by pure mathematical interest, in an earlier paper \cite{mygdalas2026spacetimequasicrystals} we showed how self-similar quasicrystals may be naturally generalized from Euclidean {\it space} to Minkowski {\it spacetime}. In that work, we constructed the first examples of such {\it Lorentzian quasicrystals}, which are essentially the spacetime analogues of Euclidean quasicrystals like the Penrose tiling or the Ammann-Beenker tiling.\footnote{We draw the reader's attention to previous references \cite{Flicker:2018tkr, vacaru2018space, friedland2024spacetime, he2025experimental} that have studied dynamical/physical models that they dub ``time quasicrystals'' or ``spacetime quasicrystals.''  These are something different from the spacetime patterns constructed in present paper, in which quasi-periodicity, non-crystallographic orientational symmetry, and discrete self-similarity are intrinsically knit together as in the Penrose or Ammann-Beenker tilings.} We pointed out key novel features of these objects (compared to their Euclidean cousins), and ended with some speculations about how such spacetime quasicrystals might relate to reality. The present thesis is a substantial extension of this work; even though all the material from \cite{mygdalas2026spacetimequasicrystals} is included (and revised, to some extent) here as well, the thesis goes beyond the original paper as mentioned in the Statement of Contributions and in the Outline.  

\section{Outline of the Thesis}
\label{sec:ThesisOutline}

This thesis contains two related -- under the umbrella of quantum gravity -- but highly disproportional parts. In the first shorter part of the thesis, we give a conceptual argument towards the necessity of quantization of gravity. The second longer part is entirely devoted to spacetime quasicrystals, which we view as the main topic of this work. The two parts can be read separately and could be of independent interest. The outlines may be used to navigate through the content and contain most of the results of the two parts; they can be found just after the beginning of each Part.

\chapter*{\centering \vfill PART I: \\ Semiclassical Gravity Efficiently \\ Solves \cc{NP}-Complete Problems \vfill}
\addcontentsline{toc}{chapter}{PART I: Semiclassical Gravity Efficiently Solves \cc{NP}-Complete Problems}
\vspace{-15mm}
\begin{center}
    \includegraphics[width=0.5\linewidth]{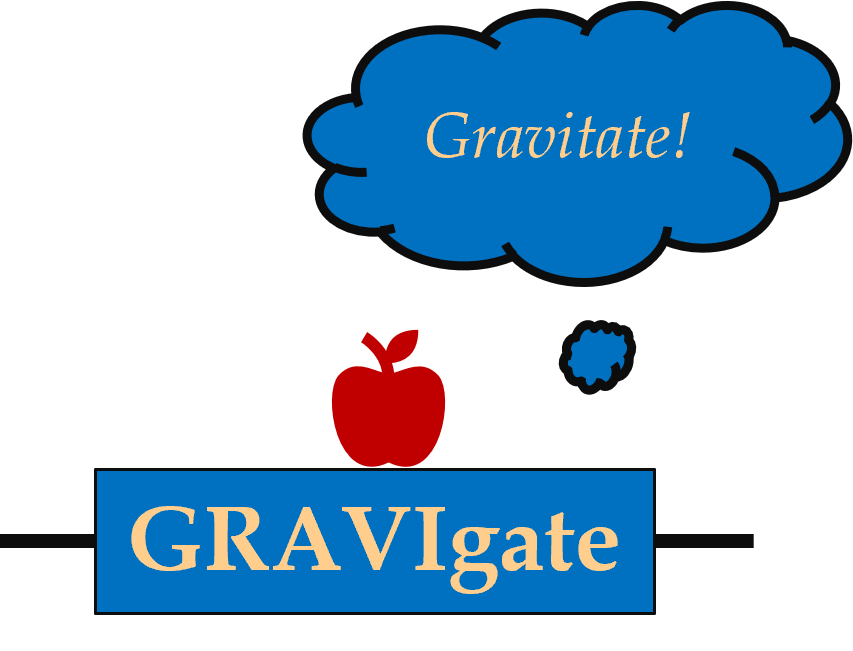}
\end{center}
\vfill

\newpage 

\begin{tikzpicture}[remember picture,overlay]
\node[
  anchor=north east,
  inner sep=0pt,
  xshift=-2.5cm,
  yshift=-2.5cm
] at (current page.north east) {
  \begin{tcolorbox}[colframe=purple,colback=orange!10,width=0.70\textwidth]
\justifying
\noindent 
Wherein the reader enters the realm of computational complexity theory and, mesmerized by its beauty, emerges convinced that gravity must be quantized.
\end{tcolorbox}
};
\end{tikzpicture}

\vspace{+5ex}
\noindent \textbf{Keywords:} Computational Complexity Theory; P vs NP; Semiclassical Gravity; Schr\"odinger--Newton Equation; Non-linearities; Quantum Gravity.

\section*{Outline of Part I}
\addcontentsline{toc}{section}{Outline of Part I}

The summary of Part I is as follows: 

Chapter \ref{ch:complexity} contains some background on computational complexity theory, in which we define certain complexity classes, the famous \cc{P} vs.~\cc{NP} problem, and its manifestation in the physical extended Church--Turing thesis, or PECTT for short. In Chapter \ref{ch:SemiclassicalGravityAndWeakField} we introduce the semiclassical Einstein equations, the most popular theory of semiclassical gravity. In particular, we are interested in its weak-field (Newtonian) limit, which we discuss in Section \ref{sec:weakfieldlimit}, and gives us the Schr\"odinger-Newton (SN) equation. In Section \ref{sec:SNQubitDynamics}, we describe how a massive qubit evolves under the SN dynamics. Chapter \ref{ch:SolvingNPSemiclassicaly} has the main result of this work, and puts together the two topics mentioned in the previous chapters. After showing in Section \ref{sec:EncodingNPinQubit} how to modify an older algorithm due to Abrams and Lloyd to encode any \cc{NP} problem into a single qubit, we show in Section \ref{sec:ALBBJ} how the SN dynamics can be exploited to efficiently solve \cc{NP} problems. We close this chapter with a discussion on these findings in Section \ref{sec:GRAVIgateDiscussion}.
\cleardoublepage
\phantomsection

\chapter{A Primer on Computational Complexity Theory}
\label{ch:complexity}

Computational complexity theory seeks to classify problems according to their intrinsic difficulty, often by quantifying the resources (e.g., time, memory, number of pages for a proof) required to solve them on a particular type of computer. 
For an introduction, see \S 3 in \cite{Nielsen_Chuang_2010}.
For example, a major open question in computational complexity theory is to lower-bound the time required to find a prime factor of a composite integer on a classical computer. Note that, typically, the ``measure'' of an amount of a resource is with respect to the number of bits it takes to encode the input.\footnote{For example, for finding a prime factor of the number $2026$, the input size is $11$ (because it takes $11$ bits to represent $2026$ in \textit{binary}; similarly, 4 base-10 `digits' in \textit{decimal}), not $2026$.} 

Even though the precise definition of a computing machine (or a computational model) is beyond the scope of this work\footnote{We do, however, implicitly assume an abstract Turing machine, as any physical computation can be efficiently simulated on one such device \cite{Nielsen_Chuang_2010}.}, what we would like to emphasize is the intricate connection between computation and physics, most eloquently put in the following slogan form (adapted from \cite{Abrams_1998} to match the spirit of Wheeler's saying about general relativity:
\begin{center}
``Computation \textit{is} a physical process, and the\\ physical laws tell us what is \textit{efficiently} computable.'' \end{center}
This is to say, a computational model leveraging classical physics -- a classical computer, like a Python program -- provides a framework that can give efficient algorithms for solving certain problems by manipulating the corresponding classical laws. A quantum computer, on the other hand, has access to a different set of laws, those of quantum theory, and comes with its own set of computational advantages and limitations. Similarly, any physical theory can, in principle, give rise to its own computational device. 

\section{How can Computational Problems be Classified?}
\label{sec:ComputationalProblems}

Given a particular resource and a type of computer, one can then discuss the class of problems that the model can solve using an amount of the given resource. There are several problems out there, but perhaps the most useful ones are those that are formulated as \textit{decision problems}, i.e.~those whose answer is emphatically `YES' or `NO'. This is because other types of problems, such as \textit{search problems} or \textit{counting problems} (i.e. those that ask `which solution?' and `how many solutions?' instead of `does it exist any solution?'), can usually be reduced to a series of decision problems. 

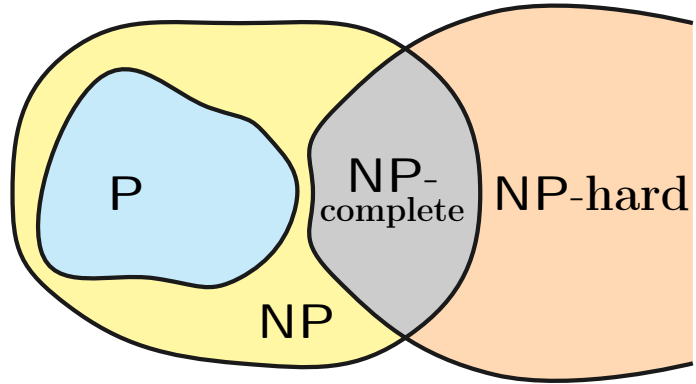
\begin{figure}[t!]
    \centering
\begin{tikzpicture}[scale=0.75, transform shape]

    \def\pathNP{plot [smooth cycle, tension=0.8] coordinates {
        (-4.5, 0) (-3.5, 2.5) (-0.5, 3) (2.5, 2.5) (3.75, 0) 
        (2.5, -2.5) (-0.5, -3) (-3.5, -2)
    }}

    \def\pathP{plot [smooth cycle, tension=0.8] coordinates {
        (-4, -1.0) (-3, 2) (-1, 1.5) (0, 1) 
        (0.5, -0.2) (-0.5, -1.5) (-2, -1.5)
    }}

    \def\pathNPHard{plot [smooth, tension=0.8] coordinates {
        (7.5, 3) (4, 3.2) (1.2, 1.5) (0.8, 0) 
        (1.2, -1.5) (4, -3.2) (7.5, -3)
    }}

    
    \fill[orange!30] \pathNPHard;

    \fill[yellow!45] \pathNP;

    \fill[cyan!20] \pathP;

    \begin{scope}
        \clip \pathNPHard -- cycle; 
        \fill[gray!40] \pathNP;
    \end{scope}

    \draw[ultra thick, color=black!90] \pathNP;
    \draw[ultra thick, color=black!90] \pathP;
    
    \draw[ultra thick, color=black!90] \pathNPHard; 

    \node at (-2.5, 0) {\Huge $\cc{P}$};
    \node at (0.5, -2.25) {\Huge $\cc{NP}$};
    
    \node[align=center] at (2.2, 0) {\huge $\cc{NP}$-\\ \Large {\bf complete}};
    \node at (5.7, 0) {\Huge $\cc{NP}$-{\bf hard}};

\end{tikzpicture}
\caption[Some relevant complexity classes and their interrelations.]{Some relevant complexity classes and their interrelations (see text for formal definitions). $\cc{P}$ contains problems that can be solved ``quickly'' (i.e.~in polynomial time), while problems in $\cc{NP}$ can have their solutions verified ``quickly'' (see Fig.~\ref{fig:PandNPComputation}). Both $\cc{P}$ and $\cc{NP}$ contain decision problems. $\cc{NP}$-hard contains problems that need not be themselves decision problems nor verifiable ``quickly''. For example, they might be decision problems that are `undecidable', like the Halting Problem \cite{Church_1936,Turing_1937}; thus \cc{NP}-hard is a strict superset of \cc{NP}. Yet, any $\cc{NP}$ problem can be mapped to an $\cc{NP}$-hard problem ``quickly''. In that sense, $\cc{NP}$-hard is at least as hard as $\cc{NP}$: if a solution to an $\cc{NP}$-hard problem is known, $\cc{NP}$ can be solved ``quickly''. Finally, the hardest problems in $\cc{NP}$ are called $\cc{NP}$-complete (i.e.~are both $\cc{NP}$-hard \textit{and} $\cc{NP}$). The sets drawn in this figure assume implicitly that $ \cc{P} \neq \cc{NP}$; for the Euler diagram corresponding to the case \cc{P} = \cc{NP}, see \cite{AB09}.}
    \label{fig:ComplexityClasses}
\end{figure}

Let's see an example: consider a logical Boolean formula like \begin{equation}
f(x_1,x_2,x_3) = (x_1 \lor x_2) \land (x_3 \lor \neg x_2),
\label{eq:SomeBooleanFormula}
\end{equation}
where $x_1, x_2, x_3 \in \{0,1\}$
are binary variables and $\lor, \land$ and $\lnot$ are the logical OR, AND and NOT operators respectively. In total, there are $2^3 = 8$ possible truth values, $$\{ 0,1\}^3 = \bigg\{ \{0,0,0\}, \{1,0,0\} \text{ (+ 3 perm.)}, \{ 1,1,0\} \text{ (+ 3 perm.)}, \{1,1,1\} \bigg\}$$ (TRUE/FALSE-assignments) for the three variables, and $f$ evaluates to either `TRUE' or `FALSE' for some input $x \in \{0,1\}^3$. Given the expression $f$, Alice could ask: 
\begin{center}
        \begin{tcolorbox}[colframe=red,colback=white,width=0.75\textwidth] \centering
        \texttt{``Does there exist any valid assignment to the variables $x_{1,2,3}$ that make $f$ yield TRUE?''}
    \end{tcolorbox}
\end{center}
This is the \textit{decision problem} for $f$ and it can be solved by brute force if, for example, one enumerates all possible eight truth values (let's say, by building a truth table) and check them one by one to see if they satisfy the expression \eqref{eq:SomeBooleanFormula}, that is, if $f(x) = 1$. Imagine that one feeds $f$ through such a decision solver and the output is $f(x) = 1$, so we now know that $f$ has a solution (i.e., that $f$ is \textit{satisfiable}). Alice's friend, Bob, now asks: 
\begin{center}
\begin{tcolorbox}[colframe=red,colback=white,width=0.80\textwidth]
\centering
    \texttt{``Which assignments of the variables $x_{1,2.3}$ satisfies $f$?''}
\end{tcolorbox}
\end{center} 
This is the \textit{search problem} for $f$ and it can be solved by a recursive procedure based on the decision problem: 
\begin{enumerate}
    \item Pick the first (or any other) variable and set it to TRUE to obtain another expression $f(x_1={\rm TRUE},x_2,x_3) = f^{(1)}$.
    \item Pass $f^{(1)}$ through the decision solver and check if it is satisfiable; if yes, lock in $x_1 = 1$ and continue recursively to determine a satisfying assignment for $f^{(1)}$ with the next variable (i.e. set $x_2 = {\rm TRUE}$ to get $f^{(2)}$ and repeat); if no, then set $x_1 = 0$ and continue similarly.
    \item For $n$ Boolean variables, the process terminates after at most $2n$ recursive queries of the decision solver (for non-binary variables, the number of calls becomes $kn$, where $k$ are the possible values for each variable); $2n$ corresponds to guessing the truth value wrongly for each single variable $x_i$ at each iterative step of the process. The total running time of the algorithm is $$\sim (\text{\# recursive calls}) \times \text{(running time it takes to solve the decision problem)},$$ so if the decision problem can be solved relatively ``quickly'', the search problem can be too (with some manageable polynomial overhead that comes from how many times we invoke the decision solver).  
\end{enumerate}
For this reason, we can focus on decision problems, which are ``easier'' to solve in the sense that, if we can solve those, we can solve more problems because of that. Indeed, decision problems are used to place problems into different computational complexity classes; some of them are shown in Fig.~\ref{fig:ComplexityClasses} and are defined more formally in the next Section \ref{sec:NPandChurchTuring}.

Let us use this example to comment on several important points which are used implicitly in this thesis; here, we focus on the intuition, while we refer the reader to a standard textbook (e.g. \cite{AB09}) in computational complexity theory for a formal treatment. 
\begin{itemize}
    \item \textbf{Algorithmic Efficiency:} when the evaluation time scales polynomially with the input-size $|x|$, we say that the solution is found efficiently (or ``quickly''), as opposed to an exponential scaling, which becomes intractable very soon once $|x|$ gets bigger and bigger. For instance, consider two algorithms that take some input $x$ and achieve the same goal, performing one step per one millisecond; one scales as $\sim |x|^5$ and the other as $2^{|x|}$. At $|x| = 10$, $10^5 = 100000$ and $2^5 = 1024$, so the exponential algorithm performs better at first and yields a result at around 1 second, whereas the polynomial algorithm needs about 100 seconds. But already at $|x|=50$, the exponential algorithm will need about 35 thousand years to give an answer, whereas the polynomial one will only need about half a week! ``Quickly'' is quoted, here, because the associated polynomial might be of a very high degree, but in any case, it would still have a more manageable growth than any exponential function with over the same input-size.
    
    \item \textbf{Problem Reducibility:} using an instance of some problem to solve another problem is a more general phenomenon in computational complexity theory. Similar to how the decision problem of $f$ was used above to solve its associated search problem, many other problems can be \textit{reduced} to some alternative problems in the sense that, \textit{if} a solution to the latter is known or can be obtained efficiently, it can be used to solve the original problem efficiently as well. Reductions are a very important tool, as they offer methods to map problems of unknown complexity onto ones of known. For example, consider a new Problem A, for which no polynomial algorithm is known; if it can be reduced efficiently to some Problem B with a known polynomial-time solver, this means that Problem A is at least as hard as Problem B, so they have similar complexity and our new problem can also be solved in polynomial time. There are several different types of reductions and their proper definitions are not entirely relevant for us here (again, see \cite{AB09}), but let us briefly mention two key ideas: 
    \begin{enumerate}
        \item \textit{Karp-reductions} \cite{karp2009reducibility} allow us to reduce Problem A to Problem B using a polynomial-time algorithm that preserves the structural difficulty of the problem. Solving Problem A amounts to writing a function that transforms the entire instance of Problem A into an instance of Problem B efficiently, and then querying the solver for Problem B exactly once to get an answer to the original question for Problem A without any further computation. Because this reduction process tell us that Problem A is at least as hard as Problem B, it is used to define the \cc{NP}-complete class (see discussion below and Fig.~\ref{fig:ComplexityClasses}).
        \item \textit{Oracle-reductions} (or \textit{Cook/Turing-reductions)} tell us if Problem A is computable by Problem B no matter how complicated the latter might be. Solving Problem A amounts to writing a function that includes a subroutine for solving Problem B, which we are allowed to call as many times as needed. This subroutine is treated as a \textit{black-box ``oracle''}, whose inner workings are irrelevant for solving Problem A. In fact, solving Problem B might be an exponentially hard task, but, provided our black-box delivers a solution, the original Problem A can be solved as well. These type of reductions are used to quantify ``oracle complexity'' (how many times you need to query a black box to get an answer).   
    \end{enumerate}

    \item \textbf{Completeness}, or the ``\textbf{One-Problem-to-Rule-`Em-All}'' idea: reductions, in particular of the polynomial Karp-type, are very useful as they allow us to study only certain problems that are representatives of entire complexity classes (are ``complete''). For example, a known decision problem, the \textit{Boolean satisfiability problem} $\cc{SAT}$, is shown to be \cc{NP}-complete, so any $\cc{NP}$ problem can be reduced to it ``quickly'' \cite{cookNPthm1971,Levin1973UniversalSorting}. Crucially, this means that all problems in \cc{NP} are structurally equivalent to $\cc{SAT}$; finding an efficient algorithm for \cc{SAT} would then imply that an efficient algorithm exists for all \cc{NP} problems (which is not expected; see next Section \ref{sec:NPandChurchTuring}).\footnote{In fact, \cc{SAT} was the first problem to be classified as \cc{NP}-complete, which gave rise to the concept, the complexity classes $\cc{P},\cc{NP}$ and the \cc{P} versus \cc{NP} problem.} 
    
    The problem \cc{SAT} is very much related to our toy expression \eqref{eq:SomeBooleanFormula} and can be formulated as follows: imagine you are handed a massive logical formula $f = f(x_1,\cdots,x_n)$ over $n$ binary variables and your goal is to find a valid assignment of TRUE/FALSE values for each of them (i.e.~`the needle' $x_0$) that satisfies the entire formula, $f(x_0) = 1$. The catch, of course, is that the `haystack' is exponentially large: the needle is hidden among \(2^{n}\) possible variable assignments, and no known algorithm improves upon merely trying all of the (exponentially many) possibilities.
\begin{center}
\begin{tcolorbox}[colframe=red,colback=white,width=0.83\textwidth]
\centering
    \texttt{``The Needle in an Exponential Haystack''}: for some function $$f: \{ 0,1\}^n \to \{0,1\},$$ does there exist a satisfying assignment $x_0$ such that $f(x_0) = 1$? 
\end{tcolorbox}
\end{center} 
Nonetheless, once the `needle' is found, a classical computer can verify that $x_0$ indeed makes $f$ evaluate to TRUE efficiently. Expression \eqref{eq:SomeBooleanFormula} is an instance of the \cc{SAT} problem, in fact of the \cc{2-SAT} problem, where each \textit{clause} of $f$ (i.e.~the logical sub-expressions that are connected with the $\land$ operator) consists of exactly two \textit{literals} (i.e.~variables $x_i$ or their negations $\lnot x_i$), which are themselves connected only by a $\lor$ operator.\footnote{This is known as the \textit{conjunctive normal form} of a Boolean expression $f$ \cite{AB09}.} This has been shown to be in $\cc{P}$, but its cousin, \cc{3-SAT}, is in \cc{NP}. 

\end{itemize}

In this work, we are primarily concerned with how the difficulty of a problem changes when a candidate solution, or \emph{proof}, is provided. Indeed, many problems exhibit an asymmetry in the time required to \emph{find} a solution versus the time required to \emph{verify} a solution. Consider, for example, a Sudoku puzzle: checking a completed grid is far easier than constructing one from scratch. However, it is not known if this distinction reflects a genuine difference in computational difficulty. In computational complexity theory, this question is formalized in the famous $\cc{P}$ versus $\cc{NP}$ problem, which we now discuss. 

\newpage
\section{The $\cc{P} \overset{?}{=} \cc{NP}$ Problem and the Physical Extended Church--Turing Thesis (PECTT)}
\label{sec:NPandChurchTuring}

Intuitively, the complexity class $\cc{P}$ (``polynomial time'') consists of all the computational problems that can be efficiently solved (a.k.a. \emph{decided}) by a classical computer.\footnote{In fact, the set of computational problems that efficient classical computers can solve is not $\cc{P}$, but $\cc{BPP}$ (``bounded-error probabilistic polynomial time''), which is the probabilistic analogue of $\cc{P}$. However, it is generally believed that $\cc{P} = \cc{BPP}$ \cite{AB09}, so in this discussion we have chosen to only talk about $\cc{P}$.} For example, deciding if a binary string is palindromic or not is in $\cc{P}$. More formally, let $\{0,1\}^n$ be the set of all $n$-bit binary strings, $\{0,1\}^*$ the set of all finite-length binary strings (i.e., $\{0,1\}^* = \bigcup_{n \geq 1} \{0,1\}^n$), and $|x|$ the length of a string $x$. Then, $\cc{P}$ consists of all the functions $L : \{0,1\}^* \rightarrow \{0,1\}$ (a.k.a. \emph{languages}) for which there exists a deterministic and polynomial time classical algorithm $M$ (i.e., a \emph{deterministic and polynomial time Turing machine}) such that for all inputs $x \in \{0,1\}^*$, $L(x) = 1$ if and only if $M(x) = 1$.

On the other hand, the complexity class $\cc{NP}$ (``non-deterministic polynomial time") consists of all of the computational problems for which a valid proof can be efficiently checked by a classical computer. For example, the \cc{3-SAT} problem we encountered before is in \cc{NP}. In particular, $L \in \cc{NP}$ if and only if there exists a deterministic and polynomial time algorithm $M$ and a polynomial $p$ such that for all inputs $x \in \{0,1\}^*$, $L(x) = 1$ if and only if there exists $y \in \{0,1\}^{p(|x|)}$ (usually called a \emph{proof}, \emph{certificate}, or \emph{witness}) such that $M(x,y) = 1$. In other words, $L \in \cc{NP}$ if and only if for all $x \in \{0,1\}^*$, if $L(x) = 1$, then there exists a proof that will cause $M$ to accept $x$, and if $L(x) = 0$, then there is no proof that will cause $M$ to accept $x$.

\begin{figure}[p]
    \centering
    \includegraphics[width=\textwidth]{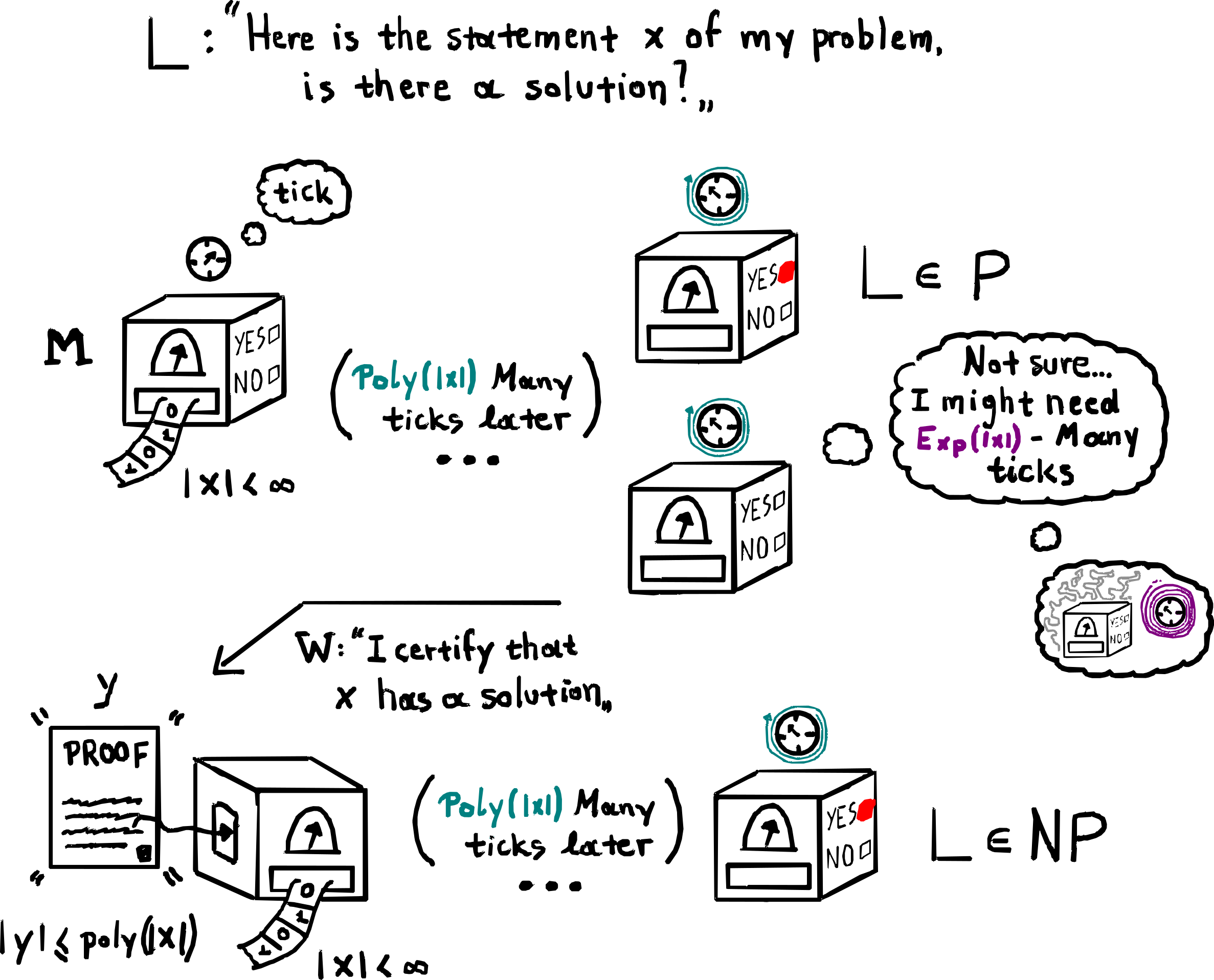}
    \caption[$\cc{P}$ vs $\cc{NP}$: an illustration.]{$\cc{P}$ vs $\cc{NP}$: an illustration. A decision problem $L$ is in $\cc{P}$ if its solution can be found (i.e.~it can be \textit{decided}) by a deterministic Turing machine $M$ in polynomial time (polynomially many ``ticks''). A decision problem $L$ is in $\cc{NP}$ if, provided a \textit{witness} $W$ supplies us with a proof $y$ that our problem has a solution, $M$ can verify the solution in polynomially many ticks. The proof $y$ must be ``written in polynomially many pages'' in terms of the size $|x|$ of the input-problem.}
    \label{fig:PandNPComputation}
\end{figure}

Evidently, $\cc{P} \subseteq \cc{NP}$, and it is a major open question (in fact, one out of the six remaining Millennium Prize Problem \cite{carlson2023millennium}; the Poincar\'{e} Conjecture is now proven) whether $\cc{P} = \cc{NP}$. Needless to say that proving this statement formally will not only earn one a substantial amount of money, but also a great deal of fame. The general consensus, though, is that $\cc{P} \neq \cc{NP}$ because the alternative suggests that proving a mathematical statement is just as easy as verifying that a given proof is correct, which is something that mathematical experience says is false \cite{Impagliazzo}. A very nice review of the \cc{P} vs \cc{NP} problem is given in \cite{aaronson2016pnp}.

Now, the entire premise of quantum computation is that it changes the narrative of what can and cannot be computed efficiently in the physical world. In particular, there exist problems widely believed to be in $\cc{NP} \backslash \cc{P}$ that can be solved efficiently on a quantum computer. The most famous such problem is integer factorization, for which Shor exhibited his eponymous polynomial-time quantum algorithm \cite{Sho99}. 

That said, it is generally expected that the hardest problems in $\cc{NP}$---the so-called \emph{$\cc{NP}$-complete} problems---do not admit efficient quantum algorithms \cite{Nielsen_Chuang_2010, aaronson2005npcompleteproblemsphysicalreality}. In part, this belief stems from the idea that \cc{NP}-complete problems appear to lack the requisite structure for there to be an exponential quantum speed-up \cite{aaronson2005npcompleteproblemsphysicalreality}. Of course, this is not to say that quantum computers offer no advantage for these problems. Grover's algorithm, for example, solves any $\cc{NP}$-complete problem quadratically faster than any classical brute-force algorithm \cite{Grover_1998}.\footnote{
\label{foot:q-computers}
Indeed, for a search-space of dimension $N$ ($= 2^n$ over binary variables, for example), classical algorithms perform at the order of $O(N)$, while Grover's algorithm scales as $O(\sqrt{N} = 2^{n/2})$, which is faster but still an exponential cost. Incidentally, this is an ``oracle complexity'' that counts Grover iterations (i.e., oracle calls) needed to find the solution. The oracle here is a bit-flip operation that, in the language of \cite{Nielsen_Chuang_2010}, ``recognizes a solution when it sees one''; see their \S 6.1 for details, specifically \S 6.1.4.} However, this advantage is modest and does not provide the exponential speed-up required to solve an $\cc{NP}$-complete problem in polynomial time on a quantum computer. It is also worth noting that if quantum computers cannot solve $\cc{NP}$-complete problems in polynomial time, then $\cc{P} \neq \cc{NP}$ \cite{Nielsen_Chuang_2010}.\footnote{The set of computational problems that quantum computers can efficiently decide is \cc{BQP} (``Bounded-Error Quantum Polynomial-Time''), the quantum analogue of \cc{BPP}. Naturally, \cc{P} sits within \cc{BQP} (since \cc{BPP} does), thus proving that $\cc{P} = \cc{NP}$ would indeed imply that $\cc{NP} \subset \cc{BQP}$ (but if $\cc{NP} \subset \cc{BQP}$, this does not necessarily show that $\cc{P} = \cc{NP}$). Shor's algorithm, implies that $\cc{BPP} \neq \cc{BQP}$, i.e.~that the quantum computer is more powerful than the classical one. But how powerful? Can it be that $\cc{NP} \subseteq \cc{BQP}$? Most quantum computing researchers \textit{conjecture} that $\cc{NP} \not\subset \cc{BQP}$, and proving this would also serve as a proof that $\cc{P}\neq\cc{NP}$. However, proving $\cc{P} \neq \cc{NP}$ with quantum computers is not merely an engineering problem; we have some theoretical evidence for this too, e.g.~\cite{bernstein1993quantum}. See \S 5.5 in \cite{aaronson2016pnp} for more discussion.} 

More generally, it is a tenet of quantum information theory, and of the theory of computation itself, that no physically reasonable model of computation (perhaps even more powerful than quantum computation) can efficiently solve an $\cc{NP}$-complete problem. This expectation is stipulated in the \emph{Physical Extended Church--Turing Thesis}, which was originally proposed by Freedman, Kitaev, Larsen, and Wang in the early 2000s \cite{freedman2002topologicalquantumcomputation} and extended an earlier ``physical'' version of the Church--Turing Thesis by Deutsch in the 1980s \cite{10.1098/rspa.1985.0070}.

\begin{thesis}[PECTT]
No physical procedure can decide an $\cc{NP}$-complete problem in polynomially many steps.
\label{thesis:PECTT}
\end{thesis}

Let us now make a comment about the wording used above in our formulation of the PECTT. One might wonder: ``what if I get lucky and guess the solution to some given instance of an \cc{NP}-problem correctly in one step? Would that violate the PECTT?'' Not so easy, no. An \cc{NP} language $L$ corresponds to an infinite collection of problems $\{ l_x \}$, one for each finite-size bit-string $x$. Each $l_x$ is of the form ``is $x$ in the set $L$''. Therefore, one might get lucky for a few of these questions, but guessing can only take you that far, because getting lucky for all $x$ must have vanishingly small probability.

To close off this section, we would like to emphasize that, while motivated from computer science, the PECTT is largely a postulate about the physical world as it articulates what can and cannot be computed efficiently by physical devices. Thus, in the context of demarcating different physical theories, the PECTT dictates that a candidate theory should be deemed implausible if it can be exploited to efficiently solve an $\cc{NP}$-complete problem. Ultimately, we will show that semiclassical gravity is such a theory.

\subsection{How \textit{Physical} is the PECTT?}
\label{subsec:physical_PECTT_comment}

At first sight, it might seem that the PECTT is against the known fact that several \cc{NP}-complete and \cc{NP}-hard problems can be rephrased in terms of some Ising model \cite{Lucas_2014, kalinin2022computational}, so that they appear to be ``solved'' by physical systems.
Usually, this is achieved by mapping a given computationally hard problem to some Ising Hamiltonian $H_{\rm Ising}$, whose ground state encodes information about the problem under consideration. For example, as explained in \cite{Lucas_2014}, an \cc{NP}-complete problem is encoded in $H_{\rm Ising}$ as the decision problem ``does the ground state have energy $\leq 0$?'', while an \cc{NP}-hard problem becomes the optimization problem ``what is the ground state energy?''. Does this mean that the PECTT is violated by physical systems, then, and wouldn't this make it... unphysical? 

The answer turns out to be negative. Even though there might exist a quantum system that could solve the \cc{NP}-complete or \cc{NP}-hard problem, this cannot be done efficiently on a quantum computer. Indeed, to achieve this, we would need to efficiently simulate the quantum system, and no such process is currently known. In fact, the estimation of the ground state of some Hamiltonian is known to be a \cc{QMA}-complete problem, which is harder than \cc{NP}-complete (see \cite{bookatz2013qmacompleteproblems} for appropriate definitions). As a result, even if we can map \cc{NP} problems onto physical systems, this is not enough to claim that \cc{NP} is efficiently solved on those systems.

\chapter{Semiclassical Gravity and its Weak-Field Limit}
\label{ch:SemiclassicalGravityAndWeakField}

Historically, the most popular theory of semiclassical gravity is described by the \textit{semiclassical Einstein field equations (semiclassical EFEs)},\footnote{We employ the natural system of units in which $c = G = \hbar = 1$. Greek indices take values from 1 to $d$, where $d$ is the spacetime dimension. Hats on top of a symbol indicate that the symbol corresponds to an operator acting on a quantum state $\Psi$ living in some appropriate Hilbert space.}
\begin{equation}
G_{\mu\nu} = R_{\mu\nu} - \frac{1}{2}g_{\mu\nu}R =  8\pi \bra{\Psi} \hat{T}_{\mu \nu} \ket{\Psi},
\label{eq:semiclassicalEFEs}
\end{equation}
where the (unquantized) metric $g_{\mu\nu}$ is sourced by the expectation value of the energy-momentum tensor $\hat{T}_{\mu \nu}$ of quantum matter on spacetime \cite{Giulini_2023}. This theory realizes M{\o}ller's \cite{Lichnerowicz:1962gnh} and Rosenfeld's \cite{ROSENFELD1963353} original proposals and, more formally, is known to arise from an action principle \cite{etde_5892929,Kibble_1980}. Despite several technical concerns (see \cite{Carlip_2008} and references therein), the expectation value $\bra{\Psi}\hat{T}_{\mu \nu}\ket{\Psi}$ can be constructed in a consistent and axiomatic way \cite{Wald:1995yp, Giulini_2023}. Moreover, the semiclassical EFEs are well-defined when the fluctuations of $\hat{T}_{\mu \nu}$ are not very large in the state $\ket{\Psi}$, i.e., when macroscopic energy-momentum configurations are not placed in superpositions \cite{DIOSI1984199,Page_1981}. As we discuss in the next subsection, the weak-field limit of these equations manifest many interesting and testable gravitational effects, thanks to the non-linearity introduced by the semiclassical EFEs.

\section{The Weak-Field Limit of Semiclassical Gravity}
\label{sec:weakfieldlimit}

As shown in \cite{Bahrami_2014, Giulini_2023}, in the Newtonian limit (i.e., assuming $g_{\mu \nu} = \eta_{\mu \nu} + h_{\mu \nu}$ with $|h_{\mu\nu}| \ll 1$ and non-relativistic velocities such that the energy density $\hat{T}_{00}$ dominates the rest of $\hat{T}_{\mu\nu}$), \eqref{eq:semiclassicalEFEs} reduces to the \emph{Schr\"odinger--Newton (SN) equation},
\begin{equation}
i\, \partial_t \Psi(r,t) = \left(-\frac{1}{2m}\nabla^2 + V_G(\Psi)\right)\Psi(r,t),
\label{eq:SNEquation}
\end{equation}
where
\begin{equation}
V_G(\Psi) = -m^2\int \frac{|\Psi(r',t)|^2}{|r - r'|} \, \dd^3 r'
\label{eq:SNpotential}
\end{equation}
is the potential due to the gravitational self-interaction of the massive particle. Consequently, as emphasized in \cite{Bahrami_2014}, \emph{if} gravity is fundamentally classical and \eqref{eq:semiclassicalEFEs} describes how gravity and quantum mechanical matter couple, \emph{then} the SN equation is the fundamental equation of motion for a massive, non-relativistic quantum particle. 

Despite the non-linear nature of the SN equation, the standard probabilistic interpretation of the wavefunction persists. Indeed, the potential due to the gravitational self-interaction \eqref{eq:SNpotential} is a function of $|\Psi|^2$, so under the usual assumptions that $\Psi$ is square-integrable and that its first derivative vanishes at infinity, $\partial_t \int |\Psi(r,t)|^2\, \dd^3r = 0$. Consequently, the norm of the wavefunction is preserved under SN evolution \cite{Carlip_2008,Giulini_2023}.

The SN dynamics yield a diversity of novel phenomena that have been extensively studied in the literature \cite{DIOSI1984199,PenroseStateCollapse,Penrose:2014nha,Grossardt_2024,hatifi2023revealingselfgravitysterngerlachhumptydumpty,Sahoo_2023,Delgado_2025}. In the context of solving hard computational problems, the most relevant aspect of the SN dynamics is the non-linearity produced by the gravitational self-interaction. Indeed, it is known that certain non-linear theories of quantum mechanics afford efficient solutions to $\cc{NP}$-complete problems (e.g., Weinberg's \cite{WEINBERG1989336ha, Abrams_1998}), so it is plausible that semiclassical gravity could as well.

\section{The Schr\"{o}dinger-Newton (SN) Dynamics of a Qubit in Weak-Field Semiclassical Gravity}
\label{sec:SNQubitDynamics}

We will now specialize the SN dynamics to the case of a massive and non-relativistic qubit. In particular, consider a non-relativistic spin-1/2 particle of mass $m$ whose overall quantum state is a time-dependent spinor $\ket{\Psi(t)}$. In the position basis, the state of the particle is a two-component wavefunction $\Psi(r,t) = \braket{r}{\Psi(t)}$ in a superposition of spin-up ($\ket{0}$) and spin-down ($\ket{1}$) states,
\begin{equation}
\Psi(r,t) = \psi_\uparrow(r,t) \otimes \alpha \ket{0} + \psi_\downarrow(r,t) \otimes \beta \ket{1},
\label{eq:SpinorSuperpositionUpDown}
\end{equation}
where $|\alpha|^2 + |\beta|^2 = 1$.\footnote{Note, it is generally unclear how entangled states evolve in a non-linear quantum theory because the superposition principle no longer holds. To mend this issue, we follow the prescription of Weinberg \cite{WEINBERG1989336ha} and Abrams and Lloyd \cite{Abrams_1998} by regarding $\{\psi_\uparrow \otimes \ket{0}, \psi_\downarrow \otimes \ket{1}\}$ as a preferred basis on the tensor product Hilbert space, relative to which the action of a non-linear map $S$ on the spin subsystem is specified by acting $S$
independently on each of the computational basis states.}

In the presence of an additional external potential $V_{\text{ext}}(r,t)$ (e.g., a magnetic field like in a Stern--Gerlach experiment \cite{Grossardt_2024,hatifi2023revealingselfgravitysterngerlachhumptydumpty,Sahoo_2023,Delgado_2025}), the time evolution of the wavefunctions $\psi_{\uparrow}$ and $\psi_{\downarrow}$, as predicted by the SN equation \eqref{eq:SNEquation}, is given by
\begin{equation}
i\, \partial_t \psi_{\uparrow\downarrow}(r,t) = \left(-\frac{1}{2m}\nabla^2 + V_G(\Psi) + V_{\text{ext}}(r,t) \right) \psi_{\uparrow\downarrow}(r,t).
\label{eq:SNforQubit}
\end{equation}
Here, the momentum term causes the wavefunction to spread out spatially in time, while the self-gravitational potential $V_G(\Psi)$ introduces a competing effect that inhibits the rate of spreading.\footnote{Incidentally, this gravitational effect was previously conjectured by Diosi and Penrose to be the mechanism behind wavefunction collapse \cite{DIOSI1984199,PenroseStateCollapse,Penrose:2014nha,Moroz1998SphericallySS}.} To see this, note that since $\ket{0}$ and $\ket{1}$ are orthonormal,
\begin{equation}
V_G(\Psi) = |\alpha|^2 V_G(\psi_\uparrow) + |\beta|^2 V_G(\psi_\downarrow).
\label{eq:potentialequation}
\end{equation}
Therefore, as $\psi_{\uparrow\downarrow}$ evolves, it is both gravitationally attracted to itself and $\psi_{\downarrow\uparrow}$, as evidenced by the first and second term in \eqref{eq:potentialequation}, respectively, thus combating the spreading \cite{Bahrami_2014}.

Ultimately, these gravitational effects manifest in the dynamics as a non-trivial, initial conditions-dependent phase difference $\Delta \phi_{\mathrm{SN}}$ between the spin-up and spin-down components of the qubit \cite{Grosssardt_2021, Grossardt_2024, Delgado_2025}. In this way, the SN dynamics mimic the dynamics of Weinberg's non-linear theory of quantum mechanics \cite{WEINBERG1989336ha, Delgado_2025}. Intriguingly, in the context of a Stern--Gerlach experiment in which the external potential models a non-uniform magnetic field that is oriented along a fixed axis, one finds that $\Delta \phi_{\mathrm{SN}} \propto (|\alpha|^2 - |\beta|^2) m^2$. Thus, at least in this case, the non-linear effects of the SN equation are maximized for very massive and non-uniform superposition states \cite{Grosssardt_2021}.\footnote{These results are based on a recursive perturbative expansion of the self-interaction potential $V_G$ and several simplifying assumptions. For detailed methodologies and numerical approaches, see \cite{Grosssardt_2021, Grossardt_2024, Delgado_2025, Sahoo_2023, hatifi2023revealingselfgravitysterngerlachhumptydumpty}.}

\chapter{Solving $\cc{NP}$-Complete Problems with Semiclassical Gravity}
\label{ch:SolvingNPSemiclassicaly}

In this Chapter, we show how the SN dynamics can be exploited to efficiently solve $\cc{NP}$-complete problems. Our idea is largely a SN-rebranding of Abrams and Lloyd's argument that Weinberg's non-linear theory of quantum mechanics can efficiently solve $\cc{NP}$-complete problems \cite{Abrams_1998}, as well as Bao, Bouland, and Jordan's \cite{Bao_2016} generalization of Abrams and Lloyd's work to other non-linear theories. To apply their ideas, however, we first need to rephrase a given $\cc{NP}$ problem as a state discrimination task, which can then be solved using the SN dynamics.

\section{Encoding an $\cc{NP}$ Problem into a Qubit}
\label{sec:EncodingNPinQubit}

Given $L \in \cc{NP}$, its corresponding algorithm $M$, its corresponding polynomial $p$, and an input $x \in \{0,1\}^*$, let 
\begin{equation}
P_{x} = \left|\left\{y \in \{0,1\}^{p(|x|)} : M(x, y) = 1\right\}\right|
\end{equation}
be the number of proofs that cause $M$ to accept. Evidently, to decide whether $P_{x} = 0$ or $P_{x} > 0$ is sufficient to decide $L$. In \cite{Abrams_1998}, Abrams and Lloyd give a quantum algorithm that can efficiently transform this question to the problem of deciding whether a particular quantum state is $\ket{0}$ or not. Specifically, the Abrams--Lloyd algorithm takes any $L \in \cc{NP}$ and prepares the single-qubit quantum state (up to normalization)
\begin{equation}
\frac{2^{p(|x|)} - P_x}{2^{p(|x|)}} \ket{0} + \frac{P_x}{2^{p(|x|)}} \ket{1}.
\label{eq:ALState}
\end{equation}
Therefore, the Abrams--Lloyd algorithm proves that to decide $L$ on a quantum computer, it suffices to be able to reliably decide if \eqref{eq:ALState} is $\ket{0}$ or not. 

Intuitively, this state discrimination task is hardest when $P_x = 1$, i.e., when the language $L$ not only has a proof, but has a \emph{unique} proof. Interestingly, as mentioned in \cite{Bao_2016}, by using ideas from computational complexity theory (namely, the Valiant--Vazirani theorem \cite{Valiant_1986}), one can formalize this intuition and refine the Abrams--Lloyd algorithm so that with sufficiently high probability, deciding $L$ does indeed reduce to this hardest case.\footnote{Intuitively, the Valiant--Vazirani (VV) theorem answers the question: ``is it easier to find the needle in the exponential haystack (i.e.~solve \cc{SAT}) if there are millions of needles, or just a single one?'' (in the language of Section \ref{sec:ComputationalProblems}). This last problem -- finding a unique solution -- is usually called a ``promise problem''; for \cc{SAT}, we denote it as \cc{Unique-SAT}. Formally, the VV theorem says that there exists a randomized polynomial-time reduction from \cc{SAT} to \cc{Unique-SAT}. Despite perhaps being counterintuitive, this tells us that we can always reduce the problem of searching for unknown solutions to searching for some particular (unique) solution without altering the structural complexity of the problem. Loosely speaking, an \cc{NP}-complete problem will still be very hard to solve even if we reduce an exponential number of needles down to just one needle. So, we can focus on \cc{Unique-SAT} and use the Abrams--Lloyd (AL) algorithm to encode \textit{that} instance of the \cc{NP} problem into a qubit. The resulting ALVV state is Eq.~\eqref{eq:languageALstate}, while the original AL state is Eq.~\eqref{eq:ALState}.}
 
\begin{theorem}[Corollary of Abrams and Lloyd \cite{Abrams_1998} and Valiant and Vazirani \cite{Valiant_1986}; see also Bao, Bouland, and Jordan \cite{Bao_2016}]
\label{thm:ALAlgorithm}
For all $L \in \cc{NP}$, there exists a polynomial time quantum algorithm $Q_L$ such that for all inputs $x \in \{0,1\}^*$, $Q_L(x)$ outputs the state (up to normalization)
\begin{equation}
\label{eq:languageALstate}
\frac{2^{p(|x|)} - L(x)}{2^{p(|x|)}} \ket{0} + \frac{L(x)}{2^{p(|x|)}}\ket{1}
\end{equation}
with constant probability.
\end{theorem}

In other words, to decide $L \in \cc{NP}$, it suffices to be able to reliably distinguish between the single-qubit states $\ket{\varphi_0} = \ket{0}$ and $\ket{\varphi_1} = N\big(\frac{2^{p(|x|)} - 1}{2^{p(|x|)}} \ket{0} + \frac{1}{2^{p(|x|)}}\ket{1}\big)$, where $N$ is the normalization factor. Notice that $\ket{\varphi_1}$ is indeed \eqref{eq:ALState} with $P_x = 1$. Now, as $\ket{\varphi_0}$ and $\ket{\varphi_1}$ are exponentially close in $|x|$, one cannot efficiently distinguish them using standard quantum mechanics \cite{Nielsen_Chuang_2010}. However, as we show in the next subsection, one can efficiently distinguish them using the SN dynamics.\footnote{Note that $\ket{\varphi_1}$ is a highly non-uniform superposition state. Thus, per our discussion in Section~\ref{sec:SNQubitDynamics}, the gravitational effects of the SN equation are large for this state, at least in a SN Stern--Gerlach experiment.}

\section{The SN Dynamics and the Abrams--Lloyd--Bao--Bouland--Jordan Algorithm}
\label{sec:ALBBJ}

Let $M(\CC^N)$ be the manifold of normalized pure states on $\CC^N$,\footnote{Therefore, $M(\CC^2)/\mathrm{U}(1) \cong \mathrm{CP}(1)$ is the Bloch sphere, where $\mathrm{U}(1)$ is the unitary group of degree $1$ and $\mathrm{CP}(1)$ is the complex projective space of complex dimension 1.} and equip it with the Fubini--Study metric $d(\ket{a},\ket{b}) = \arccos|\braket{a}{b}|$, which measures the angle between  the states $\ket{a}$ and $\ket{b}$ \cite{Nielsen_Chuang_2010}. Whereas unitary maps are isometries of $M(\CC^N)$, non-unitary maps are not and thus geometrically distort $M(\CC^N)$. An example of this is formalized in the next theorem, which is related to an older result of Mielnik \cite{Mielnik_NonlinearQM,mielnik1985phenomenon}.

\begin{theorem}[Bao, Bouland, and Jordan \cite{Bao_2016}]
Let $S : M(\CC^N) \rightarrow M(\CC^N)$ be a non-unitary diffeomorphism and let $$r = \max_{a,b \in M(\CC^N)} \frac{d(S(a), S(b))}{d(a,b)}.$$ Then, there exists a geodesic $\ell : \RR \rightarrow M(\CC^N)$ such that for all $x,y \in \mathrm{Im}(\ell)$, $\frac{d(S(x), S(y))}{d(x,y)} \geq r$.
\label{thm:BBJThm}
\end{theorem}

In other words, every non-unitary self-diffeomorphism of $M(\CC^N)$ necessarily ``stretches'' or ``magnifies'' some geodesic on $M(\CC^N)$ by a factor of at least $r$. As argued in Abrams and Lloyd \cite{Abrams_1998}, and later generalized by Bao, Bouland, and Jordan \cite{Bao_2016}, this non-isometric stretching can in principle be used to quickly drive two close states far apart, and thus assist in solving hard computational problems.

\begin{figure}[t!]
    \centering
    \includegraphics[width=0.5\linewidth]{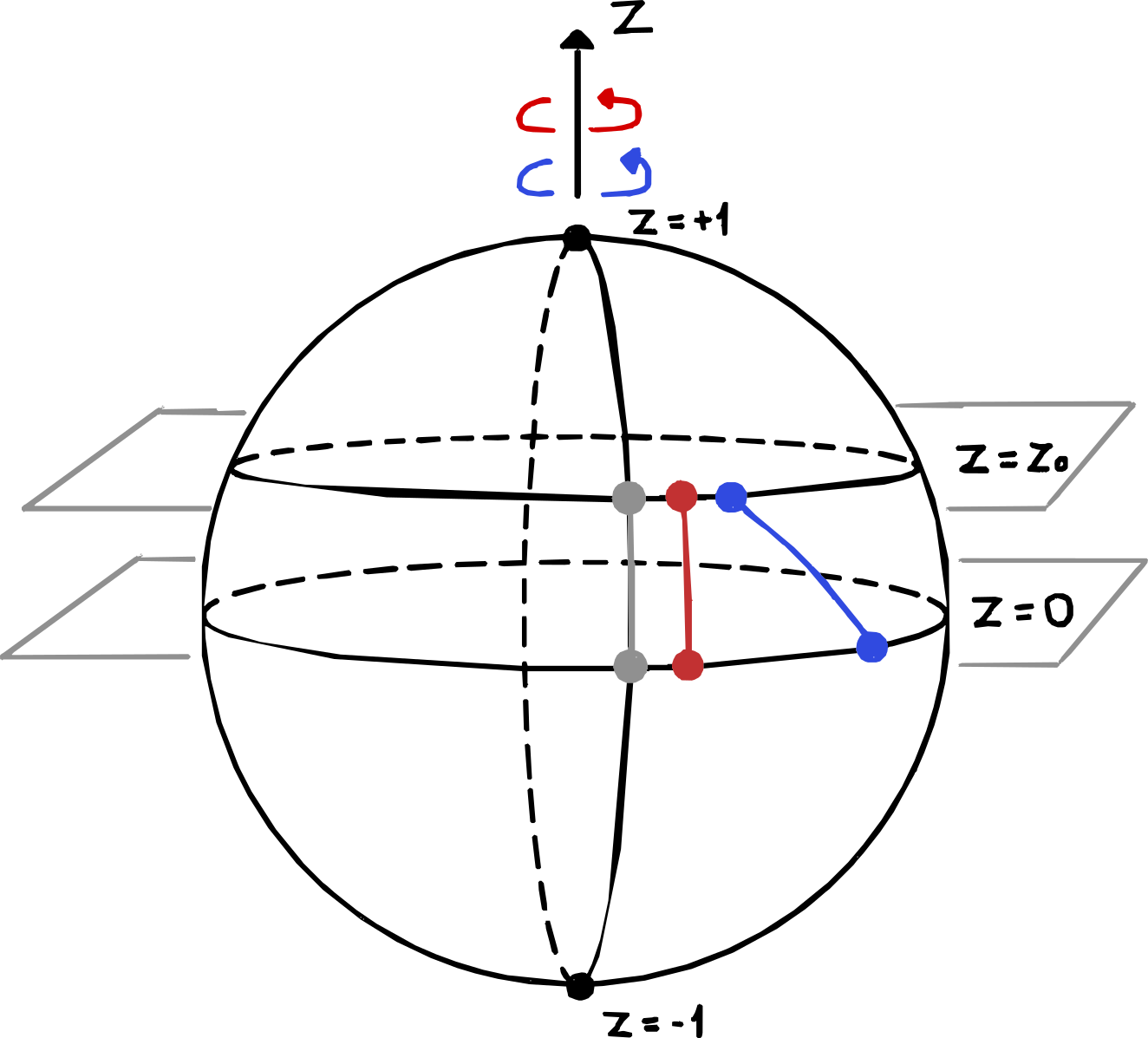}
    \caption[An example of a non-isometric diffeomorphism of the sphere $\Sph^2$: a position-dependent rotation that twists the sphere.]{An example of a non-isometric diffeomorphism of the sphere $\Sph^2$: a position-dependent rotation. This twists the sphere and stretches the distance between two grey points that were originally along the same meridian. Due to the differential rotation of $\Sph^2$, the grey point on the $z=0$ slice rotates faster than that on the $z=z_0$ slice, leading to the blue points after the map $\textcolor{blue}{\tau_{\Sph^2}}$ is applied which are now further apart. On the other hand, the isometric diffeomorphism $\textcolor{red}{\pi_{\Sph^2}}$ rotates all points at the same rate and preserves the original grey distance between the points.}
    \label{fig:UnitaryIsometryVsTwisting}
\end{figure}

An example of such a geometric distortion is shown in Fig.~\ref{fig:UnitaryIsometryVsTwisting}. Consider the sphere $\Sph^2$ in $\EE^3$, defined as $x^2 + y^2 + z^2 = 1$, and let $\rho^2 = 1-z^2$ such that $x^2 + y^2 = \rho^2 = 1-z^2$. Define polar coordinates on the $(x,y)$ plane so that a generic point on $\Sph^2$ is
$$ 
(x,y,z) \mapsto (\cos ( \theta(z) ) x - \sin (\theta(z)) y,\sin (\theta(z)) x + \cos (\theta(z)) y,z),
$$
where we let the angle $\theta$ carry a potential $z$-dependence. For example, we can let $\theta(z) = (1-z^2)\theta$ for some $\theta \in [0,2\pi)$, and, using $x = \sqrt{1-z^2}\cos \phi$ and  $y = \sqrt{1-z^2}\sin \phi$, we define a map
$$ \textcolor{blue}{
\tau_{\Sph^2}(\rho,\phi,z) = (\rho \cos (\phi + \rho(z)^2 \theta), \rho \sin (\phi + \rho(z)^2 \theta),z ) : \Sph^2 \to \Sph^2. 
}$$
Clearly, $\textcolor{blue}{\tau_{\Sph^2}}$ preserves the sphere (i.e.~the norm before and after is exactly the same) and acts as a position-dependent rotation on it which depends on $z$ through the $\rho$-factor inside the argument of the cosine and sine. Essentially, this leads to different $z$-slices of $\Sph^2$ rotating at a different rate, which is proportional to their distance from the poles located at $z=\pm 1$; the closer to the equator $z=0$, the faster the rotation happens. We also consider a rigid rotation of the sphere by the same angle $\theta$ but with no position-dependent term
$$ \textcolor{red}{
\pi_{\Sph^2}(\rho,\phi,z) = (\rho \cos (\phi + \theta), \rho \sin (\phi + \theta),z ) : \Sph^2 \to \Sph^2.
}$$
Now, consider two points on $\Sph^2$ that are originally at the same longitude and apply both maps. Because of the twisting that $\textcolor{blue}{\tau_{\Sph^2}}$ causes, the distance between the two points grows when this map is applied, while it stays the same when $\textcolor{red}{\pi_{\Sph^2}}$ acts. 

To see how this works for the SN equation and $\cc{NP}$ problems in particular, let $S_{\mathrm{SN}}$ be the map that encodes the net SN evolution of a massive qubit for some fixed amount of time, as discussed in Section~\ref{sec:SNQubitDynamics}. By the discussion in Section~\ref{sec:weakfieldlimit}, $S_\mathrm{SN}$ is a non-linear (and hence non-unitary) diffeomorphism from $M(\CC^2)$ to itself, so $S_{\mathrm{SN}}$ satisfies the conditions of Theorem~\ref{thm:BBJThm}. Therefore, with $$r_\mathrm{SN} = \max_{a,b \in M(\CC^2)} \frac{d(S_\mathrm{SN}(a), S_\mathrm{SN}(b))}{d(a,b)},$$ there exists a geodesic $\ell_\mathrm{SN} : \RR \rightarrow M(\CC^2)$ such that for all $x,y \in \mathrm{Im}(\ell_\mathrm{SN})$, $$\frac{d(S_\mathrm{SN}(x), S_\mathrm{SN}(y))}{d(x,y)} \geq r_\mathrm{SN}.$$ In particular,
\begin{equation}
\Delta_\mathrm{SN} = \min_{x,y \in \mathrm{Im}(\ell_\mathrm{SN})}\frac{d(S_\mathrm{SN}(x), S_\mathrm{SN}(y))}{d(x,y)} \geq r_\mathrm{SN} > 1,
\end{equation}
where the last inequality follows from the fact that $S_\mathrm{SN}$ is non-unitary. Here, $\Delta_\mathrm{SN}$ is the ``magnification factor'' that $S_\mathrm{SN}$ induces on $\ell_\mathrm{SN}$, i.e., it is the minimum amount by which $S_{\mathrm{SN}}$ stretches points on the geodesic $\ell_\mathrm{SN}$.

Now recall $\ket{\varphi_0}$ and $\ket{\varphi_ 1}$ from before, and let $\epsilon = d(\ket{\varphi_0}, \ket{\varphi_1})$ be the distance between them. Evidently, $\epsilon \geq c2^{-q(|x|)}$ for some real constant $c > 0$ and polynomial $q$. Since the length of $\ell_\mathrm{SN}$ is a constant (i.e., it is independent of $|x|$), if $\epsilon$ is greater than the length of $\ell_\mathrm{SN}$, then the states $\ket{\varphi_0}$ and $\ket{\varphi_1}$ can be efficiently distinguished within standard quantum mechanics by preparing a constant number of copies and then measuring them in the computational basis \cite{Nielsen_Chuang_2010}. 

However, if $\epsilon$ is less than the length of $\ell_\mathrm{SN}$, then this procedure is not guaranteed to work, at least not efficiently. Thus, in this case, apply a single-qubit unitary $U_0$ that maps $\ket{\varphi_0}$ and $\ket{\varphi_1}$ to two points $\ket{\lambda_0^{(0)}}$ and $\ket{\lambda_1^{(0)}}$, respectively, on $\ell_\mathrm{SN}$. Then, apply $S_\mathrm{SN}$ to get states $\ket{\varphi_0^{(1)}}$ and $\ket{\varphi_1^{(1)}}$, respectively, which satisfy $d(\ket{\varphi_0^{(1)}}, \ket{\varphi_1^{(1)}}) \geq \Delta_{\mathrm{SN}} \epsilon$ by Theorem~\ref{thm:BBJThm}. If $\Delta_{\mathrm{SN}}\epsilon$ is larger than the length of $\ell_\mathrm{SN}$, then again use standard techniques to distinguish the two states. However, if $\Delta_\mathrm{SN}\epsilon$ is smaller than the length of $\ell_\mathrm{SN}$, then repeat again: apply the single-qubit unitary $U_1$ that maps $\ket{\varphi_0^{(1)}}$ and $\ket{\varphi_1^{(1)}}$ back onto $\ell_\mathrm{SN}$ and reapply $S_\mathrm{SN}$. This gives states $\ket{\varphi_0^{(2)}}$ and $\ket{\varphi_1^{(2)}}$ separated by distance at least $\Delta_\mathrm{SN}^2 \epsilon$, and so on. 

By iterating this procedure, the states are separated to a constant distance with just
\begin{equation}
\log_{\Delta_{\mathrm{SN}}}(1/\epsilon) \leq \log_{\Delta_\mathrm{SN}}(2^{q(|x|)} / c) = O(q(|x|))
\end{equation}
applications of $S_\mathrm{SN}$. Since the single-qubit unitaries $U_i$ can be precomputed by knowledge of $\ket{\varphi_0}$, $\ket{\varphi_ 1}$, and how $S_\mathrm{SN}$ acts on $M(\CC^2)$,\footnote{Precomputing might feel like cheating, but in fact it is not. In particular, since $S_\mathrm{SN}$ is a fixed map, and since the initial states $\ket{\lambda_0^{(0)}}$ and $\ket{\lambda_1^{(0)}}$ on $\ell_\mathrm{SN}$ are known, the states $\ket{\varphi_0^{(1)}} = S\ket{\lambda_0^{(0)}}$ and $\ket{\varphi_1^{(1)}} = S\ket{\lambda_1^{(0)}}$ can be explicitly computed. Then, to find $U_1$, it suffices to find two points $\ket{a}$ and $\ket{b}$ on $\ell_\mathrm{SN}$ with the same inner product as $\ket{\varphi_0^{(1)}}$ and $\ket{\varphi_1^{(1)}}$, and then define $U_1$ as any unitary for which $U_1\ket{\varphi_0^{(1)}}=\ket{a}$ and $U_1\ket{\varphi_1^{(1)}}=\ket{b}$. The proof of Theorem~\ref{thm:BBJThm} in \cite{Bao_2016} is constructive and independent of $|x|$, so one can access the points on $\ell_{\mathrm{SN}}$ in constant time and easily find such an $\ket{a}$ and $\ket{b}$. Similar reasoning works for the later steps of the algorithm. Crucially, this precomputation would be intractable had we used the state \eqref{eq:ALState} instead of \eqref{eq:languageALstate}, as we would not know what initial states to evolve.} the SN map $S_\mathrm{SN}$ can be used to distinguish between $\ket{\varphi_0}$ and $\ket{\varphi_1}$ efficiently, and thus it can be used to efficiently decide any language in $\cc{NP}$.\footnote{Incidentally, this algorithm can also efficiently solve $\sharp\cc{P}$ (``sharp $\cc{P}$'') problems \cite{Abrams_1998, Bao_2016}, which count the number of proofs of $\cc{NP}$ problems. For this reason, $\sharp \cc{P}$ problems are even harder than $\cc{NP}$ problems \cite{AB09}.}

\section{Do All Non-Linear Extensions of Quantum Theory Entail PECTT Violations?}
\label{sec:other_nonlinearities}

In the previous Section, we formalized the fact that the non-linear SN qubit dynamics -- arising from taking semiclassical gravity as the fundamental starting point -- necessarily lead to efficient solutions of \cc{NP} languages.\footnote{We note that this conclusion has already been pointed out in the literature to some extent (e.g.~in \cite{Kent_2021}) before, yet it was based more on analogy with the original Abrams--Lloyd algorithm rather than on a formal treatment. Here, we believe that we showed how to put this claim on firmer grounds by combing the works of \cite{Abrams_1998,Bahrami_2014,Bao_2016,Valiant_1986}, and treating carefully which \cc{NP}-problem is to be encoded into qubit dynamics.} But do all non-linear modifications lead to the same conclusion, and thus to violations of the PECTT?   

The key is that only \textit{certain} non-linear extensions lead to such violations, namely those that satisfy Theorem \ref{thm:BBJThm}, since the corresponding map needs to be differentiable. A self-diffeomorphism on the space of all pure states for a physical system must be, by definition, smooth and invertible (with smooth inverse-map). Therefore, non-invertible discontinuous processes cannot be treated within this framework. Two particular examples of such non-invertible processes are:
\begin{itemize}
    \item \textbf{Instantaneous Wavefunction Collapse:} this is very much related to the \textit{measurement problem} in quantum theory, which is the probabilistic reduction of an initial quantum state (which could be a superposition of all possible outcomes) to a definite outcome with probability given by the Born rule \cite{Nielsen_Chuang_2010}. Even though the evolution of a quantum state happens continuously (and unitarily in standard quantum theory, with the Schr\"odinger equation), collapse occurs abruptly after interaction with some measuring device, and is clearly non-invertible (as any definite post-measurement state could have come from many pre-measurement states). Instantaneous collapse is ``non-linear'' in the sense that it does not comply with superpositions and linearity, but it is not a non-linear modification of the Schr\"odinger equation.
    \item \textbf{Objective Collapse Models:} here, the wavefunction collapse is achieved dynamically via non-linear and stochastic terms that are directly added to the usual (Schr\"odinger) quantum evolution. These lead to spontaneous collapse, which localize the wave function in space. The most well-known collapse models are the Ghirardi--Rimini--Weber (GRW) model \cite{GRW_Collapse_Model}, the Continuous Spontaneous Localization (CSL) model \cite{CSL_Collapse_Model}, and the Di\'osi--Penrose model \cite{DIOSI1984199,Diosi_CollapseModel,penrose1996gravity}; see also \cite{ReviewCollapseModels} for a modern review. Since the dynamics are stochastic, they are also inherently non-invertible. As a result, even if we measure objective collapse occurring in Nature, these models cannot be used to argue that $\cc{P} = \cc{NP}$ and thus they violate the PECTT, at least not in the sense of Theorem \ref{thm:BBJThm}.
\end{itemize}

In the context of semiclassical gravity, it has been argued that the SN dynamics together with the instantaneous collapse prescription are inconsistent. This is because the discontinuous changes lead to violations of the energy-momentum conservation ($\nabla^\mu \langle \Psi | \hat{T}_{\mu \nu} \Psi \rangle \neq 0$) on the right hand-side of the semiclassical EFEs \eqref{eq:semiclassicalEFEs}, violating the Bianchi identity ($\nabla^\mu G_{\mu \nu} = 0)$ on the left hand-side \cite{Page_1981}. This claim has been debated in the literature \cite{mattingly2005quantum,Grossardt_2022}, though, as a prescription for measurements that can be given a collapse interpretation \textit{and} is compatible with $\nabla^\mu \langle \Psi | \hat{T}_{\mu \nu} \Psi \rangle = 0$
has been worked out in \cite{Wald:1995yp} (but it restricts the possible $\Psi$ states that satisfy it). 

On the other hand, the SN dynamics are a deterministic non-linear quantum evolution equation, and as such, they also suffer from superluminal signalling \cite{GISIN19901,Polchinski_1991,czachor1991mobility}. This issue also holds for Weinberg's non-linear quantum theory \cite{WEINBERG1989336ha}, the one that the original Abrams--Lloyd paper wanted to rule out \cite{Abrams_1998}. Yet, a more careful relativistic treatment of quantum states suggests that non-linearities can exist without necessarily leading to superluminal communication with localized collapse \cite{Kent_2005}. As a result, (weak-field) semiclassical gravity in the spirit of the SN equation could still be made consistent, and a framework that employs it is the quantum readout devices of  \cite{Kent_2021} -- we'll come back to it in the next Chapter. The point is that, if SN dynamics are not unequivocally ruled out by some other condition, they imply efficient solutions to \cc{NP}-complete problems, which contradicts the PECTT.

Indeed, non-linearities, faster-than-light communication and vast computational advantages are very much related, as was established in the original Bao--Bouland--Jordan paper \cite{Bao_2016}. There, the authors point out that \textit{``the no-signalling principle is equivalent to the inability to
solve \cc{NP}-hard problems efficiently by brute force within the classes of theories analyzed''} (their wording), and generic non-linearities (that satisfy Theorem \ref{thm:BBJThm}) are included in those theories. It would be interesting to investigate whether these results can be extended to those non-invertible non-linear maps that are of interest to us here, for example in the objective collapse models. If so, one could appeal to the no-signalling condition to argue that they are unphysical, and thus the PECTT survives. In the authors' own words, 
\begin{quote}
\textit{``Thus, the quantum Church--Turing thesis appears to be remarkably robust, depending
not in a sensitive way on the complete Hilbert-space 
formalism of quantum mechanics, but rather derivable
from more foundational operational principles such as
the impossibility of superluminal signalling.''}
\end{quote}

\chapter{Discussion: Gravity from the Lens of Computational Complexity Theory}
\label{sec:GRAVIgateDiscussion}

In Part I, we showed that the weak-field dynamics of the semiclassical EFEs can efficiently solve $\cc{NP}$-complete problems. In consequence, if gravity is fundamentally classical and if the semiclassical EFEs describe the matter-gravity coupling, then one can efficiently solve $\cc{NP}$-complete problems. 

As this violates the PECTT, we are led to the conclusion that gravity is not classical. Of course, there are many ways that one may try to evade this conclusion. Another scenario, for example, is that gravity \textit{is} classical, but the semiclassical EFEs do not correctly describe the matter-gravity coupling. Alternatively, one could simply reject the PECTT altogether. Yet each such move comes at a conceptual cost and requires considerable justification.  

\section{Scenario 1: Gravity is Classical with a Different Gravity--Matter Coupling}

To avoid quantizing gravity, then, perhaps we can keep gravity classical but modify the non-linear structure of the matter-gravity coupling. 

However, given the incredibly general nature of Theorem~\ref{thm:BBJThm}, our results suggest that, in fact, \emph{any} consistent, semiclassical, and non-linear matter-gravity coupling will entail an efficient algorithm to solve $\cc{NP}$-complete problems and thus violate the PECTT. We therefore take our results not as an argument for an alternative theory of semiclassical gravity, but rather as an argument for the quantization of gravity itself.

What about a linear coupling then? Could that lead to a consistent semiclassical theory? Here, there is the example of Oppenheim's post-quantum theory of classical gravity \cite{Oppenheim_PostQuantumClassicalGravity}, where the dynamics are such that the density matrix of a quantum state evolves linearly and stochastically. Similar to the objective collapse models, the stochasticity of the dynamics evades our current argument from Chapter \ref{ch:SolvingNPSemiclassicaly}. Hence, our conclusions do not hold, and there is no obvious violation of the PECTT for this theory\footnote{\label{foot:GPTs_Galley} It is worth mentioning that, according to a more general analysis given in \cite{Galley_2023}, any consistent (definition made precise in the paper) coupling between classical gravity and quantum matter must be fundamentally irreversible. This analysis is
based on General Probabilistic Theories (GPTs), a framework that includes quantum theory but goes beyond that. This result is consistent with Oppenheim's theory, but not with the weak-field limit of the semiclassical EFEs, as those are reversible (the SN map is invertible).}. However, as is argued in \cite{Grossardt_2022} (their wording): 
\begin{quote}
\textit{``Rather than a single spacetime manifold, these models describe statistical ensembles of spacetimes. Therefore, strictly speaking, they do not qualify as models of semiclassical gravity as defined here\footnote{For semiclassical gravity, they assume a classical spacetime, i.e.~a pseudo-Riemannian 4-manifold $({\cal M}, g)$ with metric $g_{\mu \nu}$, not a superposition thereof.}, but could potentially be regarded as a theory of (semi-)classical statistical mechanics for such models.``}
\end{quote}
In other words, this linear and stochastic theory already goes beyond traditional semiclassical gravity -- in a sense, it is more towards the quantum side of gravity already, so more within the realm of quantum gravity rather than semiclassical gravity.    
\section{Scenario 2: the PECTT is Violated}

What if we just ignore the PECTT? This would mean that we believe that efficient solutions to \cc{NP}-complete problems are, in principle, attainable by some physical machine operating under the laws of a physical theory. We choose not to comment here on the possible sci-fi scenarios arising in a world where $\cc{P} = \cc{NP}$, such as the breaking of encryption and the collapse of digital privacy, or the possibility of machines ``taking over mathematics'' and automatically generate formal proofs for mathematical conjectures or simulating our hardest known computational problems\footnote{Some might argue that the latter scenario is currently unfolding; yet, as far as $\cc{P} \neq \cc{NP}$, traditional -- ``human'' -- mathematics could still be safe from the ``machines''.}. Instead, we would like to answer ``why do we believe in the PECTT?'' and what does this entail. 

First, let us rephrase the PECTT in plainer language: it states that ``whatever the correct physical description of the world is, and whatever complexity class \cc{C} of efficiently computable languages it gives rise to, \cc{NP} is not in \cc{C}''. Second, let us wonder: what could be this physical theory? As far as we know, our best contender for \cc{C} at the moment is \cc{BQP}, which can be thought of as the set of decision problems solvable on a quantum circuit with polynomially many gates \cite{AB09}. The PECTT, then, postulates that $\cc{NP} \not\subset \cc{BQP}$, which is a commonly held conjecture usually referred to as the \textit{quantum Church--Turing Thesis}; this is the statement put forward first by Deutsch \cite{10.1098/rspa.1985.0070} and then by Freedman, Kitaev, Larsen, and Wang \cite{freedman2002topologicalquantumcomputation}. As we mentioned in Footnote \ref{foot:q-computers} in Chapter \ref{sec:NPandChurchTuring}, proving the conjecture would be a proof of $\cc{P} \neq \cc{NP}$ too. And how robust is \cc{BQP} as a candidate for \cc{C}? In the words of Bao, Bouland and Jordan \cite{Bao_2016}, very:   
\begin{quote}
\textit{Physical processes which conceivably might be more computationally powerful than quantum Turing machines, such as various quantum many-body dynamics of fermions, bosons, and anyons, as well as scattering processes in relativistic quantum field theories, can all be simulated with polynomial overhead by quantum circuits'', see \cite{freedman2002topologicalquantumcomputation,aaronson2011computational,jordan2012quantum,jordan2014quantumalgorithmsfermionicquantum,Jordan:2011ci}}    
\end{quote}

We shall mention one more result in support of the claim that physical theories cannot compute \cc{NP} efficiently. First, a framework that generalizes that of quantum theory is provided by the GPTs (see Footnote \ref{foot:GPTs_Galley}). Now, in \cite{Lee_2015}, it is shown that the set of problems that a GPT can solve is \cc{BGP} (the general-probabilistic analogue of \cc{BQP}; it holds that $\cc{BQP} \subseteq \cc{BGP}$) does not include \cc{NP}.\footnote{They also show in \cite{Lee_2015} that, even allowing for \textit{post-selection} (i.e.~the imaginary ability to keep only the successful runs of a computation and throw away all other runs), $\cc{PostBGP} \subseteq \cc{PostBQP}$, and post-selection is known to give rise to extremely powerful quantum computers \cite{aaronson2004quantumcomputingpostselectionprobabilistic}. Consequently, even post-selected general-probabilistic computers are not more powerful than post-selected quantum computers.} Even in the presence of gravity, \textit{if} any consistent gravity-matter coupling (assuming the former is classical) is described by some GPT (with an irreversible interaction) as argued in \cite{Galley_2023}, this tells us that the theory cannot efficiently compute \cc{NP} either. Additionally, if the gravitational field is non-classical (and thus with a reversible interaction) and \textit{still} describable in terms of a GPT, this theory of ``quantum gravity'' would still not be able to compute \cc{NP} in polynomial time. 

In this work, we do not say that \cc{BQP} is necessarily the abstract complexity class \cc{C}, though there is good evidence that it might be. We do say, though, that when one brings in semiclassical gravity in our physical description, \cc{C} becomes so big that it includes \cc{NP}. This goes against our PECTT conjecture, which tells us that efficient computation of \cc{NP}-languages is unphysical.\footnote{See also \cite{Barrett_2019} for the computational power of physical operational theories. The PECTT conjecture would be strengthened if we show that \cc{AWPP} does not contain \cc{NP} either, as $\cc{BGP} \subseteq \cc{AWPP}$. Also, even if we end up having that $\cc{NP} \subseteq \cc{C}$ for some appropriate consistent physical theory, this would not necessarily imply that $\cc{P} = \cc{NP}$ yet; see \S 5.5 in \cite{aaronson2016pnp}.}

While preparing this manuscript, we became aware of the series of works \cite{Kent_2005,Kent_2021,Fedida_2026} by Kent et al.~concerning hypothetical ``readout devices'' capable of extracting classical information from a quantum system without disturbing it. In \cite{Kent_2021}, it was argued that semiclassical gravity may provide a physically realizable readout mechanism and, moreover, that such devices could in principle implement the original Abrams--Lloyd algorithm in \cite{Abrams_1998} for solving $\cc{NP}$-complete problems. In this respect, their framework is closely related to the gravitational mechanism studied here, but our conclusions differ primarily in their interpretation. Specifically, whereas Kent et al.~emphasize the possibility that such dynamics could lead to experimentally accessible computational advantages beyond standard quantum theory, we emphasize that there is a profound tension between such vast computational consequences and the PECTT. Thus, insofar as one regards the PECTT as a fundamental principle of physics, such readout devices cannot exist.

\section{Scenario 3: Gravity is not Classical}
\label{sec:Scenario1_GravityQuantum}

Finally, let us argue briefly that, in a world without unphysical exponential speed-ups and ad-hoc gravity-matter couplings\footnote{Indeed, the appeal of the SN equation is that it does arise directly from the weak-field limit of the semiclassical EFEs, so its foundations are known \cite{Bahrami_2014}.}, quantizing gravity might be the simplest way to respect the PECTT. 

The reason is that quantizing gravity restores linearity and thus avoids our argument altogether. For example, in the specific context of the weak-field limit of the semiclassical EFEs, quantizing the spacetime metric causes the self-gravitational potential in the SN equation to drop out, thus linearizing the dynamics \cite{Anastopoulos_2014,Giulini_2023}.\footnote{In particular, if the metric perturbation $h_{\mu \nu}$ is quantized, then the matter-gravity coupling $\hat{h}_{\mu \nu} \hat{T}^{\mu \nu}$ in the interaction Hamiltonian leads to linear weak-field dynamics (see \cite{Giulini_2023}).} Consequently, at least in this case, our argument in Section~\ref{sec:ALBBJ} does not apply, and the PECTT is not obviously violated. It would be interesting to see whether our existing quantum gravity frameworks predict linear or non-linear evolutions for the quantum state. 

Ultimately, we hope our results inspire a more computational way of thinking about demarcating fundamental physical theories. While gravitational physics and computational complexity theory are rather disparate fields, here we have demonstrated an intriguing consistency check between the two. Specifically, if the PECTT is taken seriously as a principle of physics, then any candidate theory of gravity must respect it. In this way, the PECTT, and computational complexity more broadly, provides not just a constraint, but a guiding principle in the search for a consistent theory of quantum gravity.

\chapter*{\centering \vfill PART II: \\Spacetime Quasicrystals \vfill} 
\addcontentsline{toc}{chapter}{PART II: Spacetime Quasicrystals}

\begin{center}
\includegraphics[scale=1]{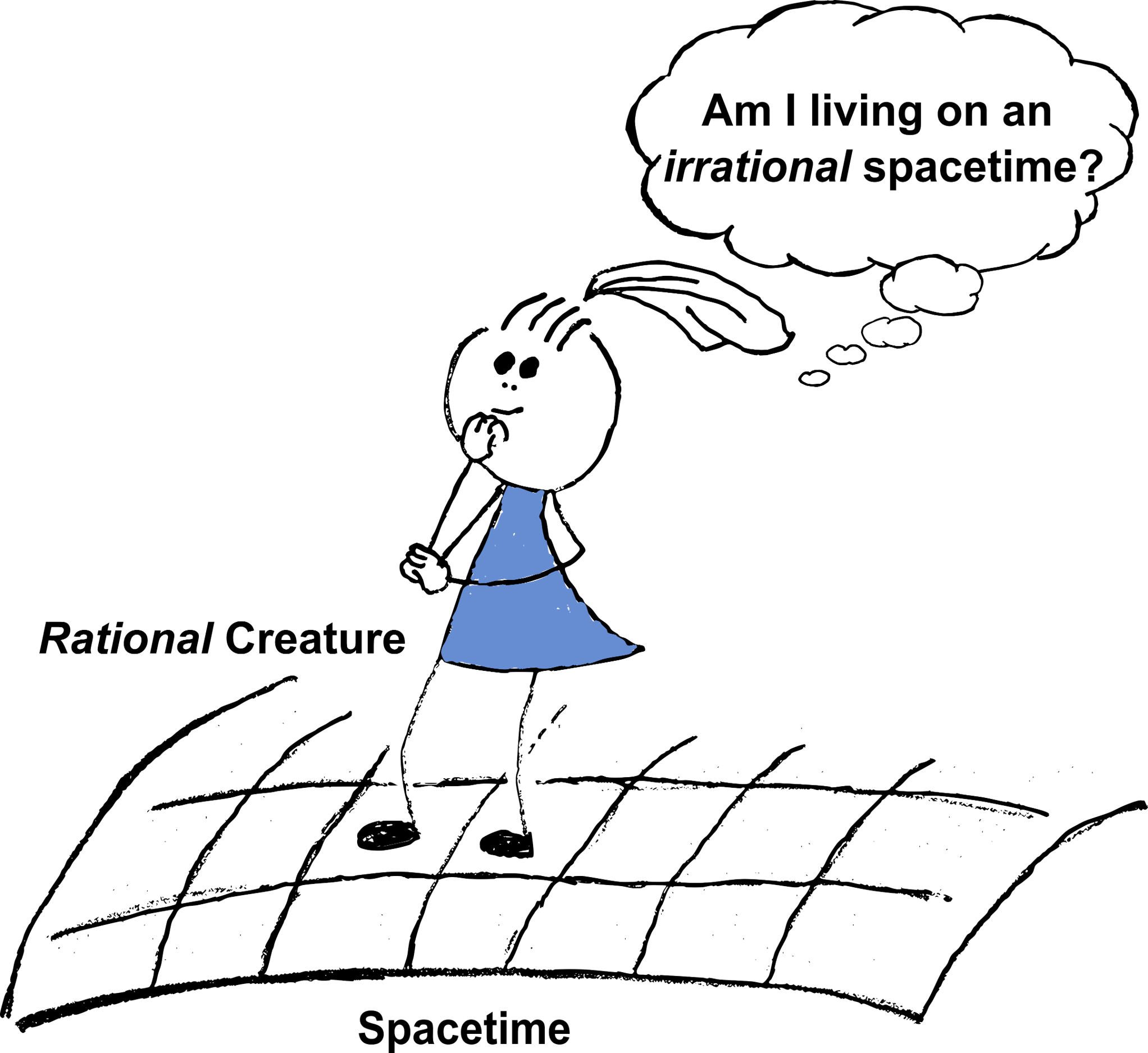}
\end{center}

\vfill

\newpage 

\begin{tikzpicture}[remember picture,overlay]
\node[
  anchor=north east,
  inner sep=0pt,
  xshift=-2.5cm,
  yshift=-2.5cm
] at (current page.north east) {
  \begin{tcolorbox}[colframe=purple,colback=blue!10,width=0.70\textwidth]
\justifying
\noindent 
Wherein the reader falls under the spell of discrete aperiodic structures, builds the first of their kind in spacetime, and uncovers their unexpected relevance to fundamental physics and quantum gravity.
\end{tcolorbox}
};
\end{tikzpicture}

\vspace{+10ex}
\noindent \textbf{Keywords:} Quasicrystals; Spacetime; Discrete Lorentz Geometry; Lorentz Symmetry; Scaling Symmetry; Aperiodic Order; Quasi-periodicity; Fractals; Self-Similarity; String Compactifications; Causal Set Theory; Hyperuniformity; Quantum Gravity.

\section*{Outline of Part II}
\addcontentsline{toc}{section}{Outline of Part II}

The summary of Part II is as follows: 

In Chapter \ref{ch:MatBackgroundPartII}, we introduce some background about lattices and their symmetries (Section \ref{sec:Lattices} and \ref{sec:Symmetries_of_Lattices}) and Coxeter's general theory of reflection groups (Section \ref{sec:Coxeter_Theory}). 

In Chapter \ref{ch:QuasicrystalsEuclidean}, as a warm-up for spacetime, we first develop the theory of quasicrystals in ordinary Euclidean space.  After reviewing the standard ``cut-and-project'' (\CNP{}) scheme (for constructing a quasicrystal by taking an irrationally-sloped ``\textit{cut}'' through a higher-dimensional lattice, and \textit{projecting} nearby lattice points onto the cut) in Section \ref{sec:CNP}, we describe a variant of particular interest, which we call the ``\textit{symmetric} cut-and-project'' (\sCNP) scheme in Section \ref{sec:sCNP}, in which quasi-periodicity, non-crystallographic symmetry, and discrete scale-invariance are intrinsically interwoven in the desired fashion (as in the Penrose or Ammann-Beenker tilings).  Whereas the most standard \CNP{} scheme uses a ``window'' to choose lattice points near the cut, we also discuss a more general \CNP{} scheme based on a ``weighting function'', which weights different points differently, according to their distance from the cut. This allows us to define ``self-dual'' and ``globally-scale-invariant'' quasicrystals, which will be important when we turn to Lorentzian quasicrystals. To illustrate all of this, we use \sCNP{} to construct the standard Ammann-Beenker tiling, as well as its self-dual variant (see Fig.~\ref{fig:AmmannBeenker}).  We end by discussing the types of (quasi)symmetries/dualities that \sCNP quasicrystals have in Section \ref{sec:CNPsymmetries}, and delineating two distinct types of scale invariance (``local'' and ``global'').  

While Chapter \ref{ch:QuasicrystalsEuclidean} is mainly a review, the reader may find certain aspects of our formulation novel or helpful, as it has been designed (i) to clarify the logical connection between quasi-periodicity, non-crystallographic symmetry and discrete self-similarity or scale-invariance; and (ii) relatedly, to make the generalization from Euclidean to Lorentzian spacetime simple and natural (and, to our knowledge, our notions of self-dual and globally-scale-invariant quasicrystals are also new.)

In Chapter \ref{ch:QuasicrystalsMinkowski}, we are finally ready to extend our discussion to spacetime. First, in Sec.~\ref{sec:MathematicalBackgroundLorentzian}, we introduce Lorentzian lattices ({\it i.e.}\ lattices in Minkowski spacetime) and their reflection symmetries, focusing on the key novel features compared to lattices in Euclidean space. Then in Sec.~\ref{sec:SpacetimeCNP}, we explain how the \sCNP{} algorithm may be naturally extended to spacetime. Crucially, the quasicrystal's non-crystallographic orientational symmetry is now a dramatically larger (in fact infinite!) group, and consequently, the \sCNP{} window (or weighting function) must also be much more symmetric. In Sec.~\ref{sec:implications}, we discuss further consequences of this fact: spacetime quasicrystals are a symmetric point set without an associated symmetric tiling; and they do not have local scale invariance, but instead can have global scale invariance or self-duality. Then, in Sec.~\ref{sec:ExamplesSpacetimeQuasicrystals}, we construct the first three examples of such spacetime quasicrystals: (i) the $1+1$ dimensional (self-dual or globally-scale-invariant) spacetime quasicrystal obtained from the $(3+1)$ dimensional odd self-dual Lorentzian lattice $\textrm{I}_{3,1}$, (ii) the three $(3+1)$-dimensional quasicrystals from the self-dual off lattice $\mathrm{I}_{7,1}$, and (iii) the four particularly interesting $3+1$ dimensional (self-dual or globally-scale-invariant) spacetime quasicrystals obtained from the remarkable $(9+1)$ dimensional even self-dual Lorentzian lattice $\textrm{II}_{9,1}$.

Chapter \ref{ch:GeneralizedSpacetimeCNP} is an addition to the paper \cite{mygdalas2026spacetimequasicrystals} and it describes a more general way of doing \CNP{} in spacetime. All quasicrystals constructed so far had three related and perhaps unsatisfying features: (i) they are only even-dimensional, (ii) they are transforming only under a 2-dimensional irreducible representation of their symmetry group, and (iii) they have a preferred splitting of spacetime into those two-dimensional planes, which are not further mixed by the symmetry. To construct quasicrystals with better symmetries, we developed a generalized \CNP{} scheme. In Section \ref{sec:RevisitingNonCrystallographicSymmetry}, we discuss aspects of non-crystallographic symmetry $G_{NC}$, in particular how it can be found as a subgroup of some crystallographic group $G_{C}$, and form what is known as a \textit{Coxeter pair}. We focus on two methods: (i) \textit{folding} $G_{C} \twoheadrightarrow G_{NC}$ and (ii) \textit{unfolding} $G_{NC} \hookrightarrow G_C$. We show that there are no more interesting foldings (than what we already have in Chapter \ref{ch:QuasicrystalsMinkowski}) if one considers the classified crystallographic symmetries, but \textit{there are} unfoldings of known non-crystallographic symmetries $G_{NC}$ into lattices that can be used to build a \CNP{}. In Section \ref{sec:SpacetimeCNPRootBasis} we develop this new method and discuss several interesting features of the resulting symmetries, specifically how they cure the issues mentioned in the beginning of the paragraph. We close the chapter in Section \ref{sec:ExamplesNewSpacetimeQuasicrystals} with the construction of the first example of an icosahedrally symmetric $(3+1)$D quasicrystal. 

After this, we dive into potential applications of our mathematical spacetime quasicrystals in physics and how they might relate to reality.

Chapter \ref{ch:StringCompactifications} contains some speculations about their relation to string theory. Section \ref{sec:FittingUniverseNutshell} is devoted to one such idea: in \cite{moore1993finitedirections}, Moore pointed out that the most symmetric toroidal compactification of the superstring was obtained by curling all 10 dimensions of spacetime (including time) into a particularly symmetric torus $\TT^{9,1}$ obtained by modding out $(9+1)$D Minkowski space by the remarkable $(9+1)$D even self-dual lattice $\mathrm{II}_{9,1}$: $\TT^{9,1}=\mathbb{R}^{9,1}/\mathrm{II}_{9,1}$.  At the time, it seemed that this 10D torus couldn't have anything to do with the real world and was of purely academic interest (as it seemed to have closed time like curves, and lack our $3+1$ macroscopic dimensions). However, we explain how the quasicrystal construction in Sec.~\ref{subsec:4DLorentzianQuasicrystal} suggests a potential way around this conclusion: the most symmetric embeddings of $(3+1)$D Minkowski spacetime $\mathbb{R}^{3,1}$ in the 10D torus $\TT^{9,1}$ are {\it irrationally-sloped} with respect to the torus, so that $\mathbb{R}^{3,1}$ wraps around and around the torus, densely filling it, without ever intersecting itself. An observer confined to this $\mathbb{R}^{3,1}$ would regard themselves as living in infinite 4D Minkowski spacetime, despite the fact that this spacetime is embedded in the completely compactified 10D spacetime torus $\TT^{9,1}$ (the universe in a nutshell!). Moreover, in Section \ref{sec:HierarchyPuzzles} we explain some (very tentative) arguments suggesting that this scenario might even be a helpful starting point for explaining the mysteriously large hierarchies between the Planck scale $M_{\rm Pl}$, the electroweak scale $M_{\rm EW}$ and the vacuum energy scale $M_{\rm vac}$ ($M_{\rm Pl}\gg M_{\rm EW}\gg M_{\rm vac}$), as well as the seesaw relationship they are observed to obey ($M_{\rm Pl}M_{\rm vac}\approx M_{\rm EW}^{2}$).

Chapter \ref{ch:sQC_CausalSets} is also an addition to the paper \cite{mygdalas2026spacetimequasicrystals}, in which we explain connections of spacetime quasicrystals to causal sets in quantum gravity. After giving a brief overview of causal set theory in \ref{sec:CSTReview}, we introduce the number-volume (N-V) correspondence in Section \ref{sec:NVCorrespondence} and the N-V test, which is a criterion to decide if a given discrete causal set approximates a continuous spacetime it lives in well. In the causal set community, the consensus is that random causal sets known as Poisson sprinklings outperform everything in the N-V test. We run this test in $(1+1)$D (Section \ref{subsec:NVTestIn1+1D}), in $(2+1)$D (Section \ref{subsec:NVTest2Plus1D}) and in $(3+1)$D (Section \ref{subsec:NVTest3Plus1D}), and we provide concrete evidence that our spacetime quasicrystals in $(1+1)$D and in $(3+1)$D could actually beat Poisson in the N-V test. Wanting to understand the results of the test from a more fundamental point of view, we were led to a concept known as \textit{spacetime hyperuniformity}, which we introduce in Section \ref{subsec:SpacetimeHyperuniformity}. In fact, in Section \ref{subsec:NVTestIsHyperuniformityTest}, we recast the N-V test as a test for spacetime hyperuniformity, and show that our $(1+1)$D spacetime quasicrystal from the $\mathrm{I}_{3,1}$ lattice is the first structure to be shown spacetime hyperuniform. We conclude this chapter in Section \ref{sec:phenoqCST} by going trough some (also very tentative) ideas about the physics one could do if they replace the traditional random causal sets with our quasi-crystalline sets.

Chapter \ref{ch:MoreConnectionsAndSpeculations} is about other interesting connections we noticed between (spacetime) quasicrystals and theoretical physics. More specifically, we point out various other suggestions for why spacetime quasicrystals may be useful as a tool for discretizing/studying a variety of (nearly) scale-invariant physical systems of interest; and how they may connect to (or be useful in) other approaches to studying quantum gravity.

Finally, there is a collection of appendices that contain more technical ideas: 

Appendix \ref{AppendixA-NumberTheory} gives an overview of (algebraic) number theory and mentions tools needed for quasicrystal-building. 
Section \ref{app:inflation}, for example, explains how the scale factors characterizing the discrete scale invariance of a quasicrystal are related to the units of an associated algebraic number field, and tabulates these number fields for the self-dual Lorentzian lattices up to dimension 10.

Appendix \ref{appendix:FoldingUnfolding} introduces our graph-origami techniques: folding and unfolding. The mathematical background is developed in Section \ref{sec:GraphOrigami} -- and we believe it to be a novel framework. In Section \ref{sec:CoxeterPairsIcosahedral}, we put the framework into work to begin a classification program for all possible (hyperbolic) Coxeter pairs. 

Appendix \ref{app:hyperbolicmirrorintersections} discusses how the different reflecting mirrors (timelike or spacelike) can intersect. Since the intersection of spacelike mirrors classifies the fundamental domain of a Coxeter group, we use this to classify the vertices of this domain. 

Appendix \ref{app:SymmetriesLorentzianLattices} summarizes Vinberg's analysis of the reflection symmetries of self-dual Lorentzian lattices (since some of Vinberg's relevant papers are hard to obtain, have not been translated into English, and also contain potentially-confusing typos in their Coxeter-Dynkin diagrams, which we correct).  

Appendix \ref{appendix:NPFrames} deals with \textit{rational} (standard) and \textit{irrational} (Coxeter) Newman-Penrose frames in spacetime, and shows that the Coxeter eigenbasis (for a hyperbolic group) is always such a frame in any dimension. 

Appendix \ref{appendix:NVTestDetails} is reserved for more details regarding the N-V test, how to sample causal diamonds off of patches and how to parametrize Lorentz boosts. We also give the data tables for the runs that led to the figures of Section \ref{sec:NVCorrespondence}.

\chapter{Mathematical Background}
\label{ch:MatBackgroundPartII}

In this Chapter, we begin by reviewing some useful background material about mathematical lattices and Coxeter's theory of reflection groups; we reserve Appendix \ref{AppendixA-NumberTheory} for a preliminary introduction to algebraic number theory.

\section{Lattices}
\label{sec:Lattices}

In this section we give a brief summary of some of the key properties of mathematical lattices we will need in this paper.  For more details see \cite{conway2013sphere}.

Let $\{{\bf b}_i\}$ ($i=1,\ldots,d$) be a basis for $V=\mathbb{R}^{d}$.  The $\mathbb{Z}$-span of $\{{\bf b}_i\}$ is a $d$-dimensional {\bf lattice} $\Lambda\in V$, {\it i.e.}\ $\Lambda$ is the set of points in $\mathbb{R}^{d}$ given by
\begin{equation}
  \Lambda=\left\{\sum_{i=1}^{d}n_{i}{\bf b}_{i} \,\, | \,\, \textrm{all } n_i\in \mathbb{Z}\right\} = \ZZ {\bf b}_1 \oplus \cdots \oplus \ZZ {\bf b}_d = \langle {\bf b}_1 , \dots, {\bf b}_d\rangle_{\ZZ}.
  \label{eq:LatticeDefinition}
\end{equation}
Also, the $\RR$-span is $V$: $\langle \Lambda\rangle_{\RR} = V$. When the basis $\{ {\bf b}_i\}$ is taken to be the \textit{roots} of the lattice (to be defined properly in the next section), we call $\Lambda$ a \textit{root-lattice}.\footnote{More formally, the basis $\{ {\bf b}_i\}$ must further satisfy the conditions $\langle {\bf b}_i,{\bf b}_i \rangle = 2$ for all $1 \leq i \leq d$ and $\langle {\bf b}_i,{\bf b}_j \rangle \in \{0,-1\}$ whenever $i \neq j$, for the lattice $\Lambda$ to be a root-lattice in the strict sense. The roots capture the reflection symmetries of $\Lambda$, as will explain shortly.} Note that the choice of lattice basis $\{{\bf b}_i\}$ is not unique: another basis $\{{\bf b}_i'\}$ generates the same lattice $\Lambda$ if and only if it is related to $\{{\bf b}_i\}$ by ${\bf b}_i'=M_{ij}{\bf b}_j$, where $M_{ij}\in{\rm GL}(d,\mathbb{Z})$ is an integer $d\times d$ matrix with determinant $\pm1$.

We let $\Lambda^{{\bf t}}$ denote the same lattice $\Lambda$ translated (relative to the origin of $\mathbb{R}^{d}$) by the vector ${\bf t}\in\mathbb{R}^{d}$:
\begin{equation}
  \Lambda^{{\bf t}}=\left\{{\bf t}+\sum_{i=1}^{d}n_{i}{\bf b}_{i}\,\, | \,\, \textrm{all } n_i\in \mathbb{Z}\right\}.
\end{equation}

The $d$-dimensional parallelotope $F_\Lambda$ with edges ${\bf b}_i$
\begin{equation}
  F_\Lambda=\left\{\sum_{i=1}^{d}x_i {\bf b}_i \,\, | \,\,  0\leq x_i<1 \textrm{ for all } i \right\}
\end{equation}
is called a {\bf unit cell} (or a \textit{fundamental domain}) for the lattice (since every point in the lattice $\Lambda$ is related to a unique point in $F_\Lambda$ by translation by an integer linear combination of the basis vectors).  In other words, $F_\Lambda=\mathbb{R}^{d}/\Lambda$.

We will regard $\mathbb{R}^{d}$ as a $d$-dimensional vector space $V$ equipped with a real-valued inner product $\langle\cdot,\cdot\rangle$.\footnote{In Section \ref{sec:MathematicalBackgroundLorentzian}, in order to pass from space to spacetime, we will generalize $\langle\cdot,\cdot\rangle$ from a (positive definite) inner product to a (indefinite) non-degenerate bilinear form.} Since the inner product is positive definite, we can choose an orthonormal basis (or ``Cartesian basis'') $\{{\bf e}_{\alpha}\}$ for $V$ with
\begin{equation}
  \langle {\bf v},{\bf w}\rangle=
  \langle v^{\alpha}{\bf e}_{\alpha},
  w^{\beta}{\bf e}_{\beta}\rangle=
  v^{\alpha}\delta_{\alpha\beta}w^{\beta}
\end{equation}
where $\delta_{\alpha\beta}$ is the Kronecker delta.\footnote{Here, and throughout the thesis, we use the ``Einstein summation convention'' in which a pair of repeated indices (one up and one down) are implicitly summed from $1$ to $d$.} Using the inner product, we define the \textbf{Gram matrix} of a lattice $\Lambda$ as the matrix of inner products of its basis vectors, 
\begin{equation}
    G_{ij} = \langle {\bf b}_i, {\bf b}_j \rangle = {\bf b}_i^\alpha \delta_{\alpha \beta} {\bf b}_j^\beta 
    \label{eq:GramMatrixDefinition}
\end{equation}
and similarly if $\langle \cdot , \cdot \rangle$ is defined with respect to some other bilinear form instead of $\delta_{\alpha \beta}$. Collecting all basis vectors $\{ \bf b_i \}$ in the columns of a $d \times d$ matrix $B$, we write $G = B^T B$. The determinant of this matrix is related to the volume of the fundamental domain of $\Lambda$ (see \S 3 in \cite{baake2013aperiodic}) 
\begin{equation}
    \det G = (\det B)^2 = \big(\mathrm{vol}(F_\Lambda) \big)^2
    \label{eq:VolumeFundamentalDomainLattice}.
\end{equation}
This is also known as the lattice determinant $\det \Lambda = \det B$; its inverse is known as the \textit{lattice density} (Example 2.6 in \cite{baake2013aperiodic})
\begin{equation}
    \mathrm{dens}(\Lambda) = \frac{1}{|\det \Lambda|} = \frac{1}{| \mathrm{vol (F_\Lambda)}|}.
    \label{eq:LatticeDensity}
\end{equation}

Given any $\Lambda$, we can define its {\bf dual lattice} $\Lambda_{\ast}$ as the set of all vectors that have integer inner product with any point in $\Lambda$:
\begin{equation}
  \Lambda_{\ast}=\{{\bf v}_{\ast}\in V|\langle {\bf v}_{\ast},{\bf v}\rangle\in\mathbb{Z}, \forall\,{\bf v}\in \Lambda\}.
\end{equation}
$\Lambda$ is called an {\bf integer lattice} if the inner product of any two lattice vectors is an integer
\begin{equation}
  \langle {\bf v},{\bf v}'\rangle\in\mathbb{Z}\quad(\forall\,{\bf v},{\bf v}'\in\Lambda).
\end{equation}
This relation ensures that an integral lattice $\Lambda$ is a sublattice of its dual $\Lambda_\ast$, i.e. that $\Lambda \subseteq \Lambda_\ast$. In fact, we can show the following inclusion of lattices (see \S 2.4 in \cite{conway2013sphere}) 
\begin{equation}
\Lambda \subseteq \Lambda_* \subseteq \frac{1}{\det \Lambda} \Lambda .
\label{eq:LatticeInclusion}
\end{equation}
We say $\Lambda$ is a {\bf unimodular lattice} if its unit cell $F_\Lambda$ has unit volume; or, in other words, if $\det \Lambda = \pm1.$
$\Lambda$ is said to be {\bf self-dual} if it is the same as its dual lattice
\begin{equation}
  \Lambda=\Lambda_{\ast}.
\end{equation}
From Eq.~\eqref{eq:LatticeInclusion}, $\Lambda$ is self-dual if and only if it is integer and unimodular.

A self-dual lattice $\Lambda$ is {\bf even} or {\bf Type II} if the inner product of any two points in $\Lambda$ is an {\it even} integer; otherwise, it is {\bf odd} or {\bf Type I}. Self-dual lattices are relatively rare: Euclidean self-dual lattices have been classified up to dimension $d=25$ and are summarized in Table 2.2 in \cite{conway2013sphere}.  

Type I lattices exist in all dimensions: up to $d=8$, the only Type I lattice is the hypercubic lattice $\mathbb{Z}^{d}$ (the analogue of the square grid or the cubic lattice in $d \geq 4$); for $d\geq 9$ a second lattice $E_{8}\oplus\mathbb{Z}^{d}$ appears; in $d=12$ there is a third lattice $D_{12}^{+}$;\footnote{The root lattice $D_n$, also known as the \textit{checkerboard lattice}, is obtained by $\ZZ^n$ by colouring its vertices black and white and keeping only one colour. These are the root lattices of the $\mathfrak{so}(2n)$ Lie algebras. The lattices $D_n^+$ contain two copies of $D_n$, one sitting in-between the points of the other (see \S 7.3 in \cite{conway2013sphere}). In lower dimensions, we do not get anything new: $D_4^+ \simeq \ZZ^4$ and $D_8^+ \simeq E_8$, but in higher-dimensions we do.} and the number increases in higher dimensions.  

Type II lattices have especially remarkable properties and are even rarer: they only occur when the dimension $d$ is a multiple of 8.  In the lowest dimension ($d=8$), the unique Type II lattice is the $E_8$ root lattice.  Next, in $d=16$, there are two such lattices: $E_{8}\oplus E_{8}$ and $D_{16}^{+}$. Then, in $d=24$, there are 24 such lattices, including one particularly remarkable lattice with no roots (the Leech lattice).  And in higher dimensions ($d=32,40,\ldots$) the number grows rapidly.

In
Chapter \ref{ch:QuasicrystalsMinkowski}, we will see there is a much simpler classification of self-dual lattices in spacetime (with signature $(d-1,1)$ and, more generally, in spaces with indefinite signature $(p,q)$ (with $p$ spatial and $q$ temporal directions, whose dimension is $d = p + q$).

\section{Coxeter Theory}
\label{sec:Coxeter_Theory}

In this subsection, we briefly summarize some basic notions of Coxeter's theory of reflection groups. Our main references are \cite{coxeterSimplices, coxeterDiscreteGroups, coxeterCompleteEnumeration, humphreys_1990, mcmullen2002coxeter}.

Consider the vector space $V=\mathbb{R}^{d}$ equipped with the (positive-definite) inner product $\langle\cdot,\cdot\rangle$.  In the basis $\{{\bf v}_{\mu}\}$ ($\mu=1,\ldots,d$), the metric $g: V \times V \to \RR$ is a symmetric, positive-definite matrix $g_{\mu\nu}=\langle {\bf v}_{\mu}|{\bf v}_{\nu}\rangle$. 

Now consider a codimension-one hyperplane through the origin in $V$ (a mirror $M$), with normal vector ${\bf r}=r^{\mu}{\bf v}_{\mu}$ (its root).  Reflection through $M$ corresponds to the linear map
\begin{equation}
  {\bf x}\to{\bf x}-2
  \frac{\langle {\bf x},{\bf r}\rangle}
  {\langle {\bf r}, {\bf r}\rangle}{\bf r}.
  \label{eq:ReflectionMap}
\end{equation}
In the basis $\{{\bf v}_{\mu}\}$, this is described by $x^{\mu}\to R^{\mu}_{\;\;\nu}x^{\nu}$ with the reflection matrix
\begin{equation}
  \label{R_eq}
  R^{\mu}_{\;\;\nu}=\delta^{\mu}_{\;\;\nu}-2\frac{r^{\mu}r_{\nu}}{r^{2}}
\end{equation} 
where, throughout this thesis, we use the standard physics notation where Greek indices are lowered and raised using the metric $g_{\mu\nu}$ and its inverse $g^{\mu\nu}$, respectively, and 
$r^{2}\equiv \langle{\bf r},{\bf r}\rangle=r^{\mu}r_{\mu}$.  Note that the matrix $R^{\mu}_{\;\;\nu}$ squares to the identity matrix
\begin{equation}
  R^{2}=1.
\end{equation}
If two different mirrors $M_i$ and $M_j$ meet at an angle $\pi/m_{ij}$ ($m_{ij}\in\mathbb{Z}^{+})$, so that their roots $\mathbf{r}_{i}$ and $\mathbf{r}_{j}$ satisfy 
\begin{equation}
  \frac{\langle {\bf r}_i,{\bf r}_j\rangle}{(r_i^2 r_j^2)^{1/2}}
  =-{\rm cos}\frac{\pi}{m_{ij}},
\end{equation}
the corresponding reflection matrices $R_i$ and $R_j$ satisfy
\begin{equation}
  (R_i R_j)^{m_{ij}}=1.
  \label{eq:CoxeterRelations}
\end{equation}
A group that is generated by successive reflections in such mirrors is a {\bf reflection group}. Since in any $n$-dimensional vector space with a symmetric bilinear form any rotation can be described as the composition of at most $n$ reflections according to the \textit{Cartan–Dieudonn\'{e}} theorem \cite{stillwell2008naive}, reflection groups are the fundamental objects to study.  

A {\bf Coxeter group} $G$ is an abstraction and generalization of a reflection group: it is a group generated by $n$ generators $S=\{R_1,\ldots,R_n\}$, subject to the relations $(R_i R_j)^{m_{ij}}=1$, where $m_{ij}$ is a symmetric integer $n\times n$ matrix $(m_{ij}=m_{ji},m_{ij}\in \mathbb{Z})$, with $m_{ii}=1$ along the diagonal, and $m_{ij}>1$ off the diagonal ({\it i.e.}\ when $i\neq j$).  The pair $(G,S)$ is called a {\bf Coxeter system}.  The matrix $m_{ij}$ is called the {\bf Coxeter matrix}, and the corresponding reflections $R_{i}$, mirrors $M_i$, and roots ${\bf r}_i$ ($i=1,\ldots,n)$ are called the {\bf fundamental reflections}, {\bf fundamental mirrors}, and {\bf fundamental roots}, respectively. The number $n = |S|$ of fundamental roots is called the {\bf rank} of $G$. 

\begin{figure}[t!]
\captionsetup{justification=justified}

\subfloat[Spherical Diagrams]{%
        \includegraphics[width=.49\linewidth]{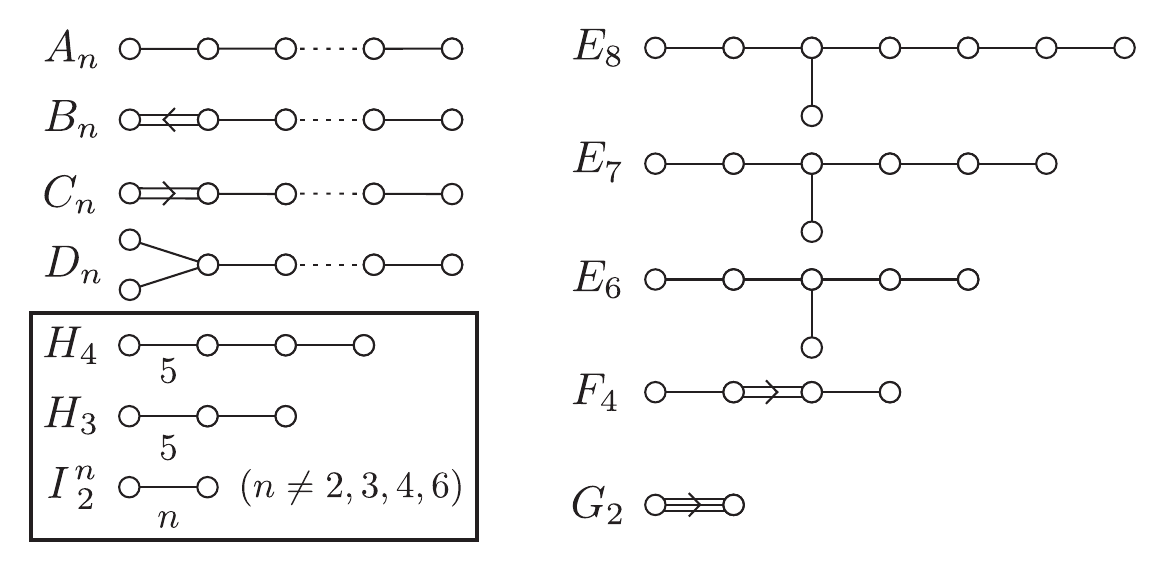}
        \label{fig:spherical_diagrams}
    }
    \hspace{+1ex}
    \subfloat[Affine Diagrams]{%
        \includegraphics[width=.49\linewidth]{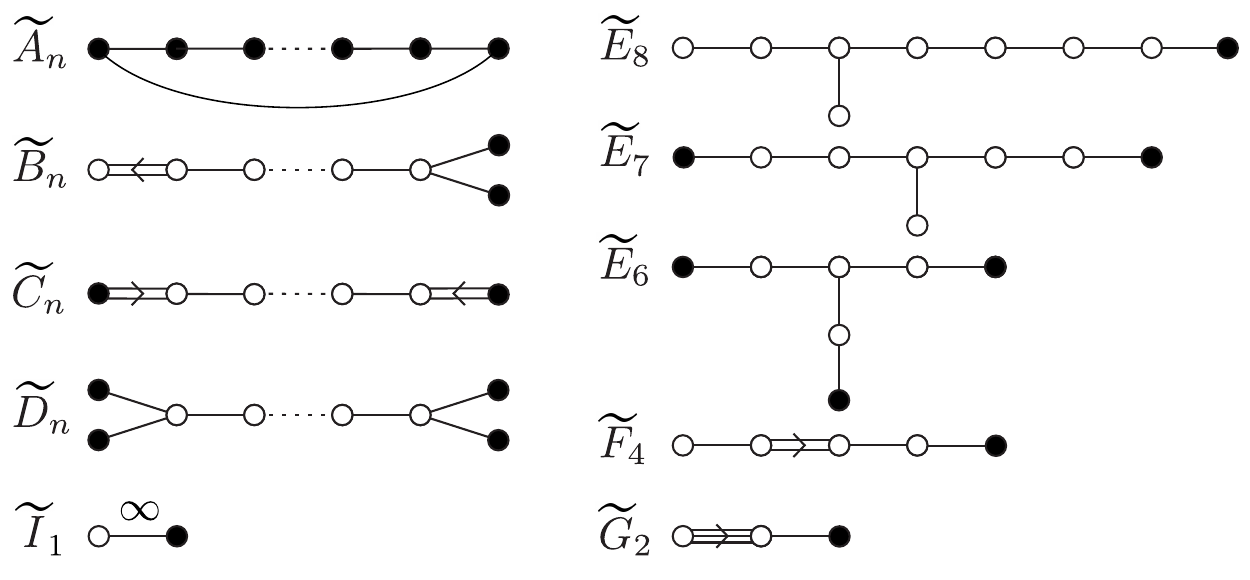}%
        \label{fig:affine_diagrams}%
    }
	\caption[Spherical and Affine Coxeter-Dynkin Graphs.]{The Coxeter-Dynkin diagrams for the irreducible Coxeter groups of finite/spherical type (left panel) and affine/planar type (right panel). The non-crystallographic cases ($H_{4}$, $H_{3}$ and $I_{2}^{n}$) are grouped in the box in the lower left.  Each affine diagram (at right) may be obtained by adding a single ``extending root'' to the corresponding finite crystallographic diagram (at left); or, alternatively, by removing a single extending node (any of the black nodes) from any affine diagram (at right) we obtain the corresponding finite crystallographic diagram (at left).}
	\label{fig:Irreducible_Root_Systems}
\end{figure}

A Coxeter group is specified entirely by its Coxeter matrix $m_{ij}$, which is conveniently summarized by a {\bf Coxeter diagram} (or {\bf Coxeter-Dynkin diagram}) in which we draw $n$ dots -- one for each fundamental reflection (or mirror, or root) -- and then draw a line between each pair of distinct dots $i$ and $j$, labelled by the corresponding integer $m_{ij}$.  For convenience, we then adopt the further standard conventions: if $m_{ij}=2$ we omit both the line and its integer label; if $m_{ij}=3$ we draw a single line and suppress the label; if $m_{ij}=4$ we draw a double line and suppress the label; and if the line connects a long root to a short root, we indicate this by adding an arrow to the line pointing to the short root. Some examples are shown in Fig.~\ref{fig:Irreducible_Root_Systems}, and the notation in the figure will be explained shortly.

A Coxeter group is {\bf irreducible} (resp.\ {\bf reducible}) if its Coxeter diagram is {\it connected} (resp.\ {\it disconnected}).  Any reducible Coxeter group naturally decomposes into a product of irreducible Coxeter groups (corresponding to the connected pieces of its Coxeter diagram).   
A Coxeter group has an associated {\bf geometric representation} in the vector space $V=\mathbb{R}^{n}$ spanned by the fundamental roots $\{{\bf r}_1,\ldots,{\bf r}_n\}$: if we take $V$ to have a symmetric bilinear form $B({\bf v},{\bf w}): V\times V \to \RR$ such that
\begin{equation}
  B({\bf r}_i,{\bf r}_j)=-{\rm cos}\frac{\pi}{m_{ij}},
  \label{eq:BilinearFormGraph}
\end{equation}
then the generator $R_{i}$ is represented by the reflection 
\begin{equation}
  R_i({\bf v})={\bf v}-2 B({\bf v},{\bf r}_i){\bf r}_i.
  \label{eq:ReflectionInRootBasis}
\end{equation}
Note that vector ${\bf r}_{i}$ is reversed by $R_i$, while the hyperplane orthogonal to ${\bf r}_{i}$, the its mirror $M_i$ is fixed pointwise by $R_i$. This natural action of the Coxeter group $G$ on $V$ preserves the form $B$ (i.e.~the metric on $V$), which means that $G$ is a discrete subgroup of the orthogonal group $O(V,B)$. 

The $n\times n$ {\bf Schl\"{a}fli matrix} $S_{ij}$ is twice the bilinear form $B$, expressed in the root basis:
\begin{equation}
  S_{ij}=-2\,{\rm cos}\frac{\pi}{m_{ij}}.
  \label{eq:SchlafliMatrix}
\end{equation}
Let $S_{ij}$ have $p$ positive and $q$ negative eigenvalues: we say it has signature ${\rm sig}(S) = (p,q)$. Note that $p+q\leq n$ since $S_{ij}$ may also have some zero eigenvalues. Using the terminology conventions in \cite{mcmullen2002coxeter}, irreducible Coxeter groups may be classified into one of four types according to the signature of $S_{ij}$
\begin{itemize}
  \item {\bf Spherical}: if $(p,q)=(n,0)$ ({\it i.e.}\ if all eigenvalues of $S_{ij}$ are positive);
  \item {\bf Affine}: if $(p,q)=(n-1,0)$ ({\it i.e.}\ if all eigenvalues of $S_{ij}$ are positive, except one which vanishes);
  \item {\bf Hyperbolic}: if $(p,q)=(p,1)$ ({\it i.e.}\ if exactly one eigenvalue of $S_{ij}$ is negative);
  \item {\bf Higher-Rank}: if $q\geq2$ ({\it i.e.}\ if at least two eigenvalues of $S_{ij}$ are negative).
\end{itemize}  
The signature of $S$ tells us what type of metric the Coxeter group $G$ preserves, namely whether it is Euclidean (positive-definite) or Lorentzian (indefinite), and thus $G$ is a subgroup of $O(n)$ in the spherical case and of $O(n-1,1)$ in the hyperbolic case. It also tells us whether our Coxeter group contains a finite number of elements or not: the Coxeter group $G$ can only be finite when the bilinear form $B$ (or $S$ equivalently) is positive definite (see Theorem in \S 6.4 in \cite{humphreys_1990}; see also \cite{howlett1982coxeter}). As a result, the only finite Coxeter groups are the spherical ones. 

\begin{figure}[t!]
\captionsetup{justification=justified}
\centering


\begin{subfigure}{\textwidth}
    \centering

    \begin{minipage}{0.32\textwidth}
        \centering
        \includegraphics[width=.95\linewidth]
        {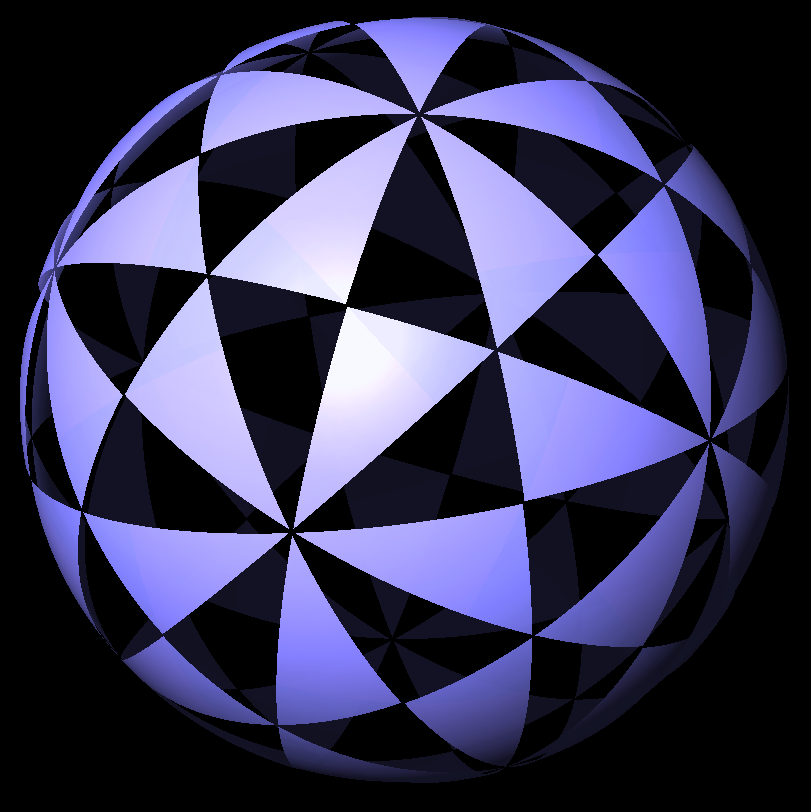}
    \end{minipage}
    \begin{minipage}{0.32\textwidth}
        \centering
        \includegraphics[width=.95\linewidth]
        {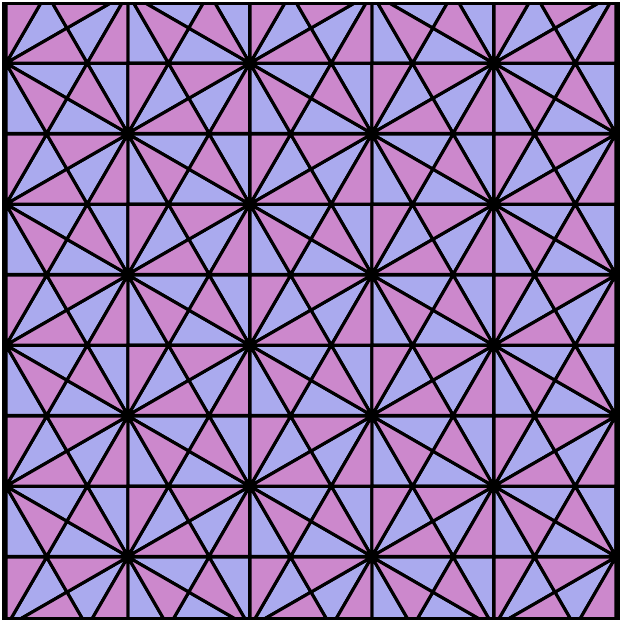}
    \end{minipage}
    \begin{minipage}{0.32\textwidth}
        \centering
        \includegraphics[width=\linewidth]
        {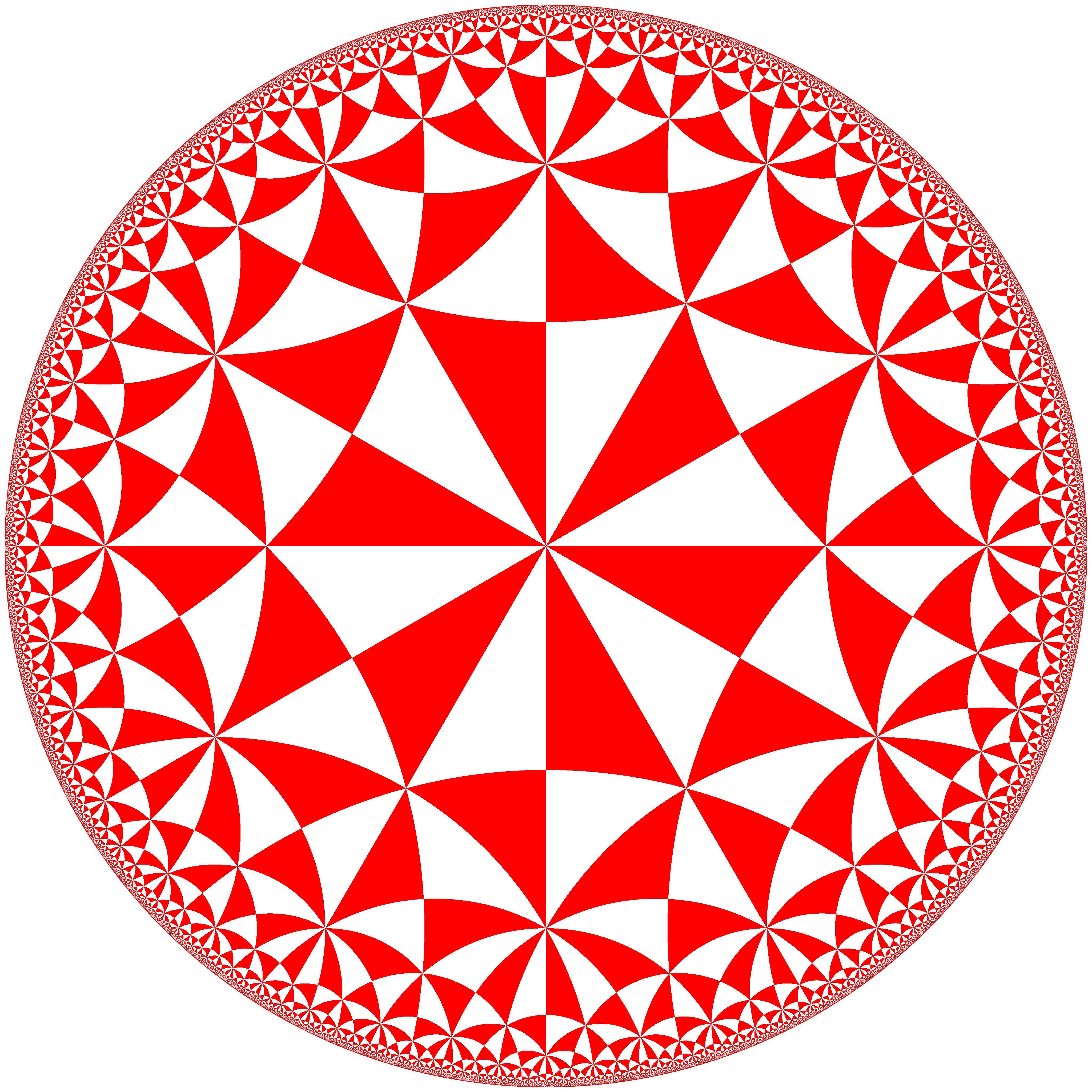}
    \end{minipage}

    \caption{The ``Coxeter complex'' showing mirrors (geodesic curves) and images of the fundamental domain (triangular regions) for various Coxeter groups:
\href{https://commons.wikimedia.org/wiki/File:Icosahedral_reflection_domains.png}{left}: the spherical (finite) $[5,3]$ (icosahedral) group;
\href{https://commons.wikimedia.org/wiki/File:Tiling_Dual_Semiregular_V4-6-12_Bisected_Hexagonal.svg}{center}: the affine (planar) $[6,3]$ group;
\href{https://en.wikipedia.org/wiki/File:Hyperbolic_domains_642.png}{right}: the hyperbolic $[6,4]$ group (figures are reproduced from Wikipedia; see links).}
\end{subfigure}

\vspace{1em}


\begin{subfigure}{\textwidth}
    \centering

    \begin{minipage}{0.32\textwidth}
        \centering
        \includegraphics[width=\linewidth]{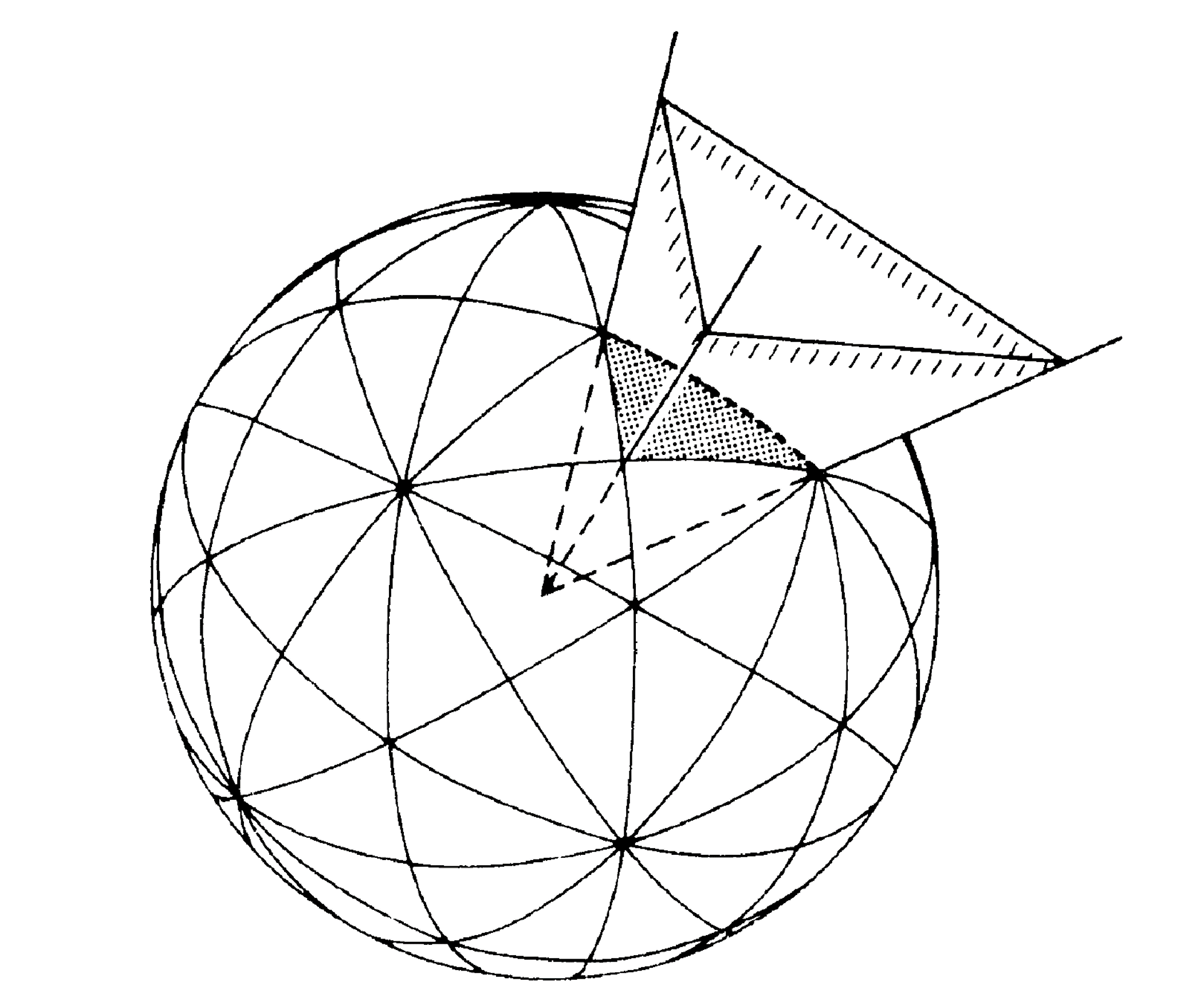}
    \end{minipage}
    \begin{minipage}{0.32\textwidth}
        \centering
        \includegraphics[width=\linewidth]{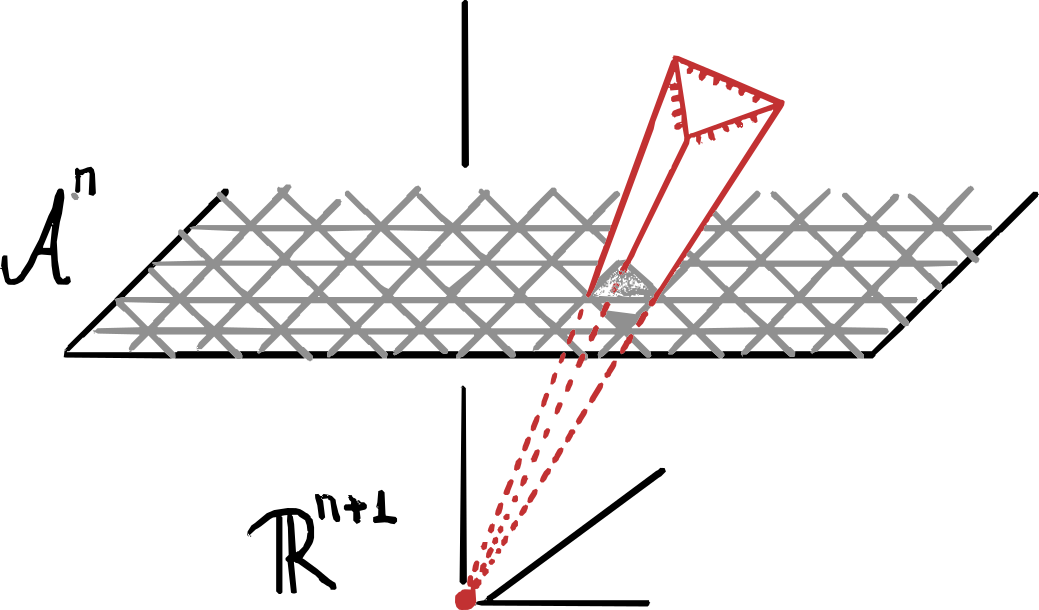}
    \end{minipage}
    \begin{minipage}{0.32\textwidth}
        \centering
        \includegraphics[width=0.85\linewidth]{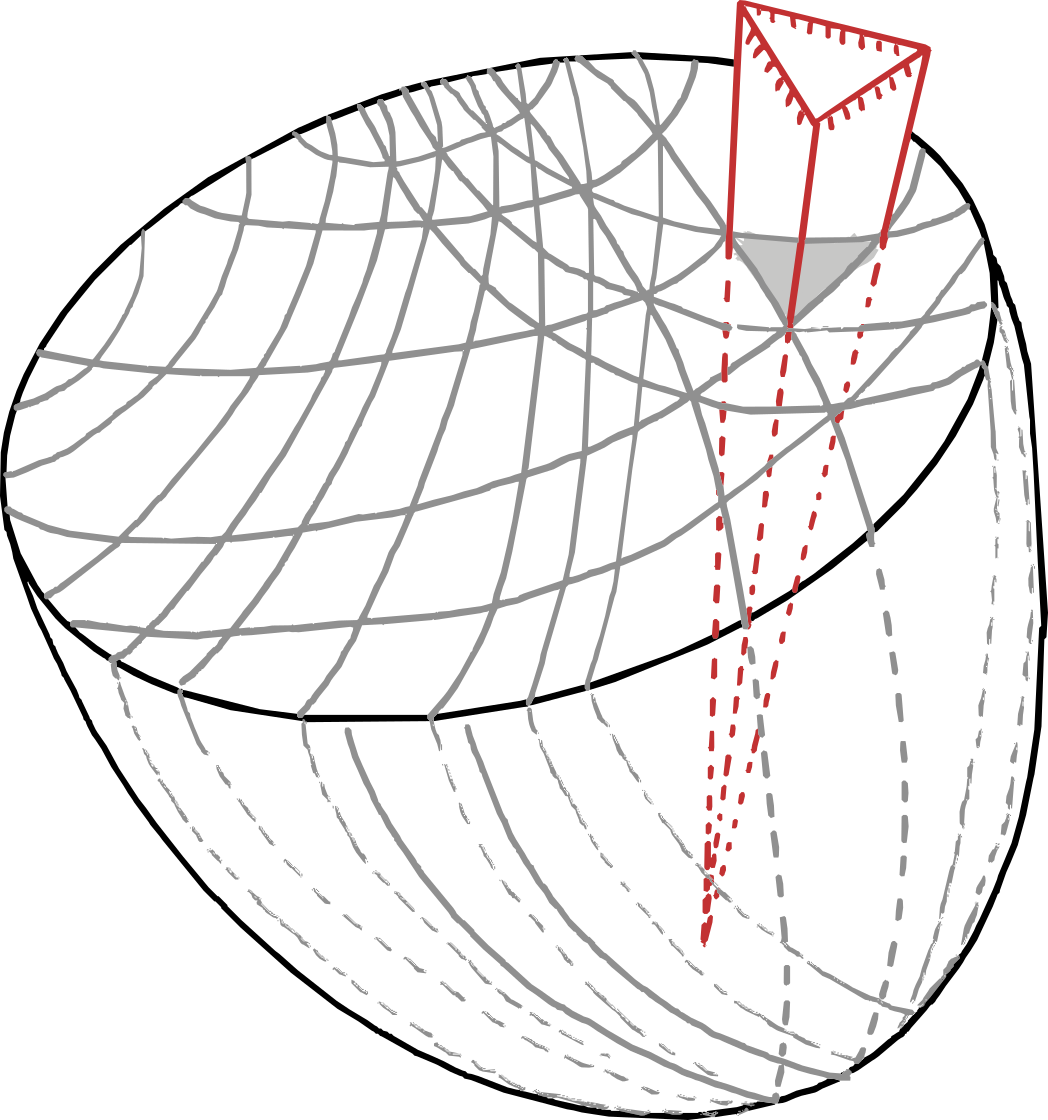}
    \end{minipage}

    \caption{Cartoon illustration of mirrors intersecting a sphere ($\mathbb{S}^2$ embedded into $\mathbb{E}^3$), a plane (thought of as the \textit{affine space} $\mathbb{A}^2$ of \textit{no origin} within the \textit{vector space} $\mathbb{E}^3$ \textit{with} an origin) and a hyperboloid (the constant-timelike-separation upper-hyperboloid $\mathbb{H}^2_+$ embedded into $\MM^3$). The iconic figure on the left is from \cite{conway2013sphere} and inspired the rest; the hyperboloid tessellation is not meant to be precise. The Euclidean triangles on the affine plane are not getting bigger and bigger as we move towards the boundaries of the plane as one might think from the illustration that has the origin of $\RR^3$ at a finite distance from $\mathbb{A}^2$; after all, the plane is infinite and the areas are calculated with an induced 2-dimensional metric and not with the metric in $\RR^3$. The hyperbolic triangles on $\HH^2_+$ \textit{are} getting smaller and smaller as one approaches the boundary of $\HH^2_+$, however, their area is all the same because of the non-trivial hyperbolic metric \cite{vinberg1993geometry}.}
\end{subfigure}
\caption{Examples of Coxeter complexes and related geometric structures.}
\label{fig:SphEucHyp}
\end{figure}

\subsection{Spherical, Affine and Hyperbolic Coxeter Groups}
\label{subsec:SphAffHypCoxeterGroups}

From a geometrical perspective, we can think of the spherical, affine, and hyperbolic Coxeter groups as reflection groups in the  $k$-dimensional maximally-symmetric space $\mathbb{Y}^k$ of positive, zero or negative curvature, respectively -- {\it i.e.}\ the $k$-dimensional sphere $\mathbb{S}^k$ (left), $k$-dimensional Euclidean space $\mathbb{E}^k$ (middle), or $k$-dimensional hyperbolic space $\mathbb{H}^k$ (right) respectively in Fig.~\ref{fig:SphEucHyp}. For spherical and affine Coxeter groups, we have $k=n-1$. In the spherical case, we can think of the $k$-dimensional sphere $\mathbb{S}^{k}$ as embedded as the surface $$(X^0)^{2}+(X^1)^{2}+\ldots+(X^k)^2=1$$ in Euclidean space $\mathbb{E}^{k+1}$; and the mirrors as $k$-dimensional hyperplanes passing through the origin of $\mathbb{E}^{k+1}$ and intersecting the $\mathbb{S}^k$ along ``great'' $(k-1)$-dimensional spheres $\mathbb{S}^{k-1}$, so that the reflections across these hyperplanes in $\mathbb{E}^{k+1}$ naturally induce corresponding reflections across the $(k-1)$-spheres $\mathbb{S}^{k-1}$ in $\Sph^{k}$. For hyperbolic Coxeter groups, we have $k=p$.  We can think of the $k$-dimensional hyperbolic space $\HH^k_+$ as the upper ($X^0>0$) sheet of the hyperboloid $$-(X^0)^{2}+(X^1)^{2}+\ldots+(X^k)^2=-1$$ in Minkowski space $\MM^{k+1}\simeq\RR^{k,1}$; and the mirrors as $n$-dimensional hyperplanes passing through the origin of $\MM^{k+1}$ and intersecting the $\HH^k_+$ along $(k-1)$-dimensional hyperbolic spaces $\HH^{k-1}_+$, so that the reflections across these hyperplanes in $\MM^{k+1}$ naturally induce corresponding reflections across the $\HH^{k-1}_+$ in $\HH^{k}_+$ (see Chapter \ref{ch:QuasicrystalsMinkowski}).

If we draw all the mirrors in a given reflection group $G$, they divide up the space $\mathbb{Y}^k$ into a set of congruent convex {\bf fundamental domains}, which together form a tessellation $T$ of $\mathbb{Y}^k$. We also say that $G$ tiles $\mathbb{Y}^k$ by reflections. If we pick one such fundamental domain (or {\bf Coxeter polytope}) $\Gamma$, the fundamental roots are the outward pointing normal vectors to its bounding mirror ``faces'' (or the hyperplanes defining them). So the corresponding Coxeter diagram specifies the shape of the fundamental domain $\Gamma$: the number of its faces, and the angle at which each pair of faces meets (see Fig.~\ref{fig:SphEucHyp}). The symmetries of the Coxeter diagram reflect symmetries of the Coxeter polytope itself: {\it e.g.}\ the $\mathbb{Z}_{2}$ symmetry of the $A_{n}$, $D_{n}$ and $E_{6}$ graphs, or the larger $S_{3}$ symmetry of the $D_{4}$ graph which gives rise to the famous {triality symmetry} of the $\mathfrak{so}(8)$ Lie algebra (see Section \ref{sec:Symmetries_of_Lattices} and Appendix \ref{appendix:FoldingUnfolding} for triality). 
Given a choice of fundamental domain $\Gamma$, successive reflections through its faces can map it to any other copy of $\Gamma$ in the tessellation $T$ of $\mathbb{Y}^k$, so that the ``tiles'' in $T$ are in one-to-one correspondence with the elements in the group $G$. 

The \textbf{spherical Coxeter groups} have been completely classified (see the left panel of Fig.~\ref{fig:Irreducible_Root_Systems}), and, moreover, Coxeter showed that these are precisely the {\it finite} Coxeter groups. They capture symmetries of regular polytopes in flat Euclidean space \cite{coxeter1973regular} and their finiteness manifests itself in the fact that the sphere they tessellate by reflections is chopped into a finite number of tiles, each of finite volume (see Fig.~\ref{fig:SphEucHyp}). 

To get some intuition about them, we introduce at this point the \textbf{Schl\"{a}fli symbol}, a shorthand for representing regular polytopes or regular tessellations by polytopes: $\{ p \}$ denotes a (convex) regular $p$-gon in 2-dimensional space;\footnote{Note, these $p$'s and $q$'s in the Schl\"{a}fli notation $\{ p,q.\dots\}$ have nothing to do with the signature of the bilinear forms we introduced earlier below Eq.~\eqref{eq:SchlafliMatrix}.} $\{ 3 \}, \{4\}$ and $\{ 6\}$, for example, would be a triangle (2-simplex), a square and a hexagon respectively. $\{ p,q\}$ represents a regular polyhedron or tessellation in which $q$-many regular $p$-gons meet at each vertex; $\{ 4,3\}$, for example, would be a cube (3 squares meet at each vertex), while $\{ 4,4\}$ would be a tiling of $\EE^2$ by squares. $\{ p,q,r\}$ would be a pattern where $r$-many shapes with Schl\"{a}fli symbol $\{ p,q\}$ meet around each \textit{edge}; $\{ 4,3,4 \}$, for example, would be the regular honeycomb of $\EE^3$ by cubes (4 cubes meet at each edge). 

The Schl\"{a}fli labels $\{ p,q,r,\dots,s\}$ become particularly illuminating when we are dealing with a Coxeter-Dynkin graph of the simplest form, 
\begin{equation*}
\begin{tikzpicture}[
    scale=0.7,
    transform shape,
    vertex/.style={
        circle,
        draw,
        fill=white,
        minimum size=5mm,
        inner sep=0pt
    }
]

\node[vertex] (v1) at (0,0) {};
\node[vertex] (v2) at (2,0) {};
\node[vertex] (v3) at (4,0) {};
\node[vertex] (v4) at (6,0) {};

\draw (v1) -- node[midway,above] {$p$} (v2);
\draw (v2) -- node[midway,above] {$q$} (v3);
\draw (v3) -- node[midway,above] {$r$} (v4);

\node at (8,0) {$\cdots$};

\node[vertex] (v5) at (10,0) {};
\node[vertex] (v6) at (12,0) {};

\draw (v5) -- node[midway,above] {$s$} (v6);

\end{tikzpicture}
\end{equation*}
because then they are the Coxeter exponents $m_{ij}$ appearing in Eq.~\eqref{eq:CoxeterRelations}. In these cases, we can see that the infinite families $A_n$, $B_n$, and $C_n$ of Coxeter groups in Fig.~\ref{fig:Irreducible_Root_Systems} describe symmetries of an $n$-simplex $\alpha_n$ ($\{ 3,3, \dots 3\}$; $n-1$ entries, all equal to $3$), of the $n$-cube $\gamma_n$ ($\{ 4,3,\dots,3\}$; $n-2$ entries of 3's) and of the \textit{$n$-cross-polytope} $\beta_n$ ($\{ 3 ,\dots, 3,4\}$; the ``reverse'' $\gamma_n$)\footnote{This last observation is not coincidental. The polytopes $\gamma_n$ (the hypercube) and $\beta_n$ (the cross-polytope, orthoplex or hyperoctahedron) form a pair of dual polytopes, in a sense that (centres of) faces of one corresponds to vertices of another \cite{coxeter1973regular}. If one polytope has Schl\"{a}fli symbol equal to $\{p,q, \dots s\}$, then its dual has the reverse Schl\"{a}fli symbol $\{ s , \dots , q,p\}$. If the Schl\"{a}fli symbol is palindromic, the polytope is \textit{self-dual}. Dual polytopes have the same reflection symmetries. A familiar example is in three dimensions: the cross-polytope is the octahedron (6 vertices, 8 faces), which is dual to the cube (8 vertices, 6 faces).} respectively. Similarly, from the ``exceptional'' cases (the ones that do not fit in any of the infinite families), $G_2$ stabilizes a regular hexagon $\{ 6\}$ and $F_4$ fixes the so-called 24-cell $\{ 3,4,3\}$, the unique self-dual regular 4-dimensional polytope in Euclidean space. The rest of the cases in Fig.~\ref{fig:Irreducible_Root_Systems} whose Coxeter-Dynkin graph has loops or more branches have complicated Schl\"{a}fli symbols and we omit them from this discussion, but the interested reader can find them in \cite{coxeter1973regular}; we will come back to the non-crystallographic boxed cases in the next subsection.

The \textbf{affine Coxeter groups} have also been completely classified (see the right panel of Fig.~\ref{fig:Irreducible_Root_Systems}). While the spherical cases describe symmetries of regular polytopes in Euclidean space, the affine ones are closely related to patterns in Euclidean spaces whose reflection symmetry is that of those polytopes. Consequently, each affine Coxeter group will contain a spherical Coxeter group as a subgroup, the one that describes the symmetries of the underlying polytope. Contrary to the spherical ones, the affine groups are \textit{infinite}: they chop an infinite hyperplane into infinitely many finite-volume fundamental domains $\Gamma$, as shown in Fig.~\ref{fig:SphEucHyp}. This infiniteness can be understood intuitively as follows: a copy of the underlying finite Coxeter group sits over each of the infinitely many points of the Euclidean pattern; this pattern is a lattice and the statement will be made more precise in the next subsection, once we introduce the idea of a crystallographic Coxeter group. To remind ourselves of this connection, the affine groups are denoted with the same letter with a tilde above: for example, $\Tilde{A}_n$ is the affine group that describes a pattern with reflection symmetry $A_{n-1}$ over each point. 

To see some of the regular tessellations, we appeal again to the Schl\"{a}fli symbols. In $\EE^2$, the sum of internal angles of $\{ p \}$ is $(p-2)\pi$ and a single angle is $(1-2/p)\pi$, so to fit $q$ such $p$-gons around a single vertex to form a tiling $\{ p,q\}$ \textit{without leaving any gaps or overlaps} we need $q(1-2/p)\pi = 2\pi$; from this, it follows that $(1/p) + (1/q) = 1/2$.\footnote{Alternatively, one could arrive at this expression by constructing the Schl\"{a}fli matrix \eqref{eq:SchlafliMatrix} for the 3-node graph with Coxeter labels $m_{12} = m_{21} = p$, $m_{23}=m_{32} = q$ and $m_{13} = m_{31} = 2$, and setting its determinant to zero, since it is an affine graph.} The only integer solutions $\{ p,q\}$ to this equation are: $\{ 4,4\}$ (square tiling), $\{3,6\}$ (triangular tiling) and $\{6,3\}$ (hexagonal tiling). These are captured by the affine graphs $\widetilde{B}_2 = \widetilde{C}_2$ (`extending' the square $B_2$) and $\widetilde{G}_2$ (`extending' the hexagon) respectively in Fig.~\ref{fig:Irreducible_Root_Systems}. Similarly, in $\EE^3$, the same condition for a tessellation $\{ p,q,r \}$ now reads $(1/p) + (1/q) + (1/r) = 1$, whose only integer solution is: $\{4,3,4\}$ (the cubic honeycomb); this is represented by the affine group $\widetilde{C}_3$ (`extending' the cube). In $\EE^4$, something sort of `miraculous' happens\footnote{The two regular polytopes $\beta_n$ and $\gamma_n$ have, in general, $2n$ and $2^n$ vertices respectively. $\gamma_n$ can be decomposed into two shapes each with $2^{n-1}$ vertices: in 3D, these shapes are tetrahedra, but in higher dimensions are more complicated; they do, however, form representations of $D_n$. Only in $n=4$, $2 \cdot 4 = 8$ and $2^{4-1} = 8$; these three shapes, each with 8 vertices, are permuted by the triality symmetry of $D_4$. Putting all of them together, we get the 24-cell; using also its dual polytope, we get the full $F_4$-root system of 48 roots, that is comprised of a 24-cell and its dual, a smaller 24-cell (hence the two shorter roots of $F_4$).} and there are exactly three regular honeycombs: one by the tesseract $\gamma_4$, or 4-cube, $\{ 4,3,3,4\}$, an \textit{exceptional} tiling by the 24-cell $\{ 3, 3, 4 ,3\}$, and one by the 16-cell $\beta_4$, or 4-cross-polytope, $\{ 3,4,3,3\}$. The corresponding affine graphs describing these are $\widetilde{C}_4$ and $\widetilde{F}_4$. In $\EE^{n\geq 5}$, though, there is again a unique regular honeycomb: the hypercubic one $\widetilde{C}_n$ \cite{Lyndon_1985}. 

The rest of the affine cases do not give rise to regular tessellations of $\EE^n$, but more generally, to \textit{Voronoi tessellations} \cite{conway2013sphere}. These are tessellations by polytopes that might not be regular, but they still exhibit all the reflection symmetries of the underlying spherical Coxeter groups. For example, $\widetilde{B}_3$ gives a tessellation of $\EE^3$ by the \textit{rhombic dodecahedron}. This polytope is not regular, but it can be built from taking the convex hull of a cube and its dual, the octahedron, and thus enjoys octahedral $B_3$ symmetry. Ultimately, each such Voronoi cell can be built from simplices. 

Note that this reasoning can be extended to positively and negatively curved spaces to find the allowed regular tessellations. One simply needs to notice that the sum of the internal angles will be larger or smaller than $2\pi$ in spherical and hyperbolic spaces respectively, as demanded by the Gauss-Bonnet theorem (see \cite{Boyle:2024qzn} and references therein). Doing so, we end up with the following criteria: 
\begin{equation}
\textbf{2D: }    
\frac{2\pi}{p} + \frac{2\pi}{q} \, 
    \left\{\begin{aligned}
        & > \pi & \text{ tessellates } \mathbb{S}^2 \\
        & = \pi & \text{ tessellates } \EE^2 \\
        & < \pi & \text{ tessellates } \HH^2_+
    \end{aligned}\right. \qquad
    \textbf{3D: } \frac{2\pi}{p} + \frac{2\pi}{q} + \frac{2\pi}{r} \, 
    \left\{\begin{aligned}
        & > 2\pi & \text{ tessellates } \mathbb{S}^3 \\
        & = 2\pi & \text{ tessellates } \EE^3 \\
        & < 2\pi & \text{ tessellates } \HH^3_+
    \end{aligned}\right.
\end{equation}
We can check that in 2 dimensions, there are only five regular tilings of $\mathbb{S}^2$ corresponding to (the projections of, in the sense of Fig.~\ref{fig:SphEucHyp}) the five Platonic solids: tetrahedron $\{3, 3\}$, octahedron
$\{3, 4\}$ and its dual cube $\{4, 3\}$, and icosahedron $\{{3, 5}\}$ and its dual dodecahedron
$\{5, 3\}$. These correspond to the spherical Coxeter groups $A_3$, $B_3/C_3$ and $H_3$. The possible tessellations of $\HH^2_+$, on the other hand, are infinite, as we can arbitrarily increase $p$ and $q$ to squeeze an arbitrarily large number $q$ of $\{ p \}$'s with arbitrarily small internal angles around each vertex.

The \textbf{hyperbolic Coxeter groups} have not been classified in the most general case \cite{guglielmetti2017commensurability} and it still a topic of active research \cite{belolipetsky2016arithmetichyperbolicreflectiongroups}, but some partial classifications are known (see \cite{allcock2010reflectivelorentzianlatticesrank,allcock2021alternativevinbergsalgorithm,Bogachev_2019}; a nice website that contains many more useful references on these topics can be found here \cite{Felikson2026HyperbolicCoxeterPolytopes}). 

Nonetheless, the hyperbolic reflection groups whose fundamental domains are simplices {\it have} been classified (see Secs.~6.8 \& 6.9 and Figs.~2 \& 3 in \cite{humphreys_1990}) and exist only in Minkowski spacetime $\MM^d$ of dimension $d \leq 10$. 
Since a $n$-simplex has precisely $n+1$ vertices and, given that the vertices of any Coxeter polytope $\Gamma$ are defined by the intersections of the bounding reflecting mirrors (the facets of $\Gamma$), the primary condition for $\Gamma$ to be a simplex is that its Coxeter-Dynkin graph must have exactly $n+1$ nodes (i.e.~simple roots). All graphs we encountered so far fall in this category, but Vinberg \cite{vinberg1967discrete,vinberg1972groups,vinberg1975some,kaplinskaja1978groups} found additional hyperbolic reflection groups with non-simplicial fundamental domains -- and thus more simple roots than the dimension of the space those groups live in -- by studying the reflection symmetries of Lorentzian lattices, which we will encounter in Section \ref{ch:QuasicrystalsMinkowski} and Appendix \ref{app:SymmetriesLorentzianLattices}. 

Hyperbolic groups are infinite as well: they naturally tessellate the infinite-volume hyperboloid $\HH^{d-1}_+$ and chop it into an infinite number of finite-volume fundamental domains as seen in Fig.~\ref{fig:SphEucHyp}. The terminology around hyperbolic Coxeter groups is usually author-dependent and can lead to confusion. For this reason, we give here a further classification based on the finiteness of the volume of the Coxeter polyhedron\footnote{We refer to the fundamental domain as \textit{polyhedron} instead of a \textit{polytope} when it can be unbounded (i.e.~not enclosed in all sides).}. 

For the fundamental domain to be a hyperbolic simplex, we need two extra things: (1) the Sch\"{a}fli matrix \eqref{eq:SchlafliMatrix} must imply a Lorentzian geometry, that is, it must have signature $(d-1,1)$ in $d$-dimensions, so that the group acts naturally on $\HH^{d-1}_+ \subset \MM^d$ and (2) the vertices of the simplex must not ``spill out'' beyond the boundary of the hyperbolic space, that is, the region bounded by the mirrors must close up \textit{within} the boundary of hyperbolic space, $\partial \HH_+^{d-1}$. The last condition guarantees that the Coxeter polytope will have a finite volume in $\HH^{d-1}_+$ and it translates into the following rule that its Coxeter-Dynkin diagram must satisfy \cite{humphreys_1990}: every proper connected subdiagram must be either spherical (positive definite) or affine (positive semi-definite). Geometrically, this rule governs the types of vertices of the fundamental polyhedron in hyperbolic space and placed hyperbolic Coxeter groups into three categories:
\begin{figure}[t!]
\centering
    \begin{tabular}{cc}
        \subfloat[Ordinary Vertices 
        \label{fig:OrdinaryVertices}
        ]{
            \includegraphics[width=0.3\linewidth]{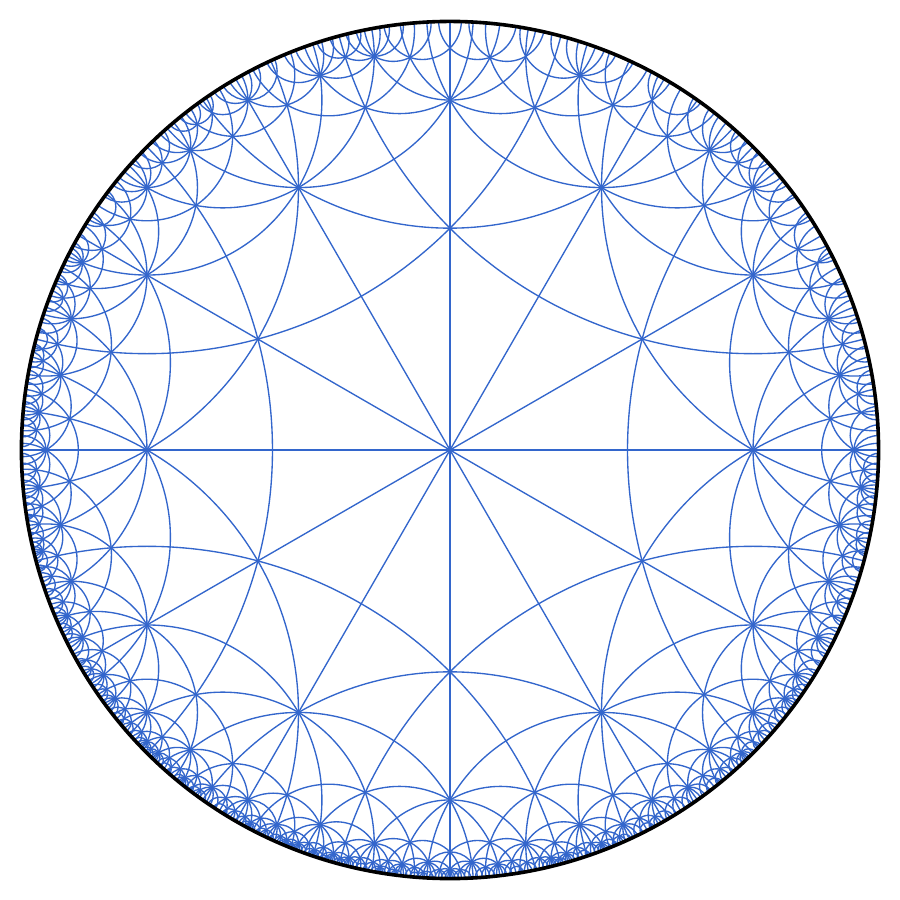}
            \hspace{1em}}
        \subfloat[Ideal Vertices
        \label{fig:IdealVertices}]{    
            \includegraphics[width=0.3\linewidth]{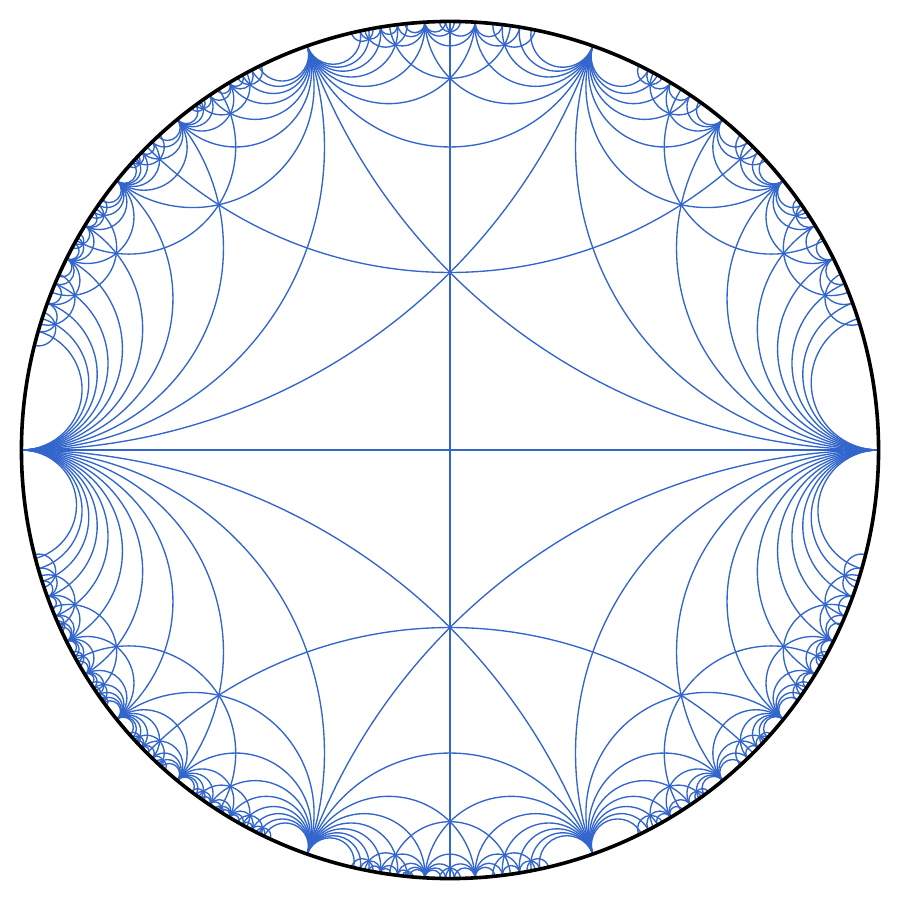}
        } \hspace{1em}
        \subfloat[Ultra-Ideal Vertices
        \label{fig:UltraIdealVertices}]{\includegraphics[width=0.3\linewidth]{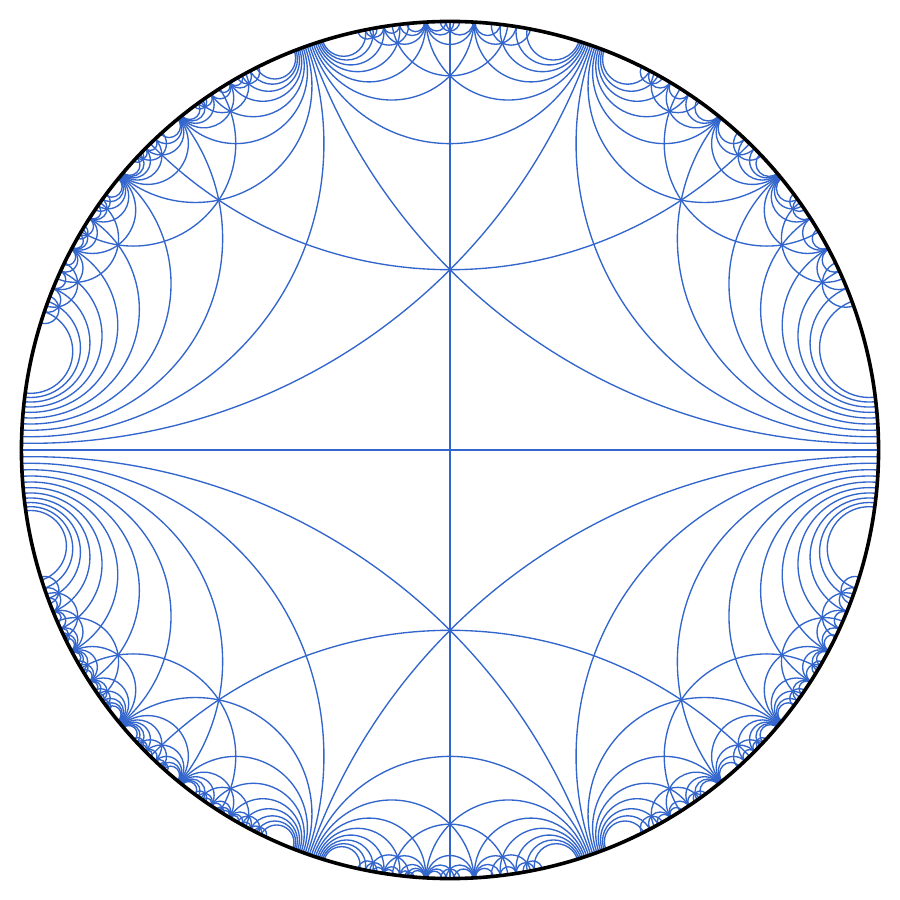}}
    \end{tabular}
\caption[Examples of ordinary, ideal and ultra-ideal vertices in hyperbolic tessellations of $\HH^2_+$]{Examples of ordinary, ideal and ultra-ideal vertices in hyperbolic tessellations of $\HH^2_+$. On the left, we show the compact hyperbolic group $[6,4]$ (the triangle group $[6,4,2]$) with only ordinary vertices. In the middle, we have the triangle group $[\infty,4,2]$ corresponding to the odd self-dual lattice $\mathrm{I}_{2,1}$ (see Sec.~\ref{sec:MathematicalBackgroundLorentzian} and Fig.~\ref{fig:VinbergRootsI1nUpto10}) that has an ideal vertex because of its affine subgraph $\widetilde{I}_1$. On the right, we give an example of an ultra-ideal vertex by considering the $B_2$ group with an ultra-parallel third side at distance $d = 0.15$. For each pattern, only a finite number of mirrors have been drawn, enough to cover the central region of $\HH^2_+$ but the region close to the boundary $\partial \HH^2_+$.}
\label{fig:OrdinaryIdealUltraIdealVertices}
\end{figure}
\begin{itemize}
    \item \textbf{Compact Hyperbolic}: if they only contain spherical subdiagrams \cite{vinberg1967discrete}; these correspond to \textit{ordinary vertices} that lie properly inside the interior of $\HH^{d-1}_+$ (see Fig.~\ref{fig:OrdinaryVertices}). All mirrors converge cleanly within $\HH^{d-1}_+$ giving polytopes of finite volume, and the mirror intersections occur at a timelike vectors in $\MM^d$. Compact hyperbolic Coxeter groups exist only in ranks 3,4 and 5 and have been classified by Lann\'{e}r \cite{Lannr1950OnCW}. 
    \item \textbf{Non-Compact Hyperbolic} (or just \textit{`Hyperbolic'}): if they contain both spherical and affine subdiagrams \cite{humphreys_1990}; the latter correspond to \textit{ideal vertices} that lie exactly on the boundary $\partial \HH^{d-1}_+$ (see Fig.~\ref{fig:IdealVertices}). In these cases, we can have mirrors that become parallel within $\HH^{d-1}_+$ and they only meet at infinity, giving non-compact, but finite-volume, polyhedra. Such mirrors intersect at a null vector in $\MM^d$, and the vertex sits exactly on the lightcone, hence the name ``ideal''. The non-compact hyperbolic Coxeter groups whose Coxeter polyhedron is a simplex are the ones given in \cite{humphreys_1990}.
    \item \textbf{Ultra Non-Compact Hyperbolic}: if they also contain hyperbolic subdiagrams; these correspond to \textit{ultra-ideal} vertices that exist outside of $\HH^{d-1}_+$ even beyond its boundary $\partial \HH_+^{d-1}$ (see Fig.~\ref{fig:UltraIdealVertices}). In these cases, there are mirrors that become ``ultra-parallel'' and literally diverge from one another within hyperbolic space. However, they are still flat planes in the ambient Minkowski space $\MM^d$, so if we extend those divergent planes outwards, they do eventually cross. But because they were diverging inside the lightcone, their mathematical point of intersection gets pushed completely outside the lightcone to a spacelike vector. Therefore, the ``vertex'' where those faces meet physically exists in the one-sheeted spacelike hyperboloid—the de Sitter region. Because the fundamental polyhedron has ultra-ideal vertices, it is geometrically ``truncated'' or ``open''. It never closes off, either in the interior or at the boundary of hyperbolic space, which means its volume is strictly infinite \cite{Felikson2026HyperbolicCoxeterPolytopes}.
\end{itemize}
Classifying infinite-volume hyperbolic Coxeter groups is not a well-posed mathematical classification problem because one can generate an infinite-volume hyperbolic graph simply by attaching a new node to any existing hyperbolic graph, which are themselves notoriously difficult to classify. We will encounter some examples in Chapter \ref{ch:GeneralizedSpacetimeCNP}. In Appendix \ref{app:hyperbolicmirrorintersections} we discuss how to understand ultra-ideal vertices. 

\subsection{Crystallographic vs Non-Crystallographic Coxeter Groups}
\label{subsec:CrystallographicVsNon}

Another very important distinction of Coxeter groups, in particular in terms of the theory of lattices and quasicrystals, is the following. A Coxeter group is {\bf crystallographic} if its geometric representation in $V=\mathbb{R}^{n}$ (the space spanned by the $n$ fundamental roots) can stabilize a lattice $\Lambda$ in $V$ ({\it i.e.}\ the $\mathbb{Z}$-span of some basis of $V$). Otherwise it is {\bf non-crystallographic}. A necessary and sufficient condition for a Coxeter group  to be crystallographic is that: (i) all of the edges in its Coxeter diagram are labelled by $2$, $3$, $4$, $6$, or $\infty$; and (ii) for each circuit (closed loop) in the Coxeter diagram, the number of edges labelled 4 (resp. 6) around the circuit is even (see Sec.~6.6 in \cite{humphreys_1990}). When either (i) or (ii) is not met, the group is non-crystallographic; in the remainder of this work, we take this to be the definition of a non-crystallographic group. 

This result is also related to the \textbf{crystallographic restriction theorem} \cite{coxeter1973regular,humphreys_1990,baake2013aperiodic}, which tell us that the numbers $A_{ij} = 2\langle {\bf r_i,r_j}\rangle/ \langle {\bf r_j, r_j} \rangle$ appearing in the reflection operations \eqref{eq:ReflectionMap} -- also known as the entries of the Cartan matrix $A$ for the simple roots $\{ \bf r_i\}$ and very much related to the entries of the Scl\"{a}fli matrix $S_{ij}$ (up to a norm-dependent scale) -- must be strictly integers. This restricts the possible angles between pairs of simple roots to only $2\pi/m_{ij}$ with $m_{ij} = 2,3,4,6, \infty$. The reason for this constraint is that if one takes the $\ZZ$-linear combination of the simple roots to define the \textit{root lattice} $\Lambda_{\rm root} = \mathrm{span}_\ZZ \big( \{ \bf r_i\}\big)$, then this will be invariant under the full Coxeter group \eqref{eq:CoxeterRelations}, as any reflection \eqref{eq:ReflectionMap} will map a simple root to another $\ZZ$-linear combination of simple roots. This is shown in more detail in \S 6.6 in \cite{humphreys_1990}, where the lattice preserved by the Coxeter group is constructed explicitly. The important point here is that this integrality\footnote{We take the word `integrality' to mean `the property of being integer-valued'. The term is more common in ring theory and algebraic number theory (i.e.~see Eq.~\eqref{eq:integral_basis_element}).} condition restricts the type of orientational symmetries a periodic pattern can exhibit to only $2-,3-,4-$ or $6-$fold rotational invariance.       

The {\it crystallographic} finite/spherical groups are closely related to affine groups: each affine diagram may be obtained by adding a single ``extending node'' to one of the crystallographic finite groups (see Fig.~\ref{fig:Irreducible_Root_Systems}).\footnote{This node corresponds to adding one more \textit{affine mirror}, which is translated and does not pass through the origin, and considering this additional affine reflection together with those already in the spherical Coxeter group. This node is the root that belongs to the 1-dimensional nullspace of the positive semi-definite bilinear form $B$ (the eigenvector corresponding to its zero eigenvalue). The method is explained more in 6.5 in \cite{humphreys_1990} and is related to the highest root construction in Lie algebras (see \S 13/14 in \cite{francesco2012conformal}).} The affine diagrams (with ${\rm sig}(S) = (n-1,0)$) describe the full reflection symmetries of some periodic pattern in Euclidean space $\mathbb{R}^{n-1}$, while the {\it crystallographic} finite/spherical diagrams arise as the \textbf{point subgroups} of those periodic patterns that fix a point where $n-1$ of the mirrors cross. Mathematically, we write $\widetilde{G} = G_{\rm sph}^{({\bf 0})} \ltimes \ZZ^{n-1}$ \cite{mcmullen2002coxeter,humphreys_1990}, where $G_{\rm sph}^{({\bf 0})}$ is the spherical Coxeter subgroup that fixes the origin $\bf 0$ in $V$, which is now translated over every lattice point in $\ZZ^{n-1}$.\footnote{$\widetilde{G}$ is just the \textit{affine extension} of a spherical Coxeter group, similarly to how the full group of isometries of Minkowski space $\MM^{d}$ is not just the Lorentz group $SO(d-1,1)$ but the Poincar\'{e} group that includes the translations, yielding the full group of isometric transformations ${\rm Isom}(\MM^d) = SO(d-1,1) \ltimes \RR^{d-1,1}$. Here, $\ltimes$ is the \textit{semi-direct} product, which is a generalization of the \textit{direct product}. At the set-level, the latter is just the Cartesian product $\times$. At the group-level, the semi-direct product $\ltimes$ uses the Cartesian product to build a bigger group $H$ from two of its subgroups $K$ and $N$, where $N$ is a \textit{normal subgroup} of $H$; we write $N \vartriangleleft H$. The symbol $\ltimes$ (or $\rtimes$)  always points towards the normal subgroup, which defines the quotient group $H/N$, to which $K$ is isomorphic; see \href{https://en.wikipedia.org/wiki/Semidirect_product}{this Wikipedia article} for more info.} Pictorially, one can think of this as follows: in Fig.~\ref{fig:SphEucHyp} (bottom left), we draw all the mirrors passing through the origin that tessellate a sphere centered there; now, one must go over every other lattice point too and draw all \textit{those} mirrors as well, as in the affine plane $\mathbb{A}^{n-1}$ (Fig.~\ref{fig:SphEucHyp}, bottom middle) there is no preferred origin. This way, we get the full translational and rotational symmetries of the periodic lattice, which is an infinite group but only because we had to draw infinitely many mirrors over infinitely many lattice points. Fixing any point on the pattern and calling it the origin is equivalent to taking the quotient $\widetilde{G}\backslash\ZZ^{n-1}$, which yields just the point group $G_{\rm sph}^{({\bf 0 })}$. 

Finally, we note that the crystallographic finite  Coxeter groups (and their associated Coxeter-Dynkin diagrams) also play a fundamental role in the classification of the finite-dimensional semi-simple Lie algebras (and their associated Lie groups) \cite{Humphreys1973}. Thus, the list of finite crystallographic Coxeter-Dynkin diagrams in Fig.~\ref{fig:Irreducible_Root_Systems} (namely $A_{n}$, $B_{n}$, $C_{n}$, $D_{n}$, $E_{6}$, $E_{7}$, $E_{8}$, $F_{4}$ and $G_{2}$) is also famous as the list of simple finite-dimensional Lie algebras. More generally, crystallographic Coxeter-Dynkin diagrams (including affine, hyperbolic and higher-rank diagrams) play an analogous role in the infinite Kac-Moody algebras \cite{MOODY1968211,kac1968simple,kac1990infinite}. Their general classification is still an open problem, but the \textit{``hyperbolic''} ones (not ultra non-compact) \textit{have} been classified in \cite{Carbone_2010}; they exist up to dimension 10 again and the nicest one is $E_{10}$, which has been shown to contain all the other ones as its subalgebras \cite{Felikson_2011}. The ultra non-compact ones fall under the general umbrella of \textit{Lorentzian} Kac-Moody algebras, for which there are several classes and classifications proposed \cite{nikulin1998theorylorentziankacmoodyalgebras, Gritsenko_2002,GABERDIEL2002403,Felikson_2008}; an example is the affine $E_{10}$, known as $E_{11}$. Interestingly, both $E_{10}$ and $E_{11}$ have been highly studied in high energy theory, string theory, cosmological singularities and the standard model of particle physics \cite{damour2002cosmologicalbilliards,DAMOUR_2006,Damour:2000hv,Damour:2002cu,Damour:2002et,Kleinschmidt:2015sfa, Meissner:2025qrx}.

Kac-Moody algebras are defined, similarly to their finite cousins the Lie algebras, by generators and relations -- the \textit{Chevalley-Serre} relations \cite{kac1990infinite} -- starting with a ``generalized Cartan matrix $A$'' with integer entries subject to the conditions that all $A_{ii} = +2$, $A_{ij} \leq 0$ and if $A_{ij} = 0$, then $A_{ji} = 0$ ($A_{ij}$ need not be symmetric, but \textit{symmetrizable}). This algebra has a \textit{root system}, a very symmetric configuration of vectors that is left invariant under the so-called \textit{Weyl group}; this also stabilizes the root lattice and turns out to be a Coxeter group. The relation is a follows (keeping the same labels): $m_{ij} = 2,3,4,6,\infty$ when $A_{ij}A_{ji} = 0,1,2,3, \geq 4$ respectively. The integrality condition is necessary for the definition of the algebras: the defining relations include nested commutators whose number is given precisely by the Cartan integers. In this sense, the difference between Lie and Kac-Moody algebras is that the commutators produce a finite or an infinite number of new generators, corresponding to the finite or infinite number of roots in their root systems. 

By contrast, the {\it non-crystallographic} finite/spherical groups are those that {\it do not} arise as the point groups within a larger periodic pattern and do not satisfy the integrality condition. From Fig.~\ref{fig:Irreducible_Root_Systems}, the only finite irreducible non-crystallographic Coxeter groups occur in two dimensions ($I_{2}^{n}$, the symmetries of a regular $n$-gon $\{ n\}$, with $n\neq 2,3,4,6$), three dimensions ($H_{3}$, the symmetries of the icosahedron $\{3,5\}$, or its dual, the dodecahedron $\{ 5,3 \}$) or four dimensions ($H_{4}$, the symmetries of the 600-cell \{3,3,5\} and its dual, the 120-cell $\{5,3,3\}$, the four-dimensional analogues of the icosahedron and the dodecahedron respectively). 

To see why these rank-$n$ Coxeter groups cannot stabilize a lattice in $\EE^n$, consider constructing the root lattice of $I_2^5$ as the integer span of its simple roots whose relative angle is $\pi/5$. Its Schl\"{a}fli matrix is 
$$ S_{I_{2}^5} = 
\begin{pmatrix}
2 & -\phi \\
-\phi & 2
\end{pmatrix}
$$
where $\phi = \tfrac{1}{2}(1+\sqrt{5})$ is the \textbf{golden ratio}. We can ask: ``is $\Lambda_{\rm root} = \mathrm{span}_\ZZ \big( {\bf r}_1, {\bf r_2} \big)$ invariant under $I_2^5$?''. Using the reflections \eqref{eq:ReflectionInRootBasis}, we can see that the new root we obtain is \textit{not} in the lattice,  
${\bf r}_k = R_i ({\bf r}_j) = {\bf r}_j + \phi {\bf r}_i  \notin \Lambda_{\rm root}$ because $\phi$ is an irrational number. Therefore, even though $\Lambda_{\rm root}$ is an ordinary lattice in $\EE^2$, it is not symmetric under the reflection group $I_2^5$. To make it symmetric, we will need to consider the integer span of all ten roots of $I_2^5$ instead, $ \Lambda_{\rm all \, roots} = \mathrm{span}_\ZZ \big( {\bf r}_1, {\bf r}_2, \dots , {\bf r}_{10} \big)$, which is closed under the action of the reflections\footnote{Note that for a crystallographic lattice, the two lattices $\Lambda_{\rm root}$ and $\Lambda_{\rm all \, roots}$ considered here are the same; this is because all roots are given as a $\ZZ$-linear combination of the simple roots for these lattices.}; but this will lead to a \textit{dense} set of points in $\EE^2$ that now includes these roots like ${\bf r}_k$ that are given as an irrational linear combinations of the simple roots. Thus, at the cost of making it symmetric under the reflection group, we ended up with a non-lattice structure according to Eq.~\eqref{eq:LatticeDefinition}. Note that the ten roots of $I_2^5$ point to the vertices of the regular decagon $\{ 10\}$ in the plane. The fact that we cannot make a lattice out of $I_2^5$ is a manifestation of the fact that we cannot tile the plane $\EE^2$ with decagons or pentagons. Similarly, all other regular polygons $\{ n \}$ with $n\neq 2,3,4,6$ do not have the tiling property, and the same intuition applies to the higher-dimensional non-crystallographic cases as well. 

At this point, it will be useful to introduce two different concepts of the ``rank'' of a Coxeter group; definitions are from \cite{Boyle_2022}. So far, we have been referring to the \textbf{ordinary rank} of a Coxeter group $G$ as the number of its fundamental roots. Equivalently, this is also the dimension of the vector space $V$ (over $\RR$) on which $G$ has a geometric (reflection) representation. Using the definitions of the previous paragraph, we can write that $V = \Lambda_{\rm root} \otimes \RR = \mathrm{span}_\RR \{ \mathbf{r}_i\}_{i=1,\dots,n} = \RR^n$. As a result, the ordinary rank $n$ of $G$, then, is the dimension of the \textit{real} vector space $V$ generated by taking all \textit{real} linear combinations of roots of $G$ (over $\RR$, $\Lambda_{\rm root}$ and $\Lambda_{\rm all \, roots}$ are obviously the same sets, equal to the the full $V$). Over $\QQ$, though, things change: we refer to the \textbf{rational rank} $n_{\QQ}$ of $G$ as the dimension of the vector space (over $\QQ$) generated by all \textit{rational} linear combinations of roots. We write $V_\QQ = \Lambda_{\rm all \, roots} \otimes \QQ = \QQ^{n_\QQ}$. When $G$ is crystallographic, $n=n_\QQ$ always; in the non-crystallographic case, $n_\QQ > n$ always. For the example of $I_2^5$ we just saw, the ordinary rank is $n=2$, but its rational rank is $n_\QQ = 4$, as it takes two rational numbers to express the golden ratio $\phi$. Thus, as mentioned poetically in \cite{Boyle_2022}, ``a non-crystallographic Coxeter group lives a double life: in one sense, its roots live in the lower-dimensional space $\RR^n$; but in another sense, they live in the higher-dimensional space $\QQ^{n_\QQ}$.''

Remarkably, we can make a different structure that \textit{does} have one of the forbidden (for lattices) non-crystallographic symmetries at the cost of losing translational symmetry (i.e. periodicity). These non-crystallographic groups play an important role in the theory of quasi-periodic tilings and quasicrystals \cite{baake2013aperiodic}, where they govern the orientational symmetries of such structures: {\it e.g.}\ the 2D Penrose tiling (with decagonal orientational symmetry $I_{2}^{10}$) \cite{penrose1974role, gardner1997penrose}, the 2D Ammann-Beenker tiling (with octagonal symmetry $I_{2}^{8}$) \cite{ammann1992aperiodic, beenker1982algebraic}, various 3D quasicrystals and tilings (with icosahedral symmetry $H_{3}$) \cite{baake2013aperiodic}, and the 4D Elser-Sloane quasicrystal (with symmetry $H_{4}$) \cite{ElserSloane4D}. We will see several examples of quasicrystals and their symmetries in the next chapters of this thesis.

\subsection{The Spherical Coxeter Pairs through Folding}
\label{subsec:SphCoxPairsIntro}

As we will see shortly in Chapter \ref{ch:QuasicrystalsEuclidean}, quasicrystals can ultimately be obtained from appropriate lattices by a delicate projection method in which they inherit the lattice's order but not their periodicity. Their non-crystallographic symmetry, which by itself cannot exist in any periodic pattern, turns out to be a subgroup of a crystallographic group that stabilizes a perfectly periodic lattice in higher dimensions. This is an example of a \textbf{Coxeter pair}: a pairing of two Coxeter groups out of which the parent group is higher-dimensional and crystallographic, while the daughter group is lower-dimensional and non-crystallographic. 

For the spherical Coxeter groups, all the possible Coxeter pairs have been identified in 
\cite{Boyle_2022}. The most interesting cases are the fivefold symmetric ones related to the golden ratio $\phi$ (also called the \textit{fivefold hierarchy} \cite{baake2013aperiodic}), which are the 2D Penrose, the 3D icosahedral and the 4D Else-Sloane quasicrystals associated with the groups $I_2^5 = H_2$, $H_3$ and $H_4$ respectively. These can be found as subgroups of the crystallographic groups $A_4$, $D_6$ and $E_8$ respectively. Two more cases are the $(B_4,I_2^8)$ pairing of octagonal symmetry and the $(F_4,I_2^{12})$ of dodecagonal symmetry. The fact that the parent groups exists in twice the dimensions of the daughter groups is no coincidence and will be explained in more detail in Appendix \ref{appendix:FoldingUnfolding}; the reason is a deep connection between algebraic number theory, geometry and aperiodic order, further explored in Appendix \ref{AppendixA-NumberTheory}. Ultimately, the rational rank $n_\QQ$ of the non-crystallographic group always coincides with the ordinary rank $n$ of the crystallographic group.

\begin{figure}[!t]
\centering
\includegraphics[width=0.55\linewidth]{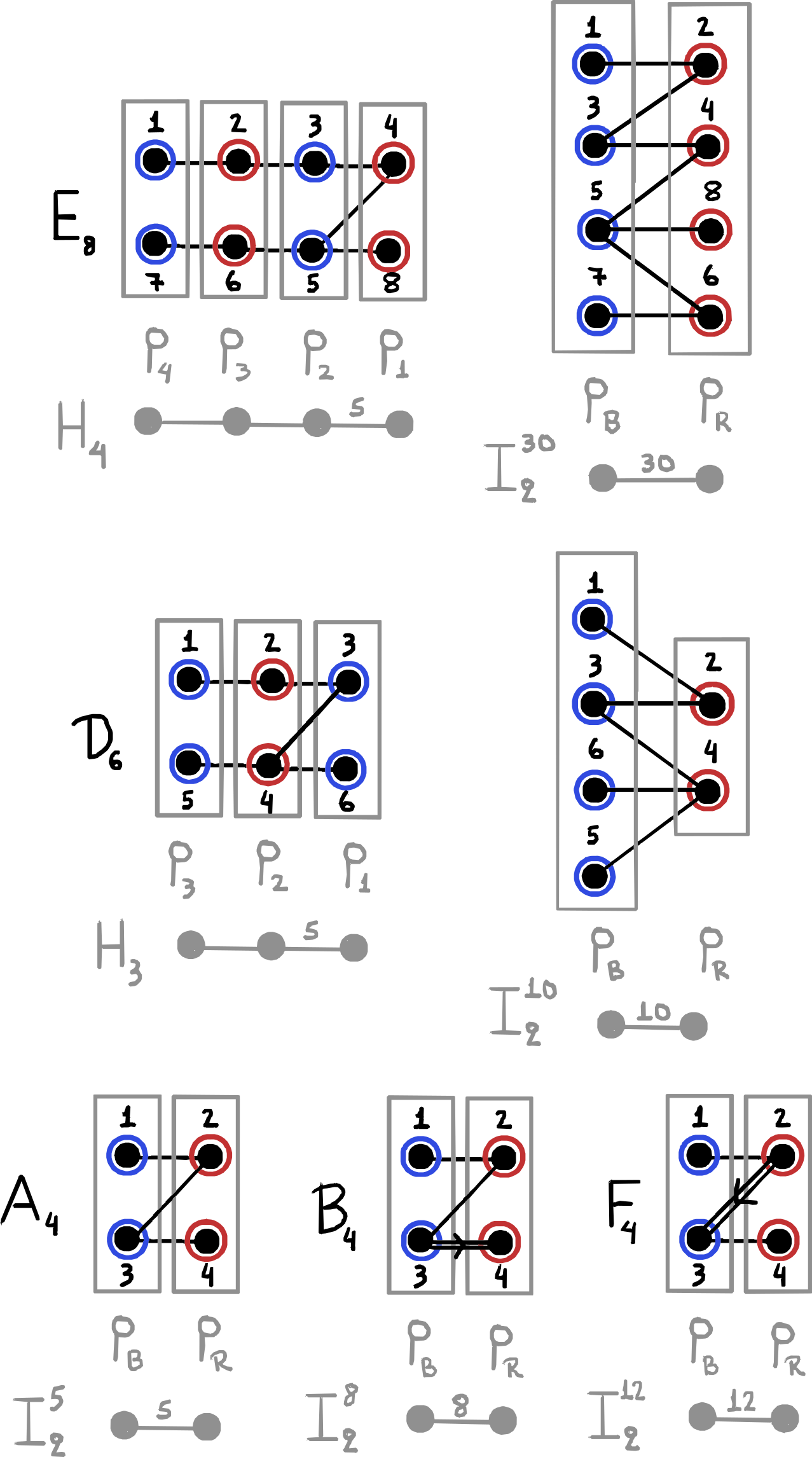}
\caption[Obtaining non-crystallographic subgroups of finite Coxeter groups by folding.]{Obtaining non-crystallographic subgroups of finite Coxeter groups by folding. Top row: the foldings yielding the $H_{4}$ and $I_{2}^{30}$ subgroups of $E_{8}$.
Middle row: the foldings yielding the $H_{3}$ and $I_{2}^{10}$ subgroups of $D_{6}$. Bottom row: the foldings yielding the $I_{2}^{5}$, $I_{2}^{8}$ and $I_{2}^{12}$ subgroups of $A_{4}$, $B_{4}$ and $F_4$ respectively.}
\label{fig:non-crystallographicEuclideanFoldings}
\end{figure}

One way to obtain a non-crystallographic Coxeter subgroup $H$ of a crystallographic Coxeter group $G$ is by {\bf folding} its Dynkin diagram. This is a process where we partition the fundamental generators of $G$ into subsets, where all the generators in a given subset commute with one another; then, for each subset, we take the product of all the generators in that subset and regard that product as a new fundamental generator (see Appendix \ref{appendix:FoldingUnfolding} and references therein). For example, see the upper left of Fig.~\ref{fig:non-crystallographicEuclideanFoldings}: if we partition the 8 reflection generators $\{R_{1},\ldots,R_{8}\}$ of $E_{8}$ into four subsets $\{R_{4},R_{8}\}$, $\{R_{3},R_{5}\}$, $\{R_{2},R_{6}\}$, $\{R_{1},R_{7}\}$, and use these subsets to define the new generators $\rho_{1}=R_{4}R_{8}$, $\rho_{2}=R_{3}R_{5}$, $\rho_{3}=R_{2}R_{6}$ and $\rho_{4}=R_{1}R_{7}$, we find that the $\rho_{i}$'s obey the Coxeter relations of the non-crystallographic Coxeter subgroup $H_{4}\in E_{8}$. 

Fig.~\ref{fig:non-crystallographicEuclideanFoldings} illustrates some important examples in which finite Coxeter groups are folded to obtain their non-crystallographic subgroups. These are the most important foldings from the standpoint of constructing ordinary Euclidean quasicrystals.~All foldings that lead to the 2D non-crystallographic $I_2^n$ arise from grouping all blue and red generators together to form $\rho_B$ and $\rho_R$, and correspond geometrically to a projection onto the 2D Coxeter planes (see Subsections \ref{subsec:Coxeter_Elements} and \ref{subsec:Hyperbolic_Coxeter_Elements}). These geometric connections are explained more in Appendix \ref{appendix:FoldingUnfolding}. 

\subsection{Coxeter Elements}
\label{subsec:Coxeter_Elements}

The product of all of a Coxeter group's fundamental reflections (in any order) gives a special type of element in the group called a {\bf Coxeter element} $C$:
\begin{equation}
  C=R_1R_2\ldots R_n.
  \label{eq:CoxeterElementProductOfGens}
\end{equation}
If the Coxeter-Dynkin graph is a tree or a forest of trees (i.e.~no closed loops, as is always the case for finite and affine Coxeter groups), the different Coxeter elements (obtained by multiplying the fundamental roots in different orders) are all equivalent up to symmetry ({\it i.e.}\ in the same conjugacy class in the Coxeter group \cite{humphreys_1990}), so we may speak of {\it the} Coxeter element. However, we will later see that for some hyperbolic Coxeter groups, the diagrams do have closed loops, and the various Coxeter elements are {\it not} equivalent to one another ({\it e.g.}\ they have different eigenvalues). 

Another representation for the Coxeter element,  more intrinsic to its Coxeter-Dynkin graph, was given in \cite{howlett1982coxeter}. Consider the Sch\"{a}fli matrix $S$ \eqref{eq:SchlafliMatrix} and split it into its upper- and lower-triangular components $T_S$ and $T_S^T$ respectively, such that 
$$ (T_S)_{ij} = \left\{ 
\begin{aligned}
    & +2, & \text{ if } i=j\\
    & -2\cos(\pi/m_{ij}), & \text{ if } i<j\\
    & 0, & \text{ elsewhere}.
\end{aligned}\right.
$$
Subtracting the identity matrix $\mathbb{I}$ from $T_S$ so that its diagonal elements are now $+1$, and calling this new matrix $U_S$, the Coxeter element $C$ will be given as
\begin{equation}
    C = -(U_T - \mathbb{I})^{-1} (U_T - \mathbb{I})^T = -U_S^{-1} U_S^T,
    \label{eq:AlternativeCoxeterElementGraph}
\end{equation}
where $S = 2B = U_S + U_S^T$. This holds for any arbitrary Coxeter-Dynkin graph, including those with loops, and always gives a Coxeter element that depends on the labelling of the generators, which is encoded in the Coxeter matrix $m_{ij}$. 

The order $h$ of a Coxeter element, {\it i.e.} the smallest positive integer $h$ such that
\begin{equation}
  C^{h}=1,
\end{equation}
is called the {\bf Coxeter number}, and will be finite (resp. infinite) for a finite (resp. infinite) Coxeter group. Note that Coxeter pairs will always share their Coxeter numbers \cite{Boyle_2022,zuber1998generalized}; interestingly, though, according to Theorem 2.7 in \cite{Kuzmanovich_FiniteOrder} (see also \cite{Hanson1990,Koo2003}), the non-crystallographic Coxeter groups \textit{cannot}
have a Coxeter element \textit{with rational entries} of order equal to their Coxeter number $h$ while keeping their size equal to their rank.\footnote{Indeed, let's focus for concreteness on $I_2^5, H_3$ and $H_4$ of rank $n=2,3$ and $4$. $\rm{GL}(n,\QQ)$ can only have an element of order $h = 5$ if its dimensions is $n \geq 4$, so no $2\times 2$ rational Coxeter element exists in $n=2$. Similarly, $H_3$ would demand at least $n = 4$ for an element of order $h = 10$, and $H_4$ would need at least $n = 6$ for its Coxeter element of order $h = 30$. 
} This can be taken as another evidence that non-crystallographic groups must ``hide'' within crystallographic ones if they are to exist over the rationals and act on a lattice.   

For our purposes here, the spectrum of the Coxeter element will turn out to be very important. Below, we give some known general results about the Coxeter eigensystems:
\begin{itemize}
    \item \textbf{Seeing the Forest \textit{and} the Tree:} in \cite{a1976valeurs}, it was shown that, for a finite Coxeter system $(G,S)$ (i.e.~with finite number of fundamental generators $|S|$), whose Coxeter-Dynkin graph is a \textit{tree or a forest}, the eigenvalues of the Coxeter element either belong on the unit circle or on the positive real line, $\mathrm{spec}(C) \subset \mathbb{S}^1 \cup \mathbb{R}^+$. 
    \item \textbf{The Bicoloured Case:} this result was extended to all graphs that are \textit{bicolourable} or \textit{bipartite} (i.e.~their vertices can be coloured red and blue such that no two vertices of the same colour are connected with an edge, like those in Fig.~\ref{fig:non-crystallographicEuclideanFoldings}; note that these might contain loops, but these must not contain an odd number of edges) in \cite{BERMAN1989339}, where also a relation between the spectrum of the Schl\"{a}fli matrix $S$ and the Coxeter element $C$ was established. In these cases, there is a preferred Coxeter element, the \textit{bicoloured} one, where the generators of a given colour are all gathered together in $C_{\rm colour} = C_{\rm red} C_{\rm blue} = C_{\rm red} C_{\rm blue} $; we will see this in Section \ref{subsec:Hyperbolic_Coxeter_Elements}. 
    \item \textbf{Generic Graphs:} in \cite{howlett1982coxeter}, it was established that (i) spherical/finite Coxeter graphs have all their eigenvalues on $\Sph^1$ ($S$ positive definite), (ii) affine Coxeter graphs have all their eigenvalues on $\Sph^1$, but also have a repeated eigenvalue $+1$ which does not appear for the spherical ones ($S$ positive semidefinite), (iii) hyperbolic Coxeter graphs have a real eigenvalue larger than 1 ($S$ indefinite). 
    \item \textbf{The Bicoloured Eigenvalue:} it has been shown in \cite{mcmullen2002coxeter} that for hyperbolic groups, the bicoloured Coxeter element $C_{\rm colour}$ is the one that achieves the lowest positive and larger-than-one eigenvalue, known as the \textit{bicoloured eigenvalue}, amongst all possible Coxeter elements for a given Coxeter graph.  
    \item \textbf{Eigensystem Always Pairs Up:} for any finite Coxeter system, the Coxeter element's \textit{non-trivial} eigenvalues appear in pairs, i.e.~if $\lambda \in \mathrm{spec}(C)$ then $\lambda^{-1} \in \mathrm{spec}(C)$ as well; this excludes the potential trivial eigenvalues $\pm 1$ that act as their own reciprocal and do not need to be paired with a distinct counterpart \cite{HORN1999411}. By extension, the eigenvectors also appear in pairs.  
\end{itemize}

Concretely, for finite Coxeter groups $G$, the eigenvalues of $C$ are the {\bf primitive $h$-th roots of unity} $\zeta_h^{m_i}$, where $\zeta_h\equiv{\rm exp}(2\pi i/h)$, and $m_i$ runs over the {\bf totatives} of $h$ (i.e. the positive integers less than $h$ and coprime to $h$).  These eigenvalues occur in complex conjugate pairs $\zeta_{h}^{m_i}$ and $(\zeta_{h}^{m_i})^{\ast}=\zeta_{h}^{h-m_i}$. The exponents $m_i$ are invariants of the characteristic polynomial of the Coxeter element and thus invariants of $G$ \cite{humphreys_1990}. Letting $\mathbf{z}_{m_i},\mathbf{z}_{m_i}^*$ be the corresponding pair of complex-conjugate eigenvectors of $C$, combinations of the form $\alpha \mathbf{z}_{m_i} + \alpha^*\mathbf{z}_{m_i}^*$ span a real, two-dimensional eigenspace called a \textbf{Coxeter plane}, and the Coxeter element rotates this plane by an angle equal to $2\pi m/h$. For hyperbolic Coxeter groups, the eigensystem is more involved and we will discuss it in more detail in Section \ref{subsec:Hyperbolic_Coxeter_Elements}.

Since the Coxeter elements $C$ are the highest-order elements in $G$, the Coxeter planes are particularly symmetric ({\it i.e.}\ they are the planes preserved by the largest subgroup of $G$), and hence can be used to construct particularly symmetrical quasicrystals. We show in Appendix \ref{eq:AlexCriterionForCoxeterElement} how the Coxeter element's eigenbasis can guide us when we try to fold a crystallographic group into a non-crystallographic subgroup.

\section{Symmetries of Lattices}
\label{sec:Symmetries_of_Lattices}

Given a lattice $\Lambda$ in a vector space $V$, we can define two notions of symmetry transformations for $\Lambda$. First, thinking of $\Lambda$ as just the underlying $\ZZ$-module that it is according to the definition \eqref{eq:LatticeDefinition}, we consider the group of linear (invertible) transformations from $V\to V$ that map $\Lambda$ to itself, 
\begin{equation}
    {\rm GL}(\Lambda) = \left\{ M \in {\rm GL} (V) \, | \, M \Lambda = \Lambda \right\}.
    \label{eq:AutomorphismLattice_Module}
\end{equation}
These are the \textbf{automorphisms} of $\Lambda$ as a $\ZZ$-module. After choosing a basis for $\Lambda$, this group is isomorphic to ${\rm GL}(d,\ZZ)$ (see discussion about bases below Eq.~\eqref{eq:LatticeDefinition}). Its elements
preserve the lattice but need not preserve the Euclidean or Lorentzian metric that $V$ might be equipped with. 

Now, if we also consider the inner product $\langle\, , \, \rangle$ on $V$ and we ask which transformations in ${\rm GL}(V)$ \textit{also} preserve the inner product between vectors, we arrive at the second symmetry group. We call this the {\bf automorphism group} of $\Lambda$ (thought of as a $\ZZ$-module that \textit{also} lives in a space with a metric), and we denote it as ${\rm Aut}(\Lambda)$.
In other words, 
\begin{equation}
    {\rm Aut}(\Lambda) = {\rm GL}(V) \cap O(V).
    \label{eq:AutomorphismsLattice_Metric}
    \end{equation}
and ${\rm Aut}(\Lambda)$ is a discrete subgroup of the orthogonal group $O(V)$. The reason why we give the name ${\rm Aut}(\Lambda)$ to this second group rather than the first one is because the Coxeter group lives within ${\rm Aut}(\Lambda)$, as reflections `care' about the metric too (as can be understood by looking at the definition of reflections \eqref{eq:ReflectionMap}).  

A {\bf root} of the lattice $\Lambda$ is a vector ${\bf r}$ of positive norm $N({\bf r})=\langle {\bf r},{\bf r}\rangle > 0$ for which the associated reflection is a symmetry of $\Lambda$. If $\Lambda$ is self-dual, its roots are just the lattice vectors ${\bf r}\in \Lambda$ whose norm $N({\bf r})$ equals $1$ or $2$. In particular, the lattice vectors with $N({\bf r})=1$ or $N({\bf r})=2$ are called the {\bf short roots} and {\bf long roots}, respectively \cite{conway2013sphere}. 

The group generated by reflections in the mirrors corresponding to these roots is the {\bf reflection subgroup} $\Gamma_\Lambda$ of ${\rm Aut}(\Lambda)$, a Coxeter group. An important subclass of lattices are {\bf reflective lattices} -- these are lattices for which the reflection subgroup $\Gamma_\Lambda$ has finite index in ${\rm Aut}(\Lambda)$  \cite{vinberg1975some,bogachev2018vinberg}. In other words, in a reflective lattice, nearly all of the automorphisms are generated by reflections in mirrors. In these cases, we let\footnote{For hyperbolic cases, where $\Lambda \subset \MM^d$, instead of $\mathrm{Aut}(\Lambda)$ we usually consider its index-2 subgroup, ${\rm Aut}_+(\Lambda)$ that contains the transformations that preserve $\HH^{d-1}_+$ and does not exchange it with the lower hyperboloid $\HH_-^{d-1}$. In other words, we consider the orientation-preserving automorphisms that do not reverse time and do not exchange the future and past lightcone. This is similar to how we usually consider the \textit{proper orthochronous} Lorentz group $SO_+(1,3)$ in relativistic theories instead of the full $SO(1,3)$.} 
\begin{equation}
\mathrm{Aut}(\Lambda) = \Gamma_\Lambda \rtimes \mathrm{Sym}(\Gamma),
\label{eq:ReflectiveLatticeAutomorphisms}
\end{equation}
where $\mathrm{Sym}(\Gamma)$ describe the symmetries of the fundamental convex polyhedron for the lattice $\Lambda$, the Coxeter polytope. Geometrically, $\Gamma_\Lambda$ maps each Coxeter polyhedron $\Gamma$ to one another, whereas ${\rm Sym}(\Gamma)$ fixes a polyhedron and describes its symmetries, or equivalently, the symmetries of the Coxeter-Dynkin graph of $\Gamma_\Lambda$ (see Fig.~\ref{fig:SphEucHyp}). The index of the reflection group inside ${\rm Aut}(\Lambda)$ is denoted by $[{\rm Aut}(\Lambda) : \Gamma_\Lambda]$, and by Lagrange's theorem (the orbit-stabilizer theorem) \cite{fraleigh2003first}, it is equal to the order of the group ${\rm Sym}(\Gamma)$, $|{\rm Sym}(\Gamma)|$. If $\Gamma$ has finite volume, the number of its geometric symmetries must be finite, meaning the index is finite. This is an alternative definition for reflective lattices, as those whose fundamental polyhedra have finite volume\footnote{A more formal treatment of this can be found in \cite{vinberg1993geometry} and in \S 12.8 of \cite{ratcliffe2006foundations} and is based on an earlier result \cite{venkov1937arithmetic} on quadratic forms; indeed, mathematicians usually classify lattices in terms of their associated quadratic forms \cite{conway1983automorphism}.}.

\subsection{Vinberg's Algorithm for Reflective Hyperbolic Lattices}
\label{subsec:vinbergsAlgorithm}

Reflective hyperbolic lattices exist only up to dimension $d \leq 21$ in $\MM^{d}$  
\cite{ESSELMANN1996103} and there are only finitely many of them in each dimension \cite{Nikulin_2007}. Consequently, the classification of reflective lattices is theoretically possible and amounts to coming up with an algorithm to produce a finite list. For more information, the reader is further directed to \cite{bogachev2018vinberg,Bogachev2019Reflective,Bogachev_2019,belolipetsky2016arithmetichyperbolicreflectiongroups}.

Indeed, Vinberg \cite{vinberg1967discrete,vinberg1972groups,vinberg1975some} found a systematic way to enumerating all \textit{spacelike} roots for a Lorentzian lattice of type $\mathrm{I}_{1,d-1}$ or $\mathrm{II}_{1,d-1}$ (automorphisms of the Minkowski metric). Essentially, the algorithm starts from an arbitrary point $\mathbf{w} \in \mathbb{H}_+^{d-1}$ and proceeds to find the convex Coxeter polyhedron $\Gamma$ around $\mathbf{w}$ that serves as the fundamental region for the hyperbolic reflection group. This will be bounded by $n$ timelike mirrors $M_{j}$ perpendicular to spacelike roots $\mathbf{r}_j$, where $\mathbf{r}_j$ is taken to be the outwards normal to the mirror (see Section \ref{app:TwoMirrorsIntersecting}). In particular, consider the half-spaces $M_j^- = \{ \mathbf{x} \in \MM^d : \langle \mathbf{x},\mathbf{r}_j \rangle \leq 0\}$, i.e.~the `inner' side of the mirror, opposite to its normal; then, the fundamental polyhedron $\Gamma$ will be the intersection of all such half-spaces 
\begin{equation*}
\mathcal{P} = \bigcap_{j=1}^m M_j^-.   
\end{equation*} 
If $m < \infty$, then there exist a finite number of fundamental roots for the hyperbolic Coxeter group and the fundamental region is of finite volume; Esselmann was able to show that this will always be the case for $d \leq 22$ \cite{ESSELMANN1996103}. The most important aspects of the algorithm - first described in \cite{vinberg1967discrete} - can be summarized as follows: 
\begin{enumerate}
    \item Pick a random timelike vector $\mathbf{w} \in \mathbb{H}^{d-1}$ as the \textbf{controlling vector} for the algorithm and define the \textbf{height} of any lattice vector $\mathbf{x} \in \Lambda$ with respect to the controlling vector as \begin{equation*}
        h_\mathbf{w}(\mathbf{x}) = -2\frac{\langle \mathbf{w},\mathbf{x}\rangle_\eta}{\sqrt{\langle \mathbf{x},\mathbf{x}\rangle_\eta}}.
    \end{equation*}
    Since the choice is arbitrary (up to isometries of the hyperboloid, the resulting roots are equivalent), Vinberg uses $\mathbf{w} = (1,0,0,\dots, 0)$.
    \item Depending on the type of the lattice $\Lambda$ (either $\mathrm{I}_{1,d-1}$ or $\mathrm{II}_{1,d-1}$), the roots will be spacelike lattice vectors $\mathbf{r}$ of norm $+1$ or $+2$. In other words, we are searching for integral solutions to the (quadratic Diophantine) equation $\langle \mathbf{r},\mathbf{r}\rangle_\eta = 1 \text{ or } 2$ (subject to the additional constraint $\mathrm{Tr(\eta \mathbf{r}) \in 2\mathbb{Z}}$ if the lattice is even; see Section \ref{subsubsec:Lorentzian_Lattices}).
    \item For a candidate lattice point $\mathbf{x}$ to be accepted and added to the list of roots $\{ \mathbf{r}_j\}$, the following conditions need to be satisfied: 
    \begin{itemize}
        \item The candidate vector $\mathbf{x} \in \Lambda$ must have non-negative inner product with the controlling vector, $h_\mathbf{w}(\mathbf{x}) \geq 0$,
        \item It must have non-positive inner product with any other root $\mathbf{r}_j$ that has already been added to the list, $\langle \mathbf{x},\mathbf{r}_j \rangle \leq 0 \, \forall \, \mathbf{r}_j$, and
        \item Its vector height $h_\mathbf{w}(\mathbf{x})$ needs to be minimized subject to the above constraints. 
    \end{itemize}
\end{enumerate}

Geometrically, the vector-height $h_\mathbf{w}(\mathbf{x})$ measures how `far away' a lattice point $\mathbf{x} \in \Lambda$ is from belonging to the spacelike mirror $P_\mathbf{w}$. When the algorithm starts, additional care needs to be taken for the zero-height vectors (since then both ${\bf x}$ and ${-\bf x}$ are solutions and a consistent choice must be made about which to accept). If the algorithm terminates, we end up with a finite list of simple roots that are sorted in non-decreasing order of vector heights. Having all the simple roots, one could obtain more roots by reflecting each of the simple roots in the mirrors defined by the others. 

The algorithm is always guaranteed to return the fundamental roots of the lattice, however it can take infinitely long in certain cases. A more recent approach can be found here \cite{allcock2021alternativevinbergsalgorithm}. A useful website with more information about hyperbolic reflection groups due to Bogachev can be found \href{https://nvbogachev.netlify.app/tools/arith-hyp-refl-gr/}{here}. An implementation of the algorithm in \texttt{SageMath} due to Bogachev and Perpechenko can be found by following \href{https://github.com/aperep/vinal}{this link} and the paper \cite{bogachev2018vinberg} that explains the idea behind the program (see also references therein).

\chapter{Quasicrystals in Euclidean Space}
\label{ch:QuasicrystalsEuclidean}

{\bf Quasicrystals} are aperiodic patterns that nevertheless have a pure-point Fourier transform (diffraction pattern), like a periodic crystal \cite{senechal1996quasicrystals}. For particularly interesting quasicrystals like the Penrose tiling or Ammann-Beenker tiling, these properties (along with a type of discrete scale invariance or \textit{self-similarity}) follow from the quasicrystal's underlying orientational symmetry.  

In this section we introduce the cut-and-project (\CNP) scheme for constructing quasicrystals -- first the ordinary $\CNP$ scheme in Sec.~\ref{sec:CNP}, and then the symmetric \CNP{} (\sCNP{}) scheme in Sec.~\ref{sec:sCNP} -- before discussing the symmetries of these quasicrystals in Sec.~\ref{sec:CNPsymmetries}. First, let us introduce some terminology: given a $d$-dimensional lattice $\Lambda\in \mathbb{R}^{d}$, and a $k$-dimensional subspace $W\subset V$, we say that $W$ is:
\begin{itemize}
\item {\bf rational} if $W\cap\Lambda$ has rank $r=k$.
\item {\bf partially irrational} if $W\cap\Lambda$ has rank $0<r<k$,
\item {\bf totally irrational} if $W\cap\Lambda$ has rank $0$.
\end{itemize}
In other words, the subspace $W$ is totally irrational if it intersects the origin of $\Lambda$, but no other points in $\Lambda$.

\section{The Cut-and-Project (\CNP) Scheme}
\label{sec:CNP}

We first review the ordinary cut-and-project (\CNP) algorithm. Note that \CNP{} quasicrystals are also referred to as {\bf model sets} in the mathematical literature \cite{baake2013aperiodic} (see their \S 7).

\subsection{Construction}
\label{subsubsec:CNPconstruction}

\begin{figure}[p]
\subfloat[\CNP{} Method: Pattern $\Lambda_{{\rm ph}}^{{\bf t},{\cal W}}$]{%
        \includegraphics[width=0.48\linewidth]{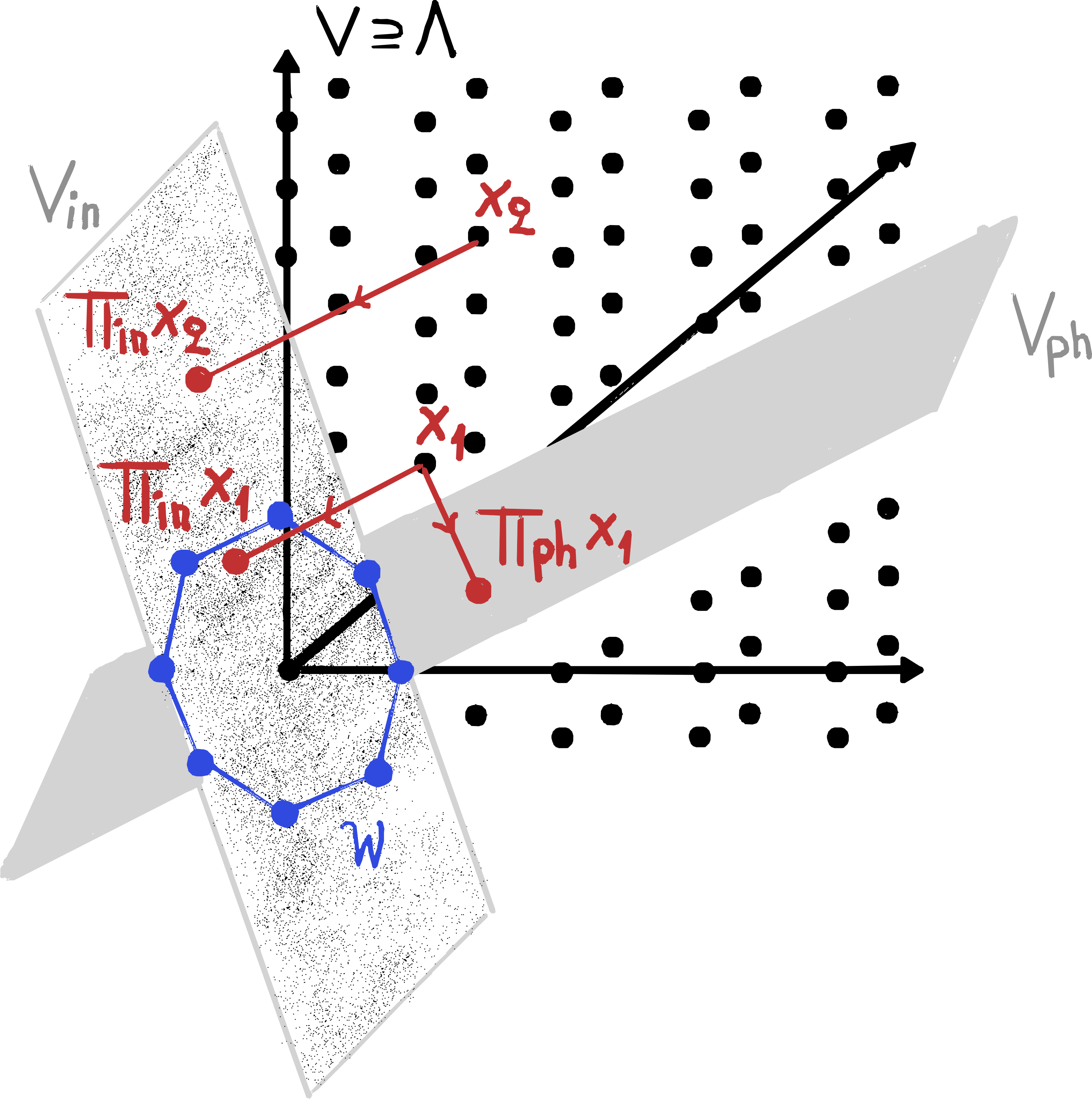}%
        \label{fig:CNPSchemeStandard}%
    }
    \hspace{+1ex}
    \subfloat[\CNP{} Method: Amplitude Pattern $\widehat{\Lambda}_{{\rm ph}}^{\,{\bf t},W}\!({\bf p}_{{\rm ph}})$]{%
        \includegraphics[width=0.48\linewidth]{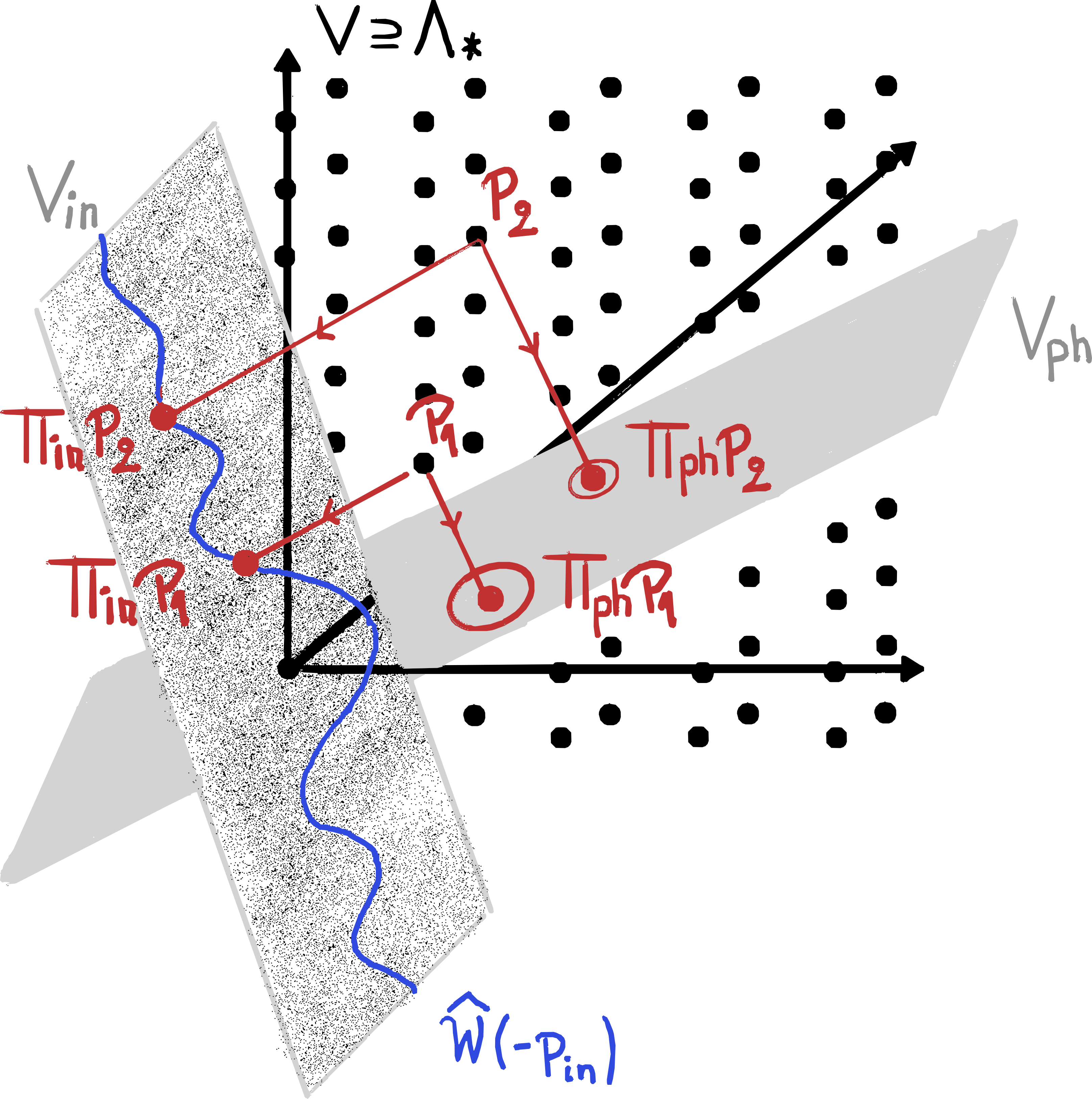}%
        \label{fig:CNPSchemeDiffraction}%
    }
    \caption[An illustration of the \CNP{} scheme explained in Sec.~\ref{sec:CNP}.]{An illustration of the \CNP{} scheme explained in Sec.~\ref{sec:CNP}. Left panel: constructing the quasicrystal $\Lambda_{{\rm ph}}^{{\bf t},W}$.  For each point ${\bf x}\in\Lambda^{{\bf t}}$, if the ``internal'' point ${\bf x}_{\rm in}=\Pi_{\rm in}{\bf x}$ lies inside the window ${\cal W}$, we include the ``physical'' point ${\bf x}_{\rm ph}=\Pi_{\rm ph}{\bf x}$ in the quasicrystal $\Lambda_{{\rm ph}}^{{\bf t},{\cal W}}$ (see Sec.~\ref{subsubsec:CNPconstruction} for details, included the weighted variant of the \CNP{} scheme). Right panel: The corresponding Fourier transform $\widehat{\Lambda}_{{\rm ph}}^{{\bf t},W}$ is produced by a closely analogous \CNP{} scheme: for each point ${\bf p}\in\Lambda_{\ast}$ (the dual lattice), the Fourier transform includes a delta function located at the ``physical'' momentum ${\bf p}_{\rm ph}=\Pi_{\rm ph}{\bf p}$, with a coefficient given by $\widehat{W}(-{\bf p}_{in})$, where $\widehat{W}$ is the Fourier transform of the ``top-hat'' or ``indicator'' function in the shape of the window ${\cal W}$, which is evaluated at the ``internal'' momentum ${\bf p}_{\rm in}=\Pi_{\rm in}{\bf p}$. As a display convention, one can represent the ``height'' of the delta function at ${\bf p}_{\rm ph}$ by a circular disk centered at ${\bf p}_{\rm ph}$, with area proportional to the coefficient $\widehat{W}(-{\bf p}_{\rm in})$, as indicated by the red circles in the right panel.} 
    \label{fig:CNPScheme}
\end{figure}

To construct a specific quasicrystal via \CNP{}, we first make three choices:
\begin{itemize}
\item {\bf Choice 1}: a $d$-dimensional lattice $\Lambda^{{\bf t}}$ in $V=\mathbb{R}^{d}$.
\item {\bf Choice 2}: a totally irrational subspace $V_{{\rm ph}}\subset V$ of dimension $0<d_{{\rm ph}}<d$, with totally irrational orthogonal complement $V_{{\rm in}}$, so that $V=V_{{\rm ph}}\oplus V_{{\rm in}}$.  We call $V_{{\rm ph}}$ and $V_{{\rm in}}$ the ``{\bf physical space}'' and ``{\bf internal space},'' respectively.  
\item {\bf Choice 3}: a subset ${\cal W}$ of $V_{{\rm in}}$ (called the ``{\bf window}'') or a function 
$W:V_{{\rm in}}\to\mathbb{C}$ (called the ``{\bf weighting function}'').
\end{itemize}  

It is convenient to let $\Pi_{{\rm ph}}$ and $\Pi_{{\rm in}}$ denote the projection operators onto the physical and internal subspaces, respectively ($V_{{\rm ph}}=\Pi_{{\rm ph}}V$ and $V_{{\rm in}}=\Pi_{{\rm in}}V$); so we have $\Pi_{{\rm ph}}^{2}=\Pi_{{\rm ph}}$, $\Pi_{{\rm in}}^{2}=\Pi_{{\rm in}}$, $\Pi_{{\rm ph}}\Pi_{{\rm in}}=\Pi_{{\rm in}}\Pi_{{\rm ph}}=0$ and $\Pi_{{\rm ph}}+\Pi_{{\rm in}}=1$.  More explicitly, we can always choose a basis $\{{\bf u}_i\}$ for $V$ that is orthogonal ($\langle {\bf u}_i,{\bf u}_j\rangle=0$ for $i\neq j$) and adapted to the decomposition $V=V_{{\rm ph}}\oplus V_{{\rm in}}$:
\begin{equation}
  \{
  \underbrace{{\bf u}_{1},\ldots,{\bf u}_{d_{{\rm ph}}}^{}}_{\text{spans}\;V_{{\rm ph}}},
  \underbrace{{\bf u}_{d_{{\rm ph}}+1}^{},\ldots,{\bf u}_{d}}_{\text{spans}\;V_{{\rm in}}}\}.
\end{equation}
In an arbitrary basis $\{{\bf v}_{\mu}\}$, where ${\bf u}_{i}=u_{i}^{\mu}{\bf v}_{\mu}$, $\Pi_{{\rm ph}}$ is then 
\begin{equation}
  \label{def_Pi}
  (\Pi_{{\rm ph}})^{\mu}_{\;\;\nu}=\sum_{i=1}^{d_{{\rm ph}}}\frac{u_{i}^{\mu}u_{i,\nu}}{u_{i}^{2}}.
\end{equation}
In general, we will use the subscripts ``{\rm ph}'' and ``{\rm in}'' to denote quantities that have been projected onto the physical or internal subspaces, respectively.

Given these three choices, we construct the corresponding quasicrystal using one of the following two schemes (as appropriate):
\begin{itemize}
\item {\bf The window scheme.} If (in Choice 3) we choose the window $\mathcal{W}\subset V_{{\rm in}}$, then the corresponding quasicrystal $\Lambda_{{\rm ph}}^{{\bf t},{\cal W}}$ is a {\it subset} of $V_{{\rm ph}}$, given by
\begin{equation}
  \label{eq:window_scheme}
  \Lambda_{{\rm ph}}^{{\bf t},{\cal W}}
  =\{{\bf x}_{{\rm ph}} \, | \, {\bf x}\in\Lambda^{{\bf t}}, \,\, {\bf x}_{{\rm in}}\in{\cal W}\}
\end{equation}
In other words, for each lattice point ${\bf x}\in\Lambda^{{\bf t}}$, if its {\it internal} position ${\bf x}_{{\rm in}}=\Pi_{{\rm in}}{\bf x}$ lies in the window $\mathcal{W}$, we include its {\it physical} position ${\bf x}_{{\rm ph}}=\Pi_{{\rm ph}}{\bf x}$ as a point in the quasicrystal $\Lambda_{{\rm ph}}$ (see Fig.~\ref{fig:CNPSchemeStandard}).

\item {\bf The weighting scheme.} If we instead choose the weighting function $W:V_{{\rm in}}\to\mathbb{C}$, the corresponding quasicrystal $\Lambda_{{\rm ph}}^{{\bf t},W}$ is a {\it distribution} on $V_{{\rm ph}}$ given by
\begin{equation}
  \label{eq:weighting_scheme}
  \Lambda_{{\rm ph}}^{{\bf t},W}({\bf v}_{{\rm ph}})=
  \sum_{{\bf x}\in\Lambda^{{\bf t}}}W({\bf x}_{{\rm in}})\delta({\bf v}_{{\rm ph}}-{\bf x}_{{\rm ph}}).
\end{equation}
In other words, for each lattice point ${\bf x}\in \Lambda^{{\bf t}}$, we add a delta function at the corresponding {\it physical} position ${\bf x}_{{\rm ph}}=\Pi_{{\rm ph}}{\bf x}$, with an amplitude given by the weighting function $W$ evaluated at the {\it internal} position ${\bf x}_{{\rm in}}=\Pi_{{\rm in}}{\bf x}$.
\end{itemize}
We note that our weighting scheme is known in the mathematical literature to produce \textit{dense and weighted Dirac combs} -- see \S 9.5 in \cite{baake2013aperiodic}.

The window scheme is essentially the special case of the weighting scheme where the weighting function $W$ is the ``indicator function'' for the region ${\cal W}$:
\begin{equation}
  W({\bf v}_{{\rm in}})=\left\{\begin{array}{ll}
  1, & \text{ if } {\bf v}_{{\rm in}}\in \mathcal{W}, \\
  0, & \text{ if } {\bf v}_{{\rm in}}\notin \mathcal{W}.\end{array}\right.
  \label{eq:WeightingFunctionAsIndicator}
\end{equation}
In the window scheme, the average {\it number density} $\bar{n}_{\rm ph}$ (in $V_{{\rm ph}}$) of points in $\Lambda_{{\rm ph}}^{{\bf t},{\cal W}}$ equals the average number density $\bar{n}$ (in $V$) of points in $\Lambda$, times the volume $\mathrm{vol}(\mathcal{W})$ (in $V_{{\rm in}}$) of ${\cal W}$ \cite{baake2013aperiodic}. Using Eq.~\eqref{eq:LatticeDensity} for the lattice density, we write
\begin{equation}
  \bar{n}_{\rm ph}= \mathrm{dens}(\Lambda_{\rm ph}^{{\bf t}, \mathcal{W}})= \bar{n}\mathrm{vol}({\cal W}) = \mathrm{dens}(\Lambda) \mathrm{vol}({\cal W}) = \frac{\mathrm{vol}({\cal W})}{|\det \Lambda |}.
  \label{eq:DensityOfCNPQuasicrystal}
\end{equation}
Hence, to obtain a quasicrystal with a finite density of points, we must have a window $\mathcal{W}$ of finite volume. Usually, the window ${\cal W}$ on $V_{\rm in}$ comes from a projection of a fundamental cell for the lattice $\Lambda$ (see, for example, Fig.~\ref{fig:sCNPWindowABTiling}), but this not necessary and any finite bounded geometric window $\mathcal{W}$ would do for \CNP{}. However, when this happens, the quasicrystal $\Lambda_{\rm ph}^{{\bf t}, \mathcal{W}}$ can be promoted to a quasi-tiling; we will come back to this in Section \ref{sec:implications} as well.  

In the weighting scheme, if the weighting function $W$ is non-zero over the entirety of $V_{{\rm in}}$, the resulting distribution $\Lambda_{{\rm ph}}^{{\bf t},W}$
will have delta function ``peaks'' living over a dense set of points in $V_{{\rm ph}}$.  In such a situation, it is less useful to talk about the number density of such delta function peaks (which is infinite); but rather, we can think about each unit delta function as the mass density distribution of a unit point mass, and talk about the average {\it mass density} of the distribution $\Lambda_{{\rm ph}}^{{\bf t},W}$ as a whole.  In particular, the average mass density $\bar{\rho}_{\rm ph}$ (in $V_{{\rm ph}}$) of points in $\Lambda_{{\rm ph}}^{{\bf t},W}$ equals the average mass density $\bar{\rho}$ (in $V$) of points in $\Lambda$, times the integral of the weighting function $W({\rm v}_{\rm in})$ over $V_{\rm in}$:
\begin{eqnarray}
  \label{mass_density_relation}
  \bar{\rho}_{\rm ph}&=&\lim_{R\to\infty}
  \frac{\int_{R}d{\bf v}_{{\rm ph}}\Lambda_{{\rm ph}}^{{\bf t},W}({\bf v}_{{\rm ph}})}{\int_{R}d{\bf v}_{{\rm ph}}} \nonumber\\
  &=&\bar{\rho}\!\int\!d{\rm v}_{{\rm in}}
  W({\bf v}_{{\rm in}}).
\end{eqnarray}
Hence, to obtain a quasicrystal with finite average mass density, the integral of $W({\bf v}_{{\rm in}})$ over $V_{{\rm in}}$ must be finite. If $W$ is just the indicator function \eqref{eq:WeightingFunctionAsIndicator}, then the integral just gives us the volume and this density reduces to the number density.

\subsection{Diffraction Pattern}
\label{subsec:CNPdiffraction}

Let us now compute the diffraction pattern (Fourier transform) of the quasicrystal $\Lambda_{{\rm ph}}^{{\bf t},W}({\bf v}_{{\rm ph}})$. In our Fourier conventions, a function $f({\bf v})$ and its Fourier transform $\widehat{f}({\bf k})$ are related by\footnote{Crystallographers would define $\langle {\bf a}, {\bf b} \rangle = 2 \pi / l$, with $l$ the lattice
spacing \cite{senechal1996quasicrystals}, but here we carry the factor of $2\pi$ in the exponential of the plane waves instead. 
}
\begin{subequations}
  \begin{eqnarray}
  \label{eq:FourierTransformDef}
    \widehat{f}\,({\bf k})&=&\int d{\bf v}\, {\rm e}^{-2\pi i\langle {\bf k},{\bf v}\rangle} f({\bf v}), \\
    f({\bf v})&=&\int d{\bf k}\, {\rm e}^{+2\pi i\langle{\bf k}, {\bf v}\rangle}\widehat{f}\,({\bf k}).
  \end{eqnarray}
\end{subequations}
The Fourier transform of $\Lambda_{{\rm ph}}^{{\bf t},W}({\bf v}_{{\rm ph}})$ is then given by 
\begin{subequations}
  \begin{eqnarray}
    \label{Fourier1}
    \widehat{\Lambda}_{{\rm ph}}^{\,{\bf t},W}\!({\bf k}_{{\rm ph}})
    &=&\int d{\bf v}_{{\rm ph}}{\rm e}^{-2\pi i\langle{\bf k}_{{\rm ph}},{\bf v}_{{\rm ph}}\rangle}\Lambda_{{\rm ph}}^{{\bf t},W}({\bf v}_{{\rm ph}}) \\
    \label{Fourier2}
    &=&\sum_{{\bf x}\in\Lambda^{{\bf t}}}\!{\rm e}^{-2\pi i\langle{\bf k}_{{\rm ph}},{\bf x}_{{\rm ph}}\rangle}W({\bf x}_{{\rm in}}) \\
    \label{Fourier3}
    &=&\frac{{\rm e}^{-2\pi i \langle {\bf k}_{{\rm ph}},{\bf t}\rangle}}{{\rm vol}(V/\Lambda)}\sum_{{\bf p}\in\Lambda_{*}}
    {\rm e}^{-2\pi i\langle{\bf p}_{{\rm in}},{\bf t}\rangle}\widehat{W}(-{\bf p}_{{\rm in}})\times\delta({\bf k}_{{\rm ph}}-{\bf p}_{{\rm ph}})\qquad\nonumber\\
  \end{eqnarray}
\end{subequations}
where, in (\ref{Fourier2}) we have substituted (\ref{eq:weighting_scheme}) and used the delta function $\delta({\bf v}_{{\rm ph}}-{\bf x}_{{\rm ph}})$ to evaluate the ${\bf v}_{{\rm ph}}$ integral, and in (\ref{Fourier3}) we have used the Poisson summation formula $\sum_{{\bf x}\in\Lambda}f({\bf x})\!=\!\frac{1}{{\rm vol}(V/\Lambda)}\sum_{{\bf p}\in\Lambda_{\ast}}\!\widehat{f}({\bf p})$ \cite{baake2013aperiodic} and then simplified using $d{\bf x}=d{\bf x}_{{\rm ph}}d{\bf x}_{{\rm in}}$ and $\langle{\bf k},{\bf x}\rangle = \langle{\bf k}_{{\rm ph}}, {\bf x}_{{\rm ph}}\rangle+\langle{\bf k}_{{\rm in}},{\bf x}_{{\rm in}}\rangle$.  

In other words, (\ref{Fourier3}) says that $\widehat{\Lambda}_{{\rm ph}}^{\,{\bf t},W}({\bf k}_{{\rm ph}})$ is a pure-point diffraction spectrum that, apart from the pre-factor $\frac{1}{{\rm vol}(V/\Lambda)}{\rm e}^{-2\pi i\langle {\bf k}_{{\rm ph}},{\bf t}\rangle}$, is given by a \CNP{} procedure closely analogous to the one for $\Lambda_{{\rm ph}}^{{\bf t},W}({\bf v}_{{\rm ph}})$: for each dual lattice point ${\bf p}\in\Lambda_{*}$, we add a delta function at the corresponding {\it physical} momentum ${\bf p}_{{\rm ph}}=\Pi_{{\rm ph}}{\bf p}$, with an amplitude determined by the {\it internal} momentum ${\bf p}_{{\rm in}}$ and given, in particular, by ${\rm e}^{-2\pi i\langle {\bf p}_{{\rm in}},{\bf t}\rangle}\widehat{W}(-{\bf p}_{{\rm in}})$ (see Fig.~\ref{fig:CNPSchemeDiffraction}).

\subsection{Self-Dual Quasicrystals}
\label{sec:SelfDualQuasicrystals}

First, consider periodic crystals.  By the Poisson summation formula, the distribution
\begin{equation}
  \gamma({\bf v})=\sum_{{\bf x}\in\Lambda}\delta({\bf v}-{\bf x})
\end{equation}
has Fourier transform
\begin{equation}
  \widehat{\gamma}({\bf k})=\sum_{{\bf p}\in\Lambda_{*}}
  \delta({\bf k}-{\bf p}).
\end{equation}
So if the lattice $\Lambda$ is self-dual ($\Lambda=\Lambda_{*}$), the distribution $\gamma$ is also self-dual in the sense of invariance under Fourier transform ($\gamma=\widehat{\gamma}$).  We say that such a distribution $\gamma$ is a {\bf self-dual crystal}.

Analogously, if the \CNP{} quasicrystal $\Lambda_{{\rm ph}}^{{\bf t},W}$ is invariant under Fourier transform ($\Lambda_{{\rm ph}}^{{\bf t},W}=\widehat{\Lambda}_{{\rm ph}}^{\,{\bf t},W}$) we say it is a {\bf self-dual quasicrystal}.  Comparing Eqs.~(\ref{eq:weighting_scheme}) and (\ref{Fourier3}), we see this happens when the higher dimensional lattice $\Lambda$ is self-dual ($\Lambda=\Lambda_{*}$), untranslated (${\bf t}=0$ or, more generally ${\bf t}\in\Lambda$), and the weighting function is invariant under Fourier transform combined with inversion through the origin ($W({\bf x})=\widehat{W}(-{\bf x})$).

Note that this {\it linear} definition of self-duality is stronger than the {\it quadratic} one defined in Section \S 9.5.2 of \cite{baake2013aperiodic}: {\it i.e.}\ whereas our linear condition (invariance of the distribution under Fourier transform) implies the  quadratic condition (equality of the autocorrelation and the diffraction measure, both of which are quadratic in the weighting function $W$), the converse is not true (since the quadratic condition discards all phase information).

Also note that the {\it window} scheme, where $W$ is the indicator function (\eqref{eq:WeightingFunctionAsIndicator}) for some compact window $\mathcal{W}$, does {\it not} yield a self-dual quasicrystal.  In particular, $W$ is non-vanishing on a compact region in $V_{{\rm in}}$, implying that $\Lambda_{{\rm ph}}^{{\bf t},W}$ has delta functions on a finite-density (non-dense) set of points in $V_{{\rm ph}}$; while $\widehat{W}$ is non-vanishing on a non-compact region of $V_{{\rm in}}$, implying that $\widehat{\Lambda}_{{\rm ph}}^{{\bf t},W}$ has delta functions on a dense set of points in $V_{{\rm ph}}$ ({\it i.e.}\ there are delta functions arbitrarily close to any point in $V_{{\rm ph}}$).  By contrast, in a self-dual quasicrystal, $W$ and $\widehat{W}$ are both non-vanishing on a non-compact region of $V_{{\rm in}}$, implying that $\Lambda_{{\rm phys}}^{{\bf t},W}$ and $\widehat{\Lambda}_{{\rm phys}}^{\,{\bf t},W}$ both have delta functions on a dense set of points in $V_{{\rm ph}}$.  For illustration, we present examples of both types (the 8-fold symmetric Ammann-Beenker tiling and its self-dual variant) in  Section \ref{subsec:AmmannBeenkerTilingsExample} and Fig.~\ref{fig:AmmannBeenker}.

\section{The Symmetric \CNP{} (\sCNP) Scheme}
\label{sec:sCNP}

In the general \CNP{} scheme of Section \ref{sec:CNP}, a particular quasicrystal is specified by making three choices.  But the most interesting quasicrystals ({\it e.g.}\ the Penrose tiling or the Ammann-Beenker tiling) are produced by a more refined version of this scheme, which we will call the \textbf{``symmetric cut-and-project''} (or \sCNP{} scheme).  

\subsection{Construction}
\label{subsec:sCNPconstruction}

In the \sCNP{} scheme, we again make three choices:
\begin{itemize}
  \item {\bf Choice 1}: a $d$-dimensional lattice $\Lambda^{{\bf t}}$ in $V=\mathbb{R}^{d}$ with automorphism group ${\rm Aut}(\Lambda)$.
  \item {\bf Choice 2}: a non-crystallographic subgroup $G\subset {\rm Aut}(\Lambda)$, and a decomposition $V=V_{{\rm ph}}\oplus V_{{\rm in}}$ that is invariant under $G$: $\Pi_{{\rm ph}}=g\Pi_{{\rm ph}}g^{-1}$ (or equivalently $[\Pi_{{\rm ph}},g]=0$), $\forall g\in G$.
  \item {\bf Choice 3}: a window ${\cal W}\subset V_{{\rm in}}$ or weighting function $W:V_{{\rm in}}\to\mathbb{C}$ invariant under $G$: {\it i.e.}\ ${\cal W}=g{\cal W}$ or $W({\bf v}_{{\rm in}})=W(g\,{\bf v}_{{\rm in}})$, $\forall g\in G$. 
\end{itemize}
This produces a quasicrystal with orientational symmetry $G$ (in the sense discussed below, Sec.~\ref{sec:CNPsymmetries}).  Note that the total irrationality of $V_{{\rm ph}}$ and $V_{{\rm in}}$ now follows from the fact that $G$ is non-crystallographic.  

\begin{figure}[t!]
\centering
    \begin{tabular}{cc}
        \subfloat[\label{fig:sCNP-ABPattern}]{
            \includegraphics[width=0.3\linewidth]{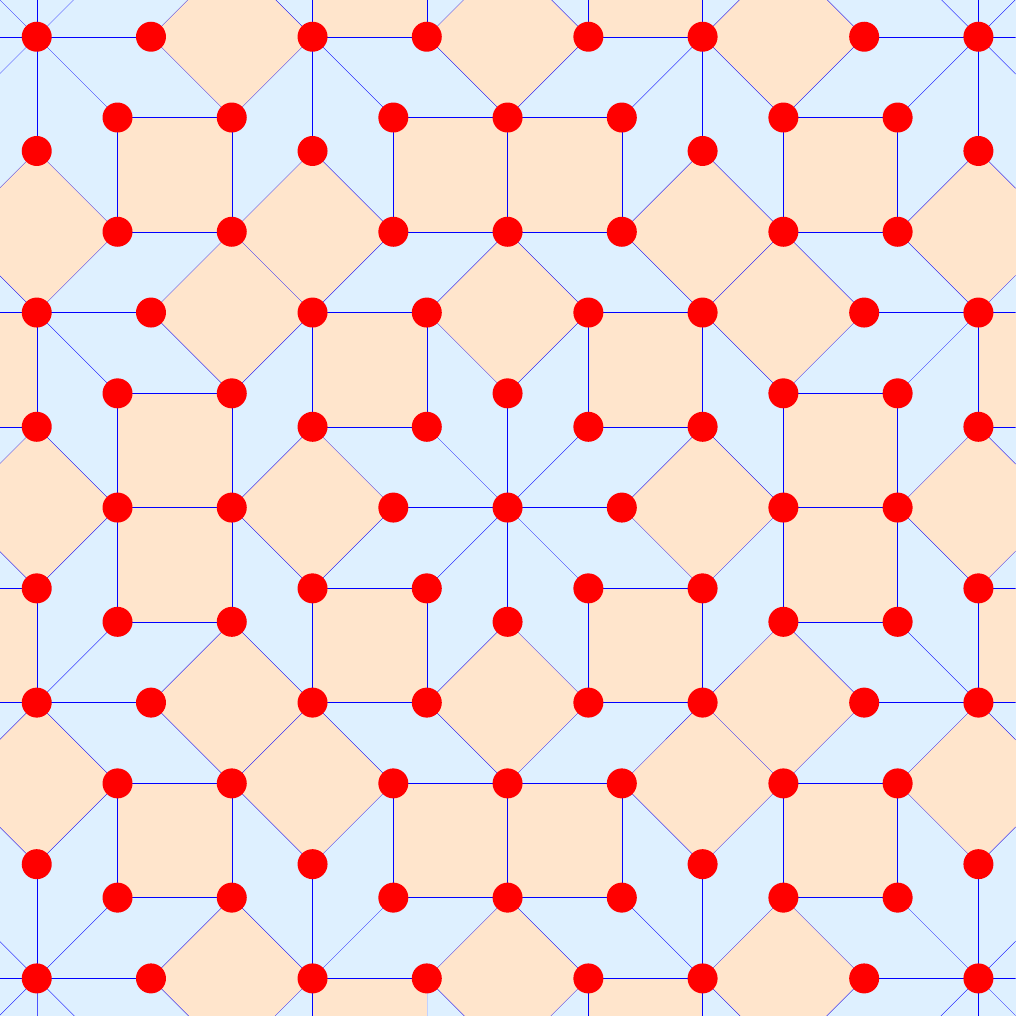}
            \hspace{1em}}
        \subfloat[\label{fig:sCNP-ABApproxAmplitudes}]{    
            \includegraphics[width=0.3\linewidth]{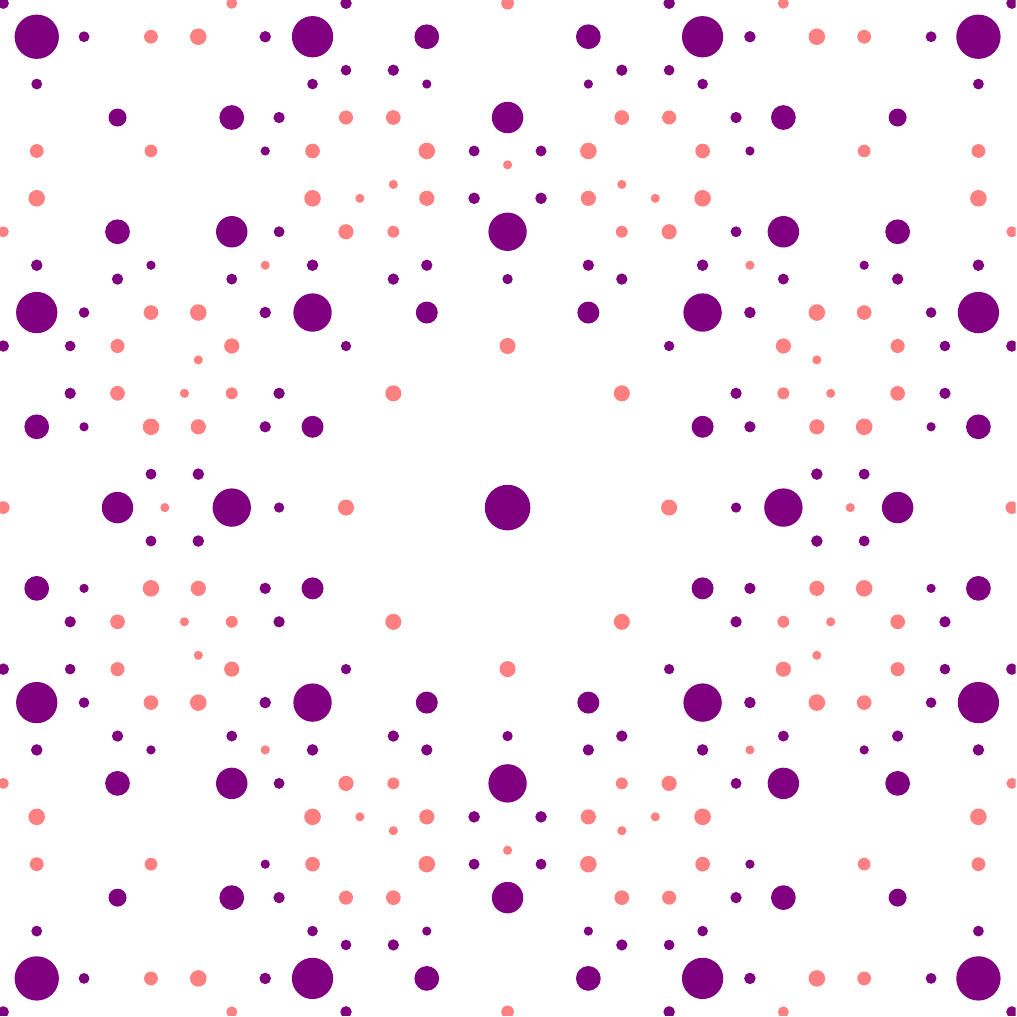}
        } \hspace{1em}
        \subfloat[\label{fig:sCNP-ABSelfDualAmplitudes}]{\includegraphics[width=0.3\linewidth]{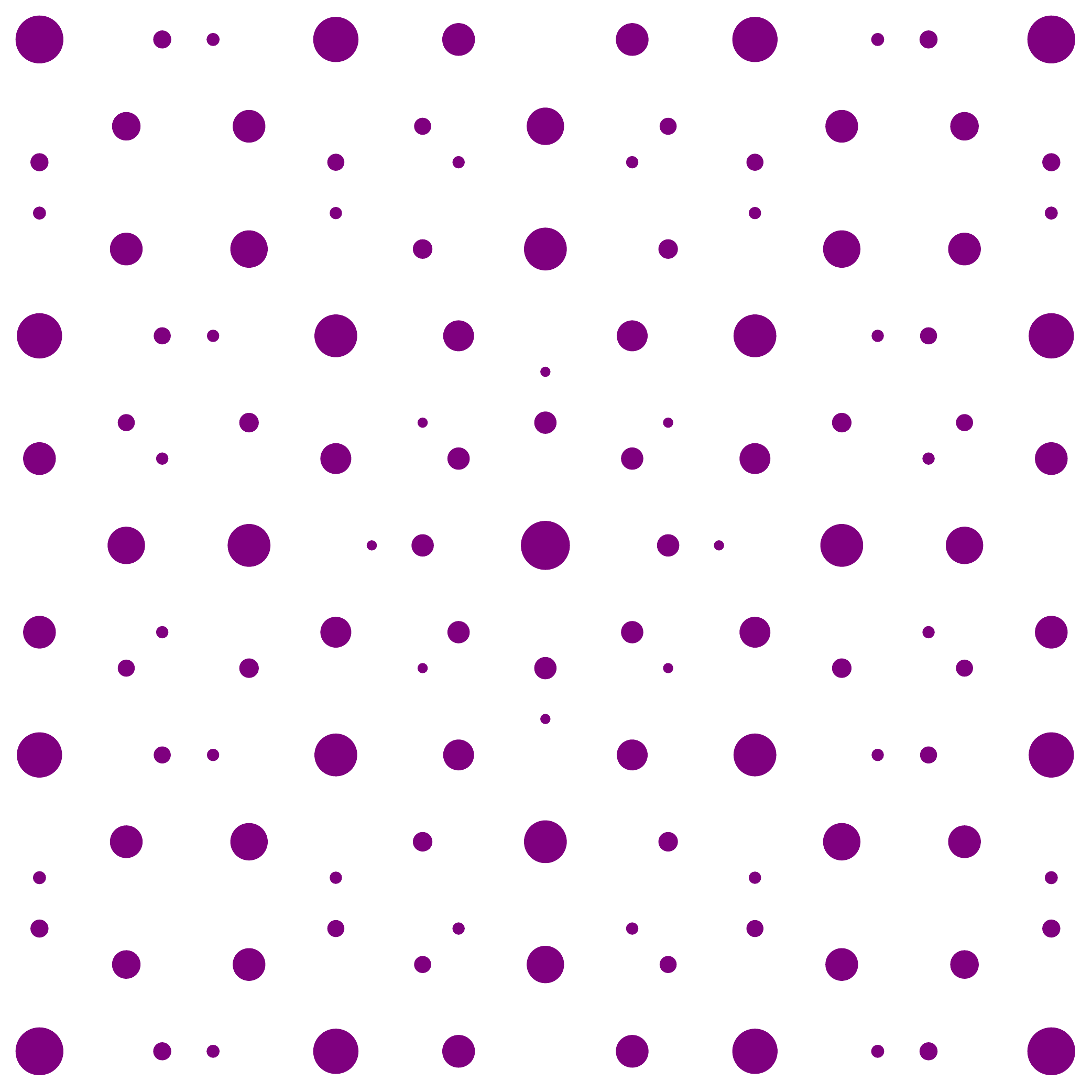}}
    \end{tabular}
\caption[Illustrating the symmetric cut-and-project (\sCNP{}) scheme (Sec.~\ref{sec:sCNP}) for Euclidean quasicrystals with 8-fold orientational symmetry.]{Illustrating the symmetric cut-and-project (\sCNP{}) scheme (Sec.~\ref{sec:sCNP}) for Euclidean quasicrystals with 8-fold orientational symmetry.  Left: a patch of the  Ammann-Beenker tiling, obtained via \sCNP{} as explained in  Sec.~\ref{subsec:AmmannBeenkerTilingsExample}. Middle: a corresponding patch of its Fourier transform. Right: the self-dual variant of the Ammann-Beenker quasicrystal (which is the same as its Fourier transform) -- see Sec.~\ref{subsec:self-dual_AB}.  In the middle and right panels, each circle's area represents the coefficient of the delta function located at its centre; purple (pink) circles indicate positive (negative) amplitudes; and only amplitudes with absolute value $>10^{-1.4}$ of the central peak's amplitude are shown (since, if all peaks are included, they densely fill the plane, but most have tiny coefficients).  Note that all three patches/panels are 8-fold symmetric; and are ``the same patch'' (but with different weightings/coefficients assigned to the points in the patch).}
\label{fig:AmmannBeenker}
\end{figure}

\subsection{Example 1: Ammann-Beenker Tiling}
\label{subsec:AmmannBeenkerTilingsExample}

As a first example of this procedure, we construct the 8-fold-symmetric Ammann-Beenker tiling.

{\bf Choice 1}: As our lattice $\Lambda^{{\bf t}}$, we choose the standard integer lattice ($\Lambda=\mathbb{Z}^{4}$) with no translation (${\bf t}={\bf 0}$).  ${\rm Aut}(\Lambda)$ is the Coxeter group $B_{4}$, whose fundamental roots ${\bf r}_{i}$ are the columns of the matrix 
\begin{equation*}
B = \left(\begin{array}{cccc}
\,1\; & -1 & 0 & 0\\
0 & 1 & -1& 0 \\
0 & 0 & 1 & -1\\
0 & 0 & 0 & 1
\end{array}\right).
\end{equation*}
From each ${\bf r}_i$ we construct the corresponding reflection matrix $R_i$ from Eq.~(\ref{R_eq}), and hence the corresponding Coxeter element
\begin{equation*}
C = R_1 R_2 R_3 R_4 = 
\left(\begin{array}{cccc}
0 & 0 & 0 & -1\\
\,1\; & 0 & 0 & 0 \\
0 & \,1\; & 0 & 0 \\
0 & 0 & \,1\; & 0
\end{array}\right).
\end{equation*}
$B_{4}$ has Coxeter number $h=8$, so $C^{8}=1$, and $C$'s eigenvalues are primitive 8th roots of unity; in particular 
\begin{equation}
  \zeta_{8},\; \zeta_{8}^\ast ,\; \bar{\zeta}_8 ,\;\bar{\zeta}_{8}^\ast,
\end{equation}
where $\zeta_{8}={\rm e}^{2\pi i/8}=(1+i)/\sqrt{2}$, $\ast$ denotes complex conjugation $i\to-i$, and bar denotes Galois conjugation
$\sqrt{2}\to-\sqrt{2}$.  The corresponding eigenvectors are
\begin{equation}
  {\bf v}, {\bf v}^\ast, \bar{\bf v}, \bar{{\bf v}}^\ast.
\end{equation}

{\bf Choice 2}: For our decomposition $V=V_{{\rm ph}}\oplus V_{{\rm in}}$, we choose $V_{{\rm ph}}=\{z{\bf v}+{z}^\ast{{\bf v}}^\ast,z\in\mathbb{C}\}$ (the real subspace spanned by ${\bf v}$ and ${\bf v}^\ast$) and $V_{{\rm in}}=\{\bar{z}\bar{{\bf v}}+\bar{z}^\ast\bar{{\bf v}}^\ast,z^\ast\in\mathbb{C}\}$ (the real subspace spanned by $\bar{\bf v}$ and $\bar{{\bf v}}^\ast$).  This decomposition is preserved by the cyclic subgroup $\mathbb{Z}_{8}$ generated by $C$ (which acts on both $V_{{\rm ph}}$ and $V_{{\rm in}}$ as the rotational symmetries of a regular octagon) and, more generally, by the non-crystallographic Coxeter group $I_{2}^{8}$ (which acts on both $V_{{\rm ph}}$ and $V_{{\rm in}}$ as the full symmetries of a regular octagon, including both rotations and reflections).  Then, from Eq.~(\ref{def_Pi}), we obtain the corresponding projectors
\begin{equation*}
\Pi_{\rm ph} = 
\frac{1}{2}
\begin{pmatrix}
1 & \frac{\sqrt{2}}{2} & 0 & -\frac{\sqrt{2}}{2}\\
\frac{\sqrt{2}}{2} & 1 & \frac{\sqrt{2}}{2} & 0 \\
0 & \frac{\sqrt{2}}{2} & 1 & \frac{\sqrt{2}}{2} \\
-\frac{\sqrt{2}}{2} & 0 & \frac{\sqrt{2}}{2} & 1
\end{pmatrix} \text{ and } \Pi_{\rm in} = \Pi_{\rm ph}'.
\end{equation*}

{\bf Choice 3}: To define the window ${\cal W}$, we start with the tesseract $\gamma_4$, {\it i.e.}\ the unit hypercube in four dimensions (centered at the origin) whose vertices are all 16 combinations of $\left( \pm \tfrac{1}{2}, \pm \tfrac{1}{2}, \pm \tfrac{1}{2}, \pm \tfrac{1}{2} \right)$. The polytope has 32 edges, 24 faces and 8 three-dimensional cells (or \textit{facets}).  The projection $\Pi_{\rm in}(\gamma_4)$ leads to the regular octagon $\mathcal{W}_{AB}$ shown in Fig. \ref{fig:sCNPWindowABTiling}, which is invariant under the non-crystallographic subgroup $G=I_{2}^{8}$ from Choice 2.

We show the resulting 8-fold-symmetric ($I_{2}^{8}$-symmetric) quasicrystal $\Lambda_{{\rm ph}}^{{\bf 0},{\cal W}}$ (the vertices of the Ammann-Beenker tiling) in Fig.~\ref{fig:AmmannBeenker} (left panel), and its Fourier transform $\widehat{\Lambda}_{{\rm ph}}^{\,{\bf 0},W}$ in Fig.~\ref{fig:AmmannBeenker} (centre panel).

\subsection{Example 2: Self-Dual Variant}
\label{subsec:self-dual_AB}

Alternatively, we can replace our previous Choice 3 by:

\begin{figure}[t!]
\centering
\begin{tabular}{cc}
        \subfloat[]{
            \includegraphics[width=0.3\linewidth]{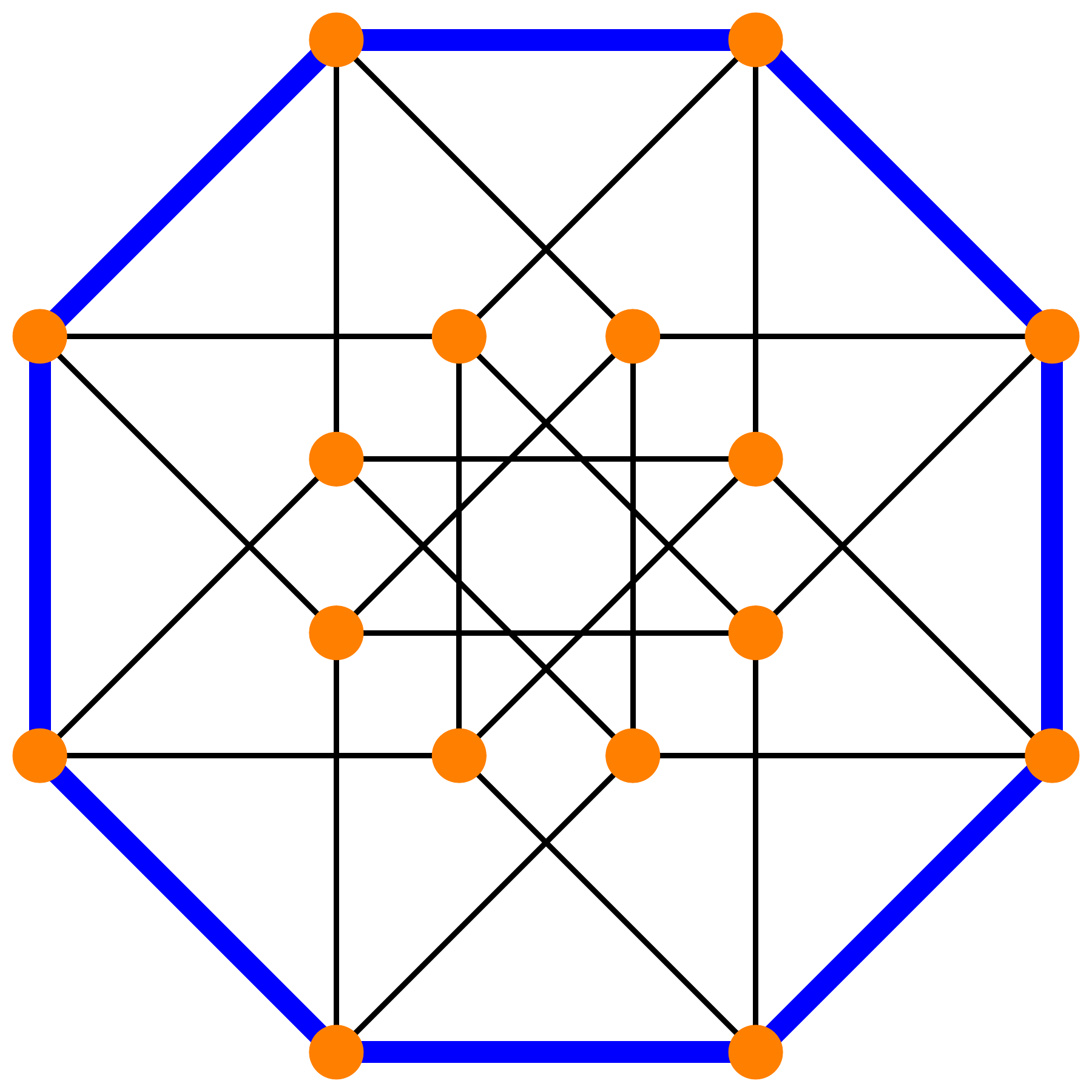}
            \hspace{1em}}
        \subfloat[]{    
            \includegraphics[width=0.3\linewidth]{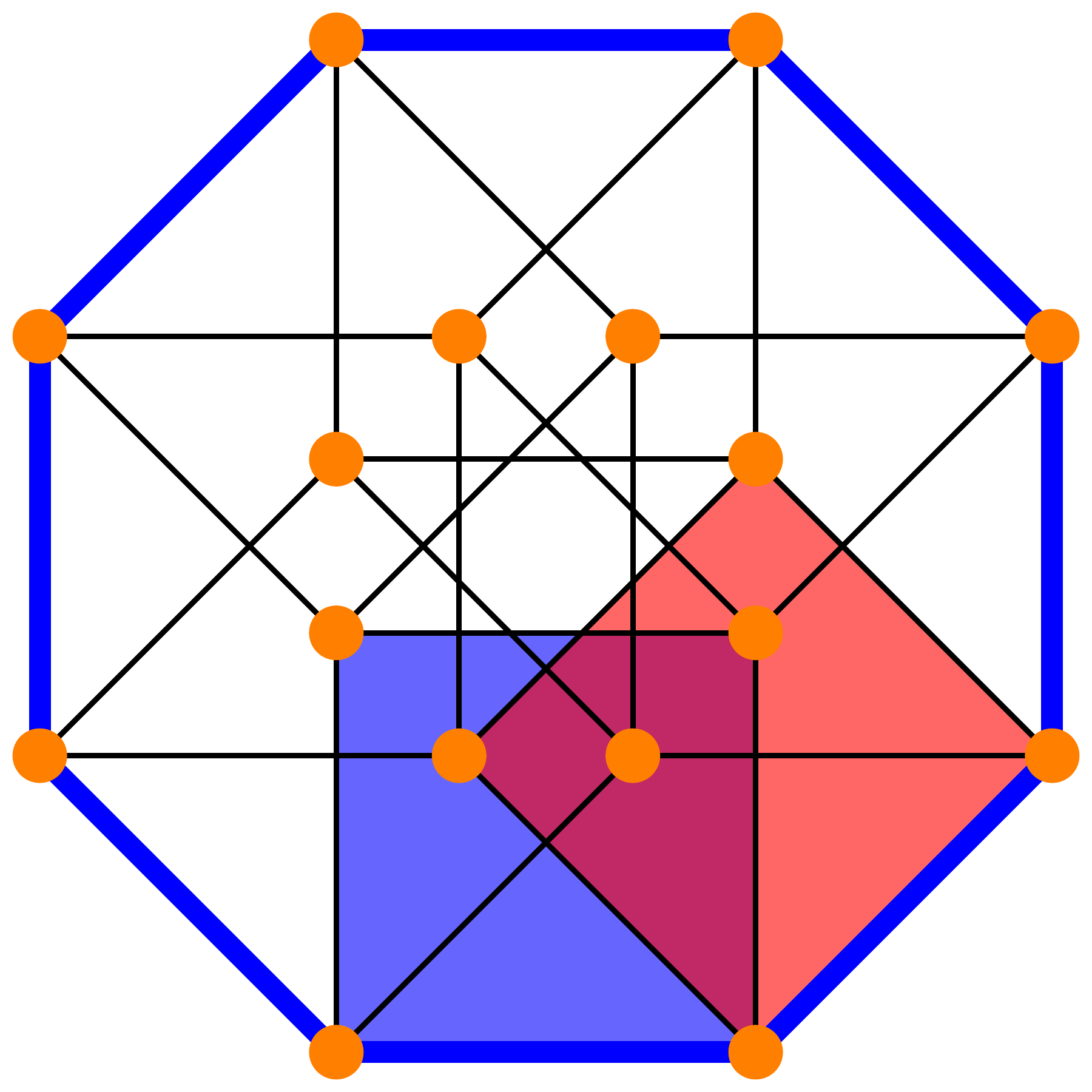}
        } \hspace{1em}
        \subfloat[]{\includegraphics[width=0.3\linewidth]{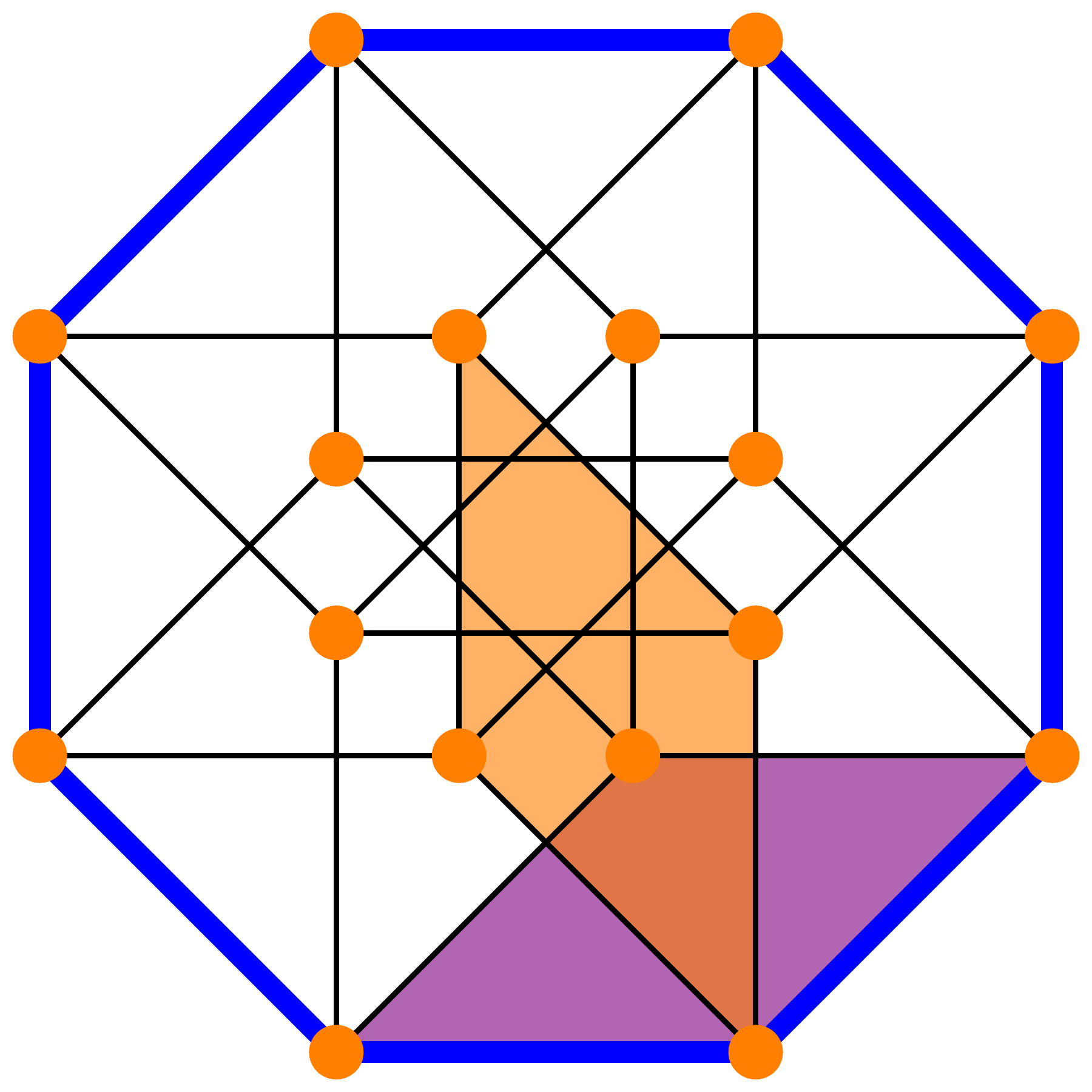}}
    \end{tabular}
\caption[The projection of the 4D unit hypercube $\gamma_{4}$ onto the internal space $V_{{\rm in}}$.]{The projection of the 4D unit hypercube $\gamma_{4}$ onto the internal space $V_{{\rm in}}$ with maximal symmetry. Left: the convex hull (blue octagon) defines the window $\mathcal{W}_{AB}$ used to generate the 8-fold-symmetric Ammann Beenker tiling shown in Fig.~\ref{fig:AmmannBeenker}. The faces of $\gamma_4$ become the tiles upon projection: in the middle, we see two such square faces and on the right we have two rhombic faces. Squares appear in eight possible orientations and rhombi in sixteen, all permuted by the action of the Coxeter element.} 
\label{fig:sCNPWindowABTiling} 
\end{figure}

{\bf Choice 3'}: For the weighting function, we now choose
\begin{equation}
  W({\bf v}_{{\rm in}})={\rm exp}(-\pi\, v_{{\rm in}}^{2}), \qquad v_{\rm in}^2 = \langle {\bf v}_{\rm in},{\bf v}_{\rm in}\rangle.
\end{equation}
Since this satisfies $W({\bf v}_{{\rm in}})=\widehat{W}(-{\bf v}_{{\rm in}})$, and we have also chosen a self-dual lattice $\Lambda=\mathbb{Z}^{4}=\Lambda_{\ast}$ with no translation (${\bf t}=0$), the resulting quasicrystal is self-dual (identical to its Fourier transform). This self-dual 8-fold-symmetric quasicrystal (the natural self-dual variant of the Ammann-Beenker quasicrystal) is given (for the first time, we believe) in Fig.~\ref{fig:AmmannBeenker} (right panel).

\subsection{To Tile or Not To Tile?}
\label{subsec:Tile_Or_Not}

Ever since the beginning of this thesis, already from Chapter \ref{sec:QC_Journey_Intro}, we were mentioning tilings and quasicrystals without being too careful defining what we mean by these terms. Before we dive into symmetries, let us give some very broad definitions regarding structures in a $d$-dimensional space $V$ (of either Euclidean -- in this chapter -- or Lorentzian signature in the next chapter).

Following \cite{baake2013aperiodic}, we call a non-empty set of non-empty subsets of $V$ a \textit{pattern} ${\cal P}$ in $V$, and the elements of ${\cal P}$ its \textit{fragments}. We want this to be sufficiently general on purpose, so that it can accommodate for two types of patterns that are of most interest to us here:
\begin{itemize}
    \item Any \textit{point set} $\mathcal{S}$ can be turned into a pattern in the obvious way: ${
    \cal P}_{\cal S} = \big\{ \{ s \} \, | \, s \in {\cal S} \big\}$. Our quasicrystals so far are such patterns. We usually demand the point set to be \textit{locally finite}, meaning that all its intersections with any compact region $K \subset V$ contain a finite number of points (i.e.~the cardinality of any $K \cap {\cal P}_{\cal S}$ is finite). This concept will become useful in Chapter \ref{ch:sQC_CausalSets}. Note that only quasicrystals arising from the window scheme are locally finite; any variant coming from the weighting scheme (including the self-dual and scale-invariant ones) have infinite points inside any finite region, thus cannot be locally finite.
    \item A \textit{tiling} (or a \textit{tessellation)} ${\cal T}$ in $V$ is the pattern 
    \begin{equation}
        \mathcal{T} = \big\{ T_i \,\, | \,\, i \in I \big\},
        \label{eq:tiling_definition}
    \end{equation}
    where $I$ is an index set (usually taken to be countable, but not restricted to be finite). The fragments $T_i$ are non-empty closed subsets of $V$, called the \textit{tiles} of the tiling ${\cal T}$, and satisfy the conditions that (i) they cover the full space $V$, $\cup_{i \in I} T_i = V$ (i.e.~no gaps), and (ii) they only intersect at their boundaries, $T_i^\circ \cap T_j^\circ = \varnothing$,   (i.e.~no overlaps); here, $T_i^\circ$ denotes the interior of the tile $T_i$. Usually, the tiles are convex sets; frequently, they are just some polytopes. We refer the reader to \S 5.1 in \cite{baake2013aperiodic} for more information.
\end{itemize}

It turns out that the Ammann--Beenker (AB) \textit{quasicrystal} (which is locally finite) can be promoted to the quasi-periodic Ammann--Beenker \textit{tiling} (or quasi-tiling) of Fig.~\ref{fig:sCNP-ABPattern}. The AB tiling (Fig.~\ref{fig:sCNP-ABPattern}) is a non-periodic tiling of $\EE^2$ with two tiles, a square and a rhombus, which (i) stem from (projections of) the faces of a cell for the lattice $\ZZ^4$ -- the tesseract $\gamma_4$, and (ii) each appears in 8 possible orientations, mapped onto one another by the action of $I_2^8$, the window's $\mathcal{W}_{AB}$ -- and, thus, the full tiling's -- symmetry group (see Fig.~\ref{fig:sCNPWindowABTiling}). In fact, this holds for any quasi-tiling: if it contains a tile, it must also contain all instances of that tile that are related to the original tile by the symmetry group action \cite{baake2013aperiodic}. 

When we say, then, that the AB tiling consists of only two tiles, we implicitly assume two \textit{equivalence classes} of tiles
\begin{equation}
\begin{aligned}
& \big[ \, \ABsquare{orange!40}{0} \,\big]_{I_2^8}
=
\big\{
\ABsquare{orange!40}{0},
\ABsquare{orange!40}{-45},
\ABsquare{orange!40}{-90},
\ABsquare{orange!40}{-135},
\ABsquare{orange!40}{-180},
\ABsquare{orange!40}{-225},
\ABsquare{orange!40}{-270},
\ABsquare{orange!40}{-315}
\big\} = \big\{ C^k \big( \ABsquare{orange!40}{0} \big) : C \in \mathrm{I}_{2}^8, \, 0 \leq k \leq 7  \big\}\\
&
\big[ \, \ABrhombus{cyan!30}{0} \big]_{I_2^8} = 
\big\{
\ABrhombus{cyan!30}{0}, 
\ABrhombus{cyan!30}{45},
\ABrhombus{cyan!30}{90},
\ABrhombus{cyan!30}{135},
\ABrhombus{cyan!30}{180},
\ABrhombus{cyan!30}{225},
\ABrhombus{cyan!30}{270},
\ABrhombus{cyan!30}{315}, + \dots
\big\} \, 
= 
\big\{ C^k \big( \ABrhombus{cyan!30}{0} \big) \cup C^k \big( \ABrhombus{cyan!30}{45} \big): C \in I_2^8, \, 0 \leq k \leq 7\big\}\\
\end{aligned}
\end{equation}
each with a finite number of tiles $\big| \big[ \, \ABsquare{orange!40}{0} \,\big]_{I_2^8} \big| = \big| \big[ \, \ABrhombus{cyan!30}{0} \big]_{I_2^8} \big|/2 = 8$; here, $C$ is the Coxeter element in $I_2^8$ of order $h=8$. Note that there are two times more orientations of rhombi than of squares in Fig.~\ref{fig:sCNPWindowABTiling}, which changes the cardinality of the two equivalence classes above. Consequently, there are still two tiles, but only if we ``mod-out'' by the symmetries of $I_2^8$ and identify all the possible orientations a tile can have with one another as a single tile. We call the representative tiles of the different equivalence classes the \textit{prototiles} \cite{baake2013aperiodic}. Note that, if any orientations of a tile were to be omitted, that would imply that the tiling is no longer symmetric under its own full symmetry group $I_2^8$, so it is not maximally symmetric. 

Given a set of vertices for a quasicrystal, there are perhaps several methods to tile it \cite{baake1990planar,kramer1989dualisation} (see also \S 7.4 in \cite{baake2013aperiodic}). A simple algorithm proceeds as follows: after picking a window ${\cal W}$ -- let's say the octagon from Fig.~\ref{fig:sCNPWindowABTiling} for the AB tiling for concreteness -- and finding the point set $\Lambda_{\rm ph}^{{\bf t}, {\cal W}}$, one can determine which points from this set have their pre-images connected at the level of the lattice $\Lambda$ (under the projector $\Pi_{\rm ph}$; irrationality guarantees a unique lifting of any ${\bf x}_{\rm ph}$ back to some ${\bf x}\in \Lambda$). Those points will belong to the same or to neighbouring fundamental cells for the lattice, and by connecting them with an edge on $V_{\rm ph}$, one eventually will form all tiles. Then, one realizes that the vertices that belong to the same tile actually did form a facet of the fundamental cell as well. So an alternative construction projects appropriate faces of the higher-dimensional polytope that belong exclusively inside the window ${\cal W}$ upon projection.\footnote{This can be done systematically and one can write a computer program that achieves it -- our quasicrystals were produced in \texttt{Mathematica} using our package, called \texttt{CutNProject}, which will be released on \texttt{GitHub} soon.} 

We note that, currently, we do not know of any algorithm to promote the non-locally-finite self-dual Ammann--Beenker quasicrystal of Fig.~\ref{fig:sCNP-ABSelfDualAmplitudes} 
into a tiling too. Partly, the issue seems to be the infinite density of the pattern: since there are infinite points (the full $\Lambda$), drawing edges between all points that were connected at the level of $\Lambda$ should lead to an infinitude of overlapping tiles. The smallest intersections of those tiles would be of arbitrarily small volume (since for each vertex of some tile, there should be another vertex arbitrarily close to it that will belong to a different tile that overlaps the original one). So the self-dual Ammann-Beenker quasicrystal (and its scale-invariant cousins that we did not construct explicitly here) are \textit{not} tilings in the usual sense.

\section{Symmetries of Quasicrystals}
\label{sec:CNPsymmetries}

In this section, we discuss the symmetries and transformations that naturally act on our quasicrystals.

\subsection{The Torus Parametrization}
\label{subsec:Torusparametrization}

In Sec.~\ref{subsubsec:CNPconstruction}, we explained that the \CNP{} quasicrystal $\Lambda_{{\rm ph}}^{{\bf t},W}$ is determined by three choices: the lattice $\Lambda^{{\bf t}}$, the decomposition $V=V_{{\rm ph}}\oplus V_{{\rm in}}$ ({\it i.e.}\ the projector $\Pi_{{\rm ph}}$), and the window ${\cal W}$ (or weighting function $W$). 

Let us describe three important relationships between quasicrystals described by different values of the parameter ${\bf t}$:
\begin{itemize}

\item {\bf Torus parametrization \cite{baake1997torus}.}  First note that if we shift ${\bf t}\to{\bf t}'={\bf t}+\bm{\lambda}$, with $\bm{\lambda}\in\Lambda$, the two quasicrystals are {\it identical}:
\begin{equation}
  \Lambda_{{\rm ph}}^{{\bf t}+\bm{\lambda},W}({\bf v}_{{\rm ph}})=\Lambda_{{\rm ph}}^{{\bf t},W}({\bf v}_{{\rm ph}})\qquad({\rm for}\;\bm{\lambda}\in\Lambda).
\end{equation}
So two parameters ${\bf t}'$ and ${\bf t}$ that differ by an element $\bm{\lambda}\in\Lambda$ are equivalent, and hence the inequivalent values of ${\bf t}$ form the $d$-dimensional torus $\TT^{d}=V/\Lambda$.  

\item {\bf Translational equivalence.}  Next note that if we shift ${\bf t}\to{\bf t}'={\bf t}+\Delta{\bf t}$, with $\Delta{\bf t}\in V_{{\rm ph}}$, the two quasicrystals are {\it equivalent up to translation}:
\begin{equation}
  \Lambda_{{\rm ph}}^{{\bf t}+\Delta{\bf t},W}({\bf v}_{{\rm ph}})
  =\Lambda_{{\rm ph}}^{{\bf t},W}({\bf v}_{{\rm ph}}+\Delta{\bf t})\quad({\rm for}\;\Delta{\bf t}\in V_{{\rm ph}}).
\end{equation}

\item {\bf Local indistinguishability.}  Two quasicrystals $\Lambda_{{\rm ph}}^{{\bf t},W}$ with same $\Lambda$ and ${\cal W}$ (or $W$), but different ${\bf t}$ are {\it locally indistinguishable} \cite{baake1999guidemathematicalquasicrystals, baake2013aperiodic}\footnote{The term ``locally indistinguishable'' used here is synonymous with the earlier term ``locally isomorphic'' introduced in \cite{levine1986quasicrystals}. See \S 5.1.1 in \cite{baake2013aperiodic}, their Definition 5.5, and the discussion immediately after that.}. Informally, this means that any finite patch found somewhere in one quasicrystal can also be found somewhere in the other.

More precisely, in the window scheme, it means that, for any finite region $R\subset V_{{\rm ph}}$, $\exists$ a region $R'\subset V_{{\rm ph}}$ (a translation of $R$) such that $\Lambda_{{\rm ph}}^{{\bf t'},{\cal W}}$ in region $R'$ is identical to $\Lambda_{{\rm ph}}^{{\bf t},{\cal W}}$ in region $R$; or, in other words, $\exists$ a vector $\Delta{\bf v}_{{\rm ph}}\in V_{{\rm ph}}$ such that 
$V_{{\rm ph}}^{{\bf t},W}({\bf v}_{{\rm ph}})=V_{{\rm ph}}^{{\bf t}',W}({\bf v}_{{\rm ph}}+\Delta{\bf v}_{{\rm ph}})$, for ${\bf v}_{{\rm ph}}\in R$.  

And in the weighting scheme, it means that, for any finite region $R\subset V_{{\rm ph}}$, $\exists$ a vector $\Delta{\bf v}_{{\rm ph}}\in V_{{\rm ph}}$ s.t.~$V_{{\rm ph}}^{{\bf t},W}({\bf v}_{{\rm ph}})$ and $V_{{\rm ph}}^{{\bf t}',W}({\bf v}_{{\rm ph}}+\Delta{\bf v}_{{\rm ph}})$ are arbitrarily close (as distributions) for ${\bf v}_{{\rm ph}}\in R$.  

Quasicrystals $\Lambda_{{\rm ph}}^{{\bf t},W}$ with the same $\Lambda$ and ${\cal W}$ (or $W$), but different ${\bf t}$, are in the same {\bf local isomorphism class}, and the quasicrystals in this equivalence class are in one-to-one correspondence with the points on the torus ${\bf t}\in T^{d}=V/\Lambda$.

Note that, for the same $\Lambda$ and ${\cal W}$ (or $W$), but different translation parameters ${\bf t}$ and ${\bf t}'$, where ${\bf t}'$ {\it cannot} be written as ${\bf t}+\Delta{\bf t}+\bm{\lambda}$ (with $\Delta{\bf t}\in V_{{\rm ph}}$ and $\bm{\lambda}\in\Lambda$), the two quasicrystals $\Lambda_{{\rm ph}}^{{\bf t},W}$ and $\Lambda_{{\rm ph}}^{{\bf t}',W}$ are {\it not} equal up to translation: so they are \emph{locally} isomorphic but {\bf globally distinct}.  The local isomorphism class contains an uncountable infinity of locally isomorphic yet globally distinct quasicrystals.

\end{itemize}

\subsection{(Quasi-)Translational Invariance}
\label{subsec:TranslationalInvariance}

Since $V_{{\rm ph}}$ is totally irrational relative to $\Lambda$ ({\it i.e.}\ $V_{{\rm ph}}$ does not intersect any points in $\Lambda$, besides the origin), the resulting \CNP{} pattern is {\it aperiodic}: {\it i.e.}\ it has no non-zero vector $\Delta{\bf v}_{{\rm ph}}\in V_{{\rm ph}}$ s.t.~$\Lambda_{{\rm ph}}^{{\bf t},W}({\bf v}_{{\rm ph}})=\Lambda_{{\rm ph}}^{{\bf t},W}({\bf v}_{{\rm ph}}+\Delta{\bf v}_{{\rm ph}})$.  

However, although $\Lambda_{{\rm ph}}^{{\bf t},W}({\bf v}_{{\rm ph}})$ has no {\it exact} translational invariance, it still has a kind of {\bf quasi-translational invariance}. 

This means that, for any region $R\subset V_{{\rm ph}}$, $\exists$ a non-zero vector $\Delta{\bf v}_{{\rm ph}}\in V_{{\rm ph}}$ s.t.~$\Lambda_{{\rm ph}}^{{\bf t},W}({\bf v}_{{\rm ph}})$ and $\Lambda_{{\rm ph}}^{{\bf t},W}({\bf v}_{{\rm ph}}+\Delta{\bf v}_{{\rm ph}})$ are equal (in the window scheme) or are arbitrarily close as distributions (in the  weighting scheme) for ${\bf v}_{{\rm ph}}\in R$.  

\subsection{(Quasi-)Orientational Invariance}
\label{subsec:OrientationalInvariance}

In the \sCNP{} scheme (where the lattice $\Lambda$ has automorphism group ${\rm Aut}(\Lambda)$, while the projector $\Pi_{{\rm ph}}$ and the weighting function $W$ are both invariant under a subgroup $G\subset {\rm Aut}(\Lambda)$): for ${\bf t}={\bf 0}$, the resulting quasicrystal $\Lambda_{{\rm ph}}^{{\bf 0},W}$ also has {\it exact} orientational invariance under $G$:
\begin{equation}
  \Lambda_{{\rm ph}}^{{\bf 0},W}({\bf v}_{{\rm ph}})
  =\Lambda_{{\rm ph}}^{{\bf 0},W}(g{\bf v}_{{\rm ph}}),\quad
  \forall g\in G.
\end{equation}

For ${\bf t}\neq0$, $\Lambda_{{\rm ph}}^{{\bf t},W}$ may no longer have {\it exact} $G$-invariance; but (since it is locally indistinguishable from $\Lambda_{{\rm ph}}^{{\bf 0},W}$) it  still has a type of {\bf quasi-orientational invariance} (or {\bf quasi $G$ invariance}).  

That is, for any region $R\subset V_{{\rm ph}}$ and any $g\in G$, $\exists$ a vector $\Delta{\bf v}_{{\rm ph}}\in V_{{\rm ph}}$ s.t.~$\Lambda_{{\rm ph}}^{{\bf t},W}({\bf v}_{{\rm ph}})$ and $\Lambda_{{\rm ph}}^{{\bf t},W}(g{\bf v}_{{\rm ph}}+\Delta{\bf v}_{{\rm ph}})$ are equal (in the window scheme) or are arbitrarily close as distributions (in the weighting scheme) for ${\bf v}_{{\rm ph}}\in R$.  

\subsection{(Quasi-)Scale Invariance}
\label{subsec:ScaleInvariance}

Quasicrystals can have a type of discrete scale invariance that periodic crystals lack.  

In particular, such symmetry arises in the following context.  In the \sCNP{} scheme, we consider the non-crystallographic subgroup $G\subset{\rm Aut}(\Lambda)$ that preserves both the splitting $V=V_{{\rm ph}}\oplus V_{{\rm in}}$ ({\it i.e.}\ $[g,\Pi_{{\rm ph}}]=0$) and the window ${\cal W}=g{\cal W}$ or weighting function $W({\bf v}_{{\rm in}})=W(g{\bf v}_{{\rm in}}), \forall g\in G$.  But there is also a larger subgroup $\widetilde{G}$ 
\begin{equation}
  G\subset\widetilde{G}\subset{\rm GL}(\Lambda)
\end{equation}
whose elements $\tilde{g}\in \widetilde{G}$ still preserve the splitting $V=V_{{\rm ph}}\oplus V_{{\rm in}}$ and the lattice $\Lambda$, but they do {\it not} (necessarily) preserve the inner product on $V$, the window ($\tilde{g}{\cal W}\neq{\cal W}$) or weighting function ($W(\tilde{g}\,{\bf v}_{{\rm in}})\neq W({\bf v}_{{\rm in}})$). Instead, the group $\widetilde{G}$ also includes elements $\tilde{g}$ that act to rescale the physical and internal subspaces 
\begin{subequations}
  \label{g_tilde}
  \begin{eqnarray}
  \tilde{g}\,{\bf v}_{{\rm ph}}&=&\lambda_{{\rm ph}}{\bf v}_{{\rm ph}}\qquad({\bf v}_{{\rm ph}}\in V_{{\rm ph}}) \\
  \tilde{g}\,{\bf v}_{\;\!{\rm in}\,}&=&\lambda_{\;\!{\rm in}\,}{\bf v}_{\;\!{\rm in}\,}\qquad({\bf v}_{\;\!{\rm in}}\in V_{\;\!{\rm in}\,}).
  \end{eqnarray}
\end{subequations}

It is straightforward to show that ${\rm Aut}(\Lambda)$ cannot contain any non-trivial real such numbers. For any ${\bf v}_{{\rm ph}}, {\bf w}_{{\rm ph}} \in V_{\rm ph}$, demanding the preservation of the metric leads to
$$
\langle {\bf v}_{{\rm ph}},{\bf w}_{{\rm ph}}\rangle = \langle \widetilde{g}{\bf v}_{{\rm ph}}, \widehat{g}{\bf w}_{{\rm ph}}\rangle = \lambda_{\rm ph}^2\langle {\bf v}_{{\rm ph}},{\bf w}_{{\rm ph}}\rangle
\Rightarrow \lambda_{\rm ph}^2 = 1,$$
where we used the non-degeneracy of the induced metric on $V_{\rm ph}$ in the last step (similarly for $\lambda_{\rm in}$). Thus, the metric-preserving automorphism group ${\rm Aut}(\Lambda)$ cannot have elements that act as a dilation by $|\lambda_{\rm ph}|\neq 1$ or $|\lambda_{\rm in}|\neq 1$. This is why $\widetilde{G}$ exists within ${\rm GL}(\Lambda)$ and \textit{not} within ${\rm Aut}(\Lambda)$ (see Definitions \eqref{eq:AutomorphismLattice_Module} vs \eqref{eq:AutomorphismsLattice_Metric}).\footnote{We thank the referee from Physical Review Research for raising this issue in their report and for motivating us to be clearer about these dilations.} Letting $\widetilde{M}$ represent the matrix of the element $\widetilde{g} \in {\rm GL}(\Lambda)$, its action on the physical and internal spaces takes the form
\begin{equation}
\left\{
\begin{aligned}
  & \widetilde{M}|_{V_{\rm ph}} = \lambda_{\rm ph} 1 \\[1ex]
  & \widetilde{M}|_{V_{\rm in}} = \lambda_{\rm in} 1
\end{aligned}
\right.
  \Rightarrow
    \widetilde{M} = 
    \setlength{\arraycolsep}{0.50pt}
    \begin{pmatrix}
    \inflblock{red}{
    \lambda_{\rm ph} & \, \dots \, & 0 \\
    \vdots & \ddots & \\
    0 & \, \dots \,  & \,\, \lambda_{\rm ph}} 
    & 
    \inflblock{white}{
    0 & \, \dots \, & 0 \\
    \vdots & \ddots & \\
    0 & \, \dots \,  & \, 0}\\
        \inflblock{white}{
    0 & \, \dots \, & 0 \\
    \vdots & \ddots & \\
    0 & \, \dots \,  & 0} & \inflblock{red}{
    \lambda_{\rm in} & \, \dots \, & \, 0 \\
    \vdots & \ddots & \\
    0 & \, \dots \,  & \,\, \lambda_{\rm in}}
    \end{pmatrix} ,
\label{g_tilde_ActionMatrix}
\end{equation}
where $1$ denotes the $d_{\rm ph} \times d_{\rm ph}$ or the $d_{\rm in} \times d_{\rm in}$ identity matrix. Since $\widetilde{M}\in{\rm GL}(\Lambda)\simeq {\rm GL}(d,\ZZ)$ has determinant $\pm1$, the numbers $\lambda_{\rm ph}$ and $\lambda_{\rm in}$ satisfy $\lambda_{{\rm ph}}^{d_{{\rm ph}}}\lambda_{{\rm in}}^{d_{{\rm in}}}=\pm 1$ with $d_{{\rm ph}}+d_{{\rm in}}=d$.

The special numbers $\lambda_{\rm ph}$ are called the {\bf scale factors} associated to the \sCNP{} scheme. These scale factors may, in turn, be written as $\lambda_{\rm ph}=\pm \varepsilon_1^{m_1} \dots \varepsilon_{r}^{m_r}$, in terms of a finite list of $r$ {\bf fundamental scale factors} $\{\varepsilon_1,\ldots,\varepsilon_r\}$ ($\varepsilon_i>1$) raised to integer powers ($m_i\in\mathbb{Z}$); and these fundamental scale factors are, in turn, the {\bf fundamental units} of an {\bf algebraic number field} $K'$ associated with the non-crystallographic group $G$ (see Appendix \ref{app:inflation} for more details).  For example, for the 2D Ammann-Beenker tiling, there is a single fundamental scale factor -- the {\bf silver ratio} $1+\sqrt{2}$; and for the 2D Penrose tiling, or the 4D Elser-Sloane quasicrystal, there is again a single fundamental scale factor -- the {\bf golden ratio} $(1+\sqrt{5})/2$.

Such transformations may lead to two types of scale invariance -- local scale invariance (associated with the window scheme) and global scale invariance (associated with the weighting scheme).  We introduce these in turn. 

\subsection{Local Scale Invariance}
\label{subsec:LocalScaleInvariance}

\begin{figure}
\includegraphics[width=0.95\linewidth]{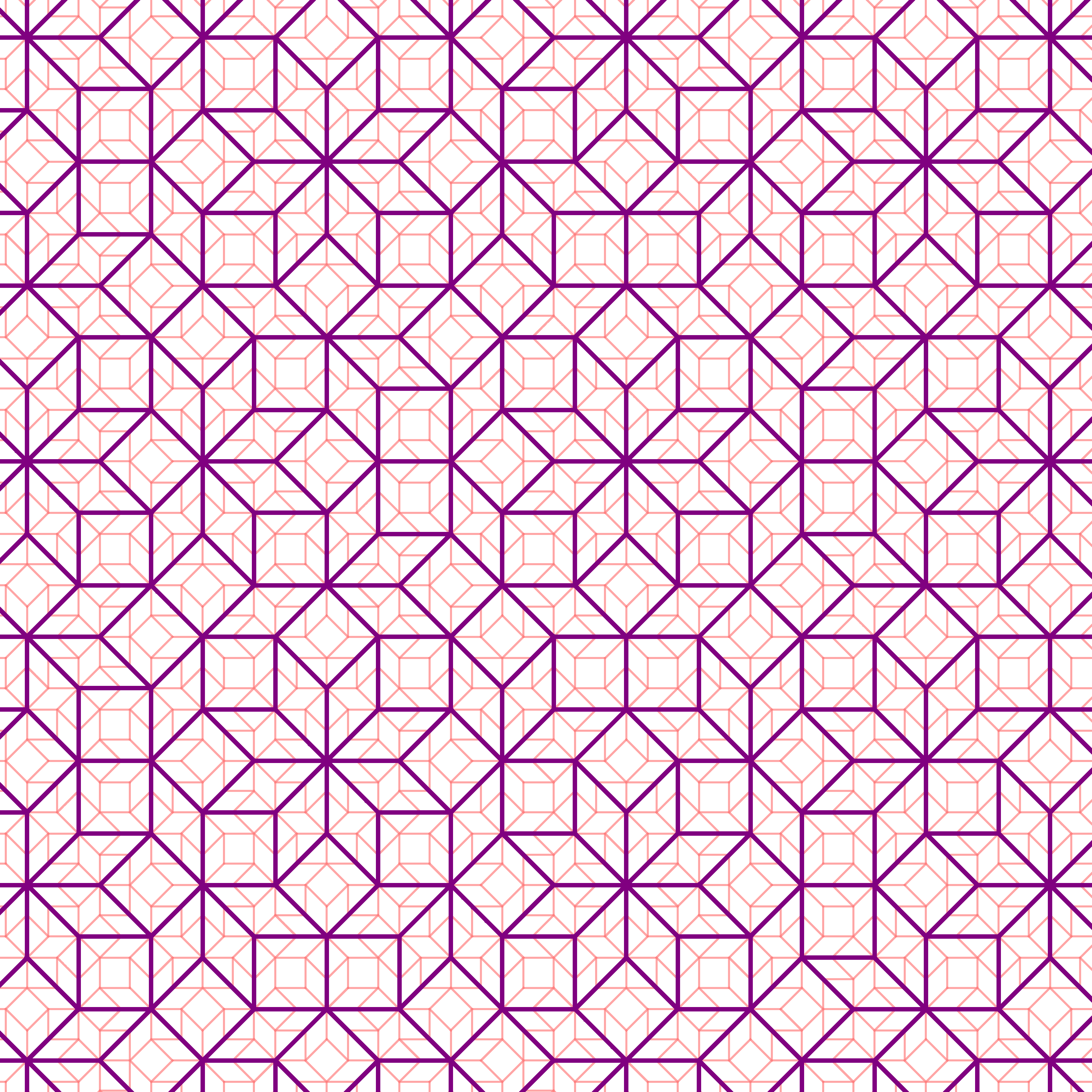}
\caption{A patch of the Ammann-Beenker tiling in purple, underlaid by its inflation in pink. Figure is reproduced from \cite{Boyle_2022}.} 
\label{fig:AmmannBeenkerInflation} 
\end{figure}

Informally, local scale invariance (also known as local inflation/deflation symmetry -- LIDS \cite{baake2013aperiodic}) refers to the inflation/deflation phenomenon exhibited by self-similar aperiodic tilings like the Penrose tiling or Ammann-Beenker (AB) tiling. Using the AB tiling for illustration, we define an {\bf inflation rule}, where we cut each large (purple) rhombus or square into a certain fixed pattern of smaller (pink) rhombi and squares, as shown in Fig.~\ref{fig:AmmannBeenkerInflation}.\footnote{The inflation rules for the Penrose tiling are shown in Fig.~\ref{fig:penroseTiling}.}  By cutting up all the tiles of any AB tiling (the ``parent'' AB tiling) in this way and rescaling it by the ``silver ratio''
$1+\sqrt{2}$, we obtain another legal AB tiling (the ``offspring'' AB tiling). This inflation process may be iterated arbitrarily many times.

We can also consider the inverse process, {\bf deflation}, in which we glue the tiles of an offspring AB tiling into ``supertiles,'' again according to the pattern in Fig.~\ref{fig:AmmannBeenkerInflation}, to recover the parent AB tiling from which it descended.  In a legal AB tiling, there is a \emph{unique} way to group all the tiles into supertiles in this way, so one can unambiguously determine the parent AB tiling from its offspring, and this deflation process may {\it also} be iterated indefinitely. 
 
To introduce local scale invariance more abstractly, we first review another idea: local equivalence. 

\begin{itemize}
    
    \item {\bf Local equivalence.}~Two patterns ${\cal P}$ and ${\cal P}'$ (which could be tilings, quasicrystals etc.) are {\it locally equivalent} -- also called {\it mutually locally derivable} (MLD) \cite{baake2013aperiodic} -- if ${\cal P}$ can be obtained from ${\cal P}'$ in a unique way by local rules, and vice versa, such that any symmetry of ${\cal P}'$ is also a symmetry of ${\cal P}$, and vice versa.

    For example (see Fig.~\ref{fig:AmmannBeenkerInflation}): the Ammann-Beenker tiling (in purple) is locally equivalent to its ``inflation'' (in pink) \cite{baake2013aperiodic}.
    
    \item {\bf Local scale invariance.}~A pattern ${\cal P}$ has local scale invariance -- also called local inflation/deflation symmetry (LIDS) \cite{baake2013aperiodic} -- with {\bf inflation factor} $\lambda>1$ if it is both locally equivalent to another pattern ${\cal P}'$, {\it and} locally indistinguishable from $\lambda {\cal P}'$ (where $\lambda {\cal P}'$ means the pattern ${\cal P}'$, expanded by the factor $\lambda$).  The pattern $\lambda {\cal P}'$ is the {\bf inflation} of ${\cal P}$, and the local rule for obtaining ${\cal P}'$ from ${\cal P}$ is the corresponding {\bf inflation rule}.  Or, viewed in the reverse direction, ${\cal P}$ is called the {\bf deflation} of $\lambda {\cal P}'$, and the local rule for obtaining ${\cal P}$ from ${\cal P}'$ is the corresponding {\bf deflation rule}.

    For example (see Fig.~\ref{fig:AmmannBeenkerInflation}):~the Ammann-Beenker tiling (in purple) is not only locally equivalent to its inflation (in pink), it is also locally indistinguishable from its inflation (after rescaling the pink tiling by the appropriate scale factor, $1+\sqrt{2}$).  Hence it has local scale invariance.
    
    Note that a periodic pattern {\it cannot} have local scale invariance with $\lambda>1$. Equivalently, the contrapositive: a pattern with local scale symmetry with $\lambda>1$ cannot be periodic. \textit{Proof}: if the two crystalline patterns ${\cal P}$ and ${\cal P}'$ are locally equivalent, they have the same fundamental domain; but then ${\cal P}$ and $\lambda {\cal P}'$ do {\it not} have the same fundamental domain, so are not locally indistinguishable.
\end{itemize}

To apply this to \sCNP{}, consider two patterns, ${\cal P}=\Lambda_{{\rm ph}}^{{\bf t},{\cal W}}({\bf v}_{{\rm ph}})$ and ${\cal P}'=\Lambda_{{\rm ph}}^{{\bf t},\widetilde{{\cal W}}}({\bf v}_{{\rm ph}})$, with $\widetilde{{\cal W}}=\lambda_{{\rm in}}{\cal W}$. On the one hand, using Eqs.~(\ref{eq:window_scheme}, \ref{g_tilde}) and the fact that $\tilde{g}\in{\rm GL}(\Lambda)$, it follows that 
\begin{equation}
  \Lambda_{{\rm ph}}^{{\bf t},\widetilde{{\cal W}}}=\lambda_{{\rm ph}}\Lambda_{{\rm ph}}^{\widetilde{{\bf t}},{\cal W}}
\end{equation}
where 
\begin{equation}
  \widetilde{{\bf t}}=\frac{{\bf t}_{{\rm ph}}}{\lambda_{{\rm ph}}}+\frac{{\bf t}_{{\rm in}}}{\lambda_{{\rm in}}}.
\end{equation}
In other words, the quasicrystal $\Lambda_{{\rm ph}}^{{\bf t},{\cal W}}$ (obtained using the old translation ${\bf t}$ and the new window $\widetilde{{\cal W}}$) is the {\it same} as the quasicrystal $\lambda_{{\rm ph}}\Lambda_{{\rm ph}}^{\widetilde{{\bf t}},{\cal W}}$ (obtained by using the old window ${\cal W}$ and the new translation ${\bf t}$, and then rescaling the resulting quasicrystal by $\lambda_{{\rm ph}}$). And this, in turn is locally isomorphic to the quasicrystal $\lambda_{{\rm ph}}\Lambda_{{\rm ph}}^{{\bf t},W}$ (by replacement $\widetilde{\bf t}\to {\bf t}$). Hence, ${\cal P}$ is locally isomorphic to $\lambda_{{\rm ph}} {\cal P}'$.

On the other hand, in the window scheme, two such quasicrystals, obtained using the same $\Lambda$, ${\bf t}$, and $\Pi_{{\rm ph}}$, but two different windows ${\cal W}$ and $\widetilde{{\cal W}}$, will be locally equivalent if and only if the set $\widetilde{{\cal W}}$ may be expressed as a finite number of unions and intersections of (translated copies of) ${\cal W}$, and vice versa \cite{baake1991quasiperiodic, baake2013aperiodic}.  So, iff this condition is satisfied,
${\cal P}$ is {\it also} locally equivalent to ${\cal P}'$, and hence
the quasicrystal $\Lambda_{{\rm ph}}^{{\bf t},W}$ has local scale invariance.
    
For example, in the AB tiling, the original window ${\cal W}$ is a regular octagon, while the rescaled window $\widetilde{{\cal W}}$ is a smaller regular octagon (scaled down by the factor $1+\sqrt{2}$).  Since each of these shapes may be construct from a finite number of unions and intersections of the other shape, it follows that the AB tiling has local scale invariance -- the inflation/deflation rule illustrated in Fig.~\ref{fig:AmmannBeenkerInflation}.

\subsection{Global Scale Invariance}
\label{subsec:GlobalScaleInvariance}

Whereas {\it local} scale invariance is a phenomenon that can occur in the {\it window} \CNP{} scheme, {\it global} scale invariance is a phenomenon that can occur in the {\it weighting} \CNP{} scheme.

First consider the set $\Lambda_{{\rm ph}}=\Pi_{{\rm ph}}\Lambda$ obtained by projecting all points in $\Lambda$ onto the physical subspace $V_{{\rm ph}}$.  $\Lambda_{{\rm ph}}$ forms a dense, but countable set of points in $V_{{\rm ph}}$.  

Next choose a pure-power-law weighting function 
\begin{equation}
  W({\bf v}_{{\rm in}})\propto v_{{\rm in}}^{-(d_{{\rm in}}+k)}.
\end{equation}
where $v_{\rm in}=|{\bf v}_{\rm in}|$ is the magnitude of ${\bf v}_{in}$, $d_{{\rm in}}=d-d_{{\rm ph}}$ is the dimension of the internal space, and $k>0$. Since $W$ is non-zero everywhere in $V_{{\rm in}}$, the resulting quasicrystal $\Lambda_{{\rm ph}}^{{\bf 0},W}$ has a dense set of delta function peaks -- one living over each point in $\Lambda_{{\rm ph}}$.  

From Eq.~(\ref{eq:weighting_scheme}), it follows that this quasicrystal has the following symmetry
\begin{eqnarray}
  \label{eq:global_scale_invariance}
  \Lambda_{{\rm ph}}^{{\bf 0},W}({\bf v}_{{\rm ph}})
  =\lambda_{{\rm in}}^{k}\Lambda_{{\rm ph}}^{{\bf 0},W}(\lambda_{{\rm ph}}{\bf v}_{{\rm ph}}),
\end{eqnarray}
where we have used $\tilde{g}\in{\rm GL}(\Lambda)\Rightarrow\lambda_{{\rm ph}}^{d_{{\rm ph}}}\lambda_{{\rm in}}^{d_{{\rm in}}}=1$.

In other words, Eq.~(\ref{eq:global_scale_invariance}) says that, as we rescale the physical space $V_{{\rm ph}}\to\lambda_{{\rm ph}}V_{{\rm ph}}$, the underlying dense subset $\Lambda_{{\rm ph}}\subset V_{{\rm ph}}$ is mapped to itself, one-to-one; and, moreover, as we map any point ${\bf x}_{{\rm ph}}\in\Lambda_{{\rm ph}}$ to the new point ${\bf x}_{{\rm ph}}'=\lambda_{{\rm ph}}{\bf x}_{{\rm ph}}\in\Lambda_{{\rm ph}}$, we find that the delta function peak living at the original point ${\bf x}_{{\rm ph}}$ was higher (by a factor of $\lambda_{{\rm in}}^{k}$) than the delta function peak living at the new point ${\bf x}_{{\rm ph}}'$.  We say such a quasicrystal has {\bf exact global scale invariance with weight $k$}.

So far we have focused on the case ${\bf t}=0$, since it yields exact global scale invariance.  But it is also singular since the delta function peaked at the origin (${\bf v}_{{\rm ph}}=0$) is multiplied by an infinite coefficient $W({\bf 0})=\infty$.  

If we shift from this singular (${\bf t}=0$) case to a generic case -- {\it i.e.}\ a value of ${\bf t}$ that {\it cannot} be written as $\lambda+{\bf t}_{{\rm ph}}$ (with $\lambda\in\Lambda$ and ${\bf t}_{{\rm ph}}\in V_{{\rm ph}}$) -- we obtain a new quasicrystal $\Lambda_{{\rm ph}}^{{\bf t},W}$ that is non-singular: it does not contain any delta functions multiplied by infinite coefficients, and the mass contained in any finite region is finite.  

Moreover, the non-singular quasicrystal $\Lambda_{{\rm ph}}^{{\bf t},W}$ is {\bf {\it almost} locally indistinguishable} from $\Lambda_{{\rm ph}}^{{\bf 0},W}$.  By this we mean that, for any finite region $R$ in $V_{{\rm ph}}$, we can find another finite region $R'$ in $V_{{\rm ph}}$ such that $\Lambda_{{\rm ph}}^{{\bf t},W}$ in region $R$ is arbitrarily close (as a measure) to $\Lambda_{{\rm ph}}^{{\bf 0},W}$ in $R'$; and, in the other direction, for any finite region $R$ in $V_{{\rm ph}}$ {\it that does not include the origin}, we can find another finite region $R'$ in $V_{{\rm ph}}$ such that $\Lambda_{{\rm ph}}^{{\bf 0},W}$ in $R$ is arbitrarily close (as a measure) to $\Lambda_{{\rm ph}}^{{\bf t},W}$ in $R'$.  We say that generic non-singular quasicrystal $\Lambda_{{\rm ph}}^{{\bf t},W}$ has {\bf global quasi-scale invariance}.

\chapter{Quasicrystals in Minkowski Spacetime}
\label{ch:QuasicrystalsMinkowski}

Having laid the groundwork in the previous sections, the extension to \textit{spacetime} quasicrystals is now rather straightforward, and we can focus on the key novel features, and giving the first examples of such quasicrystals.

\section{Mathematical Background}
\label{sec:MathematicalBackgroundLorentzian}

\subsection{Lorentzian Lattices}
\label{subsubsec:Lorentzian_Lattices}

In Sec.~\ref{sec:Lattices}, we introduced lattices living in Euclidean space $V=\mathbb{R}^{d}$.  To generalize the discussion, we now replace $\mathbb{R}^{d}$ (with $d$ spatial dimensions) to $\mathbb{R}^{s,t}$ (with $s$ spatial and $t$ temporal dimensions).  In other words, we replace the positive-definite inner product $\langle {\bf v},{\bf w}\rangle=v^{\alpha}\delta_{\alpha\beta}w^{\beta}$ on $\mathbb{R}^{d}$ by the non-degenerate ($\det \eta \neq 0$) but \textit{indefinite} symmetric bilinear form $\langle {\bf v},{\bf w}\rangle=v^{\alpha}\eta_{\alpha\beta}w^{\beta}$ on $\mathbb{R}^{s,t}$. This means that now the norm of some vector ${\bf x} \in \mathbb{R}^{s,t}$ can be positive (spacelike), negative (timelike) or even zero (null), and in the latter case this happens without ${\bf x}$ being trivially the ${\bf 0}$-vector (as it would happen for a positive-definite metric). In an orthonormal (Cartesian) basis, $\eta_{\alpha\beta}$ is
\begin{equation}
\eta_{\alpha\beta} = \mathrm{diag}(\underbrace{ \, -1,\dots,-1}_{\mathclap{t \text{ times}  }} \, , \, \underbrace{+ 1\dots,+1}_{\mathclap{s \text{ times }}} \,).
\label{eq:LorentzianMetricEta}
\end{equation}
The other definitions in Sec.~\ref{sec:Lattices} remain  unchanged.  

As before, the even and odd self-dual lattices are of particularly high symmetry and interest.  Perhaps surprisingly, the classification of such lattices is much simpler in {\it indefinite} spacetimes  ({\it i.e.} when $t, s$ are both $\geq 1$):
\begin{itemize}
  \item {\bf Odd self-dual lattices}: for any indefinite $V=\mathbb{R}^{s,t}$ (with $s,t$ both $\geq 1$), there is a unique {\it odd} self-dual lattice (denoted ${\rm I}_{s,t}$) which, in an orthonormal (Cartesian) basis, consists of all vectors ${\bf x}\in \mathbb{R}^{s,t}$ with integer components ${x}^{\mu}$:
  \begin{subequations}
  \begin{equation}
    {\rm I}_{s,t}=\{{\bf x} \,\, |\,\,x^{\mu}\in\mathbb{Z}\;(\mu=1,\ldots,s+t)\}.
    \label{eq:OddUnimodularLattice}
  \end{equation}
  In the Euclidean case $(s,t)=(d,0)$, this would be the usual (hyper)cubic lattice $\ZZ^d$, so $\mathrm{I}_{s,t}$ can be thought of their Lorentzian cousins. 
  \item {\bf Even self-dual lattices}: for any indefinite $V=\mathbb{R}^{s,t}$ (with $s,t$ both $\geq 1$) whose {\bf signature} $\sigma\equiv s-t$ is also a multiple of 8, we also have a unique {\it even} self-dual lattice (denoted ${\rm II}_{s,t}$)\footnote{Interestingly, the condition for the existence of even unimodular lattices \cite{serre2012arithmetic} coincides with the existence of Weyl-Majorana spinors \cite{Polchinski:1998rr}; the author would like to know \textit{why} this beautiful ``coincidence'' happens and explore the potentially intricate connection between Clifford algebras and lattice theory.} which, in an orthonormal (Cartesian) basis, consists of all vectors ${\bf x}\in \mathbb{R}^{s,t}$ with (i) components $x^{\mu}$ all in $\mathbb{Z}$ or all in $\mathbb{Z}+\frac{1}{2}$; and (ii) {\it even} inner product with the vector ${\bf n}=(1,\ldots,1$):
  \begin{equation}
    {\rm II}_{s,t} = \left\{{\bf x} \,\, | \,\,{\it all}\;x^{\mu}\in\mathbb{Z}\;{\rm or}\,{\it all}\,x^{\mu}\in\mathbb{Z}+\frac{1}{2}
    \,\,{\rm and}\;\langle {\bf n},{\bf x}\rangle\in2\mathbb{Z}\;{\rm for}\;{\bf n}=(1,\ldots,1)\right\}.
  \end{equation}
  \label{eq:EvenUnimodularLattice}
  \end{subequations}
Note that in the Euclidean case $(s,t)=(8,0)$ this also describes the usual $E_{8}$ lattice, so the family $\mathrm{II}_{s,t}$ can be thought of as Lorentzian cousins of the exceptional $E_8$ lattice. 
\end{itemize}

When $V={\rm R}^{s,1}$ (with a single time direction) we say spacetime is {\bf Lorentzian}, and $\Lambda$ is a {\bf Lorentzian lattice}. We focus on this case for the rest of this paper. The even and odd self-dual lattices are clearly not the only lattices in Lorentzian space; we will see some more exotic examples in Chapter \ref{ch:GeneralizedSpacetimeCNP}. 

\subsection{Symmetries of Lorentzian Lattices}
\label{subsec:Symmetries_of_Lorentzian_Lattices}

\begin{figure}[!t]
\subfloat
[The Coxeter-Dynkin diagrams describing the reflective symmetries of the odd self-dual lattices $\mathrm{I}_{s,1}$ for $3 \leq s+1 \leq 10$.]
[Odd self-dual lattices $\mathrm{I}_{s,1}$ for $3 \leq s+1 \leq 10$.]
{
\label{fig:VinbergRootsI1nUpto10}
\includegraphics[width=0.45\linewidth]{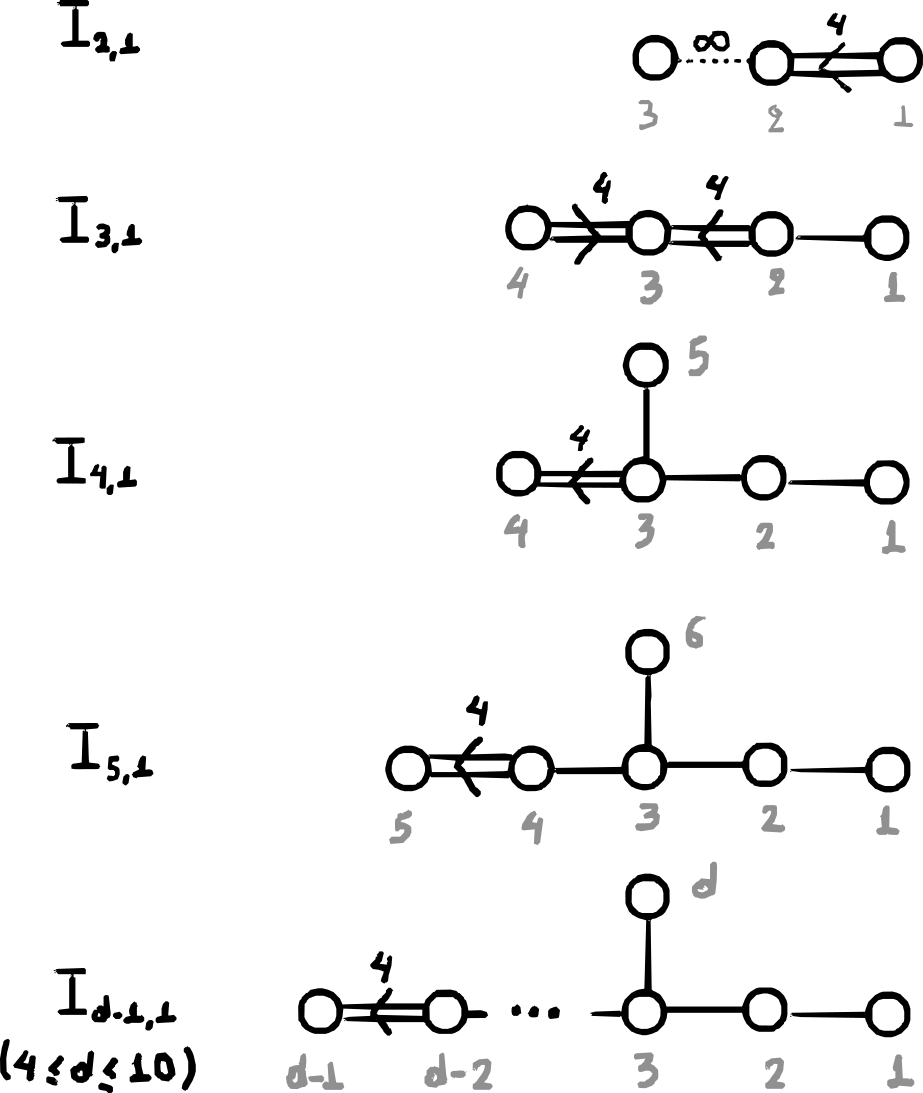}
}
\hspace{+1ex}
\subfloat
[The Coxeter-Dynkin diagrams describing the reflection groups of the even self-dual lattices ${\rm II}_{9,1}$ and ${\rm II}_{17,1}$.]
[Even self-dual lattices ${\rm II}_{9,1}$ and ${\rm II}_{17,1}$.]
{
\label{fig:CoxeterDynkinDiagramsEvenSelfDual}
\includegraphics[width=0.46\linewidth]{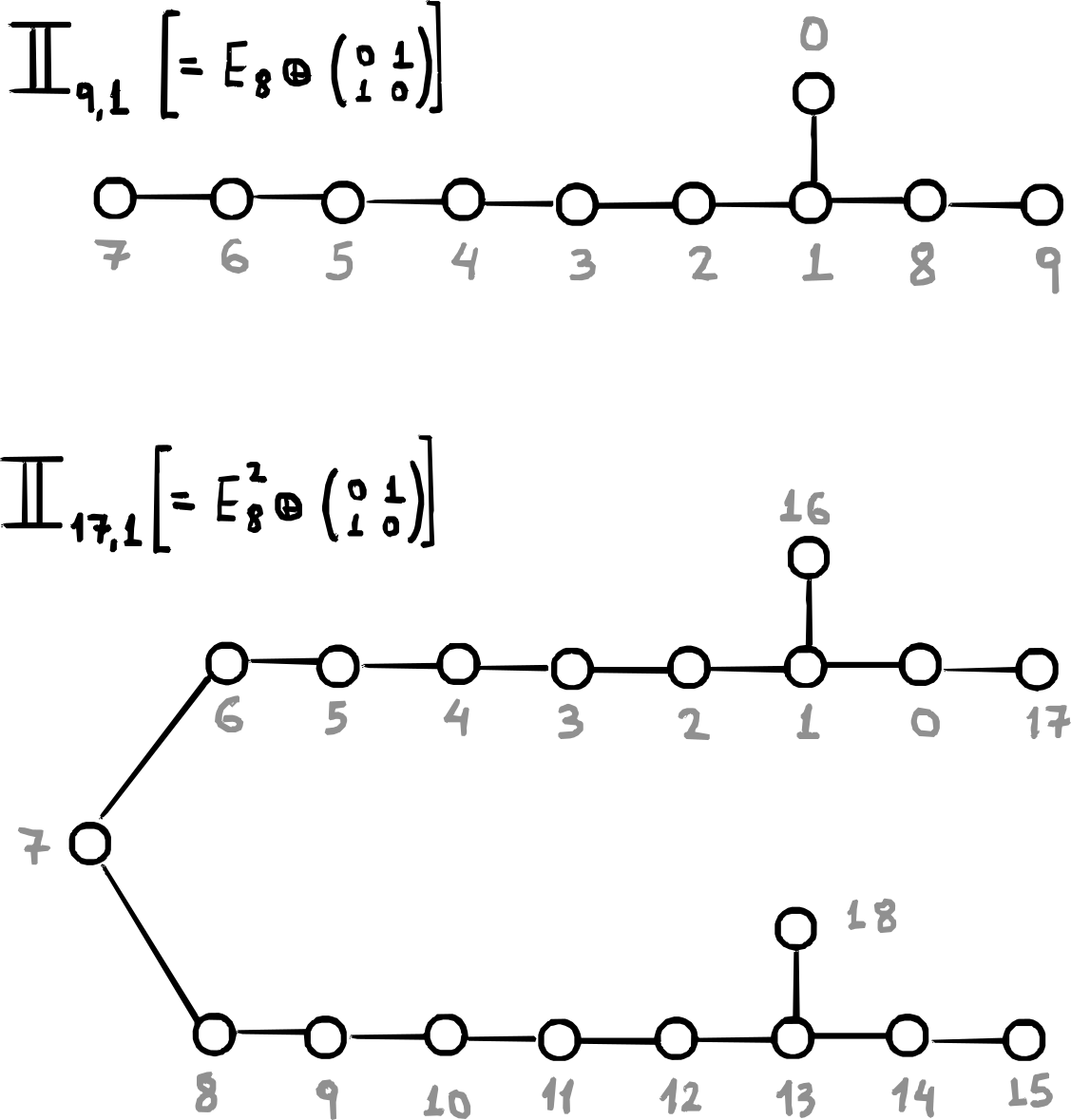}
}
\caption[The Coxeter-Dynkin diagrams describing the reflection groups of the odd and even self-dual lattices $\mathrm{I}_{s,1}$ for $3 \leq s+1 \leq 10$ and ${\rm II}_{9,1}$, ${\rm II}_{17,1}$ respectively.]{The Coxeter-Dynkin diagrams describing the reflection groups of the odd and even self-dual lattices $\mathrm{I}_{s,1}$ for $3 \leq s+1 \leq 10$ and ${\rm II}_{9,1}$, ${\rm II}_{17,1}$ respectively.
In Cartesian coordinates, and numbered as in the figures, the corresponding fundamental roots for the diagrams of Fig.~\ref{fig:VinbergRootsI1nUpto10} are given in Table \ref{tab:SimpleRootsReflectiveOddLatticesN2to17}.~The fundamental roots for ${\rm II}_{9,1}$ are: $i:(0|0^i,+1,-1,0^{7-i})$ for $0\leq i\leq 7$; $8:\,(0|-1^{2},0^{7})$, $9:\,(\frac{1}{2}|(\frac{1}{2})^{9})$. And the fundamental roots for ${\rm II}_{17,1}$ are: $i:(0|0^i,+1,-1,0^{15-i})$ for $0\leq i\leq 15$, $16:(0|-1^2,0^{15})$, $17:(\frac{3}{2}|-\frac{1}{2},(\frac{1}{2})^{16})$, $18:(1|0^{14},1^{3})$}
\label{fig:OddEvenLorentzianLatticesUpToDim10}
\end{figure}

{\bf Vinberg's results.}~For a Lorentzian lattice $\Lambda$, the reflection subgroup of ${\rm Aut}(\Lambda)$ is a hyperbolic Coxeter group (see Sections \ref{sec:Symmetries_of_Lattices} and \ref{sec:SingleMirrorIntersections}).  In dimension $s+1\geq3$, the odd and even self-dual Lorentzian lattices $I_{s,1}$ and (for $s-1=0$ mod 8) ${\rm II}_{s,1}$ have infinite automorphism groups, with infinite reflection subgroups (which are hyperbolic Coxeter groups). Vinberg \textit{et al.}~\cite{vinberg1967discrete, vinberg1972groups, vinberg1975some, kaplinskaja1978groups} studied the symmetries of these lattices and: (i) showed that they were reflective iff $s+1\leq20$; and (ii) in these reflective cases, determined the Coxeter diagrams of the reflection subgroups (using his eponymous algorithm from Section \ref{sec:Symmetries_of_Lattices}). Their results for the odd self-dual lattices ${\rm I}_{s,1}$ are shown in Fig.~\ref{fig:VinbergRootsI1nUpto10} for dimensions $3\leq s+1\leq 10$, and in Appendix \ref{app:SymmetriesLorentzianLattices} for dimensions 
$11\leq s+1\leq 20$.  Their results for the even self-dual lattices ${\rm II}_{s,1}$ ({\it i.e.} ${\rm II}_{9,1}$ and ${\rm II}_{17,1}$) are given in Fig.~\ref{fig:CoxeterDynkinDiagramsEvenSelfDual}.\footnote{Subsequently, Conway succeeded in also working out the automorphism group of the even self-dual lattice ${\rm II}_{25,1}$ and its reflection subgroup: in this case, even the number of {\it fundamental} roots is infinite and, in fact, the fundamental roots are a copy of the remarkable 24-dimensional Leech lattice \cite{conway1983automorphism}.}

\textbf{Generalized Graphical Notation}.~To describe hyperbolic Coxeter groups, we must slightly generalize the graphical notation for Coxeter diagrams introduced Sec.~\ref{sec:Coxeter_Theory}. As before, for two generators $R_i$ and $R_j$ with Coxeter exponent $m_{ij}$, the corresponding vertices $i$ and $j$ in the Coxeter graph are connected by an edge labelled by the integer $m_{ij}$ (and, as before, we use the short hand that, for an edge with $m_{ij}=2$, $3$, $4$ or $6$, we draw no edge, a single edge, a double edge, or a triple edge, respectively). Geometrically, the two corresponding roots ${\bf r}_{i}$ and ${\bf r}_{j}$ have inner product $$\frac{\langle {\bf r}_{i},{\bf r}_{j}\rangle}{(r_{i}^{2}r_{j}^{2})^{1/2}}=-{\rm cos}(\pi/m_{ij}) = \frac{S_{ij}}{2},$$ corresponding to two $s$-dimensional mirrors in $\mathbb{R}^{s,1}$ intersecting at angle $\pi/m_{ij}$ and hence -- from restricting to the intersection of these mirrors with the upper unit hyperboloid $\HH_{+}^s$ in $\mathbb{R}^{s,1}$ -- two $(s-1)$-dimensional hyperbolic mirrors in $s$-dimensional hyperbolic space $\HH_{+}^s$ (also intersecting at angle $\pi/m_{ij}$). For hyperbolic Coxeter groups, though, two more possibilities appear for the exponents $m_{ij}$:
\begin{itemize}
    \item First, two roots can define mirrors that \textit{meet only asymptotically at infinity} in the hyperbolic plane, corresponding to the Coxeter exponent $m_{ij}=\infty$. We called such vertices \textit{ideal} in Sec.~\ref{subsec:SphAffHypCoxeterGroups} and Fig.~\ref{fig:IdealVertices}. To reflect the fact that their mirrors meet only asymptotically, we connect such nodes with a \textbf{dotted line} (instead of Vinberg's notation which is to use a thick line \cite{vinberg1975some}). Ideal vertices give a Schl\"{a}fli entry exactly equal to $S_{ij} = -2$, while the ordinary vertices have strictly $-2 < S_{ij} \leq 0$. 
    \item Second, two roots can define mirrors that {\it do not intersect at all} (even at infinity) in the hyperbolic plane; these are the \textit{ultra-ideal vertices} in Sec.~\ref{subsec:SphAffHypCoxeterGroups} and Fig.~\ref{fig:UltraIdealVertices}. Such pairs of roots start to appear in $\mathbb{R}^{s,1}$ with $s\geq 18$ for the $\mathrm{I}_{s,t}$ lattices, and we connect them with a \textbf{dashed line} (following Vinberg -- see their table below Theorem 2 in \cite{vinberg1975some}). Ultra-ideal vertices give a Schl\"{a}fli entry that is even more negative, $S_{ij} < -2$, which clearly cannot be interpreted as an angle between mirrors for any real label $m_{ij}$ anymore (see Appendix \ref{app:hyperbolicmirrorintersections}).  
\end{itemize}

{\bf Remarks.}  Let us add a few remarks about the symmetries of the self-dual lattices:
\begin{itemize} 
\item For ${\rm I}_{s,1}$, the first $s$ fundamental roots form a sub-diagram of (spherical/finite) type $B_s$, which corresponds to the fact that the odd Lorentzian lattices $\mathrm{I}_{s,1}$ are the Lorentzian analogues of the $B_s$ Euclidean root lattice, which is isomorphic to the hypercubic lattice $\mathbb{Z}^{s}$.
\item In dimension $s+1\leq 10$, the lattices ${\rm I}_{s,1}$ or ${\rm II}_{s,1}$ have $n=s+1$ fundamental roots; these fundamental roots are a (linearly-independent) basis for the lattice (or the Minkowski space $\mathbb{R}^{s,1}$ in which it is embedded); and the fundamental domain of the corresponding hyperbolic Coxeter group is a {\it simplex} in hyperbolic space $H^{s}$ (so these are among the complete list of such hyperbolic Coxeter groups given in Figs. 2 \& 3 of Section 6.9 in \cite{humphreys_1990}).
\item By contrast, in dimension $s+1>10$, the lattices ${\rm I}_{s,1}$ or ${\rm II}_{s,1}$ have $n>s+1$ fundamental roots; so these fundamental roots are {\it over-complete} as a basis for the lattice (or the Minkowski space $\mathbb{R}^{s,1}$ in which it is embedded); and the fundamental domain of the corresponding hyperbolic Coxeter group is {\it not} a simplex, but rather a shape with more codimension-one faces (bounding mirrors).  Note that, for a simplex, any pair of faces intersects in the boundary of the simplex; but for a more complex fundamental domain, a given pair of mirror faces might only intersect {\it outside} the shape ({\it i.e.} if the faces are extended beyond the shape); see Appendix \ref{app:hyperbolicmirrorintersections}. 
\end{itemize}

\subsection{Hyperbolic Coxeter Eigensystems: Salem Numbers and Irrational Newman--Penrose Frames}
\label{subsec:Hyperbolic_Coxeter_Elements}

Given a hyperbolic Coxeter group with a finite number of fundamental roots ${\bf r}_{i}$, we can take the product (in some order) of all the fundamental $d\times d$ reflection matrices $R_{i}$ in Minkowski space to obtain a $d\times d$ Coxeter element $C$.  

If the group's Coxeter diagram is {\bf bicolourable} ({\it i.e.}\ if the vertices in the diagram can be coloured red and blue, such that no two vertices of the same colour are joined by an edge; see Fig.~\ref{fig:non-crystallographicEuclideanFoldings} or Fig.~\ref{fig:HyperbolicNonCrystallographicFoldingsToCoxeterPlane} for some hyperbolic cases), then we can define $C_{{\rm red}}=\prod_{i\in{\rm red}}R_{i}$ as the product of all red fundamental reflections, $C_{{\rm blue}}=\prod_{i\in{\rm blue}}R_{i}$ as the product of all blue fundamental reflections, and the Coxeter element as $C_{\rm colour}=C_{{\rm red}}C_{{\rm blue}}$. We refer to this $C_{\rm colour}$ simply as $C$ below.

For any hyperbolic Coxeter group whose Coxeter-Dynkin diagram has a finite number of nodes and is bicolourable, the eigensystem of $C$ has the following properties (the discussion here is an extension of Sec.~\ref{subsec:Coxeter_Elements} for the hyperbolic case):

\begin{itemize}
\item \textbf{Eigenvalues:} When the dimension is even ($d=2k$), the spectrum of a hyperbolic Coxeter element \cite{a1976valeurs} consists of $k$ pairs of reciprocal eigenvalues:
\begin{equation}
  \lambda_{j}^{\pm}=1/\lambda_{j}^{\mp}\quad(j=0,\ldots,k-1).
\end{equation}
The first pair ($\lambda_{0}^{+}$ and $\lambda_{0}^{-})$ are real and positive with 
\begin{equation}
  \lambda_{0}^{+} = e^{+\vartheta_0}>1,\quad \lambda_{0}^{-} = e^{-\vartheta_0}<1,
\end{equation}
while the remaining $k-1$ pairs are complex-conjugates lying on the unit circle in $\mathbb{C}$
\begin{equation*}
  \lambda_{j}^{\pm}={\rm e}^{\pm i\vartheta_{j}},
  \quad (j=1,\ldots,k-1).
\end{equation*}
If the dimension is odd ($d=2k+1$) there is also an additional eigenvalue $\lambda_{2k+1}=-1$. While this is true for all groups up to dimension $10$ where the fundamental domain is a simplex, it is not for the higher-dimensional cases, where more $\pm 1$ eigenvalues can appear. Here, we are primarily interested in these cases where the fundamental domain is a simplex.

The largest eigenvalue $\lambda_{0}^{+}$ is a {\bf Salem number} (a real algebraic number $\xi > 1$, whose conjugate roots all have absolute value no greater than 1, and at least one of which has absolute value exactly 1).\footnote{If, on the other hand, all conjugate roots of $\xi$ lie strictly within the unit circle (i.e.~no eigenvalue is allowed to have magnitude equal to 1) would make $\xi$ a closely related object known as a \textit{Pisot-Vijayaraghavan number} (or PV for short); see \S 2 in \cite{baake2013aperiodic}.} Salem numbers and their connections to hyperbolic Coxeter elements are of number-theoretic interest (see {\it e.g.}\ \cite{mcmullen2002coxeter, mukunda2007pisot}).  

In particular, {\bf Lehmer's conjecture} is that the {\it smallest} real Salem number is {\bf Lehmer's number} $\lambda_{{\rm Lehmer}}\approx 1.17628$, which is the largest real eigenvalue of the Coxeter element obtained from the first non-trivial even self-dual Lorentzian lattice $\mathrm{II}_{1,9}$ \cite{mcmullen2002coxeter}. For more about the Coxeter polynomial for $\mathrm{II}_{9,1}$, the so-called {\bf Lehmer's polynomial}, see \cite{GROSS20091034}.

\item \textbf{Eigenbasis:} When the dimension is even ($d=2k$), the corresponding eigenvectors are a pair of real vectors $\{ \mathbf{v}_{0}^{+}, \mathbf{v}_{0}^{-}\}$ and $k-1$ pairs of complex-conjugate vectors $\{ \mathbf{v}_j^{+},\mathbf{v}_j^{-}\}$ with ${\bf v}_{j}^{-}=({\bf v}_{j}^{+})^{\ast}$ $(j=1,\dots,k-1)$; and when the dimension is odd ($d=2k+1$), there is an additional real (in fact, rational) eigenvector $\mathbf{v}_{2k+1}$ (corresponding to eigenvalue $\lambda_{2k+1}=-1$).  (We henceforth restrict attention to the case of even dimension $d=2k$, as all statements have trivial extensions to the odd case.)  

Any vector ${\bf w}\in \mathbb{R}^{s,1}$ in $\MM^{s+1}$ may be expressed in this Coxeter eigenbasis as
\begin{equation}
  {\bf w}=\alpha_{0}^{+}{\bf v}_{0}^{+}+
  \alpha_{0}^{-}{\bf v}_{0}^{-}+\sum_{j=1}^{k-1}
  (\alpha_{j}{\bf v}_{j}^{+}+\alpha_{j}^{\ast}{\bf v}_{j}^{-})
  \label{eq:VectorsInCoxeterEigenbasis}
\end{equation}
where the first two coefficients are real ($\alpha_{0}^{\pm}\in\mathbb{R}$) while the other coefficients are complex 
($\alpha_{j}\in\mathbb{C}$). It is straightforward to show (see Appendix \ref{appendix:NPFrames}) that the Coxeter eigenbasis is a special type of basis called a {\bf Newman-Penrose frame} \cite{Newman:1961qr}; that is, in terms of Minkowski bilinear form $\langle {\bf v}|{\bf w}\rangle=\eta_{\mu\nu}v^{\mu}w^{\nu}=v^{\mu}w_{\mu}$, they have the following properties:
\begin{itemize}
\item \textit{Nullness} of each eigenvector
\begin{equation*}
  \langle {\bf v}_{j}^{\pm},{\bf v}_{j}^{\pm}\rangle=0\quad
  (j=0,\ldots,k-1),
\end{equation*}
\item \textit{Normalization} of each eigenvector pair: 
\begin{eqnarray*}
\langle {\bf v}_{0}^{+},{\bf v}_{0}^{-}\rangle&=&-1 \\
\langle {\bf v}_{j}^{+},{\bf v}_{j}^{-}\rangle&=&+1\quad(j=1,\ldots,k-1),
\end{eqnarray*}
\item \textit{Orthogonality} of different eigenvector pairs:
\begin{equation*}
\langle {\bf v}_{j}^{\pm},{\bf v}_{k}^{\pm}\rangle=
\langle {\bf v}_{j}^{\pm},{\bf v}_{k}^{\mp}\rangle=0
\quad (j\neq k).
\end{equation*}
\end{itemize}

Let $G$ denote the group generated by $C$ (which, in the bicolourable case, is a subgroup of the group $G'$ generated separately by $C_{\rm red}$ and $C_{\rm blue}$).  Minkowski space $\mathbb{R}^{s,1}$ transforms as a reducible representation of $G$ (or $G'$): we can split it into $k$ different two-dimensional real irreducible representations (or ${\bf 2}$'s in particle physics notation) called {\bf Coxeter planes}.  The $j$-th Coxeter plane $V_{j}$ is the plane spanned by the two real basis vectors
${\bf u}_{j}^{\pm}=({\bf v}_{j}^{+}\pm{\bf v}_{j}^{-})/\sqrt{2}$.  The Coxeter plane $V_{0}$ has an induced metric of signature $(1,1)$, while the other Coxeter planes $V_{j}$ ($j=1,\ldots,k-1$) have induced metrics of signature $(2,0)$, so we have 
\begin{equation}
\begin{array}{ccccccccc}
V & = & V_0 & \oplus & V_1 & \oplus & \dots & \oplus & V_{k-1} \\
d & = & \mathbf{2} & \oplus & \mathbf{2} & \oplus & \dots & \oplus & \mathbf{2} \\
(s,1) & = & (1,1) & \oplus & (2,0) & \oplus & \dots & \oplus & (2,0). 
\end{array}
\label{eq:VDecompositionHyperbolicCoxeter}
\end{equation}
The Coxeter element $C$ acts as an \textit{irrational rotation} on each Coxeter plane: on $V_0$, it is a rotation by a hyperbolic angle $\vartheta_0 = \mathrm{cosh}^{-1}\left[(\lambda_{0}^{+} + \lambda_{0}^{-})/2\right]$ (i.e.~\textit{a discrete Lorentz boost}); on each of the $V_j$'s, it is an ordinary (Euclidean) discrete rotation by $\vartheta_j = \cos^{-1}\left[{(\lambda_j^{+} + \lambda_j^{-})/2}\right]$. In other words, in the basis $\{{\bf u}_{0}^{+},{\bf u}_{0}^{-},\ldots,{\bf u}_{k-1}^{+},{\bf u}_{k-1}^{-}\}$, the Coxeter element takes the 2-by-2 block-diagonal form 
\begin{equation}
  \setlength{\arraycolsep}{0pt}
  \setlength{\delimitershortfall}{0pt}
  \newcommand*{\myfbox}[1]{%
    \fcolorbox{red}{red!20!white}{$#1$}%
  }
  C = 
   \begin{pmatrix}
    \,\myfbox{\mathrm{B}\big(\vartheta_0\big)}\, & 0  & \,\cdots \,  \\
     0 & \,\myfbox{\mathrm{R}\big(\vartheta_1\big)}\, &  & \\
    \, \vdots \,& & \, \myfbox{\ddots} \, & \\
    & & & \,\myfbox{\mathrm{R}\big(\vartheta_{k-1}\big)}\,
  \end{pmatrix}
  \label{eq:CoxeterActionLorentzianCase}
\end{equation}
where
\begin{equation*}
    \mathrm{B}\big(\vartheta_0\big) = \begin{pmatrix}
        \cosh \left( \vartheta_0 \right) & \sinh \left( \vartheta_0 \right) \\[2ex]
        \sinh \left( \vartheta_0 \right) & \cosh \left( \vartheta_0 \right)\\
    \end{pmatrix}
\end{equation*}
are the 2-by-2 discrete Lorentz boosts
\begin{equation*}
    \mathrm{R}\big(\vartheta_j\big) = \begin{pmatrix}
        \cos \left( \vartheta_j\right) & -\sin \left( \vartheta_j\right) \\[2ex]
        \sin \left( \vartheta_j\right) & \cos \left( \vartheta_j\right)\\
    \end{pmatrix}
    \label{eq:CoxeterElement2by2Blocks}
\end{equation*}
are the 2-by-2 discrete Euclidean rotations. 
\end{itemize}
Note that, since the Coxeter number is infinite for any of the hyperbolic groups (see Sec.~\ref{subsec:Coxeter_Elements}), one can act repeatedly with $C$ given by Eq.~\eqref{eq:CoxeterElement2by2Blocks} and never get the identity. This has the following implications:
\begin{itemize}
    \item \textbf{Infinite Set of Discrete Boosts:} on the $V_0$ plane, $C$ generates any Lorentz boost by some $\vartheta_0^k$ (for any $k \in \ZZ^+$); the possible hyperbolic angles form a discrete set on the real line. Each action of $C$ translates a point on the real line by $F_0 = \ln \lambda_0^+ = \vartheta_0$. As a group, it is isomorphic to $\ZZ$, a \textit{discrete, yet countably infinite, subgroup} of $SO(1,1)$. Physically, one can say that there exist an infinite family of preferred observers that are connected to one another by one of these discrete Lorentz boosts.  
    \item \textbf{Dense Euclidean Rotations:} similarly, on any of the $V_j$ planes of Euclidean signature, $C$ generates a dense set of possible rotation angles on the unit circle $\Sph^1$; essentially, this means that we can approximate any continuous transformation in $U(1) \simeq SO(2)$ with a Coxeter action (to an appropriate integer power). Each Coxeter action rotates a point on the circle by $F_j = (1/2\pi i)\ln \lambda_j^+ = \vartheta_j/2\pi$. The group is again $\ZZ$, which now closes/``compactifies'' into the  circle $\Sph^1$; equivalently, it is an infinite cyclic group and forms a \textit{dense, yet countably infinite, subgroup} of $SO(2)$.   
\end{itemize}
With these observations, the group generated by the coloured Coxeter element is an \textit{infinite dihedral group}, for which we propose the following presentation 
\begin{equation}
    \rm{I}_2^{\infty}({\lambda_0^+}) = \rm{Dih}_{\infty}(\lambda_0^+) = \big\langle \rho_{\rm blue}^2 = \rho_{\rm red}^2 = 1 = C_{\rm colour}^{\infty} : \max \big(\rm{spec}(C_{\rm colour})\big) = \lambda_0^+\big\rangle
    \label{eq:ColoureCoxeterGroup}
\end{equation}
In principle, we get a different group $I_2^{\infty}(\lambda_0^+)$ for each hyperbolic Coxeter graph (or even for each Coxeter element for a graph with loops), which is why we indicate the Salem number $\lambda_0^+$ in the notation. We propose that the fundamental domains for this group are $F_0$ in its $(1,1)$-representation and $F_j$ in each of its $(2,0)$-representations (see Eq.\eqref{eq:VDecompositionHyperbolicCoxeter}).

Note that this group is structurally different from $\widetilde{I}_1^\infty$ of Fig.~\ref{fig:Irreducible_Root_Systems} despite also having an infinite Coxeter number and be defined by two reflection generators.

\section{Spacetime \CNP{} Scheme}
\label{sec:SpacetimeCNP}

In the previous section, we first introduced the ordinary cut-and-project (\CNP) scheme, followed by the more refined Symmetric \CNP{} scheme, in which quasi-periodicity, orientational symmetry and discrete scale symmetry (inflation/deflation symmetry) are tightly interconnected.  The most interesting quasicrystals are produced by the \sCNP{} scheme. As discussed in Sec.~\ref{sec:sCNP}, an \sCNP{} quasicrystal is determined by three choices, so in this section we review these three choices (in blue boxes), and discuss any novelties associated with generalizing each one to the Lorentzian spacetime context.

\subsection{The Lattice $\Lambda$}

\begin{center}
\begin{tcolorbox}[colframe=blue,colback=white,width=0.99\textwidth]
\textbf{Choice 1:}
a $d$-dimensional lattice $\Lambda^{\mathbf t}$ in
$V=\mathbb{R}^d$ with automorphism group
${\rm Aut}(\Lambda)$.
\end{tcolorbox}
\end{center}

Now $\Lambda$ is a {\it Lorentzian} lattice in $V=\mathbb{R}^{s,1}$.  Whereas ${\rm Aut}(\Lambda)$ for a Euclidean lattice was a {\it finite} group, ${\rm Aut}(\Lambda)$ for a Lorentzian lattice in dimension $s+1\geq3$ can be an {\it infinite} group: in particular, 
\begin{enumerate} 
  \item for an odd or even self-dual lattice $I_{s,1}$ and $II_{s,1}$ (see Figs.~\ref{fig:VinbergRootsI1nUpto10} and \ref{fig:CoxeterDynkinDiagramsEvenSelfDual});
  \item for any {\bf Lorentzian root lattices}, {\it i.e.}\ the integer span of the fundamental roots of any of the crystallographic hyperbolic Coxeter groups enumerated in Secs.~6.7--6.9 and Figs.~2 \& 3 of Ref.~\cite{humphreys_1990}.
\end{enumerate}
In this Chapter, we will mostly focus on the self-dual lattices; after all, the nicest quasicrystals in Euclidean space are constructed from either a hypercubic or the $E_8$ lattice, which are themselves self-dual. In Chapter \ref{ch:GeneralizedSpacetimeCNP} we will tackle more general Lorentzian lattices. 

\subsection{The Non-Crystallographic Symmetry}

\begin{center}
\begin{tcolorbox}[colframe=blue,colback=white,width=0.99\textwidth]
\textbf{Choice 2:} a non-crystallographic subgroup $G\in{\rm Aut}(\Lambda)$, and a decomposition $V=V_{{\rm ph}}\oplus V_{{\rm in}}$ that is invariant under $G$: $\Pi_{{\rm ph}}=g\Pi_{{\rm ph}}g^{-1}$ (or, equivalently $[\Pi_{{\rm ph}},g]=0$), $\forall g\in G$. 
\end{tcolorbox}
\end{center}

Whereas for Euclidean quasicrystals, the relevant (finite/spherical) non-crystallographic subgroups have all been enumerated (see Fig.~\ref{fig:Irreducible_Root_Systems}), for Lorentzian quasicrystals, the relevant (hyperbolic) non-crystallographic subgroups have not been, in general, although the non-crystallographic groups with simplicial fundamental domains {\it have} been enumerated in Figs. 2 \& 3 of Section 6.9 in \cite{humphreys_1990}. 
It would be interesting to determine the non-crystallographic subgroups of ${\rm Aut}(\Lambda)$ for each Lorentzian lattice $\Lambda$ mentioned above -- we present the first steps towards that goal in Chapter \ref{ch:GeneralizedSpacetimeCNP}.

In the present Chapter, we focus for concreteness on a particular non-crystallographic subgroup $G\in{\rm Aut}(\Lambda)$ which can be constructed for any Lorentzian lattice $\Lambda$ listed above: namely, the group $G$ generated by a Coxeter element $C$ of $\Lambda$ ({\it i.e.}\ the product of the fundamental reflections corresponding to the fundamental roots of $\Lambda$). When the associated Coxeter graph of $\Lambda$ is bicolourable, which is the case for all self-dual lattices (up to $d \leq 10$), we implicitly assume the coloured Coxeter element $C_{\rm colour}$.

As explained in Sec.~\ref{subsec:Hyperbolic_Coxeter_Elements}, the action of $G$ splits $V$ into 2D subspaces $V_{j}$ ($j=0,\ldots,k-1$) that transform as 2D irreducible representations of $G$.  To obtain a {\it spacetime} quasicrystal, we must take the physical subspace $V_{\rm ph}$ to consist of the subspace $V_{0}$ which has Lorentzian signature $(1,1)$, and some (but not all) of the remaining subspaces $V_{j}$ with Euclidean signature $(2,0)$. This amounts to choosing the projector $\Pi_{\rm ph}$ to be
\begin{eqnarray}
  (\Pi_{\rm ph})^{\mu}_{\;\;\nu}&=&
  \sum_{j=0}^{k \leq k_{\rm ph}}\left(\frac{(v_{j}^{+})^{\mu}(v_{j}^{-})_{\nu}}{\langle {\bf v}_{j}^{+},{\bf v}_{j}^{-}\rangle}+\frac{(v_{j}^{-})^{\mu}(v_{j}^{+})_{\nu}}{\langle {\bf v}_{j}^{-},{\bf v}_{j}^{+}\rangle}\right)  
\end{eqnarray}
This induces a decomposition $V=V_{\rm ph}\oplus V_{\rm in}$ that is invariant under $G$ and, in the case where the original Coxeter group is bicolourable, also under the larger subgroup 
$G'=\cCG$ \eqref{eq:ColoureCoxeterGroup} (which includes $G$ as a subgroup).

Note that this subgroup $G$ generated by the Coxeter element (or, in the bicolourable case, the larger group $G'$), is the Lorentzian generalization of the group that is used in the Euclidean case to generate well-known quasicrystals like the Ammann-Beenker tiling (starting from $\Lambda=\mathbb{Z}^{4}$) or the Penrose tiling (starting from $\Lambda=\mathbb{Z}^{5}$ or $\Lambda=A_{4}$). But, whereas $G$ (or $G'$) is a {\it finite} group in the Euclidean case, it is an {\it infinite} group in the Lorentzian case (since the Coxeter number $h$ is finite for finite/spherical Coxeter groups, but infinite for hyperbolic Coxeter groups).  So Lorentzian quasicrystals have infinitely more orientational symmetry than their Euclidean cousins!

\subsection{The Window}
\label{subsec:SpacetimeCNPWindow}

\begin{center}
\begin{tcolorbox}[colframe=blue,colback=white,width=0.99\textwidth]
\textbf{Choice 3:} a window ${\cal W}\subset V_{{\rm in}}$ or weighting function $W:V_{{\rm in}}\to\mathbb{C}$ invariant under $G$: {\it i.e.}\ ${\cal W}=g{\cal W}$ or $W({\bf v}_{{\rm in}})=W(g\,{\bf v}_{{\rm in}})$, $\forall g\in G$.
\end{tcolorbox}
\end{center} 

In the Euclidean case, since the non-crystallographic group $G$ was a {\it finite} group, the window ${\cal W}\in V_{\rm in}$ or weighting function $W:V_{\rm in}\to\mathbb{C}$ only needed to have finite/discrete rotational symmetry.  For example, the 8-fold symmetric Ammann-Beenker tiling was produced by an 8-fold symmetric window ${\cal W}$ (see Fig.~\ref{fig:sCNPWindowABTiling}).

\begin{figure}[p]
\centering
    \begin{tabular}{cc}
        \subfloat[Apeirogon on $\HH^1_+$
        \label{fig:HyperbolaSeedsApeirogon}]{
        \includegraphics[width=0.45\linewidth]{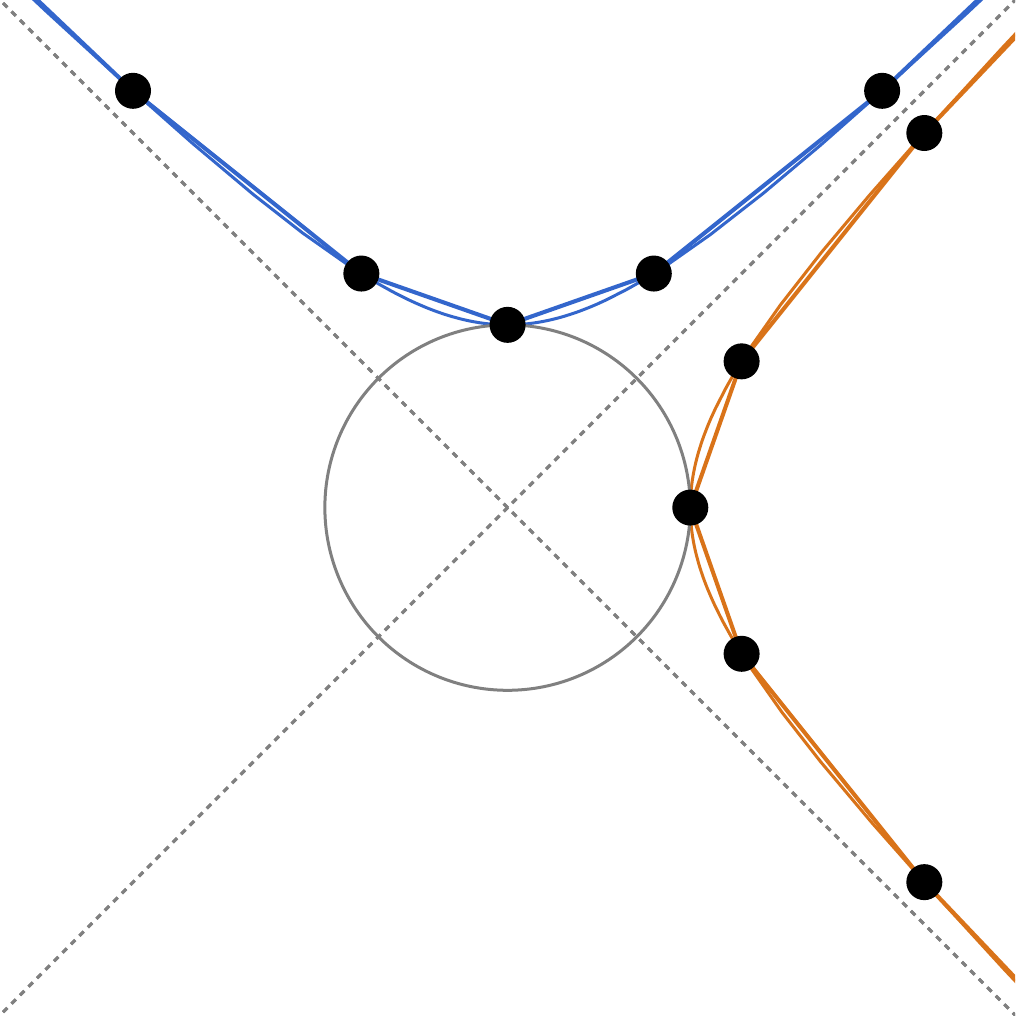}
            \hspace{1em}}
        \subfloat[Apeirogram on $\Sph^1$
        \label{fig:CircleSeedsStellatedApeirogon}]{    
            \includegraphics[width=0.47\linewidth]{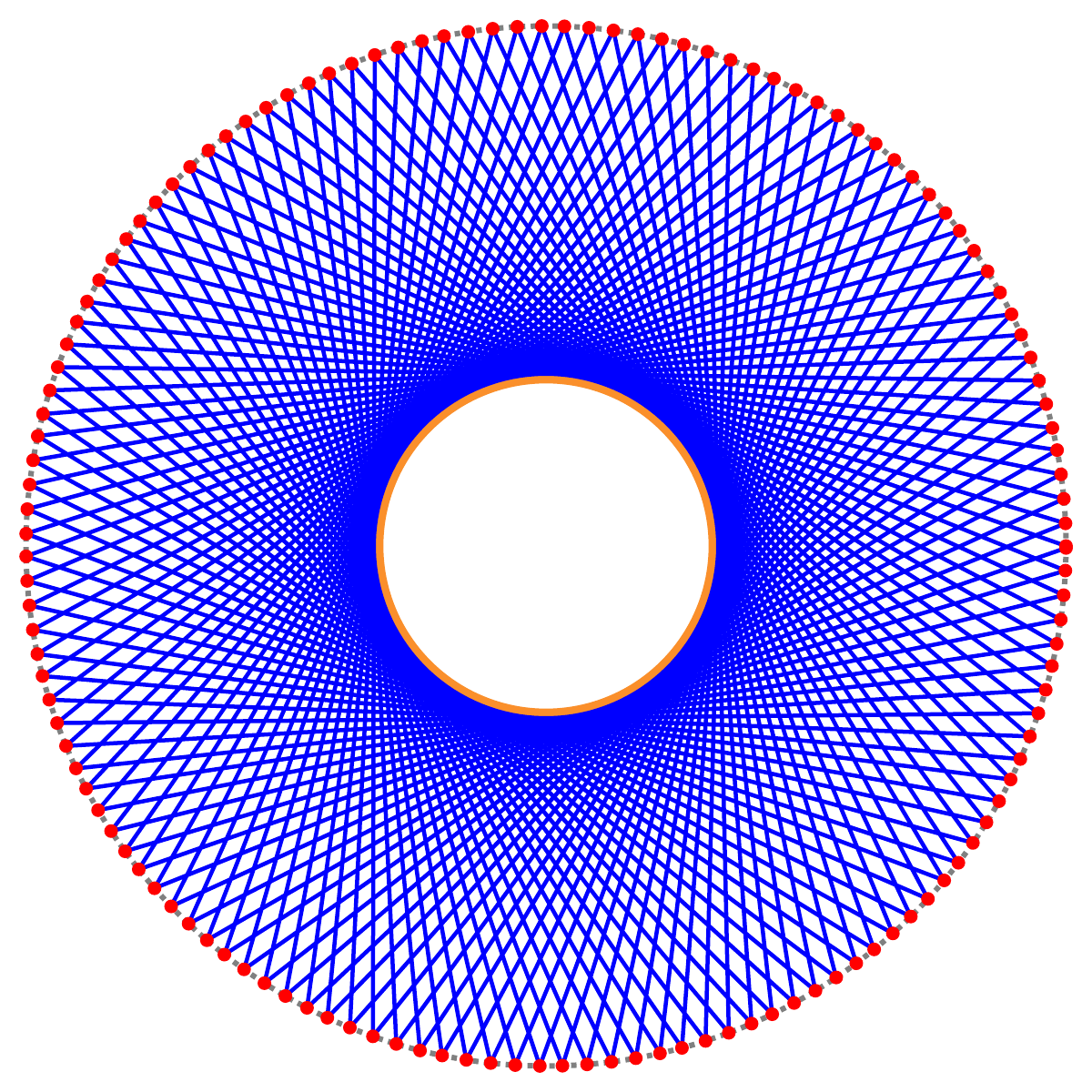}
        } \\
         \rule{0pt}{3ex}\\   
       \subfloat[Rolled-out Apeirogram on $\RR$
        \label{fig:HyperbolaUnrolledOnLine}]{
    \raisebox{9mm}{%
        \includegraphics[width=0.47\linewidth]{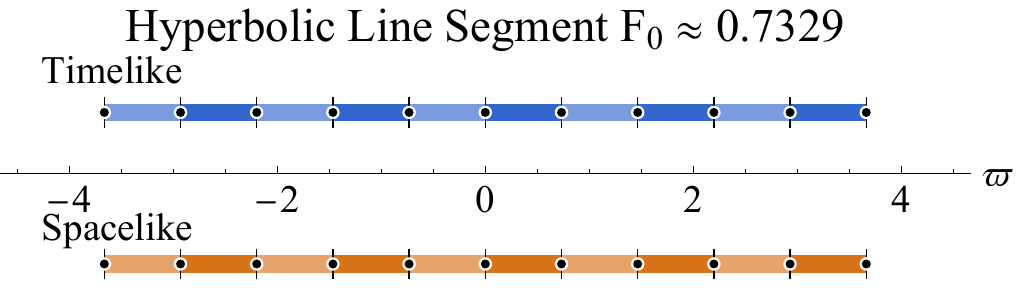}
    }
}
        \hspace{1em}
        \subfloat[Rolled-out Apeirogram on $\RR$. 
        \label{fig:CircleUnrolledOnLine}]{\includegraphics[width=0.53\linewidth]{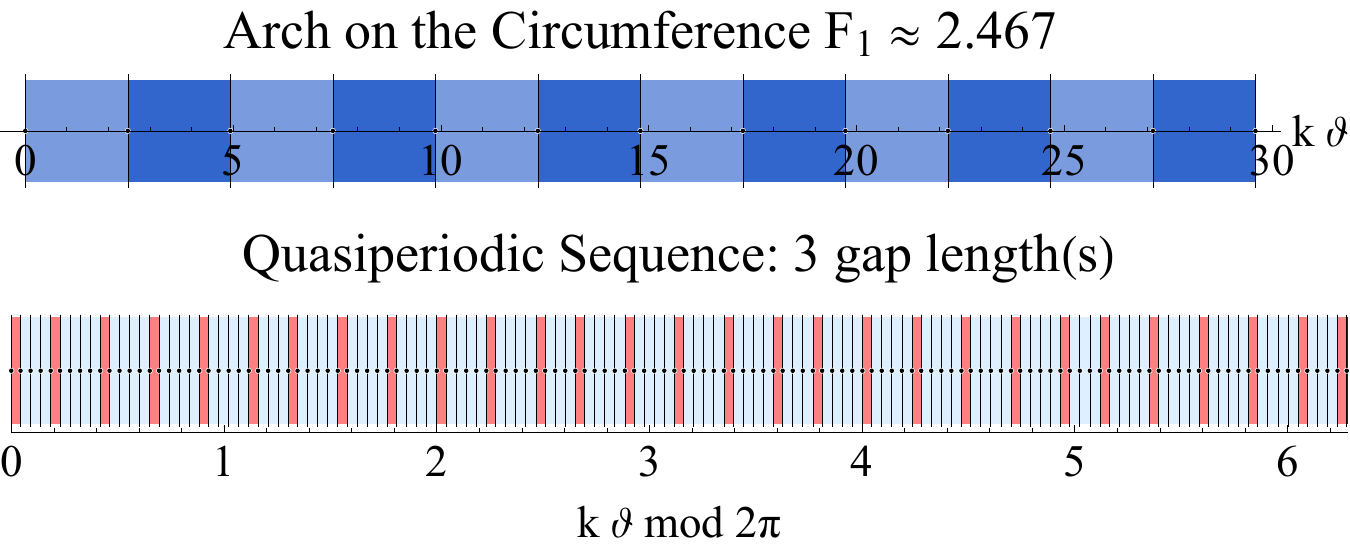}}
    \end{tabular}
\caption[
Apeirogon and Apeirogram for the infinite dihedral group generated by $\mathrm{I}_{3,1}$]{Apeirogon and Apeirogram for the infinite dihedral group generated by $\mathrm{I}_{3,1}$. On the left, we see how the Coxeter action divides the timelike and spacelike hyperbolas into segments of the same (affine) length. Below, the hyperbolas are unrolled into lines to see that the segments are indeed the same. On the right, one can see the apeirogram. For concreteness, we truncated the irrational Coxeter rotation after 135 applications, which is exactly how many are needed to come `very close' to the starting point (here, $(1,0)$). Each chord joins two points separated by central angle $\vartheta_1$, so every such line is tangent to a circle of radius $R = R_0 |\cos \vartheta_1/2|$ ($R_0 = 1$ here for our choice of the initial point). Below, we see the 12 first arches on $\Sph^1$ (each of length $F_1$), and the quasi-periodic sequence of successive differences of points on $\Sph^1$ (the three-gap theorem guarantees there are exactly three distances; the smallest one is not visible but is the last one, corresponding to the point of closest approach to $(1,0)$; see Footnote \ref{foot:FailedStars}).} 
\label{fig:ApeirogonAndApeirogram}
\end{figure}

By contrast, in the Lorentzian case, since the non-crystallographic group $G$ has infinite order, a symmetric window ${\cal W}$ or weighting function must also have a infinite rotational symmetry. Since the group $\cCG$ exists within all (bicolourable, otherwise just take its subgroup generated by a Coxeter element) hyperbolic Coxeter groups and it acts as an irrational rotation on each of the Coxeter planes within the internal space, our window
$\mathcal{W}$ must be a polytope with a countably infinite number of edges. This is a polytope known as
an \textbf{apeirogon} \cite{coxeter1973regular}. A regular apeirogon can be defined as a partition of the Euclidean line $\EE^1$ into infinitely many equal-length segments.\footnote{It generalizes the regular $n$-gon $\{ n \}$, which may be defined as a partition of the circle $\Sph^1$ into finitely many equal-length segments. For this reason, the apeirogon is denoted as $\{ \infty \}$.} Going back to our discussion in Section \ref{subsec:Hyperbolic_Coxeter_Elements}, we can now see that this is indeed the case for the boost-plane $V_0$: under the action of $C$, $\EE^1$ splits into segments of length $F_0$. We can see that on the left of Fig.~\ref{fig:ApeirogonAndApeirogram}.

The situation on the internal Coxeter planes is a bit more interesting, as we can see on the right hand-side of Fig.~\ref{fig:ApeirogonAndApeirogram}. The irrational Coxeter rotation chops the circle $\Sph^1$ into different overlapping segments of arch-length $F_j$. Connecting successive points under this rotation, one gets a \textbf{star apeirogon} or \textbf{apeirogram} \cite{coxeter1934discrete}. A \textit{star polygon} (or \textit{stellated polygon)} is a geometric shape created by extending the edges of a convex regular polygon outwards until they intersect to form a new, star-like closed -- but non-convex -- boundary. This process, known as \textit{stellation}, produces self-intersecting shapes. One can think, for example, of the regular pentagon as the shape we get by connecting each vertex by an edge to another vertex which is $2\pi/5$ away on $\Sph^1$. This way we get a partition of $\Sph^1$ onto 5 equal-length fundamental domains, which are permuted by a rational rotation (by $2\pi/5$). On the other hand, if one connects each vertex to another vertex which is now $4\pi/5$ away, we get a regular star pentagon, or \textit{pentagram}; for this shape, it looks as if the 5 fundamental domains are each $4\pi/5$ wide and overlap one another. While all such \textit{star polygons} have a finite number of vertices, edges and winding number (i.e.~the number of rotations needed to close the star), an irrational rotation produces a \textit{failed star polygon}, which has infinitely many vertices and edges, and an infinite winding number, since the star continuously wraps itself infinitely many times. This is precisely the type of apeirogram we are dealing with here, as shown in Fig.~\ref{fig:ApeirogonAndApeirogram}.\footnote{
\label{foot:FailedStars} We note that there is nothing bad about being a ``failed star polygon''. After all, the theory of music is based on one: the \href{https://en.wikipedia.org/wiki/Three-gap_theorem}{chromatic circle}. If a failed star polygon is behind all harmonious musical notes, it cannot be that detrimental for a star polygon to never close! We also note that the three-gap theorem (see the previous link) guarantees that if one places $N$ points on $\Sph^1$ at angles $\Theta, \dots, N\Theta$ from the starting point, then there will be at most \textit{three distinct gaps} between pairs of points in adjacent positions around $\Sph^1$ (and when there are three distances, the largest of the three always equals the sum of the other two).} 

In practice, then, we smoothen out the polytope to the full circle. We note that by doing so, we lose information about the irrational number that generates the rotations, since, instead of considering the infinite dihedral group, we take the full $U(1)$.\footnote{It would be interesting to see if we can avoid this and use a window that ``knows'' about the irrationality of each Coxeter group; at the time of writing, we do not know how to achieve this. Forgetting about this irrationality might be another reason why we cannot promote the spacetime quasicrystals into tilings, as explained in Section \ref{sec:implications}.} This implies that, in each ``internal'' Coxeter plane $V_{j}$ ($k_{\rm ph}<j\leq k$), the window must be a {\it circularly-symmetric disk} of radius $\rho_{j}$, also called a \textit{polydisk}, or the weighting function must be circularly-symmetric -- {\it i.e.}\ a function of the form
\begin{equation}
  W\left(\frac{r_{(k_{\rm ph}^{}+1)}}{\rho_{(k_{\rm ph}^{}+1)}},\ldots,\frac{r_{k}^{}}{\rho_{k}^{}}\right)
\end{equation}
where $r_{j}$ is the radial coordinate in the $j$th Coxeter plane $V_{j}$ ({\it i.e.}\ the coordinate that measures the distance from the origin in that plane, according to the induced metric).

\section{Other implications}
\label{sec:implications}

Let us mention some other novel features of Lorentzian quasicrystals (compared to their Euclidean cousins).

\subsection{Point Set, But No Tiling} 
 
So far have described the \sCNP{} algorithm with symmetric window ${\cal W}$ as a way to generate a $G$-symmetric quasi-crystalline point set (for some non-crystallographic group $G$).  But in the Euclidean case (where $G$ is a finite group), one can often (as in the Penrose tiling, or Ammann-Beenker tiling) regard these points as the endpoints of edges; these edges as the boundaries of faces, etc, so that the $G$-symmetric point-set is promoted to a $G$-symmetric {\it tiling} or {\it tessellation} (see Section \ref{subsec:Tile_Or_Not}).

By contrast, in the Lorentzian case, one cannot promote a $G$-symmetric Lorentzian \sCNP{} point set to a $G$-symmetric tiling (with a finite number of polygonal tiles). As mentioned in Section \ref{subsec:Tile_Or_Not}, if the $G$-symmetric tiling contains any tile $T$, it must also contain the ``rotated'' tile $gT$ ($\forall g\in G$). But if $G$ is an infinite group, this will either imply that the tile $T$ is accompanied by an infinite number of other, differently-oriented tiles, or else, $T$ must be invariant under a finite-index subgroup of $G$ (in which case the tile itself is ``spherically symmetric''). We summarize this result in the following no-go theorem:

\begin{notile} One cannot promote a $G$-invariant spacetime quasicrystal to a $G$-invariant spacetime quasi-tiling with a finite number of polygonal tiles for any infinite non-crystallographic symmetry group $G$.
\label{thm:notiletheorem}
\end{notile}

In fact, having an infinite number of tiles might not be too restricting at all. For example, the \textit{pinwheel tilings} \cite{Pinwheel_Radin_1994,radin1997aperiodic,Radin1999MilesOT} (see also \S 6.6 in \cite{baake2013aperiodic}) have also an infinite number of tiles; the simplest such tiling has infinitely many rotated copies of a single triangular tile, all congruent to the original. Thus, there is only a single equivalence class of tiles, but it contains infinitely many members. Such tilings, though, lack a property that the rest of the tilings we considered so far had: they do not have \textit{finite local complexity} (FLC). Loosely speaking, tilings with FLC have their prototiles arranged only in a finite number of possible configurations (or types of adjacencies). More formally, we say that a pattern ${\cal P} \subset V$ has FLC (with respect to translations) if, for every compact region $K \subset V$, the collection of $K$ and all its translated copies, $\left\{ ({\bf t} + K) \cap {\cal P} \,\, | \,\, {\bf t} \in V \right\}$, contains only a finite number of configurations of fragments inside it \cite{baake2013aperiodic}. The pinwheel tilings do not have FLC according to this definition. However, one can define FLC more generally with respect to the full group of isometries of $V$, ${\rm Iso}(V)$; this includes translations and rotations, if $V$ is Euclidean (these are also called \textit{Euclidean motions}), and the full Poincar\'e transformations (the latter + Lorentz boosts), if $V$ is Lorentzian. Then, the pinwheel tiling considered in \S 6.6 of \cite{baake2013aperiodic} \textit{has} FLC with respect to Euclidean motions, which explains its high degree of order over long scales.   

The pinwheel tilings are part of a family of structures with \textit{infinite local complexity} (ILC) instead -- with respect to translations -- and their theory has been understood to some extent in the works of   \cite{frank2007topologysometilingspaces,frank2013tilingsinfinitelocalcomplexity,frank2018fusiontilingsinfinitelocal} (see references therein). More strictly, then, what we have shown above in our \textit{No-Tile Theorem} is not that we cannot tile our spacetime quasicrystals at all, but even if we could, they would have ILC instead of FLC.\footnote{We thank our external examiner, Felix Flicker, for pointing out this possibility to us during the defence.} Having said that, it would be interesting to explore whether it is still possible to have a $G$-invariant spacetime quasi-tiling with a finite number of equivalence classes of tiles, each of them having infinitely many members related by the infinite symmetry group. 

If such a $G$-symmetric spacetime quasi-tiling exists, then, could it still be obtained via \sCNP{}? That we do not know. The reason is that the pinwheel tilings are not known to arise via the \CNP{} method because of their dense, circular rotational symmetry (similar to our spacetime quasicrystals, their rotational symmetry is generated by some irrational rotation). Instead, they are usually defined in terms of substitution rules (inflation and subdivision of tiles). Within the \sCNP{} framework, we can understand why we cannot tile our spacetime quasicrystals with the traditional methods as follows: the circular (or poly-disk) window ${\cal W}$ is simply too smooth (with ``infinite edges'') to lift to some higher-dimensional cell $F_{\Lambda} \in V$ with a finite number of faces that can give the lower-dimensional tiles (see Section \ref{subsec:Tile_Or_Not}).
To define the tiling with the inflate-subdivide-inflate prescription used in the pinwheel tiling, on the other hand, we would need to know at least some prototiles, which we do not know either. We are led to conclude, then, that if such a tiling exists, it would demand new methods for its construction.   

Finally, it is also worth mentioning that traditional metric-based approaches that tessellate Euclidean point sets -- such as Voronoi's tessellation algorithm, or its dual, Delone's\footnote{As we discovered with my advisor Latham after a certain funny misunderstanding during a meeting, the soviet mathematician Boris Nikolayevich Delaunay lives a double life in the literature as both \textit{Delaunay} (the surname given to him from his French ancestor, an officer in the French Army captured in Russia during Napoleon's invasion of 1812) and \textit{Delone} (a transliteration from Cyrillic of the translated French surname into the Russian language).} triangulation algorithm (see \S 2.2 in \cite{baake2013aperiodic} and \cite{conway2013sphere})-- would not work in our case. This is because of the indefiniteness of the Minkowski metric, which allows both spacelike and timelike distances.\footnote{Because of the irrationality of the physical space, our spacetime quasicrystals admit no null separated vertices that would lead to an infinite number of nearest neighbours exactly distance-0 away.} On a spacetime quasi-crystalline pattern, this issue manifests in an inherent \textit{non-locality}: for example, for any given point, its nearest neighbour (the point closest to being spacetime-distance zero away from it) might lie very far away from it on the pattern. We will come back to this in Section \ref{sec:BridgeQG}.

\subsection{No Preferred Window Size}

In the Euclidean case, if the window ${\cal W}\in V_{\rm in}$ is chosen to have a certain natural size, it uniquely lifts to a cell ${F_\Lambda}\in V$ whose vertices have rational coordinates in the lattice basis, and whose $n$-faces project onto $V_{\rm ph}$ to give the $n$-tiles in the quasi-crystalline tiling (as in Fig.~\ref{fig:sCNPWindowABTiling}).  

Is there a natural size for the spherically-symmetric window ${\cal W}$ in the Lorentzian case?  The answer appears to be negative: the larger the radius, the denser the projected set $\Lambda_{\rm phys}({\cal W})$, but there seems to be no natural or intrinsic way to choose a particular radius as the best one.  Relatedly, a spherically-symmetric window does not have a unique lift from $V_{\rm ph}$ to $V$.

\subsection{No Local Scale Invariance}

In Section \ref{subsec:ScaleInvariance}, we explained two notions of scale invariance: local and global.  And, in particular, in \ref{subsec:LocalScaleInvariance} we explained that a \CNP{} quasicrystal has local scale invariance -- also known as ``local inflation/deflation symmetry'' (LIDS) -- if and only if the inflated window $\widetilde{{\cal W}}$ can be obtained via a finite number of unions and intersections of translated copies of the original window ${\cal W}$; and vice versa \cite{baake1991quasiperiodic, baake2013aperiodic}.  For a spherically-symmetric window this is impossible (a rescaled ball cannot be expressed as a finite number of unions and intersections of translated copies of the original ball).  So a $G$-symmetric spacetime quasicrystal (with infinite non-crystallographic symmetry group $G$) cannot have local inflation invariance.  

\subsection{Self-Duality and Global Scale-Invariance} 

So far in Sec.~\ref{sec:implications} we have focused on spacetime \sCNP{} quasicrystals produced via the {\it window} scheme; and we have seen that (in the spacetime context) there is no $G$-invariant tiling associated with the resulting quasi-crystalline point set, no preferred size of the window ${\cal W}$ (and hence no preferred density of the point set), and no local inflation-deflation symmetry. 

Let us turn to consider spacetime \sCNP{} quasicrystals produced via the {\it weighting} scheme:
\begin{itemize}
\item If we take the weighting function $W$ to be a Gaussian, there {\it is} a preferred width of the Gaussian -- namely the width that makes the quasicrystal self-dual, just as explained in Sec.~\ref{sec:SelfDualQuasicrystals}.
\item If we take the weighting function $W$ to be a power law, we obtain a quasicrystal with global scale invariance, just as explained in Sec.~\ref{subsec:GlobalScaleInvariance}.  
\item In particular, if we take the particular case of a trivial power law, $W({\bf v}_{\rm in})=1$, so we project each point in the lattice $\Lambda$ with equal weight, we effectively obtain the dense point set $\Lambda_{\rm ph}=\Pi_{\rm ph}\Lambda$.  This does indeed have global scale invariance: when this dense set is rescaled by certain special numbers, it is precisely mapped to itself, one-to-one.  These special numbers are the real units of the algebraic number field associated with the eigenvalues of the Coxeter element (see Appendix \ref{app:inflation})  
\end{itemize}
Given all of these observations, we see that in the spacetime context, the weighting scheme is more natural than the window scheme; and the self-dual or globally-scale-invariant quasicrystals produced by the weighting scheme are most natural of all.

\section{Examples of Spacetime Quasicrystals}
\label{sec:ExamplesSpacetimeQuasicrystals}

In this section, we finally apply the machinery developed so far to construct the first examples of quasicrystals in Lorentzian spacetime. 

\subsection{Example 1: $(1+1)$-dimensional Quasicrystals from $\mathrm{I}_{3,1}$}
\label{subsec:2DLorentzianQuasiCrystal}

\begin{figure}[p]
\centering
\begin{threeparttable}
    \begin{tabular}{cc}
        \subfloat[Self-dual Variant
        \label{fig:SpacetimeCNP-I13SelfDualPattern}]{
        \includegraphics[width=0.35\linewidth]{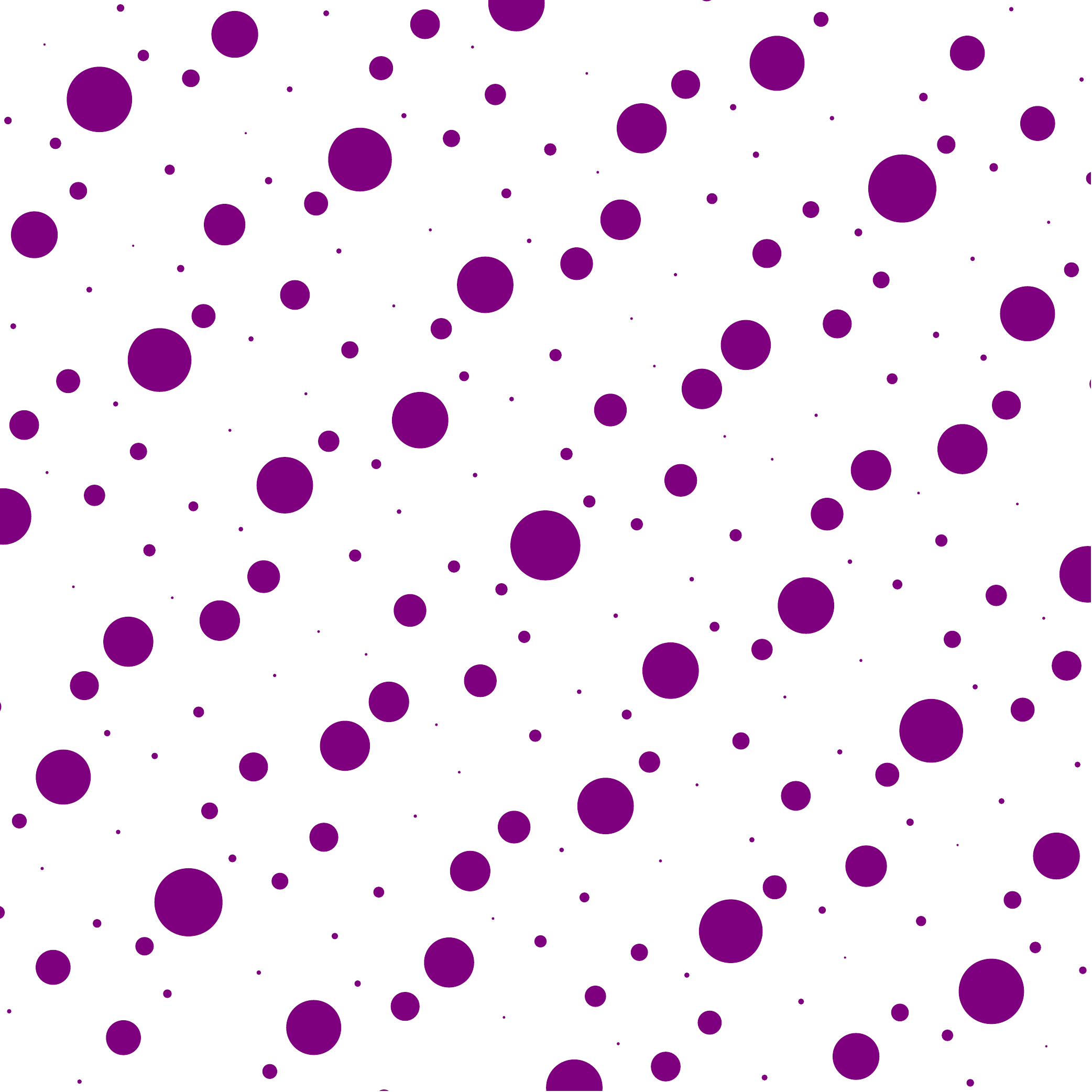}
            \hspace{1em}}
        \subfloat[Finite-Density Variant
        \label{fig:SpacetimeCNP-I13SphericalWindow}]{    
            \includegraphics[width=0.35\linewidth]{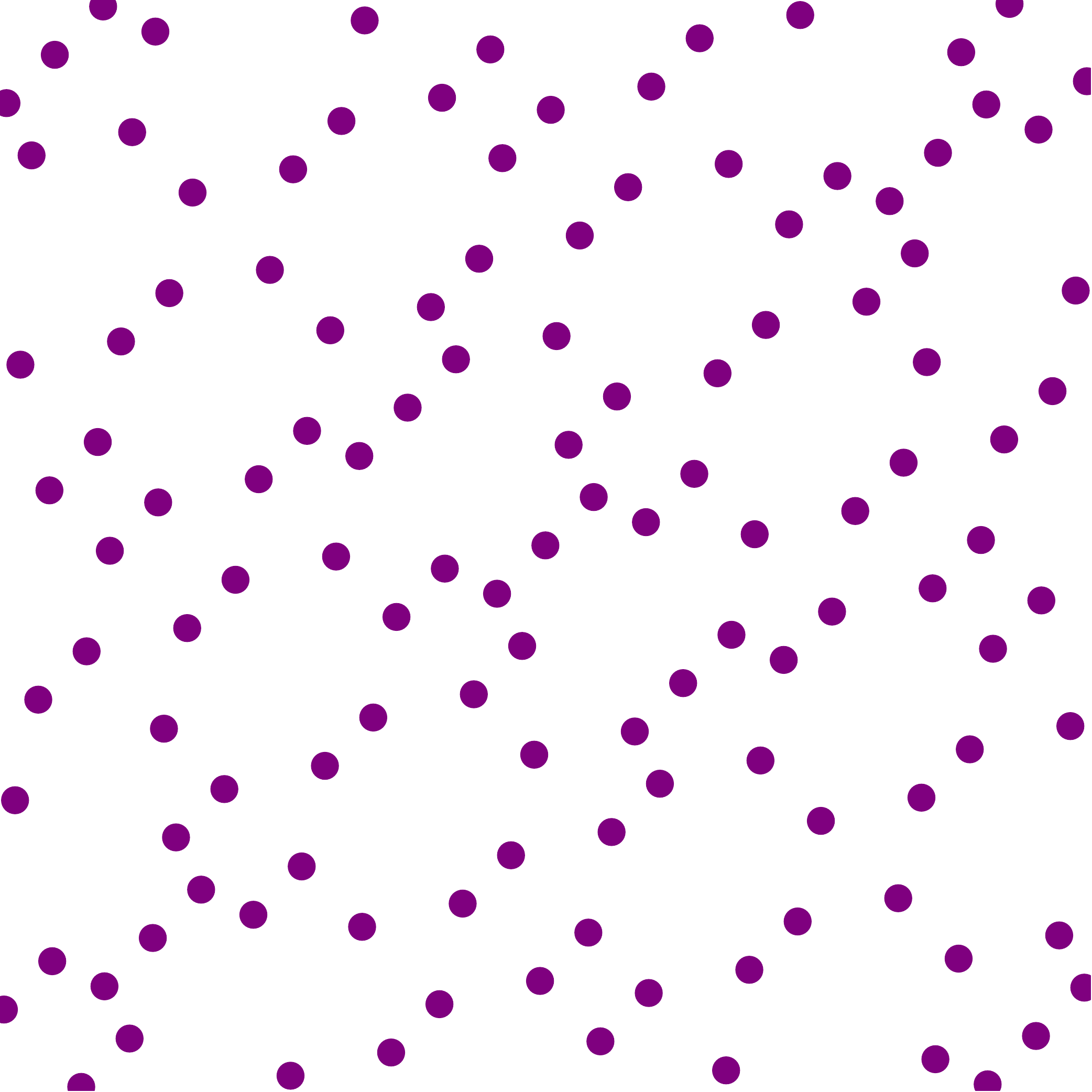}
        } \\
         \rule{0pt}{3ex}\\   
        \subfloat[Scale-Invariant ($k=1.01$)
        \label{fig:SpacetimeCNP-I13ScaleInvPattern}]{\includegraphics[width=0.35\linewidth]{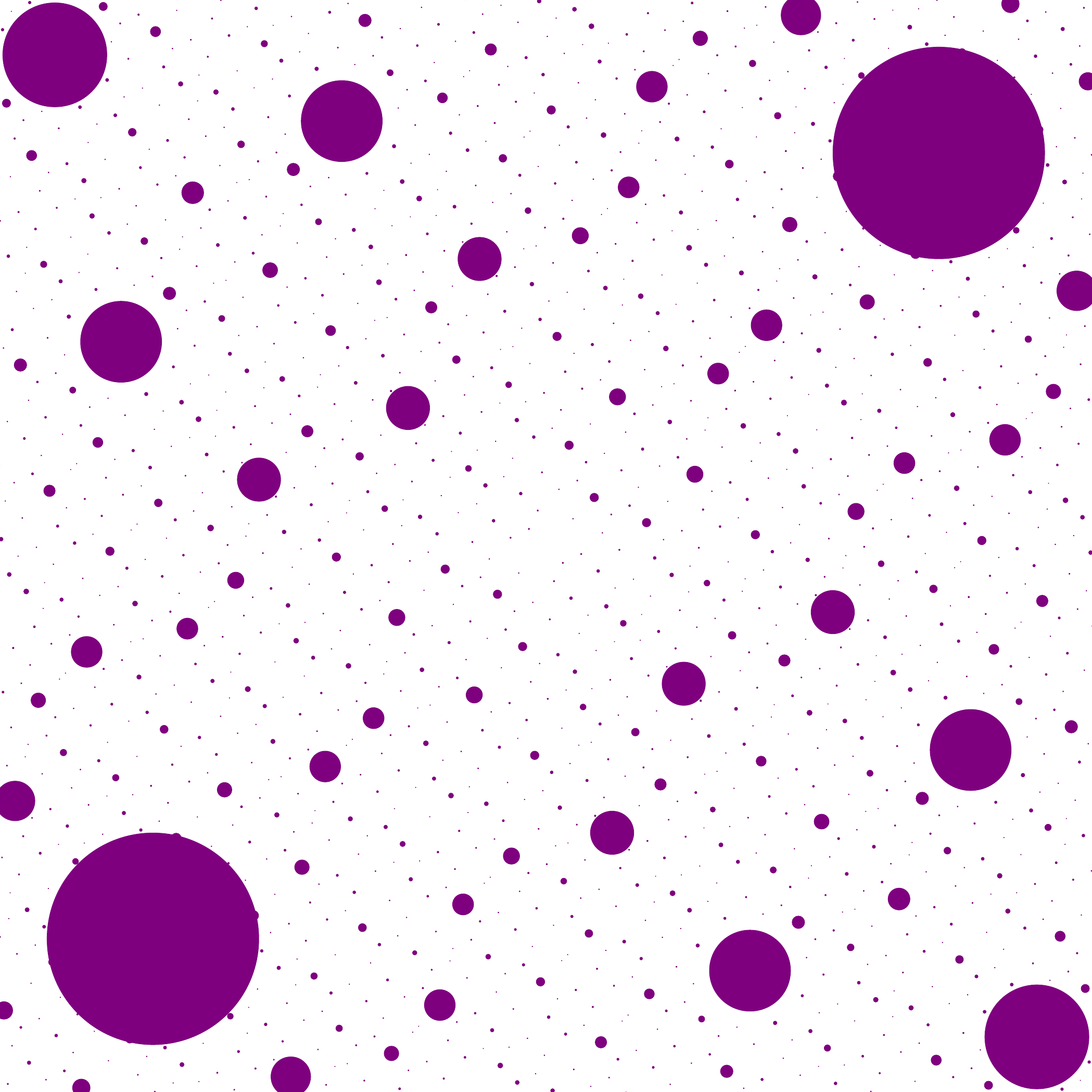}}
        \hspace{1em}
        \subfloat[Scale-Invariant ($k=0.25$)
        \label{fig:SpacetimeCNP-I13ScaleInvPattern2}]{\includegraphics[width=0.35\linewidth]{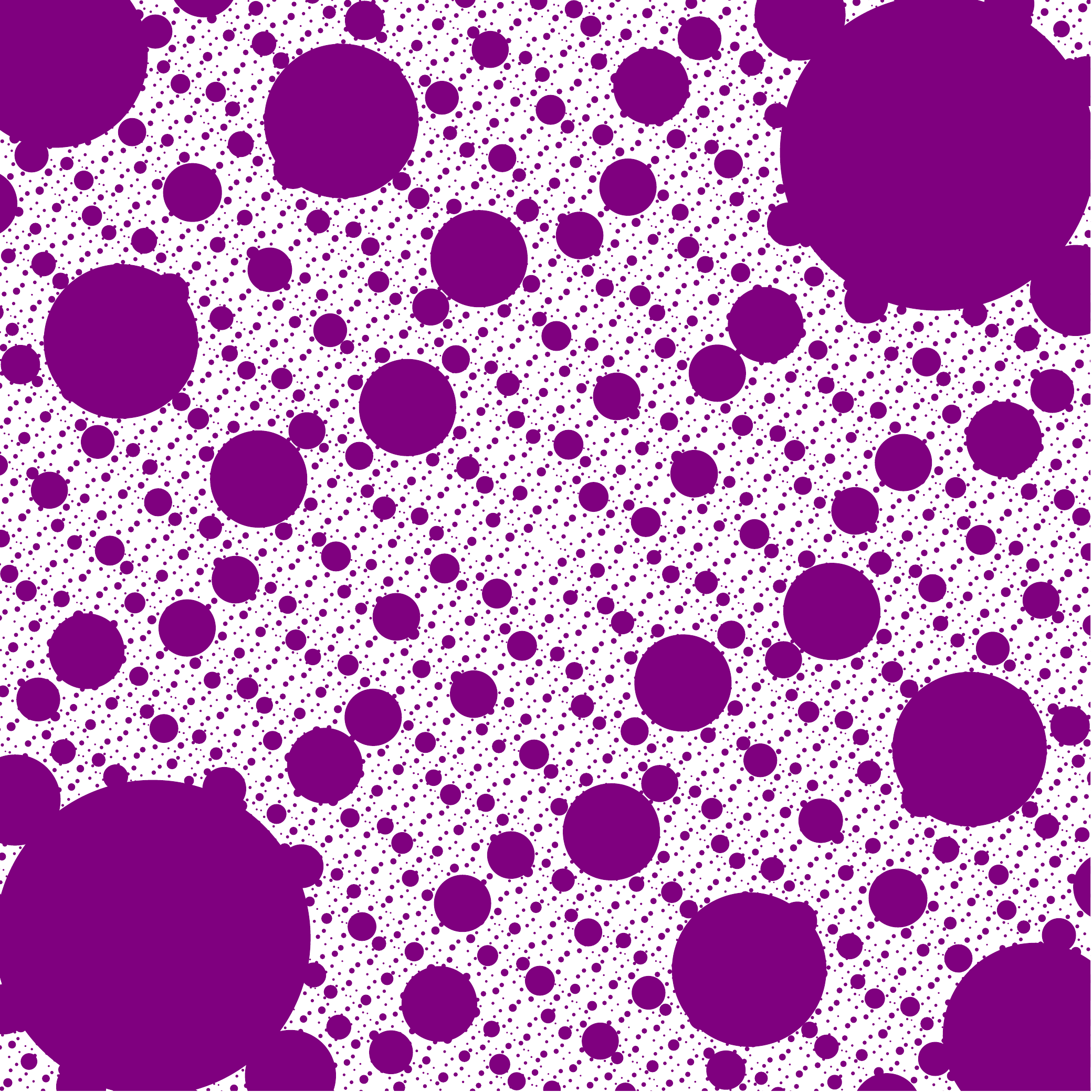}}
    \end{tabular}
\end{threeparttable}
\caption[Four variants of the $(1+1)$D spacetime quasicrystal constructed from the odd self-dual lattice $\textrm{I}_{3,1}$ in Sec.~\ref{subsec:2DLorentzianQuasiCrystal}]{
\captionsetup{width=0.99\linewidth}
The same patch of four variants of the $(1+1)$D spacetime quasicrystal constructed from the odd self-dual lattice $\textrm{I}_{3,1}$ in Sec.~\ref{subsec:2DLorentzianQuasiCrystal} with the following choices of weights. \textbf{Upper left}: Gaussian weighting function $W_{\rm Gauss}({\bf v}_{\rm in})\propto {\rm exp}(-\pi v_{in}^{2})$ with $v_{\rm in}^2\equiv\langle {\bf v}_{\rm in},{\bf v}_{\rm in}\rangle$). 
\textbf{Upper right}: a disk $\mathcal{W}$ of radius $3\sigma$ ($\sigma=(2\pi)^{-1/2}$ is the standard deviation of $W_{\rm Gauss}$). \textbf{Lower left and right:} a weighting function $W({\bf v}_{\rm in})\propto v_{in}^{-(2+1.01)}$ and $W({\bf v}_{\rm in})\propto v_{in}^{-(2+0.25)}$ respectively. In both lower panels, the central singular peak is removed; it would be absent for any shifted lattice $\Lambda^{\bf t}$ with ${\bf t}\notin\Lambda$. The power laws ($k=1.01,\,0.25$) decay sufficiently fast to yield finite mass on bounded regions, but only $k=1.01$ produces a translation-bounded complex measure (Lemma 9.5 of \cite{baake2013aperiodic}). Note that the upper left panel only includes peaks with amplitudes $>10^{-3}$; for the two lower panels, we chose peaks $>10^{-6}$. Without these cutoffs, if all peaks were included, they would densely fill the plane, though most would have negligibly tiny coefficients.
}
\label{fig:I13quasicrystal}
\end{figure}

For simplicity, we will start with a low dimensional example: a $(1+1)$ dimensional quasicrystal, obtained from the lattice $\mathrm{I}_{3,1}$. We construct three variants of this quasicrystal: ``windowed'' (of finite-density), self-dual, and scale-invariant of weight one (both of infinite density). This low dimensional example has the advantage that we can display the resulting quasicrystals explicitly (see Fig.~\ref{fig:I13quasicrystal}).

The Coxeter diagram for $\mathrm{I}_{3,1}$ is shown in Fig. \ref{fig:VinbergRootsI1nUpto10}, and the corresponding Schl\"{a}fli matrix reads
\begin{equation*}
S_{\mathrm{I}_{3,1}} = \begin{pmatrix}
2 & -1 & 0 & 0\\
-1 & 2 & -\sqrt{2} & 0 \\
0 & -\sqrt{2} & 2 &-\sqrt{2} \\
0 & 0 & -\sqrt{2}&2 
\end{pmatrix}.
\end{equation*}
Expressed in an orthonormal basis (Cartesian coordinates), the fundamental roots (from Table \ref{tab:SimpleRootsReflectiveOddLatticesN2to17}) are:
\begin{equation*}
\begin{aligned}
& \mathbf{r}_1 = (0,-1,1,0), \quad \mathbf{r}_2 = (0,0,-1,1), \\
& \mathbf{r}_3 = (0,0,0,-1), \quad \mathbf{r}_4 = (1,1,1,1).
\end{aligned}.
\end{equation*}
and the Coxeter element $C=R_1 R_2 R_3 R_4$ is 
\begin{equation*}
C_{\mathrm{I}_{3,1}} = 
\left(\begin{array}{cccc}
2 & -1 & -1 & -1 \\
-1 & 1 & 1 & 0 \\
1 & 0 & -1 & -1 \\
1 & -1 & 0 & -1 
\end{array}\right).
\end{equation*}
Its eigenvalues are solutions to
the characteristic Coxeter polynomial
$$ 
\chi_C(x) = x^4 - x^3 - 2x^2 - x + 1 . 
$$
Rounded to 5 digits, they are:
\begin{eqnarray*}
\lambda_{0}^{\pm}&=&\frac{1\!+\!\sqrt{17}}{4} \pm \frac{1}{2}\sqrt{\frac{\sqrt{17}\!+\!1}{2}} \approx (2.08102)^{\pm1}, \\ 
\lambda_1^{\pm}&=& \frac{1\!-\!\sqrt{17}}{4} \pm \frac{i}{2}\sqrt{\frac{\sqrt{17}\!-\!1}{2}} \approx -0.78078\!\pm\!0.62481\,i,
\nonumber
\end{eqnarray*}
and the corresponding projection operators are
\begin{eqnarray*}
  (\Pi_{\rm ph})^{\mu}_{\;\;\nu}&=&\frac{(v_{0}^{+})^{\mu}(v_{0}^{-})_{\nu}}{\langle{\bf v}_{0}^{+},{\bf v}_{0}^{-}\rangle}
  +\frac{(v_{0}^{-})^{\mu}(v_{0}^{+})_{\nu}}{\langle{\bf v}_{0}^{-},{\bf v}_{0}^{+}\rangle} 
  = \tfrac{+1}{\sqrt{17}}
  \left(
  \begin{array}{cccc} 
  \tfrac{7+\sqrt{17}}{2} & 0 & -2 & -2 \\
  0 & \tfrac{3+\sqrt{17}}{2} & 1 & -1 \\
  2 & 1 & \tfrac{-5+\sqrt{17}}{2} & -1 \\
  2 & -1 & -1 & \tfrac{-5+\sqrt{17}}{2}
  \end{array}\right), \\[1.5ex]
  (\Pi_{\rm in})^{\mu}_{\;\;\nu}& = &\frac{(v_{1}^{+})^{\mu}(v_{1}^{-})_{\nu}}{\langle{\bf v}_{1}^{+},{\bf v}_{1}^{-}\rangle}
  +\frac{(v_{1}^{-})^{\mu}(v_{1}^{+})_{\nu}}{\langle{\bf v}_{1}^{-},{\bf v}_{1}^{+}\rangle} 
  =\tfrac{-1}{\sqrt{17}}
  \left(
  \begin{array}{cccc} 
  \tfrac{7-\sqrt{17}}{2} & 0 & -2 & -2 \\
  0 & \tfrac{3-\sqrt{17}}{2} & 1 & -1 \\
  2 & 1 & \tfrac{-5-\sqrt{17}}{2} & -1 \\
  2 & -1 & -1 & \tfrac{-5-\sqrt{17}}{2}
  \end{array}\right).
\end{eqnarray*} 
Clearly, $(\Pi_{\rm in})^{\mu}_{\;\;\nu}=\delta^{\mu}_{\;\;\nu}-(\Pi_{\rm ph})^{\mu}_{\;\;\nu}$. Notice that, if we define the Galois-automorphism $\sigma : \sqrt{17} \to -\sqrt{17}$, then we can understand $\lambda_1^\pm$ as the images of $\lambda_0^\pm$ under $\sigma$ (in the spirit of Appendices \ref{AppendixA-NumberTheory} and \ref{appendix:FoldingUnfolding}).

In the (unnormalized) Coxeter eigenbasis, the Gram matrix $G_{ij} = \langle {\bf v}_i^\pm , {\bf v}_j^\pm \rangle$ of inner products (with respect to the sesquilinear form -- see Appendix \ref{appendix:NPFrames}) becomes 
\begin{equation}
\begin{aligned}
    G_{{\rm I}_{3,1}} &= 
    \setlength{\arraycolsep}{0.50pt}
    \begin{pmatrix}
    \GramBlock{red}{0 & \langle {\bf v}_0^+,{\bf v}_0^- \rangle\\ \langle {\bf v}_0^-,{\bf v}_0^+ \rangle & 0} & \fixedblock{white}{0 & 0\\ 0 & 0}\\
    \fixedblock{white}{0 & 0\\ 0 & 0} & \GramBlock{red}{\langle {\bf v}_1^+,{\bf v}_1^- \rangle & 0\\ 0 & \langle {\bf v}_1^-,{\bf v}_1^+ \rangle}
    \end{pmatrix} = 
    \setlength{\arraycolsep}{0.50pt}
    \begin{pmatrix}
    \fixedblock{red}{0 & G_0^{+-}\\ G_0^{-+} & 0} & \fixedblock{white}{0 & 0\\ 0 & 0}\\
    \fixedblock{white}{0 & 0\\ 0 & 0} & \fixedblock{red}{G_1^{+-} & 0\\ 0 & G_1^{-+}}
    \end{pmatrix}\\[+1.5ex]
    &= 
    \BigGramBlock{red}{0 & -\tfrac{1}{2}\big(17 + 5 \sqrt{17} \big) \\ -\tfrac{1}{2}\big(17 + 5 \sqrt{17}\big) & 0} 
    \bigoplus
    \BigGramBlock{red}{0 & -\tfrac{1}{2}\big(17 - 5 \sqrt{17} \big) \\ -\tfrac{1}{2}\big(17 - 5 \sqrt{17}\big) & 0 } 
    \end{aligned}
    \label{eq:GramMatrixI13CoxeterEigenbasis}
\end{equation}
The first and second blocks are the induced metrics on $V_{\rm ph}$ and on $V_{\rm in}$ respectively; note that one block is the Galois-conjugate of the other (see Appendix \ref{appendix:FoldingUnfolding}). The density of the windowed pattern is (see Eq.~\eqref{eq:DensityOfCNPQuasicrystal} and use $\det \mathrm{I}_{3,1} = 1$)
\begin{equation}
\rho_{\rm sQC}^{\mathrm{I}_{3,1}} = \pi R_{\rm int}^2 = 2\pi G_{1}^{+-} R^2_{\rm disk} \approx 11.3585 \qquad \text{(with $R_{\rm disk} = 1$)}.
    \label{eq:DensitysQCI13}
\end{equation}
The radius in the internal space is calculated with the induced metric together with a factor of 2; we find that $R_{\rm int} = \sqrt{2} R_{\rm disk}$. Letting $  V_{\rm in} \ni {\bf w}_{in} = \alpha_1 {\bf v}_1^+ + \alpha_1^* {\bf v}_1^-$ be an arbitrary internal vector, we can understand this factor-of-2 contribution as follows: using the real bilinear form (see Appendix \ref{appendix:NPFrames}) and the above induced metric $G_{\mathrm{I}_{3,1}}$, we get
$$ \langle {\bf w}_{\rm in}, {\bf w}_{\rm in} \rangle = |{\bf w}_{\rm in}|^2 = 2 |\alpha_1|^2 \langle {\bf v}_1^+, {\bf v}_1^- \rangle = 2 G_{1}^{+-} |\alpha_1|^2$$
where $|\alpha_1|^2$ is the ``naive'' Euclidean metric on $V_{\rm in}$ (the sum-of-squares of the real and imaginary part of $\alpha_1 \in \CC$). 

Now, in Fig.~\ref{fig:I13quasicrystal} we illustrate the resulting quasicrystal that arise from several natural choices of the weighting function (or window): first, the self-dual quasicrystal obtained by choosing a Gaussian weighting function $W({\bf v}_{\rm in})$ with the self-dual width, {\it i.e.}\ $W({\bf v}_{\rm in})\propto {\rm exp}(-\pi v_{\rm in}^{2})$ with $v_{\rm in}^{2}\equiv\langle{\bf v}_{\rm in},{\bf v}_{\rm in}\rangle$ (top left panel), the quasicrystal obtained by taking a circular-disk-shaped window ${\cal W}$ with radius $3\sigma$ (top right panel) where $\sigma=(2\pi)^{-1/2}$ is the standard deviation of the self-dual Gaussian, and two globally-scale-invariant quasicrystals (bottom panel) of weight $k=1.01$ (left) and $k=0.25$ (right), obtained by taking a weighting function $W({\bf v}_{\rm in})\propto v_{\rm in}^{-(2+k)}$ for these two values of $k$. All figures are in the Coxeter eigenbasis (i.e.~null coordinates, the lightcone is the vertical and horizontal axes).

\subsection{Example 2: $(3+1)$-dimensional Quasicrystals from $ \mathrm{I}_{7,1}$ }
\label{subsec:4DsQCFromI17}

Here, we move to a higher-dimensional example and construct a $(3+1)$D spacetime quasicrystal from the odd self-dual lattice $\mathrm{I}_{7,1}$. Its Coxeter graph is shown in Fig.~\ref{fig:VinbergRootsI1nUpto10}, and its simple roots are given in Tab.~\ref{tab:SimpleRootsReflectiveOddLatticesN2to17}. Its Schl\"afli matrix reads
\begin{equation*}
    S_{\mathrm{I}_{7,1}} = 
    \begin{pmatrix}
        2 & -1 & 0& 0& 0& 0& 0& 0\\
        -1 & 2 & -1 & 0 &0 &0 &0 &0 \\
        0 & -1 & 2 & -1 & 0&0 &0 &-1 \\
        0 & 0 & -1 & 2& -1 & 0& 0& 0\\
        0 & 0 & 0 & -1 & 2& -1 & 0 & 0\\
        0 & 0 & 0 & 0 & -1 & 2 & -\sqrt{2} & 0\\
        0&  0& 0& 0& 0& -\sqrt{2} & 2 & 0 \\
        0 & 0 & -1&0 &0 & 0& 0& 2\\
    \end{pmatrix}
\end{equation*}
The Coxeter element $C$ becomes: 
$$ C = 
\begin{pmatrix}
    2 & -1 & -1 & -1 & 0 & 0 & 0 & 0\\
    0 & 0 & 0 & 0 & 0 & 0 & 0 & -1\\
    1 & 0 & -1 & -1 & 0 & 0 & 0 & 0 \\
    1 & -1 & 0 & -1 & 0 & 0& 0 & 0\\
    1 & -1 & -1 & 0 & 0& 0&0&0\\
    0&0&0&0&1&0&0&0\\
    0&0&0&0&0&1&0&0\\
    0&0&0&0&0&0&1&0
\end{pmatrix}
$$
Its eigenvalues are solutions to the characteristic polynomial
$$ 
\chi_C(x) = x^8 - x^6 - x^5 - x^3 - x^2 + 1
$$
and are too cumbersome to be displayed analytically (but not impossible -- see Appendix \ref{app:inflation}). Numerically, they are (truncated to 5 digits)
$$
\begin{aligned}
 & \lambda_0^\pm = (1.45799)^\pm, \quad \lambda_1^\pm = 0.429221 \pm  0.9032 i,\\ & \lambda_2^\pm = -0.609155 \pm  +0.793051 i, \quad 
 \lambda_3^\pm = -0.891998 \pm  0.452039 i .
\end{aligned}
$$

The Gram matrix in the normalized Coxeter eigenbasis is given by Eq.~\eqref{eq:GramMatrixNormalizedCoxeterEigenbasisSesquilinear}. The \sCNP{} pattern's density is (again, Eq.~\eqref{eq:DensityOfCNPQuasicrystal} and $\det \mathrm{I_{7,1}}=1$)
\begin{equation}
\rho_{\rm sQC}^{\mathrm{I}_{7,1}} = \big( \pi R_{\rm int 1}^2 \big) \big( \pi R_{\rm int 2} ^2 \big) = \big( 2\pi R_{\rm disk1}^2\big) \big( 2\pi R_{\rm disk 2}\big) \approx 39.4784 \quad \text{(with $R_{\rm disk 1} = R_{\rm disk 2} = 1$)}.
\label{eq:DensitysQCI17}
\end{equation}
The factors of 2 appear for the same reason as explained below Eq.~\eqref{eq:DensitysQCI13}, but now the Gram matrix contains only $\pm 1$ and there is no additional factor coming from the induced metric.

\begin{figure}[t!]
    \centering
    \subfloat[Lorentzian plane $V_0$: TX slice]{
    \includegraphics[width=0.48\linewidth]{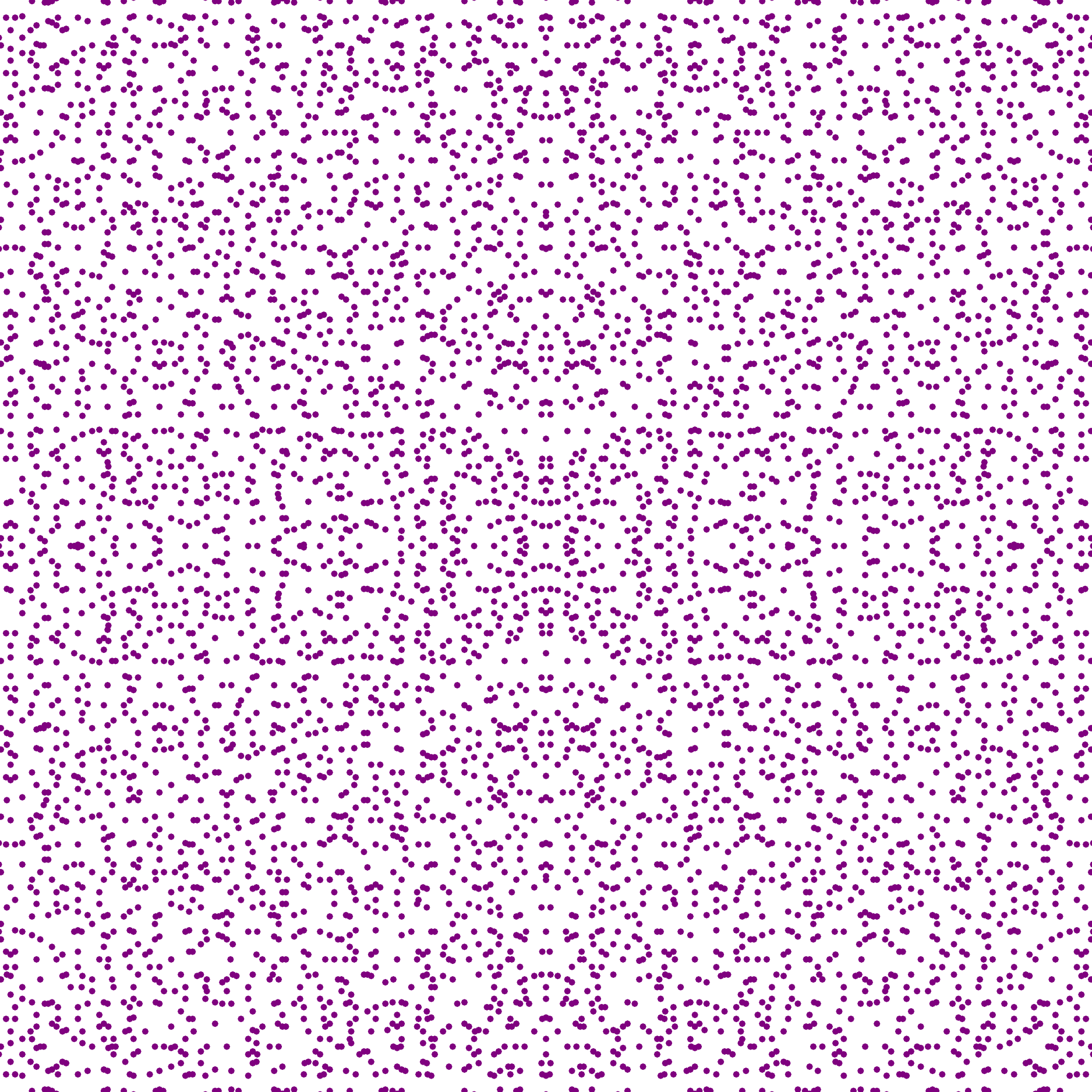}
    }
    \hfill
    \subfloat[Euclidean plane $V_1$: YZ slice]{
    \includegraphics[width=0.48\linewidth]{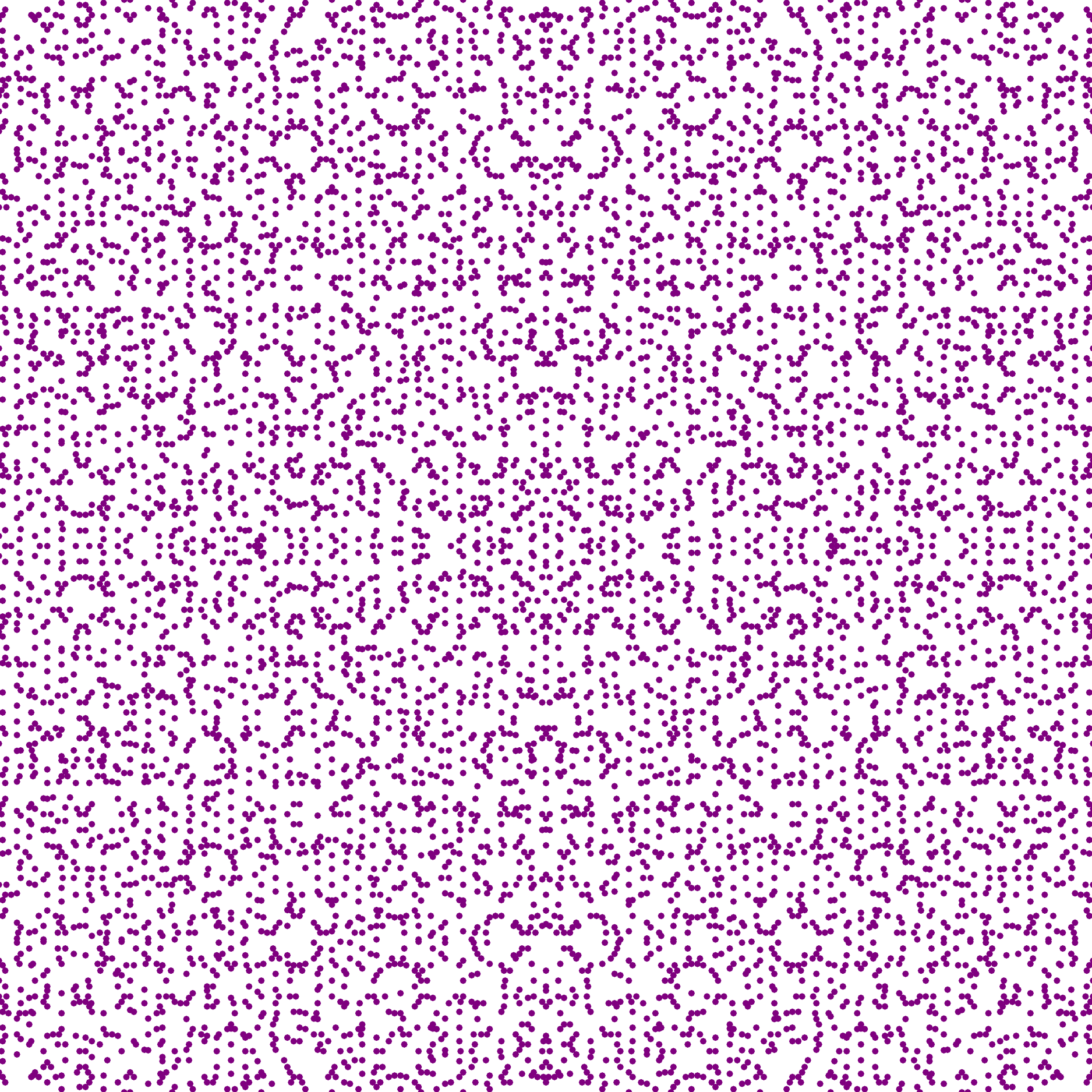}
    }
    \caption[Two-dimensional slices of the $(3+1)$-dimensional spacetime quasicrystal from $\mathrm{I}_{7,1}$.]{Two-dimensional slices of the $(3+1)$-dimensional spacetime quasicrystal from $\mathrm{I}_{7,1}$. On the left, we see the Lorentzian Coxeter plane $V_0$ by itself, where our pattern enjoys discrete Lorentz symmetry; note the hyperbolas appearing in the pattern. On the right is the Euclidean $V_1$-plane, where the spacetime quasicrystal has dense rotational symmetries; note the various circular rings appearing in the pattern.}
    \label{fig:2DSlicesOf4DsQCFromI17}
\end{figure}

In Fig.~\ref{fig:2DSlicesOf4DsQCFromI17}, we show two 2-dimensional slices of the $\mathrm{I}_{7,1}$ spacetime quasicrystal, constructed as a finite-density windowed \sCNP{} structure. The physical space $V_{\rm ph}$ contains the first two Coxeter planes, $V_0$ and $V_1$, while the rest of them are in the internal space $V_{\rm in}$. Clearly, one could have constructed two more different spacetime quasicrystals by grouping $V_0$ with the other two Coxeter planes $V_2$ and $V_3$. For each such choice, one could also construct the corresponding dense patterns by choosing appropriate weighting functions.  

\newpage
\subsection{Example 3: $(3+1)$-dimensional Quasicrystals from $\mathrm{II}_{9,1}$}
\label{subsec:4DLorentzianQuasicrystal}

Now we turn to a higher-dimensional example that will discuss in Chapter \ref{ch:StringCompactifications} when we discuss its possible physical interest.

As our lattice $\Lambda$, we now choose the remarkable 10-dimensional lattice $\mathrm{II}_{9,1}$ (the first non-trivial even unimodular lattice) and the $E_{10}$ root lattice. Its Coxeter diagram is shown in Fig.~\ref{fig:CoxeterDynkinDiagramsEvenSelfDual}, and the corresponding Schl\"{a}fli matrix reads
\begin{equation*}
S_{E_{10}} = \begin{pmatrix}
2 & -1 & 0 & 0 & 0 & 0 & 0 & 0 & 0 & 0 \\
-1 & 2 & -1 & 0 & 0 & 0 & 0 & 0 & -1 & 0 \\
0 & -1 & 2 & -1 & 0 & 0 & 0 & 0 & 0 & 0 \\
0 & 0 & -1 & 2 & -1 & 0 & 0 & 0 & 0 & 0 \\
0 & 0 & 0 & -1 & 2 & -1 & 0 & 0 & 0 & 0 \\
0 & 0 & 0 & 0 & -1 & 2 & -1 & 0 & 0 & 0 \\
0 & 0 & 0 & 0 & 0 & -1 & 2 & -1 & 0 & 0 \\
0 & 0 & 0 & 0 & 0 & 0 & -1 & 2 & 0 & 0 \\
0 & -1 & 0 & 0 & 0 & 0 & 0 & 0 & 2 & -1 \\
0 & 0 & 0 & 0 & 0 & 0 & 0 & 0 & -1 & 2 \\
\end{pmatrix}.
\end{equation*}
Expressed in an orthonormal basis (Cartesian coordinates), the fundamental roots are given explicitly in the caption of Fig.~\ref{fig:CoxeterDynkinDiagramsEvenSelfDual}, and the Coxeter element $C=R_{0}R_{1}\ldots R_{9}$ is
\begin{equation*}
C_{E_{10}} = \frac{1}{4}\left(\begin{array}{rrrrrrrrrr}
5 & -1 & -1 & -1 & -1 & -1 & -1 & -1 & -1 & -1 \\
1 & -1 & -1 & -1 & -1 & -1 & -1 & -1 & -1 & 3 \\
-1 & 1 & -3 & 1 & 1 & 1 & 1 & 1 & 1 & 1 \\
-1 & -3 & 1 & 1 & 1 & 1 & 1 & 1 & 1 & 1 \\
1 & -1 & -1 & 3 & -1 & -1 & -1 & -1 & -1 & -1 \\
1 & -1 & -1 & -1 & 3 & -1 & -1 & -1 & -1 & -1 \\
1 & -1 & -1 & -1 & -1 & 3 & -1 & -1 & -1 & -1 \\
1 & -1 & -1 & -1 & -1 & -1 & 3 & -1 & -1 & -1 \\
1 & -1 & -1 & -1 & -1 & -1 & -1 & 3 & -1 & -1 \\
1 & -1 & -1 & -1 & -1 & -1 & -1 & -1 & 3 & -1 \\
\end{array}\right).
\end{equation*}
Its characteristic polynomial is Lehmer's polynomial, 
$$ 
\chi_C(x) = x^{10} + x^9 - x^7 - x^6 - x^5 - x^4 - x^3 + x + 1. 
$$
The eigenvalues are (rounded to 5 digits)
$$
\begin{aligned}
    & \lambda_{0}^{\pm}=(1.17628)^{\pm 1}, \quad \lambda_{1}^{\pm}=-0.29233 \pm 0.95632\,i, \\
    & \lambda_{2}^{\pm}=-0.94331 \pm 0.33193\,i, \quad
    \lambda_{3}^{\pm}=+0.45687 \pm 0.88954\,i, \quad 
    \lambda_{4}^{\pm}=-0.73444 \pm 0.67868\,i .
\end{aligned}
$$
To construct a $(3+1)$-dimensional quasicrystal that is invariant under the group $G$ generated by $C$, we must choose the physical space to be spanned by the real pair of eigenvectors ${\bf v}_{0}^{\pm}$ and one of the four pairs of complex conjugate eigenvectors. For example, if we choose the first pair of complex conjugate eigenvectors ${\bf v}_{1}^{\pm}$ we obtain the projection operators
\begin{eqnarray*}
  (\Pi_{\rm ph})^{\mu}_{\;\;\nu}&=&\frac{(v_{0}^{+})^{\mu}(v_{0}^{-})_{\nu}}{\langle{\bf v}_{0}^{+},{\bf v}_{0}^{-}\rangle}
  +\frac{(v_{0}^{-})^{\mu}(v_{0}^{+})_{\nu}}{\langle{\bf v}_{0}^{-},{\bf v}_{0}^{+}\rangle} \nonumber\\
  &+&\frac{(v_{1}^{+})^{\mu}(v_{1}^{-})_{\nu}}{\langle{\bf v}_{1}^{+},{\bf v}_{1}^{-}\rangle}
  +\frac{(v_{1}^{-})^{\mu}(v_{1}^{+})_{\nu}}{\langle{\bf v}_{1}^{-},{\bf v}_{1}^{+}\rangle}, \\
  (\Pi_{\rm in})^{\mu}_{\;\;\nu}&=&\delta^{\mu}_{\;\;\nu}-
  (\Pi_{\rm ph})^{\mu}_{\;\;\nu}.
\end{eqnarray*} 
But, instead of the pair ${\bf v}_{1}^{\pm}$, we could alternatively have chosen the pair ${\bf v}_{2}^{\pm}$, ${\bf v}_{3}^{\pm}$ or ${\bf v}_{4}^{\pm}$.  So there are four inequivalent $G$-invariant ways to decompose $V=\mathbb{R}^{9,1}$ into $V_{\rm ph}=\mathbb{R}^{3,1}$ and $V_{\rm in}=\mathbb{R}^{6,0}$, with each splitting giving rise to a different type of quasicrystal.

Finally, for any of these four inequivalent decompositions, our final choice is the weighting function: once again, the two most natural choices are the Gaussian-self-dual weighting function, or the power-law weighting function which is globally scale invariant with weight $w$. Although these $(3+1)$ dimensional quasicrystals are too high dimensional to display explicitly, we can construct them numerically in exactly the same way as we did in the previous $(1+1)$ dimensional examples. 


\chapter{Generalized Spacetime $\CNP$ Scheme}
\label{ch:GeneralizedSpacetimeCNP}

In Chapter \ref{ch:QuasicrystalsMinkowski}, we chose a particular non-crystallographic symmetry group -- the one generated by the Coxeter element -- and very specific lattices -- the integer self-dual ones -- in order to produce the first examples of spacetime quasicrystals. Under the current spacetime \sCNP{} construction of Section \ref{sec:SpacetimeCNP}, all $d_{\rm ph}$-dimensional spacetime quasicrystals transform under a discrete subgroup of 
$$SO(1,1)\times \underbrace{SO(2)\times \cdots \times SO(2)}_{\mathclap{(d_{\rm ph} - 2)/2 \text{ factors }}
}$$ which is dense in the rotational part, so spacetime quasicrystals have dramatically more orientational symmetry than their Euclidean counterparts. However, there are several somewhat unsatisfying and deeply interrelated facts about such constructed spacetime quasicrystals:
\begin{itemize}
    \item \textbf{The Curse of Being Even:} they can only have even dimension; to go to odd-dimensional structures while keeping the symmetric Coxeter planes intact (i.e.~respecting the pairing of the Coxeter eigenbasis), one would need to include the rational direction (eigenvector $\mathbf{v}_{2k+1}$, see Section \ref{subsec:Hyperbolic_Coxeter_Elements}) to the physical space $V_{\rm ph}$, but this would make $V_{\rm ph}$ only a partially irrational subspace (see beginning of Chapter \ref{ch:QuasicrystalsEuclidean}).  
    \item \textbf{The Only Building Block is $\mathbf{2}$}: the symmetry group generated by the Coxeter element, or the bigger infinite dihedral group \eqref{eq:ColoureCoxeterGroup}, only possess $2$-dimensional irreducible representations $\mathbf{2}$. Any higher-dimensional quasicrystal, like the $(3+1)$D examples of Section \ref{sec:ExamplesSpacetimeQuasicrystals}, does not transform in a irreducible $\mathbf{4}$ representation but in a direct sum of appropriate numbers of $\mathbf{2}$'s (see Eq.~\eqref{eq:VDecompositionHyperbolicCoxeter}). 
    \item \textbf{Preferred Spacetime Splitting:} as a result, there is always a \textbf{preferred splitting} of spacetime into two-dimensional Coxeter planes, which are crucially \textbf{not mixed by the symmetry group}. There is the Lorentzian $V_0$ plane where one can boost (and there is an infinite set of discrete boosts that are symmetries of the pattern), and $(d_{\rm ph}-2)/2$ Euclidean planes where one could rotate (and one could practically do any planar rotation), but one cannot mix and match. Physically, one cannot mix the time direction hiding inside $V_0$ with any of the spatial directions that are not inside $V_0$, and one could not similarly rotate two spatial directions of one Euclidean plane into those of another such plane. 
\end{itemize}
In short, the symmetry group of a spacetime quasicrystal is \textit{not} a subgroup of $SO(1,d_{\rm ph}-1)$. Therefore, a question arises naturally: can we do better? Can we make our spacetime quasicrystals have even more intricate symmetries and capture a dense subgroup in the full $SO(1,d_{\rm ph}-1)$ underlying continuous group?


Building off of the ideas in \cite{mygdalas2026spacetimequasicrystals}, the first step towards answering this question happened in an MSc thesis \cite{Thorne2026}, for which the author served as a co-supervisor (together with Latham Boyle; see Statement of Contributions). In Section \ref{subsec:NoFoldingsForSimplices}, we review the results of the thesis, while in Section \ref{subsec:UnfoldingNonCrystallographicGroups} we present our own ongoing work to extend these results. The remaining of this chapter is devoted to adapting the \CNP{} scheme to our new findings. In Section \ref{sec:SpacetimeCNPRootBasis} we give a more general \sCNP{} scheme performed in the root basis (instead of the Cartesian one). Finally, in Section \ref{sec:ExamplesNewSpacetimeQuasicrystals} we put the algorithm to work to produce the first-ever icosahedrally-symmetric $(3+1)$D spacetime quasicrystal.  

\section{Revisiting the Non-Crystallographic Symmetry}
\label{sec:RevisitingNonCrystallographicSymmetry}

Constructing \sCNP{} quasicrystals can be algebraically understood as searching for non-crystallographic subgroups within crystallographic parent groups. Once we know that such a\textit{ Coxeter pair} exists (see Section \ref{subsec:SphCoxPairsIntro}), then we can build a \sCNP{} scheme starting from the lattice whose symmetries are described by that parent group and obtain the quasi-crystalline pattern with the desired non-crystallographic symmetry. In this chapter, we refer to a (non-)crystallographic group as $G_{C}$ ($G_{NC}$). A Coxeter pair is then denoted as $(G_{C},G_{NC})$.

As mentioned in that Section, all the possible Coxeter pairs are known for the Euclidean cases. Interestingly, it is also known that for all these cases, the non-crystallographic symmetry group $G_{NC}$ can be obtained by \textit{folding} the crystallographic graph; to indicate folding, we write $G_{C} \twoheadrightarrow G_{NC}$. Indeed, in Fig.~\ref{fig:non-crystallographicEuclideanFoldings} we encountered all the interesting non-crystallographic foldings of the spherical Coxeter groups, such as $E_8 \twoheadrightarrow H_4$, $D_6 \twoheadrightarrow H_3$, and the three 4D-to-2D \sCNP{} schemes, $A_4 \twoheadrightarrow I_2^5$, $B_4 \twoheadrightarrow I_2^8$ and $F_4 \twoheadrightarrow I_2^{12}$ (the three most relevant symmetries of natural quasicrystals). Knowing the lattice (i.e.~$G_C$), the symmetry $G_{NC}$ and, by extension, the window invariant under $G_{NC}$ (i.e.~a 600-cell, an icosahedron, a decagon, an octagon or a dodecagon respectively, see Section \ref{fig:SphEucHyp}), one can just follow the \sCNP{} prescription of Section \ref{sec:sCNP} to build the spatial quasicrystals.  

On the other hand, even though several examples of non-crystallographic hyperbolic Coxeter groups \textit{are} known in the literature, it is \textit{not} known if any of them can be embedded into a higher-rank crystallographic group of the same nature to form a \textbf{hyperbolic Coxeter pair}. Indeed, there are around 16 non-crystallographic graphs in Figs. 2 \& 3 in \cite{humphreys_1990} (most due to Lann\'{e}r \cite{Lannr1950OnCW}), which exist only up to dimension $d = 5$ (and thus tessellate the hyperboloid of dimension up to $d-1 = 4$), but none of them is known to exist as a subgroup of some appropriately higher-rank $G_C$. This inhibits the construction of any spacetime \sCNP{} scheme to produce quasicrystals with these symmetry groups..  

\begin{figure}[t!]
    \centering
    \includegraphics[width=0.75\linewidth]{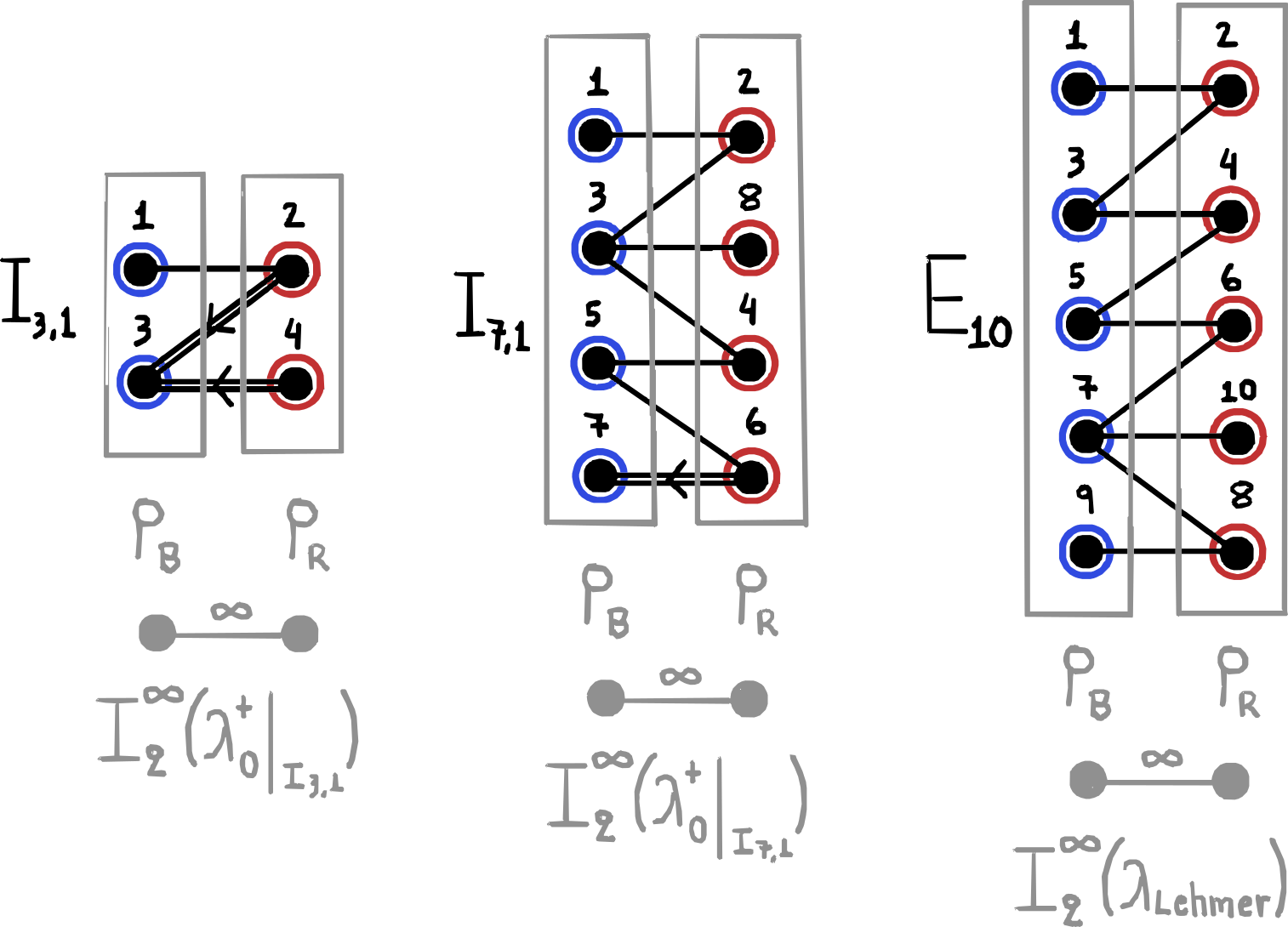}
    \caption[Some Lorentzian non-crystallographic foldings of self-dual integer lattices onto their rank-2 infinite dihedral subgroup.]{Some Lorentzian non-crystallographic foldings of self-dual integer lattices onto their rank-2 infinite dihedral subgroup. These hyperbolic Coxeter pairs were used in Section \ref{sec:ExamplesSpacetimeQuasicrystals} to produce the first examples of spacetime quasicrystals.}   \label{fig:HyperbolicNonCrystallographicFoldingsToCoxeterPlane}
\end{figure}

Despite these challenges, to make some progress  and actually construct a spacetime quasicrystal, we chose in Section \ref{sec:SpacetimeCNP} integer self-dual lattices and their $G_C$'s as the starting point (Section \ref{subsubsec:Lorentzian_Lattices}). The reasoning was simple: the nicest spatial quasicrystals are constructed from such self-dual lattice, like $E_8$ or $B_4 \simeq \ZZ^4$ and $A_4 \hookrightarrow \ZZ^5$ for the Elser-Sloane quasicrystal or the Ammann-Beenker and Penrose tilings respectively. Thus, mathematical elegance led us to the self-dual Lorentzian lattices $\mathrm{I}_{d-1,1}$ and $\mathrm{II}_{d-1,1}$ for $d \leq 10$.  All of these graphs are bicolourable (see Fig.~\ref{fig:OddEvenLorentzianLatticesUpToDim10}), which allowed us to consider the simplest $G_{NC}$ inside them, or equivalently, the simplest non-crystallographic folding: grouping all red and all blue generators together, one obtains an infinite dihedral group \eqref{eq:ColoureCoxeterGroup} characteristic of each $G_C$, as shown in Fig.~\ref{fig:HyperbolicNonCrystallographicFoldingsToCoxeterPlane}. Geometrically, these foldings correspond to projections onto the Coxeter planes and the resulting structures suffer from the issues mentioned in the beginning of this chapter.

A natural question, then, is: can we fold the graphs of the self-dual integer lattices in some other way to obtain more general $G_{NC}$'s? The answer turns out to be negative, and we now turn to this topic in the next subsection. We reserve Appendix \ref{appendix:FoldingUnfolding} for the details regarding graph folding.  

\subsection{No Non-Crystallographic Foldings of Simplicial Hyperbolic Coxeter Graphs}
\label{subsec:NoFoldingsForSimplices} 

Central to the theory of spherical Coxeter pairs are the Coxeter elements: for the cases identified in \cite{Boyle_2022}, for example, it is noted that the Coxeter number is shared between a Coxeter pair $(G_{C},G_{NC})$. Additionally, in any $d$D-to-2D \sCNP{} scheme, the Coxeter eigenbasis indicates the subspaces on which to project a higher-rank lattice $\Lambda$ (stabilized by $G_C$) to produce the quasi-crystalline pattern of lower-rank with symmetry $G_{NC}$. In fact, this generalizes to any higher-dimensional \sCNP{} scheme, as explained in Appendix \ref{appendix:FoldingUnfolding}: for the \sCNP{} scheme based on the spherical Coxeter pair $(E_8,H_4)$, for instance, the bicoloured Coxeter eigenbasis tell us how to carry out the projection to get the Elser-Sloane $H_4$-quasicrystal \textit{and} realize $H_4$ as a subgroup of $E_8$. As a result, one would reasonably expect that the Coxeter elements would hold the key to understanding hyperbolic Coxeter pairs too. After all, this led to all spacetime quasicrystals constructed so far, based on a Coxeter pair $(G_C = \mathrm{I}_{d-1,1} \text{ or } \mathrm{II}_{d-1,1}, \mathrm{Dih}_{\infty})$.

Given a Coxeter pair $(G_C,G_{NC})$, which may be spherical or hyperbolic, there are two equivalent ways to express this relationship between the two members: (1) the \textit{folding way} $G_{C} \twoheadrightarrow G_{NC}$, and (2) its ``inverse'', the \textit{embedding way} $G_{NC} \hookrightarrow G_C$, or the \textit{``unfolding''} way. Here, we focus on viewpoint (1); viewpoint (2) will be discussed in the next subsection.

In the thesis \cite{Thorne2026}, we formulated a necessary -- but crucially \textit{not sufficient} -- condition for some Coxeter graph $H$ to arise as a folding of some other Coxeter graph $G$ when both of the graphs are irreducible and their graphs are trees (i.e.~no loops). The idea is that if $H$ embeds in $G$ by graph folding, then the eigenvalues of their respective Coxeter elements $C_H$ and $C_G$ must be related as $\mathrm{spec}(C_H) \subseteq \mathrm{spec}(C_G)$. We focus on trees because then it makes sense to talk about \textit{the} Coxeter element and \textit{the} spectrum. In other words, if $\mathrm{rank}(G) = r_G$ and $\mathrm{rank}(H) = r_H$, the $r_H$ eigenvalues of the $r_H \times r_H$ folded Coxeter element $C_H$ must be drawn from the pool of the $r_G$ eigenvalues of the original $r_G \times r_G$ Coxeter element $C_G$. We call this the \textbf{eigenvalue-sharing condition} \cite{Thorne2026}; we give a more precise statement in Lemma \ref{lemma:EigenvalueSharing} in Appendix \ref{appendix:FoldingUnfolding}, together with some examples. 

Using this condition, we were able to show the following: 
\begin{itemize}
    \item \textbf{No Hyperbolic Coxeter Pairs Between Simplicial Trees \cite{Thorne2026}:} considering all simplicial hyperbolic graphs from \cite{humphreys_1990,mcmullen2002coxeter} that do not contain any closed circuits, we were unable to find any two of them that satisfy the eigenvalue-sharing condition. This shows that it is impossible to embed any known hyperbolic tree-like\footnote{In \cite{Humphreys1973} there are exactly 9 -- out of the total 16 -- such non-crystallographic graphs; the rest contain loops and we did not include them in our calculations. } $G_{NC}$ into some crystallographic partner $G_C$ of the same properties, i.e.~tree-like, simplicial and with only spherical or affine subgraphs. This theorem is proven by exhaustion using the known classification of simplicial hyperbolic Coxeter groups and a Python script that checked the eigenvalue-sharing condition against all these graphs. The script and more details about it are available in $\cite{Thorne2026}$.
    \begin{itemize}
    \item A particular subcase of this theorem is that our integer self-dual lattices $G_C = \mathrm{I}_{d-1,1} \text{ or } \mathrm{II}_{d-1,1}$ ($d\leq10$) contain no known $G_{NC}$ either. 
    \end{itemize}
    \item \textbf{A Non-Simplicial Candidate for a Hyperbolic Coxeter Pair \cite{Thorne2026}:} relaxing the condition for simplicial fundamental domains, we also considered four more tree-like graphs, those of $\mathrm{I}_{d-1,1}$ for $11 \leq d \leq 14$ shown in Fig.~\ref{fig:OddGraphsN10to13} from Vinberg's classification \cite{vinberg1975some}. We found that a non-crystallographic hyperbolic Coxeter group from the simplicial classification \cite{humphreys_1990}, known as $L_{353}$ \cite{mcmullen2002coxeter}, whose graph is 
    \begin{equation*}
\begin{tikzpicture}[
    scale=0.7,
    transform shape,
    vertex/.style={
        circle,
        draw,
        fill=white,
        minimum size=5mm,
        inner sep=0pt
    }
]

\node[vertex] (v1) at (0,0) {};
\node[vertex] (v2) at (2,0) {};
\node[vertex] (v3) at (4,0) {};
\node[vertex] (v4) at (6,0) {};

\draw (v1) -- node[midway,above] {} (v2);
\draw (v2) -- node[midway,above] {5} (v3);
\draw (v3) -- node[midway,above] {} (v4);
\end{tikzpicture}
\end{equation*}
shares its eigenvalues with the crystallographic $\mathrm{I}_{13,1}$, though we were \textbf{unable to explicitly fold} the $\mathrm{I}_{13,1}$-graph in some way to get the $L_{353}$-graph (this was also checked thoroughly by trying all possible partitions of red and blue generators with a similar computer program). This \textbf{rules out this candidate Coxeter pair}, which is to be expected, since the eigenvalue-sharing condition for one
group to be a folding of another is necessary but not sufficient, so two groups can still share eigenvalues without forming a Coxeter pair. 
\end{itemize}
As a natural next step to extend these results, it would be interesting to include those hyperbolic diagrams from \cite{humphreys_1990,vinberg1975some} that have loops but are bicolourable, and check whether the eigenvalue-sharing condition is met for the bicoloured Coxeter elements. Note that the classification of hyperbolic Kac-Moody algebras \cite{Carbone_2010} (and thus crystallographic Coxeter groups; see Section \ref{subsec:SphAffHypCoxeterGroups}) cannot, unfortunately, add much, since for algebras the root-lengths matter but for Coxeter graphs they do not; as a result, several graphs in \cite{Carbone_2010} can be mapped to a graph already in \cite{humphreys_1990} and we do not get any more diagrams to try. 

Another idea would be to check if a given non-crystallographic graph -- from either the simplicial \cite{humphreys_1990} or Vinberg's \cite{vinberg1975some} classification -- admits other non-crystallographic foldings not included in the known classifications. We have tried certain examples with no luck and we discuss these findings in Appendix \ref{appendix:FoldingUnfolding}. Before thinking of ways to extend the search deeper into the unknown, though, we have found another method to define hyperbolic Coxeter pairs, as we know explain.

\subsection{Unfolding Non-Crystallographic Simplicial Hyperbolic Coxeter Graphs Into Crystallographic Graphs}
\label{subsec:UnfoldingNonCrystallographicGroups}

We now take viewpoint (2) of understanding Coxeter pairs: instead of trying to fold a given $G_C$ into some also known $G_{NC}$, we start from a desired non-crystallographic member $G_{NC}$ and attempt to lift it to some higher-rank lattice whose stabilizer is $G_C$. Stated differently, for a given $G_{NC}$, our goal it to embed -- or ``unfold'' -- it into some $G_C$; we write $G_{NC} \hookrightarrow G_C$.

To that end, we devised a method capable of unfolding all spherical non-crystallographic groups into their corresponding crystallographic counterparts; the details are explained in Appendix \ref{appendix:FoldingUnfolding} and a sketch of the method is given below. We then applied the method to unfold certain simplicial $G_{NC}$ of particular interest to their crystallographic parent groups. This way, we constructed a first list of the long-sought hyperbolic Coxeter pairs shown in Fig.~\ref{fig:UnfoldingHyperbolicGroupsRank4} and \ref{fig:UnfoldingHyperbolicGroupsRank5} in Appendix \ref{subsec:HyperbolicCoxeterPairsAppendix}.

The idea is very simple, but it requires some amount of algebraic number theory (see Appendix \ref{AppendixA-NumberTheory}). In its simplest form, it goes as follows: consider a Coxeter graph and construct its Schl\"{a}fli matrix $S$ from it according to Eq.~\eqref{eq:SchlafliMatrix}. The entries will belong to some number field $K$: if the graph is crystallographic, the number field will be just the rationals $K = \QQ$. \footnote{This is not entirely true because when we have Coxeter labels 4 and 6 we get Schl\"afli entries $\sqrt{2}$ and $\sqrt{3}$ respectively, so the statement is true for simply-laced graphs (i.e.~only Coxeter labels 3). But this does not change the argument, it simply makes the base field and the target field slightly different. For example, $\mathrm{I}_{2}^8$ is defined over a rank-4 extension by $\sqrt{1+\sqrt{2}}$, so we would unfold it to $B_4$, which is defined over $\QQ(\sqrt{2})$, a rank-2 extension over $\QQ$. For simplicity, here only consider icosahedral $G_{NC}$'s with only labels of 5. We will extend the formalism in a future work.}; if the graph is non-crystallographic, we get some rational extension $R$ by some irrationality. When the degree of $K$ over $R$ is 2, we can represent any element of $S_{NC}$ by a $2 \times 2$ rational matrix, leading to a graph in twice as many dimensions that is crystallographic (see Appendix \ref{appendix:FoldingUnfolding}).

\begin{figure}[t!]
    \centering
    \includegraphics[width=0.99\linewidth]{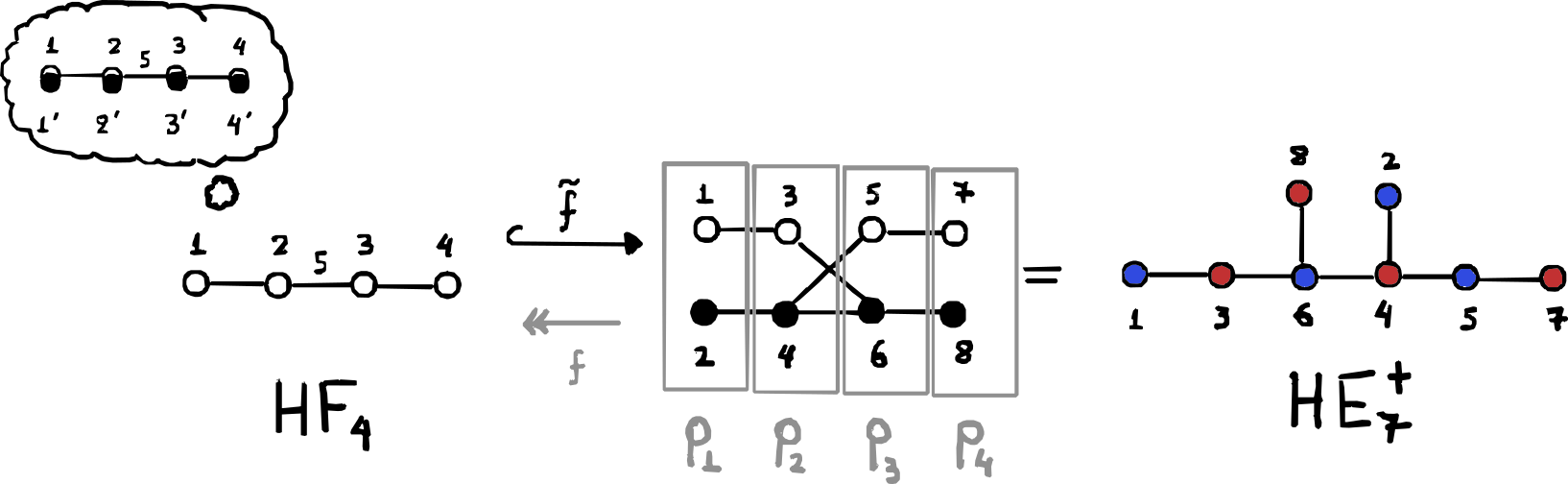}
    \caption[Unfolding the non-crystallographic group $HF_4$ into a crystallographic group in eight dimensions.]{Unfolding the non-crystallographic group $HF_4$ into a crystallographic group $HE_7^+$ in eight dimensions. One could think of the 4 nodes of $HF_4$ as being comprised of two nodes stacked on top of each other, which get unfolded into the full $HE_7^+$ graph after the unfolding $\widetilde{f}$-map. On the other hand, one could explicitly perform the folding $f$-map $HE_7^+ \twoheadrightarrow HF_4$ by grouping the corresponding nodes into pairs (shown in gray boxes in the middle) to define the new folded generators $\rho_i$.}
    \label{fig:unfoldingHF4}
\end{figure}

A very interesting non-crystallographic group from the simplicial classification \cite{humphreys_1990} is the group $L_{353}$ we encountered in the previous subsection, the one that shares eigenvalues with $\mathrm{I}_{13,1}$. Its graph resembles $F_4$ from the Euclidean classification of Fig.~\ref{fig:Irreducible_Root_Systems}, but instead of the label 4 in its middle two nodes it has a 5. Because of this, its Schl\"{a}fli matrix becomes indefinite; we call it $HF_4$ (hyperbolic $F_4$). This makes $HF_4$ very interesting because it contains a maximal finite icosahedral Coxeter subgroup $H_3$, as one can verify by removing either node from its ends. Using our unfolding techniques, we can embed the rank-4 $HF_4$ group into a lattice in twice as much dimensions whose crystallographic symmetry group is the 8-node graph $HE_7^+$, as shown in Fig.~\ref{fig:unfoldingHF4}. We call it $HE_7^+$ (hyperbolic extended $E_7$) because it is an extended (by one node) version of the finite $E_7$ graph of Fig.~\ref{fig:Irreducible_Root_Systems}: removing node $2$ or $8$ gives back the familiar $E_7$. 

This resulting rank-8 crystallographic graph belongs to no known classification \cite{humphreys_1990,vinberg1975some}.~The reason for this is that it contains hyperbolic subgraphs and thus ultra-ideal vertices (see Section \ref{subsec:SphAffHypCoxeterGroups}), which makes its fundamental domain acquire infinite volume. For example, removing node 4 from $HE_7^+$, one gets back a simplicial rank-7 hyperbolic graph that \textit{is} included in the list of \cite{humphreys_1990}. Nevertheless, it is a valid graph -- its Schl\"{a}fli matrix has the correct $(7,1)$ signature -- and defines a perfectly valid root lattice in $\MM^8$. 

\section{Spacetime $\CNP$ in the Root Basis}
\label{sec:SpacetimeCNPRootBasis}

Consequently, now that we know of the hyperbolic Coxeter pair $(HE_7^+,HF_4)$ -- and the rest of them identified in \ref{subsec:HyperbolicCoxeterPairsAppendix} -- can we design a \sCNP{} scheme to produce a spacetime quasicrystal with a full rank-4 non-crystallographic symmetry? 

The spacetime \sCNP{} of Section \ref{sec:SpacetimeCNP} is
phrased in explicit Cartesian (Minkowski) coordinates. This is because the integer self-dual lattices $\mathrm{I}_{d-1,1}$ and $\mathrm{II}_{d-1,1}$ acquire a particularly simple description in terms of these coordinates (Eqs.~\eqref{eq:OddUnimodularLattice}, \eqref{eq:EvenUnimodularLattice}), as they act as integral automorphisms of the Minkowski metric \eqref{eq:LorentzianMetricEta}. Besides, Vinberg has worked out the simple roots for all these cases and found their representation in terms of the Cartesian basis. However, we now need to work with lattices, like the $HE_7^+$ root lattice, whose simple roots are not tabulated in the Cartesian basis. Here we reformulate the entire \CNP{} algorithm so that it runs \emph{intrinsically in the simple-root basis}, where the roots themselves \textit{look like} a Cartesian basis, but crucially with the
Schl\"afli matrix $S_{ij}$ playing the role of the spacetime metric instead of the Minkowski metric \eqref{eq:LorentzianMetricEta}.

\subsection{The Lattice and the Schl\"{a}fli Matrix as the Metric}

\begin{center}
\begin{tcolorbox}[colframe=blue,colback=white,width=0.99\textwidth]
\textbf{Choice 1:}
a $d$-dimensional lattice $\Lambda^{\mathbf t}$ in
$V=\mathbb{R}^d$ with automorphism group
${\rm Aut}(\Lambda)$.
\end{tcolorbox}
\end{center}

Let $G_C$ be a crystallographic Coxeter group of rank $d$ with simple roots $\{ \mathbf{r}_i\}$.
We identify ${\bf r}_i = \bf{e}_i$, the standard Cartesian basis of $\MM^d$,\footnote{We chose here to talk about the spacetime \CNP{} scheme so we identified the $d$-dimensional vector space $V$ with $\MM^d$ from the beginning. This \CNP{} method in the root basis, though, can be trivially modified to work with $\EE^d$ too, as the signature is encoded in the Schl\"afli matrix and depends on the graphs under consideration.} so that the root lattice resembles $\Lambda_{\rm root}=\ZZ^d$. The bilinear form is the \emph{Schl\"afli matrix} \eqref{eq:SchlafliMatrix}, which now defines the inner products in $\MM^d$
\begin{equation}
\langle {\bf v}, {\bf w} \rangle = {\bf v}^T S {\bf w} = v^\mu S_{\mu \nu} w^\nu .
\label{eq:SchlafliInnerProduct}
\end{equation}
Importantly, $G_C$ is a subgroup of $\mathrm{GL}(\MM^d) \cap O(\MM^d,S)$, namely it is in the orthogonal group preserving the $S_{\mu\nu}$-metric, \textit{not} $\eta_{\mu \nu}$. The fundamental reflection generators $R_i$ in the root basis are given by Eq.~\eqref{eq:ReflectionInRootBasis} and the Coxeter element $C$ is either calculated directly from $S$ according to Eq.~\eqref{eq:AlternativeCoxeterElementGraph} or by taking the product of all $R_i$'s as in Section \ref{subsec:Coxeter_Elements}. 
 
\subsection{The Non-Crystallographic Symmetry}
\label{subsec:NCSymmetryGeneralsCNP}

\begin{center}
\begin{tcolorbox}[colframe=blue,colback=white,width=0.99\textwidth]
\textbf{Choice 2:} a non-crystallographic subgroup $G\in{\rm Aut}(\Lambda)$, and a decomposition $V=V_{{\rm ph}}\oplus V_{{\rm in}}$ that is invariant under $G$: $\Pi_{{\rm ph}}=g\Pi_{{\rm ph}}g^{-1}$ (or, equivalently $[\Pi_{{\rm ph}},g]=0$), $\forall g\in G$. 
\end{tcolorbox}
\end{center}

For our non-crystallographic group $G_{NC}$, we pick one from the hyperbolic Coxeter pairs identified in the Appendix \ref{subsec:HyperbolicCoxeterPairsAppendix}. Since the crystallographic parent $G_C$ is hyperbolic, the structure of the eigensystem of the Coxeter element for $G_C$ is exactly as described in Section~\ref{subsec:Hyperbolic_Coxeter_Elements}: there is one real Salem eigenvalue pair ${\lambda_0^+, \lambda_0^-}$ with $\lambda_0^+>1$, and $k-1$ unit-circle pairs $\lambda_j^\pm$. The eigenbasis still forms an irrational Newman-Penrose frame, and any vector ${\bf w} \in \MM^d$ can be expressed in the Coxeter eigenbasis as in  Eq.~\eqref{eq:VectorsInCoxeterEigenbasis}. 

The action of the Coxeter element $C$ of $G_C$ still splits $V$ into mutually orthogonal $2D$ Coxeter planes as in Eq.~\eqref{eq:VDecompositionHyperbolicCoxeter}, and its action onto each of them is still given by Eq.~\eqref{eq:CoxeterActionLorentzianCase}: on the Lorentzian plane $V_0$ it acts as an irrational boost, and on each of the $k-1$ Euclidean planes $V_j$ as an irrational rotation. The crucial difference from the previous discussion is that now these planes may be \textit{elegantly combined to produce irreducible representations of $G_{NC}$}. For concreteness, and since the most interesting examples of hyperbolic non-crystallographic $G_{NC}$ are rank-4 (see Section \ref{subsec:HyperbolicCoxeterPairsAppendix}), consider the case where $G_C$ is rank-8; then, the different decompositions of $V$ are shown below:    
\begin{equation}
\begin{array}{cccccccccc}
\textbf{under } I_2^\infty(\lambda_0^+) : \\[1ex]
& V & = & V_0 & \oplus & V_1 & \oplus & V_2 & \oplus & V_3 \\
& d & = & \mathbf{2} & \oplus & \mathbf{2} & \oplus & \mathbf{2} & \oplus & \mathbf{2} \\
& (7,1) & = &
\multicolumn{3}{c}{\underbrace{(1,1)\oplus(2,0)}} &
\oplus &
\multicolumn{3}{c}{\underbrace{(2,0)\oplus(2,0)}} \\[2ex]

\textbf{under } G_{NC} : \\[1ex]
& V & = &
\multicolumn{3}{c}{V_{01}=V_{\rm ph}} &
\oplus &
\multicolumn{3}{c}{V_{23} = V_{\rm int}} \\
& d & = &
\multicolumn{3}{c}{\mathbf{4}} &
\oplus &
\multicolumn{3}{c}{\mathbf{4}} \\
& (7,1) & = &
\multicolumn{3}{c}{(3,1)} &
\oplus &
\multicolumn{3}{c}{(4,0)}
\end{array}
\label{eq:VDecompositionUnderHyperbolicCoxeterPair}
\end{equation}
In defining $V_{\rm ph}$ from the first two Coxeter eigenplanes, we implicitly assumed that the first four eigen-directions of the Coxeter element have been chosen in such a way that if we restrict $C$ onto $V_{\rm ph}$, its eigenvalues will be exactly those of $G_{NC}$ (see Lemma \ref{lemma:EigenvalueSharing}). Without loss of generality, we always assume that the eigenvalues are sorted in this way.

In the spirit of Appendix \ref{appendix:FoldingUnfolding}, this would correspond to unfolding the dihedral group $I_2^\infty(\lambda_0^+)$ into $G_{NC}$, which in turn gets unfolded into $G_C$. Each of the $\mathbf{2}$'s that go into the $\mathbf{4}$'s are Galois-conjugates to one another under the Galois group acting at the level of $I_2^\infty(\lambda_0^+)$; this group will be of order $4$, since the rank-2 $I_2^\infty(\lambda_0^+)$ unfolds into 8-dimensions, and all $\mathbf{2}$'s are related to each another by some element in that group. The pair of irreducible $\mathbf{4}$'s of $G_{NC}$ are Galois-conjugate with respect to the Galois group acting at the level of $G_{NC}$ instead. This group is of order 2 (since the rank-4 $G_{NC}$ unfolds into 8-dimensions) and contains only one field automorphism $\sigma$. For the cases we will consider in this Chapter, this will be the conjugation in the golden field $\QQ(\phi)$, $\sigma: \sqrt{5} \mapsto - \sqrt{5}$, as we are dealing strictly with quadratic field extensions. 

\subsubsection{Algebraic Wick Rotation via a Galois Automorphism?}

Intriguingly, the Galois automorphism $\sigma$ relating the two $\mathbf{4}$ representations of our non-crystallographic group $G_{NC}$ inside $V$ acts as an discrete algebraic analogue of \textbf{Wick rotation}. In an ordinary (analytic) Wick rotation, one sends the Minkowski (real) time $t_M$ to the Euclidean (imaginary) time $it_E$, effectively replacing the indefinite metric $\eta_{\mu\nu}$ by the positive-definite $\delta_{\mu\nu}$. This allows one to treat the Lorentzian geometry as a Euclidean one. Here the same signature flip of the metric occurs, not by analytic continuation but by applying $\sigma$. We therefore call this an \textit{algebraic Wick rotation}: the Galois automorphism $\sigma: \phi\mapsto\bar{\phi}$ maps $V_\text{ph}$ (Lorentzian, physical spacetime) onto $V_\text{in}$ (Euclidean, internal space).

More concretely, focus on $HF_4$ from Fig.~\ref{fig:unfoldingHF4} and construct its Schl\"afli matrix: 
$$
S_{HF_4} = 
\begin{pmatrix}
2 & -1 & 0 & 0\\
-1 & 2 & -\phi & 0\\
0 & -\phi & 2 & -1\\
0 & 0 & -1 & 2
\end{pmatrix}.
$$
Under the unfolding $HF_4 \hookrightarrow HE_7^+$, we get (in the folded-root basis)
$$ 
S_{HE_7^+} = S_{HF_4} \oplus \sigma(S_{HF_4}) = S_{HF_4} \oplus S_{HF_4}^\sigma
$$
with two blocks corresponding to the induced metric on $V_{\rm ph}$ and $V_{\rm in}$ respectively. While $\det S_{HF_4} = 3 - 2\sqrt{5}< 0$ (3 positive, 1 negative eigenvalues), the Galois automorphism $\sigma$ changes the signature of $S_{HF_4}^\sigma$, which has $\det S_{HF_4}^\sigma = \sigma(\det S_{HF_4}) = 3 + 2\sqrt{5} > 0$ (all positive eigenvalues). As a result, the physical space $V_{\rm ph}$ carries a \textit{Lorentzian} representation $\mathbf{4}$, whereas its Galois-conjugate subspace $V_{\rm in}$ carries a \textit{Euclidean} representation $\mathbf{4}$. 

\subsubsection{Block-Decomposition in the Coloured Coxeter Eigenbasis}

While the Coxeter element still acts separately on its own Coxeter eigenplanes $\mathbf{2}$, we can show that those are further mixed by the rest of the infinite elements of $G_{NC}$, thus producing a rank-$m$ symmetry group that is now \textit{dense in the full} $SO(m-1,1)$ instead of just in $SO(1,1) \times SO(2)^{\times {(m-2)/2}}$. 

In particular, choosing the bicoloured Coxeter element $C_{\rm colour}$ for $G_C$ -- as explained in Appendix \ref{app:FoldingisProjection} -- and forming a change-of-basis matrix $B_C$ by placing the Coxeter eigenbasis in the rows of that matrix (ensuring, as in Eq.~\eqref{eq:VDecompositionUnderHyperbolicCoxeterPair}, that we chose the appropriate ordering), we can verify that $G_{NC}$ sits within $G_C$. This is because all $2m$-by-$2m$ block-reflections $\rho_I$ can be simultaneously put into $m$-by-$m$ block-diagonal form by a similarity transformation with $B_C$: 
\begin{equation}
    \rho_I \longmapsto B_C^{-1} \rho_I B_C = 
    \setlength{\arraycolsep}{0.50pt}
    \begin{pmatrix}
        \fixedblock{red}{\rho_I|_{V_{\rm ph}}} & \\
        & \fixedblock{red}{\rho_I|_{V_{\rm in}}}
    \end{pmatrix}
\end{equation}
with the two blocks being conjugate of one another; here, the restrictions $\rho_I|_{V_{\rm ph}}$ generate the non-crystallographic symmetry $G_{NC}$ on $V_{\rm ph}$. 

This is drastically different to our earlier construction in the following way: in Section \ref{sec:SpacetimeCNP}, apart from the already-diagonal Coxeter element (since we are working it its eigenbasis), the only other two reflection generators we could block-diagonalize were the bicoloured ones $\rho_{\rm blue}, \rho_{\rm red}$. But those were generating $C_{\rm colour}$ once taken together. As a result, the only transformations we could perform that preserved our Coxeter splitting were two reflections, and the infinite set of Coxeter boosts/rotations. On the other hand, now we can perform all $m$ reflections $\rho_I|_{V_{\rm ph}}$ \textit{and} all combinations of them, including the Coxeter element. The full infinite symmetry $G_{NC}$ allows us to mix and match the Coxeter planes that are included in $V_{\rm ph}$ (and, by extension in $V_{\rm in}$) freely: we can boost essentially along any direction in the $m$-dimensional $V_{\rm ph}$ space and perform essentially any spatial rotation. In short, $G_{NC}$ is dense in $SO(m-1,1)$ and avoids issues of defining preferred spacetime splitting in $\MM^m$.

\subsubsection{Projectors}

Given the $G_{NC}$-invariant decomposition \eqref{eq:VDecompositionUnderHyperbolicCoxeterPair}, we choose our projectors similarly to Section \eqref{sec:SpacetimeCNP}, with the only difference that the inner products are calculated with the Schl\"afli matrix as in Eq.~\eqref{eq:SchlafliInnerProduct}. Specifically, we have
\begin{eqnarray}
  (\Pi_{\rm ph})^{\mu}_{\;\;\nu}&=&
  \sum_{j=0}^{k \leq k_{\rm ph}}
  \left(\frac{(v_{j}^{+})^{\mu}
  (v_{j}^{-})^{\kappa}S_{\kappa \nu}}
  {\langle {\bf v}_{j}^{+},{\bf v}_{j}^{-}\rangle}
  +\frac{(v_{j}^{-})^{\mu}(v_{j}^{+})^{\kappa}S_{\kappa \nu}}{\langle {\bf v}_{j}^{-},{\bf v}_{j}^{+}\rangle}\right)  
  \label{eq:SProjectorGeneralsCNP}
\end{eqnarray}
with the Coxeter eigenbasis now forming a Newman-Penrose irrational frame with respect to $S$ -- the normalization conditions from Section \ref{subsec:Hyperbolic_Coxeter_Elements} remain the same but with $S$ replacing $\eta$. With these definitions, one can check that the usual projector properties hold 
\begin{equation*}
\Pi_{\rm ph}^2 = \Pi_{\rm ph}, \quad  \Pi_{\rm ph} \Pi_{\rm in} = 0, \quad \Pi_{\rm ph} + \Pi_{\rm in} = 1,
\end{equation*}
and that the projectors are $S$-self-adjoint ($S \Pi_{\rm ph} = \Pi_{\rm ph}^T S$).

\subsection{The Window}

\begin{center}
\begin{tcolorbox}[colframe=blue,colback=white,width=0.99\textwidth]
\textbf{Choice 3:} a window ${\cal W}\subset V_{{\rm in}}$ or weighting function $W:V_{{\rm in}}\to\mathbb{C}$ invariant under $G$: {\it i.e.}\ ${\cal W}=g{\cal W}$ or $W({\bf v}_{{\rm in}})=W(g\,{\bf v}_{{\rm in}})$, $\forall g\in G$.
\end{tcolorbox}
\end{center} 

Since our non-crystallographic groups $G_{NC}$ are infinite, we still cannot use a polytope with a finite number of faces. The hyperbolic Coxeter element acts as an irrational rotation on each of the Coxeter planes within the internal space, but contrary to Section \ref{subsec:SpacetimeCNPWindow}, now the internal planes can get mixed by the infinite rotational symmetry $G_{NC}$. Consequently, similar to how the irrational rotation densely filled the circle $\Sph^1$ for the rank-2 group $I_2^\infty(\lambda_0^+)$, its rank-$m$ cousins $G_{NC}$ will densely fill a sphere in $V_{\rm in} \simeq \EE^m$. As a result, the natural window $\mathcal{W}$ to use is an $m$-dimensional ball, or a spherically-symmetric weighting function $W$ on the internal space, i.e.~$W ({\bf v}_{\rm in}) \sim f(\langle {\bf v}_{\rm in},{\bf v}_{\rm in} \rangle_{V_{\rm in}})$ (for some function $f$). This is to be compared to how we generated (3+1)D dimensional spacetime quasicrystals in Section \ref{sec:ExamplesSpacetimeQuasicrystals}. Back then, to preserve the symmetry $I_2^\infty(\lambda_0^+)$, we used a \textit{polydisk}: independent disks of some (in principle, independent) radii in each Coxeter 2D plane. 

\section{More Examples of Spacetime Quasicrystals} 
\label{sec:ExamplesNewSpacetimeQuasicrystals}

In this final section, we will apply the generalized spacetime \sCNP{} algorithm developed in this chapter to produce more spacetime quasicrystals in $(3+1)$-dimensional Minkowski space $\MM^4$. The 4D examples shown here have an honest rank-4 symmetry compared to the ones shown in Section \ref{sec:ExamplesSpacetimeQuasicrystals}, which were transforming under a dense subgroup of $SO(1,1)\times SO(2)$.

Before we turn on to the examples, let us point out that the implications from Section \ref{sec:implications} remain true in this for these quasicrystals as well: there is no $G_{NC}$-symmetric tiling associated with them, no preferred size of the window, and no local scale invariance, however self-duality and global scale-invariance are still natural properties for these structures to have too.

\subsection{The First $(3+1)$-dimensional Spacetime Quasicrystal with Spatial Icosahedral Symmetry}
\label{sec:IcosahedrallySymmetricsQC}

A remarkably beautiful spacetime quasicrystal can be constructed from the Coxeter pair $(HE_7^+,HF_4)$: the \sCNP{} scheme yields a quasi-crystalline set in $\MM^4$ whose spatial sections show the symmetry of the finite icosahedral group $H_3$, the symmetry class of physical icosahedral quasicrystals (\cite{bindi2009natural}). 

\subsubsection{The Setup}

As our lattice, we take the root lattice of the extended hyperbolic $E_7$ graph shown on the right hand-side of Fig.~\eqref{fig:unfoldingHF4} (obtained by attaching a node -- node 8 -- to node 6 of the underlying $E_7$ diagram). Its Schl\"afli matrix is shown in Eq.~\eqref{eq:unfoldingHF4toHE7Plus}; it is hyperbolic ($S_{HE_7^+}$ has 1 negative and 7 positive eigenvalues) and it contains hyperbolic subdiagrams as explained in Section \ref{subsec:UnfoldingNonCrystallographicGroups}, so its fundamental domain on the hyperboloid $\HH^7$ it tessellates by reflections does not have finite volume. 

The folding of $HE_7^+ \twoheadrightarrow HF_4$ has been shown in Appendix \ref{subsec:UnfoldingAppendix}. The explicit form of the eigenvalues of the Coxeter element is again very messy and we avoid showing them directly; nevertheless, they satisfy: 
\begin{equation}
\begin{aligned}
    & \lambda_0^+ + \lambda_0^- = -\frac{\bar{\phi}}{2} + \frac{1}{2}\sqrt{\frac{5 + 7(1+2\phi)}{2}}, \\
    & \lambda_0^+ - \lambda_0^- = \sqrt{-\frac{1}{2}(1+3\bar{\phi}) +\frac{1}{2}\sqrt{5+\phi}}\\
    & \lambda_1^+ + \lambda_1^- = 2\mathrm{Re}(\lambda_1^+) = -\frac{\bar{\phi}}{2} - \frac{1}{2}\sqrt{\frac{5 + 7(1+2\phi)}{2}}, \\
    & \lambda_1^+ - \lambda_1^- = 2i\mathrm{Re}(\lambda_0^+) = i\sqrt{+\frac{1}{2}(1+3\bar{\phi}) +\frac{1}{2}\sqrt{5+\phi}}
\end{aligned}
\label{eq:eigenvaluesHF7Plus}
\end{equation}
and the Coxeter polynomial is
$$\chi_C(x) = x^8 + x^7 - x^6 - 4x^5 - 5x^4 - 4x^3 - x^2 + x + 1. 
$$
The Coxeter element (in the folded root-basis) for $HF_4$ becomes
$$ 
C|_{V_{NC}} = 
\begin{pmatrix}
0 & \phi & 0 & -\phi \\
1 & \phi & 0 & - \phi \\
0 & \phi & 0 & -1 \\
0 & 0 & 1 & -1
\end{pmatrix},
$$
with characteristic polynomial:
$$ \chi_{C|_{V_{NC}}}(x) = x^4 - \bar{\phi} x^3 - (\phi - \bar{\phi}) x^2 + \bar{\phi} x + 1. 
$$
The factorization of $\chi_C(x)$ in the field extension $\QQ(\phi)$ reads
$$ 
\chi_C(x) = \chi_{C|_{V_{NC}}}(x) \overline{\chi_{C|_{V_{NC}}}(x)} = \big( x^4 - \bar{\phi} x^3 - (\phi - \bar{\phi}) x^2 + \bar{\phi} x + 1 \big) \big( x^4 - {\phi} x^3 + (\phi - \bar{\phi}) x^2 + {\phi} x + 1\big) .
$$

Notice that $\lambda_0^+ + \lambda_0^-$ and $\lambda_1^+ + \lambda_1^-$ are related by some  Galois automorphism that flips the overall sign of the large square root (but does not flip $\phi$ individually). Similarly, the differences are related by a flip of the square-root term inside the large square root, which turns the radicand of $\lambda_0^+-\lambda_0^-$ in Eq.~\eqref{eq:eigenvaluesHF7Plus} to negative, thus bringing an imaginary unit in front of the large square root of $\lambda_1^+ -\lambda_1^-$.\footnote{Descriptively, if $\lambda_0^+ - \lambda_0^-$ is of the form $\sqrt{a + \sqrt{b}}$ for some $a,b \in \QQ(\phi)$ with $a + \sqrt{b}$ being the radicand of the large square root, then it becomes $\sqrt{ a - \sqrt{b}}$ under the sign flip of the smaller square root, which is equal to $ \sqrt{- (-a + \sqrt{b})} = i \sqrt{-a + \sqrt{b}} = \lambda_1^+ - \lambda_1^-
$ in Eq.~\eqref{eq:eigenvaluesHF7Plus}.
} 
The rest of the eigenvalues can be found by considering the Galois-conjugated Coxeter element $\sigma(C|_{V_{NC}})$ (i.e.~replacing $\phi$ with $\bar{\phi}$ in the above matrix); $\sigma:\phi \mapsto \bar{\phi}$ relates the eigenvalues too, as shown below
\begin{equation}
\begin{aligned}
    & \lambda_2^+ + \lambda_2^- = 2\mathrm{Re}(\lambda_2^+) = -\frac{\phi}{2} + \frac{1}{2}\sqrt{\frac{5 + 7(1+2\bar{\phi})}{2}} = \sigma(\lambda_0^+ + \lambda_0^-) \\
    & \lambda_2^+ - \lambda_2^- = 2 i \mathrm{Re}(\lambda_2^+) = i\sqrt{\frac{1}{2}(1+3\phi) +\frac{1}{2}\sqrt{5+\bar{\phi}}} = \sigma(\lambda_1^+ - \lambda_1^-)\\
    &  \lambda_3^+ + \lambda_3^- = 2\mathrm{Re}(\lambda_3^+) = -\frac{\phi}{2} - \frac{1}{2}\sqrt{\frac{5 + 7(1+2\bar{\phi})}{2}} = \sigma(\lambda_1^+ + \lambda_1^-)\\
    & \lambda_3^+ - \lambda_3^- = 2 i \mathrm{Re}(\lambda_3^+) = i\sqrt{\frac{1}{2}(1+3\phi) -\frac{1}{2}\sqrt{5+\bar{\phi}}} = \sigma(\lambda_0^+ - \lambda_0^-)
\end{aligned}.
\label{eq:eigenvaluesHF7Plus_GaloisActed}
\end{equation}

The spacetime quasicrystal is the \sCNP{} point set arising from projecting onto the subspace spanned by the Coxeter eigenbasis corresponding to the eigenvalues $\lambda_0^\pm,\lambda_1^\pm$ with a projector given by Eq.~\eqref{eq:SProjectorGeneralsCNP}. The $HF_4$-invariant window $\mathcal{W}$ is taken to be an $S$-ball in the internal space $V_{\rm in}$. Note that the sphere must be defined with respect to the induced metric $S|_{V_{\rm in}}$ on the internal space and not with the Minkowski metric. The mean density of the pattern is then (see Eq.~\eqref{eq:DensityOfCNPQuasicrystal})   
\begin{equation}
    \rho_{HF_4} = \frac{\mathrm{vol}_S(\mathcal{W})}{\sqrt{|\det S|}} \quad \big( \approx 5.951 \text{ with } R_{\rm phys \, disk} = 1 \big)
    \label{eq:sQCDensityHF4}
\end{equation}
where $\mathrm{vol}_S(\mathcal{W}) = 2\pi^2 R^4$ (the $S$-volume of the 4-ball; $R$ is its arbitrary radius), and $\sqrt{|\det S|}$ is the lattice co-volume of $\mathbb{Z}^8$ in the quadratic form $S$. For $HE_7^+$: $\det(S_{HE_7^+}) = (3+2\sqrt{5})(3-2\sqrt{5}) = -11 < 0$ (see discussion in Section \ref{subsec:NCSymmetryGeneralsCNP}). 

\subsubsection{Revealing the Icosahedral Structure on a Spatial Slice}

From a group-theoretic point of view, the fact that the icosahedral group $H_3$ (finite, of order 120) is included as the maximal finite subgroup of $HF_4$ is a consequence of the fact that there exists a $H_3$ subdiagram of $HF_4$. Indeed, if we look at Fig.~\ref{fig:unfoldingHF4}, removing either node 1 or 4 from the ends of the $HF_4$ Coxeter-Dynkin graph, we find the three-node graph of $H_3$. This is the point stabilizer of any vertex in the tiling of $\HH^3$ by $HF_4$. In fact, since there are two different such sub-diagrams, $H_3$ is only unique inside $HF_4$ up to conjugation with an element of $HF_4$. 

The icosahedral group $H_3$ acts on a rest frame: it fixes a timelike direction ${\bf w}$ and acts by ordinary $SO(3)$-type rotations on the orthogonal spatial subspace $\EE^3$. Therefore, to see the icosahedral symmetry, we should extract an appropriate time-slice of the spacetime quasicrystal. In other words, we should build a frame adapted to the icosahedral symmetry. 

To do this, we first note that the three reflections $\rho_{1,2,3}$ of $HF_4$ (with the labelling of Fig.~\ref{fig:unfoldingHF4}) fix a common 2-dimensional subspace; their nullspace is
$$
\ker (\rho_1 -\mathbb{I}) \cap \ker (\rho_2 -\mathbb{I}) \cap \ker (\rho_3 -\mathbb{I}) \subset \MM^8 .
$$
This is 2-dimensional because the eight directions in the space $\MM^8$ split as: 
$$ 
\begin{aligned}
& (\text{one $H_3$-fixed physical \textit{timelike} direction}) + (\text{one $H_3$-fixed internal \textit{spacelike} direction}) \\
&  \hspace{+35mm} +(\text{six complementary non-$H_3$-fixed directions})
\end{aligned}
$$
Applying $\Pi_{\rm ph}$ on this nullspace kills the $H_3$-fixed internal direction (the Galois-conjugate ``time'' of the physical ``time''). The survivor with the most-negative $S$-norm is the rest-frame timelike vector ${\bf w}$, which we normalize to $\langle {\bf w}, {\bf w} \rangle = {\bf w}^T S {\bf w} = -1.$

Icosahedrally symmetric patterns exhibit three types of rotational symmetries: (i) \textit{2-fold}, (ii) \textit{3-fold} and, (iii) \textit{5- or 10-fold}. Indeed, a regular icosahedron $\{ 3,3,5\}$ has: six 5-fold axes (passing through the 12 opposite vertices; rotations around vertices where 5 triangular faces meet, mapping one face to another), ten 3-fold axes (passing through the centres of the 20 opposite triangular faces; rotations map the three vertices of each face to one another), and fifteen 2-fold axes (passing through the midpoints of the 30 edges of the icosahedron; reflects the endpoints of each edge). These 31 axes generate 60 rotational symmetries, which are combined with a reflection, to yield the full $H_3$ group of order 120. 

Here, we are interested in the most exotic 5-fold rotational symmetries, so we would like to find the 5-fold axes $\bf f_5$. The order-5 element $\rho^{(5)} = \rho_{2}\rho_{3}$ is a rotation by $2\pi/5$ in the physical space: it rotates within the plane spanned by the folded roots $\widetilde{\beta}_{2,3}$ and fixes the plane to it that has signature $(1,1)$. The latter contains our timelike axis $\bf w$ (which is fixed by $H_3$) and also the 5-fold spacelike axis ${\bf f}_5$ we are after. To isolate ${\bf f}_5$, we compute $\Pi_{\text{ph}}(\ker(\rho^{(5)} - I))$ and remove the ${\bf w}$-component in the $S$-metric; for any vector ${\bf v}$, the ${\bf w}$ component is $({\bf v}^T S {\bf w}){\bf w}$. Then, $\langle {\bf w}, {\bf f_5} \rangle = 0$ by construction and we normalize ${\bf f}_5$ to $\langle {\bf f}_5, {\bf f}_5 \rangle = +1$. This yields an 8D vector within $\MM^8$ in the folded root basis. 

Holding the timelike direction $\bf w$ fixed, we would like to build an orthonormal spatial frame $\{ {\bf a}_1, {\bf a}_2, {\bf a}_3\}$ adapted to it. The spatial directions for the rest-frame of an observer with timelike vector ${\bf w}$ is the $S$-orthogonal complement of ${\bf w}$ within the physical space $V_{\textrm{ph}}$:
$$V_{\textrm{rest frame}} = \mathbf{w}^\perp \cap V_{\rm ph} = 
\{ {\bf x} \in V_{\textrm{ph}} : \langle {\bf x}, {\bf w} \rangle = {\bf x}^T S {\bf w} = 0\} \simeq \EE^3.
$$
This is what the observer at rest calls ``space'', on which $H_3$ acts as ordinary rotations. To find these directions $\{ {\bf a}_i\}$, we act with $\Pi_{\rm ph}$ on the root-basis of the lattice $HE_7^+$ (i.e.~the identity matrix), and we then perform a Gram-Schmidt process with the $S$-metric on the resulting columns of $\Pi_{\rm ph}$. Making sure to remove any ${\bf w}$-component of the physical vectors and after $S$-normalize it, we end up with the three orthogonal and normalized to $1$ vectors ${\bf a}_i$. In this basis, the induced metric is simply the 3-by-3 identity matrix. The 5-fold axis ${\bf f}_5$ belongs to the rest frame; expressing it in the $\{ {\bf a}_i\}$ basis, we find\footnote{The standard icosahedron has vertices at $(0, \pm 1, \pm \phi)$ and its cyclic permutations. The 5-fold axes pass through opposite vertex pairs, so a unit 5-fold direction is $\tfrac{1}{\sqrt{1 + \phi^2}}(0, 1, \phi).$ That is the hallmark number of icosahedral symmetry; it lives in $\mathbb{Q}(\sqrt{5})$ and it cannot arise accidentally from a crystallographic axis, which would have rational components.} 
$$
{\bf f}_5 = \frac{1}{\sqrt{\phi^2 + 1}} \left(1,\frac{1}{\sqrt{3}},\frac{\phi^2}{\sqrt{3}} \right), \qquad ({\bf f}_5^T {\bf f}_5 = +1).
$$

\begin{figure}[t!]
    \centering
    \subfloat[Looking down the ${\bf f}_5$-axis.]{
    \includegraphics[width=0.48\linewidth]{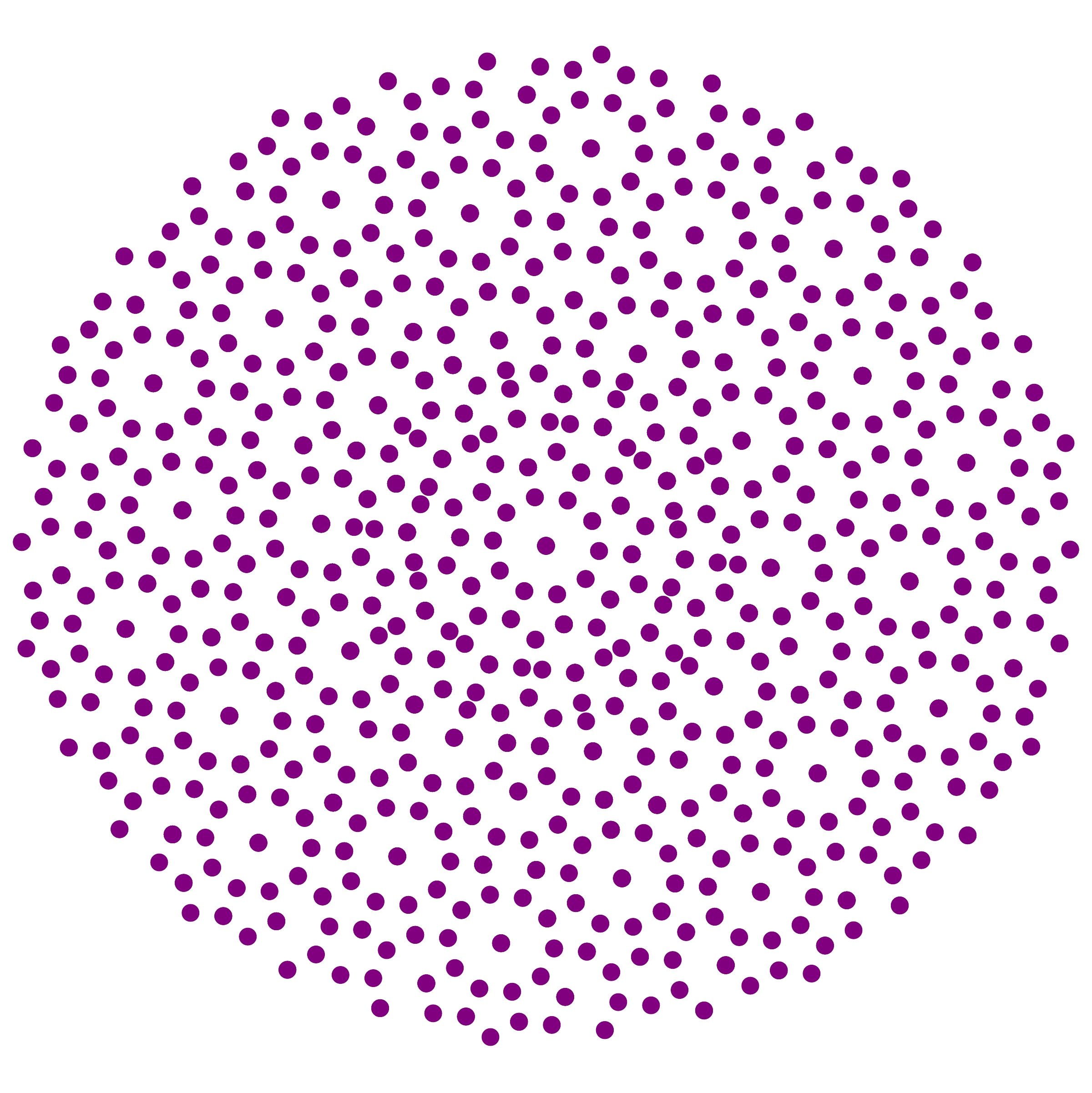}
    }
    \hfill
    \subfloat[Rotating in 3D to the 3-fold axis.]{
    \includegraphics[width=0.48\linewidth]{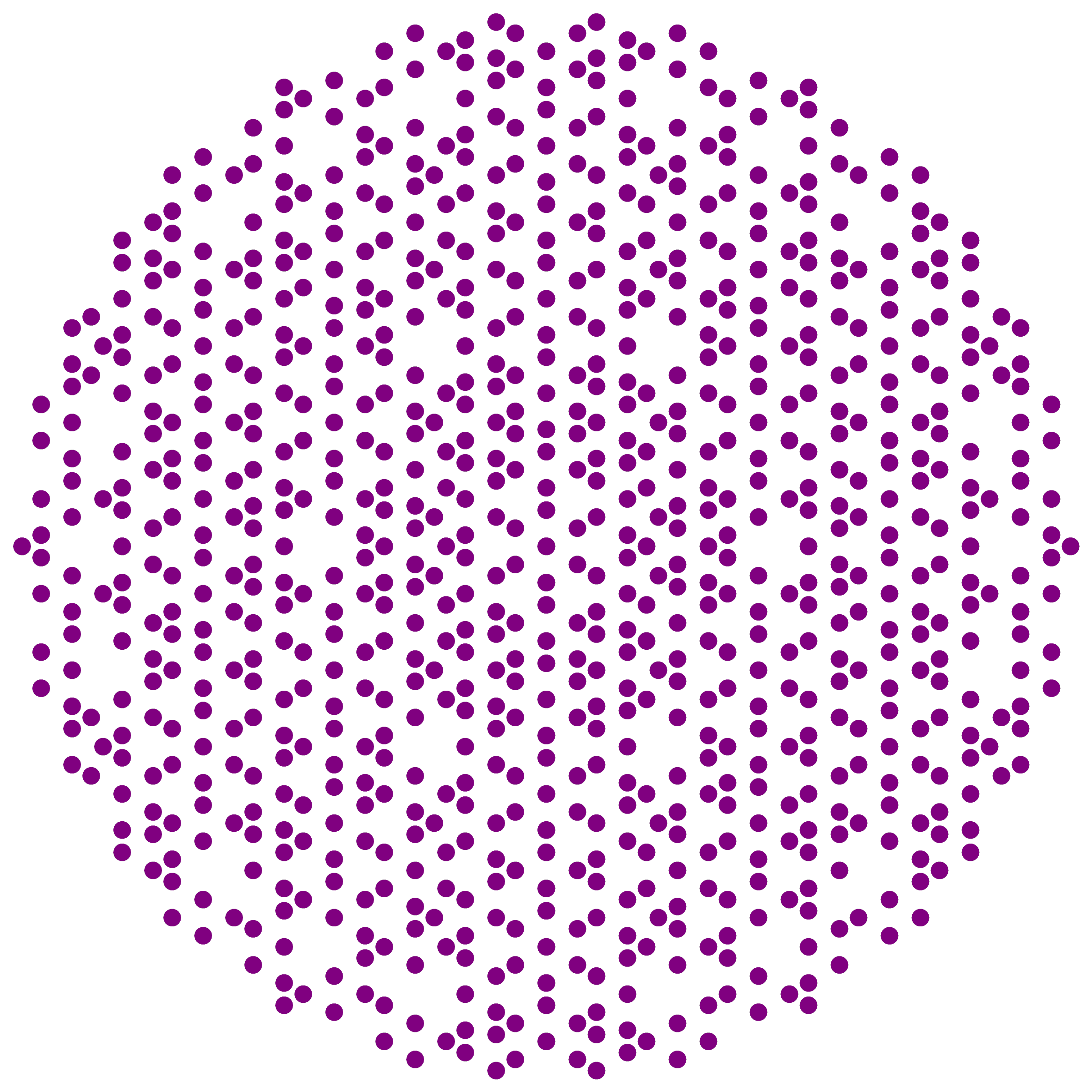}
    }
    \caption[Spatial slices of the $HF_4$ spacetime quasicrystal: the 5-fold and the 3-fold axes of symmetry.]{Spatial slices of the $HF_4$ spacetime quasicrystal: the 5-fold and the 3-fold axes of symmetry. On the left, we show the 2D projection down the ${\bf f}_5$ axis of 5-fold (or, actually, 10-fold) rotational symmetry, whereas on the right panel we rotate the full 3-dimensional pattern and find one axis of 3-fold (or, actually, 6-fold) rotational symmetry.}
    \label{fig:SpatialSlicesIcosahedralSymmetry}
\end{figure}

Having the full spacetime frame $\{ \bf w,a_1,a_2,a_3 \}$ adapted to $H_3$ at hand, the only step left is to run the \sCNP{} algorithm, obtain our spacetime quasicrystal, and look down straight at the axis $\mathbf{f}_5$. We show two slices in Fig.~\ref{fig:SpatialSlicesIcosahedralSymmetry}, one in 2-dimensions (looking down ${\bf f}_5$), and one in 3-dimensions, where we rotated manually to find a 3-fold axis. Note that the icosahedral group $H_3$ is centrosymmetric (contains inversions $-I$). The spherical window $\mathcal{W}$ is also inversion-symmetric and, therefore, the point set is centrosymmetric: if ${\bf p_{\rm ph}} \in \Lambda_{\rm ph}$ then $-{\bf p}_{\rm ph} \in \Lambda_{\rm ph}$. A 5-fold rotation plus inversion gives $2\times 5 = 10$ equivalent angular positions in the planar projection — exactly what is seen in real icosahedral quasicrystals. This is why we end up with 10-fold symmetric rings instead of just 5-fold symmetric ones. Similarly, we see hexagonal (6-fold) symmetry arising as well.

\begin{figure}[h!]
    \centering
    \subfloat[Angular distribution from 0 to $2\pi$.]{
    \includegraphics[width=0.48\linewidth]{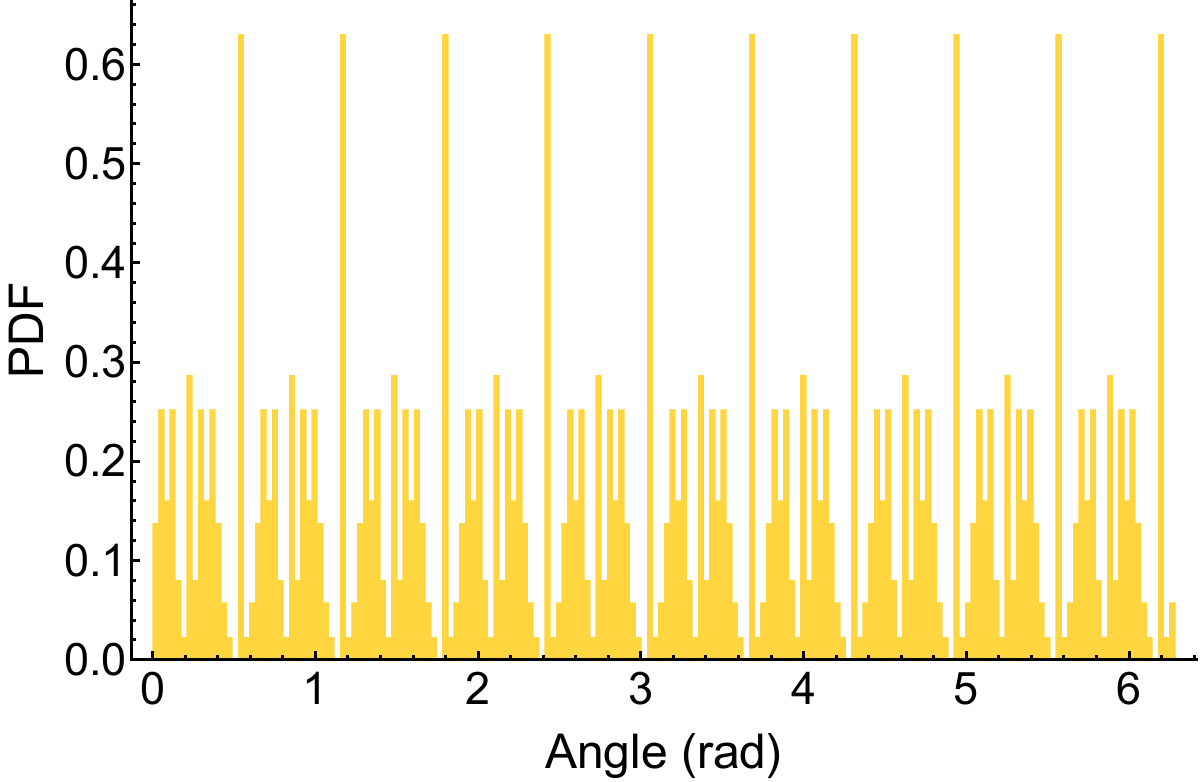}
    }
    \hfill
    \subfloat[The same angular distribution, $\mathrm{mod} \, 2\pi/5$.]{
    \includegraphics[width=0.48\linewidth]{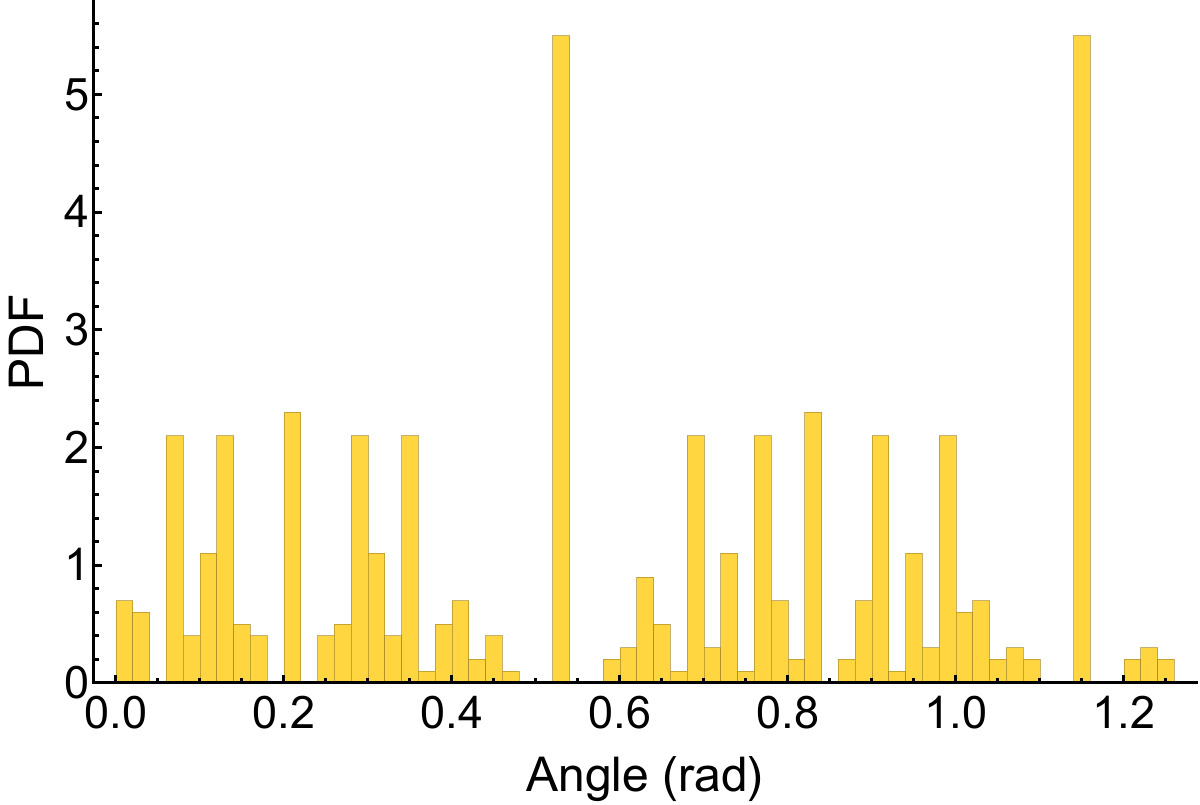}
    }
    \caption[Histograms of the angular distribution of points inside an annular ring of the 2-dimensional quasi-crystalline of the $HF_4$ spacetime quasicrystal.]{Histograms of the angular distribution of points inside an annular ring of the 2-dimensional pattern shown in Fig.~\ref{fig:SpatialSlicesIcosahedralSymmetry}. The vertical axis is the probability density function and the horizontal is the angles. On the left, we count angles from $0$ to $2\pi$ and bin them into $1^\circ$-bins. We see 10 equally-spaced peaks, confirming exact 10-fold rotational symmetry of the spatial sections of the $HF_4$ spacetime quasicrystal when viewed along the 5-fold axis. On the right, we count the same angles mod $2\pi/5$; we see sharp peaks at exactly $2\pi/10 = 36^\circ$ and $2\pi/5 = 72^\circ$, as $10$-fold symmetry demands.}
    \label{fig:IcosahedralAnglesSymmetry}
\end{figure}

Using the 2D projection on the left, we selected points in an annular ring to isolate on ``shell'' of the structure.\footnote{We calculated the radii of all projected points from the origin in that patch, found the median radius, and then selected points between 0.4 and 1.6 times that median; this is arbitrary, but we just wanted to avoid the centre of the pattern in case it was noisy.} For these points inside the ring, we computed their angular position and we show two histograms in Fig.~\ref{fig:IcosahedralAnglesSymmetry}.

After projecting the slab points down the 5-fold axis to get 2D coordinates, we selected points in an annular ring (between $0.4$ and $1.6$ times the median radius) to isolate one ``shell'' of the structure, then computed the angular position of each point in that ring. Two histograms are shown in Fig.~\ref{fig:IcosahedralAnglesSymmetry}, confirming the icosahedral orientational order of our structure.

\subsection{Enriching the Icosahedral Family of Spacetime Quasicrystals in $(3+1)$ and $(4+1)$ Dimensions}
\label{subsec:moreicosahedralsQC}

Before closing the chapter, we would like to emphasize that we have explored only the very tip of the iceberg here. The hyperbolic Coxeter pairs identified, for the first time, in Appendix \ref{subsec:HyperbolicCoxeterPairsAppendix} suggest that there are, in fact, several more icosahedrally symmetric spacetime quasicrystals one could build. 

In $(3+1)$-dimensions, the Coxeter pairs $(Ah_7^+,HH_4)$ and $(Dh_6^{++}, HD_4)$ can both give rise to such a structure, as the non-crystallographic groups all contain an $H_3$ subgroup.

In $(1+4)$-dimensions, there is the very promising case of the pair $(Eh_9^+,H_5)$, with $H_5$ being the honest-to-God 5-dimensional cousin of $H_4$. Intriguingly, the root lattice of its crystallographic counterpart $Eh_9^+$ is \textit{almost} $E_{10}$.\footnote{The idea behind spacetime quasicrystals started by trying to identify the Lorentzian analogue of the Elser-Sloane 4D quasicrystal $H_4$ inside $E_8$. Given that $E_{10}$ \textit{is} the Lorentzian cousin of $E_8$, we were naturally led to this crystallographic group, but our efforts to find the $H_4$ analogue were to no avail. Only now are we able to see that $H_5$ actually exists inside a group very closely related to $E_{10}$ but also very different at the same time (because it contains hyperbolic subgraphs).}. Indeed, it has 10 nodes -- see Fig.~\ref{fig:UnfoldingHyperbolicGroupsRank5} -- but the single extending node that would take us from the affine graph $\widetilde{E_8} = E_9$ of 9 nodes (shown in Fig.~\ref{fig:affine_diagrams}) to the hyperbolic graph $\mathrm{II}_{9,1} = E_{10}$ (shown in Fig.~\ref{fig:CoxeterDynkinDiagramsEvenSelfDual}) lies in the ``wrong'' side: node 2 in Fig.~\ref{fig:UnfoldingHyperbolicGroupsRank5} should have been appended to node 1 of the same graph to get $E_{10}$ instead of $Eh_9^+$.\footnote{Note that this is reminiscent of how we get the affine extension of the $E_7$ graph in Fig.~\ref{fig:affine_diagrams}: we add a node to the ``wrong'' direction of the $E_7$ graph. A node in the ``correct'' direction would give $E_8$ instead.} The rest of the hyperbolic Coxeter pairs, namely ($Ah_9^+, HH_5$) and ($Dh_9^+,HD_5$) also have their non-crystallographic counterpart contain $H_4$ as their maximal finite subgroup, thus they can also produce 5D quasicrystals with $H_4$ spatial symmetry.    

Interestingly, there is no known non-crystallographic group in rank 6, at least none whose fundamental Coxeter polytope has finite volume. Could it be that in spacetime one can have icosahedrally symmetric quasicrystals only up to dimension 5 (contrary to the Euclidean case, where the highest is dimension 4)? Could one form some hyperbolic Coxeter pair where both $G_C$ \textit{and} $G_{NC}$ are outside of any known classifications? And if so, would these cases be of the same interest? Answering such questions is an interesting future direction that is to be explored beyond the scope of this thesis.

\chapter[Spacetime Quasicrystals: String Compactifications and Phenomenology]{Spacetime Quasicrystals in String Compactifications and Phenomenology}
\label{ch:StringCompactifications}

So far, this part has focused on developing the mathematical formalism necessary to construct spacetime quasicrystals, and constructing the first examples. In this Chapter, we can't resist offering some speculations about how such spacetime quasicrystals might relate to the real world.  

\section{Fitting the Universe in a Nutshell}
\label{sec:FittingUniverseNutshell}

Superstring theory \cite{Green:1987sp, Green:1987mn, Polchinski:1998rq, Polchinski:1998rr} has been much-studied as a candidate unified theory; but mathematical consistency requires the superstring to live in a $(9+1)$ dimensional target space to cancel the worldsheet Weyl anomaly.  

To reconcile this with the $(3+1)$ dimensional universe we observe, the standard approach is to imagine (a modern variant of) the Kaluza-Klein picture in which four of the ten dimensions are large and non-compact, while the remaining six ``extra'' spatial dimensions are ``compactified'' ({\it i.e.}\ curled up into an unobservably tiny 6D compact manifold). Alas, this picture so far does not have any observational evidence in its favour, and arguably creates more problems than its solves. In particular, stabilizing the six extra dimensions in a way that agrees with observations has been a notoriously difficult, and has driven theorists to consider awkwardly complex mechanisms that are often compared to Rube-Goldberg machines, and lead to a vast ``string landscape'' of possible vacua for 4D spacetime.

Here we propose a new geometric picture for how our 4D universe might be embedded in a higher-dimensional spacetime, completely different from the usual Kaluza-Klein picture. Although our proposal is logically independent of string theory, superstrings provides an intriguing motivation, as we now explain.

\subsection{Classical Picture: Irrational Spacetime}
\label{subsec:IrrationalityOfSpacetime}

The most well studied compactifications of string theory are the {\it toroidal} compactifications: here one takes $m$ of the target-space dimensions to be curled up in an $m$-dimensional torus $\TT^{m}$, which is constructed by taking $m$-dimensional flat space $\mathbb{R}^{m}$ modulo an $m$-dimensional lattice $\Lambda_{m}$: $\TT^{m}=\mathbb{R}^{m}/\Lambda_{m}$. Due to stringy effects (in particular $T$-duality \cite{Harvey1987dd}), such compactifications can have much higher symmetry than one would naively expect based on the isometry group of the compactified target spacetime. In \cite{moore1993finitedirections}, Moore studied such toroidal compactifications in generality, including the possibility of compactifying the time direction! He found that the {\it most symmetric} toroidal compactification of the superstring was obtained when {\it all 10 dimensions} of the target space were compactified on the 10D Lorentzian torus $\TT^{9,1}$ constructed by taking 10D Minkowski  space $\mathbb{R}^{9,1}$ modulo the 10D even self-dual Lorentzian lattice $\Lambda_{9,1}$: $\TT^{9,1}=\mathbb{R}^{9,1}/\mathrm{II}_{9,1}$.  

This most-symmetric superstring compactification was assumed to be unphysical, and of purely academic interest, for two reasons: first, because if all ten dimensions are compactified at an ultra-microscopic scale, then one seems to be left without any remaining non-compact directions to serve as the $(3+1)$ macroscopic directions we observe; and second, because all directions (including time) are compactified, then the target spacetime has closed time-like curves which are usually assumed to be pathological, since they seem to be paradoxical and at odds with the causality properties of our universe.  

However, the previous section's spacetime quasicrystal construction suggests a way around this conclusion.

\begin{figure}[t!]
    \centering
    \includegraphics[scale=0.2]{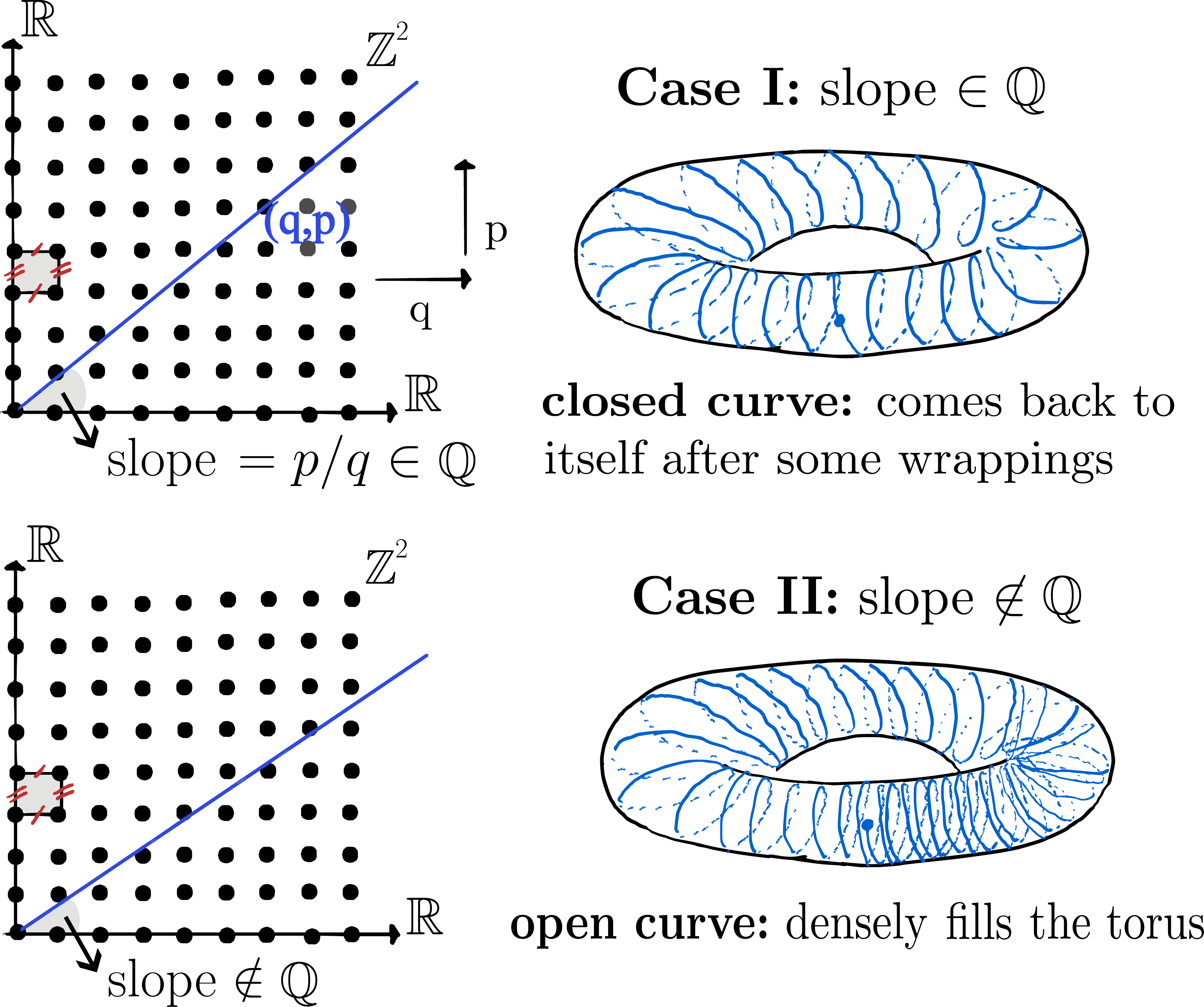}
    \caption[Toroidal compactifications with rational and irrational subspaces; an illustration.]{Case I: the line's slope is a rational number $p/q$, with $p$ and $q$ relatively prime integers. Then, if one moves $q$ steps in the $x$-direction and $p$ steps in the $y$-direction, one returns to the same point on the torus $\mathbb{R}^{2}/\mathbb{Z}^{2}$. Upon compactification (identification of the red edges on the square fundamental cell for the lattice), the origin becomes identified with any other lattice point. Thus, after compactification, the rational line will eventually come back to itself.  However, if the line is irrationally sloped (case II), then no other point of the lattice apart from the origin will belong to the line. Such a line densely fills the torus, extends indefinitely on it and avoids ever crossing itself as it wraps around it.}
\label{fig:IrrLineCompactification}
\end{figure}

To see why, let us first warm up with a simplified 2D example: consider the 2D square lattice $\Lambda_{2}$ in the 2D Euclidean plane $\mathbb{R}^{2}$, and a 1D line of slope $\sigma$ cutting through the lattice.  Now, if we mod out $\mathbb{R}^{2}$ by the 2D lattice $\Lambda_{2}$, we obtain the torus $\TT^{2}=\mathbb{R}^{2}/\Lambda_{2}$; and the 1D line in $\mathbb{R}^{2}$ is mapped to a 1D curve on the torus $\TT^{2}$ (see the blue line in Fig.~\ref{fig:IrrLineCompactification}). If the original 1D line in $\mathbb{R}^{2}$ had a \textit{rational} slope $\sigma=p/q$ (where $p$ and $q$ are relatively prime integers), it is mapped to a {\it closed} curve of {\it finite} length on $\TT^{2}$, since the curve comes back joins up with itself after wrapping $p$ times {\it around} the $\TT^2$'s hole, and $q$ times {\it through} the $\TT^2$'s hole (top panel in Fig.~\ref{fig:IrrLineCompactification}).  By contrast, if the original line in $\mathbb{R}^{2}$ had an {\it irrational} slope $\sigma$, it is mapped to an {\it open} curve of {\it infinite} length on $\TT^{2}$, since the curve never joins up with itself. Instead, it wraps around and around the $\TT^2$, densely covering it, producing a space-filling curve (bottom panel in Fig.~\ref{fig:IrrLineCompactification}). In this case, an ant confined to the line would inhabit a 1D world of infinite extent, even though it is embedded in a compact 2D torus of finite area!

\begin{figure}[!t]
    \centering
    \includegraphics[scale=0.33]{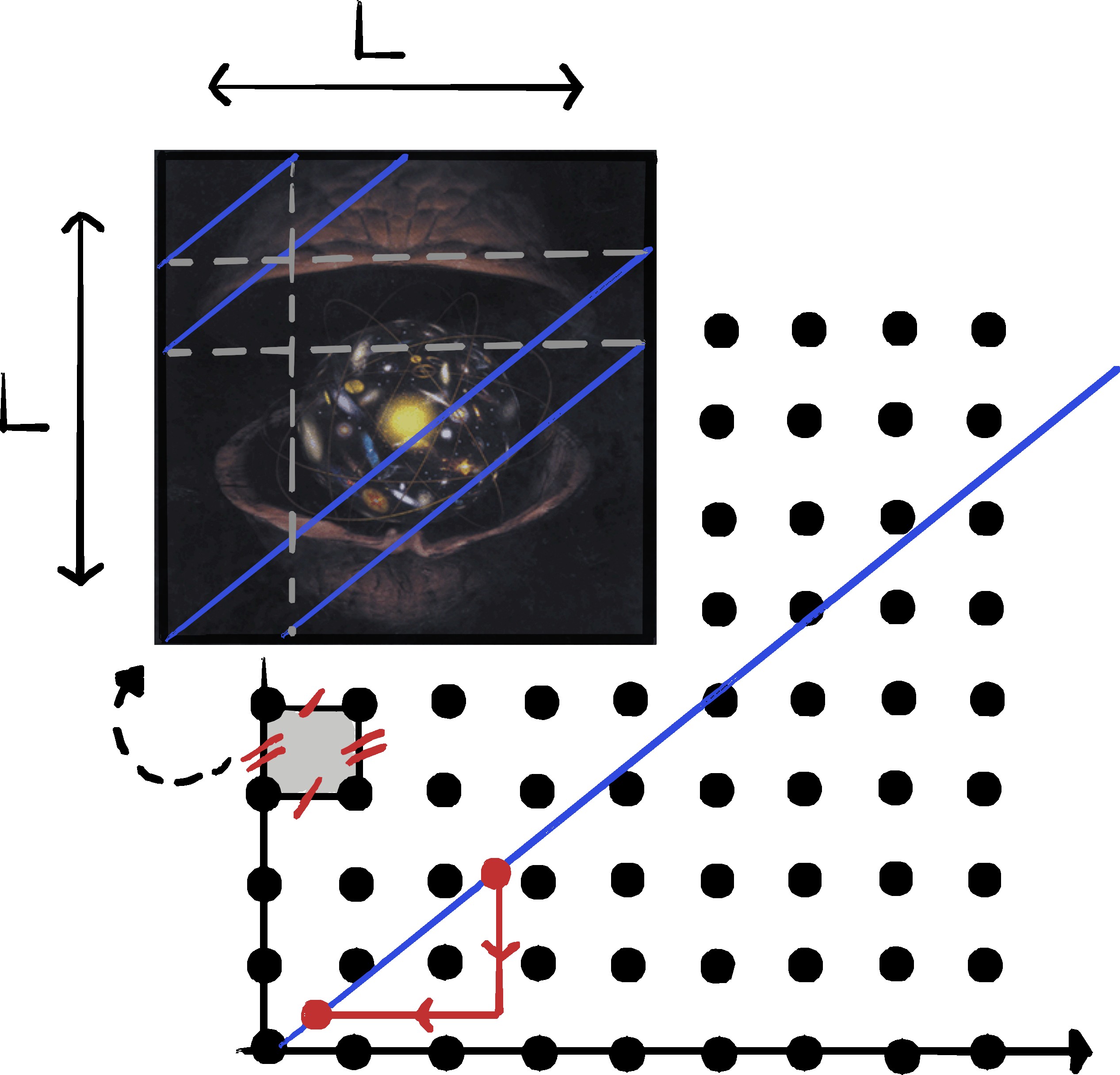}
    \caption[Fitting the Universe in a nutshell; an illustration.]{~Upon compactification, all points along the blue line can be identified with a point within the first square, which has been turned into a 2D torus.~By extension, the full infinite blue line is mapped within the square.~The game resembles ``Pac-Man'': whenever the line hits the boundary of the square, it restarts in the opposite edge from its equivalent point (up to edge-identification).~This is depicted at the background within the zoomed-in square. Thus, the whole infinite 1D blue line can fit inside a tiny 2D torus.~The no self-crossing condition follows from irrationality.~Extrapolating to the idea that the blue line represents our 4D spacetime living within 10 dimensions, the whole infinite universe can fit in a tiny nutshell! Figure adapted from Hawking's \cite{Hawking2001-HAWTUI}.}
    \label{fig:UniverseInNutshell}
\end{figure}

Now return to the 10D example of interest: consider the 10D Lorentzian lattice $\mathrm{II}_{9,1}$ in 10D Minkowski space $\mathbb{R}^{9,1}$.  We have seen in Sec.~\ref{subsec:4DLorentzianQuasicrystal} that the Coxeter element $C$ of $\mathrm{II}_{9,1}$ generates a non-crystallographic group $G$ that group induces a decomposition $V=V_{\rm ph}\oplus V_{\rm in}$, where $V_{\rm ph}=\mathbb{R}^{3,1}$ (a copy of 4D Minkowski space) and $V_{\rm in}=\mathbb{R}^{6,0}$ (a copy of 6D Euclidean space) transform under two different representations ($\rho_{3,1}$ and $\rho_{6,0}$) of $G$.  Crucially, since $G$ is non-crystallographic, this 4D Minkowski subspace $V_{\rm ph}=\mathbb{R}^{3,1}$ is {\it completely irrational} with respect to the lattice $\mathrm{II}_{9,1}$ ({\it i.e.}\ it does not intersect any points in the lattice $\mathrm{II}_{9,1}$, except for the origin).  So now, if we mod out $\mathbb{R}^{9,1}$ by the lattice $\mathrm{II}_{9,1}$, we obtain the 10D Lorentzian torus $\TT^{9,1}=\mathbb{R}^{9,1}/\mathrm{II}_{9,1}$; and the irrational 4D Minkowski subspace $V_{\rm ph}=\mathbb{R}^{3,1}$ is mapped to a $(3+1)$D surface on the $\TT^{9,1}$, which densely fills the torus, wrapping around and around it without ever intersecting itself (like the blue line in the lower panel of Fig.~\ref{fig:IrrLineCompactification}).

Wigner realized that different types of elementary particles are characterized by the way in which they transform under the fundamental symmetries of the theory.  In particular, elementary particles that transform as the representation $\rho_{3,1}$ of $G$ would be confined to the irrational $(3+1)$-dimensional surface that transforms in the appropriate way.  A being made of such particles would regard themselves as living in an {\it infinite} (non-compact) 4D Minkowski space, with {\it no closed timelike curves}, despite the fact that this 4D spacetime is embedded in a 10D torus $\TT^{9,1}$ of finite volume, with all 10 dimensions compactified (including the time direction)!  

In other words, this picture fits an infinite $(3+1)$D universe in a $(9+1)$D nutshell (see Fig. \ref{fig:UniverseInNutshell}).

\section{The Puzzles of the Hierarchy of Scales}
\label{sec:HierarchyPuzzles}

Now let us see how this picture might be helpful in explaining some of the large hierarchies between three fundamental dimensional scales in Nature: the Planck scale $M_{\rm Pl}\sim 10^{18}~{\rm GeV}$, the electroweak scale $M_{\rm EW}\sim 10^{3}~{\rm GeV}$ and the vacuum energy scale $M_{\rm vac}\sim 10^{-12}~{\rm GeV}$.

\subsection{Cartoon Picture}
\label{subsec:CartoonPicture}

To set the stage, consider the action for the standard model of particle physics coupled to gravity.  Written in its symmetric form, almost all of the terms have dimension{\it less} coefficients, and only three have dimension{\it ful} coefficients. Consider these three  dimensionful terms:
\begin{equation}  
    S =\int d^4x \sqrt{-g} \left[ \underbrace{\mspace{+5mu} \textcolor{black}{M_{\rm EW}^2} (h^{\dagger} h) \mspace{+5mu}}_{\mathclap{\text{Higgs}}}+ \underbrace{\mspace{5mu} \textcolor{black}{M_{\rm Pl}^2} R \mspace{5mu}}_{\mathclap{\text{Bending}}} \mspace{+5mu} - \underbrace{\mspace{5mu} \textcolor{black}{M_{\rm vac}^4} \mspace{5mu}}_{\mathclap{\text{Stretching}}} \mspace{5mu} \right]
    \label{eq:ActionHierarchyProblem}
\end{equation}
namely, the \textit{quadratic self-coupling} $M_{\rm EW}^{2}(h^{\dagger}h)$ of the Higgs doublet $h$ (responsible for the Higgs mechanism that gives mass to the fermions), the \textit{Einstein-Hilbert term} $M_{\rm Pl}^{2}R$ (responsible for gravity -- the ``bending'' of the spacetime fabric) and the \textit{vacuum energy term} $M_{\rm vac}^{4}$ (responsible for the universe's slowly accelerating expansion today -- the ``stretching'' of the spacetime fabric).

Historically, gravity was discovered before the standard model of particle physics; but for motivational purposes, imagine a counterfactual history where it was the other way around. To a physicist who only knew about the standard model, it would seem that there was only {\it one} dimensionful term in the action (the Higgs term), and hence only {\it one} fundamental scale in nature: $M_{\rm EW}$.  

If we then informed this physicist of the existence of gravity, and two additional dimensionful terms in the action, they would probably guess the order of magnitude of these new coefficients was set by the only scale they were aware of ($M_{\rm EW}$) -- {\it i.e.} they might guess the new terms would be of order $\int d^{4}x\sqrt{-g}(M_{\rm EW}^{2}R+M_{\rm EW}^{4})$.  They would be shocked to discover that, in reality, the coefficient of 
the Einstein-Hilbert term (proportional to the Ricci curvature $R$) was so much {\it larger} than expected (by a factor of $10^{30}$), while the coefficient of the vacuum energy term (proportional to the overall volume of spacetime) was so much {\it smaller} than expected (by a factor of $10^{60}$).  In other words, they would be shocked that there was such a {\it high cost} (in terms of action) for {\it bending} spacetime, and such a {\it low cost} for {\it stretching} spacetime.

From a ``cartoon'' standpoint, the picture presented in Sec.~\ref{subsec:IrrationalityOfSpacetime} seems like it might be a promising starting point for explaining these surprising qualitative features.  On the one hand, the fact that the irrationally-sloped 4D Minkowski slice $\mathbb{R}^{3,1}$ {\it densely} fills the 10D torus $\TT^{9,1}$ seems like it might explain why $\mathbb{R}^{3,1}$ is so {\it hard} to bend: because any given portion of the $\mathbb{R}^{3,1}$ is infinitely tightly ``hemmed in'' in the extra dimensions between other infinitely-close neighbouring portions of the $\mathbb{R}^{3,1}$, like the blue curve in the lower panel of Fig.~\ref{fig:IrrLineCompactification}. Effectively then, any local deformation that attempts to curve/bend some portion of spacetime gets suppressed by the natural tendency of the different wrappings of $\RR^{3,1}$ to repel one another and avoid crossing. On the other hand, the fact that the $\mathbb{R}^{3,1}$ slice is infinite in extent (like the blue curve in the bottom panel of Fig.~\ref{fig:IrrLineCompactification}) rather than rejoining itself after a finite distance (like the blue curve in the top panel of Fig.~\ref{fig:IrrLineCompactification}) might explain why it is so {\it easy} to stretch: since it has no intrinsic length scale, there is nothing to obstruct or ``push back against'' rescaling it.  Moreover, we have seen that the associated \CNP{} quasicrystals -- and in particular the relevant $(3+1)$D quasicrystals constructed in Sec.~\ref{subsec:4DLorentzianQuasicrystal} -- have a global discrete scale invariance (which, again, corresponds to low cost in action for rescaling spacetime).

\subsection{Seesaw Mechanism}
\label{subsubsec:SeesawMechanism}

We now attempt to go beyond the previous section's cartoon picture, to suggest a (semi-) quantitative argument for how this picture might help explain the observed seesaw relation between $M_{\rm vac}$ and $M_{\rm Pl}$:
\begin{equation}
  \label{M_seesaw}
  M_{\rm EW}^{2}\approx M_{\rm vac}M_{\rm Pl}.
\end{equation}

\begin{figure}[!t]
    \centering
    \includegraphics[width=0.5\linewidth]{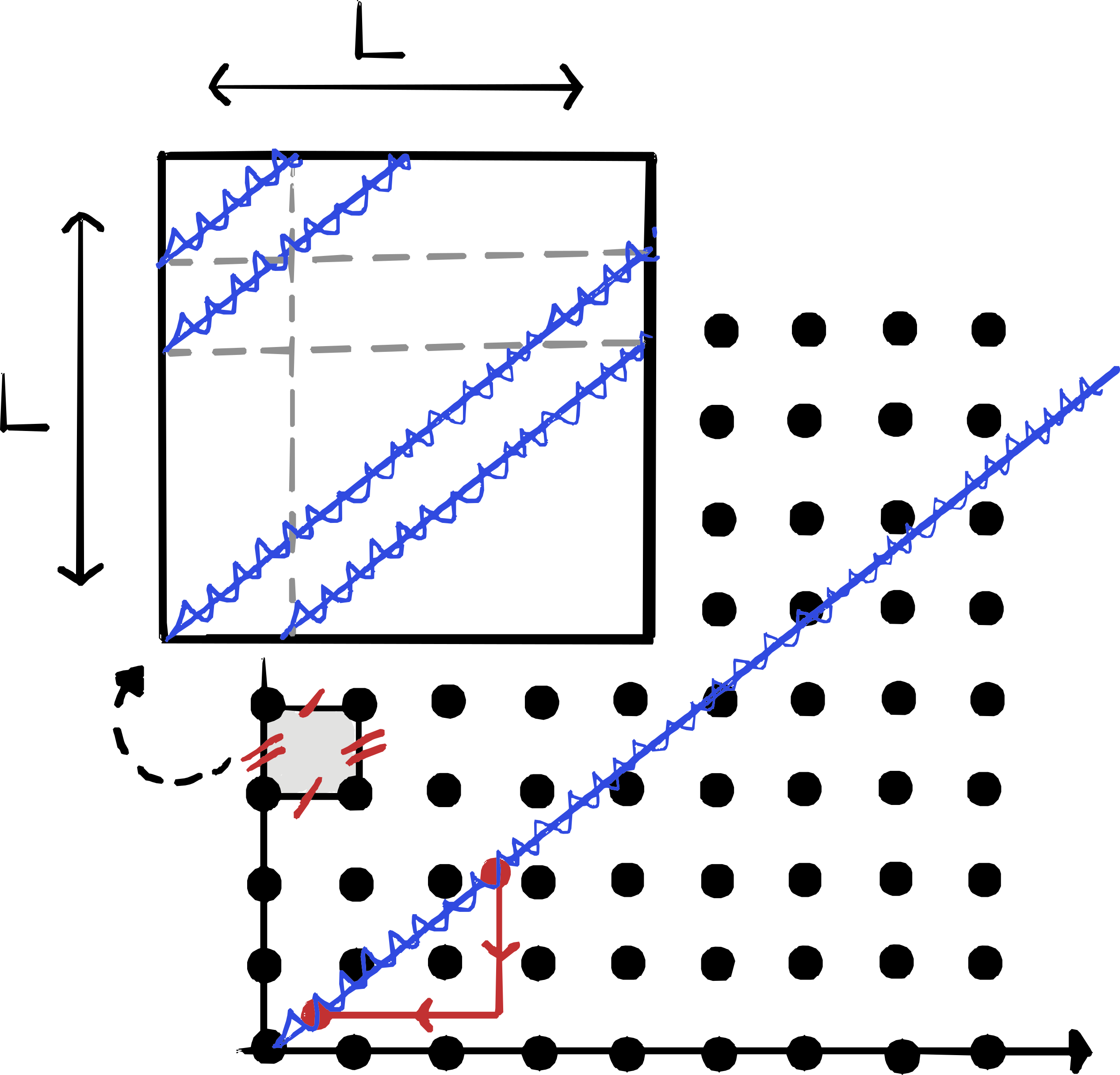}
    \caption[A fluctuating irrational spacetime around a mean classical background; an illustration.]{Enter Quantum Fluctuations. Now the line is not infinitely thin so it will eventually touch another point of the higher-dimensional lattice, and hence the corresponding curve on the torus will eventually intersect/overlap itself.}
    \label{fig:FiniteThicknessLine}
\end{figure}

The discussion in Secs.~\ref{subsec:IrrationalityOfSpacetime}, \ref{subsec:CartoonPicture} considered an idealized, classical picture, in which the 4D Minkowski slice (densely filling the 10D spacetime torus) was taken to be infinitely ``thin'' in the six extra dimensions.  But a real object won't be infinitely thin; it will have some finite thickness due to quantum fluctuations (see Fig.~\ref{fig:FiniteThicknessLine}).

To analyze the impact of such fluctuations, consider the 10D torus at some moment in time: then we have a 3D irrationally-sloped spatial slice (with 6 extra/transverse directions) densely filling a 9D spatial torus. We will refer to this 3D slice as a ``3-brane'' for short.\footnote{Let us add two comments here about why we chose a fixed time-slice. First, in a universe that is asymptotically de Sitter in the future with asymptotic Hubble rate of expansion $H_{\rm dS}$ (like ours), the horizon 3-volume approaches a finite physical volume $H_{\rm dS}^{-3}$ but the horizon 4-volume does not (since it is basically the 3-volume times an infinite physical time elapsed in the future). In other words, such universe has an infinite lifetime and expands forever and ever. Second, we are applying the uncertainty principle to argue that our 3-brane cannot be infinitely thin. This really applies to the Hamiltonian formulation of quantum mechanics, which picks some spatial slice.} And instead of assuming 6 extra dimensions from the outset, it will be instructive to initially analyze a more general picture with $n$ extra dimensions, so the 3-brane densely fills a $(3+n)$D spatial torus. The volume of the extra dimensions, ${\cal V}_n$, is governed by the transverse thickness of the 3-brane. 

Quantum fluctuations effectively give the 3-brane a finite non-zero thickness in the extra dimensions.  Let us imagine this transverse quantum thickness to be the Planck length $l_{pl}=M_{\rm Pl}^{-1}$; and hence, since the slice has $n$ transverse dimensions, its $n$D transverse volume ${\cal V}_{n}$ is 
\begin{equation}
  \label{V_n}
  {\cal V}_{n}\approx M_{\rm Pl}^{-n}.
\end{equation}
This 3-brane lives within a $(3+n)$D spatial torus $\TT^{3+n}$ of linear size $L$, and hence $(3+n)$D volume 
\begin{equation}
  \label{V_3+n}
  {\cal V}_{3+n}\approx L^{3+n}.
\end{equation}
Since the 3-brane is {\it densely} packed within the torus $\TT^{3+n}$, its 3-volume ${\cal V}_{3}$ (along the 3-brane) times its $n$D volume ${\cal V}_{n}$ (transverse to the 3-brane) must equal the torus's volume ${\cal V}_{3+n}$:
\begin{equation}
  \label{V_seesaw}
  {\cal V}_{3+n}={\cal V}_{3}{\cal V}_{n}.
\end{equation} 
Thus, we have a seesaw relation between ${\cal V}_{3}$ and ${\cal V}_{n}$ (for fixed\footnote{The key fact here is that to maintain a finite non-zero ${\cal V}_{3+n}$ for $\TT^{3+n}$ given by Eq.~\eqref{V_seesaw} compatible with an infinite spacetime $\RR^{3,1}$ of ${\cal V}_{3} \to \infty$, we need ${\cal V}_n \to 0$ to have any chance of making the product ${\cal V}_n {\cal V}_3$ finite. ${\cal V}_{3+n}$ is always governed by the compactification scale $L$, i.e.~the lattice spacing of the higher-dimensional $\Lambda$, which is, in a sense, the only length-scale at the 10D level.} ${\cal V}_{3+n}$): when the 3-brane is infinitely thin (${\cal V}_{n}\to0$) its 3-volume can be infinite (${\cal V}_{3}\to\infty$) -- indeed, this was our previous classical picture, where the extra dimensions occupy a measure-zero volume inside $\TT^{n+3}$. On the other hand, if quantum fluctuations give the 3-brane a finite non-zero thickness (${\cal V}_{n}>0$), its 3-volume must also become finite (${\cal V}_{3}<\infty$) in order to still fit inside the torus' compact volume ${\cal V}_{3+n}$; and the larger the transverse thickness ${\cal V}_{6}$, the smaller the spatial 3-volume ${\cal V}_{3}$ of ``our 3-brane.''  

Let us now see how this ``volume seesaw'' (\ref{V_seesaw}) might be converted into the observed ``mass seesaw'' (\ref{M_seesaw}).  The conversion requires two key physical inputs:
\begin{enumerate}
    \item For starters, how should we interpret the finite 3-volume of our 3-brane? The standard $\Lambda$CDM model of cosmology \cite{mukhanov2005physical, weinberg2008cosmology} suggests a natural and intriguing answer.  Since our universe is observed to have positive vacuum energy density $M_{\rm vac}^{4}$, Einstein's equations imply that we are surrounded by a ``de Sitter horizon'' of radius $H_{dS}^{-1}$, and hence are limited to only ever observe or communicate with a finite portion of our universe, with finite spatial 3-volume $H_{dS}^{-3}$, where $H_{dS}$ is the de Sitter expansion rate, determined by the Friedmann equation
\begin{equation}
  H_{dS}^{2}=\frac{8\pi G}{3}\rho_{\rm vac}\approx M_{\rm Pl}^{-2}M_{\rm vac}^{4}.
\end{equation} 
In other words, cosmological observations have led us to the surprising conclusion that our universe effectively has finite spatial 3-volume. In our scenario, this should the {\it same} as the finite spatial 3-volume ${\cal V}_{3}$ of our 3-brane. With this identification,
\begin{equation}
  \label{V_3}
  {\cal V}_{3}\approx H_{dS}^{-3}\approx M_{\rm Pl}^{3}/M_{\rm vac}^{6},
\end{equation}
and we can substitute Eqs.~(\ref{V_n}, \ref{V_3+n}, \ref{V_3}) into the volume seesaw (\ref{V_seesaw}) to obtain
\begin{equation}
     \label{L1}
     L^{3+n}\approx M_{\rm Pl}^{3-n}M_{\rm vac}^{-6}.
\end{equation}
\item The next crucial step is to imagine (following {\it e.g.}\ \cite{arkani1998hierarchy}), that -- in contrast to the rest of the standard model fields -- gravity is not confined to the 3-brane, so that the higher-dimensional ($4+n$ dimensional) Planck mass $M_{4+n}$, with associated higher-dimensional Einstein-Hilbert term $\int d^{4+n}x\sqrt{-g_{4+n}}M_{4+n}^{2+n}R_{4+n}$ is related to the lower-dimensional ($4$ dimensional) Planck mass $M_{4}$, with associated lower-dimensional Einstein-Hilbert term $\int d^{4}x\sqrt{-g_{4}}M_{4}^{2}R_{4}$, by integrating over the full extra-dimensional volume of the torus $L^{n}$ to obtain
\begin{equation}
  \label{L2}
  M_{4}^{2}\approx M_{4+n}^{2+n}L^{n}.
\end{equation}
Here $M_{4}$ is the observed (4D) Planck mass $M_{\rm Pl}$, but what should we take for the higher-dimensional Planck mass $M_{4+n}$?~Again following {\it e.g.}\ \cite{arkani1998hierarchy}, we imagine that the enormous hierarchy between the $M_{\rm Pl}$ and $M_{\rm EW}$ is a manifestation of the fact that we are confined to a lower-dimensional slice of spacetime, rather than perceiving the higher-dimensional spacetime. In other words, whereas the lower-dimensional Planck mass has an enormous hierarchy relative to the electroweak scale ($M_{4}=M_{\rm Pl}\gg M_{\rm EW}$), the higher-dimensional Planck mass has no such hierarchy ($M_{4+n}\approx M_{\rm EW}$). With this identification, Eq.~(\ref{L3}) becomes
\begin{equation}
  \label{L3}
  M_{\rm Pl}^{2}\approx M_{\rm EW}^{2+n}L^{n}.
\end{equation}
\end{enumerate}
Finally, we can combine Eqs.~(\ref{L1}) and (\ref{L3}), to eliminate $L$ and obtain
\begin{equation}
    M_{\rm vac}M_{\rm Pl}^{\frac{(n^{2}-n+6)}{6n}}\approx M_{\rm EW}^{\frac{(2+n)(3+n)}{6n}}.
\end{equation}
Strikingly, this reduces to the observed seesaw relation (\ref{M_seesaw}) precisely when there are $n=6$ extra dimensions, just as in the scenario proposed here! Of course, this is just a post-diction, rather than a prediction. After we have already noticed the geometric mean relation \eqref{M_seesaw} based on just pure numerology, we searched for a geometric reason as to why this could be the case. It would be far more satisfying if this seesaw relation arose from first principles, let's say as the condition for the stability of the torus or from some other condition (like the optimal winding number of the 3-brane around $\TT^{3+n}$ so that it avoids non-local effects between neighbouring slices).

\chapter{Spacetime Quasicrystals: a New Form of \textit{Ordered} Causal Sets}
\label{ch:sQC_CausalSets}

\setlength{\epigraphwidth}{0.45\textwidth}
\epigraph{\justifying{
``Quasicrystals: a new form of matter''}}{Paul Steinhardt's review paper \cite{STEINHARDT1990112}, 1990}

We now have a new method to generate \textit{highly symmetric}, \textit{discrete} and \textit{ordered} point sets in spacetime, our spacetime quasicrystals. A natural question arises: are there any possible applications of them in the context of quantum gravity? 

Perhaps the most straightforward such application arises in \textit{causal set theory} (CST), which is an approach to (discrete) quantum gravity, as we will explain in Section \ref{sec:CSTReview}. Similarly to how quasicrystals turned out to be a new form of ordered matter, here we lay the groundwork so that spacetime quasicrystals could eventually become a new form of ordered ``spacetime matter''. We call such a theory \textit{quasi-crystalline CST} (or qCST).

\section{Causal Set Theory: a Brief Overview}
\label{sec:CSTReview}

\textbf{Causal Set Theory} (CST) is an approach to quantum gravity where spacetime discreteness and causality are taken as fundamental axioms of the theory \cite{Bombelli:1987aa}. For a comprehensive review, see \cite{Surya_2019}; here, we only mention those aspects that are relevant for our purposes. 

\begin{figure}[t!]
\centering
\subfloat[][]
{
\label{fig:CartoonCausalSet}
    \includegraphics[width=0.61\textwidth]{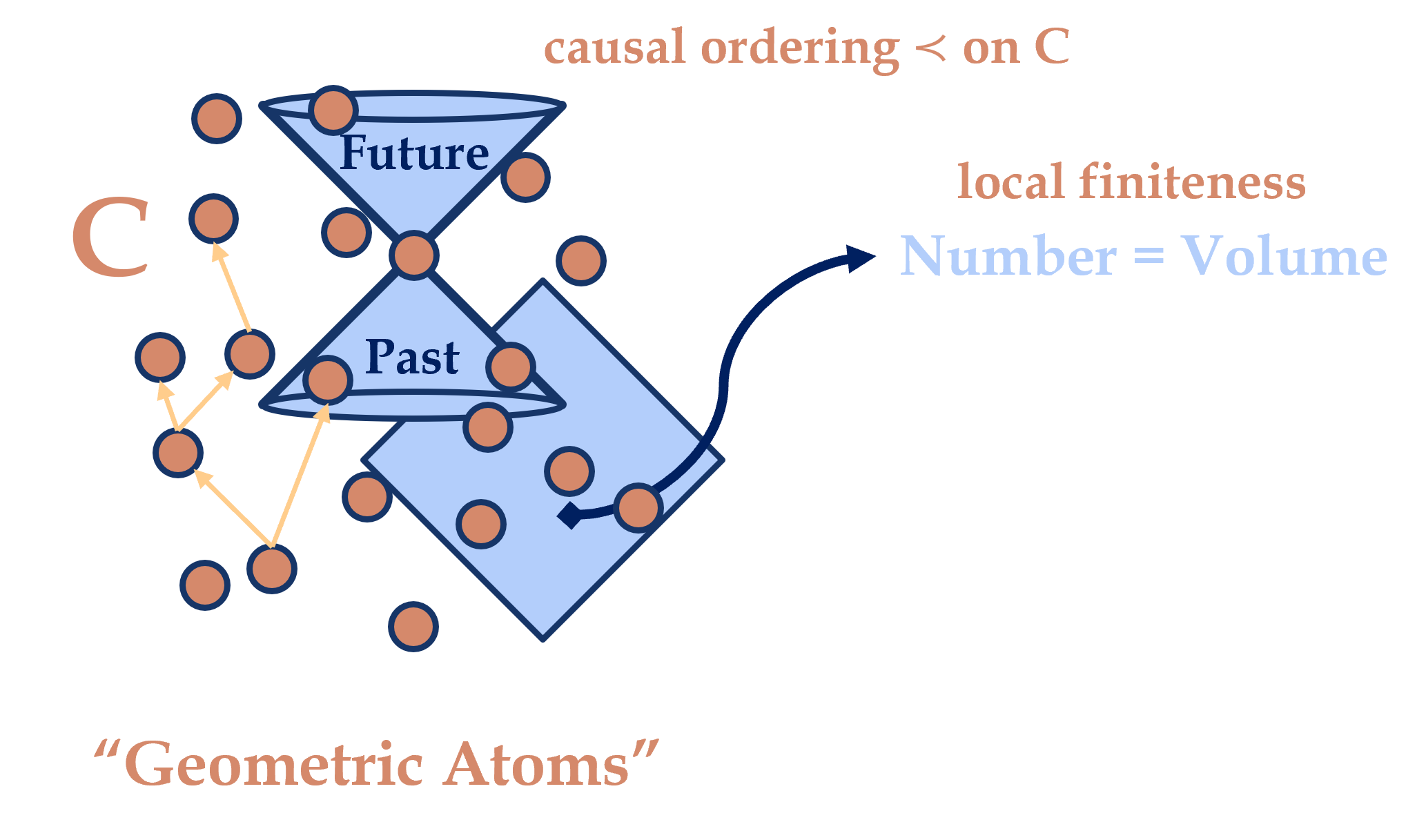}
}
\hspace{-1ex}
\subfloat[][]
{
\label{fig:CartoonCausalSetAndSpacetime}
\includegraphics[width=0.32\textwidth]{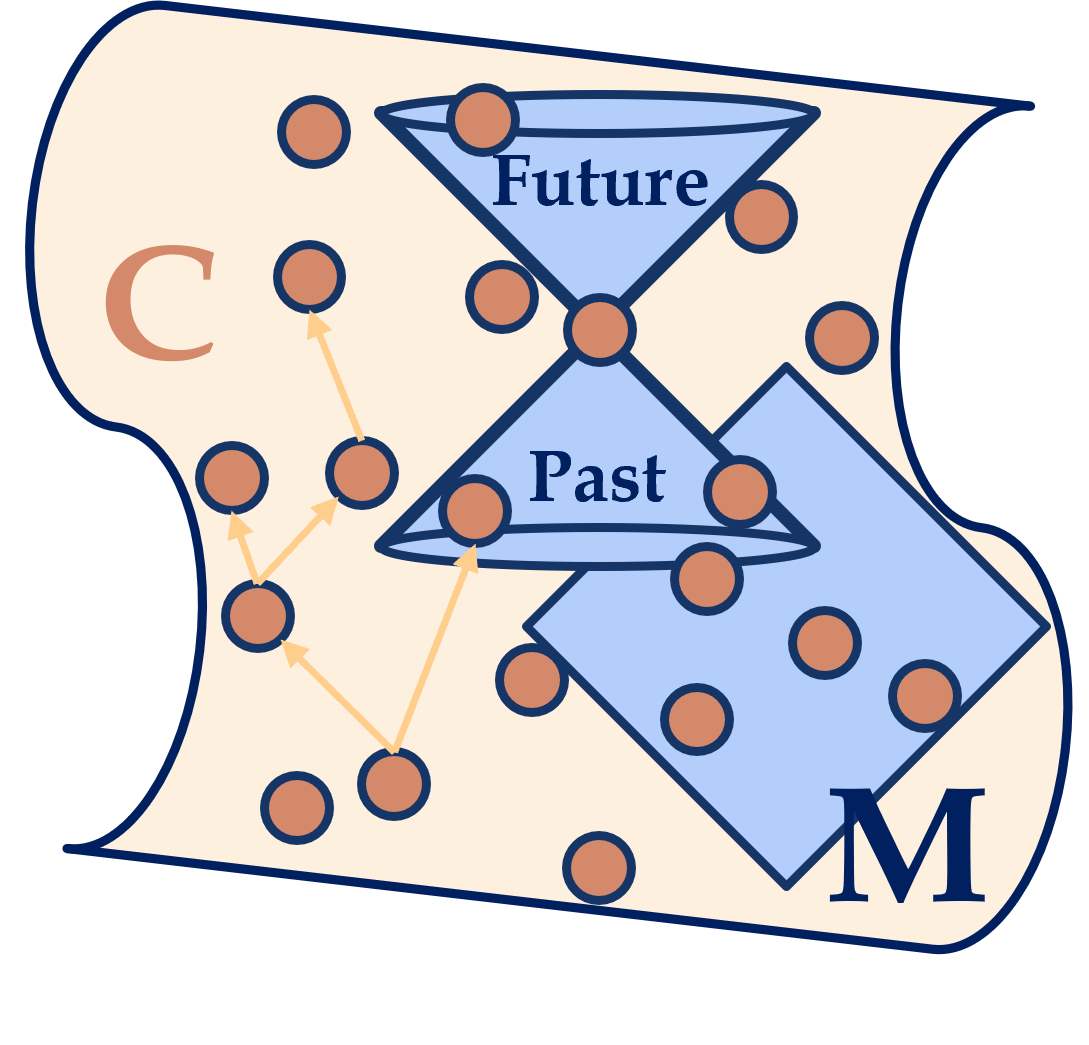}
}
    \caption[Cartoon of a causal set]{Cartoon of a causal set $\mathcal{C}$. On the left, we see a collection of geometric atoms; once $\prec$ has been defined, one can draw edges connecting the ``spacetime atoms''; because of transitivity, we avoid drawing edges implied by others. Here, we schematically drew only a couple edges. At the continuous limit, $\prec$ gives rise to the lightcone, which is not intrinsic to $\mathcal{C}$ but to the approximating spacetime $\mathcal{M}$ (shown on the right).}
    \label{fig:CartoonCauset}
\end{figure}

A \textbf{causal set} $\mathcal{C}$ (or \textit{causet}, for short) is a point set, which can be thought of as a collection of \textit{``geometric atoms''}, endowed with an order relation $\prec$, the \textit{causal ordering}, that allows one to speak about the causal future and past of a given point in $\mathcal{C}$. An illustration is given in Fig.~\ref{fig:CartoonCauset}. The causal ordering $\prec$ satisfies the properties that it is (i) \textit{acyclic}: $x \prec y $ and $ y \prec x$ $ \Rightarrow x=y, \, \forall x,y \in \mathcal{C}$ (i.e. ``no closed timelike curves in $\mathcal{C}$'') and (ii) \textit{transitive}: $ x \prec y$ and $y \prec z$ $\Rightarrow x \prec z, \, \forall x,z,y \in C$ (i.e. ``if $x$ is in causal contact with $y$ and $y$ is in causal contact with $z$, then this implies that $x$ must also be causally related to $z$''). These two properties turn $\mathcal{C}$ into a partially ordered set (or \textit{poset} for short). The fact that $\prec$ encodes causal relations between the elements of $\mathcal{C}$ is the reason why causality is fundamental in the theory; it is baked-in axiomatically\footnote{Questions like ``how this causal relation came to be'', though very interesting, are not addressed in the CST framework. Other inquiries, such as ``can we have causality without a fixed background spacetime'' have a positive answer within CST: at least mathematically, causality is defined abstractly through $\prec$.}.

To encode discreteness into the formalism, we also say that (iii) any causal set ${\cal C}$ is \textit{locally finite}. Mathematically, $\forall x,y \in C, | \Cint{x}{y} |  < \infty$, where $\Cint{x}{y}$ is the \textit{order interval} -- or \textit{causal diamond}, or \textit{Alexandrov interval} in the continuum -- defined as 
\begin{equation}
    \Cint{x}{y} = \mathrm{Fut}(x) \cap \mathrm{Past}(y), \qquad 
\left\{
    \begin{aligned}
        \mathrm{Fut}(x) &= \{ w \in {\cal C} \, | \, x \prec w , x \neq w\}  \\[1ex]
        \mathrm{Past}(x) &= \{ w \in {\cal C} \, | \, w \prec x , x \neq w\}
    \end{aligned}
    \right.
    \label{eq:CausalInterval} .
\end{equation}
Physically, this means that any causal diamond contains a finite number of points inside it, that is, $\Cint{x}{y}$ has finite cardinality for any $x,y \in {\cal C}$. This is important because we would like the number of elements inside any region of the causal set to accurately reflect the continuous volume of that region. The reason why one demands this comes from the \textit{Hawking--King--McCarthy--Malament (HKMM) Theorem} (see \S 2 in \cite{Surya_2019} and references therein), which sits at the core of the CST proposal. It tells us that, up to an overall conformal scale-factor captured by the local volume element, the causal structure $\prec_{\mathcal{M}}$ of a continuous spacetime\footnote{Taken to be future- and past-distinguishing, i.e. no closed timelike curves exist, for example.} $\mathcal{M}$ entirely determines its conformal geometry, i.e. it specifies 9 out of the total 10 metric components of $g_{\mu\nu}$ in $d=4$. The theorem can be stated as
\begin{equation*}
\mathbf{HKMM:} \textrm{ Causal Structure } + \textrm{ Volume Element  } = \textrm{ Lorentzian Geometry }.
\end{equation*}
At the discrete level, the causal structure is captured by the causal order $\prec$ (with properties (i)-(ii)) on our point set ${\cal C}$. One further assumes a fundamental correspondence between the number of elements inside a region of ${\cal C}$ and the volume element it represents. Local finiteness (property (iii)), then, guarantees that \textit{any such region will be assigned a finite volume} in the continuous approximation, where $({\cal C},\prec)$ gives rise to $(\mathcal{M},g_{\mu \nu})$. Discreteness, thus, provides the missing $(1/10)^{\rm th}$ of the metric $g_{ \mu \nu}$ and the HKMM theorem turns into the ``CST slogan'' \cite{Surya_2019}
\begin{equation*}
\textbf{CST Slogan:} \textrm{ Causal Order } + \textrm{ Discreteness } \sim \textrm{ Lorentzian Geometry }.
\end{equation*}

But what is exactly the \textit{continuous approximation} to a causal set? To make this precise, we should comment on what it actually means for a continuous spacetime to approximate a discrete causal set ${\cal C}$ ``well''. For this, we need to answer ``when does $({\cal C},\prec) \sim (\mathcal{M},g_{\mu\nu})$ hold'' and also quantify this behaviour between the perhaps many causets that satisfy the approximate relation $\sim$. To that end, we consider a faithful (injective) embedding $\Phi$ of a causal set ${\cal C}$ into a spacetime $\mathcal{M}$ that preserves causal relations,
\begin{equation}
    \Phi : ({\cal C},\prec_C) \hookrightarrow (\mathcal{M},g_{\mu\nu} \leftrightarrow \, \prec_\mathcal{M}) \quad \textrm{ such that } x \prec_C y \Leftrightarrow \Phi(x) \prec_\mathcal{M} \Phi(y). 
    \label{eq:CausalSetEmbedding}  
\end{equation}
We demand the embedding to be \textit{uniform} with respect to the spacetime volume measure of $(\mathcal{M},g_{\mu\nu})$, that is, the \textit{density} $\rho$ of the embedding is the same throughout $\mathcal{M}$ and $\Phi({\cal C})$ is a uniform distribution of points (spacetime events) in $\mathcal{M}$. This ensures that every finite spacetime volume $V$ will be accurately reflected by a finite number of elements $N \simeq \rho V$ in the causal set. Uniformity guarantees that this assignment would be the same no matter where the region is in the spacetime $\mathcal{M}$. Given the central role of causality, it is natural to take these spacetime regions to be causal diamonds $\cdiamond = \Cint{x}{y}$. Then, we say that \textbf{$\mathcal{M}$ approximates ${\cal C}$ well} if the following relation holds 
\begin{equation}
    \forall \cdiamond \subset C, \quad N_\cdiamond = {\cal C} \cap \cdiamond \simeq \rho \mathrm{vol} (\cdiamond) = \rho \int_\cdiamond \sqrt{-\det g} d^{D} x, \quad D = \dim \mathcal{M} .
    \label{eq:NumberVolumeCorrespondence}
\end{equation}
In other words, we would like the number of elements $N_\cdiamond$ inside any $\mathlarger{\cdiamond}$ of the causal set that we get by counting -- ``the discrete'' -- to match the volumes $V_\cdiamond$ calculated with the spacetime metric -- ``the continuum''. This statement is known as the \textbf{number-volume (N-V) correspondence}.

Finally, we would also like the assignment $N_\cdiamond \simeq \rho V_{\cdiamond}$ to be \textit{Lorentz covariant}, or frame-independent. After all, the goal is to recover the covariant structure of spacetime (local Lorentz invariance) from the more fundamental starting point: the underlying causal set. For this reason, $\Phi({\cal C})$ must be a uniform distribution in \textit{any} frame. We will come back to this point in far greater detail in the next section, after we introduce a test to check whether or not a given pattern $\Phi({\cal C})$ satisfies the NV correspondence. 

\section{The Number-Volume (N-V) Correspondence}
\label{sec:NVCorrespondence}

So far we said that $({\cal C},\prec) \sim (\mathcal{M},g_{\mu\nu})$ is expected to hold when the N-V correspondence in the form of Eq.~\eqref{eq:NumberVolumeCorrespondence} is satisfied. However, we have not yet quantified how well a given spacetime $\mathcal{M}$ is approximated by a discrete causal set ${\cal C}$. This is achieved through the following test, first proposed in \cite{Saravani_2014}.

\begin{statement}{\textbf{The N-V Test \cite{Saravani_2014}}} for a given causet ${\cal C}$ embedded into some spacetime $\mathcal{M}$ at a density $\rho$, sample randomly $M$ causal diamonds $ \{ \cdiamond \} = \left\{ \cdiamond^{(1)}, \dots , \cdiamond^{(M)} \right\} $ of some \textit{fixed volume} $V = V_\cdiamond$, but allowing for different shapes so that we can capture the different Lorentz boosts one could perform.~Let $N_\cdiamond^{(1)}, \dots, N_\cdiamond^{(M)}$ be the number of elements in these regions, found by intersecting ${\cal C}$ and the corresponding test region $\cdiamond^{(\rm I)}, {\rm I}=1,\dots, M$. For the number-volume correspondence to hold, we require that (as $M \rightarrow \infty$)
\begin{equation}
    \langle N_\cdiamond \rangle = \rho V_\cdiamond \quad \text{ and } \quad \frac{\delta N_\cdiamond}{\langle N_\cdiamond \rangle } = \frac{\sqrt{\langle(N_\cdiamond - \langle N_\cdiamond \rangle )^2\rangle}}{\langle N_\cdiamond \rangle } \ll 1
    \label{eq:NumberVolumeCorrespondenceTest}.
\end{equation}
\label{statement:NVTest}
\end{statement}
The first equation tells us that the N-V correspondence holds on average. The second quantifies how well this is happening: this is because the fluctuations $\delta N_\cdiamond$ around the mean behaviour $\langle N_\cdiamond \rangle$ are given by the standard deviation 
\begin{equation}
\sigma_{N_\cdiamond} = \sqrt{\langle ( N_\cdiamond - \langle N_\cdiamond \rangle    )^2\rangle} = \sqrt{\mathrm{Var}(N_\cdiamond)}
\label{eq:StandardDeviationOfNumberVariable}
\end{equation}
of the random variable $N_\cdiamond$. We would like our causal set ${\cal C}$ to \textbf{realize the N-V correspondence with the minimum possible fluctuations} because the least the fluctuations $\delta N_\cdiamond$ are, the closest ${\cal C}$ is to the continuum. 

Given this formulation \eqref{eq:NumberVolumeCorrespondenceTest} of the N-V test, the question becomes: which embedding $\Phi: {\cal C} \hookrightarrow \mathcal{M}$ achieves the best NV correspondence, i.e. the correct mean behaviour $\langle N_\cdiamond \rangle$ with the \textit{least noise} $\sigma_{N_\cdiamond}$? 

\subsection{The CST Consensus: Poisson Rules Them All!}

The consensus among the CST community seems to be that a random process that follows Poisson statistics -- known as \textit{sprinkling} -- is the best among all possible uniform distributions $\Phi({\cal C})$ in any given $\mathcal{M}$. 

A \textit{Poisson sprinkling} works as follows: we first fix the target spacetime $\mathcal{M}$ and then reverse-engineer a causal set ${\cal C}(\mathcal{M})$ by sampling Poisson-random points from $\mathcal{M}$ at density $\rho$, endowing them with their causal relations (i.e. pulling $\prec_\mathcal{M}$ that comes from $g_{\mu\nu}$ to $\prec_{\cal C})$. The probability of finding $N$ elements in any spacetime region with volume $V$ is 
\begin{equation}
    P(N) = \frac{(\rho V)^N}{N!} e^{-\rho V}
    \label{eq:PoissonProbability}
\end{equation}
which yields the known expressions of Poisson statistics for the average number of elements and its standard deviation within any region $\cdiamond$ of volume $V_\cdiamond$, 
\begin{equation}
    \langle N_\cdiamond \rangle_{\textrm{Poisson}} = \rho V_\cdiamond = \left(\sigma_{N_\cdiamond}^2\right)_{\textrm{Poisson}} = \mathrm{Var}(N_\cdiamond)_{\textrm{Poisson}}.
    \label{eq:PoissonStatistics}
\end{equation}
This leads to 
\begin{equation}
    \left.\frac{\delta N_\cdiamond}{\langle N_\cdiamond \rangle}\right|_{\textrm{Poisson}} = \frac{1}{\sqrt{\langle N_\cdiamond \rangle}} = \frac{1}{\sqrt{\rho V_{\cdiamond}}}
    \label{eq:PoissonFluctuationsAroundMean}
\end{equation}
which is believed to be the least fluctuations one could achieve in general. In fact, one goes even beyond that and formulates the following statement\footnote{To the best of the author's knowledge, this remains a conjecture and has not yet received the status of a theorem.}: 
\begin{statement}{\textbf{The Causal Set--Continuum Correspondence \cite{Saravani_2014,Surya_2019,Surya_2026}}} a causal set ${\cal C}$ is called \textit{manifold-like} (i.e.~having a \textit{good} continuous approximation) \textit{if and only if} it can be generated through a Poisson sprinkling from some continuous Lorentzian spacetime $\mathcal{M}$ (with high probability).   \label{statement:CS-C-Correspondence}
\end{statement}
Therefore, the Poisson process, and the discrete random geometries it carries with it, is central to the CST hypothesis. But is this as good as it gets? 

\subsubsection{Why Random Instead of Ordered?}

\noindent Before we answer, let us briefly mention some desirable features of the Poisson process that one would reasonably expect also for any other point process. 

\begin{itemize}
    \item \textbf{Lorentz Invariance.} Going back to the discussion about Lorentz covariance, it has been shown \cite{BOMBELLI_2009} that any random causet coming from a Poisson sprinkling \textit{does not break Lorentz symmetry} in Minkowski spacetime\footnote{The author was unable to find literature showing this formally for more general spacetimes.}, as one cannot consistently pick a direction throughout spacetime that singles out any particular frame (see also \S 3.2 in \cite{Surya_2019}). This implies that, if $\Phi_{\textrm{Poisson}}({\cal C})$ is a uniform distribution in $(\mathcal{M},\eta_{\mu \nu})$ in some frame, then it will be also uniform with respect to any other frame. In a sense, the Poisson process respects Lorentz symmetry, not by actually preserving it at the level of the discrete causal set, but by having \textit{randomness prevent the selection of preferred frames}. This way one recovers (local) Lorentz invariance at the continuous approximation, which is a property we would want for any point process $\Phi$, as Lorentz invariance seems to withstand any experimental test so far \cite{PhysRevD.91.052003,Liberati_2013,Mattingly_2005}.

\item \textbf{Generality.} The Poisson process can be used in any arbitrarily curved background spacetime (such as de Sitter, anti-de Sitter and beyond) to sample randomly points and build causal sets from it, making it a versatile mechanism to generate causets. 
\end{itemize}

\begin{figure}[!t]
    \centering
\subfloat[Unboosted frame; regular pattern]{
\includegraphics[width=0.40\linewidth]{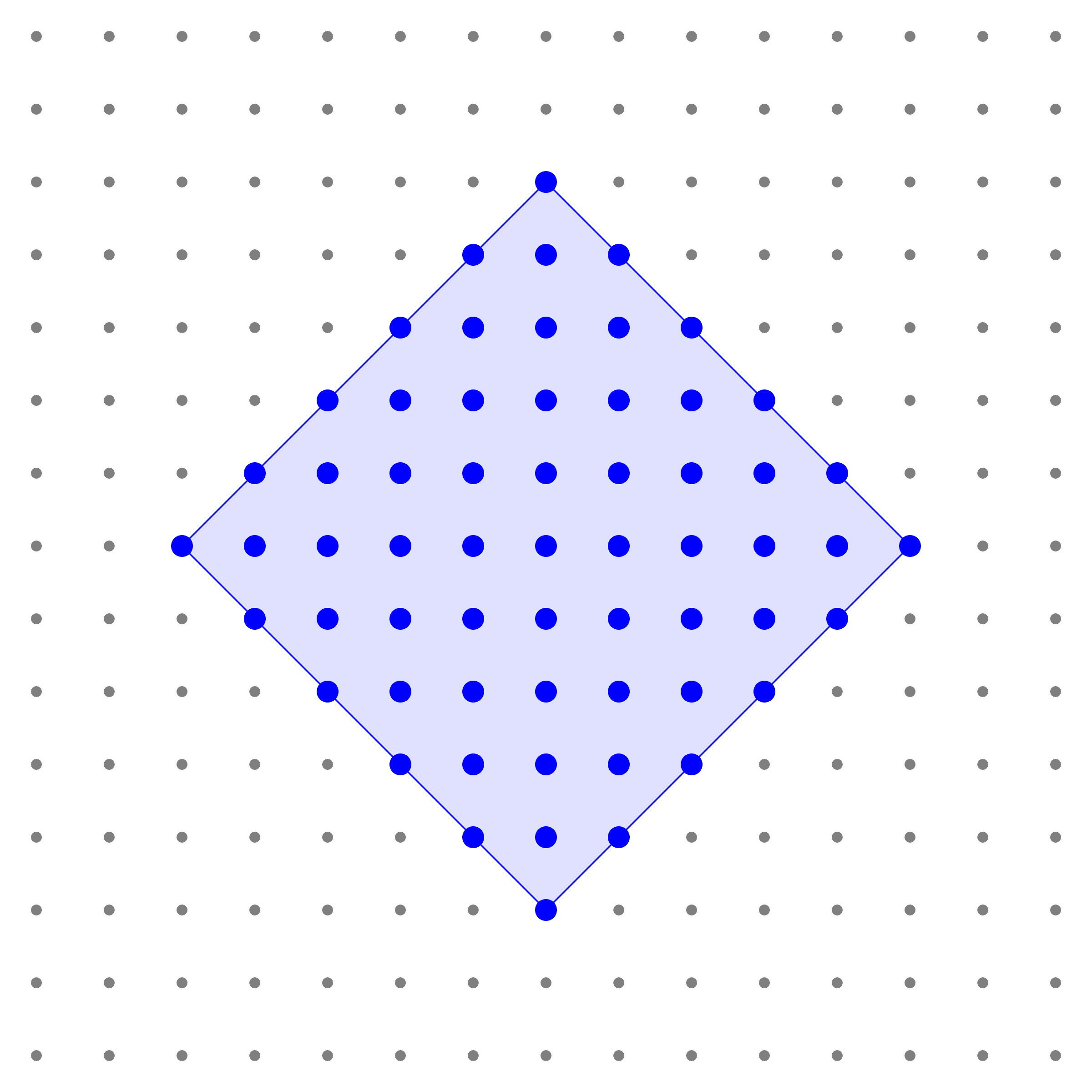}
}
\hspace{+1ex}
\subfloat[Boosted frame; irregular pattern]{
\includegraphics[width=0.39\linewidth]{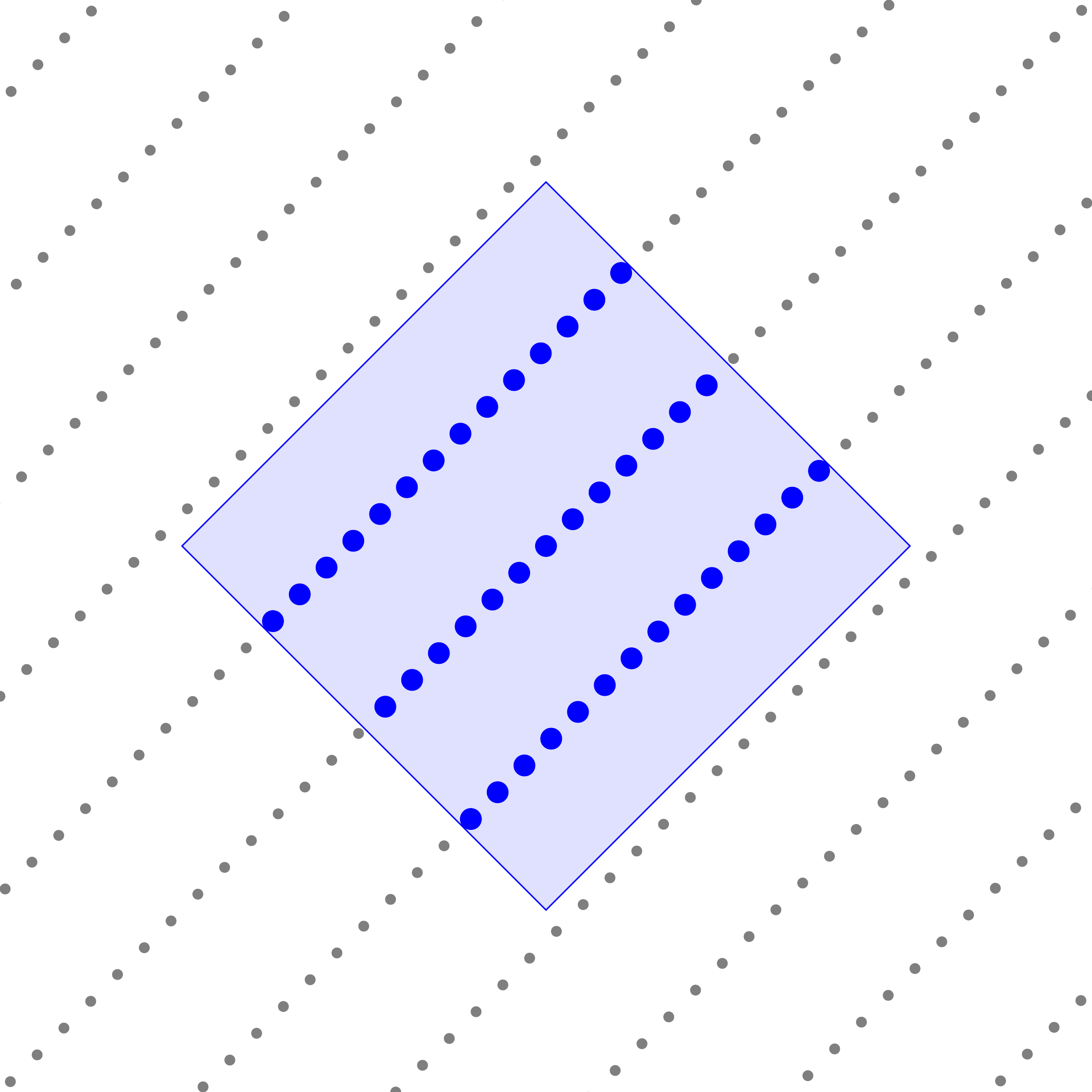}
}
    \caption[A regular lattice-distribution of points in Minkowski space is not uniform in all frames]{The integer lattice $\mathrm{I}_{1,1} \simeq \mathbb{Z}^2$ in $(1+1)$-dimensional Minkowski space ($(x,t)$ plane). On the left, the pattern looks perfectly regular and uniform in its rest frame; the causal diamond of volume $V_\cdiamond = 50$ centered at the origin contains 61 points ($\rho_{\rm Int} = 1)$. On the right, the lattice becomes irregular in a boosted frame (moving with velocity $v/c \simeq 0.7616$). The pattern loses its uniformity and the same causal diamond now contains 41 points (the rest 20 points that were sitting originally on its boundary are now outside of it).}
\label{fig:BoostedVsUnboostedIntegerLattice}
\end{figure}

There is one more result specifically about the Poisson distribution that must be mentioned, which is the fact that \textbf{Poisson beats everything in the sub-Planckian regime.} A theorem from \cite{Saravani_2014} tell us that the NV correspondence is best realized by the Poisson sprinkling \textbf{for arbitrarily small -- sub-Planckian -- volumes} (in any dimension and any background; the theorem is statistical in nature). Indeed, as the authors mention, for their proof to go through, one must take $\langle N_{\cdiamond} \rangle < 1$ (they consider $\rho = V_c^{-1} = 1$, where $V_c$ is the discreteness scale -- the fundamental spacetime volume cut-off -- thought to be at the Planck scale). This theorem, together with some suggestive results in $(1+1)$D and $(2+1)$D Minkowski spacetime (that we will reproduce in the next subsection) and an \textit{unproven conjecture} in $(3+1)$D from the same paper, are usually taken as evidence that Poisson sprinklings will outperform any other point process in any possible setting.

Letting aside questions about generality for the moment, let us first focus on the following: can a different uniform distribution $\Phi$ achieve a frame-independent assignment of numbers to volumes in Minkowski spacetime except for Poisson?

Perhaps a regular -- \textit{ordered} -- structure, a lattice in Minkowski spacetime, can do the trick. After all, its translational symmetry is enough to make it uniform in its rest frame; the whole pattern is governed by a single repeating fundamental cell, what else can look more uniform that this? Fig.~\eqref{fig:BoostedVsUnboostedIntegerLattice} shows what goes wrong with the integer lattice $\mathrm{I}_{1,1} = \langle \mathbf{e}_0 = (1,0),\mathbf{e}_1 = (0,1)\rangle_{\mathbb{Z}}$ in $(1+1)$D Minkowski spacetime. Even though the lattice appears regular and uniform in its rest frame, it becomes irregular and non-uniform for a boosted observer (squeezed in the boost-direction and stretched in the perpendicular one). This implies that such structures lead to a poor number-volume realization, and we will revisit this to see exactly what fails in the next subsection. 

If randomness beats regularity then, can we extend the theorem from \cite{Saravani_2014} and show that Poisson sprinklings \textit{actually} beat everything across all ranges of scales? The answer turns out to be no: for $\langle N_\cdiamond \rangle > 1$, the proof fails, as is noted in their work. And before one starts scheming of other methods to make Poisson stand out again, the authors have found a counterexample: perhaps surprisingly, a lattice in $(1+1)$-dimensional Minkowski spacetime with discrete Lorentz symmetry has been found to realize the N-V correspondence much better than Poisson at large scales (effectively, for any macroscopic volume $\rho V_\cdiamond \geq 1$) \cite{Saravani_2014}. Given that we have found a new class of \textit{ordered}\footnote{Here, and throughout this work, when we refer to a pattern as ``ordered'', we use this as a synonym to ``structured'', as opposed to ``random'' or ``amorphous'', i.e. one that lacks structure. This is not to be confused with the causal order relation $\prec$ inherent in \textit{any} causet, structured or not, which turns them into partially ordered sets with respect to $\prec$.} causal sets in this thesis, our spacetime quasicrystals, which also have discrete Lorentz symmetry, we should test them against the traditional random causal sets and the lattices and quantify if and how well they satisfy the N-V correspondence.  

In the following subsections, we run the N-V test for several point sets in $(1+n)$-dimensional Minkowski spacetime for $n=1,2 \text{ and } 3$. We replicate the results from \cite{Saravani_2014} in $(1+1)$D, which is our main reference on this matter, while also adding some of our spacetime quasicrystals in the analysis. Then, we continue on higher dimensions and we test their conjecture in $(3+1)D$. We reserve the Appendix \ref{appendix:NVTestDetails} for some details regarding the tests. As we shall see, we provide evidence that their conjecture does not hold, and our spacetime quasicrystals could actually beat Poisson in the N-V test.

\subsection{Challenging the Consensus in $(1+1)$D: a Lost Battle for Poisson}
\label{subsec:NVTestIn1+1D}

We start from the simplest case: we consider four discrete point processes in $(1+1)$ dimensions. The structures under test are: 
\begin{enumerate}
    \item \textbf{A Poisson Causal Set:} this is generated by a sprinkling onto the Minkowski plane at a fixed density $\rho_{\rm Poisson}$ (taken to be $\rho_{\rm Poisson} = 1$) and serves as the reference point against which all other ordered structures are measured. Poisson statistics, described by Eqs.~\eqref{eq:PoissonStatistics}, \eqref{eq:PoissonFluctuationsAroundMean}, are expected to hold for $\langle N_\cdiamond \rangle$ and $\sigma_{N_\cdiamond}$.  
    \item \textbf{The Integer Lattice} $\mathrm{I}_{1,1} \simeq \mathbb{Z}^2 = \left\langle \mathbf{e}_0 = (1,0),\mathbf{e}_1 = (0,1)\right\rangle_{\mathbb{Z}}$: this is perhaps the simplest regular discrete structure in $(1+1)$D, the set of all integer pairs $(t,x) \in \mathbb{Z}^2$ with density $\rho_{\rm Int.} = 1$. This lattice belongs to the same family as the Lorentzian lattices mentioned in Section \ref{subsubsec:Lorentzian_Lattices}: it is an odd unimodular lattice in $\mathbb{R}^{1,1}$ arising as the integral automorphisms of the Minkowski quadratic form $\eta(t,x) = -t^2 + x^2$. The reason we silently ignored it from our discussion then is that it possesses \textit{no non-trivial discrete Lorentz boost symmetry.} This is because the only possible integer solutions to $\eta(t,x) = \pm 1, \pm 2$ (the condition that gives the roots of the lattice; see Section \ref{sec:Symmetries_of_Lattices}) are $(t,x) = (\pm 1, 0)$ or $(t,x) = (0 ,\pm 1)$. Stated differently, its only non-trivial reflection symmetry is described by the root $\mathbf{r}_1 = (0,1)$ (spatial reflections) and its Coxeter group is of finite order.\footnote{The temporal reflection exchanges the upper and lower hyperbola in $\RR^{1,1}$ and is not contained in the Coxeter group (see Appendix \ref{app:hyperbolicmirrorintersections} and Section \ref{ch:MatBackgroundPartII}).} 
    \item \textbf{The ``Fibonacci'' Lattice} $\Lambda_{\rm Fib.} = \langle \xi_{t} = \left(\tfrac{\sqrt{5}}{2},\tfrac{1}{2}\right) , \xi_s = (0,1) \rangle_{\mathbb{Z}}$: this is another lattice in $\RR^{1,1}$ whose timelike basis element $\xi_t$ contains the irrational number $\sqrt{5}$, while the spacelike element $\xi_s$ is the same as ${ \bf e_1 }$ from the $\mathrm{I}_{1,1}$ lattice. The lattice was introduced in \cite{Saravani_2014} as a member of a whole family of lattices with discrete Lorentz symmetry in $(1+1)$D (see also \cite{schild1948discrete} for an example in $3+1$D, related to $\mathrm{I}_{3,1}$). In their paper, the authors referred to it as a ``Lorentzian lattice''. Here, we call it the ``Fibonacci'' lattice because of the $\sqrt{5}$-term, since the integer lattice $\mathrm{I}_{1,1}$ is also a perfectly valid Lorentzian lattice.\footnote{For us, \textit{any} lattice in Lorentzian space is a Lorentzian lattice (Section \ref{subsubsec:Lorentzian_Lattices}). For example, one can map $\mathrm{I}_{1,1}$ onto \cite{Saravani_2014}'s parametrisation by using the following values for their parameters: $\epsilon = \delta = \gamma = 1$, $\psi = \theta = \chi = 0$ which turns the Lorentz boost into the identity map.} Since there is no $GL(2,\ZZ)$-transformation between the two bases, the Fibonacci lattice $\Lambda_{\rm Fib.}$ is \textit{not isomorphic} to $\mathrm{I}_{1,1}$. The density of $\Lambda_{\rm Fib.}$ is $\rho_{\rm Fib.} = 2/\sqrt{5} \simeq 0.8944$ (see Section \ref{sec:Lattices}). It also enjoys discrete Lorentz symmetry: a nice way to see it  (see also Appendix B in \cite{Saravani_2014}) is to pass on to \textit{lightlike coordinates} (Appendix \ref{appendix:NVTestDetails}) 
    $${\rm u} = t+x \qquad \text{and} \qquad {\rm v} = t-x, $$
    where the Fibonacci lattice takes an interesting form
    \begin{equation}
        \Lambda_{\rm Fib.} = \left\{ ({\rm u},{\rm v}) = (m \phi + n, -m\Bar{\phi} - n) \, | \, m,n \in \ZZ \right\}
        \label{eq:FibonacciLatticeNullSpace},
    \end{equation}
    with $\phi = (1+\sqrt{5})/2$ being the golden ratio and $\Bar{\phi} = (1-\sqrt{5})/2$ its Galois conjugate. One can now verify that the following operation $T_L$ (since $\phi^2 = \phi+1, -\Bar{\phi} = 1/\phi$)
    $$ ({\rm u},{\rm v}) \overset{T_L}{\longmapsto} ( \phi^2 {\rm u}, \phi^{-2} {\rm v}) = \bigg( (2m + n) \phi + (m+n), -(2m + n)\Bar{\phi} - (m+n) \bigg)$$
    leaves the lattice exactly invariant. This corresponds to a Lorentz boost \eqref{eq:LorentzTransformation1+1D} with
    $$ \left\{
    \begin{aligned}
        & \cosh{\varpi} = \frac{e^{\varpi} + e^{-\varpi}}{2} = \tfrac{1}{2}\big(\phi^2 + \Bar{\phi}^2\big) = \tfrac{1}{2}\big(\phi + \Bar{\phi} + 2\big) = \tfrac{3}{2}\\
        & \sinh{\varpi} = \frac{e^{\varpi} - e^{-\varpi}}{2} = \tfrac{1}{2}\big(\phi^2 - \Bar{\phi}^2\big) = \tfrac{1}{2}\big(\phi - \Bar{\phi}\big) = \tfrac{\sqrt{5}}{2}
    \end{aligned}
    \right. \Rightarrow \boxed{\tanh\varpi = \frac{\sqrt{5}}{3}},
    $$
    or to an observer moving with velocity $v/c = \sqrt{5}/3 \approx 0.745356$. However, this is not the only symmetry of the pattern: \textit{any integer power of the map $T_L$} will actually preserve $\Lambda_{\rm Fib.}$.
    As a result, there is an \textbf{infinite set of boosts} that are exact symmetries of the pattern (i.e. it is mapped onto itself under their actions). They yield the velocities
    $$ \tanh\varpi = \frac{\phi^{2k} - \Bar{\phi}^{2k}}{\phi^{2k} + \Bar{\phi}^{2k}} = \frac{(\phi^k + \Bar{\phi}^k)(\phi^k - \Bar{\phi}^k)}{(\phi^k + \Bar{\phi}^k)^2 - 2(-1)^k}, \quad k \in \ZZ .$$
    In other words, there is an \textbf{infinite family of preferred observers}, with respect to which the pattern looks identical, but other observers (moving with some other velocities) will see a different pattern. 

    It is also instructing from the number-theoretic point of view to note why the ``square-root'' of the transformation $T_L$ is \textit{not} a Lorentz transformation, despite \textit{being} a lattice automorphism. Let this map be ``$T_{\sqrt{L}}$'', acting as
    $$ ({\rm u},{\rm v}) \overset{T_{\sqrt{L}}}{\longmapsto} (\phi {\rm u}, \Bar{\phi} {\rm v}) = \big( (m+n)\phi + m, -(m+n)\Bar{\phi} - m \big)$$
    such that $(T_{\sqrt{L}})^2 = T_L$. Clearly, it preserves the lattice \eqref{eq:FibonacciLatticeNullSpace}. However, a Lorentz boost in null coordinates \textit{must} take the form $( {\rm u} , {\rm v}) \mapsto ( \lambda {\rm u}, \lambda^{-1} {\rm v})$ for some $\lambda > 0$, but in the $T_{\sqrt{L}}$-map the scaling of ${\rm v}$ is $\Bar{\phi} = -1/ \phi <0$. This is why $T_{\sqrt{L}}$, and by extension, the odd powers of $\phi$ (the unit of the field extension $\mathbb{Q}(\sqrt{5})$) do not produce valid Lorentz transformations. One can see that, as a consequence, $T_{\sqrt{L}}$ does not preserve the Minkowski metric, but rather flips it: 
    $$\eta({\rm u,v}) = - {\rm uv} \overset{T_{\sqrt{L}}}{\longmapsto} - (\phi \Bar{\phi}) {\rm uv} = + {\rm uv} = -\eta({\rm u,v})$$ 
    This emphasizes the importance of \textit{totally positive} generators like $\phi^2$ for generating Lorentz transformations (generators of number fields whose Galois conjugates are all positive -- see Appendices \ref{AppendixA-NumberTheory} and \ref{appendix:FoldingUnfolding}).
    \item \textbf{The Spacetime Quasicrystal} (sQC) from Section \ref{subsec:2DLorentzianQuasiCrystal}: this is an aperiodic structure coming from the $\CNP$ method from the integer lattice $\mathrm{I}_{3,1}$. This point process is a different type of sprinkling: not random, but a very ordered one, as we ``sprinkle'' points from the higher-dimensional lattice onto an irrational slice according to the $\CNP$ procedure. The pattern's number density - given by Eq.~\eqref{eq:DensitysQCI13} -- is $\rho_{\rm sQC} \approx 11.3585$ (with $R_{\rm disk} = 1$). 
    This pattern also enjoys discrete Lorentz symmetry: its infinite set of boosts are given by the powers of the eigenvalues of the Coxeter element (see Section \ref{subsec:Hyperbolic_Coxeter_Elements} and \ref{subsec:2DLorentzianQuasiCrystal}); in fact, we have 
    $$\tanh\varpi = \sqrt{1 - \frac{4}{(\lambda_0^+ + \lambda_0^-)^2}} = \sqrt{\frac{2}{1+\sqrt{17}}} \approx 0.624811$$ and are related to the irrational number $\sqrt{17}$.
\end{enumerate}

\begin{figure}[!t]
    \centering
    \includegraphics[width=0.90\linewidth]{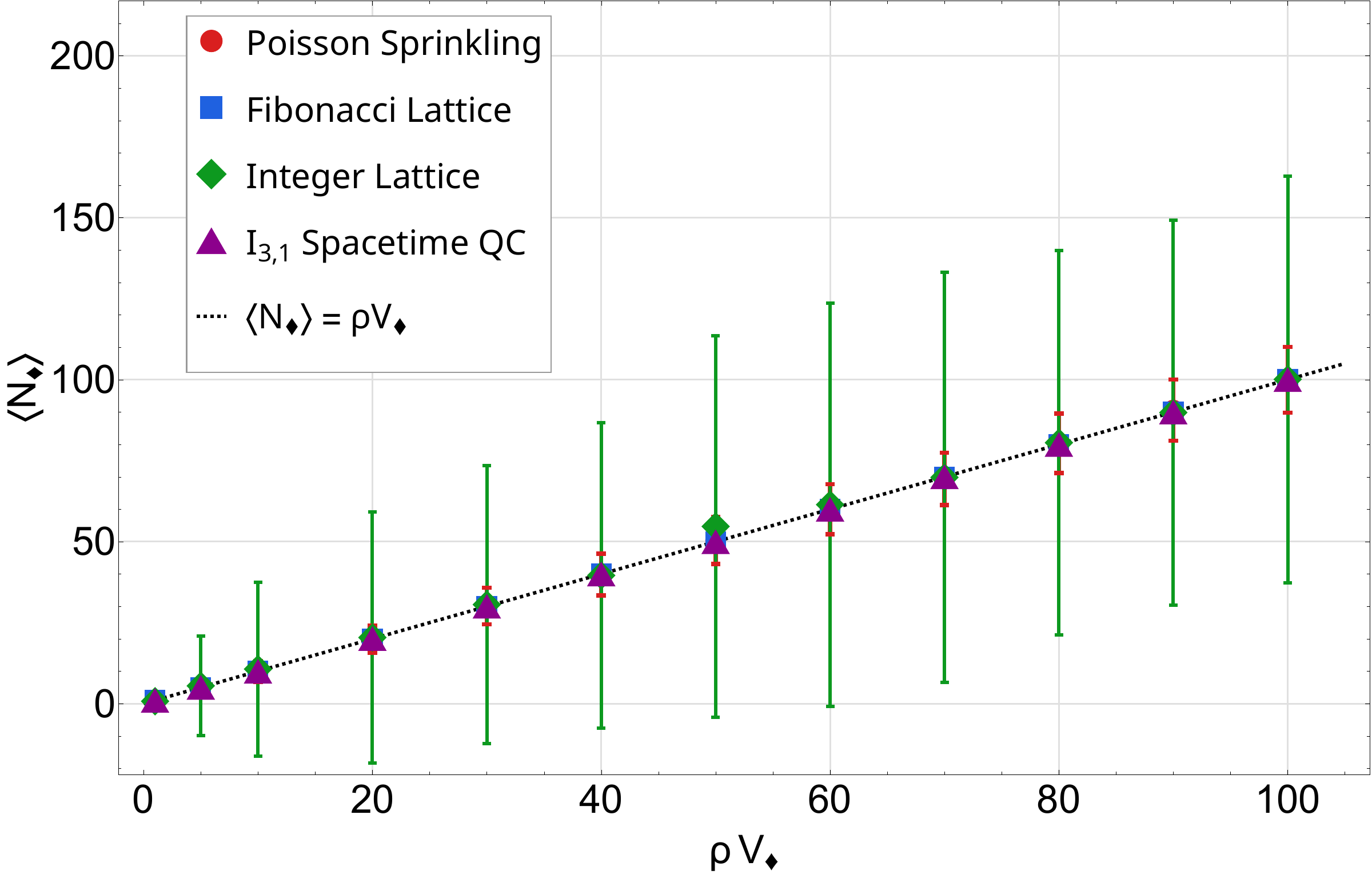}
    \caption[The N-V test for four $(1+1)$-dimensional point sets: the mean behaviour.]{The N-V test for four $(1+1)$-dimensional point processes, where we plot the average number of points $\langle N_\cdiamond \rangle$ inside each $\cdiamond$ against the physical volumes $\rho_{\rm pattern} V_\cdiamond = V_{\rm sample}$. The error-bars are given by the standard deviation $\sigma_{N_\cdiamond}$ for each process. Notice that the errors for the Fibonacci lattice (in blue) and for the spacetime quasicrystal (in purple) are small enough to be covered by the size of the markers.}
\label{fig:NVTest1Plus1DMean}
\end{figure}

Running the N-V test for these four point sets, we obtain the results shown in Fig.~\ref{fig:NVTest1Plus1DMean}. The setup is similar\footnote{The authors there do not give many details of how \textit{exactly} they sample their causal diamonds. Attempting to reconstruct their method, we came up with our own setup, which is described in more detail in Appendix \ref{appendix:NVTestDetails}.} to that of \cite{Saravani_2014}: we consider $M = 1000$ different causal diamonds $\cdiamond$ with the same volume $V_\cdiamond$, whose centres and shapes vary randomly throughout the pattern patches. To make sure we are actually comparing the same volumes across our four patterns that have different densities, we sample $V_{\rm sample} = \rho_{\rm pattern} V_\cdiamond$; clearly, if $\rho=1$, those two volumes agree. For each realization, the number of lattice points $N_\cdiamond$ inside $\cdiamond$ is counted, leading to a distribution of the number of points for a given volume $V_{\cdiamond}$. From this, we calculate the mean $\langle N_{\cdiamond} \rangle$ and the standard deviation $\sigma_{N_{\cdiamond}}$. This procedure is then repeated for different volumes. The proposed volume levels for our causal diamonds are $V_{\rm sample} = \{ 1,5,10,20,30,40,50,60,70,80,90,100\}$ in order to match \cite{Saravani_2014}. We then plot our data points in Fig.~\ref{fig:NVTest1Plus1DMean} with the same horizontal axis given by $\rho V_\cdiamond$ for better comparison.\footnote{One could instead rescale all patterns to adjust their density so that all patterns have $\rho = 1$. Here, we chose not to do that so that we can keep the patterns ``pristine'' (i.e.~the Fibonacci lattice with only $\sqrt{5}$ and the spacetime quasicrystal with only $\sqrt{17}$ as their irrationalities).}

The result that follows from Fig.~\ref{fig:NVTest1Plus1DMean} is clear: \textbf{all four point sets satisfy the N-V correspondence}, as they do track the dashed diagonal $\langle N_{\cdiamond} \rangle = \rho V_\cdiamond $ describing the average -- \textit{desirable} -- behaviour. However, the do so with very different fluctuations around the mean, with the integral lattice dominating the errors. 

\begin{figure}[!t]
    \centering
    \includegraphics[width=0.95\linewidth]{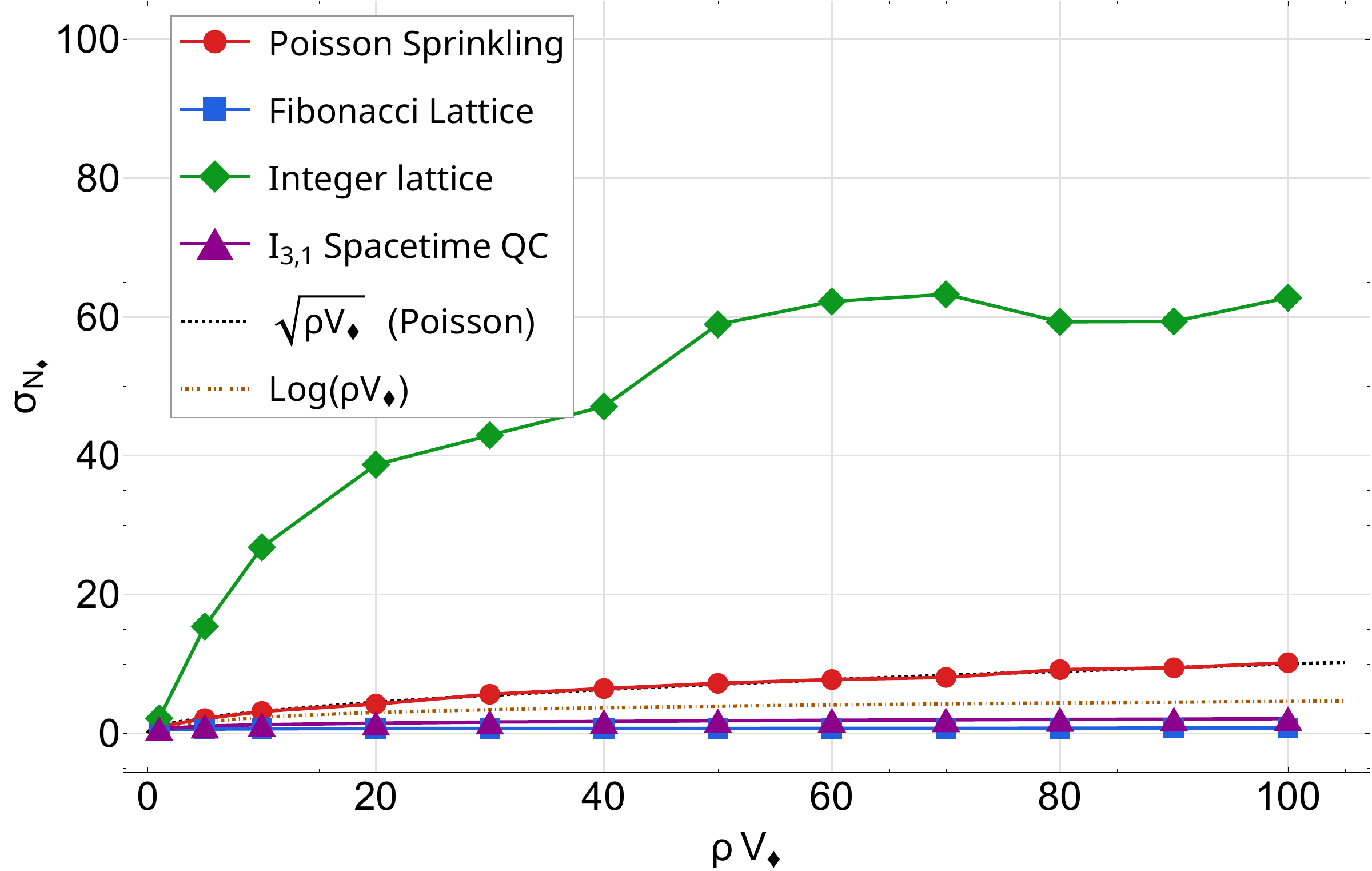}
   \caption[The N-V test for four $(1+1)$-dimensional point sets: the standard deviation.]{The N-V test for four $(1+1)$-dimensional point processes, where we plot the standard deviation $\sigma_{N_\cdiamond}$ (the fluctuations) against the physical volumes $\rho_{\rm pattern}V_\cdiamond= V_{\rm sample}$. The Poisson sprinkling (in red) tracks the dashed diagonal $\sqrt{\rho_{\rm pattern} V_\cdiamond}$, while the integer lattice (in green) is super-Poissonian. The Fibonacci lattice (in blue) and the spacetime quasicrystal (in purple) are sub-Poissonian and sub-logarithmic (in dashed orange)}
    \label{fig:NVTest1Plus1DStandardDeviation}
\end{figure}

This can be seen better in Fig.~\ref{fig:NVTest1Plus1DStandardDeviation}, where we now plot the standard deviation $\sigma_{N_\cdiamond}$ (i.e. the error-bars of Fig.~\ref{fig:NVTest1Plus1DMean}) against the volumes $V_\cdiamond$. Here, the reference curve is that of the Poisson statistics \eqref{eq:PoissonStatistics}: we see that the red curve coming from the Poisson sprinkling exactly tracks the expected $\sqrt{\rho V_\cdiamond}$-behaviour. We call the regions above and below the reference curve as \textit{super-Poissonian} and \textit{sub-Poissonian} respectively. The new result of the paper \cite{Saravani_2014} was that \textbf{the Fibonacci lattice realizes the N-V correspondence much better than the Poisson sprinkling for macroscopic volumes} $V_\cdiamond \geq 1$, since its fluctuations remain in the sub-Poissonian regime throughout the range of tested volumes. In fact, we see that its standard deviation (in blue) remains almost constant and notably even below the logarithm (shown in dashed orange). On the other hand, our other candidate, the integer lattice shown in green, is super-Poissonian throughout the tested range, as expected from our earlier discussion in Fig.~\ref{fig:BoostedVsUnboostedIntegerLattice}. This confirms the statement from \cite{Saravani_2014}, which never really shows the test results for $\mathrm{I}_{1,1}$.  

The new result of this thesis is that \textbf{the spacetime quasicrystal} -- built from the $\mathrm{I}_{3,1}$ lattice -- \textbf{also significantly outperforms the Poisson sprinkling in the N-V test for the same test range.} Indeed, we see in Fig.~\ref{fig:NVTest1Plus1DStandardDeviation} that the purple curve remains in the sub-Poissonian region as well, however it stays consistently above the blue curve of the Fibonacci lattice. This means that our aperiodic structure beats Poisson in the number-volume realization, but it still cannot beat the periodic Fibonacci lattice. From Fig.~\ref{fig:NVTest1Plus1DStandardDeviation}, though, we see that both structures have a similar behaviour: their standard deviations are held almost constant for the full range, exhibiting a slow, sub-logarithmic growth. 

Before moving on to higher dimensions, it will be instructive to understand why our ordered structures behave the way they do, which was not explicitly mentioned in \cite{Saravani_2014}.

\subsubsection{Why the Integer Lattice Fails:}

\begin{figure}[!t]
    \centering

    \subfloat[Stretched $\cdiamond$ hitting or missing entire rows.]{
        \begin{minipage}[c]{0.3\linewidth}
        \centering

        \includegraphics[scale=0.175]{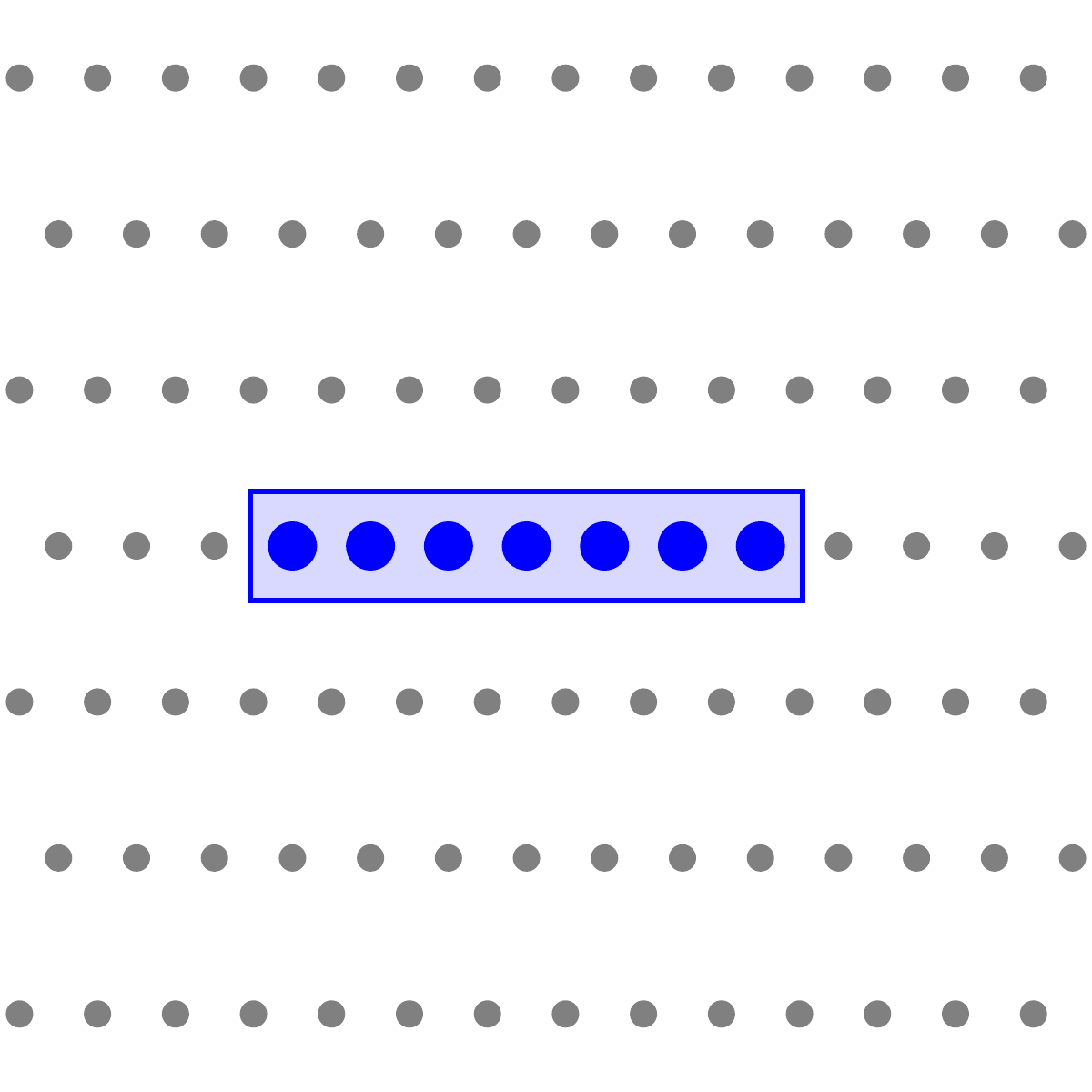}

        \vspace{0.5em}

        \includegraphics[scale=0.175]{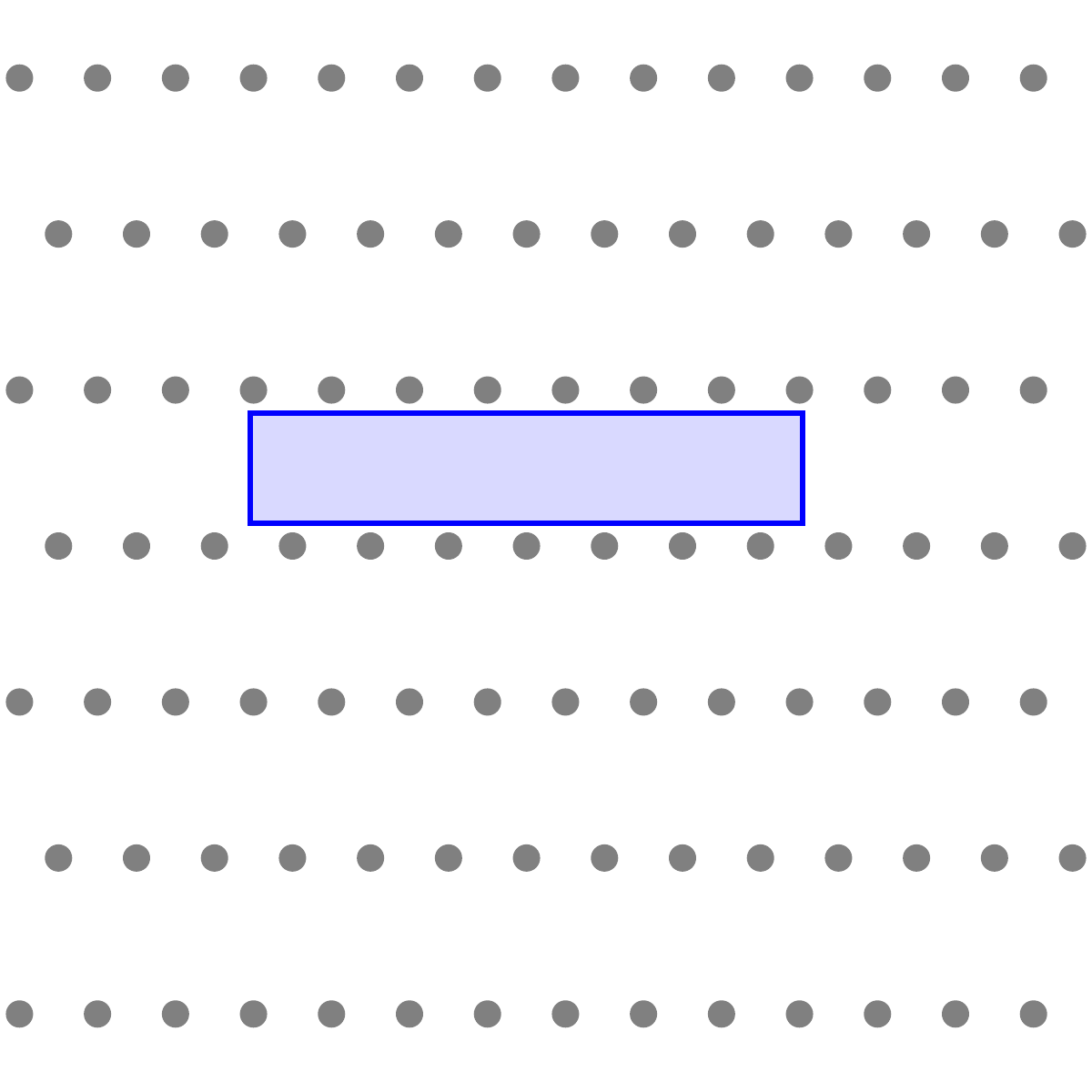}
    \end{minipage}
    \label{fig:StretchedDiamondBetweenRows}
    }
    \subfloat[The boost catastrophe of the integer lattice]{
        \includegraphics[width=0.67\linewidth]{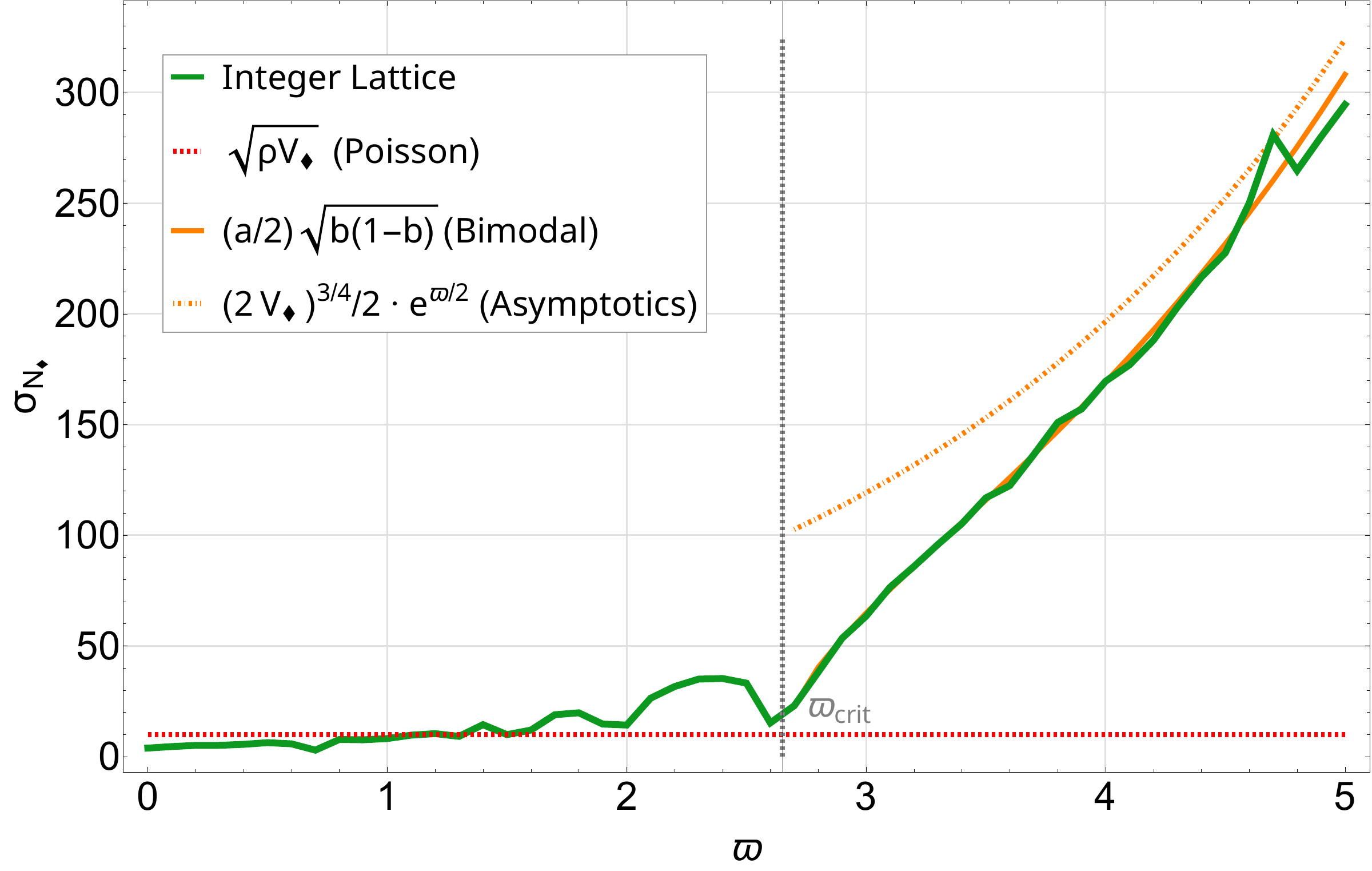}
    }
    \caption[The boost catastrophe of the integer lattice $\mathrm{I}_{1,1}$.]
    {On the left we see a diamond of volume $V_\cdiamond = 5$ $(\rho = 1)$, stretched by a boost $\varpi = 1.5 >\varpi_{\rm crit}(5) \simeq 1.16$; as its centre varies, it either hits or misses entire rows. On the right, we see this \textit{boost catastrophe} of the integer lattice $\mathrm{I}_{1,1}$ for $V_\cdiamond = 100$. The lattice remains around the Poisson curve for small boosts. After $\varpi > \varpi_{\rm crit}(100) \simeq 2.65$, the more we boost, the more we break the regularity and the variance grows with no limit. For this \textit{boost-stratified} N-V test, we sampled $M=500$ causal diamonds $\cdiamond$ at each boost level in the interval $\varpi \in [0,5]$ with step $\delta \varpi = 0.1$. For each such distribution, we calculated $\langle N_\cdiamond \rangle$ and $\sigma_{N_{\cdiamond}}$, and plotted the latter against the values of $\varpi$.}

    \label{fig:BoostCatastopheIntegerLattice}
\end{figure}

\noindent The reason turns out to be that $\mathrm{I}_{1,1}$ is simply \textbf{too regular in the nullspace} ${\rm (u,v)}$. In these coordinates, the integer lattice becomes 
$$ \mathrm{I}_{1,1} = \{ ({\rm u},{\rm v}) = (m+n,m-n) \, | \, m,n \in \ZZ \}, $$
which means that, for any given value of ${\rm u} = {\rm u}_0 \in \ZZ$ (similarly for fixing ${\rm v}$), there is an entire row of infinite ${\rm v}$'s, which share the same parity as ${\rm u}_0$, actually, since ${\rm v} = {\rm u}_0 - 2n$ (that is, in a given ${\rm u}$-row, the ${\rm v}$'s are equally-spaced, two integers away from one another; the null $\mathrm{I}_{1,1}$ lattice \textit{is} the $\mathrm{II}_{1,1}$ lattice -- see Section \ref{subsubsec:Lorentzian_Lattices}). Adjacent ${\rm u}$-rows are one integer apart. As a result, there \textit{exists a minimum spacing} in the null space between adjacent rows, which is 
$$\delta {\rm v}_{\min} = 1 = \delta {\rm u}_{\min}. $$ 
In the ${\rm (u,v)}$-plane, causal diamonds $\cdiamond$ become rectangles (see Appendix \ref{appendix:NVTestDetails}), and for a fixed volume $V_\cdiamond$, their null extents ${(a,b)}$ in the ${\rm u}$- and ${\rm v}$-direction are given by Eq.~\eqref{eq:NullExtentsCDiamond1+1D} and are boost-dependent. As we increase the boost parameter, one side of the rectangle gets squeezed and the other gets stretched, leading to thinner and thinner rectangles. Because of the regularity along the null directions, for large enough boosts, \textbf{a very thin $\cdiamond$ could slip through entire rows}, thus missing the lattice points altogether. This situation is depicted in Fig.~\ref{fig:StretchedDiamondBetweenRows}. Consequently, such $\cdiamond$ of non-zero volume would intersect no points from the causal set, leading to a bad realization of the N-V correspondence. 

In fact, one can analytically calculate a threshold in the boost parameter $\varpi$ and predict when this behaviour will kick in. For a given volume $V_\cdiamond$, the condition for $\cdiamond$ to have a non-zero chance of falling in-between adjacent rows is that its null extent (let's say, in the ${\rm v}$-direction) becomes smaller than $\delta {\rm v}_{\min}$. Letting the critical boost parameter that achieves that be $\varpi_{\rm crit}$, we obtain (using results from Appendix \ref{appendix:NVTestDetails})
\begin{equation}
  \sqrt{2V_{\cdiamond}}e^{-\varpi_{\rm crit}} = b = \delta {\rm v}_{\min} \Rightarrow \boxed{\varpi_{\rm crit} = \frac{1}{2}\ln{2V_\cdiamond} - \ln{\delta {\rm v}_{\min}}}.
  \label{eq:BoostCatastropheCondition}
\end{equation}
For the integer lattice, the second term vanishes because $\delta {\rm v}_{\min} = 1$ and $\varpi_{\rm crit}$ is governed by the volume $V_{\cdiamond}$ under consideration. We call $\varpi_{\rm crit}$ the \textbf{catastrophe threshold}. 

For example, for $\rho V = 100$ (with $\rho = 1$, so $V = 100$), we find $\varpi_{\rm crit} = \frac{1}{2}\ln 200 \approx 2.65$, thus any boost with $\varpi > \varpi_{\rm crit}$ is expected to break the regularity of the lattice and lead to large volumes $V_\cdiamond$ containing no points. In fact, we can also calculate the standard deviation $\sigma_{N_{\cdiamond}}$ assuming a \textit{bimodal distribution} at the regime where $\varpi > \varpi_{\rm crit}$ ($\cdiamond$ will either hit an entire row and intersect points, or miss it completely). We find -- see Appendix \ref{appendix:NVTestDetails} -- that 
$$ \sigma_{N_{\cdiamond}} = \frac{(2V_\cdiamond)^{3/4}}{2}  e^{\varpi/2}, \qquad \text{ for } b \ll 1 \text{ (very thin diamonds)}$$ 
deep into the catastrophic regime. We call this behaviour the \textbf{boost catastrophe} of the lattice, because the variance grows without limit, as shown in Fig.~\ref{fig:BoostCatastopheIntegerLattice}.

\subsubsection{Why the Fibonacci Lattice and the Spacetime Quasicrystal Win:}

\noindent We can summarize our discussion about the $\mathrm{I}_{1,1}$ lattice as follows: at every $\rm u$-value there is an entire row of infinitely many points (spaced 2 apart in $\rm v$). When a boosted diamond (with null extents $(a,b)$) captures one such $\rm v$-slice, it encloses $\sim a/2$ points all at once; if it misses it, we get zero. This mechanism drives the boost catastrophe.

A key property that the Fibonacci lattice and the spacetime quasicrystal share, but the integer lattice crucially lacks, is that \textbf{their null directions are not regular}: for any fixed value of $\rm u$ (similarly for $\rm v$) there is \textit{not} an entire null-row of infinite $\rm v$'s (and vice versa). This is a consequence of the fact that the \textbf{null-projections are now dense} in $\RR$. 

For concreteness, focus on the Fibonacci lattice \eqref{eq:FibonacciLatticeNullSpace}. Its $\rm (u,v)$-coordinates are elements of $\ZZ(\phi)$ and thus dense in the reals $\RR$. One can show directly that no two distinct points of the Fibonacci lattice share the same $\rm u$- or $\rm v$-coordinate: if $m_1\phi + n_1 = m_2\phi + n_2$ for integers $m_i, n_i$, then $(m_1-m_2)\phi = n_2 - n_1$, which forces $m_1 = m_2$ and hence $n_1 = n_2$ by irrationality of $\phi$. Consequently, the lattice has \textbf{no null rows}: every lattice point occupies a unique position in both null directions. However, the denseness of $\ZZ(\phi)$ guarantees that, for any given value of ${\rm u}_0 \in \ZZ(\phi)$ (or ${\rm v}_0$), one can find two integers $m,n$ such that the resulting ${\rm u} = m \phi + n$ approximates ${\rm u}_0$ within some small $\epsilon<1$; of course, the resulting ${\rm v}$ might actually be further away from ${\rm v}_0$. Given that the constant $\rm u$- and $\rm v$-lines define the boundaries of causal diamonds, we conclude that, for any point in $\Lambda_{\rm Fib.}$, one can find another point in the pattern that comes arbitrarily close to the lightcone of the original point, but does not belong to it (i.e. that has the same $\rm u-$ or $\rm v-$ coordinate, but differs in the other; of course, for a given $\epsilon$, we might need to travel a long distance on the pattern to find this other point) \cite{kronecker1884naherungsweise, WeissteinKroneckerApproximation}.

We view this as an important property, so we decided to give it a name: 
\begin{statement}{Quasi-Null Property (QNP)} 
a discrete pattern is said to be \textit{quasi null} if it contains no null-separated points, but admits sequences of separations approaching any lightcone arbitrarily closely.
\label{statement:QuasiNullity}
\end{statement}
Such a pattern has no null-rows, because then, for any given point, we would have an infinite number of null neighbours. 

The contrast with $\mathrm{I}_{1,1}$ is now sharp. The integer lattice does not have the QNP: its null directions are \textit{rational} relative to $\ZZ^2$, giving $\delta{\rm v}_{\min}=1=\delta{\rm u}_{\min}$ and a finite $\varpi_{\rm crit}$ according to Eq.~\eqref{eq:BoostCatastropheCondition} for any volume $V_\cdiamond$. On the other hand, the Fibonacci lattice has \textit{irrational} null directions with respect to $\ZZ^2$ and there is no global minimum spacing.\footnote{The minimum gap in any finite patch vanishes in the thermodynamic limit where the volume of the patch -- and the number of points it contains -- is taken to be very large while its density remains constant.} Therefore, $\delta {\rm v}_{\min}, \delta {\rm u}_{\min} \rightarrow 0$, which drives the catastrophe threshold to $\varpi_{\rm crit} \rightarrow + \infty$. This means that there is \textit{no finite boost} at which a causal diamond can reliably slip between all ${\rm v}$-values: for any $\varpi$ however large, the $\cdiamond$ window of width $b > 0$ will always intersect lattice ${\rm v}$-values. This fact, together with the discrete Lorentz boost-symmetry that the Fibonacci lattice has, implies that no finite boost can completely empty a $\cdiamond$, and \textbf{the structure avoids the boost catastrophe. }

The spacetime quasicrystal has the exact same behaviour, with the difference that its algebraic structure is governed by a different irrational number, $\sqrt{17}$, and the field $\ZZ(\sqrt{17})$. It still has the QNP \textit{and} discrete Lorentz symmetry, but its structure is more complicated to write down than that of the Fibonacci lattice. Note that, for \textit{any} \CNP{} pattern, the QNP follows from the irrationality of the \CNP{} slice. As a result, \textbf{any spacetime quasicrystal is guaranteed to have no null separated vertices.}

Before we move on to higher dimensions, let us offer some insights about \textit{why} the standard deviation for the Fibonacci lattice and the spacetime quasicrystal remains low. The actual explanation will have to wait till Section \ref{subsec:SpacetimeHyperuniformity}, when we introduce the concept of \textit{hyperuniformity}. The reasoning is that a boosted $\cdiamond$ gains and loses points \textit{individually} as its centre is displaced, rather than capturing or missing parts of rows of points simultaneously, as is the case for the integer lattice $\mathrm{I}_{1,1}$. On top of that, due to the Lorentz boost symmetry, both patterns eventually come back to themselves after (integer multiples of) certain boosts ($\varpi_{\rm Fib.} = \tanh^{-1}(\sqrt{5}/3) \approx 0.962424 $ and $\varpi_{\rm sQC} = \cosh^{-1}{\left(\frac{1+\sqrt{17}}{4}\right)} \approx 0.732858 $), so the patterns absorb boosts better (e.g.~they do not change drastically). These two properties push the standard deviation to lower-than-Poisson values.

Inspired by these properties, we formulate the following conjecture:
\begin{conjecture}{(The ``Poisson's Demise'') Conjecture} a point set that has discrete Lorentz boost-symmetry and the QNP can beat Poisson sprinklings in the N-V test in \textit{any number of spacetime dimensions}.
\label{conj:PoissonDemise}
\end{conjecture}

As was argued in \cite{Saravani_2014}, it is expected that \textit{no} structure, lattice or other, even if it has discrete Lorentz symmetry, could realize the N-V correspondence better than Poisson in higher than 2 spacetime dimensions. Their reasoning is that the freedom to perform boosts independently in more than one plane will distort the structure much more drastically than in $(1+1)$D where there is only a single plane, \footnote{In $(1+1)$D, there is a single boost parameter $\varpi_x$ along the t-x plane (but an infinite number of directions along the plane!), while in $(1+3)$D there are three, $\varpi_x,\varpi_y$ and $\varpi_z$, including boosts in the t-y and t-z planes as well.} and this will quickly drive any structure into the super-Poissonian regime. On the contrary, a boosted Poisson structure will still satisfy Poissonian statistics, even though it will change, in general, and never get mapped exactly onto itself.  

More specifically, we saw that in $(1+1)$D our structures with discrete Lorentz boost symmetry only change within some rapidity interval from $|\varpi| = 0$ to some $|\varpi_{\rm sym} | > 0$, after which they come back to themselves (as boosting by $\varpi_{\rm sym}$ is a symmetry of the pattern). If $\varpi \in [k \varpi_{\rm sym}, (k+1)\varpi_{\rm sym}]$ for some $\ZZ \ni k \geq 1  $, then the previous changes of the pattern repeat: two boosted patterns by rapidities $\varpi$ and $\varpi+\varpi_{\rm sym}$ are the same (up to labelling of points). Now, imagine that some higher-dimensional pattern has independent boost symmetries along, let's say, all three directions $x,y$ and $z$, given by some parameters $\varpi_{\rm sym, x},\varpi_{\rm sym, y}$ and $\varpi_{\rm sym, z}$. Even if the pattern does not change `too much' for a boost in the t-x plane, there is still `plenty of room' to boost in any of the other planes and distort the pattern till we reach the values $\varpi_{\rm sym ,y}$ or $\varpi_{\rm sym ,z}$. An even better scenario would be that the pattern under consideration has also high rotational symmetries that can mix the three independent planes, leading to even more symmetric boost directions. For a Lorentzian lattice, the rotational symmetries are limited by the crystallographic restriction theorem, but a quasicrystal avoids this and can show more intricate symmetries. In these cases, a \textit{dense set of symmetric boost directions} is expected to exist, like for the $HF_4$-quasicrystal of Section \ref{sec:IcosahedrallySymmetricsQC}, and the spacetime quasicrystals will absorb most boosts.       
  
Since our spacetime quasicrystals satisfy the requirements of the Poisson's demise conjecture in any dimension, they are promising candidates to outperform sprinklings in higher dimensions as well. 

\subsection{The N-V Test in $(2+1)$D: Poisson Shines Again}
\label{subsec:NVTest2Plus1D}

Climbing up the ladder of dimensionality, in $(2+1)$D we test the following structures:

\begin{enumerate}
    \item \textbf{A Poisson Causal Set:} generated by a sprinkling in $\RR^{1,2}$ at density $\rho_{\rm Poisson} =1$, which will serve again as our reference. 
    \item \textbf{The Integer Lattice} $\mathrm{I}_{2,1} \simeq \ZZ^3$: in $(1+1)$D, the integer lattice $\mathrm{I}_{1,1}$ has only a very modest symmetry (one root, finite discrete Coxeter group). Perhaps this is the reason why it underperformed in the N-V test? But in $(2+1)$D, $\mathrm{I}_{2,1}$ has an infinite symmetry group described by the first Coxeter-Dynkin graph of Fig.~\ref{fig:VinbergRootsI1nUpto10}. Could it behave better now? We picked the integer lattice to answer this question. As a root lattice, it is the unique self-dual lattice in $(2+1)D$, it is isomorphic to $\ZZ^3$ and it has density $\rho_{\rm Int.} = 1$. It has discrete boost symmetry but it does not have the QNP. 
    \item \textbf{The Spatially Hexagonal Lattice with Integer Time Steps} $\ZZ \oplus A_2$: this lattice candidate describes a tesselation by triangles on the spatial slice (i.e.~the $A_2$ root lattice; the densest two-dimensional packing \cite{conway2013sphere}, with 6-fold rotational symmetry vs 4-fold for $\mathbb{Z}^2$)), which is repeated over integer time steps (we scale the temporal direction by $\tfrac{2}{\sqrt{3}}$ to make the lattice unimodular). The lattice is 
    \begin{equation}
    \Lambda_{\rm Hex.} = \left\langle (\tfrac{2}{\sqrt{3}},0,0),(0,1,0),(0,\tfrac{1}{2},\tfrac{\sqrt{3}}{2})\right\rangle_\ZZ = \left\{ \big(\tfrac{2t}{\sqrt{3}}, m+\tfrac{1}{2}n, n\tfrac{\sqrt{3}}{2} \big) \, | \, t,m,n \in \ZZ \right\}
    \end{equation}
    with density $\rho_{\rm Hex.} = 1$. Despite its lack of discrete Lorentz boost-symmetry, we chose this lattice because it still has some irrationality (through the $\sqrt{3}$). As such, it will allow us to check whether having the QNP without Lorentz boost symmetry (but with higher rotational symmetry) is enough to beat Poisson. 
    \item \textbf{The 3D Fibonacci Lattice:} inspired by the results of the previous section, we construct a lattice which is a 2D Fibonacci lattice in its $(t,x)$ or $(t,y)$ plane, but not in any arbitrary combination of the $x,y$-directions. The lattice is 
    \begin{equation}
    \Lambda_{\rm Fib}^{3D} = \left\langle (\tfrac{\sqrt{5}}{2},\tfrac{1}{2},\tfrac{1}{2}),(0,1,0),(0,0,1)\right\rangle_\ZZ = \left\{ \big(\tfrac{t\sqrt{5}}{2}, \tfrac{m}{2}+n, m\tfrac{1}{2}+n \big) \, | \, t,m,n \in \ZZ \right\}
    \end{equation}
    with density $\rho_{\rm Hex.} = 2/\sqrt{5} \simeq 0.8945$. We chose this lattice to test if having discrete Lorentz symmetry on 2D subspaces is still enough to beat Poisson; note that this pattern also has the QNP. 
\end{enumerate}

\begin{figure}[t!]
    \centering
    \includegraphics[width=0.95\textwidth]{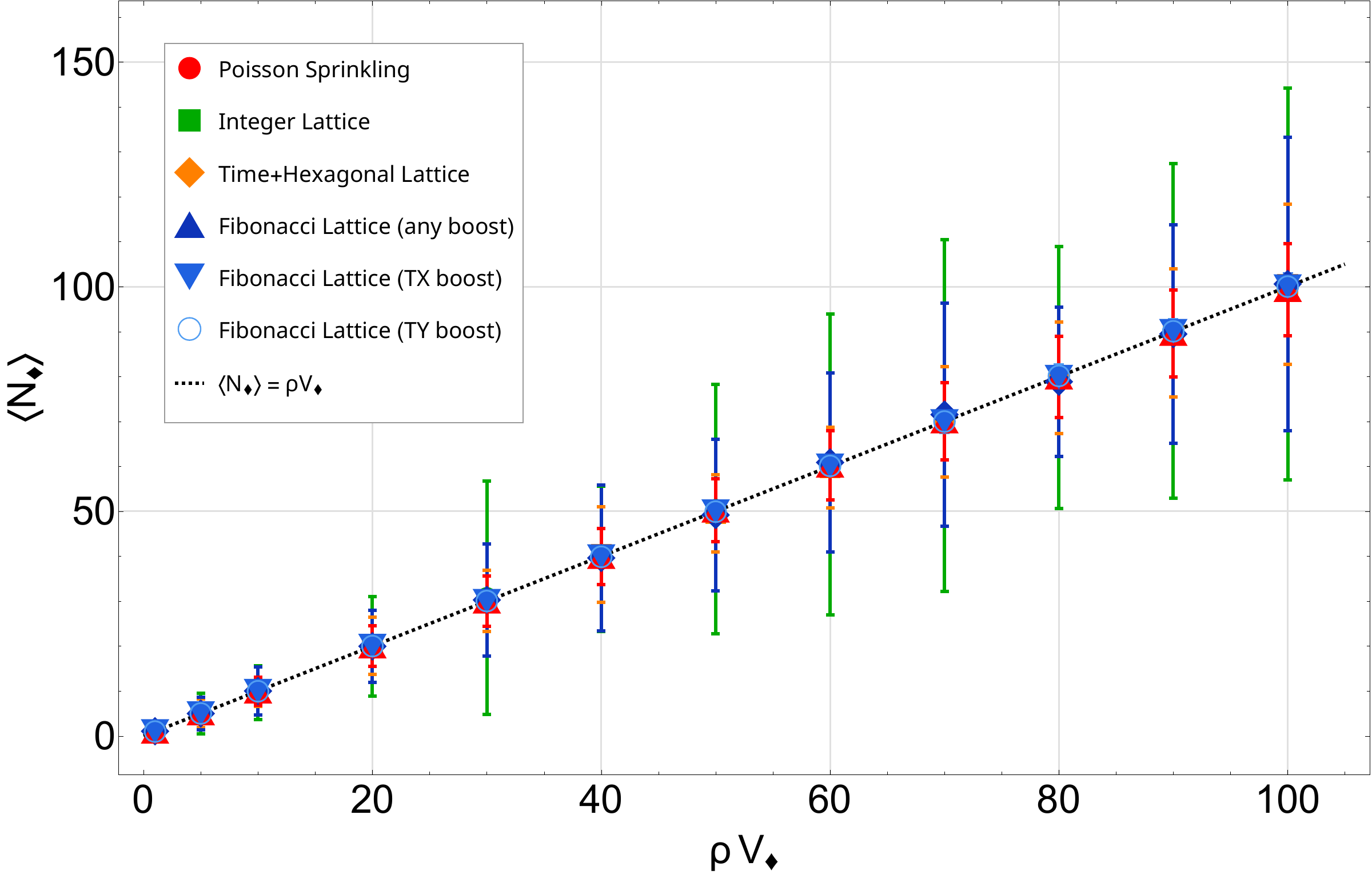}
    \caption[The N-V test for four $(2+1)$-dimensional point sets: the mean behaviour.]{The N-V test in $(2+1)$D, where we plot the average number of points $N_{\cdiamond}$ inside each $\cdiamond$ against the physical volumes $\rho_{\rm pattern} V_{\cdiamond} = V_{\rm sample}$; the error-bars are given by $\sigma_{N_{\cdiamond}}$. For the 3D Fibonacci lattice, we consider three cases: (1) generic boosts (in navy blue), (2) TX boosts (in royal blue), and (3) TY boosts (in steel blue). In particular, the error bars for the last two cases are too small and are covered completely by the markers.}
    \label{fig:1Plus2DNVTestComparisonPlots}
\end{figure}

\begin{figure}[t!]
    \centering
    \includegraphics[width=0.95\linewidth]{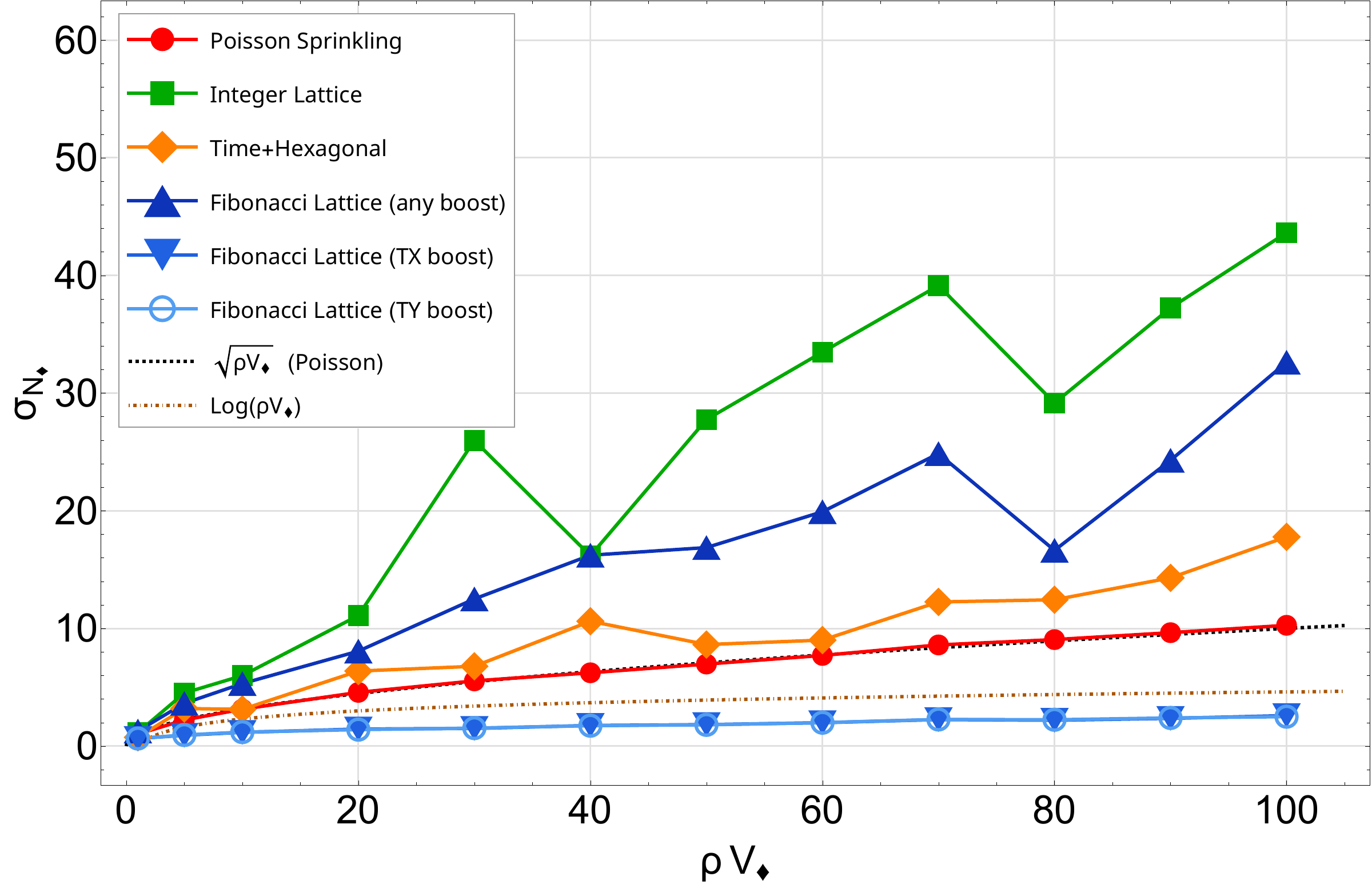}
    \caption[The N-V test for four $(2+1)$-dimensional point sets: the standard deviation.]{The N-V Test for four $(2+1)$-dimensional point processes, where we plot the standard deviation $\sigma_{N_\cdiamond}$ against the physical volumes $\rho_{\rm pattern} V_\cdiamond = V_{\rm sample}$. The Poisson sprinkling (red) tracks the dashed square-root curve, while all other lattices are generically in the super-Poissonian region. Notably, the 3D Fibonacci lattice is sub-Poissonian \textit{only} if we consider boosted $\cdiamond$ along either the $(t,x)$ or the $(t,y)$ plane.}
    \label{fig:1Plus2DNVTestStandardDeviationComparisonPlots}
\end{figure}

Running the N-V test with the same setup as in the $(1+1)$D case (see Appendix \ref{appendix:NVTestDetails} for the details), we obtain the results shown in Fig.~\ref{fig:1Plus2DNVTestComparisonPlots}. We see that all structures satisfy the N-V correspondence again, but \textbf{no structure can beat Poisson in its standard deviation}. An important lesson from the $(2+1)$D-analysis seems to be that \textit{neither} discrete Lorentz symmetry ($\mathrm{I}_{2,1}$) nor irrational null directions ($\ZZ \oplus A_2$ or $\Lambda_{\rm Fib}^{3D}$) \textit{by themselves} are enough to guarantee a better N-V realization than Poisson sprinklings. Interestingly, the 3D Fibonacci lattice can beat Poisson \textit{if} we confine the boosts on either the $(t,x)$- or the $(t,y)$-plane, but \textit{not} for any generic boost that generally mixes these two planes. As explained in Appendix \ref{appendix:NVTestDetails}, in $(2+1)$D one can parametrize a boost by specifying a hyperbolic angle $\varpi$ \textit{and} a direction on the circle $\Sph^1$, given by a unit vector ${\bf n } = (\cos \theta, \sin \theta)$. For $\theta = 0,\pi$ or for $\theta = \pi/2, 3\pi/2$, we find that the 3D Fibonacci lattice is actually sub-Poissonian. This can be seen better in Fig.~\ref{fig:1Plus2DNVTestStandardDeviationComparisonPlots}, where the 3D Fibonacci lattice is shown to have a sub-logarithmic growth as in Fig.~\ref{fig:NVTest1Plus1DStandardDeviation}). 

These results enhance our belief in the Poisson's demise conjecture: no point structure that meets only half of the requirements of the conjecture was able to beat Poisson in $(2+1)$D. However, the larger the (spatial) symmetry group of the lattice, the better the structure behaves on average in the N-V test. Compare for example, the generic 3D Fibonacci curve with that of the hexagonal spatial lattice with integer (in fact, irrational) time steps. While the former has only a modest 4-fold rotational symmetry on its spatial plane (which looks like the $\ZZ_2$ lattice), the latter enjoys the higher 6-fold rotational symmetry. We see that the orange curve stays always below the navy blue curve in Fig.~\ref{fig:1Plus2DNVTestStandardDeviationComparisonPlots}, suggesting that the hexagonal spatial lattice behaves better in the N-V test, on average, than the 3D Fibonacci lattice.   

We have also confirmed that all lattices suffer from boost catastrophe, though an analytical model for how the variance grows after $\varpi_{\rm crit}$ has not been yet found. We refer the reader to the Appendix \ref{appendix:NVTestDetails} for the plots and the associated discussion. The boost-stratified tests shed more light on where exactly each lattice starts to fail, which is hidden in Fig.~\ref{fig:1Plus2DNVTestStandardDeviationComparisonPlots}, because the result is an average behaviour over all performed boosts.

\subsection{The N-V Test in $(3+1)$D: Poisson is \textit{not} Invincible}
\label{subsec:NVTest3Plus1D}

The real test, though, is in $(3+1)$D, where we do have a natural quasicrystal to use in our simulations. The point processes we test now are
\begin{enumerate}
    \item A \textbf{Poisson Causal Set:} generated by a sprinkling in $\RR^{3,1}$ at density $\rho_{\rm Poisson} =1$, which will serve again as our reference. 
    \item \textbf{The Integer Lattice} $\mathrm{I}_{3,1}$: as a root lattice, it is the unique self-dual lattice in $(3+1)D$; it is isomorphic to $\ZZ^4$ and it has density $\rho_{\rm Int.} = 1$. Its infinite discrete group of symmetries is described by the second Coxeter-Dynkin graph in Fig.~\ref{fig:VinbergRootsI1nUpto10}. 
    \item \textbf{The Spacetime Quasicrystal} from Section \ref{subsec:4DsQCFromI17}: this is a $\sCNP$ quasicrystal coming from the lattice $\mathrm{I}_{7,1}$. The pattern's number density is given by Eq.~\eqref{eq:DensitysQCI17} as
    $ \rho_{\rm sQC \, I_{7,1}} \approx 39.4784 $
    (see Section \ref{sec:ExamplesSpacetimeQuasicrystals}). The pattern enjoys discrete Lorentz boost symmetry \textit{only in its first Lorentzian $\mathbf{2}$-plane} (see Section \ref{subsec:Hyperbolic_Coxeter_Elements}). There is \textit{no boost symmetry} if one performs a different boost that mixes the temporal direction of the first $\mathbf{2}$-plane with any of the spatial directions of the second (Euclidean) $\mathbf{2}$-plane. The boosts that are symmetries of the pattern are (integer multiples of) the following velocity 
    $$ \tanh \varpi = \sqrt{1- \frac{4}{(\lambda_0^+ + \lambda_0^-)^2}} \approx 0.360144$$ (see Table.~\ref{tab:FieldExtensionsHyperbolicCases} for the calculation of $\lambda_0^+ + \lambda_0^-$ for this lattice). 
    On the second spatial $\mathbf{2}$-plane, the quasicrystal has dense orientational symmetry (i.e.~\textit{almost} the full $SO(2)$ group); a discrete rotation by any integer multiple of 
    $$ \vartheta_1 = \cos^{-1} \left( \frac{\lambda_{1}^+ + \lambda_1^-}{2} \right) \approx 1.1272 $$
    maps the pattern onto itself.
    \item \textbf{The Icosahedral Spacetime Quasicrystal} from Section \ref{sec:IcosahedrallySymmetricsQC}: this is a \sCNP{} quasicrystal from the root-lattice $HE_7^+$; the pattern's density is given by Eq.~\eqref{eq:sQCDensityHF4} as $\rho_{\rm sQC \, HF_4} \approx 5.951$. $HF_4$ contains the spatial icosahedral symmetry group $H_3$ (so it enjoys at least the icosahedral rotational symmetries), but it has much more rotational symmetries than that, as one can combine each of these $H_3$ elements with the additional generator of $HF_4$ to generate infinitely many rotations. Similarly, the boost symmetry is not only along the $V_0$ plane: we have boost symmetry along at least every icosahedral direction as well, which mixes the different spatial directions. The Coxeter-boost is given by
    $$ \tanh\varpi = \sqrt{1-\frac{4}{(\lambda_0^+ + \lambda_0^-)^2}} \approx 0.5476.
    $$
\end{enumerate}

\begin{figure}[t!]
    \centering
    \includegraphics[width=\linewidth]{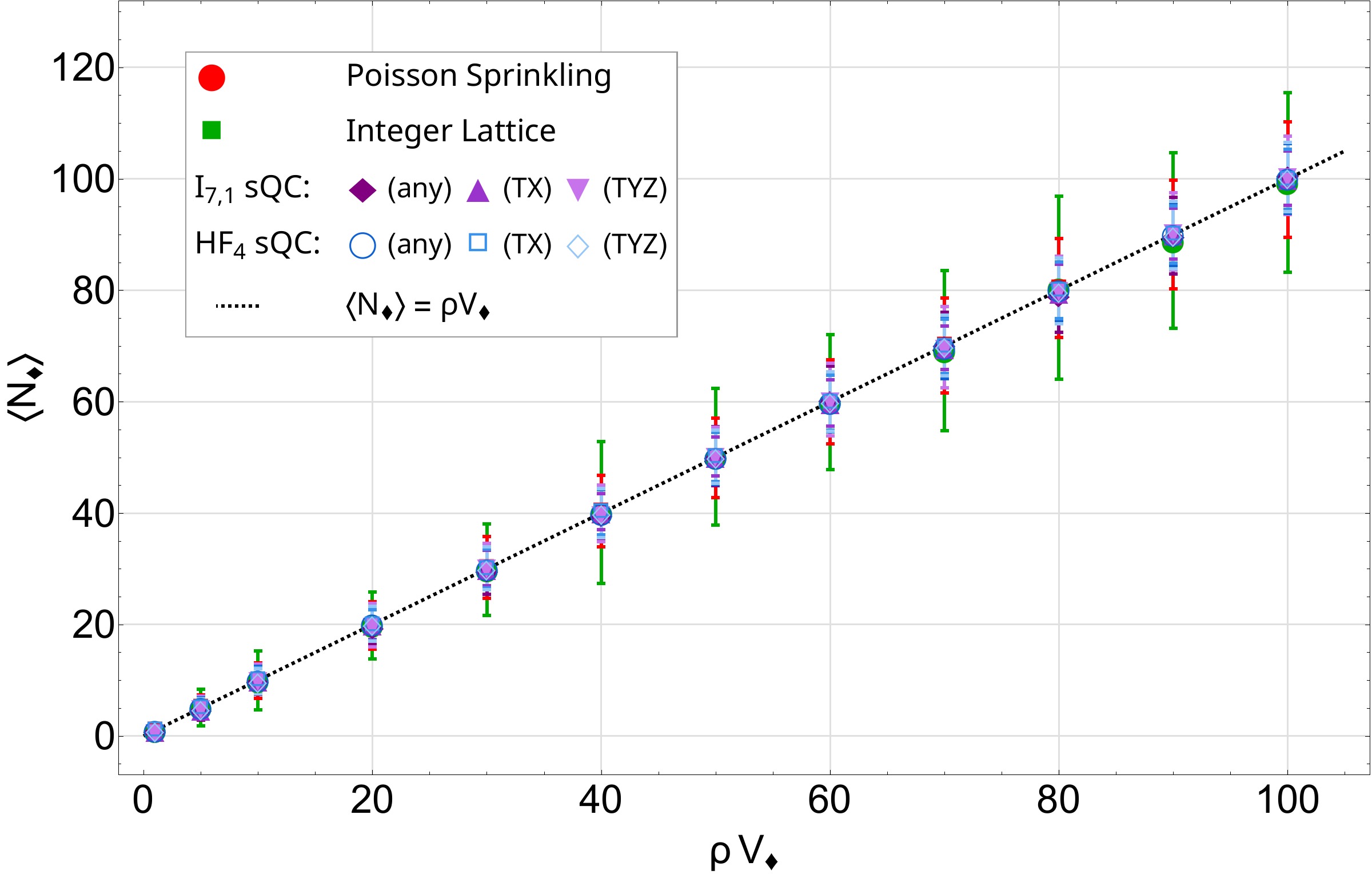}
    \caption[The N-V test for five $(3+1)$-dimensional point sets: the mean behaviour.]{The N-V test in $(3+1)$D, where we plot the average number of points $N_{\cdiamond}$ inside each $\cdiamond$ against the physical volumes $\rho_{\rm pattern} V_{\cdiamond} = V_{\rm sample}$; the error-bars are given by $\sigma_{N_{\cdiamond}}$. For both $\mathrm{I}_{7,1}$ and $HF_4$ sQCs, we consider three cases: (1) generic boosts (in deep purple and blue), (2) TX boosts (in medium purple and blue), and (3) TYZ boosts (in light purple and blue). In particular, the error bars for the last case are too small and are covered completely by the markers.}
    \label{fig:NVTest1Plus3DMean}
\end{figure}

\begin{figure}[t!]
    \centering
    \includegraphics[width=0.95\linewidth]{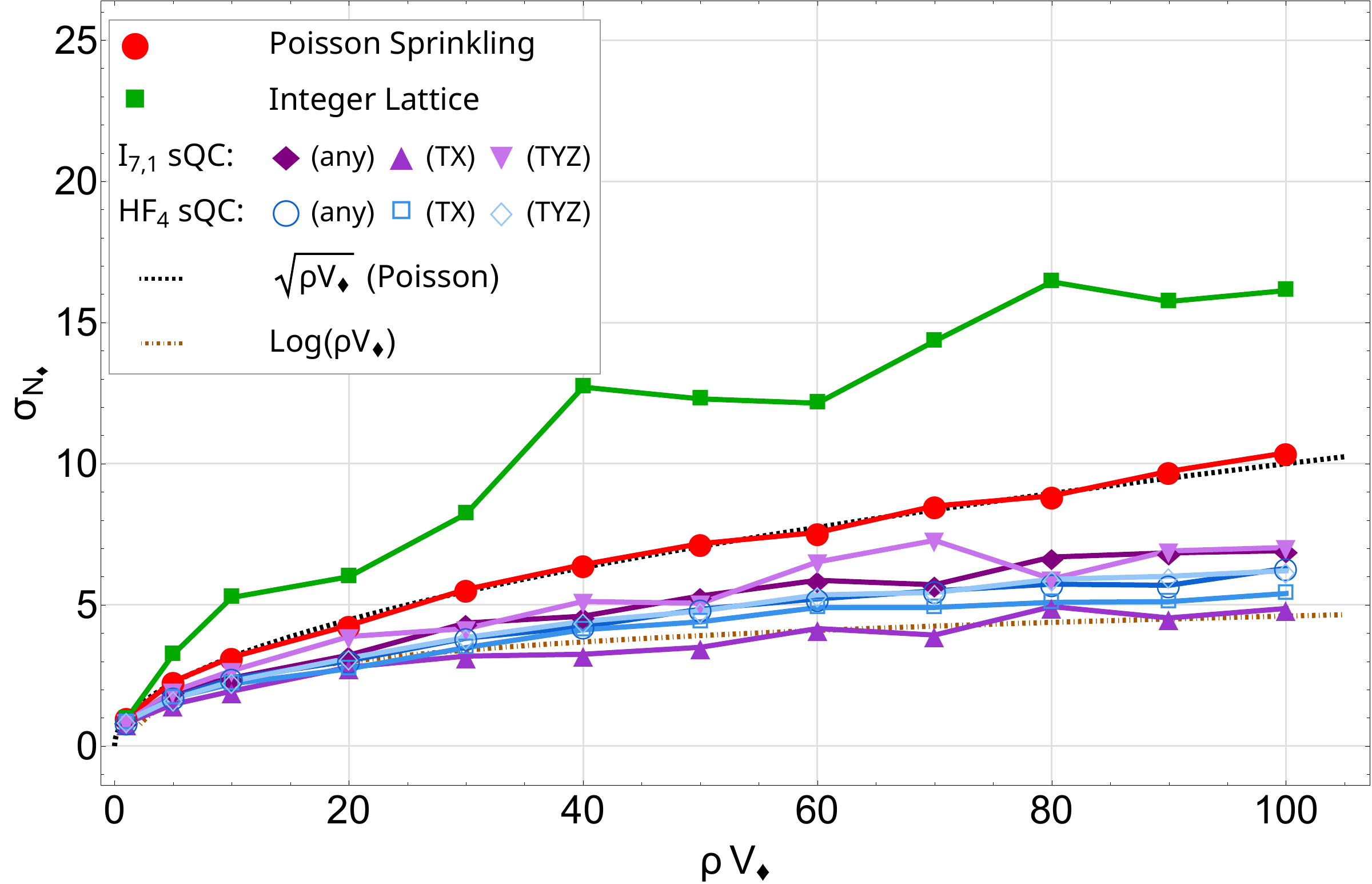}
    \caption[The N-V test for five $(3+1)$-dimensional point sets: the standard deviation.]{The N-V Test for five $(3+1)$-dimensional point processes, where we plot the standard deviation $\sigma_{N_\cdiamond}$ against the physical volumes $\rho_{\rm pattern} V_\cdiamond = V_{\rm sample}$. The Poisson sprinkling (red) tracks the dashed square-root curve and the integer lattice is consistently above it (super-Poissonian). Both spacetime quasicrystals appear sub-Poissonian, and the $HF_4$ sQC has, on average, less fluctuations than the $\mathrm{I}_{7,1}$ sQC.}
\label{fig:NVTest1Plus3DStandardDeviation}
\end{figure}

Running the N-V test for these patterns, we obtain the results of Fig.~\ref{fig:NVTest1Plus3DMean} (see Appendix \ref{appendix:NVTestDetails} for more details). All patterns do realize the number-volume correspondence well and the integer lattice belongs to the super-Poissonian regime, as expected. The boost catastrophe of $\mathrm{I}_{3,1}$ is explored further in Appendix \ref{appendix:NVTestDetails} (see Fig.~\ref{fig:BoostCatastrophe1Plus3DInteger}). In Fig.~\ref{fig:NVTest1Plus3DStandardDeviation} we see the standard deviations. Interestingly, the $\mathrm{I}_{7,1}$ quasicrystal stays close to the Poissonian reference curve throughout the range of tested volumes and boosts, and so does the $HF_4$ quasicrystal. Thus, both structures appear to be sub-Poissonian. 

To investigate this behaviour of the spacetime quasicrystals a bit more, we perform two independent N-V tests with their results shown also in Figs.~\ref{fig:NVTest1Plus3DMean} and \ref{fig:NVTest1Plus3DStandardDeviation}. The setup is the same: we sample the same number of $\cdiamond$ of the same volumes $V_\cdiamond$ and with the same range in the boost parameter $\varpi$. The difference lies in the direction the boost is happening. For the general test -- shown in deep purple and blue in Fig.~\ref{fig:NVTest1Plus3DMean} for the $\mathrm{I}_{7,1}$ and $HF_4$ sQCs respectively -- we allow for a generic direction vector ${\bf \hat{n}} = (n_x,n_y,n_z) \in \mathbb{S}^2$, sampled uniformly over the sphere. For the other two tests -- shown in medium and light purple and blue in Fig.~\ref{fig:NVTest1Plus3DMean} -- the boost is constrained along either the TX-plane with ${\bf \hat{n}} = (1,0,0) \in \mathbb{S}^2$, or the YZ-plane with ${\bf \hat{n}} = (0,n_y,n_z) \in \mathbb{S}^2$ respectively. 

For a TX-boost, we find that both spacetime quasicrystals realizes the N-V correspondence better than the Poisson sprinkling: the fluctuations $\sigma_{N_{\cdiamond}}$ seem to exhibit only a slow logarithmic growth, remaining consistently below the Poisson reference for all tested volumes and boosts. We attribute this to the fact that both structures have a genuine discrete Lorentz-boost symmetry in their $V_0$ Coxeter plane. On the other hand, for a TYZ-boost, the spacetime quasicrystals still remain sub-Poissonian but their fluctuations are slightly larger than in the TX-case. On this plane, the $\mathrm{I}_{7,1}$ has no boost symmetry, while the $HF_4$ sQC \textit{has}; this is why its $TYZ$-curve in Fig.~\ref{fig:NVTest1Plus3DStandardDeviation} appears slightly lower than that of the other quasicrystal. On average, when we sample generic boosts, it happens that the ``good'' and the ``bad'' directions average out, producing a sub-Poissonian behaviour. In other words, if more ``good'' directions are sampled, $\sigma_{N_\cdiamond}$ drops to its lower end and the structure beats Poisson in the N-V test. 

As explained in Appendix \ref{subsec:TestedPatched1Plus3D}, for the $\mathrm{I}_{7,1}$ sQC we used only 200 samples to produce these results (because of the increased computational cost), while for the $HF_4$ sQC we used 1000 samples (but with a much smaller patch). With only 200 samples, we cannot guarantee that the ``good'' and the ``bad'' directions are sampled equally for the $\mathrm{I}_{7,1}$ sQC. Additionally, since the size of the patch effectively constrains the maximum boost of $\cdiamond$, the boosts performed for the N-V test of the $HF_4$ sQC are lower, on average, than those of the $\mathrm{I}_{7,1}$ sQC (see Tables in Appendix \ref{subsec:TestedPatched1Plus3D}). The boosts do, however, cover at least 4-5 ``periods'' of the Coxeter boost $\varpi_{\rm sQC \, I_{7,1}} \approx 0.3856$ and 3-4 ``periods'' of $\varpi_{\rm sQC \, HF_4} \approx 0.6149$. To be able to definitively say that both spacetime quasicrystals actually do beat Poisson in $(3+1)$D, we would need to increase the number of samples for the $\mathrm{I}_{7,1}$ to at least 1000 and generate a larger for the $HF_4$ sQC to sample also at least 1000 diamonds off of it. Both of these tasks are computationally expensive, though they will be addressed thoroughly in a subsequent article about these findings.

Nevertheless, we view our results as a partial confirmation of the Poisson's Demise Conjecture \ref{conj:PoissonDemise}, and, by extension, as \textit{evidence against} the conjecture from \cite{Saravani_2014} -- that Poisson is invincible in higher than $(1+1)$D. It is interesting to notice that the generic-boost curve for the $HF_4$ sQC remains always below that of the $\mathrm{I}_{7,1}$ sQC. We believe that this happens because of the higher symmetries the $HF_4$ sQC has (rank-4 compared to rank-2). Had we been able to perform more exhaustive tests on the $\mathrm{I}_{7,1}$ sQC, we expect that eventually it would 
become super-Poissonian, as the number of its ``good'' directions is too low to save it (in fact, there is only one such direction, the TX-plane, along which it shines according to Fig.~\ref{fig:NVTest1Plus3DStandardDeviation}). On the other hand, the $HF_4$ sQC has far more ``good'' directions (at least icosahedrally many) and we would expect its symmetries to save it from ever becoming super-Poissonian. These preliminary results might not be a definitive answer to whether any other structure can outperform Poisson in $(3+1)$D, but definitely challenges this consensus and makes the waiting time for better tests even more exciting.

\section{Spacetime Hyperuniformity}
\label{subsec:SpacetimeHyperuniformity}

Let's try to understand the results of the previous section from a more fundamental point of view. In particular, we would like to know from first principles \textit{why} we see the following progression of fluctuations in the N-V test (focusing on the $(1+1)$D case for concreteness)
$$ \underbrace{\left(\sigma_{N_\cdiamond}\right)_{\textrm{Integer}}}_{\mathclap{\textrm{\bf super-Poissonian}}} 
>
\overbrace{\left(\sigma_{N_\cdiamond}\right)_{\textrm{Poisson}}}^{\mathclap{= \sqrt{\rho V_\cdiamond} \textbf{ : reference} } }
> 
\underbrace{ \left(\sigma_{N_\cdiamond}\right)_{\textrm{sQC}}
> \left(\sigma_{N_\cdiamond}\right)_{\textrm{Fibonacci}}}_{\mathclap{\textrm{\bf sub-Poissonian}}}.
$$
Since the number variance $\mathrm{Var}(N_\cdiamond)$ is the square of the standard deviation $\sigma_{N_\cdiamond}$, and using Eqs.~\ref{eq:PoissonStatistics} for Poisson statistics, the above becomes
$$ \underbrace{\left(\mathrm{Var}(N_\cdiamond)\right)_{\textrm{Integer}}}_{\mathclap{\textrm{\bf super-Poissonian}}} 
>
\left(\rho V_\cdiamond \right)_{\textrm{Poisson}}
> 
\underbrace{ \left(\mathrm{Var}(N_\cdiamond)\right)_{\textrm{sQC}}
>
\left(\mathrm{Var}(N_\cdiamond)\right)_{\textrm{Fibonacci}}}_{\mathclap{\textrm{\bf sub-Poissonian -- hyperuniform!}}}
.$$
In condensed-matter physics, structures whose number variance grows slower than that of Poisson-random structures (i.e. slower than the volume of some observation window, which for our purposes is given by $\cdiamond$) are called \textbf{hyperuniform} \cite{Torquato_2018}. This gives us an entry point in understanding more formally the results of the N-V test, together with an opportunity to extend known techniques from condensed matter (and statistical mechanics in general) to the realm of spacetime structures. 

\subsection{Direct-Space Variance, Euclidean Spatial Statistics and the First Definition of Hyperuniformity}
\label{subsec:EuclideanVariance}

First, we need to understand to a certain extent what hyperuniformity is in the Euclidean context, which is where the concept was originally developed. The main reference here is the review \cite{Torquato_2018} and the seminal work \cite{Torquato_2003}, to which we point the reader for details; we also found \href{https://mitia.duerinckx.web.ulb.be/workshop%2720-Torquato-slides.pdf}{this presentation material} by Torquato useful in writing this section.  

\textit{Hyperuniformity} is a property that led to a new classification and structural characterization of matter, including ordered structures -- crystals, quasicrystals -- but also more exotic disordered point configurations \cite{Torquato_2003}. States of matter with this property are \textit{correlated systems} characterized by strong \textit{suppression of large-scale (i.e.~long-wavelength) density fluctuations}, which does not occur for usual disordered systems such as ordinary fluids (e.g. ideal gases, liquids) and amorphous solids (e.g. glasses). The origins of this behaviour can be traced back to how the variance grows with the geometric properties of some observational patch.

Imagine a point pattern $P$ in $d$-dimensional Euclidean space $\RR^d$; the points could represent atoms of a physical material, points of a mathematical structure, like a periodic lattice or an aperiodic quasicrystal, or stars \textit{in some galaxy far, far away.} Let $\Omega$ be some observation window over the pattern, i.e.~we probe $P$ by placing a patch $\Omega$ on it and looking only at those points inside $\Omega$. The shape of $\Omega$ could be any convex set, and if the point process has certain symmetries, $\Omega$ can be chosen to be adapted to those. A convenient choice, though, especially for statistically isotropic point sets in Euclidean space, is a (hyper-)spherical window whose geometric parameter is only its radius $R$ \cite{Torquato_2003,Torquato_2018}. By sampling many such patches $\Omega$ over $P$ and collecting the statistics, one can calculate the average number of points inside $\Omega$, $\langle N_\Omega (R) \rangle$, and the associated local number variance $\sigma_\Omega^2 (R) = \langle N_\Omega^2(R) \rangle - \langle N_\Omega(R) \rangle^2$ as functions of $R$.\footnote{In \cite{Torquato_2003,Torquato_2018}, they use an \textit{ensemble-average} formulation in which they fix some $\Omega$ and generate many realizations of $P$; then, take the average over the full ensemble. We comment on how this relates to our formalism in Section \ref{subsec:NVTestIsHyperuniformityTest}.} We distinguish the following cases:
\begin{itemize}
    \item \textbf{Random Poisson Structures:} it follows from Poisson statistics (Eq.~\ref{eq:PoissonStatistics}) that $$\sigma_{\Omega}^2(R) = \langle N_\Omega (R) \rangle = \rho \mathrm{vol}(\Omega) \propto R^d$$
    and the variance grows with the volume of $\Omega$ ($\rho$ is the number density of the pattern $P$).~This is the standard scaling for typical disordered systems, like liquids and structural glasses.~The fact that $\langle N_\Omega \rangle$ is given by the volumes independently of where we sample $\Omega$ (i.e. of where its centroid is on $P$) is a consequence of statistical homogeneity, i.e. uniformity of the density over $P$. 
    \item \textbf{Hyperuniform Structures:} a hyperuniform system is one in which the number variance grows slower than $\mathrm{vol}(\Omega)$, that is, $ \sigma_{\Omega}^2(R) < R^d$ in $d$-dimensions. Three different asymptotic scalings for $\sigma_\Omega^2$ have been found \cite{Zachary_2009}, which place hyperuniform systems into three distinct classes (the slower the growth, the more hyperuniform, so we get the sequence I>II>III): 
    \begin{equation}
    \sigma_{\Omega}^2(R) \propto 
    \left\{ 
    \begin{aligned}
        & R^{d-1}, & \quad \textbf{(Class I)}\\
        & R^{d-1} \log R, & \quad \textbf{(Class II)}\\
        &  R^{d-\alpha}, (0<\alpha<1) & \quad \textbf{(Class III)}
    \end{aligned}
    \right. \qquad \text{ (large-$R$ limit: $R \to \infty$)}
    \label{eq:HyperuniformVarianceScalings}
    \end{equation}
    All crystals \cite{Zachary_2009}, some quasicrystals \cite{Oguz_2017} and certain exotic disordered states of matter (``stealthy disordered systems'' \cite{Torquato_2003,Torquato_2018}) belong to Class I, which is shown to be the minimal scaling exponent the variance can exhibit (see \cite{Zachary_2009} and references therein). Class II-hyperuniform structures show a slow logarithmic growth modifying the Class I-behaviour. This class includes several interesting objects: certain quasicrystals \cite{Oguz_2017}, zeros of the Riemann function and eigenvalues of random matrices \cite{Torquato_2018}, the Harrison-Zeldovich scale-invariant power-spectrum of density fluctuations in the early universe \cite{Harrison_1970_Powerspectrum,zeldovich1972hypothesis} and more (see \cite{Torquato_2018} for references).
\end{itemize}

For a generic point configuration with a well-defined average number density $\rho$ (i.e.~uniform structure), the variance $\sigma_\Omega^2(R)$ is shown to be specified entirely by \textit{pair correlations} and $\Omega$-related geometric factors \cite{Torquato_2003}; specifically, we obtain
\begin{equation}
    \sigma_\Omega^2(R) = \langle N_{\Omega}(R) \rangle \left(1 + \rho \int_{\RR^d} h({\bf r}) \alpha_{\Omega}({\bf r},R) {\rm d}^d{\bf r} \right), \qquad \langle N_\Omega(R) \rangle 
    = \rho \mathrm{vol}(\Omega),  
    \label{eq:VarianceGenericPointSet}
\end{equation}
Here, $\alpha_{\Omega}({\bf r},R)$ is a non-negative geometric quantity -- known as the \textit{scaled intersection volume }-- that describes the intersection of two spherical windows $
\Omega$ of radius $R$ (at a fixed orientation) with their centroids separated at a distance $r = |{\bf r}|$ (normalized by $\mathrm{vol}(\Omega)$). The second factor, $h({\bf r})$, is the \textit{total correlation function}, defined as 
\begin{equation}
    h({\bf r}) = g_2(\mathbf{r}) - 1, \quad {\bf r} \in \RR^d 
    \label{eq:TotalCorrelationFunction}
\end{equation}
where $g_2({\bf r})$ is the \textit{pair correlation function}. For \textit{statistically homogeneous} point processes, the argument $\mathbf{r}$ refers to relative displacements between the points instead of their absolute positions.~If the pattern is also \textit{statistically isotropic} (rotationally invariant), one can integrate out the angular part and the argument becomes just the radial part $r = |\bf r|$. Then, one talks about the \textit{radial distribution function} $g_2(r)$.  

Mathematically, $\rho g_2({\bf r}) {\rm d}^d {\bf r}$ measures the conditional probability of finding a point (``a neighbour'') within a volume ${\rm d}^d {\bf r}$ around $\bf r$, given there is one at the origin $\mathbf{r}$. This volume is usually taken to be (hyper)spherical, but one could consider some angle-dependent volumes as well. Intuitively, $g_2(\mathbf{r})$ answers the question `is there a probability of finding a neighbour at exactly distance $\mathbf{r}$ away?'. For completely \textit{uncorrelated point sets}, like random Poisson structures without any long-range order, $g_2({\bf r}) = 1$ for all arguments and $h({\bf r})$ vanishes identically \cite{Torquato_2003}.\footnote{Formally, this holds exactly only as $r = |{\bf r}| \to \infty$, as in small scales one could still be finding deviations of $g_2({\bf r})$ from unity that lead to a non-zero $h({\bf r})$, which average out once we consider larger observational windows.} The explanation here is that for a random structure there is always a probability that one will find a neighbour at any scale, as any distance can appear on a random pattern. On the other hand, for an ordered structure, \textit{only certain distances can appear}; for example, for a lattice point configuration, the nearest neighbour will be a lattice-spacing away, so $g_2(\mathbf{r})$ vanishes for distances lower than that, but picks up a sharp peak at exactly the lattice-spacing (and, in fact, for any higher-than-nearest neighbour; see Fig.~\ref{fig:RipleysFunction}). Based on this, we note the following cases; the interpretations are to be taken against a random structure:
\begin{itemize}
    \item \(g(r) = 1\): equal probability of finding a neighbour at \textit{any} distance \(r\) (\textit{randomness}),
    \item \(g(r) > 1\): higher probability of finding a neighbour at distance \(r\); pairs of points at distance $r$ are \textit{more} abundant than the spatial average (\textit{clustering}).
    \item \(g(r) < 1\): lower probability of finding a neighbour at distance \(r\); pairs of points at distance $r$ are \textit{less} abundant than the spatial average (\textit{dispersion/exclusion})
\end{itemize}
Adapting from \cite{law2009ecological}, ``the function $g(r)$ is a normalized
measure of how, on average, a point perceives the density of
other points as it `looks out' over increasing distances $r$ into
the point configuration it is living''.
Therefore, deviations of $g_2(\bf r)$ from unity -- or, equivalently, $h({\bf r})$ from zero -- is an indication for the existence of spatial correlations on the pattern.

While $g_2({r})$ captures spatial correlations \textit{specifically at a given distance} ${r}$ away from a certain origin-point, there are two other measures of spatial statistics related to $g_2(r)$, whose role it to \textit{cumulatively consider correlations across all scales} up to some distance $\mathbf{r}$. First, we have the \textit{cumulative coordination number} $ Z(r)$, which gives us the expected number of points found in a sphere of radius $r$ centered at
an arbitrary point of the point process \cite{Torquato_2018}. This is a monotonically increasing function of $r$ defined as
\begin{equation}
    Z(r) = \rho s_d \int_{0}^r x^{d-1} g_2(x) {\rm d}x
    \label{eq:CumulativeCoordNumber}
\end{equation}
where $s_d = d \pi^{d/2}/\Gamma(1+d/2)$ is the surface-area of the $d$-dimensional unit sphere. For a totally random point configuration with $g_2(r) = 1$ everywhere, the integration simply yields $Z(r) = \rho v_d(r) \approx N(r)$, where $v_d(r) = \pi^{d/2} r^d/\Gamma(1+d/2)$ is the volume of the $d$-sphere. For an ordered structure, $g_2(r)$ consists of sharp peaks (i.e.~is a sum of delta functions) at shell distances $r_1 < r_2 < \cdots$, so $Z(r)$ is flat between shells and `jumps' discretely whenever it hits a shell, producing a ladder-like behaviour. In fact, we can quantify the height of the `step': the jump at shell $r_k$ is the \textit{shell coordination number} $z_k$ — the average number of neighbours at exactly that distance from a typical point (see Fig.~\ref{fig:RipleysFunction}), 
  $$\Delta Z(r_k) = \rho \quad \times \underbrace{ \left(\text{area under the }g(r)\text{ peak at }r_k\right)}_{\sim \text{(height}\times\text{d}r) \text{, so
  d}r\text{-independent}; \text{ height } \sim 1/{\rm d} r, \text{ unphysical}  } \times \quad  s_d(r_k) $$
A second -- very related -- measure is \textit{Ripley's K-function} $ K(r)$\footnote{\textit{Believe it or not (!)}, this is due to the British statistician Brian D. Ripley (1952--) and \textit{not} due to the American anthropologist and cartoonist LeRoy Robert Ripley (1890--1949).} \cite{dixon2001ripley,law2009ecological,stoyan1993estimation,stoyan1994fractals}, which is defined as the density-normalized $Z(r)$, 
\begin{equation}
    K(r) = s_d \int_0^r g_2(x)  x^{d-1} {\rm d} x = \frac{Z(r)}{\rho}
    \label{eq:RipleysKFunction}
\end{equation}
and it always has units of volume, as we are accumulating the surface area swept over certain radial distance $r$ to get the enclosed volume. Eq.~\eqref{eq:RipleysKFunction} offers itself for another definition of $g_2(r)$, 
\begin{equation}
    g_2(r) = \frac{1}{v_d(r)}\frac{{\rm d} K(r)}{{\rm d} r} 
\end{equation}
Intuitively, $K(r)$ counts the ``effective volume'' one would need to assign to a ball of radius $r$ to reproduce the observed mean neighbour-count $Z(r)$ at the background density $\rho$, but as a measure per se, it does not depend on the density. For a random distribution, we get $K(r) = v_d(r)$. The deviation $K(r) - v_d(r)$ tells us directly whether the structure packs more neighbours than Poisson ($>0$, \textit{clustering}) or fewer ($<0$, \textit{dispersion}). 

\subsection{Fourier-Space Variance, Spectral Functions and the Second Definition of Hyperuniformity}
\label{subsec:FourierVariance}

The best way to see how the variance is related to the suppression of fluctuations is to consider the Fourier representation\footnote{\label{foot:NewFourierDefs}
Following a slightly different notation than our previous Eq.~\eqref{eq:FourierTransformDef}, where  now the Fourier transform is $\widehat{f}({\bf k}) = \int f({\bf r}) \exp{[-(i {\bf k \cdot r})]}{\rm d}{\bf r}$ and its inverse is $ f({\bf r}) = (1/2\pi)^d \int \widehat{f}({\bf k}) \exp{[(i {\bf k \cdot r})]}{\rm d}{\bf k}$ to match \cite{Torquato_2018}.} of $\sigma_\Omega^2(R)$
\begin{equation}
    \sigma_\Omega^2(R) = \langle N_\Omega(R) \rangle \left( \frac{1}{(2\pi)^d} \int_{\RR^d} S({\bf k}) \widehat{\alpha}_{\Omega}({\bf k},R) {\rm d}^d {\bf k}\right).
\label{eq:VarianceGenericPointSetFourierRep}
\end{equation}
Here, $\widehat{\alpha}_{\Omega}({\bf k},R)$ is the Fourier transform of the scaled intersection volume and is always non-negative. $S({\bf k})$ is a non-negative spectral function called the \textbf{structure factor} (or \textit{power-spectrum)}, and is defined as 
\begin{equation}
S({\bf k}) = 1 + \rho \widehat{h}({\bf k})
\label{eq:StructureFactor}
\end{equation}
where $\widehat{h}({\bf k})$ is the Fourier transform of Eq.~\eqref{eq:TotalCorrelationFunction}. It provides a mathematical description of how incident radiation scatters off the point pattern, as well as a measure of which wavelengths can be excited for the material under question, and is related to the diffraction measure \cite{baake2013aperiodic,senechal1996quasicrystals}. It follows immediately from our previous discussion that $S({\bf k}) = 1, \,\, \forall {\bf k }$ if the pattern is random/disordered, while it is a non-trivial function in the presence of correlations for an ordered configuration. 

Taking the large-volume ($R \to \infty$), or infinite-wavelength, limit of Eq.~\eqref{eq:VarianceGenericPointSetFourierRep}, and using Eq.~\eqref{eq:StructureFactor} and the Fourier transform definitions (Footnote \ref{foot:NewFourierDefs}), we obtain \cite{Torquato_2003}
\begin{equation}
    \lim_{{\rm vol}(\Omega) \to \infty} \frac{\sigma_{\Omega}^2(R)}{\langle N_{\Omega}(R) \rangle} = \lim_{{\rm vol}(\Omega) \to \infty} \frac{\sigma^2_{\Omega}(R)}{{\rm vol}(\Omega)} = \lim_{\bf |k| \to 0} S ({\bf k}) = 1 + \rho \int_{\RR^d} h({\bf r}) {\rm d} {\bf r}.
    \label{eq:HyperuniformityCondition}
\end{equation}
This equation contains much information: first, it tell us that the hyperuniformity condition we discussed in the beginning of the section, namely that the variance grows more slowly than the window volume, makes the first limit of $\sigma_{\Omega}^2(R)$ equal to zero. Secondly, this translates into the vanishing of the second limit for the structure factor $S({\bf k})$, 
\begin{equation}
    \lim_{|\bf k| \to 0} S({\bf k}) = 0 
    \label{eq:StructureFactorGoesToZero}
\end{equation}
implying that a hyperuniform system is, \textit{equivalently}, one in which the (normalized) density fluctuations are completely suppressed at very large length-scales (very small wave-numbers). Finally, Eq.~\eqref{eq:HyperuniformityCondition} indicates the reason \textit{why} a point pattern is hyperuniform: the last term must vanish, yielding 
\begin{equation}
    \int_{\RR^d} h({\bf r}) {\rm d} {\bf r} \overset{\text{Eq.~\eqref{eq:TotalCorrelationFunction}}}{=} \int_{\RR^d} [g_2({\bf r}) -1 ] {\rm d} {\bf r} = -\frac{1}{\rho} < 0
    \label{eq:HyperuniformityNegativeCorrelations}
\end{equation}
This means that, \textbf{for a hyperuniform point process, $h({\bf r})$ must exhibit
negative correlations} (i.e., anti-correlations) for some values of $r$, which in turn implies that $g_2(r) < 1$ and thus the pattern exhibits dispersion instead of clustering.\footnote{We shall note that for a lattice or quasicrystal, $g_2(r)$ is a sum of delta functions and it is not true pointwise that $g_2(r) < 1$. In fact, $g_2(r) = 0$ almost everywhere (between shells), so it is trivially less than 1 there, but it diverges to infinity on the peaks at the shell distances. The weighted integral, however, in Eq.~\eqref{eq:VarianceGenericPointSet}, takes care of the infinities and leads to a negative contribution on average.}

Going back to Eq.~\eqref{eq:VarianceGenericPointSet}, we can now appreciate how the existence of spatial correlations in our point configuration is responsible for driving the local number variance to the sub-Poissonian regime. For a random point process, $h(\mathbf{r})$ vanishes and the second term contributes nothing to $\sigma_{\Omega}^2(R)$; what we get, then, is that the variance is given by the average number count $\langle N_{\Omega}(R) \rangle$, exactly as in Poisson statistics \eqref{eq:PoissonStatistics}. On the other hand, when a pattern has anti-correlations, $h({\bf r}) < 0$ and the second term brings negative contributions to the variance. This happens, for example, for our ordered structures, where we have $g(r) < 1$ at short range (anti-correlation/exclusion zone). As a result, this negative correction lowers the average $\langle N_{\Omega}(R) \rangle$ leading to a sub-Poissonian $\sigma_{\Omega}^2(R)$. Finally, if we have clustering in the pattern with $h({\bf r}) > 0$, then the positive correlations similarly drive the variance $\sigma_{\Omega}^2(R)$ to the super-Poissonian regime. 

Given that a hyperuniform structure is one in which Eq.~\eqref{eq:StructureFactorGoesToZero} holds, one could quantify how fast the structure factor $S({\bf k})$ goes to zero, when it approaches it continuously. To recover the variance scalings we mentioned when we discussed the three hyperuniformity classes in Eq.~\eqref{eq:HyperuniformVarianceScalings}, a simple power-law $S({\bf k}) \propto k^\alpha \,\,(\alpha > 0)$ suffices, as it can be verified from Eq.~\eqref{eq:VarianceGenericPointSetFourierRep}; we have
\begin{equation}
    S({\bf k}) \propto k^\alpha \, : \,
    \left\{ 
    \begin{aligned}
        & \alpha > 1, & \quad \textbf{(Class I)}\\
        & \alpha = 1, & \quad \quad \textbf{(Class II)}\\
        & 0 < \alpha < 1, & \quad \textbf{(Class III)}\\
    \end{aligned}
    \right. \qquad \text{(small-$k$ limit; $|{\bf k}| \to 0$)}
    \label{eq:StructureFactorPowerLaw}
\end{equation}
The limiting cases $\alpha = 0$ and $\alpha = \infty$ correspond to a random Poisson structure and a crystalline lattice respectively \cite{Torquato_2018}. Interestingly, our universe's matter power-spectrum is in Class II, if we take the Harisson-Zeldovich, exactly scale-invariant power-spectrum \cite{zeldovich1972hypothesis,Harrison_1970_Powerspectrum}, or in Class III if we consider its spectral tilt. 

For a quasicrystal, however, $S({\bf k})$ consists of \textit{a dense set of Bragg peaks} separated by gaps of arbitrarily small size \cite{levine1986quasicrystals,senechal1996quasicrystals}. This makes $S(\bf k)$ highly discontinuous with dense support, and Eq.~\eqref{eq:StructureFactorPowerLaw} gets difficult to interpret. For this reason, in \cite{Oguz_2017} a new criterion to quantify hyperuniformity for quasicrystals was established. Instead of looking at the behaviour of $S(\bf{k})$ as $|{\bf k}| \to 0$, one turns their attention to the \textit{cumulative intensity function} $Z(k)$ (or \textit{integrated spectral density}), defined similarly to Eq.~\eqref{eq:RipleysKFunction} by
\begin{equation}
    Z(k) = \int_{0}^k S({\bf q}) q^{d-1} s_q {\rm d} q.
    \label{eq:CumulativeIntensityFunction}
\end{equation}
with the role of $g_2(r)$ played by $S(k)$ now. Since 
$$ S({k}) = \frac{1}{s_d(k)} \frac{{\rm d} Z(k)}{{\rm d} k}$$
and the surface-term $s_d(k)$ is killed by the integration in Eq.~\eqref{eq:VarianceGenericPointSetFourierRep}, we get (after integration by parts and handling the boundary term)
\begin{equation}
    \sigma_{\Omega}^2(R) = - \langle N_{\Omega}(R) \rangle \left( \frac{1}{(2\pi)^d} \int_{0}^{\infty} Z(k) \frac{\partial \widehat{\alpha_{\Omega}}(k,R)}{\partial k} {\rm d} k\right)
    \label{eq:VarianceFourierRepForQuasicrystals}
\end{equation}
$Z(k)$ is a smoother function than $S(k)$, it is monotonically increasing and, as such, can be used to extract the scaling exponent $\alpha$ in Eq.~\eqref{eq:StructureFactorPowerLaw}. The criterion for hyperuniformity developed in \cite{Oguz_2017} is that $Z(k)$ is bounded both above and below by functions of the form $c_\pm k^{1+\alpha}$ for small $k$
\begin{equation}
    c_- k^{\alpha+1} < Z(k) < c_+ k^{\alpha+1} \qquad \text{(as $k \to 0$)}.
    \label{eq:HyperuniformityQuasicrystals1}
\end{equation}
In other words, if the ladder $Z(k)$ is contained entirely within two lines of potentially different slope $c_{\pm}$ but of the same power-law behaviour $\propto k^{\alpha+1}$ as $k$ approaches $0$, we say
\begin{equation}
    Z(k) \propto k^{\alpha+1} \qquad \text{(as $k \to 0$)}.
    \label{eq:HyperuniformityQuasicrystals2}
\end{equation}
Hyperuniformity corresponds to $\alpha > 0$, and one can see that this definition agrees with Eq.~\eqref{eq:StructureFactorPowerLaw}, since $S(k) \sim \partial_k Z(k) \propto k^\alpha$. 

\subsection{Sub-Poissonian Variance via Non-Trivial Correlations}
\label{subsec:NonTrivialCorrelationsAndVariance}

Before we turn into the variance for our spacetime structures, let us first look into the cumulative spatial-statistics measures we just defined. In direct space, we have $Z(r)$ and $K(r)$, given by Eqs.~\eqref{eq:CumulativeCoordNumber} and \eqref{eq:RipleysKFunction} respectively. In Fourier space, we have $Z(k)$ given by Eq.~\eqref{eq:CumulativeIntensityFunction}, which is interesting on its own because it has already shown to give a criterion for hyperuniformity of quasicrystals, i.e.~Eq.~\eqref{eq:HyperuniformityQuasicrystals1}. The catch is that all those measures are heavily Euclidean, as they all deal with integration over spherical windows which do not make much sense in the Lorentzian case. Nevertheless, they will give some intuition about the connection of the variance to (spatial) correlations, so let us try to apply this Euclidean intuition into our Lorentzian point processes. 

Ignoring the subtleties that come from the indefiniteness of the metric, let us calculate the pair correlation function $g_2(r)$ by following exactly the same prescription as we would do in Euclidean space. This is to say, we first fix a patch of volume $V_{\rm patch}$ for each of the point configurations and estimate $g_2(r)$ as a \textit{centre-averaged shell count}. For each reference point ${\bf x}_i$ drawn from the interior of the patch, we count the number of neighbours $n_i(r)$ falling in the annular shell of radius $r$ and width $\dd r$; then, we average over all $N_{\mathrm{ref}}$ reference centres. Normalizing by the count expected for a uniform Poisson field of the same density $\rho$ -- namely $\rho s_d \dd r$ and $\rho (2\pi r) \dd r$ in two dimensions -- yields 
$$
    g_2(r) = \frac{1}{N_{\rm ref}} \sum_i \frac{n_i(r)}{\rho \, 2\pi r \, \dd r}
$$
so that $g_2(r)$ measures the observed shell occupancy relative to the uncorrelated expectation. Equivalently, letting every point serve in turn as a centre, this is a histogram of all pairwise separations within a disk of radius $r_{\max}$ binned by distance (from $r = 0$ to $r = r_{\max}$, with each bin having width $\dd r$). It is therefore a two-point statistic averaged over all points, not a property of any single point. The average over centres is essential: a single reference point yields only a noisy sample of its local environment, whereas averaging over many centres suppresses the shot noise as $N_{\mathrm{ref}}^{-1/2}$ without introducing bias, since the centres are drawn independently and uniformly. To avoid edge effects, reference centres are restricted to an interior guard zone lying at least $r_{\max}$ from the boundary of the patch, ensuring that every shell is fully contained within the patch, while all points remain admissible as neighbours.

\begin{figure}[p]
    \centering

    \begin{subfigure}{0.35\textwidth}
        \centering
        \includegraphics[width=\linewidth]{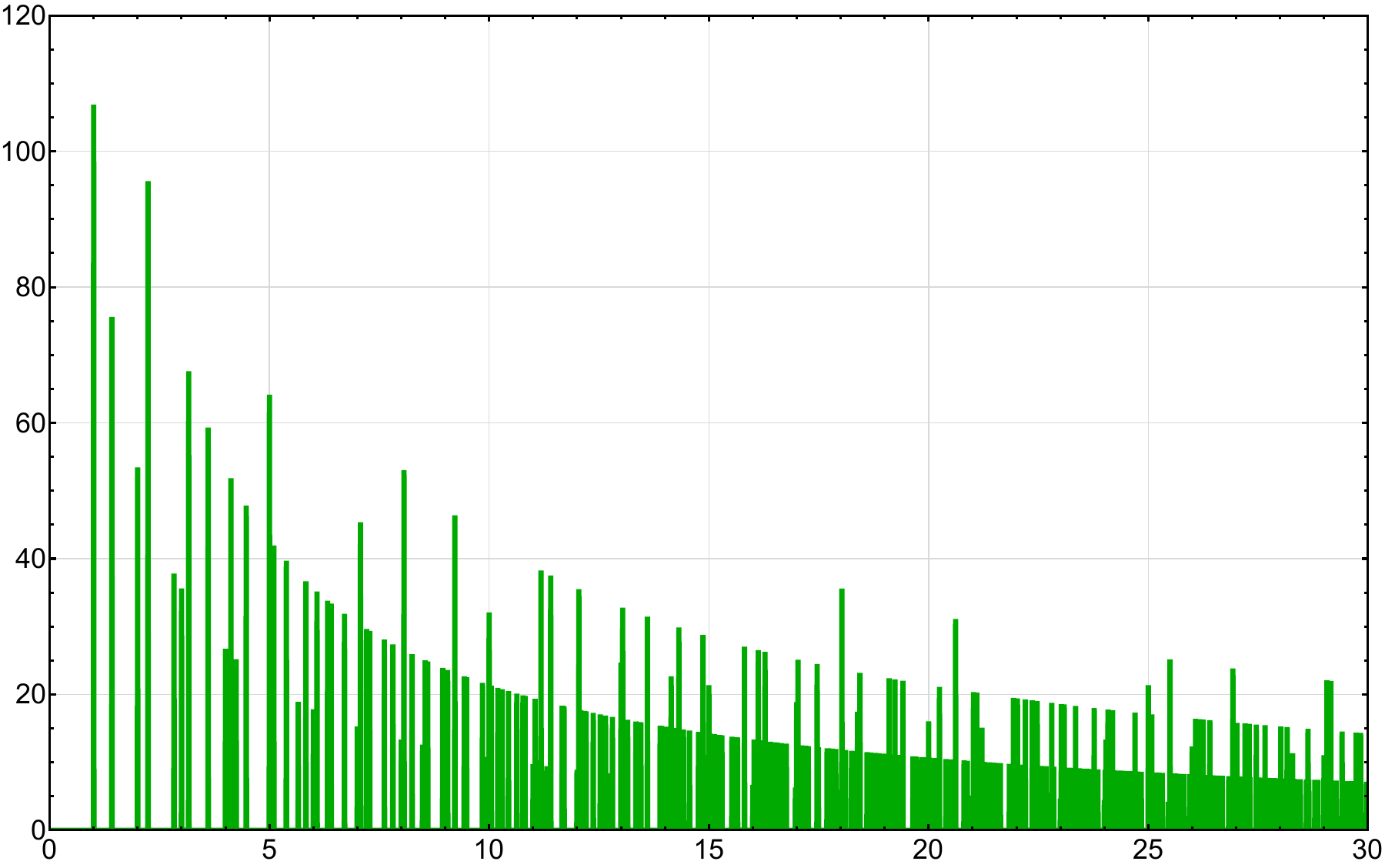}
        \caption{Integer Lattice}
        \label{fig:IntegerLatticePairCorrelations}
    \end{subfigure}
    \hfill
    \begin{subfigure}{0.35\textwidth}
        \centering
        \includegraphics[width=\linewidth]{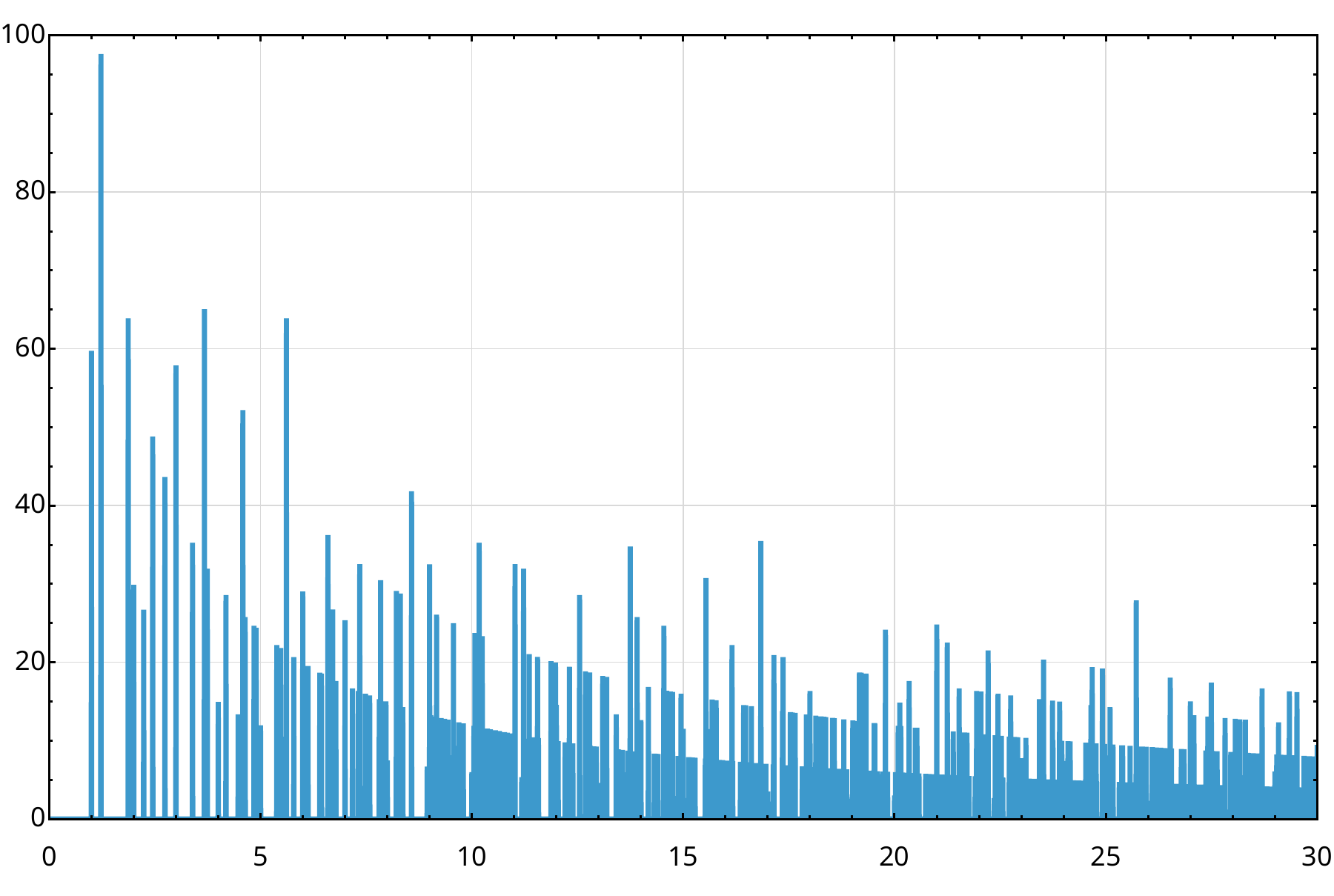}
        \caption{Fibonacci Lattice}
         \label{fig:FibonacciLatticePairCorrelations}
    \end{subfigure}
    \hfill
    \begin{subfigure}{0.35\textwidth}
        \centering
        \includegraphics[width=\linewidth]{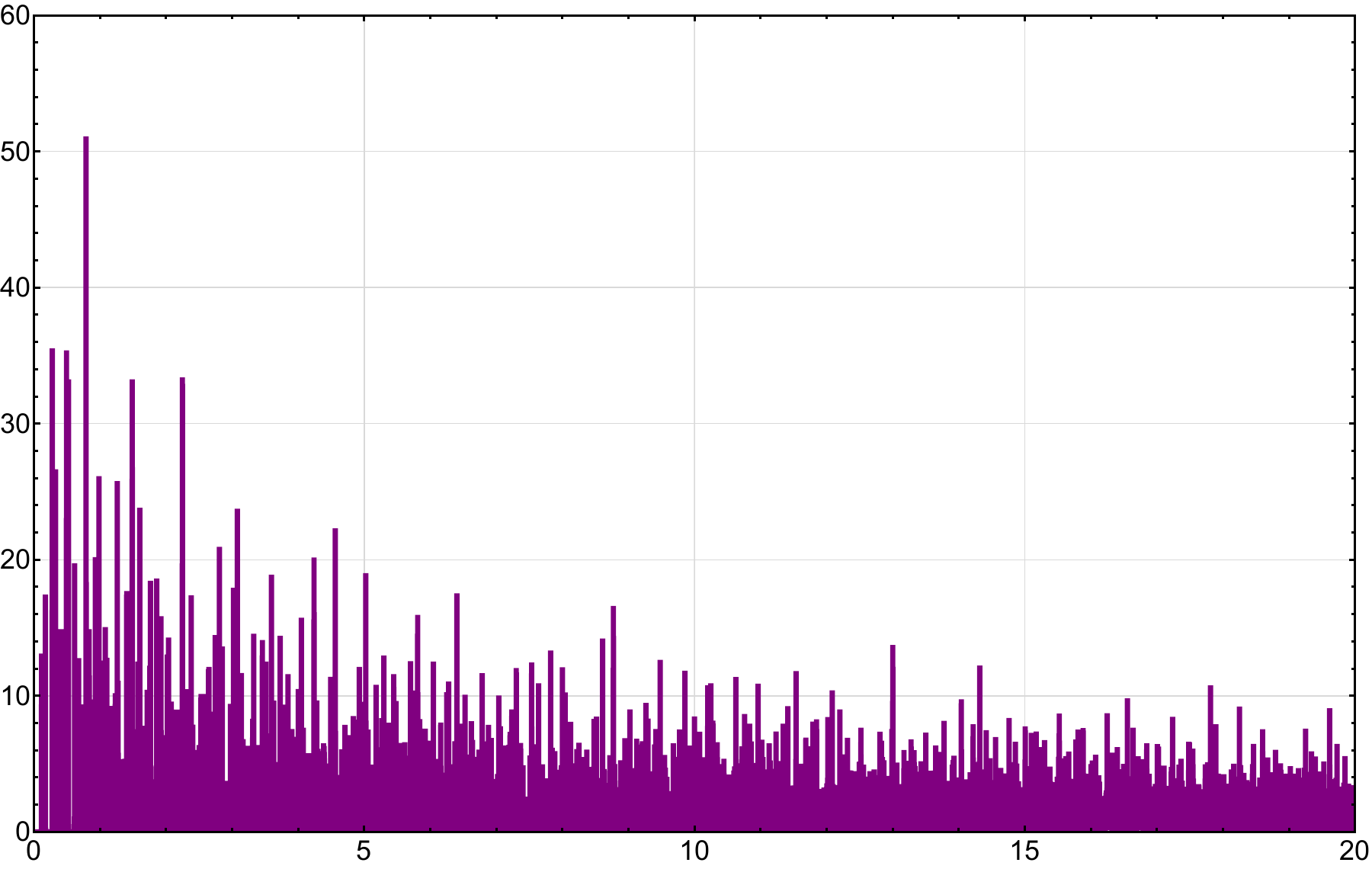}
        \caption{Spacetime Quasicrystal}
         \label{fig:SpacetimeQuasicrystalPairCorrelations}
    \end{subfigure}

    \vspace{0.5em}

    \begin{subfigure}{0.75\textwidth}
        \centering
        \includegraphics[width=\linewidth]{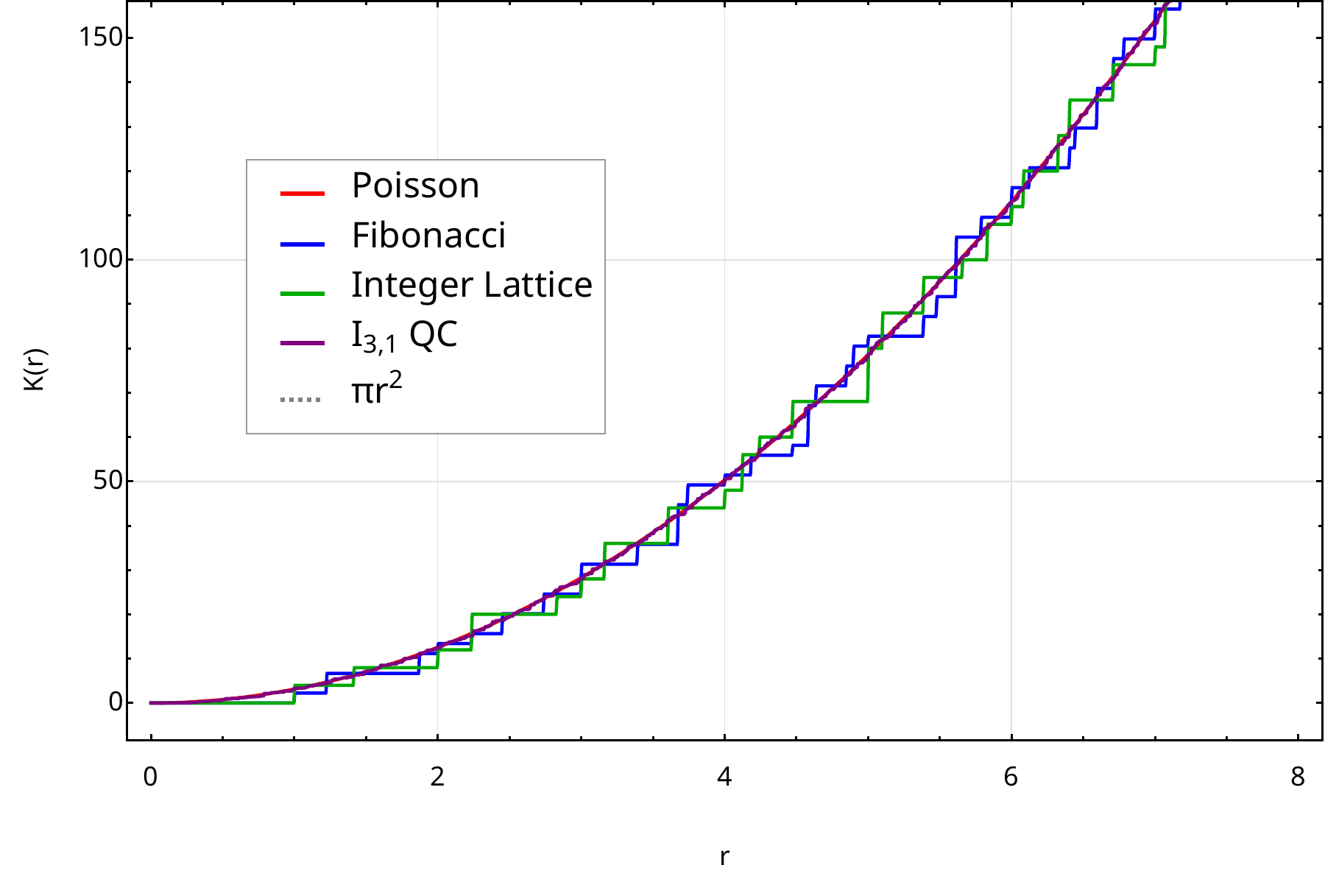}
        \caption{Ripley's Function $K(r)$ for all four point sets at zero boost.}
        \label{fig:RipleysLadder}
    \end{subfigure}

    \caption[Pair correlations $g_2(r)$ and their Ripley's functions $K(r)$ for the integer lattice, the Fibonacci lattice and the spacetime quasicrystal in $(1+1)$D.]{In the first row, we see the pair correlation function $g_2(r)$ (vertical axis) for three ordered structures; the horizontal axis is the Euclidean distance $r$. In the second row, their Ripley's functions $K(r)$ have been calculated.}
    \label{fig:RipleysFunction}
\end{figure}

In Figs.~\ref{fig:IntegerLatticePairCorrelations}, \ref{fig:FibonacciLatticePairCorrelations} and \ref{fig:SpacetimeQuasicrystalPairCorrelations} we see the resulting pair correlation functions $g_2(r)$ for our three ordered structures in $(1+1)$D: the integer lattice $\mathrm{I}_{1,1}$, the Fibonacci lattice and the $\mathrm{I}_{3,1}$-spacetime quasicrystal respectively. For all three structures, $g_2(r)$ is not a smooth curve but rather is comprised entirely of sharp peaks, i.e.~Dirac deltas, at certain \textit{Euclidean distances} $r$. \footnote{In fact, we can calculate them easily. For the integer lattice, the distances are given by $\sqrt{a^2 + b^2}$ for $a,b \in \ZZ$, whereas for the Fibonacci lattice we find $\sqrt{3 m^2/2 + m n + n^2}$ for $m,n \in \ZZ$. For both structures, the minimum Euclidean separation is 1, which is the same as the lattice spacing. For the quasicrystal, though, we do not have such an analytic expression.} This is because for an ordered point set, only a discrete set of pair-separations is possible. For all patterns, there exist a minimum separation below which the pair correlation function $g_2(r)$ vanishes as $r \to 0$; the same occurs in-between the different peaks as well as. The apparent amplitude of the peaks carries no physical information: when $g_2(r)$ is estimated as a histogram with finite bin width $\dd r$, each (infinite) delta peak is smeared across a single bin and acquires a spurious height of order $\sim \text{(weight)}/\dd r$ (which is why we did not show the axes for the delta peaks in Fig.~\ref{fig:RipleysFunction} properly). Its weight, on the other hand, is physically meaningful: it equals to the fraction of pairs at separation given by the distance $r$. The peak heights are therefore an artifact of the binning resolution and should not be interpreted quantitatively. Only the peak positions, which mark the discrete shell radii, and the peak weights, which can be recovered by integration, are robust. 
  
Cumulative measures, on the other hand, like Ripley's $K$-function \eqref{eq:RipleysKFunction} yield a bin-independent measure of the correlations, in which the integration sums the delta weights and cancels the $1/\dd r$ scaling. In Fig.~\ref{fig:RipleysFunction} we show the results of the integration for all four point processes. For the ordered structures, integrating the spiky $g_2(r)$ turns its sequence of delta weights into the monotonically increasing staircase we described earlier (see discussion at Eq.~\eqref{eq:CumulativeCoordNumber}). For the Poisson random causal set, $K(r) = \pi r^2$, and all three structures track this parabola closely, confirming the correct mean behaviour in the N-V test. On average each ordered structure lies marginally below $\pi r^2$ -- a sub-Poissonian signature of suppressed density fluctuations according to Eq.~\eqref{eq:HyperuniformityNegativeCorrelations} -- with the staircase occasionally jumping above the parabola at a strongly occupied shell before falling back. 

We see that the spacetime quasicrystal tracks the Poisson curve more tightly than the two lattices. There are two related reasons for this: one is because of its higher density ($\rho_{\rm sQC \, I_{3,1}} \approx 11.85 \gg \rho_{\rm Int.} = 1 > \rho_{\rm Fib.} \approx 0.89$), the other because of its richer spatial correlations. While for all three ordered structures only certain distances can appear, the quasicrystal has a more intricate structure and a richer set of possible distances between pairs of points. Now, if we also account for the higher density of the sQC, we get the denser set of Dirac peaks in Fig.~\ref{fig:SpacetimeQuasicrystalPairCorrelations}. Upon integration, $K(r)$ cannot escape much to the right of Fig.~\ref{fig:RipleysLadder} (to higher $r$-values) before hitting a peak and gets driven upwards, therefore always remaining very close to the Poisson parabola. 

\begin{figure}[t!]
    \centering
    \includegraphics[width=0.99\linewidth]{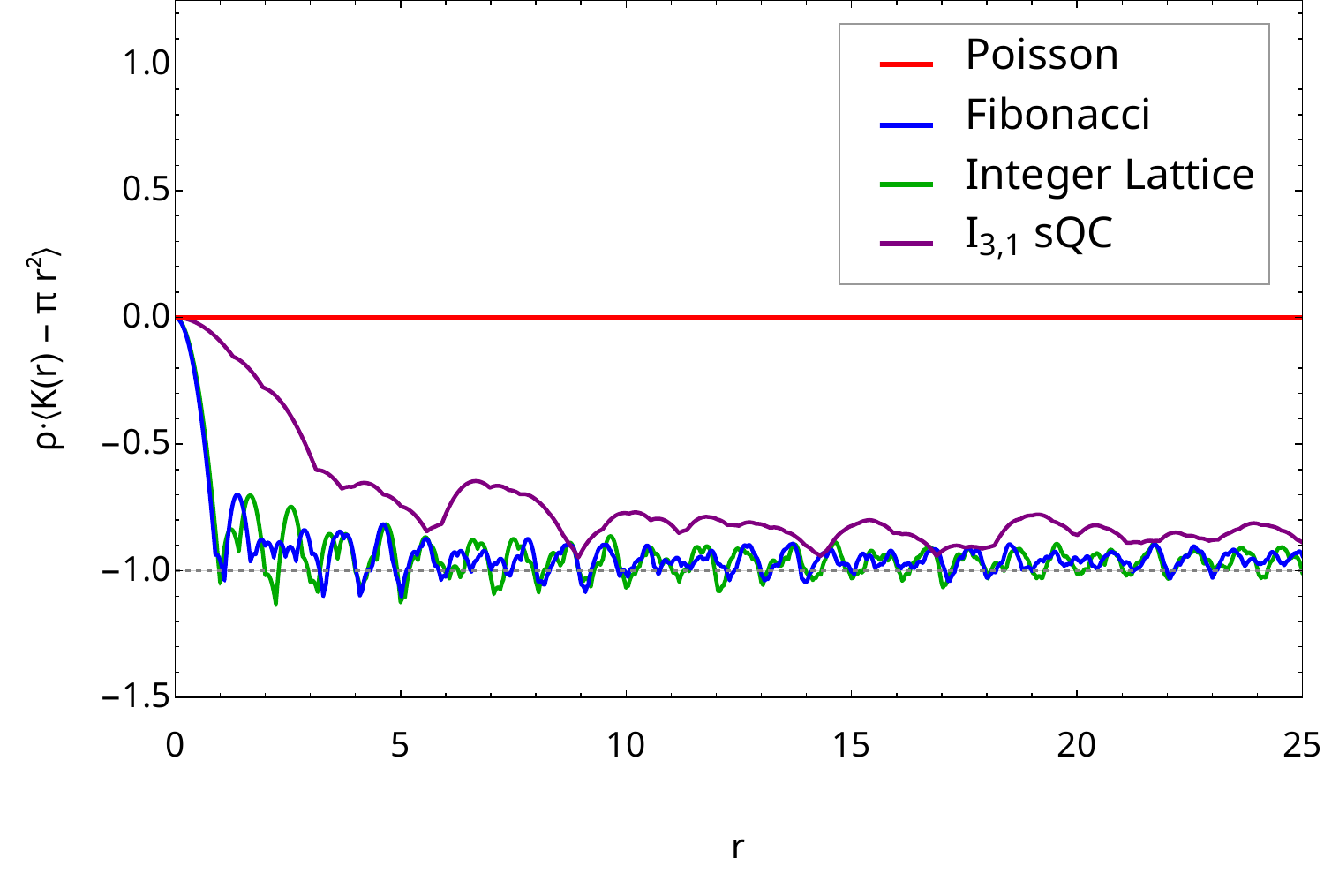}
    \caption[ Ripley's deficit $\rho_{\rm pattern} \langle K(r)-\pi r^2 \rangle $ for the four point sets of Fig.~\ref{fig:RipleysFunction}.]{Ripley's deficit $\rho_{\rm pattern} \langle K(r)-\pi r^2 \rangle $ for the four point sets of Fig.~\ref{fig:RipleysFunction}. All patterns eventually converge to $-1$, which is a signature of hyperuniformity (see Eq.~\eqref{eq:HyperuniformityNegativeCorrelations}).}
    \label{fig:RipleysDeficitHyperuniformity}
\end{figure}

To remove the dependence on the density $\rho_{\rm pattern}$, we show in Fig.~\ref{fig:RipleysDeficitHyperuniformity} the Ripley's deficit/excess $\rho_{\rm pattern}(K(r) - \pi r^2)$ for all four point sets.\footnote{In practice, this would look like a spiky sawtooth; for this reason, we first find the deficits $D(r) = K(r) - \pi r^2$ and then calculate the cumulative (running) mean $(1/r)\int_0^r D(s)\dd s$ that smoothens the sawtooth, producing the smooth curves in Fig.~\ref{fig:RipleysDeficitHyperuniformity} (after multiplied by the density of each pattern).} This is precisely the content of Eq.~\eqref{eq:HyperuniformityNegativeCorrelations}: a hyperuniform structure has a sub-Poissonian variance because of its negative correlations driving the integral $\rho \int (g_2({\bf r})-1)\dd {\bf r}$ to -1. We see that all three structures eventually converge to -1 (signature of hyperuniformity), but the spacetime quasicrystal shows a longer transient period. We believe this is because the dense, incommensurately-spaced peaks for the quasicrystal in Fig.~\ref{fig:SpacetimeQuasicrystalPairCorrelations} introduce oscillations at many different length scales, which take longer to average out. Consequently, the integral for the sQC contributes less negatively to the variance formula \eqref{eq:VarianceGenericPointSet} than that of the Fibonacci lattice, which makes $(\sigma_{N_\cdiamond})_{\rm Fib}$ lower than $(\sigma_{N_\cdiamond})_{\rm sQC}$

\subsubsection{Future Work}

Of course, all this analysis is just schematics: we know that applying a Euclidean intuition to Lorentzian point sets is not meant to lead to a proper understanding about why the variances are what they are. However, even within this scheme, we could try several more things to get a better intuition about how correlations drive the variance to a sub-Poissonian regime:
\begin{itemize}
    \item \textbf{Ripley's Boosted Ladder}: in Fig.~\ref{fig:RipleysFunction} we did not consider any boosts on our patterns. We could repeat this process for different choices of boosts and see if the behaviour shown in Fig.~\ref{fig:RipleysLadder} persists. Since we know that the integer lattice is not spacetime hyperuniform from our N-V tests, we would expect, for example, that the ladder-like behaviour around the parabola would eventually break for this structure. If this intuition is correct and the ladder is, on average, above the parabola for high enough boosts, this would mean that there is a change in the correlations from negative to positive, which could be attributed to a Lorentz contraction of the lattice spacing (see Fig.~\ref{fig:BoostedVsUnboostedIntegerLattice}). On the other hand, since both the Fibonacci lattice and the spacetime quasicrystal have boost symmetry, we expect that the ladder will not change that much for higher boosts and it would eventually come back to itself after a full boost ``period'' has been explored.  
    \item \textbf{Ripley's Ladder in $k$-Space:} since we already know that there exist a criterion for deciding hyperuniformity of quasicrystals based on bounding Ripley's ladder in Fourier space as we approach $k \to 0$ \cite{Oguz_2017}, perhaps the most reasonable thing to try next would be to generate Fig.~\ref{fig:RipleysFunction} in Fourier-space. Of course, we would need to find a different, Lorentz-invariant way to calculate the pair correlation function -- we discuss this possibility in Section \ref{subsec:LorentzInvariantHyperuniformity}. 
\end{itemize}
We leave such explorations for future work; we now turn back to the N-V test and argue that this is indeed a Lorentz-invariant way to decide whether or not a given spacetime structure is hyperuniform or not.

\subsection{The N-V Test \textit{is} a Hyperuniformity Test in Spacetime!}
\label{subsec:NVTestIsHyperuniformityTest}

Armed with the intuition of Sections \ref{subsec:EuclideanVariance} and \ref{subsec:FourierVariance}, we can now see that the N-V test of Section \ref{sec:NVCorrespondence} is nothing but a hyperuniformity test, not in space but in spacetime instead. 

Indeed, the test consists of sampling causal diamonds $\cdiamond$ of increasing volumes as our observational windows $\Omega$, and calculating $\langle N_{\cdiamond} \rangle$ and $\sigma_\cdiamond^2$ from the collected statistics. Among all structures that realize the N-V correspondence $\langle N_\cdiamond \rangle = \rho V_\cdiamond$ on average, the best candidate is the one with the least $\sigma^2_\cdiamond$. The goal is to beat Poisson, i.e.~have sub-Poissonian variance $$\sigma^2_\cdiamond < (\sigma^2_\cdiamond)_{\textrm{Poisson}} = \langle N_\cdiamond \rangle$$ throughout all tested volumes $V_\cdiamond$ and boosts.~Equivalently, this translates into a requirement for our point configuration to remain \textit{spacetime hyperuniform} across the full test-range. For example, the integer lattice $\mathrm{I}_{1,1}$, while beginning sub-Poissonian for small boosts, it quickly becomes super-Poissonian and loses the battle against the random causal set (Fig.~\ref{fig:BoostCatastopheIntegerLattice}). Ultimately, what we are after is a Lorentz-invariant hyperuniform spacetime structure to beat the random Poisson distribution across all scales.  

Adapting everything we discussed about hyperuniformity from the purely spatial to the fully spacetime setting might be a considerably difficult task. Given that the volume of a $\cdiamond$ in a $d$-dimensional flat Minkowski space is \cite{Berthiere_2015,Gibbons_2007}
\begin{equation}
    V_\cdiamond = \frac{2 v_{d-2}}{d(d-1)} \left(\frac{T}{2}\right)^d,
    \label{eq:VolumeCausalDiamondGeneralDimensions}
\end{equation}
where $T$ is the proper time between the past- and future-tip of $\cdiamond$ in its rest-frame (T/2 is what we called the half-extent $T_\cdiamond$ in Appendix \ref{appendix:NVTestDetails}), one could expect that we could use the results we already derived by performing the substitution
$$ \mathrm{vol}(\Omega) \leftrightarrow \mathrm{vol}(\cdiamond) = V_\cdiamond \Rightarrow R \leftrightarrow T.$$
However, it is not that simple. The subtlety is that the observational windows $\cdiamond$ now have varying shapes and orientations because of the different boosts one could perform, even though they all have the same volume $V_\cdiamond$ within the particular volume levels tested. As a result, the variance Eq.~\eqref{eq:VarianceGenericPointSet} now depends on the rapidity $\varpi$ and the direction $\hat{\bf n }$ of the boost (i.e.~a unit vector on $\mathbb{S}^{d-2}$, whose volume governs the purely spatial part of Eq.~\eqref{eq:VolumeCausalDiamondGeneralDimensions}). What we would actually need to correctly describe hyperuniformity in the spacetime setting is $\langle N_\cdiamond (T,\varpi,\hat{\bf n}) \rangle$ and $\sigma_{\cdiamond}^2(T,\varpi,\widehat{{\bf n}})$. We turn to this problem in the next subsection, where we offer some ideas about how to define these quantities more properly.

\subsubsection{Setup}

For now, we work in $(1+1)$D where there is no directional dependence on the boosts in order to simplify our lives. Since $\cdiamond$ is a rectangle in the $(\rm u,v)$-space with null extents $(a,b)$ (see Appendix \ref{app:CausalDiamondsInVariousDimensions}, the maximum wavenumber (in reciprocal, Fourier space) that can be supported inside $\cdiamond$ will be $ {\bf k} = (k_{\rm u}, k_{\rm v})$,  where $k_{\rm u} \sim 1/a$ and $k_{\rm v} \sim 1/b$. Taking their geometric mean, we define a Lorentz-invariant wavenumber $k_\cdiamond$ (since $a \mapsto e^{+\varpi} a$ and $b \mapsto e^{-\varpi} b$ under a boost) 
\begin{equation}
    k_\cdiamond \sim \sqrt{\frac{1}{a b}} = \sqrt{\frac{(|\det \eta_{\rm null}|)^{1/2}}{V_\cdiamond^{\rm uv}}}
    \label{eq:MINwavenumberDiamond}
\end{equation}
where $V_{\cdiamond}^{\rm uv}$ is the (density-normalized; physical) volume of $\cdiamond$ in the $(\rm u,v)$-coordinates. 

\textbf{Size $\cdiamond$}. To test for hyperuniformity, we need to find how $\sigma_{\cdiamond}^2$ grows in the large-volume limit, but how large must $V_\cdiamond$ be? In \cite{Oguz_2017}, their tests -- based on $Z(k)$ -- started at minimum wavenumber $k_{\min} \sim 0.01$ (in units of the lattice-spacing, which here is taken to be 1; see their Figs. 3,4 and 5). For this reason, our N-V test must reach at least at $k_\cdiamond^{\min} \sim 0.01$ if we would like to claim that any structure is actually hyperuniform. For example, for our $(1+1)$D sQC from ${\rm I}_{3,1}$ with $\det \eta_{\rm null} = -(G_0^{+-})^2$ (see Section \ref{subsec:2DLorentzianQuasiCrystal}), we find that $(V_\cdiamond^{\rm uv})_{\max} \sim (G_0^{+-}) \cdot 10^{4} \sim 2 \cdot 10^5$, and since $V_\cdiamond = V_{\rm sample}/\rho_{\rm sQC \, I_{3,1}}$ (where $V_{\rm sample}$ are the sampled volume levels), we must reach up to $V_{\rm sample} \sim 2 \cdot 10^6$ ($\rho_{\rm sQC \, I_{3,1}} \sim 11.85$). Computationally speaking, this is a formidable task. For comparison, the N-V test of Fig.~\ref{fig:NVTest1Plus1DMean} went up to $V_{\rm sample} = 10^2$, so we need 4 orders of magnitude more -- and a much larger patch -- to perform such an N-V test. 

\textbf{Boost $\cdiamond$.} Now that we know up to which volumes we should run the N-V test, another question that arises is by how much we should Lorentz-boost them. Stated differently, what range of values of our rapidity $\varpi$ we should consider? Given that the boost range for a given volume $V_\cdiamond$ is heavily restricted by the volume of the patch under consideration (see Appendix \ref{appendix:NVTestDetails}), $\varpi_{\max}(\cdiamond) \sim \ln{ (L_{\rm patch} k_{\cdiamond}) }$, if we would like to test small $k_\cdiamond$ at high boosts $\varpi$, we would need to enlarge the volume of the patch considerably (here, $L_{\rm patch}$ is the null extent of the patch in the coordinates $k_\cdiamond$ is calculated; $\lambda_\cdiamond \sim 1/k_\cdiamond$ is taken to be the characteristic length of $\cdiamond$). This is another formidable computational task. For patterns that have some discrete boost symmetry, like the Fibonacci lattice and the spacetime quasicrystal, we know that they will eventually get back to themselves after we reach a certain boost $\varpi_{\rm pattern}$. For this reason, we decided to limit ourselves to at least 1 full period of boosts for the largest $(V_\cdiamond^{\rm uv})_{\max}$ in the test.  

\textbf{Sample Number $\cdiamond$.} There is one more subtle point that we shall discuss before we are ready to formulate the spacetime hyperuniformity test which has to do with \textbf{patch saturation}. 

More specifically, hyperuniformity of a point configuration is tested in an \textit{ensemble-average formulation}, where a fixed observational window $\Omega$ is sampled over different realizations of the point pattern $P$ \cite{Torquato_2003,Torquato_2018}. This guarantees that each time we randomly place $\Omega$ somewhere on $P$, it `sees' a different version of $P$; the samples are independently distributed throughout $P$ and non-overlapping\footnote{Usually, this statistical independence is denoted as i.i.d. (or iid), which stands for identical, independently distributed samples; it formalizes what we usually just indicate as ``random samples''.}. In this case, the resulting fluctuations captured by $\sigma_{\Omega}^2$ are indeed a bulk property of the underlying point process, not an artifact of any finite-patch effect. 

On the other hand, in our numerical N-V tests, we are dealing with a single point configuration from which we are sampling many -- $M_\cdiamond$, in number, for each $\cdiamond$ -- diamonds of the same volume $V_\cdiamond$. Let $V_{\rm patch}$ denote the volume of the patch. Fluctuations for a fixed $V_\cdiamond$ arise because we let $\cdiamond$ uniformly sample the pattern $P$ (i.e.~each diamond centre $c_\cdiamond$ is drawn i.i.d.~uniform from a valid region of volume 
\begin{equation}
    V_{\rm valid}(\cdiamond) = V_{\rm patch} \left( 1- \sqrt{\frac{V_\cdiamond}{V_{\rm patch}}}\right)^2 < V_{\rm patch} \quad \text{ (for any non-trivial $\cdiamond$) }
    \label{eq:ValidRegionForSamplingDiamondCenter}
\end{equation} 
such that it lies entirely within the patch; see Appendix \ref{appendix:NVTestDetails}). The resulting $\sigma_{\cdiamond}^2$ is an unbiased estimator of the positional variance of $N_\cdiamond$ over $V_{\rm valid}$, but now this is a \textit{volume-average formulation} instead.~For non-ergodic point process, such as completely deterministic, regular periodic lattices and aperiodic quasicrystals, these two formulations might not agree (see also \S 3.2 in \cite{Torquato_2018}).

The intuition is as follows: as $V_\cdiamond$ grows, $V_{\rm valid}(\cdiamond)$ shrinks and all centres crowd into a smaller and smaller central region of the patch. Consecutive diamonds overlap almost entirely, leading to $N(c_\cdiamond^{(i)}) \approx N(c_\cdiamond^{(j)})$ for nearby centres $c_\cdiamond^{(i)}$ and $c_\cdiamond^{(j)}$. As the centre moves by a small displacement $\delta \ll a$ ($a$ is one of the null-extents of the diamond), $\cdiamond$ sweeps a thin boundary slab of width $\delta$ into its interior while sweeping an equally thin slab out. The vast majority of points -- those deep inside the diamond -- are shared between successive samples and contribute nothing to the spread of $N_\cdiamond$. The measured volume-averaged variance $\sigma_\cdiamond^2$ over $V_{\rm valid}$, therefore, is dominated by the fluctuations of these thin boundary layers, not the total count fluctuations one would observe by placing the diamonds at genuinely distinct locations across the patch. This boundary effect might drive $\sigma_{\Omega}^2$ to lower values resulting in an apparent hyperuniformity signal, which is partly a geometric artifact of the finite-size patch rather than a property of the underlying point process.

Consequently, to make our test as unbiased as possible, we need to ensure \textit{ergodicity} so that volume-averages agree with ensemble-averages. Computationally speaking, generating an ensemble of realizations for our quasicrystals of large enough volumes (so we can test large enough boosts) is relatively expensive,\footnote{One would need to re-run the $\CNP$ algorithm with a shifted window in the internal space; each shift would produce a new, locally indistinguishable but globally different, quasi-crystalline patch. Alternatively, one could generate a massive patch of, let's say $\sim 10^{10}$ or more points, and then sample from it some number of sub-patches that contain the desired number of elements.} so we would like to remain within the premises of testing a single point configuration. In practice, then, ergodicity restricts the number of samples we can take for each $\cdiamond$ from a patch of given volume $V_{\rm patch}$ before we \textit{saturate} the patch and all our samples are overlapping. The maximum number of geometrically independent (non-overlapping) diamond placements is found\footnote{For simplicity, we assume that we perform no boost and both null-extents are equal; for this reason, Eq.~\eqref{eq:MaximumIndependentDiamondSamples} can be taken as a conservative upper bound.} by packing diamonds within $V_{\rm valid}$ with centre-to-centre separation exactly equal to their null-extent $a$, which leads to 
\begin{equation}
    M_{\max}(\cdiamond) = \left( \sqrt{\frac{V_{\rm patch}}{V_\cdiamond}} - 1\right)^2
    \label{eq:MaximumIndependentDiamondSamples}
\end{equation}
For a given number of samples $M_{\rm target}$, the condition $M_{\max}(\cdiamond) = M_{\rm target} $ gives the saturation threshold $V_{\cdiamond}^{\rm sat}$: for any $V_{\cdiamond} > V_{\cdiamond}^{\rm sat}$, the patch cannot supply $M_{\rm target}$ independent samples. When running the spacetime hyperuniformity test, we perform an \textit{adaptive sampling} by setting the number of samples for a given $\cdiamond$ to  
$ M(\cdiamond) = \max \{ M_{\rm target}, M_{\max}(\cdiamond)\}$
This ensures that the scaling of $\sigma_{\cdiamond}^2$ is not contaminated by patch-size effects. 

\subsubsection{Results}

With these simplifications in mind, we propose the following criterion:
\vspace{-1ex}
\begin{statement}{\textbf{Criterion for Testing Spacetime Hyperuniformity}} consider the setup \ref{statement:NVTest} for the N-V test, where we now sample $M(\cdiamond)$ volumes up to $(V_{\cdiamond})_{\max}$ that makes $(k_\cdiamond)_{\min} \sim 0.01 \sim (V_\cdiamond)_{\max}^{-1}$, with the wavenumber defined as in Eq.~\eqref{eq:MINwavenumberDiamond}. In $(1+1)$D, let the range of boosts for each volume $V_\cdiamond$ be at least $|\varpi| \in [0,\varpi_{\rm pattern}]$, where $\varpi_{\rm pattern}$ is the smallest possible boost that is an exact symmetry of the pattern. Adjust $V_{\rm patch}$ of the tested patch so that $\varpi_{\max}\big((V_\cdiamond)_{\max} \big) \geq \varpi_{\rm pattern}$ and $M(\cdiamond)$ for $(V_\cdiamond)_{\max}$ is at least 200. If after running the N-V test we find 
    $$ \sigma_\cdiamond^2 < (\sigma_\cdiamond^2)_{\rm Poisson} = \langle N_\cdiamond \rangle = \rho V_\cdiamond$$
    throughout the range of tested volumes, we call the structure under the N-V test \textit{spacetime hyperuniform}.
\label{statement:HyperuniformityCriterion}
\end{statement}

To further place it into a hyperuniformity class, we look at the scaling of $\sigma_{\cdiamond}^2$ with $V_\cdiamond$ in the regime $[(V_\cdiamond)_{\rm onset}|_{k_\cdiamond = 0.1}, (V_\cdiamond)_{\max}|_{k_\cdiamond = (k_\cdiamond)_{\min}}].$ Depending on this asymptotic behaviour, we define
    \begin{equation}
        \sigma^{2}_\cdiamond \propto 
        \left\{
        \begin{aligned}
        & \langle N_\cdiamond \rangle^{1/2}, & \quad \textbf{(Class I)} & \qquad \left(\langle N_\cdiamond \rangle^{(d-1)/d}\right) \\
        & \langle N_\cdiamond \rangle^{1/2} \frac{\log \langle N_\cdiamond \rangle}{2}, & \quad \textbf{(Class II)} & \qquad \left( \langle N_\cdiamond \rangle^{(d-1)/d} \frac{\log \langle N_\cdiamond \rangle}{d}\right)\\
        &  \langle N_\cdiamond \rangle^{1-\alpha/2}, \quad (0<\alpha<1) & \quad \textbf{(Class III)} & \qquad \left(\langle N_\cdiamond \rangle^{(d-\alpha)/d}\right)
        \end{aligned}
        \right.
\label{eq:VarianceScalingsForSpacetimeHyperuniformity}
    \end{equation}

\begin{figure}[t!]
    \centering
    \includegraphics[width=0.99\linewidth]{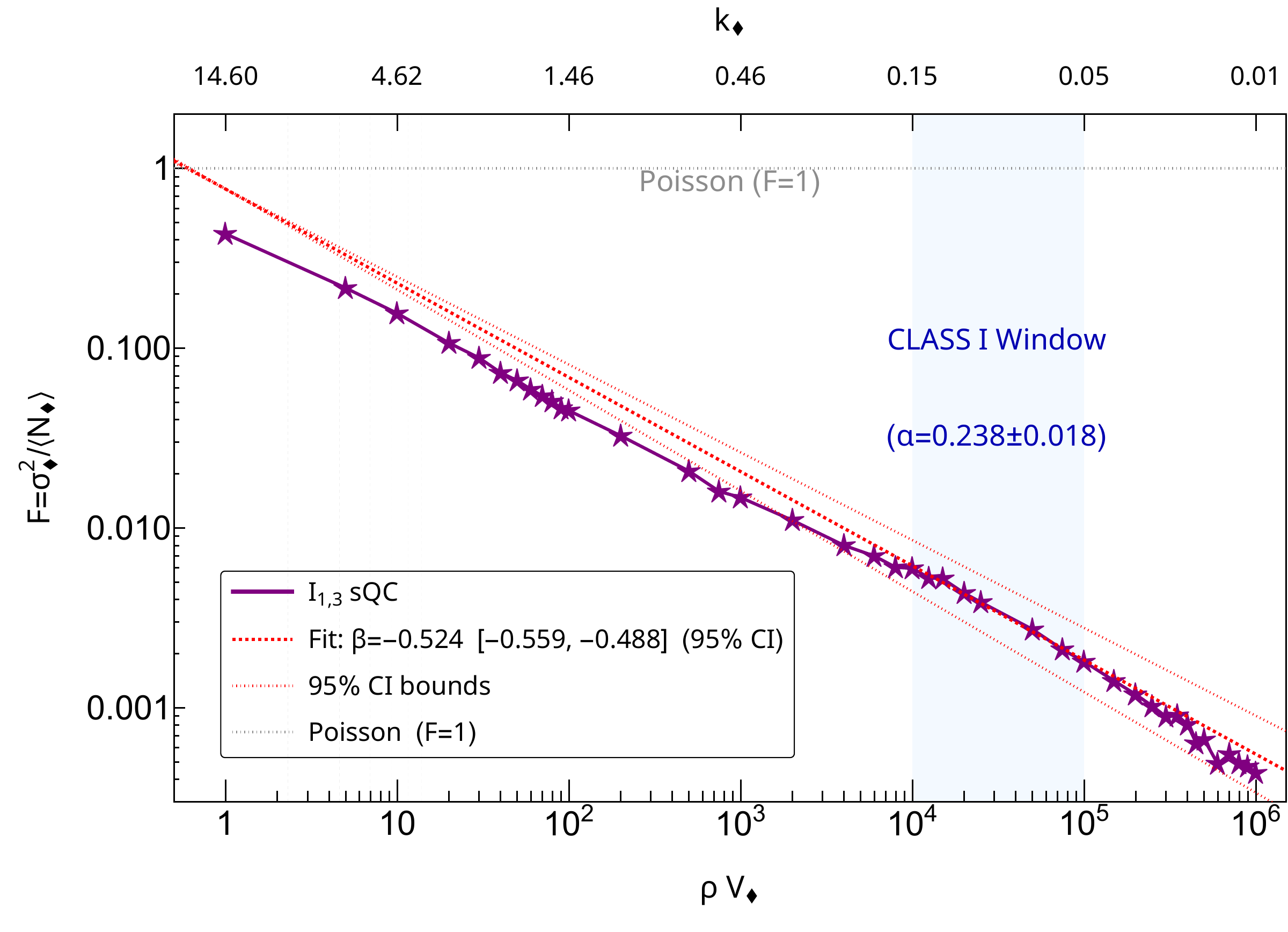}
    \caption[Spacetime hyperuniformity test]{Spacetime hyperuniformity test for the $\mathrm{I}_{3,1}$ quasicrystal, where we plot the Fano factor $F_\cdiamond = \sigma_\cdiamond^2/\langle N_\cdiamond \rangle$ against the volume levels $\rho V_\cdiamond$ (in purple) for a N-V test and fit a power-law model (best-fit shown in red).}
\label{fig:SpacetimeHyperuniformityTest}
\end{figure}

So far, we have only been able to test our $(1+1)$D spacetime quasicrystal from $\mathrm{I}_{3,1}$ and the result is presented in Fig.~\ref{fig:SpacetimeHyperuniformityTest}. We find that the $\mathrm{I}_{3,1}$ sQC \textbf{is consistent with Class I hyperuniformity} with a scaling exponent $\sigma_{\cdiamond} \propto \langle N_\cdiamond \rangle^{0.238 \pm 0.018}$ (95\% confidence interval - CI). We find this by using the results of the N-V test to plot $F_\cdiamond = \sigma_\cdiamond^2 / \langle N_\cdiamond \rangle$ (also known as the \textit{Fano factor} in statistics) for each volume $V_\cdiamond$. Spacetime hyperuniformity predicts $
F_\cdiamond = C (\rho V_\cdiamond)^\beta$ with $\beta = -1/2$. Equivalently, writing $\sigma_\cdiamond^2 \sim \langle N_\cdiamond \rangle^{2\alpha}$  defines $\alpha = (1 + \beta)/2$; Class I means $\alpha = 1/4$ and  for Poisson, $\beta = 0, F_\cdiamond = 1$ everywhere). We use the data -- results shown in Tab.~\ref{tab:NVTestForHyperuniformitysQCI31} -- to fit a power-law model for $F_\cdiamond$ (by ordinary least squares) in log-log space over the range $\rho V_\cdiamond \in [10^4, 10^5]$, where $M_{\rm samples} \geq 1000$ independent samples were available at each level (apart from the last one). The best-fit exponent $\beta = -0.535$ at 95\% CI ($[-0.565, -0.505])$ is consistent with the Class I hyperuniformity prediction. We see in Fig.~\ref{fig:SpacetimeHyperuniformityTest} that after $\rho V_\cdiamond = 10^5$, variance starts to drop: this is because of patch saturation, as mentioned earlier. As a result, with our current setup we were unable to reach up to $k_{\cdiamond} = 0.01$ (see the top axis), and a future test must be carried out.

In practice, we can use the criterion \ref{statement:HyperuniformityCriterion} to quantify hyperuniformity in $(3+1)$D as well. 
For this reason, the general $d$-dimensional scalings are given as well in the above equation.
In this case, we can still use the volume $V_{\cdiamond}$ to define a length-scale $T_\cdiamond \sim (V_{\cdiamond})^{1/d}$ (its half-extent), but this is strictly only valid in the diamond's rest-frame. For the boosted diamond, we would need to incorporate the angular-dependence ${\hat{\bf n}}$ of a general Lorentz-boost as well. We give some thoughts in the next short subsection.

We could like to conclude this section by emphasizing the following: finding a structure that beats Poisson in the N-V test in $(3+1)$D is actually \textbf{a quest for finding the first spacetime hyperuniform structure in $(3+1)D$ as well}. We have established so far that one of our spacetime quasicrystals in $(1+1)$D is spacetime hyperuniform; we did not say that this is the first such structure, though, because we expect the Fibonacci lattice in Fig.~\ref{fig:NVTest1Plus1DStandardDeviation} to also be spacetime hyperuniform, even though we have not carried out that test at the time of writing.\footnote{In \cite{Saravani_2014}, where this structure was first tested, the concept of spacetime hyperuniformity -- first developed here -- was not considered.
} Similarly, our N-V tests in $(3+1)$D provide intriguing evidence that the $\mathrm{I}_{7,1}$ and $HF_4$ spacetime quasicrystals are also sub-Poissonian. Could it be that these are the first examples of spacetime hyperuniform structures? Only more elaborate tests could settle this question once and for all.

\subsection{Towards a Lorentz-invariant Hyperuniformity Estimator?}
\label{subsec:LorentzInvariantHyperuniformity}

We close the section about spacetime hyperuniformity by mentioning a couple thoughts about how to extend the results of the previous section in a more Lorentz-invariant framework. 

First, for the spatial statistics measures of Section \ref{subsec:NonTrivialCorrelationsAndVariance}, an idea would be to calculate the pair correlation function $g_2(r)$ on a Lorentzian point set not by drawing spheres (Euclidean invariant) but causal diamonds instead. For example, we could fix the centre of some $\cdiamond$ on a point of the pattern and draw causal diamonds of increasing proper time away from that centre; then, we could calculate the spacelike and timelike distances of all points included in the causal diamonds. Having both negative and positive spacetime distances would make the interpretation of figure like Fig.~\ref{fig:SpacetimeQuasicrystalPairCorrelations} difficult, however it would be more in the spirit of Lorentz-invariance. 

Second, as mentioned before, our formulas so far -- like Eq.~\eqref{eq:MINwavenumberDiamond} for $(k_\cdiamond)_{\min}$ -- depend only on the magnitude of the wavenumber; equal volumes $V_\cdiamond$, no matter how boosted they are, will give the same $k_\cdiamond$. It is clear that, in order to properly consider a hyperuniformity estimator in ($3+1$)D, for example, where boosts carry direction dependence on $\Sph^2$, we would need to find a way to factor that in. Also, in our definition of spacetime hyperuniformity (Eq.~\eqref{eq:HyperuniformityCondition} and related formulas), what matters is the behaviour of the structure factor as the \textit{magnitude} of the wavenumber ${\bf k}$ goes to zero, independent of ${\bf k}$'s directional degrees of freedom. This has been generalized in \cite{GeneralizingHyperuniformity_Torquato} to account for anisotropic spectral densities. Perhaps this is relevant for us here; we leave this for future work.

\section{Phenomenology and Future Directions in qCST}
\label{sec:phenoqCST}

It is rather tempting to close this section on causal set theory -- or its \textit{quasi-crystalline} version -- by offering some very speculative ideas about some interesting physics one could do with spacetime quasicrystals. Normally, these would belong to Chapter \ref{ch:MoreConnectionsAndSpeculations} with the rest of the speculative ideas, but since they do refer to causal set theory more directly, we decided to give them here.

\subsection{Spectral Graph Properties}
\label{subsec:SpectralGraphProperties}

Graph observables have been shown to be very powerful in distinguishing between different causal sets \cite{Surya_2026,eichhorn2026chartingcausalsetconfiguration} (the last reference treated also our $\mathrm{I}_{3,1}$ spacetime quasicrystal). While random causal sets have no inherent structure (and thus their graph properties are not analytically understood), we can analytically study the underlying graphs of ordered structures easier. Ultimately, the goal is to get a more graph-theoretic understanding of how to build, let's say, an action for causal sets from objects inherent to the causal set, or how to describe operators on a causal set, or how to quantify the entropy of a causal set. 

\subsubsection{Propagators and Graph Laplacians}
\label{subsec:Propagators}

We know from the work of \cite{johnston2010quantumfieldscausalsets} that certain graph-observables, like the adjacency matrix of a graph, are related to field propagators (Green's functions) on a causal set (see also \S 5 in \cite{Surya_2019}). For example, in $(1+1)$D, the \textit{causal propagator} $\Delta$ is given as $i\Delta = G_R - G_A = \frac{1}{2}(C-C^T)$, where $G_{A/R}$ are the \textit{advanced/retarded propagators} of a (scalar) quantum field living on a causal set $\mathcal{C}$, and $C$ is the \textit{adjacency matrix} of $\mathcal{C}$. Recently, the eigenvalues of this object in various dimensions have captured some attention in the literature \cite{jones2026spectraldensitycausalpropagator}, and conjectures have been formulated about their asymptotic behaviours. It would be interesting to calculate the spectrum of these operators on a spacetime quasicrystal -- a quasi-crystalline causet $\mathcal{C}$ -- and compare to that of a traditional causet. 

In the spirit of understanding better how to build propagators \textit{from} a causal set (and not by taking a field \textit{on} a causal set and matching its propagators to quantities of $\mathcal{C}$, as done in \cite{johnston2010quantumfieldscausalsets}), perhaps the following techniques will be useful. In \cite{beveridge2015hittingtimeformuladiscrete,ellis2003discretegreensfunctionsproducts,CHUNG2000191}, the authors study \textit{discrete Green's functions} ${\cal G}$, solutions to the equation ${\cal L}{\cal G} = 1$, where ${\cal L}$ is the \textit{graph Laplacian} and $1$ here is a unit matrix of appropriate dimensions. This is in analogy to the continuous Green's functions, which are solutions to $\Box G_{R/A}(x,x') = \delta(x,x')$, $\Box$ being the D'Alembertian. In that sense, $G_{R/A}$ can be seen as $\Box^{-1}$ and $\mathcal{G}$ can as $\mathcal{L}^{-1}$. For an undirected graph, ${\cal L}$ becomes ${\cal L} = {\cal B} {\cal B}^T$, where ${\cal B}$ is another graph observable, known as the \textit{incidence matrix} (this keeps track of edges with which vertices are connected on a graph). Since $\Box$ is the ``square'' of the Dirac operator $D$ and ${\cal L}$ is the ``square'' of ${\cal B}$, one can try to define Green's functions for ${\cal B}$ by solving some analogous equation ${\cal B} {\cal G} = 1$. If the solutions of ${\cal L} {\cal G} = 1$ lead to \textit{bosonic} propagators (for fields satisfying the Klein--Gordon equation), ${\cal B} {\cal G} = 1$ would give \textit{fermionic} propagators (for fields satisfying the Dirac equation). It would be interesting to see what kind of $\mathcal{G}$'s we get for a spacetime quasicrystal.   

\subsubsection{Point-Set Correlations and Intrinsic Spacetime Entanglement}
\label{subsec:IntrinsicSpacetimeEntanglement}

Usually, when we consider the entanglement entropy of a system (i.e.~a quantum field) on spacetime, we do not say that the underlying spacetime is itself entangled. Similarly, when we deal with quantum fields on a fixed causal set (like in \cite{johnston2010quantumfieldscausalsets,Keseman_2022,Yazdi_2023}), we calculate the entanglement entropy of that field on some $\mathcal{C}$, and there is no contribution there from $\mathcal{C}$ itself. However, if $\mathcal{C}$ is itself a quantum description of spacetime, could it have some \textit{intrinsic spacetime entanglement}?  

In \cite{Braunstein_2006}, a density matrix was constructed for any given graph, and some of its properties were studied. Thinking of $\mathcal{C}$ as a graph, we can use these techniques to assign a density matrix to $\mathcal{C}$. We can then attempt to calculate the spacetime entanglement entropy between a spacetime region (i.e.~a subset of our causet ${\cal C}$) and its complement by tracing out the complement. Could it be that this process leads to different results for a traditional and a quasi-crystalline causet? Can we use the intrinsic non-trivial long-range correlations inherent to a spacetime quasicrystal (see Section \ref{subsec:NonTrivialCorrelationsAndVariance}) to encode this intrinsic spacetime entanglement (i.e. connectivity between its different patches)? If so, this would provide new insights into the ``entanglement builds geometry'' paradigm, usually emphasized in holographic contexts \cite{Pastawski:2015qua}. 

\subsection{Higher Everpresent Cosmological Constant $\Lambda$?}
\label{eq:LowerCosmoConstantQSCT}

In \cite{sorkin1991spacetime,Sorkin_2007,Ahmed_2004}, causal-set inspired ideas led to the model of the \textit{everpresent} $\Lambda$, which notably provided a heuristic prediction for the magnitude of the cosmological constant $\Lambda$ in our Universe -- well before the 1998 discovery of its accelerated expansion!

As described in Section \ref{sec:CSTReview}, in causal sets the spacetime continuum gets replaced by discreteness; the total spacetime volume $V_{\rm spacetime}$ is then proportional to the number of elements in the Universe, $V_{\rm spacetime} \sim N_{\rm spacetime} L_{\rm Pl}^4$, where $L_{\rm Pl}$ is the Planck-length. But the mapping from a continuous manifold to a discrete causal set is achieved via a Poisson sprinkling process, according to the CST consensus. As a result, the number of elements in a given fixed volume is not absolute, but is subject to statistical Poisson fluctuations. The fluctuation in the spacetime volume according to Poisson statistics is 
$ \Delta V_{\rm spacetime} \sim \sqrt{ N_{\rm spacetime} L_{\rm Pl}^4}$. The key idea in \cite{sorkin1991spacetime} is to propose an uncertainty relation $$\Delta V_{\rm spacetime} \Delta \Lambda \sim 1 \qquad (c= \hbar = 1)$$ between $V_{\rm spacetime}$ and its conjugate variable, the cosmological constant $\Lambda$. Substituting the Poisson fluctuation of the volume into this uncertainty relation, we get a residual fluctuation for the cosmological constant
$$ \Delta \Lambda \sim \frac{1}{L_{\rm Pl}^2 \sqrt{N_{\rm spacetime}}} \sim \frac{M_{\rm Pl}^2}{\sqrt{N_{\rm spacetime}}}.
$$
Assuming that the mean $\Lambda$ is \textit{exactly} zero\footnote{Our external examiner brought to our attention another proposal, the ``Universe in a Droplet of Helium'' argument \cite{Volovik_Helium}, which seems to predict that $\Lambda \simeq 0$ due to perfect thermodynamic equilibrium (the huge zero-point energy of the vacuum is perfectly cancelled by the macroscopic pressure of the quantum helium super-fluid). In this picture, the universe emerges out of a quantum fluid; if the droplet is finite, the droplet's surface tension can introduce a residual correction that produces a tiny $\Lambda \neq 0$ as a finite-size effect. It would be interesting to see if this proposal can be treated within the hyperuniformity paradigm, or if a spacetime quasicrystal behaves ``like a liquid'' in this sense in some coarse-grained limit.} , the above tell us that what we observe as a tiny cosmological constant today is because of its fluctuations $\Delta \Lambda$ around zero. In other words, setting $\Lambda = \langle \Lambda \rangle + \Delta \Lambda \sim 1/\sqrt{V_{\rm spacetime}}$, we get an estimate of $\Lambda$. Using the volume of the observable universe today $V_{\rm spacetime} \sim \lambda_{H_0}^4 \sim H_0^{-4}$ ($\lambda_{H_0}$ is the Hubble horizon today), this equation yields $\Lambda_0 \sim H_0^{2} \sim 10^{-122} M_{\rm Pl}^2$, close to the observational values. 

However, since the key idea depends on the model for the variance, we would expect that for a quasi-crystalline causet with sub-Poissonian variance, $$(\Delta V_{\rm spacetime})_{\rm qCST} < \sqrt{V_{\rm spacetime}} = (\Delta V_{\rm spacetime})_{\rm Poisson},$$ and thus $$\Lambda_{\rm qCST} > \Lambda_{\rm Poisson}.$$ 
How much larger depends on the variance exponent: for $d=4$ and Class I hyperuniformity, we expect $$(\Delta V_{\rm spacetime})_{\rm qCST} \sim V_{\rm spacetime}^{3/8}$$ from Eq.~\eqref{eq:VarianceScalingsForSpacetimeHyperuniformity} (compare to $(\Delta V_{\rm spacetime})_{\rm Poisson} \sim V_{\rm spacetime}^{1/2}$), thus $$\Lambda_{\rm qCST} \sim M_{\rm Pl}^2 H_0^{-3/2} \sim 10^{-93} M_{Pl}^{2}$$ (for the same ``universe atoms''). Class III hyperuniformity -- which is what the tilted matter power-spectrum seems to imply (see Section \ref{subsec:FourierVariance}) -- would give a value closer to $\Lambda_{\rm Poisson}$. Since the spectral tilt parameter ($n_s -1$) is very well measured and constrained already by \textit{Planck} \cite{Planck_2020}, it would be interesting to use this departure from Class II hyperuniformity to constraint the possible scaling exponents for a Class-III hyperuniform spacetime quasicrystal.     

\subsection{Suppression of Large-Scale Density Fluctuations}
\label{eq:SuppressionFluctuationsHyperuniformity}

Finally, recall from Section \ref{subsec:FourierVariance} that hyperuniform structures suppress large-scale fluctuations. For a physical material, this means that there exist a minimum wavenumber $k_{\min}$ and we cannot excite any modes on that material with $k < k_{\min}$. In other words, the intensity of the power-spectrum $S(k)$ at low-$k$ is heavily suppressed if the material is hyperuniform (see Eq.~\eqref{eq:StructureFactorGoesToZero}). How could that translate into a spacetime hyperuniform structure? Could it be that the existence of some $k_{\min}$ indicates quantization of spacetime excitations? 

In our Universe, we see a characteristic suppression of low-$k$ fluctuations: the low-$\ell$ end of the cosmic microwave background (CMB) has systematically less power than in higher scales ($\ell \sim k$). Could the fact that the low-$\ell$ end of the CMB has its power suppressed be a signature of some UV, quantum-gravity effect, like the one proposed here? Could it be that the granularity of spacetime is responsible for the power-spectrum of our Universe? To this author, this seems like a promising new avenue to test for quantum-gravitational signatures that spacetime discreteness could leave on the CMB.

\chapter{Further Connections of Spacetime Quasicrystals to Physics}
\label{ch:MoreConnectionsAndSpeculations}

In this last Chapter, let us end by briefly mentioning some other connections of spacetime quasicrystals to physics, as interesting topics for future investigations. 

\section{In Computational Complexity Theory}

Scale-symmetric quasicrystals, in particular the Ammann-Beenker tilings, have recently shown to be efficient in solving the general-case \cc{NP}-complete problem of finding Hamiltonian cycles (\textit{i.e.}~closed loops on a graph visiting each vertex precisely once) \cite{HamCycles_Felix2024}. Since many interesting problems in physics fall in the \cc{NP} complexity class, this result hints towards the fact that quasicrystals are essentially as powerful (if not more) as periodic lattices in simplifying certain physical problems. It would be interesting to understand the graph-theoretic properties of spacetime quasicrystals and see what kind of problems their special properties can help us solve. 

\section{In Various Discretizations}

Given their discrete scale invariance, quasicrystals provide natural mathematical structures on which to discretize and study systems in which (exact or approximate) scale or conformal invariance plays an important role ({\it e.g.}\ QFTs near their fixed points, condensed matter systems near their critical points, or our universe near the Big Bang). 

We mention two interesting ideas:
\begin{itemize}
    \item \textbf{Fermions on a Lattice:} (chiral) fermions are notoriously difficult to be placed on a lattice without doubling the degrees of freedom (thus the \textit{fermion doubling problem}). This is formalized in the \textit{Nielsen-Ninomiya} theorem, a no-go theorem telling us that, under locality, hermiticity, and translational symmetry, any attempt to place a chiral fermion on a lattice (let's say, a right-handed one), necessarily leads to an equal number of left-handed and right-handed fermions on the lattice (see \cite{NielsenNinomiya1981a,NielsenNinomiya1981b,NielsenNinomiya1981NoGo,TongGaugeTheory}). However, perhaps fermion discretizations on non-periodic (spacetime) quasicrystals could avoid the doubling problem (by going around no-go theorem since there is no translational invariance). 
    \item \textbf{Anderson Localization:} condensed-matter analogues are ubiquitous in high-energy theory (like the Higgs mechanism), and Anderson localization \cite{PhysRev.109.1492} is another very interesting one. Effectively, this refers to inhibition of wave propagation and confinement of eigenstates (as opposed to delocalized Bloch electrons shared among all atoms in a material), and can occur either because of randomness (as in the original paper \cite{PhysRev.109.1492}) or because of a quasi-random, quasi-periodic potential \cite{PhysRevB.103.L241106,aubry1980analyticity,QC_CriticalPhases_AndersonLocalization}. It is known, for example, that, while random and uncorrelated potentials perfectly localize electrons in materials (the decay of a wavefunction is exponential), quasi-periodic potentials lead to a scale-invariant, power-law localization.
     
    It would be interesting to explore how something like this manifests on a spacetime quasicrystal. Here, the role of wavefunctions in materials is played by quantum fields living on spacetime. The role of the potential is less clear, but an intriguing scenario would be to consider whether the quasi-periodicity of the geometric structure itself is enough to lead to the emergence of localized excitations (i.e.~particles) from fundamental quantum fields (delocalized states), if one discretizes the fields on a spacetime quasicrystal.  

    Relatedly, it has been shown that the \textit{return probability} of walks on a graph (starting from some initial site and returning back to it after, let's say, $N$ moves) plays a role for transient behaviours of statistical models or Anderson localization in disordered solids (see \cite{Joseph_2025} and references in their introduction). Treating spacetime quasicrystal as causal sets and considering their underlying graph structure would be a promising start to investigate if they have localization properties (by studying the return probability of walks on them), which might also relate to the Hamiltonian cycles problem mentioned before. 
\end{itemize}

\section{In String Theory, Particle Physics, and Cosmology}

There may be interesting connections between the compactification scenario we suggest in Secs.~\ref{subsec:IrrationalityOfSpacetime}, \ref{subsec:CartoonPicture}, \ref{subsubsec:SeesawMechanism} and the ``quasicrystalline string compactifications'' idea discussed in \cite{Harvey:1987da, Baykara2024Quasicrystalline}. It shall be noted, however, that the two approaches seem to have a different starting point. While the latter is an orbifold-type compactification (operating at the level of the string world-sheet where the string's left- and right-going modes live), our construction is a string target-space compactification (i.e.~it compactifies the 10D target space on a 10D lattice). For this reason, we view our method as a new class of compactifications that employ ``forbidden'' non-crystallographic symmetries distinct from \textit{quasi-crystalline orbifolds}, which could realize Moore's \cite{moore1993finitedirections} original idea.

Another approach in string theory potentially related to our work is the much-studied IKKT matrix model of Type IIB string theory. This may also be thought of as arising from D=10 super Yang-Mills theory, by compactifying all 10 dimensions of spacetime down to a point \cite{Ishibashi:1996xs, Aoki:1998bq}. Since our compactified space is a 10D torus and 
$E_{10}$ is an infinite group appearing frequently in supergravity, it is natural to explore whether the resulting infinite matrices recover the IKKT model in the limit where our torus shrinks to a point. 

Finally, our compactification scenario (based on compactifying 10D spacetime using the $\mathrm{II}_{9,1} = E_{10}$ root lattice) might have connections to other investigations of the role of $E_{10}$ in string theory, cosmology and the standard model \cite{Damour:2000hv, Damour:2002cu, Damour:2002et, Kleinschmidt:2015sfa, Meissner:2025qrx}. One potentially interesting observation in particle physics could be the following\footnote{We thank our advisor, Latham Boyle, for bringing this up during the defence.}: could the discrete symmetry of spacetime quasicrystals somehow account for the new intriguing pattern in the Yukawa couplings observed in \cite{arkanihamed2026nearlyrighttheoryflavor}? In particular, their Fig. 3 of the distribution of angles for the rephasing-invariant CP violating phase $\varphi$ with peaks at angles multiples of $\pi/8$ is reminiscent of our Fig.~\ref{fig:IcosahedralAnglesSymmetry}, which could emerge for a $\pi/8$-symmetric  (instead of $\pi/5$-symmetric) spacetime quasicrystal. On that same topic of angular distributions, there are also works showing a preference for icosahedral symmetry of CMB-correlations in the sky \cite{luminet2003dodecahedral,Roukema_2008,Kang_2016}; it would be nice if those can be explained by spacetime's underlying aperiodic order.

\section{In Quantum Error Correction}

\begin{figure}
    \centering
    \includegraphics[width=0.5\linewidth]{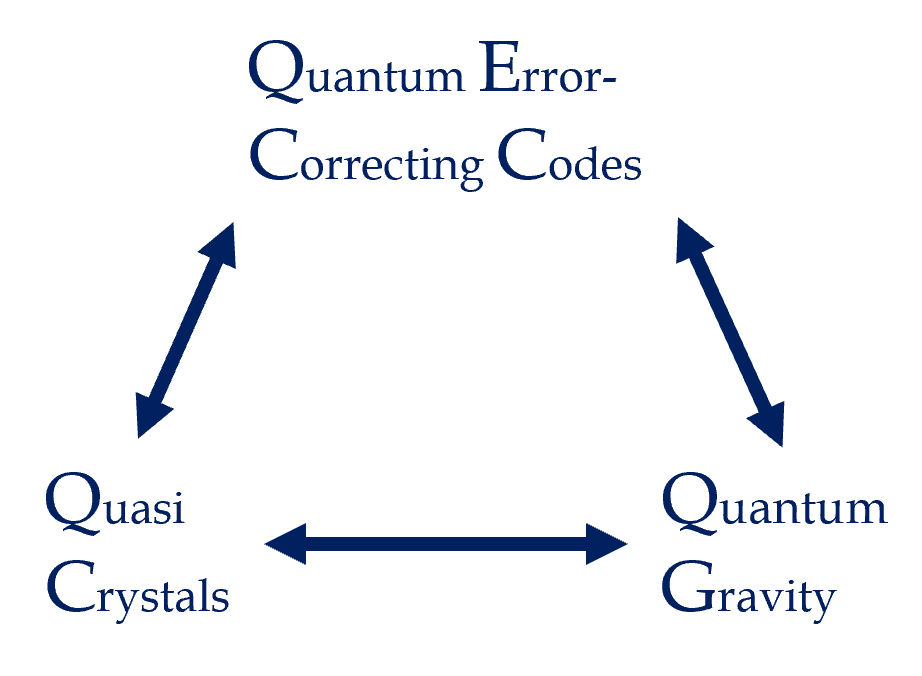}
    \caption[The Q-Triangle: Quantum Error-Correcting Codes (QECC), Quantum Gravity (QG) and QuasiCrystals (QC).]{The Q-Triangle: Quantum Error-Correcting Codes (QECC), Quantum Gravity (QG) and QuasiCrystals (QC).}
    \label{fig:QECC-QC-QG-Triangle}
\end{figure}

Spacetime quasicrystals may yield new insights into the recently noticed, but still incompletely understood, triangle of connections between (discrete) holography (an approach to quantum gravity) and quantum error correction \cite{Pastawski:2015qua}, between quantum error correction and quasicrystals \cite{li2024penrosetilingquantumerrorcorrecting}, and between quasicrystals and (discrete) holography \cite{Boyle:2018uiv, Boyle:2024qzn}. We refer to this affectionately as the Q-Triangle in Fig.~\ref{fig:QECC-QC-QG-Triangle}. 

In particular, in \cite{li2024penrosetilingquantumerrorcorrecting} it was shown that the Penrose tiling \textit{is} a particular type of quantum error-correcting code, or QECC for short. QECCs offer a way to store quantum information with some redundancy so that it can be efficiently recovered if corrupted \cite{Nielsen_Chuang_2010}. Schematically, the relation between QECCs and quasicrystals (QCs) is based on two principles: 
\begin{itemize}
    \item \textit{Local indistinguishability}, on the side of the quasicrystals (see Section \ref{subsec:Torusparametrization}), provides the needed \textit{redundancy for encoding information} (there are infinitely many quasi-crystalline patches that are locally the same but globally different, so locally one cannot say with certainty in which patch they are).
    \item \textit{Recoverability}, or \textit{reconstructability}, allows then for \textit{error-correction}. This is a property of quasicrystals that allows us to rebuild any missing patch from the pattern by inspecting the local neighbourhood only, without needing global information.
    \end{itemize}
We call this the \textit{QECC-QC} dictionary. While the first property follows almost directly from the irrationality of the \CNP{} slice for a quasicrystal, the second is less trivial to have, and indeed we are unsure how to extend it to the spacetime quasicrystals. This is because recoverability becomes possible either by using local scale invariance (see Section \ref{subsec:LocalScaleInvariance}), or the so-called Ammann-lines \cite{Boyle_2022}. Our spacetime quasicrystals lack the first property (see Section \ref{sec:implications}), and we did not yet think carefully about whether our patterns admit Ammann-lines.

Nevertheless, this does not stop us from imagining what would be the case if spacetime had quasi-crystalline structure. Could it be this gives rise to a quantum error-correcting spacetime? Perhaps before we even go there, it would undoubtedly be an interesting question to show whether and what type of QECCs spacetime quasicrystals generate.


\section{In Quantum Gravity: Bridging Different Approaches}
\label{sec:BridgeQG}

Spacetime quasicrystals may be of interest in studying quantum gravity, as explicit, mathematically natural models for discretized spacetime, with many interesting, non-trivial, and counterintuitive properties, and a high degree of (orientational and scale) symmetry (and quasi-symmetry). 

In particular, they provide a novel class of causal sets, as we discussed explicitly in Chapter \ref{ch:sQC_CausalSets}, very different from either Lorentzian lattices or random sets \cite{eichhorn2026chartingcausalsetconfiguration}. In Sections \ref{sec:NVCorrespondence} and \ref{subsec:SpacetimeHyperuniformity}, we provided concrete evidence that our quasi-crystalline causal sets are good models (in the sense of the N-V test) for discrete spacetime. In Section \ref{sec:phenoqCST}, we provided several speculative ideas about how they might relate to physics. 
    
Due to their discrete scale invariance, spacetime quasicrystals may be especially relevant to quantum gravity approaches where spacetime has a fractal, scale-invariant character ({\it e.g.}~asymptotic safety \cite{Weinberg:1976xy, Weinberg:1980gg, Niedermaier:2006wt, Dupuis:2020fhh}). Interestingly, the suggest a way to tie the ``discretum'' and the ``continuum''. Finite-density spacetime quasicrystals are discrete yet preserve much of the underlying continuous Lorentz symmetry. To recover full (global) scale-symmetry, though, we need a countable but \textit{dense} quasicrystal, which better approximates the continuous and scale-symmetric fractal spacetime of Asymptotic Safety. Indeed, the asymptotic safety community is very interested in understanding the configuration space of all possible causal sets \cite{eichhorn2026chartingcausalsetconfiguration}. This is because they are searching for a rationale for asymptotic safety, which tells us that gravity has a UV fixed-point close to which scale-symmetry kicks in. Perhaps this can be understood as arising from a second-order phase transition in the space of all possible causal sets, similar to how a quantum system exhibits scale-symmetry and universality close to its fixed points. To that end, perhaps adding the scale-invariant spacetime quasicrystals is helpful.

Another key question is whether quasi-crystalline causets admit natural (aperiodic) tessellations and what could be their ``fundamental geometric atoms''. In Section \ref{sec:implications}, we concluded that a tiling that respects the non-crystallographic symmetry is probably impossible in Lorentzian space, so we would need to consider different types of tessellations. To that end, one could leverage the graph-theoretic properties of spacetime quasicrystals to develop algorithms that connect points to form tiles, thus finding their {`geometric atoms'}. This will add a novel alternative to Euclidean distance-based tessellation methods, like Voronoi's algorithm \cite{conway2013sphere}, which fail in Lorentzian geometry because of the indefiniteness of the metric. If a tessellation exists that also preserves all (or even, some) symmetry, this could provide a new way to generate highly ordered and symmetric spacetime triangulations, which could bridge (quasi-crystalline) CST with the CDT-approach. 

All these potential applications of spacetime quasicrystals in quantum gravity enhance the known connections described at the Q-Triangle of Fig.~\ref{fig:QECC-QC-QG-Triangle}. Indeed, so far all quantum-gravity applications were holographic, as mentioned in the previous section. In this thesis, for the first time, we point out several structural parallels between different approaches to quantum gravity that are certainly worth to be pursued more in our opinion.


\chapter{Outlook}
\label{ch:Outlook}

In this final short Chapter, we take some time to remind the reader of the main two points we would like to get across from Part I and Part II of this thesis. 

\section{Computation at the Core of Physics?}

In Part I, we argued that two research fields that are perceived as rather disparate, namely computational complexity theory in computer science and (quantum) gravitational physics in theoretical physics, are actually closer than one might think. The connection, as emphasized in Chapter \ref{ch:complexity}, is simply something that no physical theory can escape: computation \textit{is} a physical process, and the physical laws tell us what is \textit{efficiently} computable.

The distinction between efficiently computable (\cc{P}) and efficiently verifiable (\cc{NP}) is the essence of the $\cc{P} \overset{?}{=} \cc{NP}$ problem. Experience suggests that these two complexity classes cannot be the same: coming up with creative new proofs must be harder than verifying one that is already available. Mathematically, this statement is the physical extended Church-Turing thesis (PECTT), which tell us that no physical device, no matter how powerful, can efficiently solve \cc{NP}. And this is exactly what all of our physical theories, including theories about gravity and the quantum world, shall respect. 

A classical or a quantum computer abides by the classical or quantum laws of physics when it processes information and manipulates (qu)bits in order to perform tasks. This is a physical process, governed by a set of physical laws, and thus must conform to the PECTT. Of course, the classical set of laws is included within the quantum set, so we expect a computational speed-up for the quantum computer. But if the PECTT is right, this would not be enough to solve \cc{NP} efficiently. Similarly, a ``(quantum) gravitational computer'' functions through a very different set of physical laws, but it remains a physical device as much as its classical or quantum version.\footnote{Of course, the author can use a classical computer to type this thesis, or visit the closest centre where a quantum computer is being built and inspect it, but holding a quantum gravitational computer that employs gravity to perform tasks is probably beyond even the wildest imagination. In any case, gravity is a physical theory, and it gives rise to systems that can use this physical theory to compute something, in the most abstract sense.} As such, it must also adhere to the PECTT, and any computational advantage from it must be modest.   

In Chapter \ref{sec:GRAVIgateDiscussion}, we propose that the PECTT shall become part of the axioms of our physical theories. We used this criterion to argue -- in Section \ref{ch:SolvingNPSemiclassicaly} -- that semiclassical gravity violates the PECTT and thus must not be a fundamental description of how gravity behaves. Since this is the most popular alternative to quantum gravity (which keeps the gravitational field completely classical instead of quantizing it in some way), we interpreted our results as an argument in favour of quantum gravity. At the end of the day, the message is clear: computational complexity theory has a lot to say about our physical theories.

\section{The Role of Aesthetics in Pure and Applied Quantum Gravity Research}

\setlength{\epigraphwidth}{0.73\textwidth}
\epigraph{\justifying{
``The Role of Aesthetics in Pure and Applied Mathematical Research''}}{Roger Penrose in his iconic paper \cite{penrose1974role} introducing the Penrose tilings}

Historically, quasicrystals are very interesting because they started from the world of recreational mathematics \cite{gardner2013undiluted,gardner1977extraordinary,gardner1997penrose}, where people were thinking of different ways to tile a plane. Imagine an infinite table and an endless supply of puzzle pieces. Your task: cover the surface perfectly—no gaps, no overlaps—and yet never forming a pattern that repeats, no matter how far you go. How many distinct shapes would you need? In 1961, Wang \cite{wang1961proving} conjectured that it is impossible, until his student, Berger, showed in 1966 \cite{berger1966undecidability} that there exist a set of 20426 such ``dominoes'' that does the trick. By 1974, Penrose had dramatically reduced the number to just two \cite{penrose1974role}, a ``kite'' and a ``dart'' (Fig.~\ref{fig:penroseTiling}). Yet it took another 50 years—and one more shape-hobbyist, the retired printing-technician David Smith—for the ``einstein'' (German for ``one stone'' and a nod to a famous physicist) to do the impossible: to find a single tile that fills the plane aperiodically all by itself \cite{Smith_2024,Smith_2024_Chiral}! 

Nice patterns have always fascinated us because they are aesthetically pleasant. Roger Penrose's original work \cite{penrose1974role} introducing the Penrose tilings contains aesthetics in its title, appreciating that aesthetics brings nice mathematics. Indeed, quasicrystals are not only pleasant to the eye but carry a lot of beautiful mathematics with them, from lattices and algebraic number theory, to discrete algebra, Fourier analysis and measure theory. In the 1980s, they even found their way to our labs, becoming the blueprints for a new form of matter (see Chapter \ref{sec:QuasicrystalsIntro}), thus moving from the mathematical to the physical world. This discovery revolutionized how we thought about crystallography and what we considered as ``impossible'' \cite{steinhardt2019second}. 

In Part II, we discussed a natural extension of quasicrystals from space to spacetime, providing the first Lorentzian cousins of the Penrose tilings: spacetime quasicrystals. And like this, quasicrystals made their way for the first time into the realm of relativistic physics. 
This discovery opens an entirely new -- and likely very rich -- area of inquiry, establishing a novel crossroads between crystallography, discrete geometry and algebraic number theory on the one hand-side, and quantum gravity, quantum error-correction and discrete conformal symmetry on the other.

Indeed, this thesis points out a vastly unexplored class of string-compactifications that employ ``forbidden'' symmetries distinct from \cite{Baykara2024Quasicrystalline}, and suggests a geometric seesaw explanation for the large hierarchies among fundamental energy scales. It also hints at several exciting applications of spacetime quasicrystals in other approaches to quantum gravity, including in causal set theory (as highly-symmetric and highly-ordered alternatives to random causal sets), in asymptotic safety (as point sets with discrete scale invariance), and as a way of making concrete the belief that `quantum gravity \textit{is} a quantum error-correction code' (since quasicrystals \textit{are} quantum error correcting codes \cite{li2024penrosetilingquantumerrorcorrecting}).

From a quasi-crystalline lens, there are several intricate connections between quantum gravity theories, as mentioned in Section \ref{sec:BridgeQG} and throughout this thesis. But perhaps even more interestingly, spacetime quasicrystals import new tools from condensed matter physics and from algebraic number theory and discrete geometry into the realm of quantum gravity. The concept of spacetime hyperuniformity, developed in Section \ref{subsec:SpacetimeHyperuniformity}, is a new way of understanding the N-V test (Section \ref{sec:NVCorrespondence}) in causal set theory, and the algebraic Wick rotation by a Galois-automorphism (Section  \ref{subsec:NCSymmetryGeneralsCNP}) is perhaps a new way to transform between a Euclidean and Lorentzian representation. 

Could aesthetics play a role in the next breakthrough in quantum gravity? We do not know, but taken all together, our work introduces spacetime quasicrystals as a qualitatively new paradigm for spacetime discreteness that may reshape how we think about geometry, information, and the quantum structure of spacetime.



\bibliographystyle{unsrt}
\cleardoublepage 
\phantomsection  
\renewcommand*{\bibname}{References}

\addcontentsline{toc}{chapter}{\textbf{References}}

\bibliography{uw-ethesis.bib}



\appendix
\chapter*{\centering \vfill APPENDICES \vfill \vfill}
\addcontentsline{toc}{chapter}{APPENDICES}
\chapter[Number-Theoretic Tools for Building Quasicrystals]{Number-Theoretic Tools for Building Quasicrystals}
\label{AppendixA-NumberTheory}

While (algebraic) number theory may initially seem like pure mathematical abstraction, it provides the rigorous machinery for classifying discrete symmetries of periodic and aperiodic lattices and related algebraic structures. This appendix outlines some necessary formalism, with an emphasis on mapping numbers to the geometric language of lattices and projections. 

\section{Some Notes on Algebraic Number Theory}
\label{sec:AlgebraicNumberTheory}

Here we briefly introduce a few basic definitions, ideas and results in {\bf algebraic number theory}, which will be useful below. For more details, see Refs.~\cite{serre2012arithmetic, lang1994algebraic, cassels1967algebraic} for standard introductions to algebraic number theory. For our purposes, the first few chapters of \cite{baake2013aperiodic} are also very useful. 

\subsection{Preliminary Definitions}

A {\bf field} is a commutative ring in which the additive identity $0$ and the multiplicative identity $1$ are distinct, and any non-zero element has a multiplicative inverse.\footnote{A \textit{ring} is a set with two binary operations, addition and multiplication, satisfying associative and distributive rules. Yet, multiplication might not be commutative and elements in the set might not have inverses. In a \textit{vector space}, its elements can also be scaled (multiplied) by numbers called scalars; every element (apart from 0) also have a multiplicative inverse. A \textit{module}, one the other hand, is a generalization of the notion of vector space in which the field of scalars is replaced by a (not necessarily commutative) ring. } Classic examples are the rational numbers $\mathbb{Q}$, the real numbers $\mathbb{R}$, and the complex numbers $\mathbb{C}$.

An {\bf algebraic number field} $K$ is a field that is also a finite dimensional vector space over $\mathbb{Q}$, which we always take as our base field here. If the base field is different from $\QQ$, we write $L$ instead and most definitions get adjusted to such a choice. In particular, if $K$ is a $d$-dimensional vector space over $\mathbb{Q}$, we say that the {\bf field extension} $K/\mathbb{Q}$ has {\bf degree} $d_K = d$. The degree of a field extension is denoted as $d_K = [K : \QQ]$ (similarly over other fields than $\QQ$).  

For example, the field $\mathbb{Q}(\sqrt{D})$ (of all numbers of the form $a+b\sqrt{D}$, where $a$ and $b$ are rational numbers, and $D$ is some fixed square-free integer) may be regarded as a two-dimensional vector space over $\mathbb{Q}$ (with basis vectors $1$ and $\sqrt{D}$), and hence is an algebraic number field of degree two. Algebraic number fields of degree two are particularly simple, and are also known as {\bf quadratic fields}.  $\mathbb{Q}(\sqrt{D})$ is a {\bf real quadratic field} when $D>0$ and an {\bf imaginary quadratic field} when $D<0$.

A {\bf monic polynomial} $P(x)$ is a polynomial in $x$ whose leading coefficient is $1$.  To any element $\alpha$ in the algebraic number field $K$ is associated a {\bf minimal polynomial} $P_{\alpha,\mathbb{Q}}(x)$, which is the lowest-order monic polynomial with coefficients in $\mathbb{Q}$, having $\alpha$ as a root.  The roots of $P_{\alpha,\mathbb{Q}}(x)$ are called the {\bf algebraic conjugates} (or just {\bf conjugates}) of $\alpha$. The degree $d_K$ of the algebraic field extension $K(\alpha)$ is also equal to the degree of the minimal polynomial $P_{\alpha,\mathbb{Q}}(x)$ of $\alpha$ over $\QQ$. $P_{\alpha,\mathbb{Q}}(x)$ is irreducible over the base field (here $\QQ$). 

If we just take one root $\alpha$ of $P_{\alpha,\mathbb{Q}}(x)$, the field extension $K = \QQ(\alpha)$ might not contain all the other roots. A \textbf{splitting field} of a polynomial is the smallest possible field extension that contains all the roots of that polynomial. In this larger field, the original polynomial can be factored completely into linear factors. If there is no repeated factors, the field extension is called \textit{separable}. If the extension is also \textit{normal}, namely $K$ itself \textit{is} the splitting field, the extension is called a \textbf{Galois extension}. 

In a Galois extension $K/\QQ$, the automorphism group of $K/\QQ$, $\mathrm{Aut}(K/\QQ)$, is the Galois group $\mathrm{Gal}(K/\QQ)$ of the extension. All elements of $\mathrm{Gal}(K/\QQ)$ fix the base field $\QQ$ and map $K$ onto itself. Then, a key idea in Galois theory is that the conjugates of some root $\alpha$ are the images of $\alpha$ under the field automorphisms of $K$. In other words, one could think of the action of $\mathrm{Gal}(K/\QQ)$ as permuting the roots of a minimal polynomial whose splitting field is $K$. The order of the Galois group is also equal to the degree of the field extension, $|\mathrm{Gal}(K/\QQ) | = d_K = [K : \QQ ]$.  

An {\bf algebraic integer} is a complex number that is the root of a monic polynomial $P(x)$ with \textit{integer} coefficients. Let ${\cal O}_{K}$ denote the {\bf ring of algebraic integers} in $K$. Intuitively, these numbers behave `like the integers $\ZZ = \mathcal{O}_\QQ$' inside the larger field $K$. $\mathcal{O}_K$ is a finitely-generated $\ZZ$-module, and as such, it has an \textit{integral basis}, that is a basis $\{ b_i \in \mathcal{O}_K \}_{i=1,\dots,d_K}$ such that each element of $\mathcal{O}_K$ can be expressed uniquely as
\begin{equation}
    x = \sum_{i=1}^{d_K} x_i b_i, \quad (x_i \in \ZZ \,\, \forall i)
    \label{eq:integral_basis_element}
\end{equation}
with $d_K$ also being the rank of $\mathcal{O}_K$. An element of ${\cal O}_{K}$ whose multiplicative inverse is also in ${\cal O}_{K}$ is called a {\bf unit} of $K$.  The units of $K$ form a group under multiplication: the {\bf unit group} ${\cal O}_{K}^{\times}$. 

If we regard $K$ as a finite-dimensional vector space over $\mathbb{Q}$, then multiplication by any element $\alpha\in K$ defines a $\QQ$-linear transformation $m_{\alpha}:K\to K$, $m_{\alpha}(x)=\alpha x$. Choosing a basis for $K$ (for example, if $K=K(\alpha)$, a convenient choice is $\{ 1, \alpha, \alpha^2, \dots, \alpha^{d_K} \}$), the transformation $m_\alpha$ can be represented by a $d_K \times d_K$ matrix $M_\alpha$ with entries in $\QQ$. Of particular interest, then, are the following algebraic quantities:
\begin{itemize}
    \item The \textbf{field trace} (or just trace) of some $\alpha \in K$, defined as the trace of the matrix $M_\alpha$: $\Tr_{K/\QQ}(\alpha) = \Tr M_\alpha$. Equivalently, 
    \begin{equation}
        \Tr_{K/\QQ}(\alpha) = \sum_{i=1}^{d_K} \sigma_i(\alpha)
        \label{eq:NumberFieldTrace}
    \end{equation}
    where $\sigma_i(\alpha)$ are the Galois-conjugates of $\alpha$ and $\sigma$ is a field isomorphism, which, for our purposes\footnote{For example, if we denote by $\alpha_i$ all the conjugates of some $\alpha$ that is used to define $K=\QQ(\alpha)$, while $\sigma_i : K \to \QQ(\alpha_i)$ will always be a field isomorphism, it will only be an automorphism of $K$ if $K$ contains all the roots. For a Galois extension, this is always the case, and these are the extensions we are mostly interest here.}, is a $K$-automorphism, i.e.~$\sigma \in \mathrm{Gal}(K/\QQ)$.  
    \item The \textbf{field norm} (or just norm) of some $\alpha \in K$, defined as the matrix determinant $\mathcal{N}_{K/\QQ}(\alpha) = \det M_\alpha$. Equivalently, 
    \begin{equation}
    \mathcal{N}_{K/\QQ}(\alpha) = \prod_{i=1}^{d_K} \sigma_i(\alpha) 
        \label{eq:NumberFieldNorm}
    \end{equation}
    \item The \textbf{field discriminant} $\Delta_K$, defined as the determinant of the $d_K \times d_K$ matrix of traces of products over some basis $\{ \alpha_i \}_{i=1,\dots,d_K} $ for $K$, 
    \begin{equation}
        \Delta_K  = \det \big( \Tr_{K/\QQ}(\alpha_i \cdot \alpha_j) \big)
        \label{eq:NumberFieldDiscriminant}
    \end{equation}
\end{itemize}
The field trace will be important later because it defines a map $K \times K \to \QQ$, known as the \textit{trace form}, taking two $x,y \in K$ to $(x,y) = \Tr_{K/\QQ}(x\cdot y) \in \QQ$, which is a symmetric bilinear form (i.e.~a metric). The norm is used to define units: the element $\alpha\in K$ is a unit of $K$ if and only if its norm is a unit in $\mathbb{Q}$, {\it i.e.}\ iff $N_{K/\mathbb{Q}}(\alpha)=\pm1$. The discriminant measures the size of the (ring of integers of the) algebraic number field. More specifically, it is proportional to the squared volume of the fundamental domain of the ring of integers.    

An algebraic number field $K$ of degree $d_K$ may be embedded in $\mathbb{C}$ in $d_K$ different ways: $r_{1}$ of these are real embeddings (of $K$ in $\mathbb{R}$), while the remaining embeddings come in $r_{2}$ complex-conjugate pairs, so 
\begin{equation}
    d_K = r_1 + 2r_2.
\end{equation}
One way to determine the numbers $r_1$ and $r_2$ is as follows: if $K$ is $\mathbb{Q}(\alpha)$ (defined as the smallest field containing both $\mathbb{Q}$ and $\alpha$), and $P_{\alpha,\mathbb{Q}}(x)$ is the minimal polynomial of $\alpha$ over $\mathbb{Q}$, then $r_1$ is the number of its roots that are real, and $2r_2$ is the number that are complex.

{\bf Dirichlet's Unit Theorem}, then tells us that the unit group $\mathcal{O}_{K}^\times$ of the number field $K$ contains exactly $r = r_1 + r_2 - 1$ generators $\{\varepsilon_i\}_{i=1,\dots,r}$, such that any unit can be written uniquely in the form
\begin{equation}
  \zeta \varepsilon_1^{m_1} \dots \varepsilon_{r}^{m_r}
\end{equation}
with $m_{i}\in\mathbb{Z}$, and $\zeta$ a root of unity in $\mathcal{O}_{K}^{\times}$. As an abstract group, $\mathcal{O}^\times_K \simeq \mu(\mathcal{O}) \times \mathbb{Z}^{r}$; $r$ is called the {\bf rank} of $\mathcal{O}_K$, $\mu(\mathcal{O})$ is called the {\bf torsion group} and the set $\{ \varepsilon_i\}_{i=1,\dots,r}$ are the {\bf fundamental units} of $K$.

\subsection{Symmetries, Subfields and Towers of Fields}
\label{subsec:symmetriesandsubfields}

Let $K$ be a splitting field over $\QQ$. An automorphism $\sigma:K\to K$ is a bijective field homomorphism that fixes the base field element-wise (i.e., $\sigma(q)=q$ for all $q\in \QQ$). The set of all such automorphisms forms the Galois group, $\mathrm{Gal}(K/\QQ)$.

The foundational result of Galois theory is a bijective correspondence between the subgroups of $\mathrm{Gal}(K/Q)$ and the intermediate subfields of $K$. For any subgroup $H \subseteq \mathrm{Gal}(K/\QQ)$, the set of elements in 
$K$ left invariant by every automorphism in $H$ forms a subfield $K^H$, known as the fixed field. Conversely, every intermediate field $\QQ\subseteq F \subseteq K$ corresponds to a symmetry group $\mathrm{Gal}(K/F)$.

Let's see some examples that will be relevant for our purposes. 

First, the \textbf{golden field} 
$$\QQ(\phi)  = \{ a+b\phi : a,b \in \QQ \}  
$$
with $\phi = (1+\sqrt{5})/2$ being the golden ratio is a quadratic extension. The minimal polynomial of $\phi$ and its conjugate root $\Bar{\phi} = (1-\sqrt{5})/2$ is $P_{\phi,\QQ}(x) = x^2 - x - 1$. The Galois group contains a single automorphism, $\sigma : \sqrt{5} \to -\sqrt{5}$, permuting the two roots $\sigma (\phi) = \bar{\phi}$; thus, $\mathrm{Gal}(\QQ(\phi)/\QQ) \simeq \ZZ_2$. For any element $x \in \QQ(\phi)$, its trace and norm are $\Tr_{\QQ(\phi)/\QQ} (x) = 2a$ and $\mathcal{N}_{\QQ(\phi)/\QQ}(x) = a^2 - 5b^2$; the condition $\mathcal{N}_{\QQ(\phi)/\QQ}(x) = \pm 1$ identifies the units in $\QQ(\phi)$, which are $\pm \phi^k$ for any $k \in \ZZ$. 

Second, consider the field $\QQ(2\cos 2\pi/30)$, with $\cos( 2\pi/30) = (\zeta_{30} + \zeta_{30}^*)/2$, $\zeta_{30} = e^{2\pi i /30}$ being a 30-th primitive root of unity. Since 
$$
2\cos 2\pi/30 = -\frac{\bar{\phi}}{2} + \frac{1}{2}\sqrt{3\phi\sqrt{5}}, 
$$
contains the square root of a square root, 
$\QQ(\cos 2\pi/30)$ is an extension of degree 4. Indeed, the minimal polynomial for $\xi = 2\cos 2\pi/30$ is
$$ P_{\xi,\QQ}(x) = x^4 + x^3 - 4x^2 - 4x + 1,
$$
and its conjugates are $\xi_7 = -\phi/2 + \sqrt{-3\bar{\phi}\sqrt{5}}/2, \xi_{11} = -\bar{\phi}/2 - \sqrt{3\phi\sqrt
5}/2$ and $\xi_{13} = -\phi/2 - \sqrt{-3\bar{\phi}\sqrt{5}}/2 $. 
The automorphisms of $\QQ(\xi)$ are 
$$
\sigma_k(\xi) = 2\cos (2k \pi/30) \quad \text{ with } k \in \{ 1,7,11,13\}
$$
i.e.~$k$ is relatively prime to $15$ and permute the roots of $P_{\xi,\QQ}(x)$. 

Now, let $K_4 = \QQ(2\cos 2\pi/30)$ and $K_2 = \QQ(\phi)$. $K_2$ is a subfield of $K_4$: notice that $\phi = 2\cos 3\pi/15$. In addition, $K_2$ is fixed by the automorphism $\sigma_{11}$: 
$$ \sigma_{11}(\phi) = \sigma_{11}(\sigma_3 (\xi) ) = 2\cos 33\pi/15 = 2 \cos 3\pi/15 = \phi
,$$ 
Consequently, $\rho = \sigma_{11}$ is a non-trivial automorphism of $K_4$ that leaves $K_2$ invariant; it belongs to $\mathrm{Gal}(K_4/K_2)$, which is of order 2 because $K_4/K_2$ is a degree-2 extension. The latter follows from the so-called \textit{tower law} (multiplicativity of the degree): if \(K\) is an intermediate field between \(F\) and \(E\) (\(F \subseteq K \subseteq E\)), then \(E\) is an extension of \(K\), and \(K\) is an extension of \(F\). The tower law governs the degree of extensions in this tower: \([E:F] = [E:K] \times [K:F]\). Moreover, $\sigma_7 \in \mathrm{Gal}(K_4/\QQ)$ fixes $\QQ$ but flips $\phi$: 
$$ \sigma_7(\phi) = \sigma_7 (\sigma_3(\phi)) = 2 \cos 21\pi/15 = 2 \cos(7\pi/5) = \bar{\phi}.$$
As a result, the fixed field by the automorphism $\sigma = \sigma_7$ inside $K_2$ is just $\QQ$. This way we have ended up with the following tower of fields: 
\begin{equation*}
    \QQ \subset K_2 = \QQ(\sqrt{5}) \subset K_4 = \QQ \big( \cos 2\pi/30 \big)
\end{equation*} 

Now this is a very important tower of fields for quasicrystal-building because it relates to the icosahedrally symmetric groups $E_8, H_4$ and $I_2^{30}$ (see Section \ref{ch:MatBackgroundPartII}) and explain why the sit so nicely inside one another. $E_8$ is crystallographic: it defines a lattice in 8 dimensions. It is a rank-8 lattice (or $\ZZ$-module) over $\QQ$. $H_4$ is non-crystallographic: it is a rank-4 module over $\QQ(\phi)$ which is itself a degree-2 extension over $\QQ$. Some highly non trivial mathematics like $4 \times 2 = 8$ tell us that $H_4$ lifts into a lattice -- the $E_8$ lattice -- in 8 dimensions. We will see this in Appendix \ref{appendix:FoldingUnfolding}. The 8D space decomposes as 
$$ \mathbf{8} = \mathbf{4} \oplus \mathbf{4}^\rho$$
under the automorphism $\rho$. Finally, $I_2^{30}$ is a rank-2 module over $K_4$, a degree-4 extension over $\QQ$; again, since $2 \times 4 = 8$, it also lifts to the same lattice in 8 dimensions. The 8D space decomposes as 
$$
\mathbf{8} = \mathbf{2} \oplus \mathbf{2}^{\sigma} \oplus \mathbf{2}^\rho \oplus \mathbf{2}^{\sigma \rho}
$$
where bolded numbers denote the dimension of a representation and the exponents are the Galois automorphisms that act on the original representation. The $\mathbf{2}$-representation corresponds to the 2D Coxeter planes, while the $\mathbf{4}$ to the space on which $H_4$ gets its reflection representation.  

We can also see the above in terms of polynomial factorizations. Start with the Coxeter polynomial for $E_8$
$$ 
\begin{aligned}
     \chi_C(x) &= \Phi_{30}(x) = x^8 + x^7 - x^5 - x^4 - x^3 + x + 1 \\[1ex]
    &= (x - \zeta_{30})(x - \zeta_{30}^{29}) (x - \zeta_{30}^{7})(x - \zeta_{30}^{23})(x - \zeta_{30}^{11})(x - \zeta_{30}^{19})(x - \zeta_{30}^{13})(x - \zeta_{30}^{17}) \\[1ex]
    & = x^4 (y^4 + y^3 - 4y^2 - 4y + 1), \qquad y = x + x^{-1}. 
\end{aligned} 
$$
and combine its linear factors pairwise in the order shown above to get 
$$
\chi_C(x) = (x^2 - f_1 x + 1) (x^2 - f_2 x + 1) (x^2 - h_1 x + 1) (x^2 - h_2 x +1)
$$
where $$f_{1,2} = \tfrac{1}{2}\big( -\phi \mp \sqrt{3\phi\sqrt{5}} \big) \text { and } h_{1,2} = \tfrac{1}{2}\big( -\bar{\phi} \mp \sqrt{3\phi\sqrt{5}} \big)$$
Note that these are the solutions of the $y$-polynomial and they all belong to the interval $[-2,2]$ -- compare that to the discussion in Section \ref{app:inflation}. The first quadratic factor in the above is the Coxeter polynomial for $\mathrm{I}_2^{30}$ and the rest are its Galois-conjugates. Combining now the quadratic terms in the order shown, we arrive (using $f_1 + f_2 = -\phi = - (2 + f_1 f_2)$ and $h_1 + h_2 = - \bar{\phi} = - 2 + h_1 h_2$)
$$
\chi_C(x) = (x^4 + \phi x^3 + \phi x^2 + \phi x + 1) (x^4 + \bar{\phi} x^3 + \bar{\phi} x^2 + \bar{\phi} x + 1)
$$
with the second quartic factor being the Coxeter polynomial for $H_4$. 

In fact, the field $K_4$ is itself a subfield of an even larger field, $K_8 = \QQ(\zeta_{30})$. For any primitive $n$-th root of unity, $\zeta_n = e^{2\pi i/n}$, the field extension $\QQ(\zeta_n)$ is the $n$-th cyclotomic field of degree $\phi(n)$ (Euler's totient function). These extensions are very well-studied within the quasicrystal community (see \S 2 in \cite{baake2013aperiodic}) because they are associated with all finite Coxeter groups whose Coxeter element is of finite order (\ref{ch:MatBackgroundPartII}). The Galois group consists of automorphisms of the form $\sigma_k : \zeta_n \to \zeta_n^k$ where $\gcd(k,n) = 1$: that is, $\mathrm{Gal}(\QQ(\zeta_n)/\QQ) \simeq (\ZZ/n\ZZ)^\times$, the multiplicative group with $\phi(n)$ elements. An immediate symmetry of this field is complex conjugation, which corresponds to the automorphism $\ast : \zeta_n \to \zeta_n^{-1} = \zeta_n^*$. By the Galois correspondence, the subgroup $H$ that contains $\ast$ and $\ast \circ \ast = \mathrm{id}$ yields a fixed field $\QQ(\zeta_n)^H$. The elements invariant under $\ast$ are purely real; this subfield is its \textbf{maximal real subfield}, whose generator is $\zeta_n + \zeta_n^\ast = 2 \cos (2\pi/n)$.   

The \textit{maximal totally real subfield} $K^+$ of any field $K$ is very important because it has the property that $K$, no matter how higher-degree extension it is over $\QQ$, it becomes a quadratic extension over $K^+$, i.e. $[K : K^+ ] = 2$ always over its maximal real subfield. As we shall see in the next sections of this appendix, the maximal real subfield turns out to be the most important one for quasicrystals, as it is associated with the Coxeter planes. For example, for the Elser-Sloane quasicrystal, we do not need to deal with $\QQ(\zeta_{30})$ of degree $\phi(30)=8$, but with its maximal real subfield, $\QQ(2\cos 2\pi/30) = K_4$.  

\subsection{Extensions by Cyclotomic or Salem Numbers: Spherical vs Hyperbolic}

As mentioned already, the finite Coxeter groups with their finite-order $h$ Coxeter element $C$ always give rise to a cyclotomic field $\QQ(\zeta_h)$. This is because the eigenvalues of $C$ are always the primitive $h$-th roots of unit, and come in complex-conjugate pairs (Section \ref{ch:MatBackgroundPartII}); these always belong on $\Sph^1$ and are all complex. On the other hand, though, a hyperbolic Coxeter element's eigenvalues also come in pairs but crucially, there is a real pair associated with a Salem number (see Section \ref{subsec:Hyperbolic_Coxeter_Elements}). This drastically changes some structural properties of the associated field extensions.

Salem numbers $\theta$ are real algebraic integers greater than 1 whose other Galois conjugates consist of $1/\theta$ and pairs of complex-conjugate eigenvalues lying on $\Sph^1$; their relevance for Coxeter theory has been pointed out several times in the literature \cite{sacépée2025salemnumbers4937,mcmullen2002coxeter,mukunda2007pisot,Smyth_2015,Smyth_SalemReview}. The field extensions they produce, though, are rather different than the cyclotomic extensions we encounter so far. The most important difference is that they are \textit{not of the Galois-type} because they fail to be normal. Consider a Salem number $\theta$ and its field extension $\QQ(\theta)$. Because $\QQ(\theta)$ is purely real, it cannot contain all the complex roots of the minimal polynomial for $\theta$. This means that the extension $\QQ(\theta)$ is not normal because the rational extension $\QQ(\theta)$ fails to be the splitting field for the characteristic polynomial of $\theta$. 

However, the rational field extension $\QQ(\theta)$ becomes normal over its maximal real subfield, $\QQ(\theta + \theta^{-1})$. This is why the term \textit{Salem extension} refers to the quadratic extension $\QQ(\theta)/\QQ(\theta+\theta^{-1})$ and not to $\QQ(\theta)$ itself \cite{ghate2001arithmetic}. And indeed, this is the relevant field for us, as we will see in Section \ref{app:inflation}. The extension $\QQ(\theta)/\QQ(\theta + \theta^{-1})$ \textit{is} Galois; its only non-trivial automorphism is the one that swaps the two elements of the pair, i.e.~$\theta \to \theta^{-1}$, thus preserving the base field and acting as an automorphism of $\QQ(\theta)$. For now, we note that there exist an automorphism, let's call it $\omega$, that achieves the following map:
\begin{equation}
    \omega  : \theta_0 = \theta + \theta^{-1} = e^{\vartheta_0} + e^{-\vartheta_0}  = 2 \cosh \vartheta_0 \longmapsto \theta_i =  e^{i \vartheta_i} + e^{- i \vartheta_i} = 2\cos\vartheta_i
    \label{eq:GaloisWickRotation}.
\end{equation}
Since $\theta + \theta^{-1} > 2$ and the other roots sit on $\Sph^1$ (they are constrained $-2 < \theta_i < 2$), the automorphism $\omega$ practically maps a real number strictly larger than 2 into some number inside $(-2,2)$. In other words, it turns a real $\theta_0$ into some imaginary $i \theta_i$, which turns the hyperbolic cosine $\cosh$ into a normal cosine $\cos$. This idea will be used in Section \ref{subsec:NCSymmetryGeneralsCNP}. 

But now the question is, where does this automorphism that exchanges the ``traces'' of the pairs of roots $\{ \theta_0,\theta_1, \theta_2 \dots \}$ live in? Is it an automorphism of $\mathrm{Gal}(\QQ(\theta)/\QQ(\theta + \theta^{-1})$? Well, it cannot be, because $| \mathrm{Gal}(\QQ(\theta)/\QQ(\theta + \theta^{-1})| = 2 $ and we already discussed its automorphism. To find where it belongs to, one must pass onto a larger field, known as the \textbf{Galois closure} of $\QQ(\theta)$, which contains both real and imaginary roots \cite{christopoulos2010galois}. This is different for the cyclotomic fields, where the automorphisms that mapped the different Coxeter planes to one another where actually living within the original rational extension (see, for example, the case of $H_4$ and the decomposition of $\mathbf{8}$ from the previous section). But nevertheless, such an automorphism is guaranteed to exist, but one needs to be careful to define in which field it lives naturally.

\section{Minkowski Embedding: Fields as Lattices}
\label{sec:GeometryOfNumbers}

Here, we emphasize a significant connection -- due to Minkowski (\S 3 in \cite{baake2013aperiodic}) -- between number fields and lattices that makes the geometry of numbers apparent.

We saw in our preliminaries section that within any number field $K$ there exist its ring of integers $\mathcal{O}_K$ which has the natural algebraic structure of a free $\ZZ$-module. A lattice -- as defined in Section \ref{ch:MatBackgroundPartII} -- is nothing but the integer span of some basis, again, a $\ZZ$-module. Mapping a number field to a lattice i appropriate dimensions is formalized by the Minkowski embedding, a construction connecting number theory and geometry. 

\subsection{The Standard One: the Diagonal Embedding}

We also mentioned that an extension $K$ of degree $d_K$ has, in total, $r_1$ real and $2r_2$ complex embeddings. For example, the golden field $\QQ(\phi)$ has $r_1 = 2$ real embeddings and no complex embeddings, a cyclotomic field $\QQ(\zeta_n)$ has $r_2 = \phi(n)$ and no real embeddings, whereas an extension by a Salem number $\QQ(\theta)$ has $r_1 = 2$ real embeddings \textit{and} $r_2 = (d_K - 2)/2$ pairs of complex-conjugate embeddings. 

One can gather all these embeddings into a map, called the \textbf{Minkowski embedding}; letting the $r_1$ real embeddings be $\sigma_i : K \hookrightarrow \RR$ and the $2r_2$ complex embeddings be $\rho_i : K \hookrightarrow \CC$, the Minkowski embedding $\mathfrak{M}$ maps an element of $K$ into a $d_K$-dimensional vector space over $\RR$ 
\begin{equation}
    \mathfrak{M}: K \ni x \mapsto \big( x , \sigma_1(x), \dots, \sigma_{r_1}(x), \mathrm
    {Re}(\rho_i(x)), \mathrm{Im}(\rho_i(x)), \dots \big)^T \in \RR^{r_1} \oplus \CC^{r_2} \simeq \RR^{r_1} \oplus \RR^{2r_2} 
    \label{eq:MinkowskiEmbeddingDiagonal}
\end{equation}
The signature of the embedding is given by $(r_1,r_2)$. In particular, inside any number field $K$ hides its ring of integers $\mathcal{O}_K$, which, under $\mathfrak{M}$, embeds into a $d_K$-dimensional discrete lattice $\Lambda_K$. 

The geometry of this lattice encapsulates the field's algebraic invariants. For instance, the volume of the fundamental domain of $\Lambda_K$ is proportional to $\sqrt{|\Delta_K|}$, where $\Delta_K$ is the discriminant of the field \eqref{eq:NumberFieldDiscriminant} Using the field trace \eqref{eq:NumberFieldTrace}, we can obtain the Gram matrix of inner products between the lattice basis' elements of $\Lambda_K$. This is because the trace gives rise to the metric on the vector space $V = \RR^{r_1} \oplus \RR^{2r_2}$. If the metric is $g : V \times V \to \RR$, we have that $g(x,y) = \Tr_{K/\QQ}(xy)$, establishing a neat connection to geometry.

There are several examples of Minkowski embeddings of different rings of integers worked out in \cite{baake2013aperiodic}, but for concreteness, and for future reference, let us work out the embedding of $\ZZ(\phi) = \mathcal{O}_{\QQ(\phi)}$ into a lattice in $\EE^2$. First, notice that $\phi \ZZ(\phi) = \ZZ(\phi)$ (as $\phi^2 = \phi + 1$), so $\ZZ(\phi)$ is a $\ZZ(\phi)$-module (of rank 1) \textit{and} a $\mathbb{Z}$-module of rank 2 (with integral basis $\{ 1, \phi \}$). It can be embedded into a lattice in 2-dimensional Euclidean space by considering the Galois automorphism $\sigma$ of $\QQ(\phi)$ and using Eq.~\eqref{eq:MinkowskiEmbeddingDiagonal}:
$$ \Lambda_K = \{ (x,\bar{x})^T \in \EE^2 : x \in \ZZ(\phi) \}
$$
Under this map, the integral basis of $\ZZ(\phi)$ is mapped to the lattice basis 
$$
1 \mapsto {\bf b}_1 = 
\begin{pmatrix}
1 \\
1 
\end{pmatrix}
\quad \text{ and }
\quad
\phi 
\mapsto 
{\bf b}_2 = 
\begin{pmatrix}
    \phi\\
    \bar{\phi} = 1 - \phi
\end{pmatrix}, \qquad B = ( {\bf b}_1 | {\bf b}_2),
$$
giving rise to the following Gram matrix 
$$
G_\Lambda = B B^T =  
\begin{pmatrix}
2 & 1\\
1 & 3
\end{pmatrix}
= \begin{pmatrix}
\Tr_{\QQ(\phi)/\QQ}({\bf b}_1^T {\bf b}_1) & \Tr_{\QQ(\phi)/\QQ}({\bf b}_1^T {\bf b}_2)\\
\Tr_{\QQ(\phi)/\QQ}({\bf b}_2^T {\bf b}_1) & \Tr_{\QQ(\phi)/\QQ}({\bf b}_2^T {\bf b}_2)
\end{pmatrix} 
= \begin{pmatrix}
g(1,1) & g(1,\phi)\\
g(\phi,1) & g(\phi,\phi)
\end{pmatrix}.
$$
While this is a lattice in $\EE^2$, it has no symmetry. There are two symmetric lattices in 2D: square and triangular (hexagonal), and our $\Lambda_K$ is neither of them. Thankfully, we can modify our embedding prescription by a bit to get a lattice with an improved symmetry.

\subsection{The One with the Twist: Embedding of Maximal Symmetry}

Imagine that we perform a relative rescaling of the two coordinates in the basis ${\bf b}_{1,2}$ by some real number $\tau$ to get the modified basis 
$$ 
B(\tau) = \begin{pmatrix}
    1 & 0 \\
    0 & \tau
\end{pmatrix} 
\begin{pmatrix}
    1 & \phi\\
    1 & 1 - \phi
\end{pmatrix} = 
\begin{pmatrix}
    1 & \phi \\
    \tau & \tau (1-\phi)
\end{pmatrix}.
$$
This leads to the modified Gram matrix
$$
G_\Lambda^{(\tau)} = 
\begin{pmatrix}
1+\tau^2 & \phi + \tau^2 (1-\phi) \\
\tau^2 (1-\phi) & (1 +\phi) + \tau^2 (2-\phi)
\end{pmatrix}.
$$
The two symmetric lattices have Gram matrices
$$ 
G_{\rm square} \sim \mathbb{I} \quad \text{and} \quad G_{\rm triangular} \sim \begin{pmatrix}
    2 & 1\\
    1 & 2
\end{pmatrix}
$$
Attempting to make the diagonal elements of $G_\Lambda^{(\tau)}$ equal, we find that for $\tau = \phi$, we obtain $G_{\Lambda}^{(\tau)} = (\phi + 2)\mathbb{I}$, thus $\Lambda^{(\tau)}$ becomes a scaled square lattice with a 4-fold rotational symmetry. This is the embedding of $\ZZ(\phi)$ with \textbf{maximal symmetry}. 

To conclude, for the ring of integers $\ZZ(\phi)$ to embed into a lattice in $\EE^2$ with the maximal possible symmetry, we need to perform a relative rescaling of the two components coming from the standard Minkowski prescription \eqref{eq:MinkowskiEmbeddingDiagonal}, which leads to an overall scaling of the resulting Gram matrix of inner products. This idea will be explored further in Section \ref{appendix:FoldingUnfolding}, where we treat more general embeddings of non-crystallographic symmetries into lattices. 

\section{Scale Factors as Algebraic Numbers}
\label{app:inflation}

In this section, we explain some points about how to determine a quasicrystal's scale factors algebraically from an associated number field; and we determine this number field for the self-dual Lorentzian lattices up to dimension $d=10$.

Consider an \sCNP{} quasicrystal obtained by starting from a lattice $\Lambda$, constructing its Coxeter element $C$, and using the non-crystallographic subgroup $G\in {\rm Aut}(\Lambda)$ generated by $C$ to decompose $V$ as $V=V_{\rm ph}\oplus V_{\rm in}$.  

The characteristic polynomial of $C$ -- called the {\bf Coxeter polynomial} $\chi_{C}(\lambda)$ -- is a polynomial with integer coefficients, which we take to be irreducible ({\it i.e.}\ not factorizable over $\mathbb{Q}$), which corresponds to the fact that $G$ is non-crystallographic.  Let $K$ denote the {\bf splitting field} of $\chi_{C}(\lambda)$ -- {\it i.e.}\ the smallest field extension of $\mathbb{Q}$ over which $\chi_{C}(\lambda)$ ``splits'' ({\it i.e.} decomposes into linear factors).  And let $K'$ denote the maximal real subfield of $K$ ({\it i.e.}\ the largest subfield of $K$ contained in $\mathbb{R}$). The (fundamental) units of $K'$ will be the quasicrystal's (fundamental) inflation factors.

For example, consider the Ammann-Beenker tiling (see Sec. \ref{subsec:AmmannBeenkerTilingsExample}): the Coxeter polynomial $\chi_C(\lambda)\in \mathbb{Z}[\lambda]$ reads
\begin{equation*}
\begin{aligned}
\chi_C(x) &= \lambda^4+1 = (\lambda - \zeta_8)(\lambda-\zeta_8^*)(\lambda-\zeta_8^3)(\lambda-\zeta_8^{3*})\\
& = (\lambda^2 - 2\mathrm{Re}(\zeta_8) \lambda + 1)(\lambda^2 - 2\mathrm{Re}(\zeta_8^3) \lambda + 1)
\end{aligned}.
\end{equation*}
where $\zeta_{8}={\rm exp}(2\pi i/8)$.  We see that $\chi_{C}(\lambda)$ cannot be factored over $\mathbb{Q}$, can be fully factored over the splitting field $K=\mathbb{Q}(\zeta_{8})$ (a degree four extension of $\mathbb{Q}$), and can be partly factored over its maximal real subfield $K^+=\mathbb{Q}(\zeta_{8}+\zeta_{8}^{\ast})=\mathbb{Q}(\sqrt{2})$.  The inflation factors of the corresponding quasicrystal will be the units of $K^+$; for the Ammann-Beenker tilings, the fundamental inflation factor -- {\it i.e.}\ the fundamental unit of $K^+=\mathbb{Z}(\sqrt{2})$ -- is the silver ratio $\epsilon=(1+\sqrt{2})$, and the general unit may be expressed as $\pm(1+\sqrt{2})^{m}$ ($m\in\mathbb{Z}$).

In fact, for all the most well-known and widely-studied Euclidean quasicrystals, the associated real algebraic number fields $K+$ are also quadratic fields $\mathbb{Q}(\sqrt{D})$ -- and for such fields, the units always have this simple structure (that, up to a sign, they are simply all integer powers of a single fundamental unit $\lambda$). Once we turn to spacetime quasicrystals, we will (usually) no longer be dealing with quadratic field extensions, but with more complicated field extensions (of higher degree, and with more than one fundamental unit); but otherwise, the above considerations still apply.

For illustration, consider the spacetime quasicrystal obtained via \sCNP{} from the odd self-dual lattice $\mathrm{I}_{3,1}$ (see Section \ref{subsec:2DLorentzianQuasiCrystal}).  The Coxeter polynomial $\chi_{C}(\lambda)$ is
\begin{eqnarray*}
\chi_C(\lambda)&=&\lambda^4 - \lambda^3 - 2\lambda^2 - \lambda + 1 \\
&=&(\lambda-\lambda_{0}^+)(\lambda-\lambda_{0}^-)(\lambda-\lambda_1^+)(\lambda-\lambda_1^-)
\end{eqnarray*}
where two of these roots ($\lambda_{0}^{\pm}$) are real, two of them ($\lambda_{1}^{\pm}$) are complex, and the larger real root $\lambda_{0}^{+}$ is a Salem number.  All four roots lie in the splitting field $K=\mathbb{Q}(\lambda_{1}^{+})$ ({\it i.e.}\ the smallest field containing both $\mathbb{Q}$ and one of the complex roots of $\chi_{C}$ -- {\it e.g.} $\lambda_{1}^{+}$).  However, finding the exact form of $\lambda_{1}^{+}$ would require solving a relatively high-order polynomial equation (in this simple example it is just a quartic equation, which is still possible to solve, but higher-dimensional examples would involve higher-order equations that are impossible to solve in closed form).  

However, if we are just interested in our quasicrystal's scale factors, then solving this higher-order equation is unnecessary since, as explained above, we don't need the full splitting field $K$, but rather its maximal real subfield $K^+$.  The subfield $K^+$ is not generated by a single root (like $\lambda_{1}^{+}$), but rather by the sum of a pair of such roots: $K^+=\mathbb{Q}(\mu_{j})$ where $\mu_{j}\equiv\lambda_{j}^{+}+\lambda_{j}^{-}$; and solving for $\mu_{j}=\lambda_{j}^{+}+\lambda_{j}^{-}$ is easier (requires solving a lower-order polynomial) than solving for either $\lambda_{j}^{+}$ or $\lambda_{j}^{-}$ separately.  

To see this, note that $\chi_{C}(\lambda)$ is (i) monic, (ii) irreducible over $\mathbb{Q}$, and (iii) palindromic ({\it i.e.}\ its coefficients $a_0,\ldots,a_4$ satisfy $a_i=a_{n-i}$), so we can write it as
\begin{equation}
  \chi_{C}(\lambda)=\lambda^{2}Q(\mu)\qquad\mu\equiv\lambda+\frac{1}{\lambda}
  \label{eq:ReducedFactoredCoxeterPoly}
\end{equation}
where
\begin{equation}
    Q(\mu)=\mu^{2}-\mu-4
\end{equation} 
is a lower order polynomial with roots
\begin{eqnarray*}
  \mu_{0}&=&\lambda_{0}^{+}+\frac{1}{\lambda_{0}^{+}}
  =\lambda_{0}^{+}+\lambda_{0}^{-}=\frac{1+\sqrt{17}}{2}>2, \\
  \mu_{1}&=&\lambda_{1}^{+}+\frac{1}{\lambda_{1}^{+}}
  =\lambda_{1}^{+}+\lambda_{1}^{-}=\frac{1-\sqrt{17}}{2}<2.
\end{eqnarray*}
If $C$ is a hyperbolic Coxeter element, $Q(\mu)$ will have exactly one root larger than 2; otherwise, if $C$ was the Coxeter element for a spherical group, for example, or roots would be contained in $(-2,2)$ \cite{Smyth_2015,ghate2001arithmetic}. This larger $\mu$, which we denote by $\mu_0$, corresponds to the real pair of eigenvalues $\lambda_0^+ + \lambda_0^-$, which are absent in the non-hyperbolic cases. To find the quasicrystal's fundamental scale factors, we simply have to find the units of $K^+(\mu_0)$. Here, it is the fundamental unit of $\mathbb{Q}[\sqrt{17}]$ (namely, $4+\sqrt{17}$).   

We can analogously identify the roots $\mu_{j}$ and field $K^+=\mathbb{Q}[\mu_{0}]$ for $I_{s,1}$ up to dimensions $s+1=10$, as summarized in Table \ref{tab:FieldExtensionsHyperbolicCases}.\footnote{When working with reduced Coxeter polynomials like the one in Eq.~\eqref{eq:ReducedFactoredCoxeterPoly}, a useful recursion relation can be worked out to find the $(k+1)$-th term, if we know the previous two terms, 
\begin{equation*}
        x^{k+1} + x^{-(k+1)} = y \big( x^{k} + x^{-k} \big) - \big( x^{(k-1)} + x^{-(k-1)} \big), \quad k\geq 1
\end{equation*}
which means knowing the $0$-th and the $1$st terms (which are 2 and $y$ respectively), we can obtain all higher terms. For the results of Table \ref{tab:FieldExtensionsHyperbolicCases}, in particular, we needed all terms up to $k=5$.} In all cases, $\mu_{0}=\lambda_{0}^{+}+\lambda_{0}^{-}>2$, while the $\mu_{j}=\lambda_{j}^{+}+\lambda_{j}^{-}<2$ for $j\geq 1$.  But as we climb up the ladder of dimensions, the relevant field extensions $K$ and $K'$ become of higher and higher degree, with more and more fundamental units.  We have only shown the units for the first three cases (the simplest, quadratic cases), but the units in the remaining cases, may be computed from the given information via standard methods of algebraic number theory.

\begin{table}[h!]
    \scriptsize
    \centering
\begin{threeparttable}    
    \begin{tabular}{c || c | c | c | c}
         $\Lambda$ & $\chi_C(\lambda)$ & $\mu_0 \equiv \lambda_{0}^{+} + \lambda_{0}^{-} > 2$ & $ \mathbb{Q}(\mu_0)$ & Fund. Units \\ \hline \hline 
         $\textrm{I}_{2,1}$ & $- (\lambda+1)(\lambda^2 - 4\lambda + 1)$ & $2+\sqrt{3} \approx 3.73205$ & $\mathbb{Q}(\sqrt{3})$ & $2\pm\sqrt
         3$  \\[+1ex]
         $\textrm{I}_{3,1}$ & $\lambda^2 (\mu^2 - \mu - 4)$ & $\frac{1}{2}(1+\sqrt{17}) \approx 2.56155$ & $\mathbb{Q}(\sqrt{17})$ & $4\pm\sqrt{17}$ \\[+1ex]
         $\textrm{I}_{4,1}$ & $-\lambda^2(\lambda+1)(\mu^2 - \mu - 3)$ & $\frac{1}{2}(1+\sqrt{13}) \approx 2.30277$ & $\mathbb{Q}(\sqrt{13})$ & $\frac{1}{2}(3 \pm \sqrt{13})$\\[+1ex]
         $\textrm{I}_{5,1}$ & $\lambda^3 (\mu^3 - 4\mu - 2)$ & $ \frac{1}{3^{1/3}}\big( \beta_6^{1/3} + \beta_6^{* 1/3} \big) \approx 2.21432 \qquad \big(\beta_6 = \frac{1}{3}(9+i \sqrt{111})\big) $ & $\cdot$ & $\cdot$ \\[1ex]
         $\textrm{I}_{6,1}$ & $- \lambda^3(\lambda+1)(\mu^3-\mu^2-3\mu+1) $ & $\frac{1}{3}\big( 1 + \beta_7^{1/3} + \beta_7^{* 1/3}\big) \approx 2.17009 \qquad \big( \beta_7 = 1 + 3i\sqrt{111}\big) $ & $\cdot$ & $\cdot$ \\[1ex]
         $\textrm{I}_{7,1}$ & $\lambda^4(\mu^4 - 5\mu^2 - \mu + 4)$ & [see below \tnote{a} $\,$] $\qquad \big( \beta_8 = \frac{1}{2}(1217+3i\sqrt{8331})\big)$ & $\cdot$ & $\cdot$ \\[1ex]
         $\textrm{I}_{8,1}$ & $ -\lambda^4(\lambda+1)(\mu^4 - \mu^3 - 4\mu^2 + 2\mu + 3)$ & [see below \tnote{b} $\,$] $\qquad \big(\beta_9 = \frac{1}{2}(853+3i\sqrt{5871})\big)$ & $\cdot$ & $\cdot$ \\[1ex]
         $\mathrm{I}_{9,1}$ & $\lambda^5(\mu^2-2)(\mu^3-4\mu-1)$ & $ \frac{1}{3}\big( \beta_{10}^{1/3} + \beta_{10}^{* 1/3}\big) \approx 2.11491, \qquad \big( \beta_{10} = \frac{3}{2}(9+i\sqrt{687})\big)$ & $\cdot$ & $\cdot$ \\[1ex]
         $\mathrm{II}_{9,1}$ & $\lambda^5(\mu^5 + \mu^4 - 5\mu^3 - 5\mu^2 + 4\mu + 3) $ & [no closed form for the quintic] = 2.02642 & $\cdot$ & $\cdot$
    \end{tabular}
    \caption[The real field extension $K^+=\mathbb{Q}(\mu_{0})$ for $I_{s,1}$, up to dimension $s+1=10$.]{The real field extension $K'=\mathbb{Q}(\mu_{0})$ for $I_{s,1}$, up to dimension $s+1=10$.  We have only given the fundamental units for the simplest (first three) cases; the fundamental units for the remaining cases may be obtained by standard computational methods in algebraic number theory.}
    \label{tab:FieldExtensionsHyperbolicCases}
    \begin{tablenotes}
    \item[a] $$\mu_0 = \frac{1}{4}\bigg( 2\sqrt{\tilde{\beta}_8^+} + \sqrt{2} \sqrt{\tilde{\beta}_8^+ + 3 \tilde{\beta}_8^- + \frac{4}{\sqrt{\tilde{\beta}_8^+}}}  \bigg)\approx 2.14386, \qquad \tilde{\beta}_8^\pm = \frac{1}{3}\big(10 \pm (\beta_8^{1/3} + \beta_8^{* 1/3} )\big)$$
    \item[b] $$ \mu_0 = \frac{1}{4}\bigg(1 + \sqrt{\tilde{\beta}_9^+} + \frac{1}{\sqrt{2}} \sqrt{\tilde{\beta}_9^+ + 3\tilde{\beta}_9^- + \frac{4}{\sqrt{\tilde{\beta}_9^+}}}\bigg) \approx 2.12676, \qquad \tilde{\beta}_9^\pm = \frac{1}{3}\big( 35 \pm 4 (\beta_9^{1/3} + \beta_9^{* 1/3})\big)$$
    \end{tablenotes}
        \end{threeparttable}
\end{table}

\chapter{Folding and Unfolding Coxeter-Dynkin Graphs}
\label{appendix:FoldingUnfolding}

This generalizes, or perhaps elucidates, techniques already developed in  \cite{Dechant2016,Mehmet_Koca_2001,Moody_1993,Fring_2006,Fring_2005}. It uses some algebraic number theory, so the reader might find Appendix \ref{AppendixA-NumberTheory} useful.

\section{Graph-Origami Techniques}
\label{sec:GraphOrigami}

Given a Coxeter-Dynkin graph, what can you do to it to produce new graphs from it? Here, we are interesting in processes that either \textit{fold} a graph down to another graph, similarly to how one would fold a piece of paper along some crease and identify opposite sides, or \textit{unfold} a graph into some bigger graph, similarly to how an already folded piece of paper can be unfolded and the identified sides are now distinct. For this reason, we call the techniques developed here \textit{graph origami}.

\subsection{Folding}
\label{subsec:FoldingDefinitions}

Consider a finite Coxeter system $(G,S)$ of rank $|S| = n$ and its Coxeter-Dynkin graph; let $V$ be the $n$-dimensional space in which $G$ has its geometric reflection representation. Here, we follow the definitions given in Section \ref{ch:MatBackgroundPartII}.

A \textbf{folding partition} is a partition of the $n$ nodes of the graph into $b$ \textbf{disjoint blocks} $B_I$ such that each block $B_I$ contains a subset of original nodes which are mutually non-adjacent (i.e.~\textbf{not connected with an edge to one another}). Let $b_I = |B_I|$ be the number of nodes in the block $B_I$ (the size of the block); we write
\begin{equation}
    S = \{ R_1,R_2,\dots R_n \} = \bigsqcup_{I=1}^b B_I \qquad \text{ where } \qquad \sum_{i=1}^b b_I = n
    \label{eq:DisjointNodePartition}
    \end{equation}
The trivial partition is one in which each $B_I$ contains a single node (i.e.~we did not group any nodes together). A non-trivial partition will generally have more than one generators in at least one block $B_I$. An equal partition would have all blocks contain the same number of nodes, i.e. $b_I = n/b \,\, \forall I=1,\dots, b $.

For each block $B_I$, define the product of generators $R_i$ included in that block as
\begin{equation}
    \rho_I = \prod_{i \in B_I} R_i = R_{i_1} R_{i_2} \cdots R_{i_{b_I}}
    \label{eq:BlockGenerators}
\end{equation}
The partition into blocks $B_I$ guarantees that all nodes are mutually disconnected, which translates into all reflection generators $R_i$ for $i \in B_I$ being mutually commuting: $R_i R_j = R_j R_i$. This is a consequence of the fact that $m_{ij} = 2$ for all pairs in $B_I$. This condition guarantees that $\rho_I$ are \textbf{involutions}: 
\begin{equation}
\begin{aligned}
    \rho_I^2 &= \big( R_{i_1} \underbrace{R_{i_2} \cdots R_{i_{b_I}}\big) \big( R_{i_1}}_{\mathclap{\text{ pass $R_{i_1}$ to the left freely, till it hits $R_{i_1}$}}} R_{i_2} \cdots R_{i_{b_I}}\big)  = 
     \underbrace{R_{i_1}^2}_{\mathclap{=1}} \big (  R_{i_2} \cdots R_{i_{b_I}} \big) \big(  R_{i_2} \cdots R_{i_{b_I}} \big) = \cdots = 1 
    \end{aligned}
\end{equation}
where $\cdots$ indicate several similar steps of using the commutation relations. Thus, $\rho_I$'s are reflection generators. Note that this holds if and only if the reflections in each block $B_I$ commute. If one groups two adjacent nodes with $m_{ij} \neq 2$ inside the same $B_I$, the resulting $\rho_I$ is a rotation, not a reflection, and there will be no reflection group generated by these $\rho_I$'s to speak of. We refer to $\rho_I$ as the \textbf{folded reflection generators}. 

By definition, reflections must reflect a vector (a root ${\bf r}_i$) and stabilize pointwise a hyperplane (the codimension-1 mirror $M_i = {\bf r}_i^\perp$ orthogonal to that root). All $R_i$'s preserve the natural metric $S$ on $V$ and have determinant $-1$, as they only reflect their corresponding root ${\bf r}_i$. On the contrary, Eq.~\eqref{eq:BlockGenerators} yields $\det \rho_I = (-1)^{b_I}$, so unless the folding partition is trivial, $\rho_I$ reflects more than one vectors in $V$ and it is not a reflection there. 

We can see that by working in the root basis (see Section \ref{sec:SpacetimeCNPRootBasis}), where the original reflections act as $R_{i_1}({\bf r}_{i_2}) = {\bf r}_{i_2} - {\bf r}_{i_1} S_{{i_1}{i_2}}$. Each block $B_I$ contains only non-adjacent original roots, so 
\textbf{all pairs of roots are mutually orthogonal}, i.e.~$\langle {\bf r}_i,{\bf r}_j \rangle = S_{ij} = 0$ and only $S_{{ii}} = 2$. Letting $V_I = \mathrm{span}({\bf r}_i : i \in B_I)$ be the subspace spanned by only the roots in the block $B_I$, it follows that the restriction of $S$ on $V_I$, denoted as $S|_{V_I}$, is just $2 \mathbb{I}_{b_I}$, where $\mathbb{I}_{b_I}$ is the $b_I \times b_I$ identity matrix. $V_I$ is, then, non-degenerate and the space $V$ decomposes as $V = V_I \oplus V_I^\perp$. It is straightforward to show that $\rho_I$ acts as $-\mathbb{I}_{b_I}$ on $V_I$ while fixes $V_I^\perp$. Since each $R_i$ reflects exactly one simple root $\mathbf{r}_i$ of the original graph, $\rho_I$ will reflect $b_I$ such roots, one for each $R_i$ included in the block $B_I$. Letting $\rho_I$ act on an arbitrary element on $V_I$, we find 
\begin{equation}
    \begin{aligned}
    \rho_I \sum_{i \in B_I} c_i {\bf r}_i &= c_{i_1} \big( R_{i_1} R_{i_2} \cdots R_{i_{b_I}} \big) {\bf r}_{i_1} + c_{i_2}\big( R_{i_1} R_{i_2} \cdots R_{i_{b_I}} \big) {\bf r}_{i_2} + \cdots + c_{i_{b_I}}\big( R_{i_1} R_{i_2} \cdots R_{i_{b_I}} \big) {\bf r}_{i_{b_I}} = \\
    & =  c_{i_1}\big( R_{i_1} {\bf r}_{i_1}\big)  +  c_{i_2}\big( R_{i_2} {\bf r}_{i_2} \big)  + \cdots + c_{i_{b_I}} \big( R_{i_{b_I}} {\bf r}_{i_{b_I}} \big) = -\sum_{i \in B_I} c_i {\bf r}_i
    \label{eq:FoldedReflectionsReflectAnythingInTheBlock}
    \end{aligned}
\end{equation}
Thus, each $\rho_I$ has a $(-1)$-eigenspace which is $V_I$. On $V$, it acts as a \textit{block reflection}, reflecting the entire space $V_I$. To make it act as a proper reflection, we must restrict $V$ onto a subspace $U$ on which it only reflects a single vector. 

For the subspace $U$ (of dimension equal to the number of blocks $b$) to be a \textit{folding subspace} of $V$, we need it to satisfy the following three reasonable conditions: 
\begin{itemize}
    \item[\textbf{(I)}] $U$ is a non-degenerate subspace and the projection is $S$-orthogonal: this means that, if $\Pi_U$ denotes the $S$-orthogonal projector from $V$ onto $U$ ($\Pi_U^2 = \Pi_U$), then $S \Pi_U  = \Pi_U^T S$, and $U$ respects the natural metric $S$ on $V$.   
    \item[\textbf{(II)}] $U$ contains precisely one vector for each block $B_I$ that is reflected by the corresponding $\rho_I$, which means that \begin{equation}
        \dim ( U \cap V_I ) = 1 \,\, \forall I = 1,\dots, b.
    \end{equation}
    This vector is a linear-combination of the roots ${\bf r}_i$ for $i \in B_I$.  
    \item[\textbf{(III)}] 
    $U$ is $\rho_I$-invariant for all $I=1,\dots, b$: this means that reflecting elements of $U$ with the generators $\rho_I$ does not take us outside of $U$, i.e.~ $ \rho_I (U) = U$ and all $\rho_I$ respect the splitting $V = U \oplus U^\perp$. 
\end{itemize}
Since all $R_i$'s preserve the metric $S$ in $V$, it follows that any products of $R_i$'s, such as our folded generators $\rho_I$, do preserve the metric as well, namely
\begin{equation}
    R_i^T S R_i = S \,\, \forall i = 1,\dots, n \Rightarrow \boxed{\rho_I^T S \rho_I = S \,\, \forall I=1,\dots, b.} 
    \label{eq:BlockReflectionsPreserveMetric}
\end{equation}
Thus all $\rho_I$ are $S$-isometries and preserve the $S$-invariant subspace $U$, so they must also leave its $S$-orthogonal complement $U^\perp$ invariant. From Condition (I) and (III), we then get that the projectors $\Pi_U$ commute with the generators (we also say $\Pi_U$ ``intertwines'' the generators)
\begin{equation}
    \Pi_U \rho_I = \rho_I|_U \Pi_U \,\, \forall I = 1,\dots,b,
    \label{eq:ProjectorOnFoldingSpaceCommutesWithGens}
\end{equation}
so the generators $\rho_I$ preserve the splitting $V = U \oplus U^\perp$. Here, $\rho_I|_U$ is the folded reflection generator $\rho_I$ restricted into the subspace $U$ in which it only has 1 negative eigenvalue. 

We can now answer the basic question: \emph{what is the \textbf{folded root} $\beta_I$ of a block $B_I$, as a vector built from the original simple roots?}. Recall that the spectrum of each $\rho_I$ in $V$ is 
$$ (\underbrace{-1,-1, \dots, -1}_{\mathclap{\text{exactly $b_I$}}}, \underbrace{+1, +1 \cdots, +1}_{\mathclap{\text{exactly $n-b_I$}}})$$ 
since its $(-1)$-eigenspace is the entire $V_I$. Every direction in $V_I$ is treated equally by $\rho_I$: it gets reflected. What we are actually after is the unique direction inside $V_I$ that is reflected by $\rho_I|_U$, which is a reflection operator in the subspace $U$ and its spectrum is 
$$ (
\underbrace{-1}_{\mathclap{\text{\textit{exactly 1}}}}, \underbrace{+1, +1, \dots, +1}_{\mathclap{\text{ exactly $b-1$}}} 
) $$
The simplest vector that does the trick is the sum of all roots in $B_I$
\begin{equation}
    \alpha_I = \sum_{i \in B_I} {\bf r}_i
    \label{eq:SumRootsInBlock}
\end{equation}
but belongs entirely to the block-space $V_I$; assuming that the projection $\Pi_U$ does not completely kill the sum, we let instead
\begin{equation}
    \alpha_I|_U = \widetilde{\alpha_I}  = \Pi_U \alpha_I = \Pi_U \sum_{i \in B_I} {\bf r}_i=  \Pi_U \big ({\bf r}_{i_1} + {\bf r}_{i_2} + \cdots + {\bf r}_{i_{b_I}} \big).
    \label{eq:BlockRoots}
\end{equation}
The projection $\Pi_U$ guarantees that the resulting direction $\widetilde{\alpha}_I$ is the unique folding direction in $U \cap V_I$, the eigenvector of the block-restricted $\rho_I|_U$ with eigenvalue $-1$. \footnote{While the direction is unique, the eigenvector $\widetilde{\alpha}_I$ is not. Note that the $(-1)$-eigenspace of $\rho_I|_U$ is the full line $\RR \widetilde{\alpha}_I$; this is because we could have considered instead any $\RR$-linear combination of the roots $\mathbf{r}_i$ included in the block $B_I$, as shown in Eq.~\eqref{eq:FoldedReflectionsReflectAnythingInTheBlock}, and not the equal-sum of Eq.~\eqref{eq:BlockRoots}. However, we will eventually normalize this $\widetilde{\alpha}_I$ to have $\widetilde{\alpha}_I^2 = 2$, thus this overall factor becomes irrelevant.} The rest $b-1$ dimensions of $U$ lie in $V_I^\perp$, on which $\rho_I$ is the identity. 

We note that these positive eigenvectors of $\rho_I|_U$ lie in $U \cap V_I^{\perp}$, which is the mirror  hyperplane normal to $\widetilde{\alpha}$ inside the subspace $U$. It can be shown that all restricted reflections $\rho_I|_U$ preserve the induced metric $S|_U$, exactly as every $\rho_I$ preserves $S$. Take any two vectors ${\bf v,w} \in U$; since $U$ is $\rho_I$-invariant, $\rho_I {\bf v}$ is already in $U$, so acting with $\Pi_U$ does not affect it; then, from Eq.~\eqref{eq:ProjectorOnFoldingSpaceCommutesWithGens} we get that $\rho_I {\bf v} = \Pi_U \rho_I {\bf v} = \rho_I|_U \Pi_U{\bf v} = \rho_I|_U {\bf v}$ (similarly for $\bf w$). Thus, 
\begin{equation}
\underbrace{\langle \rho_I|_U {\bf v}, \rho_I|_U {\bf w} \rangle}_{\mathclap{\text{${\bf v,w}$ in $U$}}} = {\bf v}^T \rho_I|_U^T S|_U \rho_I|_U {\bf w} = \underbrace{\langle \rho_I {\bf v} , \rho_I {\bf w}\rangle}_{\mathclap{\text{$ {\bf u,v}$ lifted to $V$}}} = {\bf v}^T \rho_I^T S \rho_I {\bf w} \overset{Eq.~\eqref{eq:BlockReflectionsPreserveMetric}}{=} {\bf v}^T S {\bf w} = {\bf v}^T S|_U {\bf w}
\label{eq:RestrictedReflectionsPreserveRestrictedMetric}
\end{equation}
As a result, $\rho_I|_U$ are $S|_U$-preserving involutions. The Coxeter group $\widetilde{G}$ of rank $b$ generated by the folded reflections $\rho_I|_U$ in the mirrors defined by the folded roots $\alpha_I|_U$ is the \textbf{folded Coxeter group}. The reflections are given concretely as
\begin{equation}
    \rho_I|_U {\bf v} = {\bf v} - \langle {\bf v},\widetilde{\beta_I}\rangle \widetilde{\beta_I}
    \label{eq:FoldedReflections}
\end{equation}
where we have normalized $\widetilde{\alpha}_I \mapsto \widetilde{\beta}_I$ with $\widetilde{\beta}_I^2 = \langle \widetilde{\beta}_I, \widetilde{\beta}_I \rangle = 2$. We will see how to do this shortly. 

The natural next question is: what is the \textbf{folded Coxeter graph} generated by the folded generators $\rho_I|_U$? To find this, we need to calculate the \textbf{folded Schl\"{a}fli matrix} $\widetilde{S}$, which is just the matrix of inner products between the folded roots $\widetilde{\alpha}_I$
\begin{equation}
    \widetilde{S}_{IJ} = \langle \widetilde{\alpha}_I , \widetilde{\alpha}_J \rangle = \widetilde{\alpha}_I^T S|_U \widetilde{\alpha}_J, 
\end{equation}
in the normalization chosen. $\widetilde{S}$ is just the restricted metric $S|_U$, which we can express as an $n\times n$ matrix (acting on all of $V$ but blind to $U^\perp$) as 
\begin{equation}
    S|_U = S \Pi_U = \Pi_U^T S 
    \label{eq:RestrictedSchlafliAsFoldedSchlafli}
\end{equation}
which is symmetric (since $S\Pi_U = \Pi_U^T S = (S\Pi_U)^T$ by $S$-self-adjointness), has rank $b = \dim U$, and has kernel exactly $U^\perp$ (as $\Pi_U \Pi_{U\perp} = 0$). 

First of all, let's see that we can actually enforce the normalization $\widetilde{\beta}_I^2 = 2$. Using Eq.~\eqref{eq:BlockRoots}, we get the general formula:
\begin{equation}
    \langle \widetilde{\alpha}_I, \widetilde{\alpha}_J \rangle = \alpha_I^T S \Pi_U \alpha_J= \left(\sum_{i \in B_I} {\bf r}_i^T \Pi_U^T\right) S|_U \left(\Pi_U \sum_{j \in B_J} {\bf r}_j\right) = \sum_{i \in B_I, \,\,\, j \in B_J} {\bf r}_i^T S \Pi_U {\bf r}_j 
\end{equation}
where ${\bf r}_i^T (\text{ some matrix }) {\bf r}_j$ picks out the $(i,j)$-th entry of any matrix, yielding, 
\begin{equation}
    \langle \widetilde{\alpha}_I, \widetilde{\alpha}_J \rangle = \alpha_I^T S \Pi_U \alpha_J= \sum_{i \in B_I, \,\,\, j \in B_J} (S\Pi_U)_{ij} = \sum_{i \in B_I, \,\,\, j \in B_J} S_{ij} - \sum_{i \in B_I, \,\,\, j \in B_J} (S\Pi_{U^\perp})_{ij}.
    \label{eq:InnerProductFoldedRoots}
\end{equation}
Here we have used $\Pi_U = 1 - \Pi_{U^\perp}$ for our orthogonal projectors. Setting $I=J$, we first note that $\langle \widetilde{\alpha}_I, \widetilde{\alpha}_I \rangle > 0$ because $\widetilde{\alpha}_I$ is in $U \cap V_I \subset V_I $ and $S|_{V_I}$ is positive-definite always.\footnote{Since $\alpha_I\in V_I$ (the $(-1)$-eigenspace of $\rho_I$) and $\Pi_U$ commutes with $\rho_I$ by Eq.~\ref{eq:ProjectorOnFoldingSpaceCommutesWithGens}, we have
  $$\rho_I(\Pi_U\alpha_I) = \Pi_U(\rho_I\alpha_I) = \Pi_U(-\alpha_I) = -\Pi_U\alpha_I,$$ 
  so $\Pi_U\alpha_I$ lies in the $(-1)$-eigenspace of $\rho_I$, which is $V_I$. Hence $\Pi_U\alpha_I\in U\cap V_I\subseteq V_I$.} This means that \textbf{our folded roots $\widetilde{\alpha}_I$ are always spacelike}, as demanded. Now,
\begin{equation}
    \widetilde{\alpha}_I^2 = \langle \widetilde{\alpha}_I, \widetilde{\alpha}_I \rangle = 2b_I - \alpha_I^T \Pi_{U^\perp}^T S|_{U^\perp} \Pi_{U^\perp} \alpha_I = 2b_I - \alpha_I^T S \Pi_{U^\perp} \alpha_I > 0
    \label{eq:NormFoldedRoots}
\end{equation}
where we used that $S_{{i_1}{i_2}} = 0$ for any $i_1,i_2 \in B_I$, $\Pi_{U^\perp} = \Pi_{U^\perp}^2$ and $S\Pi_U^\perp = S|_{U^{\perp}} = \Pi_{U^\perp}^T S$. Eq.~\eqref{eq:NormFoldedRoots} gives an upper bound for the norm of the $U^\perp$-component of $\alpha_I$: it cannot exceed the value $2b_I$, as then $\widetilde{\alpha}_I$ would become null or timelike. Now that we have ensured that $\widetilde{\alpha}_I$ has positive norm, we can normalize it to $\widetilde{\beta}_I^2 = 2$ by letting 
\begin{equation}
    \widetilde{\alpha}_I \mapsto \widetilde{\beta}_I = \sqrt{\frac{2}{\langle \widetilde{\alpha}_I, \widetilde{\alpha}_I \rangle}}\widetilde{\alpha}_I
\end{equation}

For the off-diagonal entries $I\neq J$, we obtain from Eq.~\eqref{eq:InnerProductFoldedRoots} that \begin{equation}
    \widetilde{S}_{IJ} = \frac{2\alpha_I^T S \Pi_U \alpha_J}{\sqrt{\langle \widetilde{\alpha}_I, \widetilde{\alpha}_I \rangle \langle \widetilde{\alpha}_J, \widetilde{\alpha}_J \rangle}}  = 
    \frac{2}{\sqrt{\langle \widetilde{\alpha}_I, \widetilde{\alpha}_I \rangle \langle \widetilde{\alpha}_J, \widetilde{\alpha}_J \rangle}}
    \left( \sum_{i \in B_I, \,\,\, j \in B_J} S_{ij} - \sum_{i \in B_I, \,\,\, j \in B_J} (S\Pi_{U^\perp})_{ij} \right) . 
    \label{eq:FoldedSchlafliMatrixEntries}
\end{equation} 
The first term in the parenthesis is already negative because all entries of $S_{ij}$ are as well. For the folding to be genuine, we would need $\widetilde{S}_{IJ} \leq 0$ and this depends crucially on the geometry, i.e.~on the projection and the splitting $V = U \oplus U^\perp$. Provided that $\widetilde{S}_{IJ}$ is negative, though, we can read the Coxeter labels $\widetilde{m}_{IJ}$ from the Sch\"afli entries since $\widetilde{S}_{IJ} = -2 \cos (\pi/\widetilde{m}_{IJ})$, where $\widetilde{m}_{IJ}$ is the order of $\rho_I|_U \rho_J|_U$. 

To fold, then, one needs to find the invariant folding subspace $U$, but how can we do that? We will see several examples below to illustrate the methods, starting from the most familiar case: folding by a graph automorphism. 

\subsubsection{Folding by Graph Automorphisms}

Folding has been mostly developed in the context of Lie algebras and crystallographic Coxeter groups, when one identifies a symmetry of the Coxeter-Dynkin graph and uses it to \textit{identify nodes that belong to the same orbit under this automorphism}. Since most of the graphs in Fig.~\ref{fig:Irreducible_Root_Systems} have either a modest $\ZZ_2$ symmetry or no symmetry at all, the method is called \textit{folding}, like folding a piece of paper after you created a crease in the middle. 

\begin{figure}[t!]
    \centering
    \includegraphics[width=0.95\linewidth]{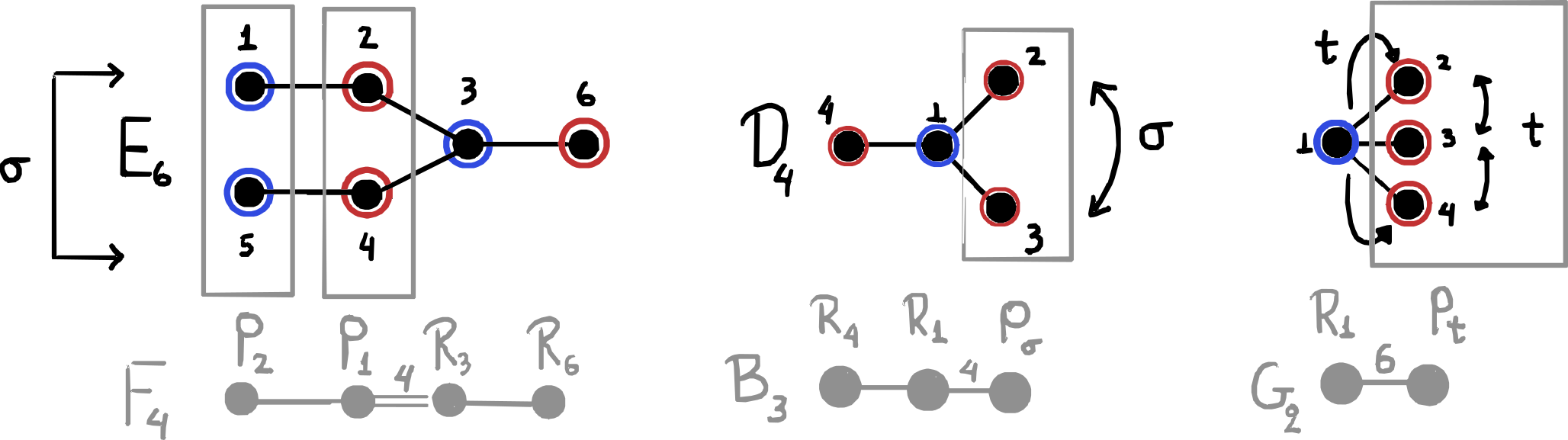}
    \caption[Three examples from the finite Coxeter groups of folding by a graph automorphism.]{Three examples from the finite Coxeter groups of folding by a graph automorphism. On the left and middle panels, we show how to use the $\ZZ_2$ symmetry (automorphism $\sigma$ of order 2) to fold $E_6$ and $D_4$ down to $F_4$ and $B_3$ respectively. On the right, we use the exceptional triality automorphism of $D_4$ to fold it down to $G_2$.} \label{fig:EuclideanExamplesFoldingAutomorphism}
\end{figure}

\textbf{Example 1:} $D_4 \twoheadrightarrow G_2$. To illustrate the machinery developed above, we consider the classic folding of the simply-laced $D_4$ Coxeter graph into the multiply-laced $G_2$ graph shown in Fig.~\ref{fig:EuclideanExamplesFoldingAutomorphism}. Let the $D_4$ graph be labelled as in Fig.~\ref{fig:EuclideanExamplesFoldingAutomorphism}; its Sch\"afli matrix then is
\begin{equation}
S_{D_4} = 
\begin{pmatrix}
2 & -1 & -1 & -1 \\
-1 & 2 & 0 & 0 \\
-1 & 0 & 2 & 0 \\
-1 & 0 & 0 & 2 
\end{pmatrix}
\end{equation}
with $\det S_{D_4} = 4$. The $D_4$ graph possesses an exceptional automorphism $t$ of order 3 that cyclically permutes the outer nodes, $t: (2 \to 3 \to 4 \to 2$, while leaving the central node fixed, $t(1) = 1$. We use this automorphism to define our folding partition, grouping the nodes into orbits of $t$. This yields $b=2$ disjoint blocks:
\begin{equation*}
B_1 = \{ 1 \} \quad (b_1 = 1) \qquad \text{and} \qquad B_2 = \{ 2, 3, 4 \} \quad (b_2 = 3).
\end{equation*}
Notice that $B_2$ is a valid block because the nodes 2,3, and 4 are mutually non-adjacent. The block generators are $\rho_1 = R_1$ and $\rho_2 = R_2 R_3 R_4$. 

The important point here is that the \textbf{graph automorphism $t$ is an $S$-isometry}: the reader can convince themselves that 
$$ S_{t(i) t(j) } = S_{ij} \,\, \forall i,j=1,\dots, 4,$$
which gives $t^T S t = S$ (here, we think of $t$ as being lifted to a transformation in $\mathrm{GL}(V)$, where $D_4$ has its reflection representation). The natural candidate, then, for the folding subspace $U$ is the space that is invariant under $t$, i.e.~the space of vectors symmetric under the permutation of the outer roots. Consequently, 
$$U = V^t = \mathrm{Fix}(t) = \ker(t-\mathrm{id}),\qquad \text{ with projector } \qquad \Pi_U = \tfrac{1}{3}(t+t^2+t^3) \quad (t^3 = \mathrm{id}).$$
(For an order $k$ automorphism $\sigma$, we take $\Pi_U =\frac{1}{k}\sum_{j=0}^{k-1}\sigma^j$.)

One checks conditions (I)–(III):
\begin{itemize}
    \item[(I)] $t$ is an $S$-isometry, so $\Pi_U^T S = S\Pi_U$, 
    \item[(II)] $\dim(U\cap V_2)=1$: the only $t$-fixed direction in $V_2 = \mathrm{span}({\bf r}_2,{\bf r}_3,{\bf r}_4)$ is the symmetric sum $\alpha_2 = {\bf r}_2+{\bf r}_3+{\bf r}_3$. And $\dim(U\cap V_1)=1$ trivially since $b_1=1$; the vector is $\alpha_1 = {\bf r}_1$.
    \item[(III)] The block generators $\rho_I$ are products of $R_i$'s that $t$ permutes among themselves, so $t\rho_I = \rho_I t$, and $U$ is preserved.
\end{itemize}
Because the block-sums $\alpha_1$ and $\alpha_2$ are strictly invariant under $t$, they lie entirely within $U$ and are automatically the folded roots, so no genuine projection is needed. Because $\alpha_I \in U$, their projections onto the orthogonal complement $U^\perp$ vanish entirely, $\Pi_{U^\perp} \alpha_I = 0$, and the orthogonal projector $\Pi_U$ acts as the identity on them. Following Eq.~\eqref{eq:BlockRoots}, the unnormalized folded roots are simply the unprojected sums themselves: $\widetilde{\alpha_I} = \alpha_I$. 

Using Eq.~\eqref{eq:NormFoldedRoots}, the $U^\perp$-subtractive term is zero, meaning the squared norm of each folded root achieves its maximal bound of $2b_I$:
\begin{equation*}
\langle \widetilde{\alpha}_1, \widetilde{\alpha}_1 \rangle = 2 \qquad \text{and} \qquad \langle \widetilde{\alpha}_2, \widetilde{\alpha}_2 \rangle = 6.
\end{equation*}
We now enforce the normalization to obtain the proper folded roots $\widehat{\beta}_I$:
\begin{equation*}
\widetilde{\beta}_1 = {\bf r}_1 \qquad \text{and} \qquad \widetilde{\beta}_2 = \frac{1}{\sqrt{3}} \big( {\bf r}_2 + {\bf r}_3 + {\bf r}_4 \big).
\end{equation*}
Finally, we construct the folded Schl\"{a}fli matrix $\widetilde{S}$ for the rank $b=2$ folded graph. The diagonal entries are 2 by construction. For the off-diagonal entry $\widetilde{S}_{12}$, since the $\Pi_{U^\perp}$ term vanishes, the inner product evaluates to:
\begin{equation*}
\widetilde{S}_{12} = \langle \widetilde{\beta}_1, \widetilde{\beta}_2 \rangle = \frac{1}{\sqrt{3}} \big\langle {\bf r}_1, {\bf r}_2 + {\bf r}_3 + {\bf r}_4 \big\rangle = \frac{1}{\sqrt{3}} (S_{12} + S_{13} + S_{14} ) = -\sqrt{3}
\end{equation*}
The folded Schl\"afli matrix is
  $$\widetilde{S}_{G_2} = 
  \begin{pmatrix}
  2 &-\sqrt{3}\\
  -\sqrt{3}&2
  \end{pmatrix},$$
  which reads off Coxeter label $\tilde{m}_{12}=6$ — exactly the $G_2$ diagram. Note that the two folded roots have different norms before
  normalization ($\widetilde{\alpha}_1^2=2$ vs $\widetilde{\alpha}_2^2=6$), producing the two inequivalent root lengths of $G_2$: $\widetilde{\alpha}_1$ is the short root and $\widetilde{\alpha}_2$ the long root with length-ratio $\widetilde{\alpha}_1^2/\widetilde{\alpha}_1^2 = 1/3$.

\textbf{Example 2:} $E_6 \twoheadrightarrow F_4$. We now consider the folding of the simply-laced $E_6$ Coxeter-Dynkin graph into the $F_4$ graph with the labelling shown in Fig.~\ref{fig:EuclideanExamplesFoldingAutomorphism}. Its Schl\"afli matrix is:
\begin{equation*}
S_{E_6} = 
\begin{pmatrix}
    2 & -1 & 0 & 0 & 0 & 0\\
    -1 & 2 & -1 & 0 & 0 & 0\\
    0 & -1 & 2 & -1 & 0 & -1\\
    0 & 0 & -1 & 2 & -1 & 0\\
    0 & 0 & 0 & -1 & 2 & 0\\
    0 & 0 & -1 & 0 & 0 & 2
\end{pmatrix}
\end{equation*}
The $E_6$ graph possesses an involutive graph automorphism $\sigma$ of order 2 that reflects the graph across its horizontal axis of symmetry in Fig.~\ref{fig:EuclideanExamplesFoldingAutomorphism}. This automorphism swaps the two identical arms of length two while fixing the central axis:
\begin{equation}
\sigma :  1 \leftrightarrow 5, 2 \leftrightarrow 4 \quad \sigma(3) = 3, \quad \sigma(6) = 6.
\end{equation}
We use this automorphism to define our folding partition, grouping the nodes into orbits of $\sigma$. This yields $b=4$ disjoint blocks:
\begin{equation*}
B_1 = \{ 1, 5 \} \, (b_1 = 2), \quad B_2 = \{ 2, 4 \} \, (b_2 = 2), \quad B_3 = \{ 3 \} \, (b_3 = 1), \quad B_4 = \{ 6 \} \, (b_4 = 1).
\end{equation*}

As before, the folding subspace $U$ is the invariant subspace of the automorphism $\sigma$, $U = V^{\sigma}$. All block-sums $\alpha_I$ are strictly invariant under $\sigma$, meaning they lie entirely within $U$. The unnormalized folded roots are simply:
\begin{equation*}
\widetilde{\alpha}_1 = {\bf r}_1 + {\bf r}_5, \quad \widetilde{\alpha}_2 = {\bf r}_2 + {\bf r}_4, \quad \widetilde{\alpha}_3 = {\bf r}_3, \quad \widetilde{\alpha}_4 = {\bf r}_6.
\end{equation*}
The conditions (I)-(III) are satisfied similarly to the previous case. The squared norm of each folded root achieves the maximal value again:
\begin{equation}
\langle \widetilde{\alpha}_1, \widetilde{\alpha}_1 \rangle = 4, \quad \langle \widetilde{\alpha}_2, \widetilde{\alpha}_2 \rangle = 4, \quad \langle \widetilde{\alpha}_3, \widetilde{\alpha}_3 \rangle = 2, \quad \langle \widetilde{\alpha}_4, \widetilde{\alpha}_4 \rangle = 2.
\end{equation}
Enforcing normalization, we get 
\begin{equation*}
\widetilde{\beta}_1 = \frac{1}{\sqrt{2}} \big( {\bf r}_1 + {\bf r}_5 \big), \quad \widetilde{\beta}_2 = \frac{1}{\sqrt{2}} \big( {\bf r}_2 + {\bf r}_4 \big), \quad \widetilde{\beta}_3 = {\bf r}_3, \quad \widetilde{\beta}_4 = {\bf r}_6.
\end{equation*}

We now calculate the entries of the folded Schl\"afli matrix $\widetilde{S}_{IJ}= \langle \widetilde{\beta_I},\widetilde{\beta}_J \rangle$. The diagonal entries are 2. For the off-diagonal entries, we evaluate the inner products between adjacent blocks from Eq.~\eqref{eq:InnerProductFoldedRoots}; after some algebra, we get
\begin{equation*}
    \widetilde{S}_{F_4}
    = 
    \begin{pmatrix}
       2 & -1 & 0 & 0 \\
       -1 & 2 & -\sqrt{2} & 0 \\
       0 & -\sqrt{2} & 2 & -1\\
       0 & 0 & -1 & 2 
    \end{pmatrix}
\end{equation*}
This matrix represents a linear graph of 4 nodes, where the first two nodes are connected by a single edge, the middle two by a double edge, and the last two by a single edge. This is precisely the Coxeter-Dynkin graph of $F_4$ (see Fig.~\ref{fig:EuclideanExamplesFoldingAutomorphism}). Again, we have long ($\widetilde{\alpha}_{1,2}$ and short ($\widetilde{\alpha}_{3,4}$) roots. 

\subsubsection{Folding by Graph Colouring}

While these foldings by graph automorphisms are generally interesting for Lie algebras, here we are primarily concerned with foldings that lead to non-crystallographic groups, like those shown in Fig.~\ref{fig:non-crystallographicEuclideanFoldings}. These foldings are not guided by a graph automorphism, but by a different graph property: bicolourability. Hence, to treat those, we would need to go beyond the standard literature, and the framework developed so far can deal with these cases in an elegant way.\footnote{We note that a very general framework for foldings has also been developed already in \cite{zuber1998generalized,Zuber_1996} which treats the non-crystallographic cases too. Our framework here is a bit more abstract and does not introduce their generalized Coxeter graph notation, but deals with the known Coxeter graphs.
}

A graph is referred to
as `\textit{bicolourable'} (or, more commonly in mathematical literature, `bipartite') if its vertices can
be separated into two disjoint and independent subsets -- we saw the definition in Sections \ref{ch:MatBackgroundPartII} and \ref{sec:MathematicalBackgroundLorentzian}. Let's dive straight into an example to see how the method works in these cases but also, and perhaps most importantly, how they differ from the folding-by-automorphisms cases we dealt before.

\begin{figure}[t!]
    \centering
    \includegraphics[width=0.60\linewidth]{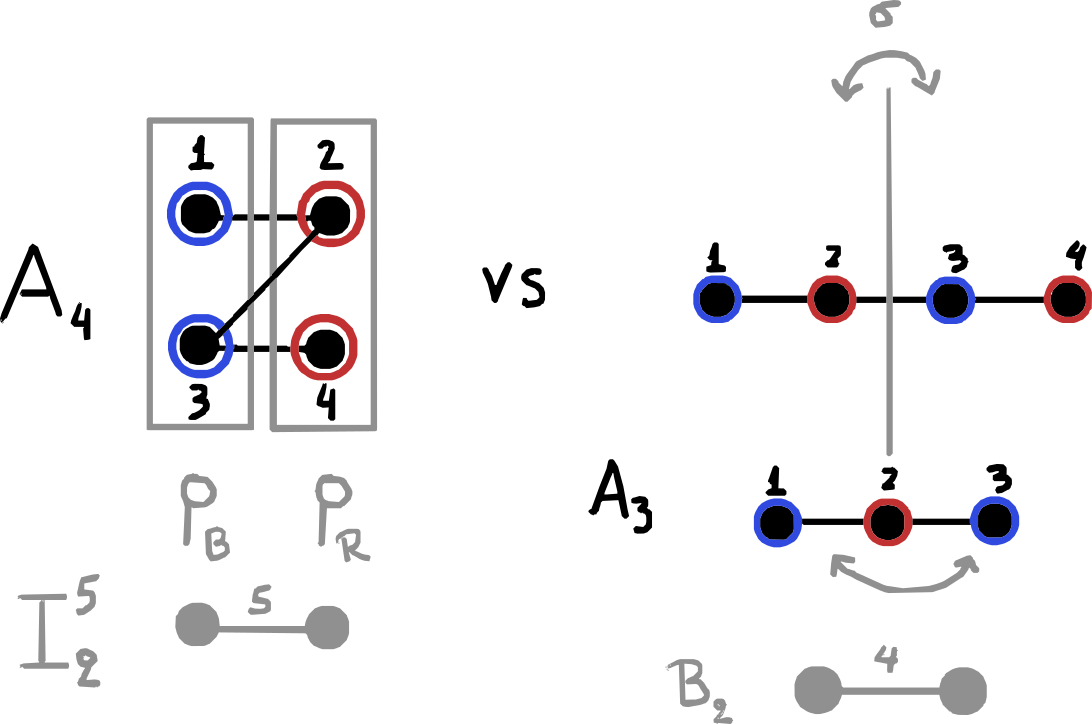}
    \caption[Folding by colouring vs folding by automorphism for $A_4$.]{Folding by colouring vs folding by automorphism for $A_4$.}
    \label{fig:A4FoldingByColouringVSAutomorphism}
\end{figure}

\textbf{Example:} $A_4 \twoheadrightarrow I_2^5$. Consider the $A_4$ graph with the labelling shown in Fig.~\ref{fig:A4FoldingByColouringVSAutomorphism}; this is bicolourable, so we can paint the vertices red and blue as shown in the figure. The blocks suggested by colouring are
$$B_1 = \{ 1,3 \} \quad \text{(blue, odd)}, \qquad B_2 = \{2,4\} \quad \text{(red, even)};$$
both blocks contain mutually non-adjacent nodes. The block generators, then, are $\rho_I= \rho_B = R_1R_3$ and $\rho_2=\rho_R = R_2R_4$ and are involutions. The block-sum roots are $\alpha_1 = \alpha_B={\bf r}_1+{\bf r}_3$, $\alpha_2= \alpha_R={\bf r}_2+{\bf r}_4$, and the $A_4$ Sch\"afli matrix is
$$S_{A_4} = \begin{pmatrix}
2 & -1 & 0 & 0 \\
-1 & 2 & -1 & 0\\
0 & -1 & 2 & -1\\
0 & 0 & -1 & 2
\end{pmatrix}.$$

We first note that the $A_4$ Coxeter-Dynkin graph has a $\ZZ^2$ symmetry $\sigma$: it's left-right reflection along the vertical axis shown in Fig.~\ref{fig:A4FoldingByColouringVSAutomorphism}. This acts as
$$ \sigma : 1 \leftrightarrow 4, \,\, 2 \leftrightarrow 3 
$$
and does not respect the colouring as it maps red to blue vertices and vice versa. The blocks that the automorphism suggests are different from $B_1,B_2$: $\sigma$ induces the partition $\{1.4\}, \{2,3\}$ instead of the coloured partition $\{ 1,3\},\{ 2,4\}$. Crucially, there is no diagram automorphism that preserves the latter. Consequently, the block-sums $\alpha_{1,2}$ are not fixed by any diagram automorphisms and are not directly in the desired folding subspace $U$. This is the structural departure from the previous folding $D_4 \twoheadrightarrow G_2$.\footnote{One could, however, follow their nose and apply the machinery to fold $A_4$ by this diagram automorphism $\sigma$. The result turns out to be $B_2$. But wait a minute, did we just identified adjacent nodes 2 and 3? Yes, but we can think of this as follows: one can first merge these nodes into a single node by contracting the line in-between to get the rank-3 graph $A_3$ (it can be checked that this line contraction does not alter the connections of nodes 2 and 3 to the outer nodes). Then, we can fold $A_3$ by the $\ZZ_2$ symmetry as before to get $B_2$.}

Let's see why the naive computation would fail. Had we treated $\alpha_I$ as already living in $U$ and identified $\widetilde{\alpha}_{I}$ with $\alpha_I$ without any projection, then we would get $\widetilde{S}_{12} = -3/2$, which is negative but gives $\widetilde{m}_{12} = \pi/\arccos{3/4} \simeq 4.37$ instead of 5, so we do not get the correct $I_2^5$ folded diagram. So here, the projection $\Pi_U$ plays a significant role and cures this issue. Note also another related difference: in the folding-by-automorphism picture, an automorphism like $\sigma$ or $t$ of Fig.~\ref{fig:EuclideanExamplesFoldingAutomorphism} acts transitively inside each block such that nodes within a block $B_I$ have the same number of neighbours with another block $B_J$. However, in the folding-by-colouring picture of Fig.~\ref{fig:A4FoldingByColouringVSAutomorphism}, this does not happen: node 1 in block $B_1$ of the $A_4$-partition has one neighbour in block $B_2$ (node 2), while node 3 in the same block has two neighbours (nodes 2 and 4). It can be verified by looking now at Fig.~\ref{fig:non-crystallographicEuclideanFoldings} that this is indeed a general phenomenon with non-crystallographic foldings. 

To identify $U$, we need the Coxeter element for the graph $A_4$. We can calculate this as the product \eqref{subsec:Coxeter_Elements}, but we shall be careful to take the \textit{bicoloured Coxeter element} 
$$ C = C_{\rm colour} = \rho_B \rho_R = R_1 R_3 R_2 R_4$$
instead of some other order in the product, as this is the natural one that `knows' about colouring. It satisfies $C^5 = 1$, since $h=5$ for $A_4$, and its characteristic polynomial -- and its factorization over $\ZZ(\phi)$, $\phi = (1+\sqrt{5})/2$ is the golden ratio -- reads  
$$ \chi_C(x) = \Phi_5(x) = x^4 + x^3 + x^2 + x + 1 = \left(x^2 +\bar{\phi} x + 1\right) \left(x^2 + \phi x +
  1\right) = \tilde{\chi}_1(x) \tilde{\chi}_2(x)
$$
The eigenvalues are $5$-th primitive roots of unity (see Section \ref{subsec:Coxeter_Elements}) and they split into two complex-conjugate pairs: $\zeta_5^\pm = e^{\pm 2\pi i/5}$ and $(\zeta_5^2)^\pm = e^{\pm 4\pi i/5}$ with 
$$\zeta_5 + \zeta_5^- = 2\cos{2\pi/5} = - \bar{\phi} = \phi^{-1} , \quad \text{ and } \quad 
\zeta_5^2 + (\zeta_5^2)^{-} = 2\cos{4\pi/5} = -\phi
$$
where $\,{}\bar{} : \sqrt{5} \to -\sqrt{5}$ denotes Galois-conjugation and $\phi \bar{\phi} = -1$ (now one can appreciate how the above factorization happened; see Section \ref{subsec:symmetriesandsubfields}). Each pair spans a 2D Coxeter plane, inducing a decomposition into invariant subspaces
$$V = U_{\rm phys}\oplus U_{\rm int}, \qquad U_{\rm phys} = \ker(C^2 +\bar{\phi}C+1),\quad U_{\rm
  int}=\ker(C^2+\phi C+1).$$
We take $U = U_{\rm phys}$ (eigenvalues $e^{\pm2\pi i/5}$), which contains whichever direction is killed by the matrix $C^2 + \bar{\phi} C + 1$ (by the Cayley-Hamilton theorem, $C$ satisfies its own $\chi_C(C) = 0$). This subspace is $\rho_I$-invariant and satisfies all three Conditions (I)–(III) from the general theory, as it can be checked.

To find the explicit projector, we can use a version of B\'{e}zout's identity for polynomials and the Cayley--Hamilton theorem:\footnote{To understand how to build our projector matrix, starts with integers. Take two coprime numbers (i.e.~they share no common factors, $\gcd(p,q)=1$) like $p=3$ and $q=5$; B\'{e}zout's identity guarantees you can find coefficients $a$ and $b$ such that $ap + bq =1$ (one such choice is $a = 2$ and $b=-1$). Now, take $bq$ ($=-5$ for concreteness) and do some modular arithmetic: $bq \mod q$ = 0 trivially, while $bq = 1 - ap = 1 \mod p$. As a result, $bq$ acts as $0$ in the $p$-domain and as $1$ in the $q$-domain. Upgrading this construction to coprime polynomials over any number field (could be $\ZZ$ or a field extension), and replacing the variable $x$ with our matrix operator $C$ from the Cayley-Hamilton theorem (this is crucial, otherwise B\'ezout's identity might not hold), we get a matrix equation to solve. } first, we use $\chi_C(C) = \tilde{\chi}_1(C) \tilde{\chi}_2(C) =0$ and the fact that the two factors do not have a common divisor, to write 
$$ 
c_1(C) \tilde{\chi}_1(C) + c_2(C)\tilde{\chi}_2(C) = \mathbb{I}
$$
Now, compare this to $\Pi_{\rm phys} + \Pi_{\rm int} = \mathbb{I}$ from Section \ref{sec:CNP}. With the above definitions for $U_{\rm phys}$ and $U_{\rm int}$, we can see that  
\begin{equation*}
\begin{aligned}
& \text{on } U_{\rm phys} = \ker \tilde{\chi}_1(C): \text{ for any ${\bf v} \in U_{\rm phys}$, $\tilde{\chi}_1(C) {\bf v} =0$, so  $c_2(C)\tilde{\chi}_2(C){\bf v} = {\bf v}$,}
\\
& 
\text{on } U_{\rm int} = \ker \tilde{\chi}_2(C): \text{ for any ${\bf v}\in U_{\rm int}$, $\tilde{\chi}_2(C){\bf v}=0$, so $c_2(C)\tilde{\chi}_2(C){\bf v} = 0$.}
\end{aligned}
\end{equation*}
Thus, $c_2(C)\tilde{\chi}_2(C)$ acts as the identity on one subspace and zero on the other — that is exactly what a projector does. For this reason, we let 
$$ \Pi_{U_{\rm phys}} = \Pi_{\rm phys} = c_2(C) \tilde{\chi}_2(C)$$
and similarly for the $\Pi_{\rm int}$. Specifying $c_2(C)$ demands some more polynomial algebra which we will not show here; the starting point is to notice that $\tilde{\chi}_2(x) - \tilde{\chi}_1(x) = (\phi - \bar{\phi})x = \sqrt{5}x$; similar to the integer case, we have to solve $c_2(x) \tilde{\chi}_2(c) = 1 \mod \tilde{\chi}_1(x)$, which yields the following projector  
$$ \Pi_{U_{\rm phys}} =  
\frac{1}{\sqrt{5}}\left(\frac{1}{\phi}{\mathbb{I}} - (C^2+C^3) \right) \overset{\chi_C(C)=0}{=} \frac{2}{5}\bigg( {\mathbb{I}} + \big( \cos 2\pi/5 \big) \big(C + C^4) + \big( \cos 4\pi/5 \big) \big(C^2 + C^3) \bigg)
$$
As a matrix it becomes,
$$\Pi_{U_{\rm phys}} = 
\frac{1}{\sqrt5}
\begin{pmatrix}
1/\phi & 0 & 1 & 0 \\ 
0 & \phi & 0 & 1 \\ 
1 & 0 & \phi & 0 \\ 
0 & 1 & 0 & 1/\phi
\end{pmatrix}
=
\frac{1}{10} 
    \begin{pmatrix} 
        5-\sqrt{5} & 0 & 2\sqrt{5} & 0 \\ 
        0 & 5+\sqrt{5} & 0 & 2\sqrt{5} \\ 
        2\sqrt{5} & 0 & 5+\sqrt{5} & 0 \\ 
        0 & 2\sqrt{5} & 0 & 5-\sqrt{5} 
    \end{pmatrix}.
$$
Notice that $\Pi_{U_{\rm phys}}$ has non-zero entries only at the pairs $1 \leftrightarrow 3$ and $2 \leftrightarrow 4$, as indicated by the folding of Fig.~\ref{fig:A4FoldingByColouringVSAutomorphism}. However, one member of the pair gets a scaling $\phi$, whereas the other $1/\phi$. 

To find the folded roots, we apply the projector $\Pi_{U_{\rm phys}}$ onto $\alpha_1 = {\bf r}_1 + {\bf r}_3$ and $\alpha_2 = {\bf r}_2 + {\bf r}_4 $
\begin{equation}
    \begin{aligned}
        & \widetilde{\alpha}_1 = \Pi_{\rm phys} \alpha_1 = \frac{\phi\sqrt{5}}{5}
        \begin{pmatrix}
           1 \\
           0 \\
           \phi \\
           0
        \end{pmatrix}
        = \frac{\phi\sqrt{5}}{5} ({\bf r}_1 + \phi {\bf r}_3)\\
        & \widetilde{\alpha}_2 = \Pi_{\rm phys} \alpha_2 = \frac{\phi \sqrt{5}}{5} 
        \begin{pmatrix}
            0 \\
            \phi \\
            0 \\
            1
        \end{pmatrix} 
        = \frac{\phi \sqrt{5}}{5} ({\bf r}_4 + \phi {\bf r}_2)
    \end{aligned}
    \label{eq:FoldedRootsA4toI25}
\end{equation}
The \textbf{golden ratio enters as a relative scaling} between the two original roots in each block as a direct consequence of the Coxeter eigenspace structure, which did not happen in the folding-by-automorphism case. 

We can now calculate the folded Schl\"afli matrix from Eq.~\eqref{eq:InnerProductFoldedRoots}. The norms of the folded roots are
$$ \langle \widetilde{\alpha}_1, \widetilde{\alpha}_1 \rangle = \left( \frac{\phi+1}{5}\right) (4 + 2\phi) = \langle \widetilde{\alpha}_2, \widetilde{\alpha}_2 \rangle
$$
Notice that the $U^\perp$-component in Eq.~\eqref{eq:InnerProductFoldedRoots} is not non-zero and contributes a $\phi$-dependent term; we recognize the first term $4$ in the second parenthesis above as the familiar $2 \times (\text{block size})$ contribution. Dividing by $\sqrt{2/\langle \widetilde{\alpha}_I, \widetilde{\alpha}_I \rangle }$), we define the properly normalized roots $\widetilde{\beta}_{1,2}$. Then, the non-diagonal Schl\"afli entry reads
$$\widetilde{S}_{12} = \frac{\phi \sqrt{5}}{(\phi+1)(\phi+2)} 
\begin{pmatrix}
    1 & 0 & 1 & 0
\end{pmatrix}
S_{A_4}
\begin{pmatrix}
    0 \\
    \phi \\
    0 \\
    1
\end{pmatrix} = -\frac{\phi \sqrt{5} (2\phi + 1)}{(\phi +1)(\phi + 2)} = -\phi
$$
The resulting matrix 
$$ S_{I_2^5} = \begin{pmatrix}
    2 & -\phi \\
    -\phi & 2
\end{pmatrix}$$
gives exactly the $I_2^5$ graph. Compare this to the naive result we got earlier without the projection term. More complicated examples, like $D_6 \twoheadrightarrow H_3$ or $E_8 \twoheadrightarrow H_4$ can be treated similarly.  

\subsubsection{Folding is Always a Geometric Projection}
\label{app:FoldingisProjection}

We have seen so far that folding always corresponds to a geometric projection onto an appropriate folding subspace $U$. In the case of the folding-by-automorphisms picture, $U$ is specified entirely by the automorphism, whereas in the folding-by-colouring picture, $U$ is singled out by the Coxeter element of the original graph. One could say that in the first case, the projection happens entirely along a rational direction of the crystallographic lattice (i.e.~the block-sum roots are already within $U$), whereas in the second case it happens along an irrational one (one would need the projectors to introduce the relative irrational scaling, as for $A_4 \twoheadrightarrow I_2^5$). But in any case, folding can always be seen as a projection: the question is always about how to find the correct folding subspace. 

As it turns out, the Coxeter element has always the answer for this. Its eigenplanes contain all the information we need to specify $U$, construct the projectors and get patterns of maximal symmetry. Familiar examples from lower dimensions include:

$B_4 \twoheadrightarrow I_2^8$ : projection of the hypercube with maximal symmetry.

$A_4 \twoheadrightarrow I_2^5$ : projection of $A_4$'s Voronoi cell (the \textit{permutohedron}) with maximal symmetry.

$F_4 \twoheadrightarrow I_2^{12} $ : projection of the 24-cell with maximal symmetry. 

In particular, in the coloured Coxeter eigenbasis, all block-generators become immediately block-diagonal. Note that this happens only for the coloured eigenbasis and not for any other Coxeter element: even though any two Coxeter element will share eigenvalues (for tree-like graphs), their eigenbasis will be different. So if one uses a different Coxeter element than the coloured one, then one needs to also find the change of basis that block-diagonalizes the generators separately.

The importance of the Coxeter element for folding is studied in the next subsection. 

\subsection{The Coxeter Element Criterion}
\label{eq:AlexCriterionForCoxeterElement}

The question here is the following: are all possible foldings we could think of for a given graph \textit{admissible}? In other words, given a Coxeter group $G$ and its graph, we can always partition its nodes in certain ways to define block reflection-generators according to the techniques of the previous section that could give the Coxeter relations for another group $H$. The question is: have we \textit{actually} showed that $H$ is a valid folding, in a sense that there exist a folding subspace $U$ with the properties described before? 

Since the Coxeter element plays an important role in the whole construction, we would expect that it will hold the key to this question as well, and, in fact, it does. In \cite{Thorne2026}, we were able to show the following:\footnote{As we were working on this topic, we found \cite{zuber1998generalized,Zuber_1996} which prove something similar about the spectra, yet in a more general context.} 

\begin{lemma}[Eigenvalue Sharing Under Folding \cite{Thorne2026}]
\label{lemma:EigenvalueSharing}
Let $G$ and $H$ be irreducible Coxeter groups of rank $n$ and $m$ respectively with tree Coxeter graphs, and suppose $H$ embeds in $G$ by graph folding. Then the $m$ eigenvalues of (a faithful matrix representation of) the Coxeter element $C_H$ form a subset of the $n$ eigenvalues of $C_G$: $\mathrm{spec}(C_H) \subset \mathrm{spec}(C_G)$.
\end{lemma}

\begin{proof}[Proof 1 \cite{Thorne2026}]
Write $C_G=R_1\cdots R_n$ (order-independent, $G$ a tree) which acts on a $n$-dimensional vector space $V$ as an irreducible representation of $G$. The folded generators $\rho_I$ express the folded Coxeter element $C_H^{(V)} = \rho_1 \cdots \rho_m$ through products of the same $R_i$'s; this way, it is also an $n \times n$ matrix on $V$ that has the same eigenvalues as $C_G$. But from the point of view of the subgroup $H$ whose natural geometric representation is on a $m$-dimensional vector space, this $n\times n$ Coxeter element $C_H^{(V)} $ must transform in a \textit{reducible} $n$-dimensional representation. In other words, the $n$-dimensional representation of $C_G$ must restrict into a direct sum of $m$-dimensional representations of $H$ (plus potential trivial or other representations that can appear if $n$ does not divide $n$) as 
  \[
    D(C_H^{(V)})\;=\;\bigoplus_{i} D^{(i)}(C_H)\ \oplus\ (\text{remainder}),
  \]
$D$ denotes a representation, i.e.~an isomorphism $G \to \mathrm{GL}(V)$; similarly for $H$. Each $D^{(i)}(C_H)$ is an $m$-dimensional representation of $H$. The direct-sum structure arises because $C_G \sim C_H^{(V)} $ splits $V$ into mutually orthogonal Coxeter planes and its action does not mix the subspaces. This tells us that the $m$-dimensional subspace $H$ acts consists of eigenvectors of $C_G$ and the restricted $C_H$ on this subspace is an $m \times m$ matrix whose eigenvalues are contained in $\mathrm{spec}(C_G)$. \end{proof}

In light of the framework we presented in the previous section, we decided to give another complimentary proof:
\begin{proof}[Proof 2]
The block Coxeter element is $C_{\rm block} = \rho_1\cdots\rho_b$, and since generators within each block commute, this is simply the product of all $n$ generators of $G$ taken in block-then-block order, so $C_{\rm block}$ is a Coxeter element for $G$.

By Condition (III), each $\rho_I$ preserves $U$. Hence their product $C_{\rm block}$ also preserves $U$: $$C_{\rm block}(U) = \rho_1\cdots\rho_b(U) \subseteq U.$$

By Condition (I), $\Pi_U$ is $S$-self-adjoint, where $S$ is the Schl\"afli matrix of $G$. Since $C_{\rm block}$ is an $S$-isometry, $S$-isometries map $S$-orthogonal complements to $S$-orthogonal complements, so $U^\perp$ is also $C_{\rm block}$-invariant. Thus $V = U \oplus U^\perp$ is an $C_{\rm block}$-invariant splitting. If the characteristic polynomial for $C_G$ is $\chi_{C_G}(x)$, then, we have 
$$\chi_{C_G}(x) = \chi_{C_{\rm block}|U}(x)\cdot\chi_{C_{\rm block}|_{U^\perp}}(x), \qquad (\ast) $$
where $C_{\rm block}|U$ denotes the restriction of $C_{\rm block}$ onto the folding subspace $U$ (similarly for $U^\perp$). 

By Condition (II), each $\rho_I|_U$ has exactly one $(-1)$-eigenvalue (the direction $U\cap V_I$) and fixes $U\cap V_I^\perp$, so it is a
genuine reflection in $U$. Their product is therefore the Coxeter element of the folded group:
$$C_{\rm block}|_U = \rho_1|_U\cdots\rho_b|_U = C_H.$$

From $(*)$, $\chi_{C_H}(x) = \chi_{C_{\rm block}|U}(x)$ divides $\chi_{C_G}(x)$. Thus, the eigenvalues of $C_H$ must be a subset of those of $C_G$.
\end{proof}

If the eigenvalues of $C_H$ are not among those of $C_G$, then no $C_{\rm block}$-invariant subspace $U$ of $V$ can have $C_{\rm block} |_U = C_H$. In this context, failing eigenvalue sharing means the system of equations
$$\underbrace{\rho_I(U) = U \,\, \forall I}_{\mathclap{\text{Condition (III)}}} \quad \text{and} \quad \underbrace{\dim(U\cap V_I)=1 \,\, \forall I}_{\mathclap{\text{Condition
  (II)}}}
 $$
has no simultaneous solution, because any $U$ satisfying (III) is forced to have $C_{\rm block}|_U$ with spectrum $\subseteq \mathrm{spec}(C_G)$, while any $U$ satisfying (II) is forced to have $C_{\mathrm block}|_U = C_H$ with spectrum $\not\subseteq \mathrm{spec}(C_G)$. (II) controls the spectrum of $\rho_I|_U$;  (III) controls the domain: $\rho_I(U)\subseteq U$. Thus, we can have the right spectrum with the wrong domain, or the right invariance with the wrong number of $(-1)$-eigenvalues. 

The condition is \emph{necessary but not sufficient}: shared spectra are forced by folding, but do not conversely guarantee one.

\subsection{Unfolding}
\label{subsec:UnfoldingAppendix}

Up to this point, we saw how to perform a folding and we developed a framework that can handle all the interesting for quasicrystals cases of folding-by-colouring. In the language of the Coxeter pairs, given a pair $(G_C,G_{NC})$ we know how to fold $G_C$ to get $G_{NC}$. We can also check whether some folding of $G_C$ is admissible. We also know that, given two known graphs, if the eigenvalues of the Coxeter element for $G_{NG}$ are not all included in those of the Coxeter element for $G_C$, then we will not be able to perform the folding $G_C \twoheadrightarrow G_{NC}$, so $G_{NC}$ and $G_C$ do not form a Coxeter pair. 

For our purposes, especially in the context of hyperbolic Coxeter groups where no Coxeter pairs are known (see Chapter \ref{ch:GeneralizedSpacetimeCNP}), it would be interesting to know the reverse process also, i.e.~starting from a non-crystallographic graph $G_{NC}$, can we \textit{unfold} it into a crystallographic higher-dimensional lattice? We call this process \textbf{unfolding} (or embedding). One could reasonably argue that this process is hopeless: if folding is a projection, how could one possibly guess the lost dimensions of the projecting space from the lower-dimensional one? As we will see, the number fields hold the memory of the higher-dimensional embedding space. The idea is inspired by \cite{Fring_2005,Fring_2006} and the Minkowski embedding for number fields (see Appendix \ref{AppendixA-NumberTheory}). If the reader has not read that Appendix, now might be a good time to do so.  

We believe this idea is best described first through a familiar example, that of $H_4 \hookrightarrow E_8$, as we now turn to explain.  

\subsubsection{Unfolding $H_4$ into $E_8$}

\begin{figure}
    \centering
    \includegraphics[width=0.95\linewidth]{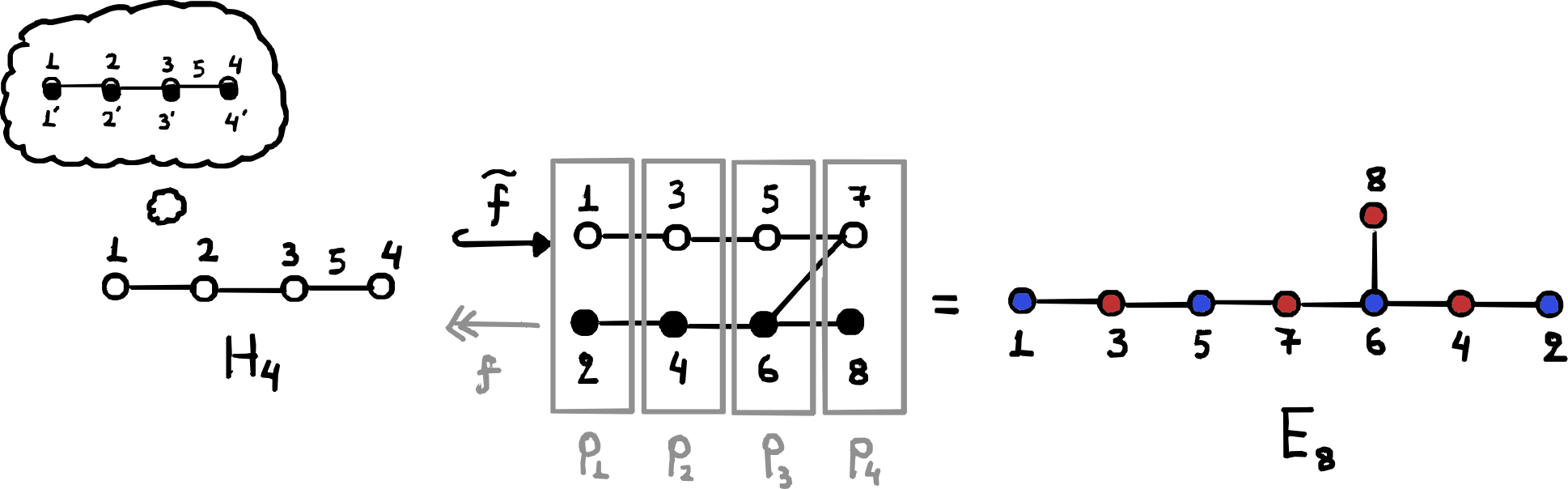}
    \caption[Unfolding $H_4$ to $E_8$]{Unfolding the non-crystallographic group $H_4$ into a crystallographic group $E_8$ in eight dimensions. One could think of the 4 nodes of $H_4$ as being comprised of two nodes stacked on top of each other, which get unfolded into the full $E_8$ graph after the unfolding $\widetilde{f}$-map. On the other hand, one could explicitly perform the folding $f$-map $E_8 \twoheadrightarrow H_4$ by grouping the corresponding nodes into pairs (shown in gray boxes in the middle) to define the new folded generators $\rho_I$.}
    \label{fig:unfoldingH4toE8}
\end{figure}

Imagine we start from the non-crystallographic $H_4$ and we did not know that it can exist within $E_8$: how could we show it? First, from its Coxeter-Dynkin graph, we build its Schl\"afli matrix $S_{H_4}$; keeping the labelling of Fig.~\ref{fig:unfoldingH4toE8} in mind, the matrix reads
\begin{equation*}
    S_{H_4} = \begin{pmatrix}
        2 & -1 & 0 & 0\\
        -1 & 2 & -1 & 0\\
        0 & -1 & 2 & -\phi\\
        0 & 0 & -\phi & 2
    \end{pmatrix}
\end{equation*}

The entries of $S_{H_4}$ live in the number field $\QQ(\sqrt{5})$ because of the irrationality $\phi$, which is a quadratic extension of the rationals $\QQ$. As such, each number in $\QQ(\sqrt{5})$ needs precisely 2 rationals to be described; we write $\QQ(\phi) = \{ a + b \phi : a,b \in \QQ\}$. $\QQ(\phi)$ is a vector space of dimension 2 over $\QQ$; this means we can represent the elements of $\QQ(\phi)$ with 2-dimensional column-vectors such as
$$ \QQ(\phi) \ni \xi = \begin{pmatrix}
    a \\ 
    b 
\end{pmatrix}$$
and the action of such elements on $\QQ(\phi)$ with 2-by-2 rational matrices. To do so, we first pick a basis for $\QQ(\phi)$, and the most natural choice is $\{ 1, \phi \}$, and then we consider the action of the basis on an arbitrary element of $\QQ(\phi)$. This leads to 
\begin{equation*}
    1 \cdot (a + b\phi) = (a + b\phi) \text{ and } \phi \cdot (a + b\phi ) = b + (a+b)\phi 
\end{equation*}
Translating this to matrices is straightforward and yields\footnote{If this feels odd, consider our familiar complex numbers $\CC$, which is also a real extension $\CC = \RR(i)$. The idea is similar to how one would represent 1 as the identity matrix $\mathbb{I}$ and the imaginary unit $i$ as the matrix $ 
J = \begin{pmatrix}
    0 & -1\\
    1 & 0
\end{pmatrix}$, which squares to $-1$.} 
\begin{equation}
    1 \mapsto M_1 = \mathbb{I} = \begin{pmatrix}
        1 & 0\\
        0 & 1
    \end{pmatrix} \quad \text{ and } \quad 
    \phi \mapsto M_\phi = \begin{pmatrix}
        0 & 1\\
        1 & 1
    \end{pmatrix}.
    \label{eq:IdentityAndPhiAsMatrices}
\end{equation}
As a result, an arbitrary element $\xi = a+b\phi$ will be mapped to the matrix
\begin{equation}
    \xi \mapsto M_\xi = \begin{pmatrix}
        a & b \\
        b & a+b 
    \end{pmatrix}
    \label{eq:AnyNumberInGoldenFieldAsMatrix}.
\end{equation}

From here, unfolding $H_4$ is a trivial substitution. We go back to $S_{H_4}$ and literally substitute every $\QQ(\phi)$-entry with its 2-by-2 block given by Eq.~\eqref{eq:AnyNumberInGoldenFieldAsMatrix}. This will give an 8-by-8 matrix, that is exactly the Schl\"afli matrix for $E_8$! The process is shown below: 
\begin{equation}
        S_{H_4} = \begin{pmatrix}
        \textcolor{olive}{2} & \textcolor{orange}{-1} & 0 & 0\\
        \textcolor{orange}{-1} & \textcolor{olive}{2} & \textcolor{orange}{-1} & 0\\
        0 & \textcolor{orange}{-1} & \textcolor{olive}{2} & \textcolor{violet}{-\phi}\\
        0 & 0 & \textcolor{violet}{-\phi} & \textcolor{olive}{2}
    \end{pmatrix}
    \hookrightarrow
    \setlength{\arraycolsep}{0.50pt}
    \begin{pmatrix}
        \fixedblock{olive}{2 & 0\\ 0 & 2} & \fixedblock{orange}{-1&0 \\ 0&-1} & \fixedblock{white}{0&0\\0&0} & \fixedblock{white}{0&0\\0&0} \\
        \fixedblock{orange}{-1&0 \\ 0&-1} & \fixedblock{olive}{2 & 0\\ 0 & 2} & \fixedblock{orange}{-1&0 \\ 0&-1} & \fixedblock{white}{0&0\\0&0}\\
        \fixedblock{white}{0&0\\0&0} & \fixedblock{orange}{-1&0 \\ 0&-1} & \fixedblock{olive}{2 & 0\\ 0 & 2} & \fixedblock{violet}{0 & -1\\ -1 & -1}\\
        \fixedblock{white}{0&0\\0&0} & \fixedblock{white}{0&0\\0&0} & \fixedblock{violet}{0 & -1\\ -1 & -1} & \fixedblock{olive}{2 & 0\\ 0 & 2}
    \end{pmatrix}
    = S_{E_8}
    \label{eq:UnfoldingH4toE8Schlaflis}
\end{equation}
Note that with this labelling of roots of $H_4$, we get that 
$$I=1 \leftrightarrow i = 1,2, \quad I=2 \leftrightarrow i =3,4, \quad I=3 \leftrightarrow i=5,6, \text{ and } \quad I=4 \leftrightarrow i=7,8.$$ In other words, the unfolded roots 1 and 2 where hiding in the first block, the unfolded roots 3 and 4 in the second and so on. Different labelling would lead to different identifications of indices, but the method works with any labelling and will always produce the correct $E_8$ diagram up to the labels. 

Now let us show a bit more explicitly why this block-substitution trick works.

\subsubsection{The General Framework}

Consider a non-crystallographic group $G_{NC}$ of rank $m$ whose geometric representation acts on an $m$-dimensional space $V_{NC}$. From its Sch\"afli matrix $S_{G_{NC}}$, we find the number field $K$ in which the Schl\"afli entries belong to. $K$ is the minimal number field needed to construct the root system. For now, consider the quadratic extension $K = \QQ(\sqrt{5})$, which will be enough for our purposes as most $G_{NC}$ contain only Coxeter labels $m=3,5$. We leave the generalization to all other cases for a future work. 

If $d_K = [K : \QQ(\sqrt{5}) ] = 2$ is the degree of the extension, $G_{NC}$ can be lifted to a lattice inside a crystallographic space $V_C$ of dimension $n = m \cdot d_K = 2m$. This $n$ is the rank of $G_C$ -- the parent to-be-determined crystallographic group that forms the Coxeter pair $(G_C,G_{NC})$ -- and coincides with the rational rank $m_\QQ$ of $G_{NC}$ (see Section \ref{subsec:CrystallographicVsNon}). We say that $G_{NC}$ is a rank-$m$ module over $\mathcal{O}_K$, so it unfolds into a lattice (module over $\ZZ$) in $n$-dimensions. Our goal it to realize this embedding via a Minkowski-like method (see Section \ref{AppendixA-NumberTheory}).

We start with an $m$-dimensional vector ${\bf v}$ with entries in $K$, which we want to map into a $2m$-dimensional real vector space. The Galois group $\Gal(K/\QQ)$ of order $|\Gal(K/\QQ)| = d_K=2$ governs this embedding. Let $\sigma = \,{} \bar{} : \sqrt{5} \mapsto -\sqrt{5}$ denote the single Galois automorphism of $K$. Then, we can embed ${\bf v}$ as 
\begin{equation}
    \mathfrak{M}: V_{NC} \ni {\bf v} \longmapsto \big({\bf v}, \sigma({\bf v})\big)^T = \big({\bf v}, \bar{\bf v} \big)^T \in  V_C = \mathbf{m} \oplus \mathbf{m}^{\sigma}
    \label{eq:NaiveMinkowskiEmbedding1}
\end{equation}
where $\sigma$ induces a decomposition of $V_C$ into $d_K = 2$ $m$-dimensional representations of $V_{NC}$; $\mathbf{m}^\sigma$ is the Galois-conjugate representation of $\mathbf{m}$. Given that the Schl\"afli matrix $S_{G_{NC}}$ is the inner products of the roots of $G_{NC}$, we our next task is to read the inner products in the embedding space $V_C$ from those in $V_{NC}$. 

To avoid cluttering the notation to much, let us use the conventions of the previous section, where $\tilde{S}_{IJ}$ and $\widetilde{\beta}_I$ denote the folded Schl\"afli matrix and the folded roots in of $G_{NC}$ respectively, while untilded quantities refer to $G_C$. We also let $V_C = V$ and $V_{G_{NC}} = \widetilde{V}$.  

Since the Schl\"afli entries of $G_{NC}$ are in the extension $K$, 
$$ \widetilde{S}_{IJ} = \langle \widetilde{\beta}_I,\widetilde{\beta}_J \rangle = a + b \phi \text{ for some choice of } a,b \in \QQ, 
$$
and those of $G_C$ are in $\QQ$, we would need a way to \textit{rationalize} the matrix $\widetilde{S}_{IJ}$ to ``get rid of'' $\phi$. The natural way to do this is by using the \textit{trace form} for the field $K$ (see Eq.~\eqref{eq:NumberFieldTrace}); what we would like to establish, then, is a dictionary like this
\begin{equation}
    \left\langle 
    \mathfrak{M}({\bf x}), \mathfrak{M}({\bf y}) 
    \right\rangle_{V} = 
    \Tr_{K/\QQ} \left( 
    \langle 
    {\bf x}, {\bf y} 
    \rangle_{\widetilde{V}} \right), \quad \forall {\bf x,y} \in \widetilde{V}, \, 
    \langle 
    {\bf x}, {\bf y} 
    \rangle
    \in K
    \label{eq:NaiveTraceForm1}
\end{equation}
where the map $\mathfrak{M}$ acts as an \textbf{isometric isomorphism} between $G_{NC}$ and $G_C$ respecting the inner products. For the specific example of $H_4 \hookrightarrow E_8$, something similar was first noted in \cite{Moody_1993} where the trace form is called the \textit{rational form}; in \cite{Fring_2005,Fring_2006} they extended those results to all icosahedrally-symmetric Coxeter pairs (i.e.~$I_2^5,H_3,H_4$). These works are the ones that inspired our method, though we view our techniques more general and applicable, in principle, to other embeddings than those three. 

We can ask: does Eq.~\eqref{eq:NaiveTraceForm1} achieve an isometric mapping? To answer this, we consider the action of the trace form $\Tr_{K/\QQ}$ on the standard $K$-basis $z = \{ 1, \phi \}$:
$$ 
\begin{aligned}
& \Tr_{K/\QQ}( 1 \cdot 1) = 2, \qquad \Tr_{K/\QQ}( 1 \cdot \phi) = \phi + \bar{\phi} = 1, \\
& \Tr_{K/\QQ}( \phi \cdot 1) = 1,
\qquad \Tr_{K/\QQ}( \phi \cdot \phi) = 2 + (\phi + \bar{\phi}) = 3
\end{aligned}
$$
Grouping these into a matrix $T_{ij} = \Tr_{K/\QQ} (z_i z_j)$, we find that its determinant is $\det T = 5$ -- equal to the discriminant $\Delta_K$ of Eq.~\eqref{eq:NumberFieldDiscriminant} -- which means that the naive trace prescription \eqref{eq:NaiveTraceForm1} we wrote down ``skews'' the $K$-basis and thus stretches volumes. We would like the trace form to give rise to an orthogonal mapping such that the basis $\{ 1, \phi \}$ corresponds to a matrix $T_{ij} = \mathbb{I}$. This is equivalent to how we asked for the Minkowski embedding of \textit{maximal symmetry} instead of the standard one in Section \ref{AppendixA-NumberTheory}. 

To achieve this, we want a \textbf{twisting factor} $\tau$ that perfectly diagonalizes the basis. This will guarantee that the volume determinant stays strictly at 1. Let the \textbf{twisted trace} 
\begin{equation}
    T_{ij}^{(\tau)} = \Tr_{K/\QQ}(\tau z_i z_j).
    \label{eq:TwistedTraceForm}
\end{equation} 
We want this to be orthonormal for the basis $z = \{1 ,\phi \}$, so we want it to satisfy
$$ \Tr_{K/\QQ} (\tau \cdot 1 \cdot 1) = \tau + \bar{\tau} = 1 \qquad \text{ and } \qquad \Tr_{K/\QQ} (\tau \cdot 1 \cdot \phi) = \tau \phi + \bar{\tau} \bar{\phi} = 0.
$$
Solving this linear system of equations, we find that the twisting factor is 
$$ 
\tau = -\frac{\bar{\phi}}{\phi - \bar{\phi}} = \frac{5 - \sqrt{5}}{10} = -\frac{\bar{\phi}\sqrt{5}}{5} = \frac{3 - \phi}{5}
$$
This is a very interesting number within $K$: first, its norm is exactly right to cancel the discriminant: $\mathcal{N}_{K/\QQ} (\tau) = \tau {\bar{\tau}} = 1/5 = 1/\Delta_K$. Second, it is a \textbf{totally positive generator} in $\QQ(\phi)$, as all its conjugates are positive: $\tau > 0 $ \textit{and} $\sigma(\tau) = \bar{\tau} > 0$, which does not happen for $\phi$, for example ($\bar{\phi} < 0$). Finally, notice that the twisting factor $\tau$, or, actually, its Galois-conjugate, appeared as an overall scaling factor in front of the folded roots in Eq.~\eqref{eq:FoldedRootsA4toI25} when we folded $A_4$ to $I_2^5$. 

With our twisted trace form in hand, we can return to our naive dictionary \eqref{eq:NaiveTraceForm1} and see how to modify it. Adding the twisting factor $\tau$ in the right hand-side of this equation and working backwards, we obtain
\begin{equation}
\begin{aligned}
    \Tr_{K/\QQ}(\tau  \langle 
    {\bf x}, {\bf y} 
    \rangle_{\widetilde{V}}) &= \tau \langle 
    {\bf x}, {\bf y} 
    \rangle_{\widetilde{V}} + \bar{\tau} \overline{\langle 
    {\bf x}, {\bf y} 
    \rangle_{\widetilde{V}}} = 
    \langle \sqrt{\tau} {\bf x}, \sqrt{\tau} {\bf y} \rangle_{\widetilde{V}} +
    \langle \sqrt{\bar{\tau}} {\bf \bar{x}}, \sqrt{\bar{\tau}} {\bf \bar{y}} \rangle_{\widetilde{V}}\\
    &= \big\langle (\sqrt{\tau} {\bf x}, \sqrt{\bar{\tau}} {\bf y} ) , (\sqrt{\tau} {\bf y}, \sqrt{\bar{\tau}} {\bf y} ) \big\rangle_{V} \equiv 
    \big\langle \mathfrak{M}_\tau ({\bf x}), \mathfrak{M}_\tau ({\bf y}) \big\rangle_V
    \end{aligned}
    \label{eq:TwistedTraceFormInnerProducts}
\end{equation}   
As a result, this tell us that, in order to use the twisted trace that is orthogonal for the $K$-basis, we have to modify the standard Minkowski embedding \eqref{eq:NaiveMinkowskiEmbedding1} into
\begin{equation}
    \mathfrak{M}_\tau: V_{NC} \ni {\bf v} \longmapsto \big(\sqrt{\tau} {\bf v}, \sqrt{\bar{\tau}}\sigma({\bf v})\big)^T = \big(\sqrt{\tau}{\bf v}, \sqrt{\bar{\tau}}\bar{\bf v} \big)^T \in  V_C = \mathbf{m} \oplus \mathbf{m}^{\sigma}
    \label{eq:ProperMinkowskiEmbedding1}
\end{equation}
Notice that the total positivity of the generator $\tau$ ensures that both radicands are positive. 

Let us note two important implications of the \textit{twisted} Minkowski embedding $\mathfrak{M}_{\tau}$ and how it relates to what we have seen already. First, notice that the \textit{relative scaling} of the two $\mathbf{m}$-components in Eq.~\eqref{eq:ProperMinkowskiEmbedding1}:
$$ 
\sqrt{\frac{\bar{\tau}}{\tau}} = \sqrt{\frac{2 + \phi}{3-\phi}} = \phi.
$$
Now, compare this to the folded roots of $A_4$ (Eq.~\eqref{eq:FoldedRootsA4toI25}): we get precisely the relative scaling equal to $\phi$ between the original $A_4$ roots ${\bf r}_{1,4}$ and ${\bf r}_{3,2}$. Restricting these roots onto the $I_2^5$ subspace $\widetilde{V}$, we can say that Eq.~\eqref{eq:FoldedRootsA4toI25} describes, equivalently, the unfolding of $I_2^5$ group into $A_4$ via the twisted Minkowski embedding of maximal symmetry. In general, 
\begin{equation}
    \mathfrak{M}_{\tau} : \widetilde{V} \ni {\bf v} \mapsto \sqrt{\tau} \big( {\bf v} \oplus \phi {\bf \bar{v}} \big) \in V = {\mathbf{m}} \oplus \mathbf{m}^\sigma
\end{equation}
The second observation is that, if we take any entry $\xi = a + b \phi \in \QQ{\phi}$, like the Schl\"afli entries $\widetilde{S}_{IJ}$ for example, we can use the twisted trace \eqref{eq:TwistedTraceForm} to evaluate the following matrix of inner products
\footnote{
In verifying this, the reader might find it useful to evaluate the traces first $\Tr_{K/\QQ}(\tau) = 1, \Tr_{K/\QQ}(\tau\phi) = 0,\Tr_{K/\QQ}(\tau \phi^2) = 1 = \Tr_{K/\QQ}(\tau \phi^3) = 1$. }  
\begin{equation}
    T_{ij}^{(\tau)}(\xi) = \begin{pmatrix}
        \Tr_{K/\QQ}(\tau \xi ) & \Tr_{K/\QQ}(\tau \xi \cdot \phi )\\ 
        \Tr_{K/\QQ}(\tau \xi \cdot \phi ) & \Tr_{K/\QQ}(\tau \xi \cdot \phi^2) 
    \end{pmatrix} =
    \begin{pmatrix}
        a & b\\
        b & a+b
    \end{pmatrix}
    \label{eq:TraceFormAndBlockSubstitution}
\end{equation}
This is precisely the block-substitution \eqref{eq:AnyNumberInGoldenFieldAsMatrix} that let us unfold $H_4$ into $E_8$ in Eq.~\eqref{eq:UnfoldingH4toE8Schlaflis}, now given a more exact number-theoretic rationale.

Let's see how this all works in practice. 

Consider the $m$-dimensional folded roots $\widetilde{\beta}_I$ of $G_{NC}$ (there are $m$ of them) and their Galois-scaled counterparts $\phi \widetilde{\beta}_I$ (another $m$ of them). These are obviously not $\QQ{(\phi)}$-linearly independent, but feeding them into the Minkowski embedding $\mathfrak{M}_{\tau}$ will generally produce $2m$ $\QQ$-linearly independent vectors that span the $2m$-dimensional crystallographic space $V$. This is because the two halves of the $2m$-dimensional vector scale by completely different, non-proportional values, so the new vector is physically ``sheared'' away from the original root. This shearing action guarantees that $\mathfrak{M}_\tau(\widetilde{\beta}_I)$ and $\mathfrak{M}_\tau(\phi \widetilde{\beta}_I)$ point towards different $2m$-directions. 

Each $\widetilde{\beta}_I$ gives rise to a pair of higher-dimensional roots
\begin{equation}
    \widetilde{\beta}_I \longrightarrow 
    {\bf r}_{2I - 1} = \mathfrak{M}_\tau (\widetilde{\beta}_I) 
    \qquad \text{ and } \qquad
    {\bf r}_{2I} = \mathfrak{M}_\tau (\phi \widetilde{\beta}_I) 
\end{equation}
such that the Schl\"afli entries $\widetilde{S}_{IJ}$ of $G_{NC}$ describe exactly 4 Schl\"afli entries $S_{ij}$ of $G_C$
\begin{equation}
    \widetilde{S}_{IJ} = \langle \widetilde{\beta}_I, \widetilde{\beta}_J \rangle 
    \longrightarrow
    \begin{pmatrix}
        \langle {\bf r}_{2I-1}, {\bf r}_{2J-1} \rangle & \langle {\bf r}_{2I-1}, {\bf r}_{2J} \rangle \\
        \langle {\bf r}_{2I}, {\bf r}_{2J-1} \rangle & \langle {\bf r}_{2I}, {\bf r}_{2J} \rangle
    \end{pmatrix} = 
    \begin{pmatrix}
        S_{{2I-1}{2J-1}} & S_{{2I-1}{2J}} \\
        S_{{2I}{2J-1}} & S_{{2I}{2J}}
    \end{pmatrix}.
\end{equation}
Even more compactly, if we let ${\bf r}_{2I - 1 + k} = \mathfrak{M}_\tau (\phi^k \widetilde{\beta}_I)$ for $k=0,1$, we get
\begin{equation}
    S_{{(2I-1+k)}{(2J-1+l)}} = \Tr_{K/\QQ} \bigg( \tau \phi^{k+l} \widetilde{S}_{IJ} \bigg)
    \label{eq:RelatingFoldedAndUnfoldedSchlaflis}
\end{equation}
which allows us to read off the entries of $S$ from $\widetilde{S}$. 

\section{Coxeter Pairs via Unfolding: Icosahedral Family}
\label{sec:CoxeterPairsIcosahedral}

Finally, in this section we give the first results of a more general program about classifying all possible Coxeter pairs and thus \sCNP{} schemes for producing quasi-crystalline patterns with these non-crystallographic symmetries. We begin with unfolding the icosahedrally symmetric Euclidean $G_{NC}$'s; then, we show how to unfold several of the simplicial non-crystallographic cases from \cite{humphreys_1990} that have their entries also in the golden field $\QQ(\phi)$. In a future work, we will deal with the rest of the non-crystallographic cases from \cite{humphreys_1990} with entries in different number fields.

Our notation closely follows that of \cite{mcmullen2002coxeter}. For each spherical Coxeter graph, one can adjoin a single extending node to obtain an affine diagram (see Fig.~\ref{fig:Irreducible_Root_Systems}). Attaching one more node to the affine cases (with a single edge), we get hyperbolic graphs: we denote them by $Ah_{n+2}, Bh_{n+2}, Dh_{n+2}$ and $Eh_{n+2}$. Note that $E_{10} = Eh_{10}$. The notation $L_{353}$ indicates a linear graph with three edges whose labels are 3, 5 and 3 respectively. Similarly, $K_{53}$ indicates a linear graph with labels 5 and 3, but with an additional node attached to the last node with a simple line. For hyperbolic extended graphs, if an underlying hyperbolic graph is known, we denote the corresponding extended graph by a superscript with as many ``+'' signs as the number of extending nodes.  

In the analysis below, we first give a figure with all the relevant unfoldings and then describe how to get the Schl\"afli matrix for $G_C$ using that of $G_{NC}$ and following the prescription of the previous section. In all figures, we first give the $G_{NC}$ graph (in a thought bubble, one can see in black the ``mirror'' nodes that are folded on top of the original nodes), then we present the unfolding (and we group in grey boxes the nodes that are to be identified if one performs the folding), and finally we give the full graph and its colouring. 

\subsection{Spherical}
\label{subsec:SphericalCoxeterPairsAppendix}

\begin{figure}[t!]
    \centering
    \includegraphics[width=0.90\linewidth]{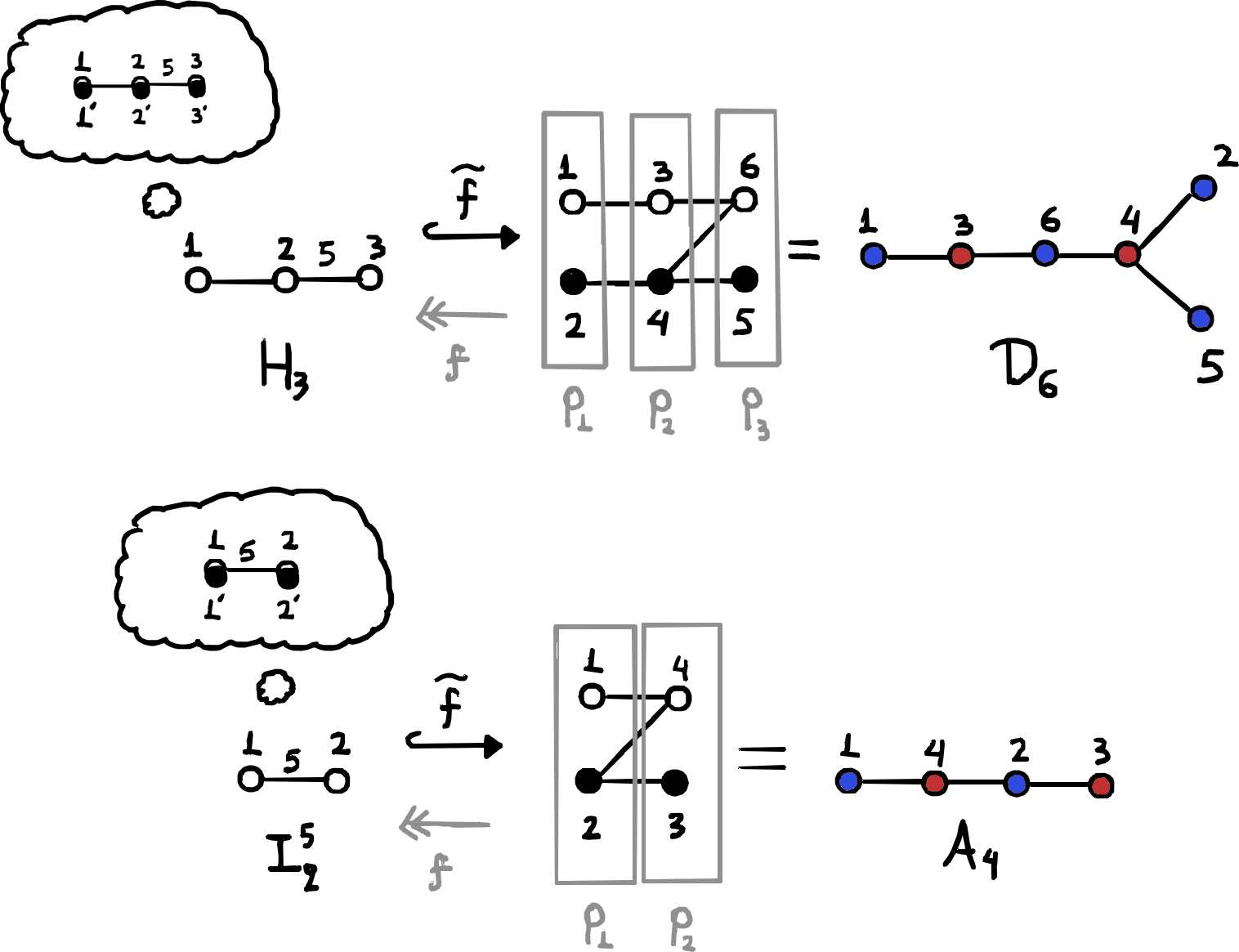}
    \caption[Unfolding the spherical non-crystallographic groups $I_2^5$ and $H_3$ into their crystallographic parent groups $A_4$ and $D_6$ respectively.]{Unfolding the spherical non-crystallographic groups $I_2^5$ and $H_3$ into their crystallographic parent groups $A_4$ and $D_6$ respectively.}
    \label{fig:UnfoldingEuclideanRank2and3}
\end{figure}

\begin{itemize}
    \item 
$I_2^5 \hookrightarrow A_4$:
\begin{equation}
        S_{I_2^5} = 
        \begin{pmatrix}
        \textcolor{olive}{2} & \textcolor{violet}{-\phi} \\
        \textcolor{violet}{-\phi} & \textcolor{olive}{2}
    \end{pmatrix}
    \hookrightarrow
    \setlength{\arraycolsep}{0.50pt}
    \begin{pmatrix}
        \fixedblock{olive}{2 & 0\\ 0 & 2} & \fixedblock{violet}{0 & -1\\ -1 & -1}\\
        \fixedblock{violet}{0 & -1\\ -1 & -1} & \fixedblock{olive}{2 & 0\\ 0 & 2}
    \end{pmatrix}
    = S_{A_4}
    \label{eq:unfoldingI25toA4}
\end{equation}

\item 
$H_3 \hookrightarrow D_6$:
\begin{equation}
        S_{H_3} = 
        \begin{pmatrix}
        \textcolor{olive}{2} & \textcolor{orange}{-1} & 0 \\
        \textcolor{orange}{-1} & \textcolor{olive}{2} & \textcolor{violet}{-\phi} \\
        0 & \textcolor{violet}{-\phi}  & \textcolor{olive}{2}\\
    \end{pmatrix}
    \hookrightarrow
    \setlength{\arraycolsep}{0.50pt}
    \begin{pmatrix}
        \fixedblock{olive}{2 & 0\\ 0 & 2} 
        & 
        \fixedblock{orange}{-1&0 \\ 0&-1} 
        & 
        \fixedblock{white}{0&0\\0&0} \\
        \fixedblock{orange}{-1&0 \\ 0&-1} 
        & \fixedblock{olive}{2 & 0\\ 0 & 2} 
        & \fixedblock{violet}{0 & -1\\ -1 & -1}\\
        \fixedblock{white}{0&0\\0&0} 
        & 
        \fixedblock{violet}{0 & -1\\ -1 & -1} & 
        \fixedblock{olive}{2 & 0\\ 0 & 2}
    \end{pmatrix}
    = S_{D_6}
    \label{eq:unfoldingH3toD6}
\end{equation}

\item 
$H_4 \hookrightarrow E_8$: see Eq.~\eqref{eq:UnfoldingH4toE8Schlaflis} and Fig.~\ref{fig:unfoldingH4toE8}

\end{itemize}

\subsection{Hyperbolic}
\label{subsec:HyperbolicCoxeterPairsAppendix}

\textbf{Rank 4:}

\begin{figure}[t!]
    \centering
    \includegraphics[width=0.99\linewidth]{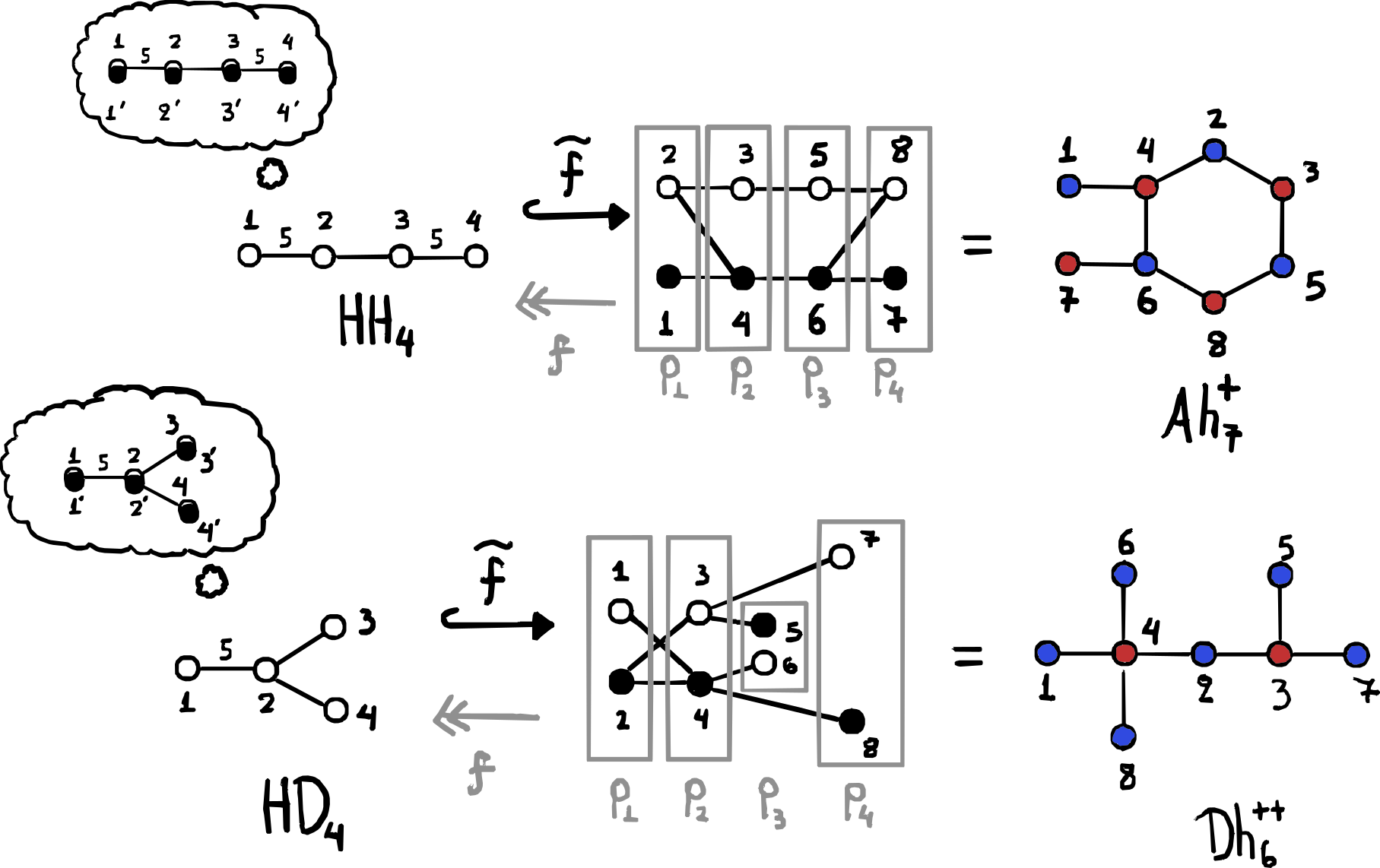}
    \caption[Unfolding hyperbolic non-crystallographic rank-4 groups into their crystallographic counterparts of rank-8.]{Unfolding hyperbolic non-crystallographic rank-4 groups of the icosahedral family into their crystallographic counterparts of rank-8.}
    \label{fig:UnfoldingHyperbolicGroupsRank4}
\end{figure}

\begin{itemize}
    \item 
$HF_4 \hookrightarrow HE_7^+$: $HF_4$ is also known as $L_{353}$ in \cite{mcmullen2002coxeter}; our notation is because of its resemblance to the spherical $F_4$. $HE_7^+$ is the extended $E_7$ graph by one node; see Fig.~\ref{fig:unfoldingHF4} and related discussion there.
\begin{equation}
        S_{HF_4} = 
        \begin{pmatrix}
        \textcolor{olive}{2} & \textcolor{orange}{-1} & 0 & 0\\
        \textcolor{orange}{-1} & \textcolor{olive}{2} & \textcolor{violet}{-\phi} & 0\\
        0 & \textcolor{violet}{-\phi} & \textcolor{olive}{2} & \textcolor{orange}{-1} \\
        0 & 0 & \textcolor{orange}{-1} & \textcolor{olive}{2}
    \end{pmatrix}
    \hookrightarrow
    \setlength{\arraycolsep}{0.50pt}
    \begin{pmatrix}
        \fixedblock{olive}{2 & 0\\ 0 & 2} & 
        \fixedblock{orange}{-1&0 \\ 0&-1} & 
        \fixedblock{white}{0&0\\0&0} & \fixedblock{white}{0&0\\0&0} \\
        \fixedblock{orange}{-1&0 \\ 0&-1} & \fixedblock{olive}{2 & 0\\ 0 & 2} & 
        \fixedblock{violet}{0 & -1\\ -1 & -1} & \fixedblock{white}{0&0\\0&0} \\
        \fixedblock{white}{0&0\\0&0} & \fixedblock{violet}{0 & -1\\ -1 & -1} & \fixedblock{olive}{2 & 0\\ 0 & 2} & \fixedblock{orange}{-1&0 \\ 0&-1}\\
        \fixedblock{white}{0&0\\0&0} & \fixedblock{white}{0&0\\0&0} & \fixedblock{orange}{-1&0 \\ 0&-1} & \fixedblock{olive}{2 & 0\\ 0 & 2}
    \end{pmatrix}
    = S_{HE_7^+}
    \label{eq:unfoldingHF4toHE7Plus}
\end{equation}

\item 
$ HH_4 \hookrightarrow Ah_7^+$: $HH_4$ is our notation for the Lann\'er graph from \cite{humphreys_1990}, which looks like a hyperbolic variant of the spherical $H_4$ graph. $Ah_7^+$ is the extended $Ah_7$ graph from \cite{mcmullen2002coxeter}; the extending node is either node 1 or 7 in Fig.~\ref{fig:UnfoldingHyperbolicGroupsRank4}. 
\begin{equation}
        S_{HH_4} = \begin{pmatrix}
        \textcolor{olive}{2} & \textcolor{violet}{-\phi} & 0 & 0\\
        \textcolor{violet}{-\phi} & \textcolor{olive}{2} & \textcolor{orange}{-1} & 0\\
        0 & \textcolor{orange}{-1} & \textcolor{olive}{2} & \textcolor{violet}{-\phi}\\
        0 & 0 & \textcolor{violet}{-\phi} & \textcolor{olive}{2}
    \end{pmatrix}
    \hookrightarrow
    \setlength{\arraycolsep}{0.50pt}
    \begin{pmatrix}
        \fixedblock{olive}{2 & 0\\ 0 & 2} & \fixedblock{violet}{0 & -1\\ -1 & -1} & \fixedblock{white}{0&0\\0&0} & \fixedblock{white}{0&0\\0&0} \\
        \fixedblock{violet}{0 & -1\\ -1 & -1} & \fixedblock{olive}{2 & 0\\ 0 & 2} & \fixedblock{orange}{-1&0 \\ 0&-1} & \fixedblock{white}{0&0\\0&0}\\
        \fixedblock{white}{0&0\\0&0} & \fixedblock{orange}{-1&0 \\ 0&-1} & \fixedblock{olive}{2 & 0\\ 0 & 2} & \fixedblock{violet}{0 & -1\\ -1 & -1}\\
        \fixedblock{white}{0&0\\0&0} & \fixedblock{white}{0&0\\0&0} & \fixedblock{violet}{0 & -1\\ -1 & -1} & \fixedblock{olive}{2 & 0\\ 0 & 2}
    \end{pmatrix}
    = S_{Ah_7^+}
    \label{eq:unfoldingHH4toAh7Plus}
\end{equation}

\item 
$HD_4 \hookrightarrow Dh_6^{++}:$ $HD_4$ is also known as $K_{53}$ in \cite{mcmullen2002coxeter}; our notation is because of its resemblance to the spherical $D_4$. $Dh_6^{++}$ is our notation for the doubly extended $Dh_6$ graph from \cite{mcmullen2002coxeter}. The two extending nodes are nodes 5 and 7 in Fig.~\ref{fig:UnfoldingHyperbolicGroupsRank4}.
\begin{equation}
        S_{HD_4} = 
        \begin{pmatrix}
        \textcolor{olive}{2} & \textcolor{violet}{-\phi} & 0 & 0\\
        \textcolor{violet}{-\phi} & \textcolor{olive}{2} & \textcolor{orange}{-1} & \textcolor{orange}{-1}\\
        0 & \textcolor{orange}{-1} & \textcolor{olive}{2} & 0\\
        0 & \textcolor{orange}{-1} & 0 & \textcolor{olive}{2}
    \end{pmatrix}
    \hookrightarrow
    \setlength{\arraycolsep}{0.50pt}
    \begin{pmatrix}
        \fixedblock{olive}{2 & 0\\ 0 & 2} & \fixedblock{violet}{0 & -1\\ -1 & -1} & \fixedblock{white}{0&0\\0&0} & \fixedblock{white}{0&0\\0&0} \\
        \fixedblock{violet}{0 & -1\\ -1 & -1} & \fixedblock{olive}{2 & 0\\ 0 & 2} & \fixedblock{orange}{-1&0 \\ 0&-1} & \fixedblock{orange}{-1&0 \\ 0&-1}\\
        \fixedblock{white}{0&0\\0&0} & \fixedblock{orange}{-1&0 \\ 0&-1} & \fixedblock{olive}{2 & 0\\ 0 & 2} & \fixedblock{white}{0 & 0\\ 0 & 0}\\
        \fixedblock{white}{0&0\\0&0} & \fixedblock{orange}{-1&0\\0&-1} & \fixedblock{white}{0 & 0\\ 0 & 0} & \fixedblock{olive}{2 & 0\\ 0 & 2}
    \end{pmatrix}
    = S_{Dh_6^{++}}
    \label{eq:unfoldingHD4toDh6++}
\end{equation}
\end{itemize}

\textbf{Rank 5:}

\begin{figure}[t!]
    \centering
    \includegraphics[width=0.99\linewidth]{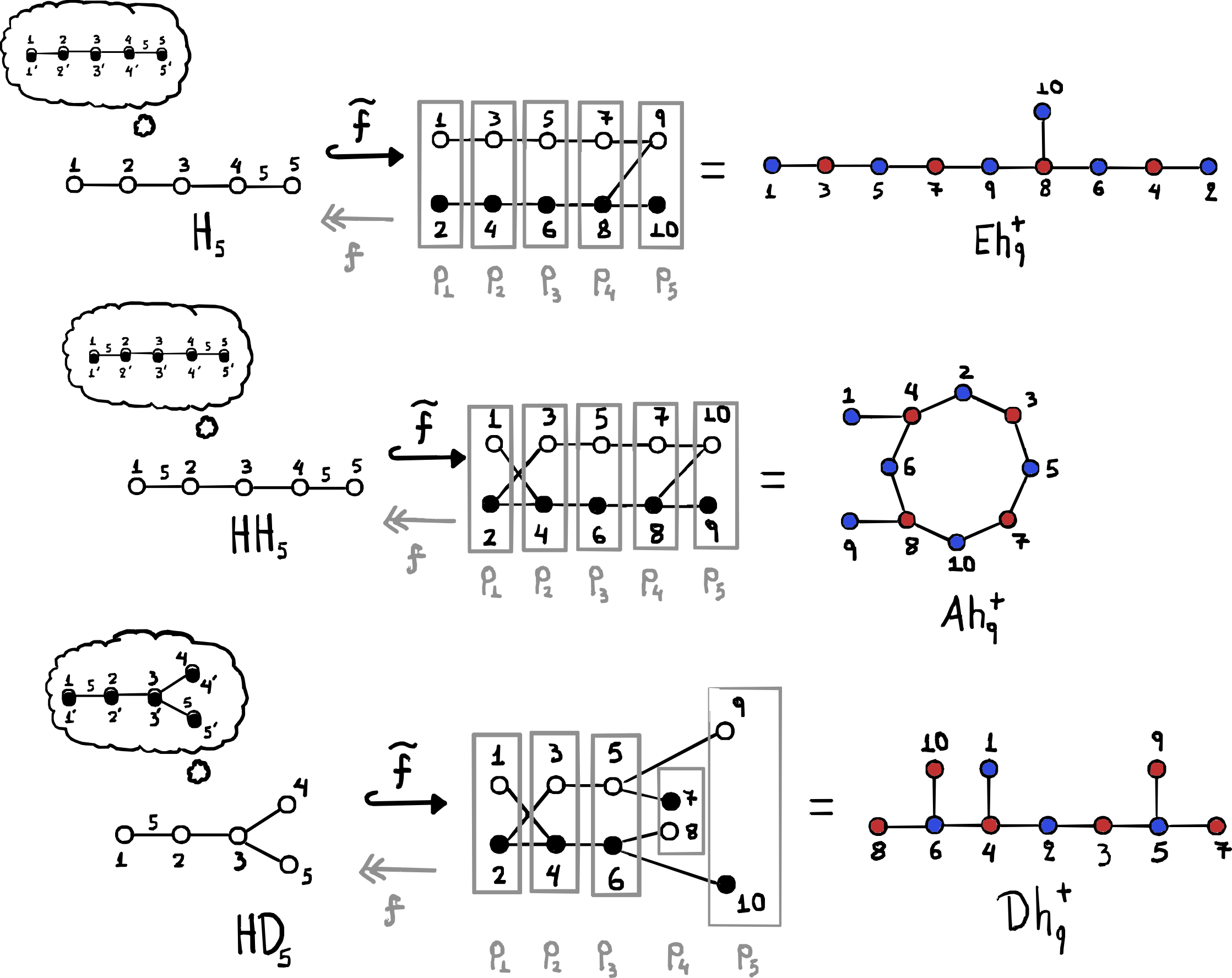}
    \caption[Unfolding hyperbolic non-crystallographic rank-5 groups into their crystallographic counterparts of rank-10.]{Unfolding hyperbolic non-crystallographic rank-5 groups of the icosahedral family into their crystallographic counterparts of rank-10.}
    \label{fig:UnfoldingHyperbolicGroupsRank5}
\end{figure}

\begin{itemize}
    \item 
$H_5 \hookrightarrow Eh_9^+$: $H_5$ is also known as $L_{5333}$ in \cite{mcmullen2002coxeter}, and $Eh_9^+$ is \cite{mcmullen2002coxeter}'s $Eh_9$, extended by another hyperbolic node (our node 1 in Fig.~\ref{fig:UnfoldingHyperbolicGroupsRank5}). 
\begin{equation}
        S_{H_5} = \begin{pmatrix}
        \textcolor{olive}{2} & \textcolor{orange}{-1} & 0 & 0 & 0\\
        \textcolor{orange}{-1} & \textcolor{olive}{2} & \textcolor{orange}{-1} & 0 & 0\\
        0 & \textcolor{orange}{-1} & \textcolor{olive}{2} & \textcolor{orange}{-1} & 0\\
        0 & 0 & \textcolor{orange}{-1} & \textcolor{olive}{2} & \textcolor{violet}{-\phi} \\ 
        0 & 0 & 0 & \textcolor{violet}{-\phi} & \textcolor{olive}{2} 
    \end{pmatrix}
    \hookrightarrow
    \setlength{\arraycolsep}{0.50pt}
    \begin{pmatrix}
        \fixedblock{olive}{2 & 0\\ 0 & 2} & \fixedblock{orange}{-1&0 \\ 0&-1} & \fixedblock{white}{0&0\\0&0} & \fixedblock{white}{0&0\\0&0} &
        \fixedblock{white}{0&0\\0&0}\\
        \fixedblock{orange}{-1&0 \\ 0&-1} & \fixedblock{olive}{2 & 0\\ 0 & 2} & \fixedblock{orange}{-1&0 \\ 0&-1} & \fixedblock{white}{0&0\\0&0} &
        \fixedblock{white}{0&0\\0&0} \\
        \fixedblock{white}{0&0\\0&0} & \fixedblock{orange}{-1&0 \\ 0&-1} & \fixedblock{olive}{2 & 0\\ 0 & 2} & \fixedblock{orange}{-1&0 \\ 0&-1} & \fixedblock{white}{0&0\\0&0} \\
        \fixedblock{white}{0&0\\0&0} &
        \fixedblock{white}{0&0\\0&0} & 
        \fixedblock{orange}{-1&0 \\ 0&-1}& 
        \fixedblock{olive}{2 & 0\\ 0 & 2} & \fixedblock{violet}{0 & -1\\ -1 & -1}\\
        \fixedblock{white}{0&0\\0&0} &
        \fixedblock{white}{0&0\\0&0} &
        \fixedblock{white}{0&0\\0&0} & \fixedblock{violet}{0 & -1\\ -1 & -1} & \fixedblock{olive}{2 & 0\\ 0 & 2}
    \end{pmatrix}
    = S_{Eh_9^+}
    \label{eq:unfoldingH5toEh9Plus}
\end{equation}

\item 
$HH_5 \hookrightarrow Ah_9^+:$ $HH_5$ is our notation for the Lann\'er graph from \cite{humphreys_1990}, as this is another hyperbolic variant of $H_5$ from above; $Ah_9^+$ is again an extended version of $Ah_9$ from \cite{mcmullen2002coxeter} that we get by attaching node 1 or 9 as shown in Fig.~\ref{fig:UnfoldingHyperbolicGroupsRank5}. 
\begin{equation}
        S_{HH_5} = \begin{pmatrix}
        \textcolor{olive}{2} & \textcolor{violet}{-\phi} & 0 & 0 & 0\\
        \textcolor{violet}{-\phi} & \textcolor{olive}{2} & \textcolor{orange}{-1} & 0 & 0\\
        0 & \textcolor{orange}{-1} & \textcolor{olive}{2} & \textcolor{orange}{-1} & 0\\
        0 & 0 & \textcolor{orange}{-1} & \textcolor{olive}{2} & \textcolor{violet}{-\phi} \\ 
        0 & 0 & 0 & \textcolor{violet}{-\phi} & \textcolor{olive}{2} 
    \end{pmatrix}
    \hookrightarrow
    \setlength{\arraycolsep}{0.50pt}
    \begin{pmatrix}
        \fixedblock{olive}{2 & 0\\ 0 & 2} & \fixedblock{violet}{0 & -1\\ -1 & -1} & \fixedblock{white}{0&0\\0&0} & \fixedblock{white}{0&0\\0&0} &
        \fixedblock{white}{0&0\\0&0}\\
        \fixedblock{violet}{0 & -1\\ -1 & -1} & \fixedblock{olive}{2 & 0\\ 0 & 2} & \fixedblock{orange}{-1&0 \\ 0&-1} & \fixedblock{white}{0&0\\0&0} &
        \fixedblock{white}{0&0\\0&0} \\
        \fixedblock{white}{0&0\\0&0} & \fixedblock{orange}{-1&0 \\ 0&-1} & \fixedblock{olive}{2 & 0\\ 0 & 2} & \fixedblock{orange}{-1&0 \\ 0&-1} & \fixedblock{white}{0&0\\0&0} \\
        \fixedblock{white}{0&0\\0&0} &
        \fixedblock{white}{0&0\\0&0} & 
        \fixedblock{orange}{-1&0 \\ 0&-1}& 
        \fixedblock{olive}{2 & 0\\ 0 & 2} & \fixedblock{violet}{0 & -1\\ -1 & -1}\\
        \fixedblock{white}{0&0\\0&0} &
        \fixedblock{white}{0&0\\0&0} &
        \fixedblock{white}{0&0\\0&0} & \fixedblock{violet}{0 & -1\\ -1 & -1} & \fixedblock{olive}{2 & 0\\ 0 & 2}
    \end{pmatrix}
    = S_{Ah_9^+}
    \label{eq:unfoldingHH5toAh9Plus}
\end{equation}

\item 
$HD_5 \hookrightarrow Dh_9^+:$ $HD_5$ is also known as $K_{535}$ in  \cite{mcmullen2002coxeter}; our notation is because of its resemblance to the spherical $D_5$. $Dh_9^+$ is an extended version of $Dh_9$ from \cite{mcmullen2002coxeter}; the additional hyperbolic node is node 10 in Fig.~\ref{fig:UnfoldingHyperbolicGroupsRank5}. 

\begin{equation}
        S_{HD_5} = \begin{pmatrix}
        \textcolor{olive}{2} & \textcolor{violet}{-\phi} & 0 & 0 & 0\\
        \textcolor{violet}{-\phi} & \textcolor{olive}{2} & \textcolor{orange}{-1} & 0 & 0\\
        0 & \textcolor{orange}{-1} & \textcolor{olive}{2} & \textcolor{orange}{-1} & \textcolor{orange}{-1}\\
        0 & 0 & \textcolor{orange}{-1} & \textcolor{olive}{2} & 0 \\ 
        0 & 0 & \textcolor{orange}{-1} &0 & \textcolor{olive}{2} 
    \end{pmatrix}
    \hookrightarrow
    \setlength{\arraycolsep}{0.50pt}
    \begin{pmatrix}
        \fixedblock{olive}{2 & 0\\ 0 & 2} & \fixedblock{violet}{0 & -1\\ -1 & -1} & \fixedblock{white}{0&0\\0&0} & \fixedblock{white}{0&0\\0&0} &
        \fixedblock{white}{0&0\\0&0}\\
        \fixedblock{violet}{0 & -1\\ -1 & -1} & \fixedblock{olive}{2 & 0\\ 0 & 2} & \fixedblock{orange}{-1&0 \\ 0&-1} & \fixedblock{white}{0&0\\0&0} &
        \fixedblock{white}{0&0\\0&0} \\
        \fixedblock{white}{0&0\\0&0} & \fixedblock{orange}{-1&0 \\ 0&-1} & \fixedblock{olive}{2 & 0\\ 0 & 2} & \fixedblock{orange}{-1&0 \\ 0&-1} & \fixedblock{orange}{-1&0 \\ 0&-1} \\
        \fixedblock{white}{0&0\\0&0} &
        \fixedblock{white}{0&0\\0&0} & 
        \fixedblock{orange}{-1&0 \\ 0&-1}& 
        \fixedblock{olive}{2 & 0\\ 0 & 2} & \fixedblock{white}{0 & 0\\ 0 & 0}\\
        \fixedblock{white}{0&0\\0&0} &
        \fixedblock{white}{0&0\\0&0} &
        \fixedblock{orange}{-1&0 \\ 0&-1} & \fixedblock{white}{0 & 0\\ 0 & 0} & \fixedblock{olive}{2 & 0\\ 0 & 2}
    \end{pmatrix}
    = S_{Dh_9^+}
    \label{eq:unfoldingHD5toDh9Plus}
\end{equation}

\end{itemize}

\chapter{Mirror Intersections and Ultra-Ideal Vertices}
\label{app:hyperbolicmirrorintersections}

This appendix collects geometric facts about how the bounding mirrors of a hyperbolic Coxeter polyhedron meet, focusing on the two distinct ways an \emph{ultra-ideal} vertex can arise. The conventions introduced in Section \ref{ch:MatBackgroundPartII} are followed throughout. 

Let $V = \MM^d = \mathbb{R}^{d-1,1}$ be a Minkowski spacetime on which a hyperbolic Coxeter group $G$ acts by reflections. Its Schl\"afli matrix $S$ has signature $(d-1,1)$, which defines the inner product $\langle {\bf x},{\bf y} \rangle$ for any two vectors ${\bf x,y} \in \MM^d$ (see Eq.~\eqref{eq:SchlafliInnerProduct}). $G$ is a discrete subgroup of $SO(d-1,1)$. Within $\MM^d$, and with respect to the inner product, we distinguish the following subspaces, all of which are preserved by an $SO(d-1,1)$ transformation:
\begin{itemize}
    \item The \textbf{future and past lightcones} $C_\pm$, defined as the locus of all null vectors in $\MM^d$:
    \begin{equation}
    \begin{aligned}
        & C_+ = \{ {\bf v} \in \MM^d : \langle {\bf v}, {\bf v} \rangle = 0 \text{ and } v^0 > 0\} \\
        & C_- = \{ {\bf v} \in \MM^d : \langle {\bf v}, {\bf v} \rangle = 0 \text{ and } v^0 < 0\}
        \end{aligned}
        \label{eq:LightconeDef}
    \end{equation}
    where $v^0$ is the temporal component of ${\bf v}$. 
    \item The \textbf{two-sheeted hyperboloid} $\HH^{d-1}$, or the inner (with respect to $C_\pm$) -- or timelike -- hyperboloid, which is the locus of all vector of constant timelike separation (negative; here taken to be unity) from the origin
    \begin{equation}
        \HH^{d-1} = \{ {\bf v} \in \MM^d : \langle {\bf v}, {\bf v} \rangle = -1 \} = \HH^{d-1}_+ \cup \HH^{d-1}_- .
        \label{eq:TimelikeHyperboloidDef}
    \end{equation}
    We call $\HH^{d-1}_\pm$ the upper/lower hyperboloids respectively, and usually we take $\HH^{d-1}_+$ (or omitting the $+$) to be a copy of hyperbolic space (as the induced metric on $\HH^{d-1}$ is positive-definite).
    \item The \textbf{one-sheeted hyperboloid} $\HH^{d-2,1}$, or outer (with respect to $C_{\pm}$ -- or spacelike -- hyperboloid, which is the locus of all vectors of constant spacelike separation (positive; here taken to be unity) from the origin
    \begin{equation}
        \HH^{d-2,1} = \{ {\bf v} \in \MM^d : \langle {\bf v}, {\bf v} \rangle = +1 \}
        \label{eq:SpacelikeHyperboloidDef}.
    \end{equation}
    This is a copy of \textbf{de Sitter spacetime} $\mathrm{dS_{d-1}}$, since the induced metric on this surfaces is of signature $(d-2,1)$.   
\end{itemize}

\section{What Can a Single Mirror Intersect?}
\label{sec:SingleMirrorIntersections}

\begin{figure}[t!]
\centering
    \begin{tabular}{cc}
        \subfloat[Mirror intersections with $\HH_+^{2}$. 
        \label{fig:MirrorIntersectionsWithHyperbolicSpace}
        ]{
            \includegraphics[scale=0.35]{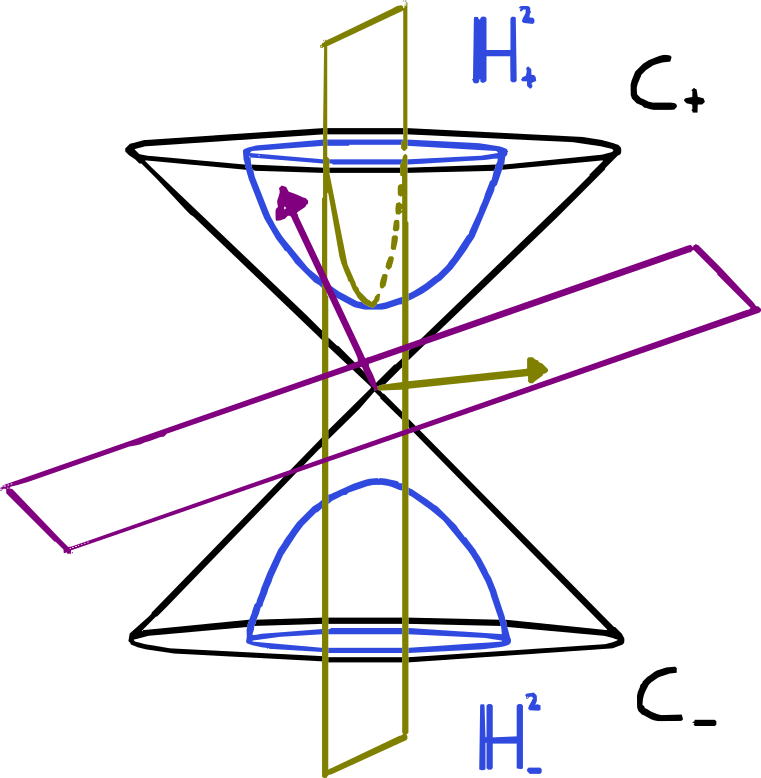}
            \hspace{1em}}
        \subfloat[
       Mirror intersections with $\dS_{2}$ \label{fig:MirrorIntersectionsWithDeSitter}]{    
            \includegraphics[scale=0.35]{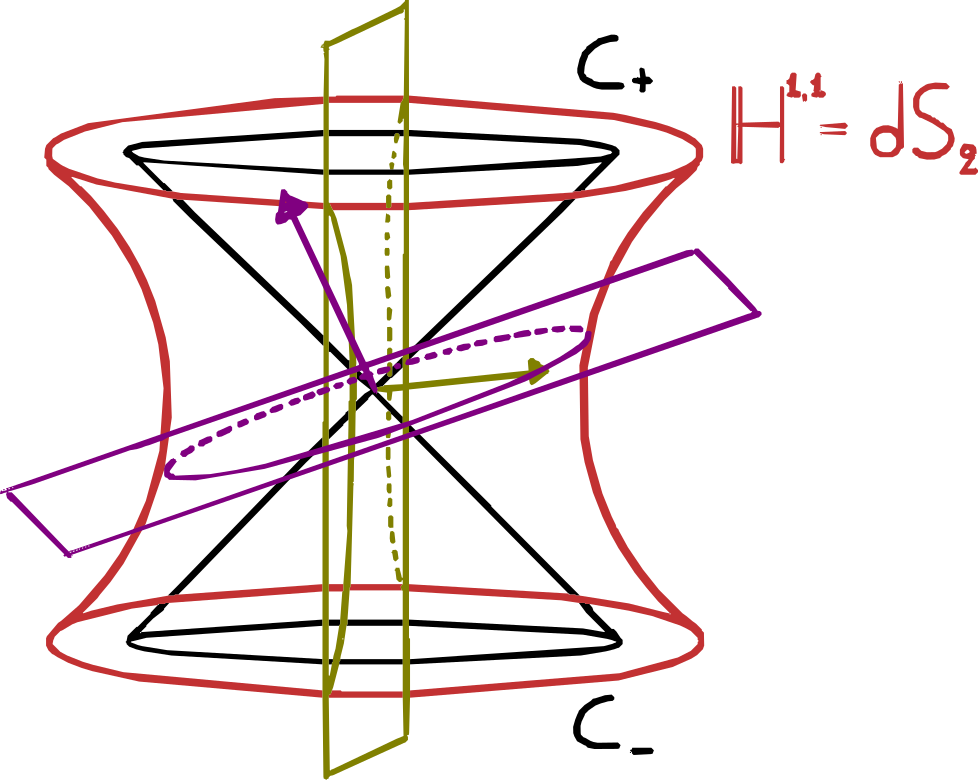}
        }
    \end{tabular}
\caption[Spacelike and timelike mirror intersections with the hyperboloids $\HH^2_+$ and $\dS_2$.]{Spacelike and timelike mirror intersections with the hyperboloids $\HH^2_+$ and $\dS_2$. The timelike hyperboloid $\HH^{2}_+$ on the left is only intersected by timelike mirrors shown in olive green colour (at hyperbolas). The spacelike hyperboloid $\dS_2$ on the right is intersected both by timelike and spacelike mirrors shown in purple colour (at hyperbolas and ellipses respectively).}
\label{fig:MirrorIntersectionsForSingleMirror}
\end{figure}

A given vector ${\bf r}$ in $\MM^d$ defines a codimension-1 hyperplane through the origin, its reflection mirror $M$. Depending on the nature of of the norm of ${\bf r}$, its mirror could intersect the following surfaces (see Fig.~\ref{fig:MirrorIntersectionsForSingleMirror}): 
\begin{itemize}
    \item \textbf{Timelike Vectors}: if ${\bf r} $ is timelike, then its mirror $M$ lies completely outside of the lightcones $C_{\pm}$ and intersects only the de Sitter hyperboloid $\dS_{d-1}$ at a spacelike surface. Such mirrors are called \textit{spacelike} (since their normal is timelike). Reflections across spacelike mirrors do not preserve the lightcones $C_\pm$ or the timelike hyperboloids $\HH^{d-1}_\pm$ but generally exchange them. This means that such transformations flip the orientation and the time (i.e.~exchange the future with the past) in $\MM^d$. However, they act as isometries on the de Sitter hyperboloid $\dS_{d-1}$.    
    \item \textbf{Spacelike Vectors:} if ${\bf r}$ is spacelike, then its mirror $M$ lies completely inside the lightcones $C_\pm$ and intersects \textit{both} the timelike $\HH^{d-1}$ \textit{and} the spacelike $\dS_{d-1}$ hyperboloids. Such mirrors are called timelike (since their normals are spacelike). Reflections across timelike mirrors are genuine isometries of $\HH^{d-1}_\pm$, i.e.~orientation-preserving, with no time-flipping occurring; they also preserve the spacelike hyperboloid $\dS_{d-1}$. Usually, if $\Lambda \subset \MM^d$ is a reflective hyperbolic lattice (see Section \ref{ch:MatBackgroundPartII}), we say that its automorphism group $\mathrm{Aut}(\Lambda)$ has an index-2 subgroup of orientation-preserving automorphisms $\mathrm{Aut}(\Lambda)$, which are generated by the spacelike roots of a lattice. The timelike roots are not considered as they only generate a flip, a $\ZZ_2$ transformation, such that the full group is $\mathrm{Aut}(\Lambda) = \mathrm{Aut}_+(\Lambda) \times \ZZ^2$ (see Section \ref{sec:Symmetries_of_Lattices}). 
\end{itemize}

Note that one can span Minkowski space $\MM^d$ entirely with linearly independent spacelike vectors, without the need of any timelike vector. The easiest example to see that is to consider just $\MM^{2}$ and two spacelike vectors in opposing sides of the lightcone $C_+$ (for instance, ${\bf v_1} = (2,1)$ and ${\bf v}_2 = (2,-1)$. We can see that by taking arbitrary $\RR$-linear combinations of them we can define a timelike vector or a null vector. However, two timelike vectors and their scalar multiples cannot generate a spacelike direction. In this sense, considering only spacelike roots is appropriate for describing the more fundamental transformations. 

\section{Facets of a Coxeter Polyhedron}
\label{app:TwoMirrorsIntersecting}

For a Coxeter polyhedron, its geometry is encoded entirely in its Coxeter-Dynkin graph, equivalently in its Schl\"afli matrix. In the root basis $\{ \bf r\}_{i=1,\dots n}$ ($n$ is the number of simple roots or the rank of $G$), $S_{ij}$ is the same as the Gram matrix of inner products $G_{ij} = \langle {\bf r}_i , {\bf r}_j \rangle = {\bf r}_i^T S{\bf r}_j = S_{ij}$. Following \cite{vinberg1967discrete}, ${\bf r}_i$ are taken to be the outwards unit normals to their corresponding reflecting mirrors, which define the bounding facets of the polyhedron.

First of all, since all mirrors are hyperplanes through the origin,\footnote{Cases such as parallel mirrors, for example, are excluded by this condition.} they must meet \textit{somewhere} in $\MM^d$. So the question is not about \textit{whether} two mirrors meet, but about \textit{where} they meet. 

Let $M_i = {\bf r}_i^\perp$ (the orthogonal complement taken with respect to $S$) be the mirror. For any subset of $k$ mirrors $\Sigma_M = \{ i_1, i_2, \dots , i_k \}$ of the total $n$, let $F_{\Sigma_M}^\perp$ be the subspace spanned by the $k$ linearly independent roots ${\bf r}_{i_1}, \dots, {\bf r}_{i_k}$, and 
$$F_{\Sigma_M}= \bigcap_{i=1}^k M_i$$
be the linear subspace on which they intersect. Since the metric $S$ is non-degenerate, $\MM^d$ decomposes as $\MM^d = F_{\Sigma_M} \oplus F_{\Sigma_M}^\perp$, which implies that 
$$ \dim \MM^d = d = \dim F_{\Sigma_M} + \dim F_{\Sigma_M}^\perp \Rightarrow \dim F_{\Sigma_M} = d-k,
$$
so the intersection is a codimension $k$ surface. In addition, the signatures of the two spaces must add up to $(d-1,1)$, so whatever signature $F_{\Sigma_M}^\perp$ has, $F_{\Sigma_M}$ must have its complement. The signature of $F_{\Sigma_M}^\perp$ is determined by the induced metric on $F_{\Sigma_M}^\perp$, which is given by the reduced Schl\"afli matrix $S|_{\Sigma_M}^\perp = (S_{ij})_{i,j \in \Sigma_M}$ we get by only considering those indices that belong to the subset $\Sigma_M$. Finding the determinant of $S|_{\Sigma_M}^\perp$ we get the signature of $F_{\Sigma_M}^\perp$ and thus the signature of the intersection $F_{\Sigma_M}$. We call the set of all $F_{\Sigma_M}$'s, namely, all the possible length-$k$ subsets $\Sigma_M$, the $(d-k)$-facets of the Coxeter polyhedron. 0-facets are vertices, 1-facets are edges, 2-facets are faces and so on. 

\subsection{Where Can Two Mirrors Intersect?}

Let's first consider the simple case of just two mirrors ($|\Sigma_M| = 2)$ in $\MM^d$. $F_{\Sigma_M}^\perp$ is 2-dimensional, so the intersection $F_{\Sigma_M}$ is $d-2$-dimensional. The induced metric is 
$$
S|_{\Sigma_M}^\perp = 
\begin{pmatrix}
    2 & S_{i_1 i_2}\\
    S_{i_1 i_2} & 2
\end{pmatrix}
$$
with eigenvalues $2 \pm S_{i_1 i_2}$, where we take all non-diagonal entries of the Schl\"afli matrix to be non-positive \cite{vinberg1967discrete}. One eigenvalue is automatically positive; $2 + S_{i_1 i_2}$, though, depends on the particular Schl\"afli entry. Hence, we distinguish between the following three cases:
\begin{center}
\begin{tabular}{lllll}
$|S_{i_1 i_2}|$ & $\mathrm{sig}(F_{\Sigma_M}^\perp)$ & $\mathrm{sig}(F_{\Sigma_M})$ & \textbf{Intersection} & \textbf{Product $R_{i_1}R_{i_2}$} \\
$< 2$ & $(2,0)$; only \textit{spacelike} & $(d-3,1)$; has timelike & inside $\mathbb{H}^{d-1}$ & Rotation \\
$ = 2$ & degenerate (\textit{null}) & tangent to light cone & on $\partial\mathbb{H}^{d-1}$ & -- \\
$ > 2$ & $(1,1)$; has \textit{timelike} & $(d-2,0)$; only spacelike & outside, on $\mathrm{dS}_{d-1}$ & Boost \\
\end{tabular}
\end{center}
Geometrically, this trichotomy records \emph{where} the mirrors meet relative to the hyperboloid $\HH^{d-1}$. The first case is the regular one where the Schl\"afli entries are negative but not \textit{too} negative to turn their space $F_{\Sigma_M}^\perp$ into a Lorentzian one. The mirrors meet at a dihedral angle of $\pi/m_{i_1 i_2}$, where $S_{i_1 i_2} = -2\cos (\pi/m_{i_1 i_2})$, they intersect within the timelike hyperboloid $\HH^{d-1}$, and the element $R_{i_1}R_{i_2}$ of the Coxeter group can be given the interpretation of a rotation on a Euclidean 2-dimensional plane $F_{\Sigma_M}^{\perp}$. When the space $F_{\Sigma_M}^\perp$ becomes degenerate, the two mirrors become parallel and they only meet asymptotically at infinity, which mathematically corresponds to the boundary $\partial \HH^{d-1}$, on the light cone of the ambient space $\MM^d$. Finally, when $|S_{i_1 i_2}| > 2$, the Sch\"afli entry becomes too negative and $F_{\Sigma_M}^\perp$ contains a timelike direction, which makes the mirror intersection $F_{\Sigma_M}$ purely spatial. This means that the two mirrors miss each other on the timelike hyperboloid $\HH^{d-1}$ and they only meet at the spacelike hyperboloid $\dS_{d-1}$. Such mirrors will appear to diverge within $\HH^{d-1}$, and the reflection product cannot be given the interpretation of a rotation on $F_{\Sigma_M}^\perp$. Instead, the angle becomes ``imaginary'', which converts the cosine into a hyperbolic cosine and the angle $\pi/m_{ij}$ becomes a divergent distance $d_{ij}$ such that $S_{ij} = -2\cos(\pi/m_{ij}) = -2\cos(id_{ij}) = -2\cosh d_{ij} < -2$. In that case, $R_{i_1}R_{i_2}$ generate a boost on the 2-dimensional plane $F_{\Sigma_M}^\perp$.    

For rank-3 groups (right-angled triangle groups $[p,q,2]$), the above translates directly into the vertex classification we mentioned in Section \ref{subsec:SphAffHypCoxeterGroups}; see in particular Fig.~\ref{fig:OrdinaryIdealUltraIdealVertices}. \textit{Ordinary vertices} corresponds to $|S_{i_1 i_2}| < 2$, \textit{ideal vertices} to $|S_{i_1 i_2}| = 0$, and \textit{ultra-ideal vertices} to $|S_{i_1 i_2}| > 2$. This is because in $\HH^2$, the intersection of two mirrors defines always a vertex of the Coxeter polyhedron, so the classification of vertices reduces to the Coxeter labels on the graph. We call this \textit{mechanism 1} for identifying ultra-ideal vertices. We also note that, if $G$ is a rank-3 Coxeter group, then the reduced 2-by-2 Schl\"afli matrix $S|_{\Sigma_M}^\perp$ is a principal submatrix of $S$, and its determinant, the principal minor, determines the signature of the intersection space and thus the type of vertices here.

\subsection{Classifying Vertices of the Coxeter Polyhedron}

More generally, a vertex of the Coxeter polyhedron on the $(d-1)$-dimensional space $\HH^{d-1}$ is a codimension $d-1$ object (i.e.~zero-dimensional), thus arising from the intersection of $d-1$ reflecting mirrors. Here, we will see a more general prescription to identify the type of vertices based on this last observation we made. The original argument is due to Vinberg \cite{vinberg1967discrete,vinberg1972groups,vinberg1975some}, and it gives a \textit{mechanism 2} for generating ultra-ideal vertices, even in the case where all Schl\"afli entries are well-behaved $|S_{i_1 i_2}|<0$ and all pairs of mirrors meet inside $\HH^{d-1}$, so we cannot identify them by just reading the Coxeter labels.

Since a vertex is a 0-facet of the Coxeter polyhedron, we must look into intersections of $d-1$ mirrors. Now, $F_{\Sigma_M}^\perp$ contains $d-1$ (out of the total $d$) directions in $\MM^d$, and the key to the classification is again hidden in the eigenvalues of the induced matrix $S|_{\Sigma_M}^\perp$. The latter is itself a principal submatrix of $S$, which we get from $S$ by removing the row and column corresponding to a particular node of the Coxeter-Dynkin graph. Denote this matrix $S^{(i)}$, where $i$ is the removed node. Based on the determinant of this matrix, we distinguish the following possibilities:
\begin{center}
\begin{tabular}{lllll}
$\det|S^{(i)}|$ & $\mathrm{sig}(F_{\Sigma_M}^\perp)$ & $\mathrm{sig}(F_{\Sigma_M})$ & \textbf{Intersection} & \textbf{$i$th Vertex Type} \\
$ > 0$ & $(d-1,0)$; only \textit{spacelike} & $(0,1)$; timelike & inside $\mathbb{H}^{d-1}$ & Ordinary \\
$ = 0$ & degenerate (\textit{null}) & tangent to light cone & on $\partial\mathbb{H}^{d-1}$ & Ideal \\
$ < 0$ & $(d-2,1)$; has \textit{timelike} & $(1,0)$; only spacelike & outside, on $\mathrm{dS}_{d-1}$ & Ultra-Ideal \\
\end{tabular}
\end{center}
As a result, we see that, even if there are no diverging mirrors within $\HH^{d-1}$ (i.e. all the rank-2 principal minors are positive), we can still generate ultra-ideal vertices by considering the subdiagrams we get from the Coxeter graph of $G$ and their determinants (i.e. a higher-rank principal minor can still get negative). This classification of vertices is similar to how we classified hyperbolic graphs in Section \ref{subsec:SphAffHypCoxeterGroups}. Ordinary vertices corresponds to spherical subgraphs, ideal vertices to affine and ultra-ideal to hyperbolic. 

Finally, let us comment on how mechanism 1 is included in mechanism 2: if a pair of mirrors is already diverging, then there is at least one rank-2 principal minor that is negative. Every larger submatrix containing this pair will also have negative determinant, and since there are always precisely one negative eigenvalue for our hyperbolic Coxeter groups \footnote{Recall, we refer to Coxeter groups as hyperbolic if they tessellate $\HH^{d-1}$, and we allow for these ``incomplete'' tessellations by non-bounded Coxeter polyhedra.}, any vertex involving an this divergent pair is automatically ultra-ideal, and the diagnosis is immediate from a single off-diagonal Schl\"afli entry.

\section{Simplicial Hyperbolic Crystallographic Groups with Ideal Vertices}
\label{subsec:UnclassifiedLatticesAppendix}

In this last section of the appendix, we would like to comment on three different types of simplices we can have according to the above vertex classification. 

First of all, a simplex in $(d-1)$-dimensional space like $\HH^{d-1}$ requires the minimum number of bounding mirrors, which is $d$; in that sense, it is ``simple''. In Section \ref{subsec:Symmetries_of_Lorentzian_Lattices}, we mentioned that these ultra-ideal vertices for the integer self-dual lattices, which correspond to having dashed lines in the Coxeter graph, start to appear in dimension 18, and we give the graphs in Appendix \ref{app:SymmetriesLorentzianLattices}. These groups have more than $d$ simple roots, thus more than $d$ fundamental mirrors, and thus their Coxeter polyhedron is way more complicated than a simplex. 

However, we will see crystallographic groups whose Coxeter graph has exactly $d$ simple roots and yet they still have ultra-ideal vertices because they contain hyperbolic subdiagrams. By the usual definitions, since the bounding mirrors are still $d$, they must bound a simplex, but this will now have infinite volume (from the point of view of $\HH^{d-1}$ which $G$ is supposed to tessellate by reflections). So how does such a simplex look like in $\MM^d$? 

\begin{figure}[h!]
\centering
    \begin{tabular}{cc}
        \subfloat[Simplex with only ordinary vertices 
        \label{fig:SimpleWithOrdinaryVertices}
        ]{
            \includegraphics[scale=0.4]{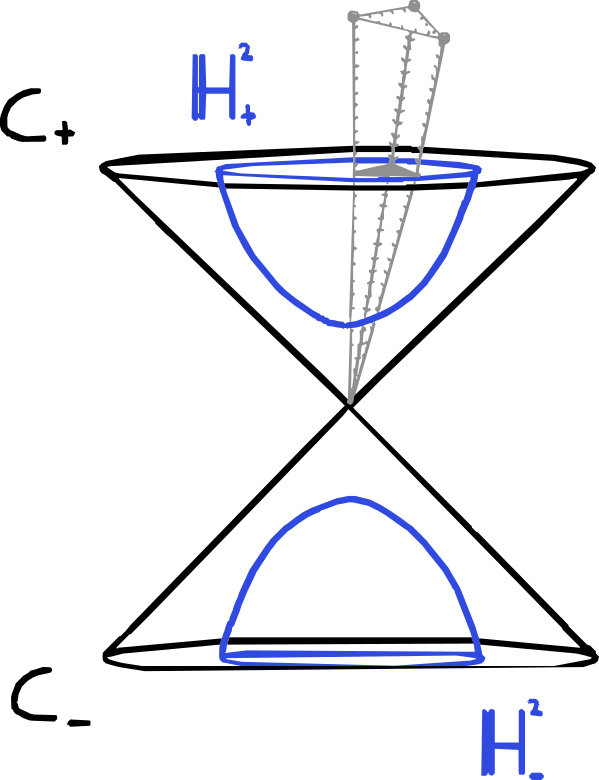}
            \hspace{5em}}
        \subfloat[Simplex with ideal vertices
        \label{fig:SimplexWithIdealVertices}]{    
            \includegraphics[scale=0.39]{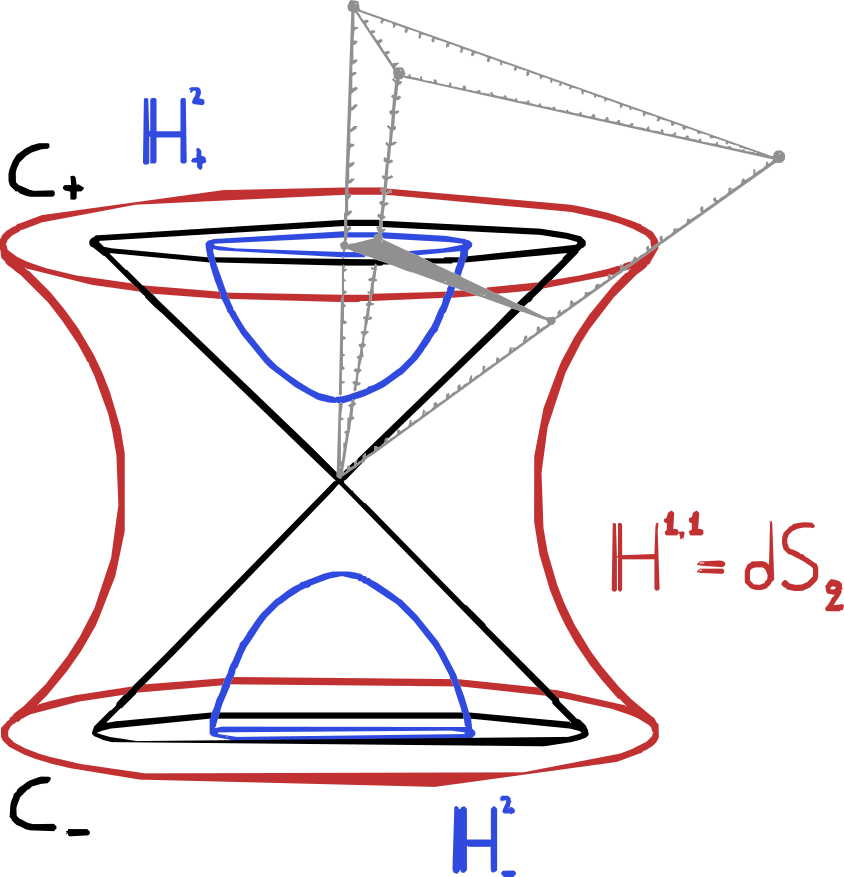}
        }
    \end{tabular}
\caption[Examples of simplices in $\MM^3$ with ordinary and ideal vertices.]{Examples of simplices in $\MM^3$ with ordinary and ideal vertices. On the left, all mirrors intersect within $\HH^2_+$ and give rise to ordinary (timelike) vertices. On the right, two pairs of mirrors intersect within $\HH_+^2$ and produce ordinary vertices, but another pair meets at the de Sitter hyperboloid $\dS_2$ instead; this intersection corresponds to an ideal vertex of the simplex.}
\label{fig:SimplicesWithVerticesOnTimelikeSpacelikeHyperboloids}
\end{figure}

Using the above classification of vertices and the intuition we got from our analysis, we end up with the simplices shown in Fig.~\ref{fig:SimplicesWithVerticesOnTimelikeSpacelikeHyperboloids}. When ultra ideal vertices are present, those must live on a spacelike point in the ambient Minkowski space $\MM^d$, which means that to close the simplex, one must look at the de Sitter hyperboloid $\dS_{d-1}$ for the missing non-ordindary vertices. From the point of view of the tessellating space $\HH^{d-1}$, such an unbounded simplex has infinite volume. But because all the mirrors see each other in the ambient space and they do intersect (because all their dihedral angles are well-defined and equal to some $\pi/m_{ij}$), they will carve out a simplex of finite volume from $\MM^d$, but with some vertices that are timelike (ordinary) and some vertices that are spacelike (ultra-ideal).

\chapter{Reflection Symmetries of Odd Self-Dual Lorentzian Lattices $I_{s,1}$}
\label{app:SymmetriesLorentzianLattices}

In this Appendix, we summarize the reflection symmetries of the odd self-dual lattices $I_{s,1}$ for dimension $s+1\leq 20$ ({\it i.e.}\ the dimensions for which the Coxeter group has a finite number of fundamental roots).

For $3\leq s+1\leq 18$, the fundamental roots and Coxeter-Dynkin diagrams for the lattices $\mathrm{I}_{s,1}$ and $\mathrm{II}_{s,1}$ were determined by Vinberg in \cite{vinberg1967discrete, vinberg1972groups, vinberg1975some} (see Table 4A and 4B in \cite{vinberg1975some} in particular).  The cases $s+1=19,20$ were then treated by Vinberg and Kaplinskaja in \cite{kaplinskaja1978groups} (see also \cite{meyer1977prasentation}).  And for $s+1>20$ there are infinitely many fundamental roots \cite{ESSELMANN1996103}. For convenience, since some of these references are only available in Russian or German, or contain errors in the printed diagrams (e.g.~in \cite{vinberg1975some}), we have carefully re-checked their results, and present all the graphs here, along with a list of the fundamental roots, for all the case with a finite number of fundamental roots ({\it i.e.}\ $s+1\leq 20$).

We follow Vinberg's approach in \cite{vinberg1975some,kaplinskaja1978groups} with the difference that our fundamental roots are denoted by $\mathbf{r}_i$ (instead of $\mathbf{e}_j$), while our standard orthonormal (Cartesian) basis vectors are denoted by ${\bf e}_{i}$ (instead of ${\bf u}_{i}$); and, to align with the physicists' notation, we label the Cartesian basis in $\mathbb{R}^{s,1}$ as $\{\mathbf{e}_j\}$ ($j=0,\dots,s$), with $\mathbf{e}_0$ corresponding to the temporal basis vector. 

For $3\leq s+1\leq 18$, the fundamental roots for the lattices $\mathrm{I}_{s,1}$ are given in Table \ref{tab:SimpleRootsReflectiveOddLatticesN2to17}.  For $s+1=19$ or $20$, the first $s+8$ fundamental roots are given in Table \ref{tab:SimpleRootsReflectiveOddLatticesN1819}; while for $s+1=19$, the remaining fundamental roots are given in Table \ref{tab:AllSimpleRootsReflectiveOddLattices18}; and for $s+1=20$, the remaining fundamental roots are given in Table \ref{tab:AllSimpleRootsReflectiveOddLattices19}.

\begin{table*}[!t]
    \centering
    \begin{tabular}{c | c | c | c }
         Root Index $i$ & Root Vector $\mathbf{r}_i$ & $\langle \mathbf{r}_i,\mathbf{r}_i \rangle$ & for which $s$ \\
         \vspace{-2ex}& & & \\
         \hline \hline 
         \vspace{-1.75ex}& & & \\
         $1 \leq i \leq s-1$ & $-\mathbf{e}_i + \mathbf{e}_{i+1}$ & $2$ & $s\geq 2$ \\[+1ex]
         \hline 
         \vspace{-1.75ex}& & & \\
         $s$ & $-\mathbf{e}_s$ & $1$ & $s\geq 2$ \\[1ex] \hline
         \vspace{-1.75ex}& & & \\
         $s+1$ & $\mathbf{e}_0 + \mathbf{e}_1 + \mathbf{e}_2$ & 1 & $s=2$ \\[+1ex]
         & $\mathbf{e}_0 + \mathbf{e}_1 + \mathbf{e}_2 + \mathbf{e}_3$ & 2 & $s\geq 3$ \\[+1ex] \hline
        \vspace{-1.75ex}& & & \\
         $s+2$ & $3\mathbf{e}_0 + \mathbf{e}_1 +  \dots + \mathbf{e}_{10}$ & 1 & $s=10$ \\[1ex]
         &$3\mathbf{e}_0 + \mathbf{e}_1 +  \dots + \mathbf{e}_{11}$ & 2 & $s\geq 11$ \\[+1ex] \hline
        \vspace{-1.75ex}& & & \\
         $s+3$ & $4\mathbf{e}_0 + 2\mathbf{e}_1 + \mathbf{e}_2 + \dots + \mathbf{e}_{14}$ & 1 & $s=14$ \\[1ex]
               & $4\mathbf{e}_0 + 2\mathbf{e}_1 + \mathbf{e}_2 + \dots + \mathbf{e}_{15}$ & 2 & $s\geq 15$ \\[+1ex] \hline
         \vspace{-1.75ex}& & & \\      
         $s+4$ & $6\mathbf{e}_0 + 2(\mathbf{e}_1 + \dots + \mathbf{e}_7) + (\mathbf{e}_8 + \dots + \mathbf{e}_{16})$ & 2 & $s=16$ \\[1ex] 
        & $4\mathbf{e}_0 + \mathbf{e}_1 +  \dots + \mathbf{e}_{17}$ & 1 & $s=17$ \\[1ex]\hline
         \vspace{-1.75ex}& & & \\
         $s+5$ & $6\mathbf{e}_0 + 2(\mathbf{e}_1 + \dots + \mathbf{e}_7) + (\mathbf{e}_8 + \dots + \mathbf{e}_{17})$ & 2  & $ s=17$ 
    \end{tabular}
    \caption[A list of the fundamental roots for all odd self-dual lattices $\mathrm{I}_{s,1}$ with $3 \leq s+1\leq 18$.]{A list of the fundamental roots for all odd self-dual lattices $\mathrm{I}_{s,1}$ with $3 \leq s+1\leq 18$. The indices in the first column match the labelling of the roots in the Coxeter-Dynkin graphs of Figs.~\ref{fig:VinbergRootsI1nUpto10}, \ref{fig:OddGraphsN10to13}, and \ref{fig:OddGraphsN14to17}.  The second column expresses each root in terms of the standard Cartesian basis $\{{\bf e}_{0},\ldots,{\bf e}_{s}\}$ for $\mathbb{R}^{s,1}$, where ${\bf e}_0$ is the unit vector along the timelike direction. The third column shows the root's norm ({\it i.e.}\ whether it is a short or a long root).  And the fourth column explains for which $s$ ({\it i.e.}\ in which number of spatial dimensions) a given fundamental root exists.  } 
\label{tab:SimpleRootsReflectiveOddLatticesN2to17}
\end{table*}

\begin{table*}[t]
    \centering
    \begin{tabular}{c | c | c }
         Root Index $i$ & Root Vector $\mathbf{r}_i$ & $\langle \mathbf{r}_i,\mathbf{r}_i \rangle$ \\
         \vspace{-2ex}& & \\
         \hline \hline 
         \vspace{-1.75ex}& & \\
         $1 \leq i \leq s-1$ & $-\mathbf{e}_i + \mathbf{e}_{i+1}$ & $2$ \\[+1ex]
         \hline 
         \vspace{-1.75ex}& & \\
         $s$ & $-\mathbf{e}_s$ & $1$ \\[1ex] \hline
         \vspace{-1.75ex}& & \\
         $s+1$ & $\mathbf{e}_0 + \mathbf{e}_1 + \mathbf{e}_2 + \mathbf{e}_3$ & 2 \\[+1ex] \hline
        \vspace{-1.75ex}& & \\
         $s+2$ &$3\mathbf{e}_0 + \mathbf{e}_1 +  \dots + \mathbf{e}_{11}$ & 2 \\[+1ex] \hline
        \vspace{-1.75ex}& & \\
         $s+3$ & $4\mathbf{e}_0 + 2\mathbf{e}_1 + \mathbf{e}_2 + \dots + \mathbf{e}_{15}$ & 2 \\[+1ex] \hline
         \vspace{-1.75ex}& & \\      
         $s+4$ & $4\mathbf{e}_0 + \mathbf{e}_1 +  \dots + \mathbf{e}_{18}$ & 2 \\[1ex]\hline
         \vspace{-1.75ex}& & \\
         $s+5$ & $5\mathbf{e}_0 + 3\mathbf{e}_1 + \mathbf{e}_2 + \dots + \mathbf{e}_{s}$ & 1,2 
         \\[1ex] \hline
         \vspace{-1.75ex}& & \\
         $s+6$ & $6\mathbf{e}_0 + 2(\mathbf{e}_1 + \dots + \mathbf{e}_7) + \mathbf{e}_8 + \dots + \mathbf{e}_{17}$ & 2  
         \\[1ex] \hline
         \vspace{-1.75ex}& & \\
         $s+7$ & $7\mathbf{e}_0 + 3\mathbf{e}_1 + 2(\mathbf{e}_2 + \dots + \mathbf{e}_9) + \mathbf{e}_{10} + \dots + \mathbf{e}_{s}$ & 1,2 
         \\[1ex]\hline
         \vspace{-1.75ex}& & \\
         $s+8$ & \, $9\mathbf{e}_0 + 3(\mathbf{e}_1 + \dots + \mathbf{e}_5) + 2(\mathbf{e}_6 + \dots + \mathbf{e}_{13}) + \mathbf{e}_{14} + \dots + \mathbf{e}_{s}$ \, & 1,2 \\[1ex]
    \end{tabular}
    \caption[A list of the first $s+8$ fundamental roots for the odd self-dual lattices $\mathrm{I}_{s,1}$ when $s+1=19$ or $20$.]{A list of the first $s+8$ fundamental roots for the odd self-dual lattices $\mathrm{I}_{s,1}$ when $s+1=19$ or $20$. The indices in the first column match the labelling of the roots in the Coxeter-Dynkin graphs of Fig.~\ref{fig:OddGraphsN18to19}.  The second column expresses each root in terms of the standard Cartesian basis $\{{\bf e}_{0},\ldots,{\bf e}_{s}\}$ for $\mathbb{R}^{s,1}$, where ${\bf e}_0$ is the unit vector along the timelike direction. The third column shows the root's norm ({\it i.e.}\ whether it is a short or a long root).} 
\label{tab:SimpleRootsReflectiveOddLatticesN1819}
\end{table*}

Using the fundamental roots in these tables, one can construct the corresponding Coxeter-Dynkin graphs for the lattice $\mathrm{I}_{s,1}$ in Minkowski space of dimension $d=s+1$, with $3\leq d \leq 20$.  In order to describe the structure of the remaining roots, we present these graphs in four batches of increasing complexity, commenting on certain properties and interesting aspects along the way.  (In any dimension $d=s+1$, the first $s$ roots form a copy of the spherical/finite Coxeter graph of type $B_{s}$.)
\begin{itemize}
\item \textbf{Dimensions $3 \leq d \leq 10$}: see Fig. \ref{fig:VinbergRootsI1nUpto10}.  For these graphs, the number of fundamental roots 
(the rank) equals the dimension, $n=d$, so the fundamental roots form a (linearly-independent) basis for $\mathbb{R}^{s,1}$, and define a fundamental domain in hyperbolic space $H^{s}$ bounded by $s+1$ mirrors (a hyperbolic {\it simplex}).

\item \textbf{Dimensions $11 \leq d \leq 14$}: see Fig. \ref{fig:OddGraphsN10to13}. Now the rank exceeds the dimension by one, $n=d+1$, so the fundamental roots are over-complete, and the fundamental domain in $H^{s}$ is {\it not} a simplex \cite{vinberg1975some}. 

\begin{figure}[!h]
\centering
\includegraphics[scale=0.25]{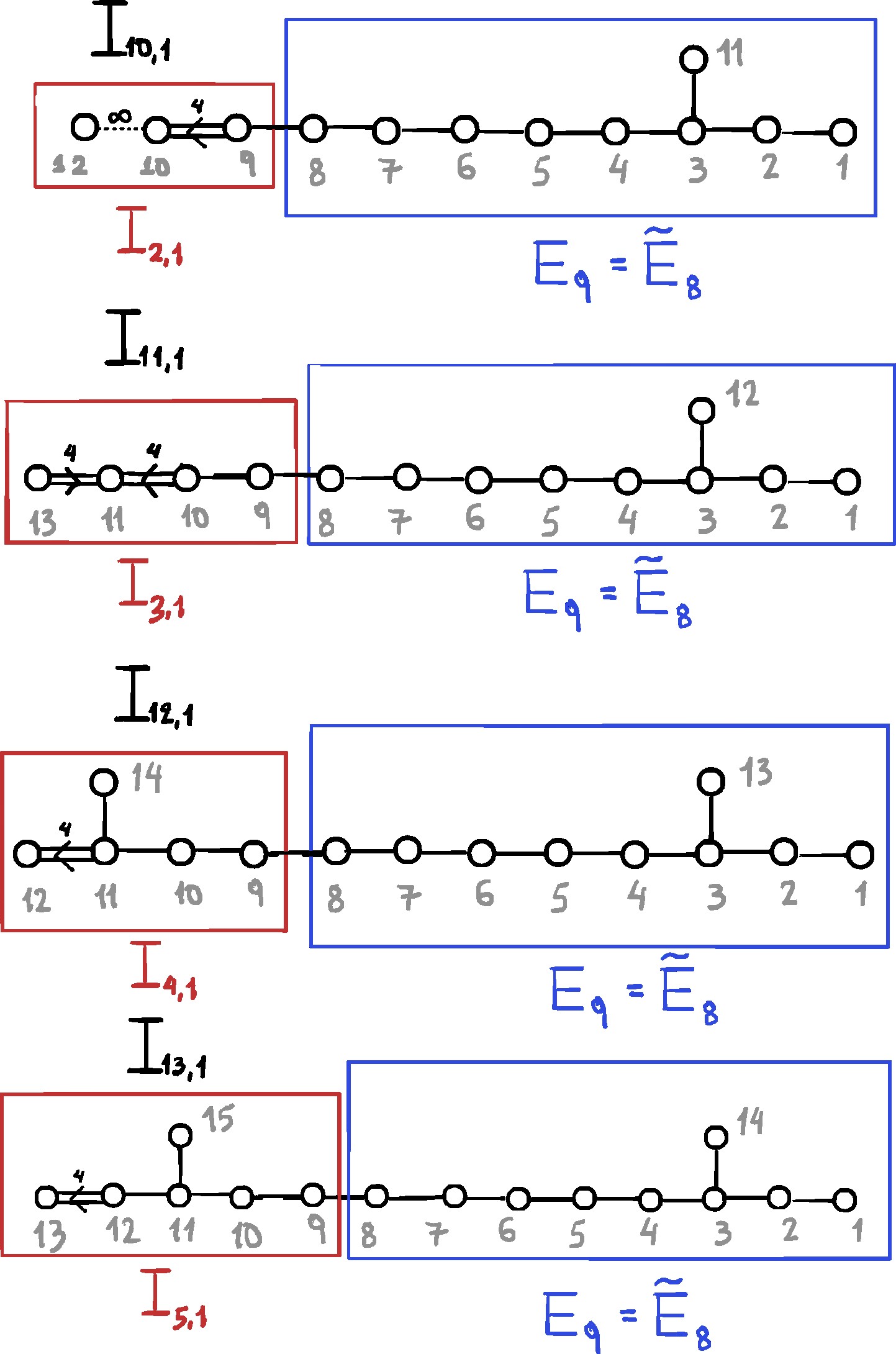}
\caption{\justifying The Coxeter-Dynkin diagrams for $I_{s,1}$ for $11\leq s+1\leq 14$ (with fundamental roots given in Table \ref{tab:SimpleRootsReflectiveOddLatticesN2to17}).}
\label{fig:OddGraphsN10to13}
\end{figure}

Note a kind of 8-fold periodicity here: the graphs for $11 \leq d \leq 14$ are obtained from those in dimensions $3 \leq d \leq 6$ by `gluing' them (see \S 4 in \cite{conway2013sphere}) to a copy of $\tilde{E}_8 = E_9$ ({\it i.e.}\ the affine extension of $E_8$).

Note a misprint in \cite{vinberg1975some}: the $\mathrm{I}_{11,1}$ graph incorrectly has a triple (rather than double) line in its $\mathrm{I}_{3,1}$ component, obscuring this 8-fold periodicity.

\item \textbf{Dimensions $15 \leq d \leq 18$}: 
see Fig.~\ref{fig:OddGraphsN14to17}. The rank continues to grow relative to the dimension : $n=d+2=s+3$ in dimension $d=15,16$; $n=d+3=s+4$ in dimension $d=17$; and $n=d+4=s+5$ in dimension $d=18$.  

Note that in dimensions $d\geq15$, the Coxeter graphs have non-trivial symmetries ({\it e.g.} the horizontal reflection symmetry of the graph for $I_{14,1}$), are no longer trees (as they contain closed loops), and are no longer bicolorable (except when $d=16$) since they contain at least one closed loop of odd length.  For $d\geq18$ they are also no longer {\it planar} graphs.

\begin{figure*}
    \centering
    \subfloat[]{%
        \includegraphics[width=0.95\linewidth]{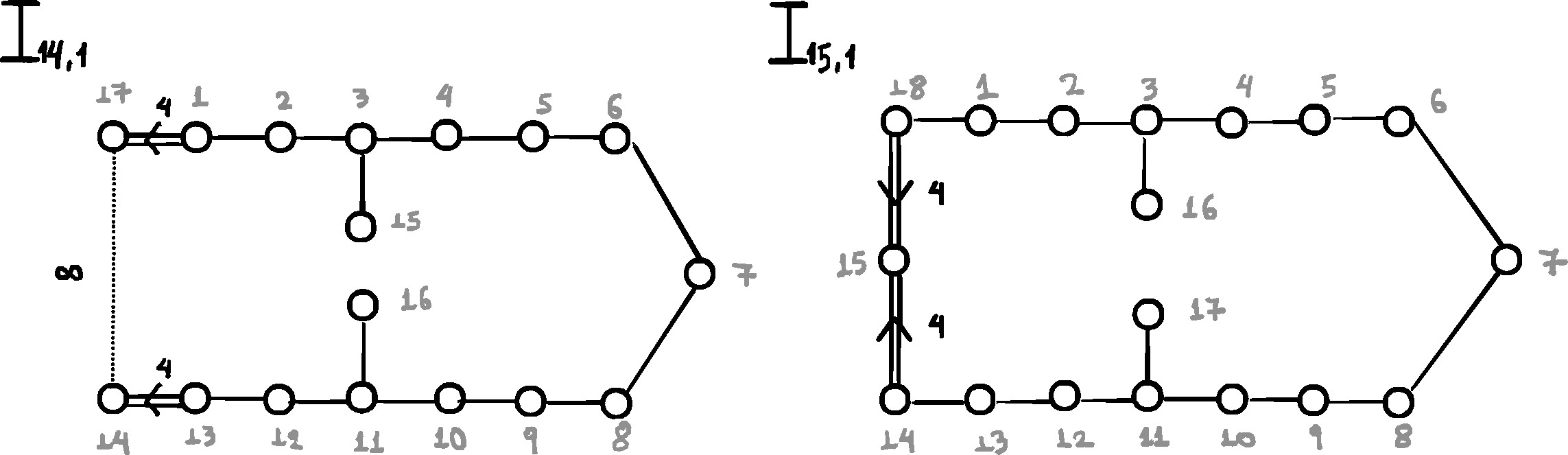}%
        \label{fig:OddGraphsN14and15}%
    }\par
    \subfloat[]{%
        \includegraphics[width=0.80\linewidth]{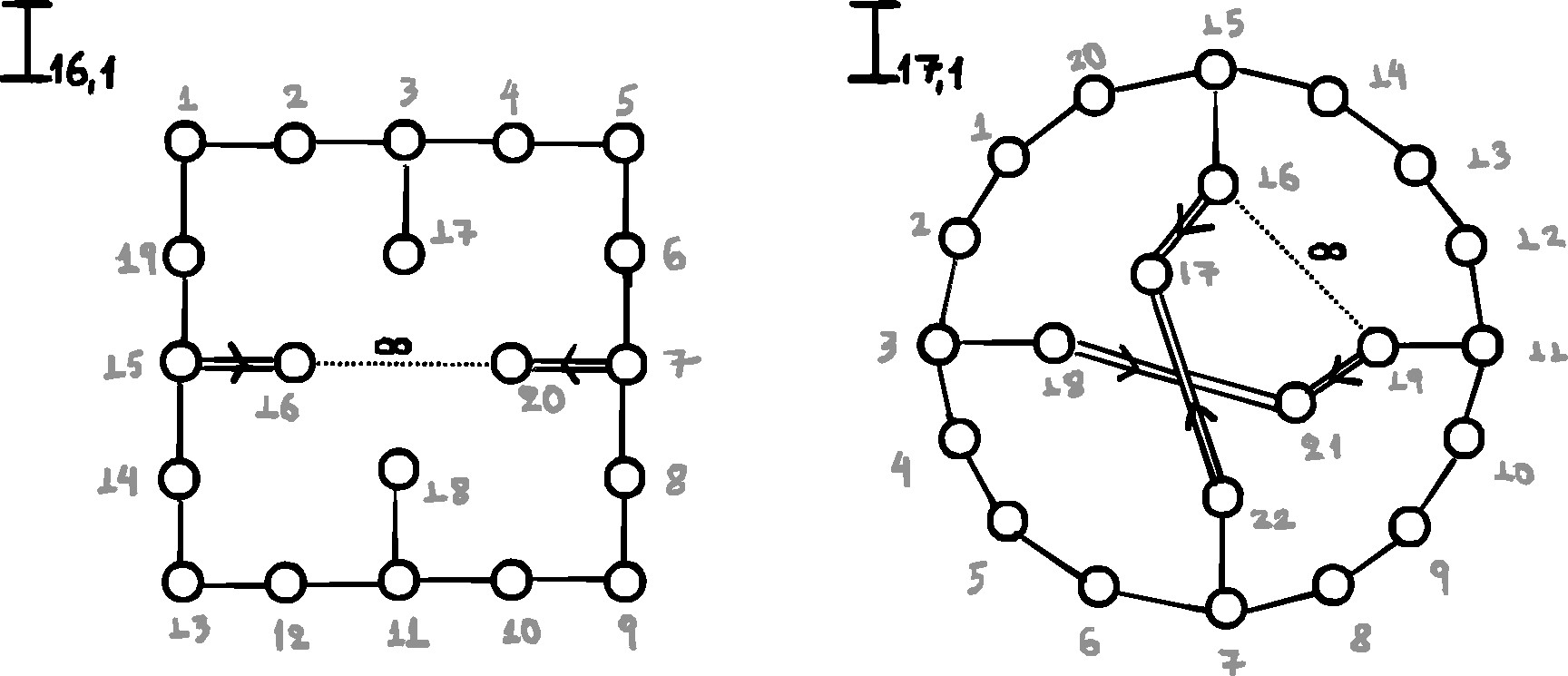}%
        \label{fig:OddGraphsN16and17}%
    }
    \caption{\justifying The Coxeter-Dynkin diagrams for $I_{s,1}$ for $15\leq s+1\leq 18$ (with fundamental roots given in Table \ref{tab:SimpleRootsReflectiveOddLatticesN2to17}).}
    \label{fig:OddGraphsN14to17}
\end{figure*} 

\item \textbf{Dimensions $19 \leq d \leq 20$}:
These last two graphs are the most complex, and were first worked out by Vinberg and Kaplinskaja in\cite{kaplinskaja1978groups}.

\begin{figure*}
    \centering
    \subfloat[\justifying The Coxeter-Dynkin diagram for $I_{18,1}$.  The left graph contains the 22 roots in the skeleton subset $\Sigma_{18}^*$ and is symmetric under the full permutation group $S_4$. The middle graph illustrates one (out of three) vertices of the 1st kind, while the right graph illustrates one (out of 12) vertices of the 2nd kind. In total, $\Sigma_{18}$ has the $22+3+12=37$ fundamental roots given in Tables \ref{tab:SimpleRootsReflectiveOddLatticesN1819},  \ref{tab:AllSimpleRootsReflectiveOddLattices18}.]{%
        \includegraphics[width=0.90\linewidth]{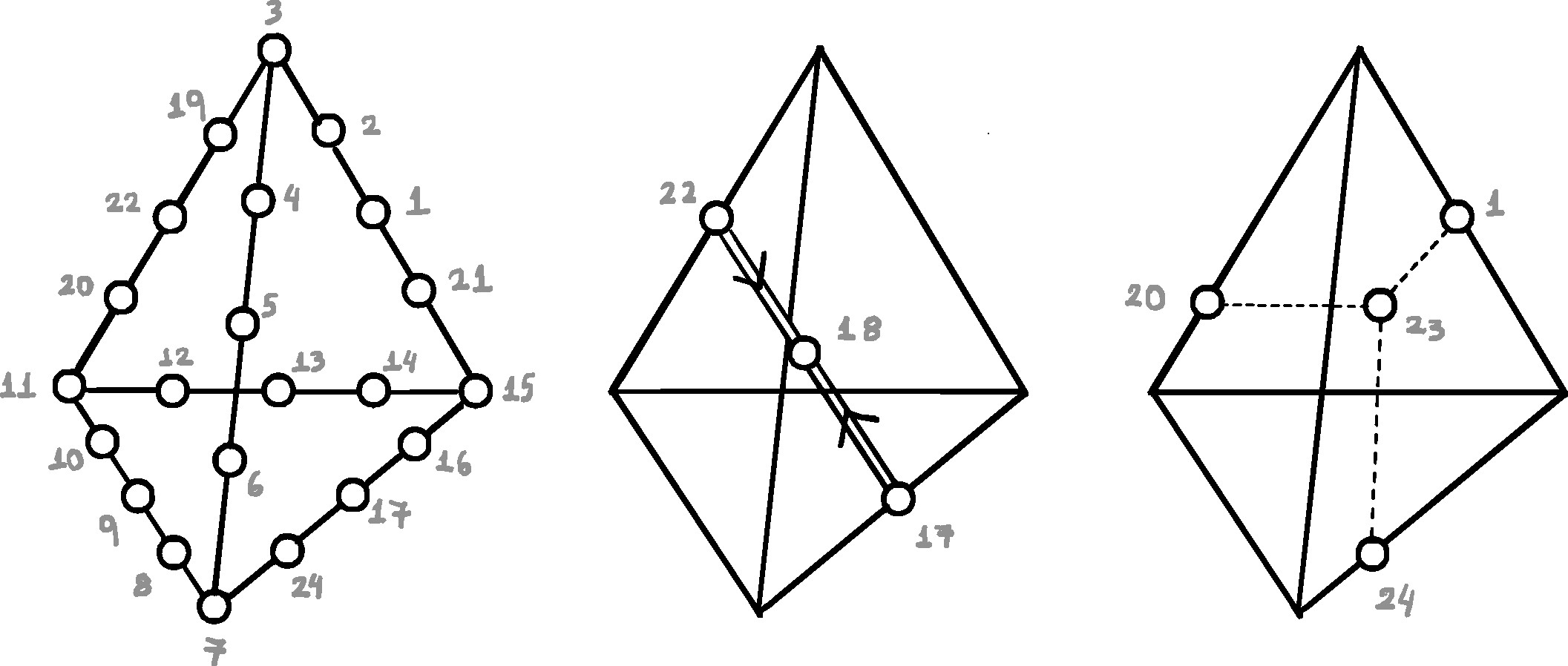}%
        \label{fig:OddGraphsN18}%
    }\par
    \subfloat[\justifying The Coxeter-Dynkin diagram for $I_{19,1}$.  The left graph contains the 25 roots in the skeleton subset $\Sigma_{19}^*$ and is symmetric under the full permutation group $S_5$. The middle graph illustrates one (out of five) vertices of the 1st kind, while the right graph illustrates one (out of 20) vertices of the 2nd kind. In total, $\Sigma_{19}$ has the $25+5+20=50$ fundamental roots given in Tables \ref{tab:SimpleRootsReflectiveOddLatticesN1819}, \ref{tab:AllSimpleRootsReflectiveOddLattices19}. ]{%
        \includegraphics[width=0.95\linewidth]{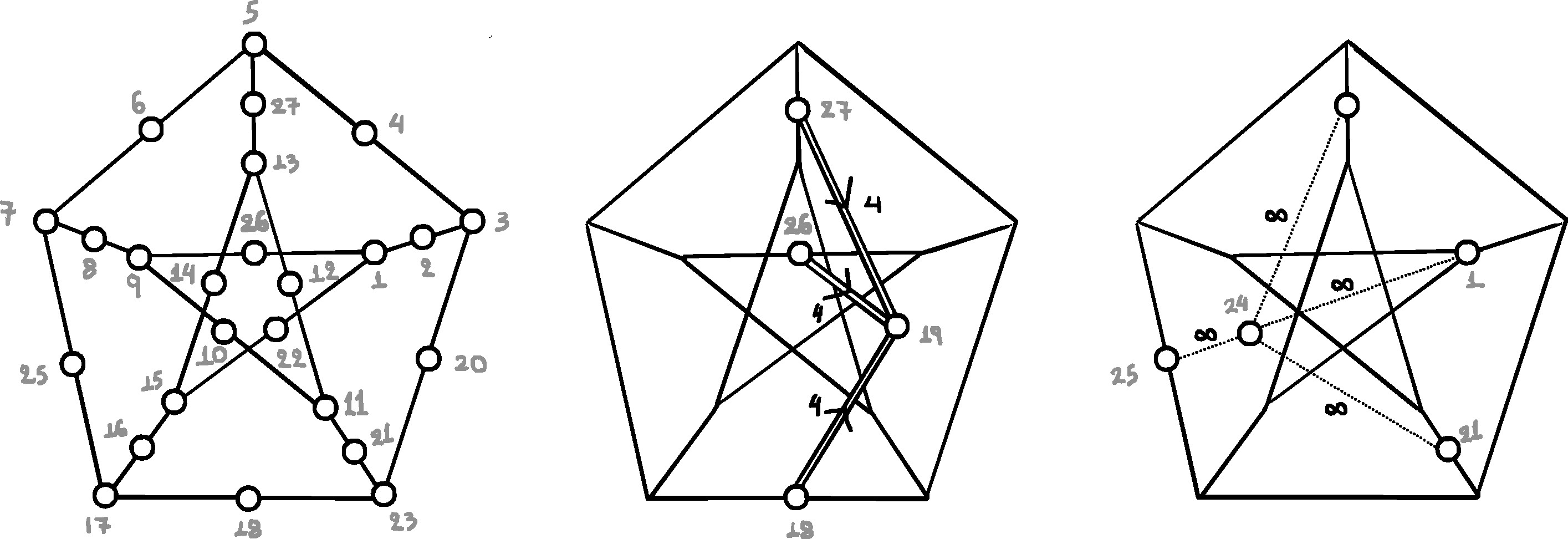}%
        \label{fig:OddGraphsN19}%
    }
    \caption{\justifying Coxeter-Dynkin diagrams for (the skeletons of) $\textrm{I}_{18,1}$ and $\textrm{I}_{19,1}$ (see text for more information).} 
\label{fig:OddGraphsN18to19}
\end{figure*}

To explain the set of fundamental roots in these two cases, first consider the $s+8$ fundamental roots given in Table \ref{tab:SimpleRootsReflectiveOddLatticesN1819} for both $\mathrm{I}_{18,1}$ and $\mathrm{I}_{19,1}$.  These are sets of 26 and 27 roots, which we call $\Sigma_{18}^{(0)}$ and $\Sigma_{19}^{(0)}$, respectively.  Then, define the subsets
\begin{equation*}
\begin{aligned}
& \Sigma_{18}^* = \{1,\dots,24\} - \{ 18,23\} \quad \text{and} \\[1ex]
& \Sigma_{19}^* = \{ 1,\dots,27\} - \{ 19,24\},
\end{aligned}
\end{equation*}
consisting of only the long roots (of norm 2).  These subsets $\Sigma_{18}^*$ and $\Sigma_{19}^*$ form, respectively, a {\it skeleton} Coxeter-Dynkin diagram in the shape of a tetrahedron (left panel of Fig.~\ref{fig:OddGraphsN18}), or a pentagon with an inscribed star (known as the \textit{Petersen graph}), left panel of Fig.~\ref{fig:OddGraphsN19}).  Note that these skeleton diagrams each have non-trivial symmetries described, respectively, by the \textit{full} permutation groups $S_4$ (of order $4!=24$) and $S_5$ (of order $5!=120$).  

Now, we obtain the {\it full} set of fundamental roots ($\Sigma_{s}$) by adding `extra' roots to the skeleton set ($\Sigma_{s}^*$) in such a way that the corresponding Coxeter diagram's ($S_{4}$ or $S_{5}$) symmetry is maintained.  In particular, the extra roots are the roots we previously excluded from $\Sigma_{s}^{(0)}$ to obtain $\Sigma_{s}^{\ast}$, as well as any additional roots needed to maintain the symmetry ($S_{4}$ or $S_{5}$, respectively) of the Coxeter-Dynkin diagram.  In particular, all the extra vertices are either \textbf{vertices of the 1st kind}, related by ($S_{4}$ or $S_{5}$) symmetry to the root $\mathbf{r}_s$, or \textbf{vertices of the 2nd kind}, related by symmetry to the root $\mathbf{r}_{s+5}$.  

For $\Sigma_{18}$, there are three vertices of the first kind (one of which is illustrated in the middle panel of Fig.~\ref{fig:OddGraphsN18}) and twelve vertices of the second kind (one of which is illustrated in the right panel of (Fig.~\ref{fig:OddGraphsN18}), yielding $22+3+12=37$ fundamental roots in total.  Finally, the full set of fundamental roots  $\Sigma_{18}$ consists of the roots ${\bf r}_{1},\ldots,{\bf r}_{26}$ given in Table \ref{tab:SimpleRootsReflectiveOddLatticesN1819}, along with the remaining fundamental roots ${\bf r}_{27},\ldots,{\bf r}_{37}$ given in Table \ref{tab:AllSimpleRootsReflectiveOddLattices18}.

For $\Sigma_{19}$, there are five vertices of the first kind (one of which is illustrated in the middle panel of Fig.~\ref{fig:OddGraphsN19}) and 25 vertices of the second kind (one of which is illustrated in the right panel of Fig.~\ref{fig:OddGraphsN19}), yielding $25+5+20=50$ fundamental roots in total.  Fundamental roots ${\bf r}_{1},\ldots,{\bf r}_{27}$ are given in Table \ref{tab:SimpleRootsReflectiveOddLatticesN1819}, while the remaining fundamental roots ${\bf r}_{28},\ldots,{\bf r}_{50}$ are given in Table \ref{tab:AllSimpleRootsReflectiveOddLattices19}.

\end{itemize}

\begin{table}[t]
\scriptsize
    \centering
        \begin{tabular}{c | c | c | c }
         Root Index $i$ & Root Vector $\mathbf{r}_i$ & $\langle \mathbf{r}_i,\mathbf{r}_i \rangle$ \\
         \vspace{-2ex}& & & \\
         \hline \hline 
         \vspace{-1.75ex}& & & \\
         27 & $9\mathbf{e}_0 + 3(\mathbf{e}_1 + \dots + \mathbf{e}_8) + \mathbf{e}_9 + \dots + \mathbf{e}_{18}$  & 1 \\[1ex]
         \vspace{-1.75ex}& & & \\
         28 & $11\mathbf{e}_0 + 3(\mathbf{e}_1 + \dots + \mathbf{e}_{13}) + \mathbf{e}_{14} + \dots + \mathbf{e}_{18}$  & 1 & \\[1ex]
         \vspace{-1.75ex}& & & \\
         29 & $15\mathbf{e}_0 + 5(\mathbf{e}_1 + \dots + \mathbf{e}_{5}) + 3(\mathbf{e}_{6} + \dots + \mathbf{e}_{16}) + \mathbf{e}_{17} + \mathbf{e}_{18} $  & 1 \\[1ex]
         \vspace{-1.75ex}& & & \\
        30 & $15\mathbf{e}_0 + 5(\mathbf{e}_1 + \dots + \mathbf{e}_{6}) + 3(\mathbf{e}_{7} + \dots + \mathbf{e}_{14}) + \mathbf{e}_{15} + \dots + \mathbf{e}_{18} $  & 1 \\[1ex]
         \vspace{-1.75ex}& & & \\
        31 & $17\mathbf{e}_0 + 5(\mathbf{e}_1 + \dots + \mathbf{e}_{9}) + 3(\mathbf{e}_{10} + \dots + \mathbf{e}_{16}) +\mathbf{e}_{17} + \mathbf{e}_{18} $ & 1 \\[1ex]
         \vspace{-1.75ex}& & & \\
        32 & $19\mathbf{e}_0 + 7(\mathbf{e}_1 + \mathbf{e}_2) + 5(\mathbf{e}_3 + \dots + \mathbf{e}_{10}) + 3(\mathbf{e}_{11} + \dots + \mathbf{e}_{17}) + \mathbf{e}_{18} $ & 1 \\[+1ex] 
        \vspace{-1.75ex}& & & \\
        33 & $21\mathbf{e}_0 + 7(\mathbf{e}_1 + \dots + \mathbf{e}_4) + 5(\mathbf{e}_5 + \dots + \mathbf{e}_{12}) + 3(\mathbf{e}_{13} + \dots + \mathbf{e}_{17}) + \mathbf{e}_{18}$ & 1 \\[+1ex]
         \vspace{-1.75ex}& & & \\
        34 & \,\, $21\mathbf{e}_0 + 7(\mathbf{e}_1 + \dots + \mathbf{e}_5) + 5(\mathbf{e}_6 + \dots + \mathbf{e}_{10}) + 3(\mathbf{e}_{13} + \dots + \mathbf{e}_{18}) \,\, $ & 1 \\[+1ex]
         \vspace{-1.75ex}& & & \\
       35 & \,\, $23\mathbf{e}_0 + 9 \mathbf{e}_1 +  7(\mathbf{e}_2 + \dots + \mathbf{e}_6) + 5(\mathbf{e}_7 + \dots + \mathbf{e}_{12}) + 3(\mathbf{e}_{13} + \dots + \mathbf{e}_{18}) \,\, $ & 1 \\[+1ex]
        \vspace{-1.75ex}& & & \\
        36 & \,\, $25\mathbf{e}_0 + 9 (\mathbf{e}_1 + \mathbf{e}_2) +   7(\mathbf{e}_3 + \dots + \mathbf{e}_8) + 5(\mathbf{e}_9 + \dots + \mathbf{e}_{13}) + 3(\mathbf{e}_{14} + \dots + \mathbf{e}_{18}) \,\, $ & 1 \\[+1ex]
         \vspace{-1.75ex}& & & \\
         37 & \, $27\mathbf{e}_0 + 9 (\mathbf{e}_1 + \dots + \mathbf{e}_4) +   7(\mathbf{e}_5 + \dots + \mathbf{e}_9) + 5(\mathbf{e}_{10} + \dots + \mathbf{e}_{14}) + 3(\mathbf{e}_{15} + \dots + \mathbf{e}_{18})$ \, & 1 
\end{tabular}
    \caption{\justifying The remaining fundamental roots for $\mathrm{I}_{18,1}$.} 
\label{tab:AllSimpleRootsReflectiveOddLattices18}
\end{table}

\begin{table}[t]
\scriptsize
    \centering
        \begin{tabular}{c | c | c | c }
         $i$ & Root Vector $\mathbf{r}_i$ & $\langle \mathbf{r}_i,\mathbf{r}_i \rangle$ \\
        \vspace{-2ex}& & & \\
        \hline \hline 
        \vspace{-1.75ex}& & & \\
        28 & $6\mathbf{e}_0 + 2(\mathbf{e}_1 + \dots + \mathbf{e}_6) + (\mathbf{e}_7 + \dots + \mathbf{e}_{19})$  & 1 \\[1ex]
         \vspace{-1.75ex}& & & \\
         29 & $9\mathbf{e}_0 + 3(\mathbf{e}_1 + \dots + \mathbf{e}_8) + (\mathbf{e}_9 + \dots + \mathbf{e}_{19})$  & 2 \\[1ex]
         \vspace{-1.75ex}& & & \\
        30 & $11\mathbf{e}_0 + 3(\mathbf{e}_1 + \dots + \mathbf{e}_{13}) + (\mathbf{e}_{14} + \dots + \mathbf{e}_{19})$  & 2 \\[1ex]
        \vspace{-1.75ex}& & & \\
        31 & $8\mathbf{e}_0 + 3(\mathbf{e}_1 + \mathbf{e}_{2}) + 2(\mathbf{e}_{3} + \dots + \mathbf{e}_{12}) + (\mathbf{e}_{13} + \dots + \mathbf{e}_{19} $  & 1 \\[1ex]
         \vspace{-1.75ex}& & & \\
        32 & $10\mathbf{e}_0 + 3(\mathbf{e}_1 + \dots +\mathbf{e}_{8}) + 2(\mathbf{e}_{9} + \dots + \mathbf{e}_{14}) + (\mathbf{e}_{15} + \dots + \mathbf{e}_{19} $  & 1 \\[1ex]
         \vspace{-1.75ex}& & & \\
        33 & $15\mathbf{e}_0 + 5(\mathbf{e}_1 + \dots + \mathbf{e}_{5}) + 3(\mathbf{e}_{6} + \dots + \mathbf{e}_{16}) + (\mathbf{e}_{17} + \mathbf{e_{18}} + \mathbf{e}_{19}) $  & 2 \\[1ex]
         \vspace{-1.75ex}& & & \\
        34 & $15\mathbf{e}_0 + 5(\mathbf{e}_1 + \dots + \mathbf{e}_{6}) + 3(\mathbf{e}_{7} + \dots + \mathbf{e}_{14}) + (\mathbf{e}_{15} + \dots + \mathbf{e}_{19}) $  & 2 \\[1ex]
         \vspace{-1.75ex}& & & \\
        35 & $12\mathbf{e}_0 + 4(\mathbf{e}_1 + \dots + \mathbf{e}_{4}) + 3(\mathbf{e}_{5} + \dots + \mathbf{e}_{10}) + 2(\mathbf{e}_{11} + \dots + \mathbf{e}_{16}) + (\mathbf{e}_{17}+\mathbf{e}_{18}+\mathbf{e}_{19}) $  & 1 \\[1ex]
         \vspace{-1.75ex}& & & \\ 
        36 & $17\mathbf{e}_0 + 5(\mathbf{e}_1 + \dots + \mathbf{e}_{9}) + 3(\mathbf{e}_{10} + \dots + \mathbf{e}_{16}) + (\mathbf{e}_{17} + \mathbf{e}_{18} +\mathbf{e}_{19}) $  & 2 \\[1ex]
         \vspace{-1.75ex}& & & \\
        37 & $19\mathbf{e}_0 + 7(\mathbf{e}_1 + \mathbf{e}_2) + 5(\mathbf{e}_3 + \dots + \mathbf{e}_{10}) + 3(\mathbf{e}_{11} + \dots + \mathbf{e}_{17}) + (\mathbf{e}_{18} + \mathbf{e}_{19}) $ & 2 \\[+1ex]
        \vspace{-1.75ex}& & & \\
        38 & $21\mathbf{e}_0 + 7(\mathbf{e}_1 + \dots + \mathbf{e}_4) + 5(\mathbf{e}_5 + \dots + \mathbf{e}_{11}) + 3(\mathbf{e}_{12} + \dots + \mathbf{e}_{19})$ & 2 \\[+1ex]
        \vspace{-1.75ex}& & & \\
        39 & $21\mathbf{e}_0 + 7(\mathbf{e}_1 + \dots + \mathbf{e}_4) + 5(\mathbf{e}_5 + \dots + \mathbf{e}_{12}) + 3(\mathbf{e}_{13} + \dots + \mathbf{e}_{17}) + (\mathbf{e}_{18} + \mathbf{e}_{19})$ & 2 \\[+1ex]
        \vspace{-1.75ex}& & & \\
        40 & $21\mathbf{e}_0 + 7(\mathbf{e}_1 + \dots + \mathbf{e}_5) + 5(\mathbf{e}_6 + \dots + \mathbf{e}_{10}) + 3(\mathbf{e}_{11} + \dots + \mathbf{e}_{18}) + \mathbf{e}_{19}$ & 2 \\[+1ex]
        \vspace{-1.75ex}& & & \\
        41 & \,\, $23\mathbf{e}_0 + 7(\mathbf{e}_1 + \dots + \mathbf{e}_7) + 5(\mathbf{e}_8 + \dots + \mathbf{e}_{12}) + 3(\mathbf{e}_{13} + \dots + \mathbf{e}_{19}) \,\, $ & 2 \\[+1ex]
        \vspace{-1.75ex}& & & \\
        42 & \,\, $23\mathbf{e}_0 + 9 \mathbf{e}_1 +  7(\mathbf{e}_2 + \dots + \mathbf{e}_6) + 5(\mathbf{e}_7 + \dots + \mathbf{e}_{12}) + 3(\mathbf{e}_{13} + \dots + \mathbf{e}_{18}) + \mathbf{e}_{19} \,\, $ & 2 \\[+1ex]
        \vspace{-1.75ex}& & & \\
        43 & \,\, $25\mathbf{e}_0 + 9 (\mathbf{e}_1 + \mathbf{e}_2) +   7(\mathbf{e}_3 + \dots + \mathbf{e}_7) + 5(\mathbf{e}_8 + \dots + \mathbf{e}_{14}) + 3(\mathbf{e}_{15} + \dots + \mathbf{e}_{19}) \,\, $ & 2 \\[+1ex]
        \vspace{-1.75ex}& & & \\
        44 & \,\, $25\mathbf{e}_0 + 9 (\mathbf{e}_1 + \mathbf{e}_2) +   7(\mathbf{e}_3 + \dots + \mathbf{e}_8) + 5(\mathbf{e}_9 + \dots + \mathbf{e}_{13}) + 3(\mathbf{e}_{14} + \dots + \mathbf{e}_{18}) + \mathbf{e}_{19} \,\, $ & 2 \\[+1ex]
        \vspace{-1.75ex}& & & \\
        45 & \,\, $27\mathbf{e}_0 + 9 (\mathbf{e}_1 + \dots + \mathbf{e}_3) +  7(\mathbf{e}_4 + \dots + \mathbf{e}_{10}) + 5(\mathbf{e}_{11} + \dots + \mathbf{e}_{14}) + 3(\mathbf{e}_{15} + \dots + \mathbf{e}_{19}) \,\, $ & 2 
        \\[+1ex]
        \vspace{-1.75ex}& & & \\
        46 & \,\, $27\mathbf{e}_0 + 9 (\mathbf{e}_1 + \dots + \mathbf{e}_4) +  7(\mathbf{e}_5 + \dots + \mathbf{e}_{8}) + 5(\mathbf{e}_{9} + \dots + \mathbf{e}_{15}) + 3(\mathbf{e}_{16} + \dots + \mathbf{e}_{19}) \,\, $ & 2
        \\[+1ex]
        \vspace{-1.75ex}& & & \\
        47 & \,\, $27\mathbf{e}_0 + 9 (\mathbf{e}_1 + \dots + \mathbf{e}_4) +  7(\mathbf{e}_4 + \dots + \mathbf{e}_{9}) + 5(\mathbf{e}_{10} + \dots + \mathbf{e}_{14}) + 3(\mathbf{e}_{15} + \dots + \mathbf{e}_{18}) + \mathbf{e}_{19} \,\, $ & 2 \\[+1ex]
        \vspace{-1.75ex}& & & \\
        48 & \,\, $29\mathbf{e}_0 + 9 (\mathbf{e}_1 + \dots + \mathbf{e}_6) +  7(\mathbf{e}_{7} + \dots + \mathbf{e}_{10}) + 5(\mathbf{e}_{10} + \dots + \mathbf{e}_{15}) + 3(\mathbf{e}_{16} + \dots + \mathbf{e}_{19}) \,\, $ & 2 \\[+1ex]
        \vspace{-1.75ex}& & & \\
        49 & \,\, $31\mathbf{e}_0 +11 (\mathbf{e}_1 + \mathbf{e}_2) + 9 (\mathbf{e}_3 + \dots + \mathbf{e}_6) +  7(\mathbf{e}_7 + \dots + \mathbf{e}_{10}) + 5(\mathbf{e}_{11} + \dots + \mathbf{e}_{15}) + 3(\mathbf{e}_{16} + \dots + \mathbf{e}_{19}) \,\, $ & 2
        \\[+1ex]
        \vspace{-1.75ex}& & & \\
        50 & \,\, $33\mathbf{e}_0 + 11 (\mathbf{e}_1 + \mathbf{e}_2 + \mathbf{e}_3) + 9 (\mathbf{e}_4 + \dots + \mathbf{e}_8) +  7(\mathbf{e}_9 + \dots + \mathbf{e}_{13}) + 5(\mathbf{e}_{14} + \dots + \mathbf{e}_{17}) + 3(\mathbf{e}_{16} + \dots + \mathbf{e}_{19}) \,\, $ & 2 
\end{tabular}
    \caption{\justifying The remaining fundamental roots for $\mathrm{I}_{19,1}$} 
\label{tab:AllSimpleRootsReflectiveOddLattices19}
\end{table}

\chapter{Rational and Irrational Newman-Penrose Frames}
\label{appendix:NPFrames}

\section{The Standard Newman-Penrose Frame}

The standard -- \textit{rational} -- \textbf{Newman-Penrose (NP)\footnote{Not to be confused with the complexity class \cc{NP}!} formalism} in general relativity \cite{Newman:1961qr} is itself a \textit{tetrad formulation} (i.e.~a choice of a $4$-frame over each point of four-dimensional Lorentzian spacetime with signature $(1,3)$). The NP formalism is based on a choice of a \textit{null tetrad} that consists of one pair of real null (co)vectors $\{l,n\}$ and one pair of complex-conjugate null (co)vectors $\{ m,\bar{m}\}$.\footnote{Thinking in terms of spherical coordinates $(t,r,\theta,\phi)$, for example, $l^\mu$ and $n^\mu$ can be taken to be the outgoing and ingoing null vectors respectively: $l^\mu = (\widehat{t} + \widehat{r})/\sqrt{2}$ and $n^\mu = (\widehat{t} - \widehat{r})/\sqrt{2}$, with $\hat{t}, \hat{r}$ being the unit vectors along the temporal and radial direction respectively. The complex null vector $m^\mu$ is constructed by combining two real, orthogonal, unit spacelike vectors; for instance, $m^\mu = (\widehat{\theta} + i \widehat{\phi})^\mu/\sqrt{2}$, where $\widehat{\theta},\widehat{\phi}$ are the angular unit vectors. Then, $\bar{m}^\mu = m^{\mu*} = (\widehat{\theta} - i \widehat{\phi})^\mu/\sqrt{2}$ completes the tetrad frame.} Norms are calculated with the global Lorentzian metric $g_{\mu \nu}(x)$, which can be written as the square of the tetrad as $g_{\mu\nu} = e_{\mu}^a(x) \eta_{a b} e_{\nu}^b(x)$, with $\eta_{ab}$ being the standard Minkowski metric. 

Using the convention $\eta = \mathrm{diag}(-1,1,1,1)$ (similarly for the mostly-negative convention, with the signs flipped), the following normalization conditions are adopted: 
\begin{itemize}
\item \textit{Self-normalization} vanishes for null frames: 
\begin{equation*}
l^\mu l_{\mu} = l^\mu g_{\mu \nu} l^\nu = n^\mu n_\mu = m^\mu m_\mu = \bar{m}^\mu \bar{m}_\mu = 0
\end{equation*}
\item \textit{Cross-normalization} within same pairs: 
\begin{equation*}
l^\mu n_\mu = -1 = l_\mu n^\mu, \quad m^\mu \bar{m}_\mu = + 1 = m_\mu \bar{m}^\mu,
\end{equation*}
\item \textit{Contraction} across different pairs vanishes:
\begin{equation*}
l_\mu m^\mu = l_\mu \bar{m}^\mu = n_\mu m^\mu = n_\mu \bar{m}^\mu = 0
\end{equation*}
\end{itemize}
The metric $g_{\mu \nu}$ used to raise/lower indices can be decomposed into products of the tetrad null vectors as 
\begin{equation}
g_{\mu \nu} = - l_\mu n_\nu - n_\mu l_\nu + m_\mu \bar{m}_\nu + \bar{m}_\mu m_\nu,
\label{eq:NPFormalismMetricDecomposition}
\end{equation}
where real vectors appear with a minus sign and complex vectors appear with a plus sign. Raising one component of the metric leads to the Kronecker symbol
\begin{equation}
\delta^\mu_\nu = g_{\alpha \nu} g^{\alpha \mu} = - l^\mu n_\nu - n^\mu l_\nu + m^\mu \bar{m}_\nu + \bar{m}^\mu m_\nu.
\label{eq:NPFormalismIdentityDecomposition}
\end{equation}

\subsection{Defining Null Projectors}

Let the tetrad be$\{ l^\mu, n^\mu, m^\mu, \bar{m}^\mu\}$ and satisfy the normalization conditions of the previous paragraph. Let's say we would like to define a projector onto the $l-m$ plane; then, the projector can be constructed as follows
\begin{equation}
(\Pi_{lm})^{\alpha}_\beta = -l^\alpha n_\beta + m^\alpha \bar{m}_\beta.
\end{equation}
Indeed, one can verify the following properties: 
\begin{itemize}
\item $\Pi_{lm}$ \textbf{squares to itself} $\Pi_{lm}^2 = \Pi_{lm}$:
\begin{equation*}
\begin{aligned}
(\Pi_{lm})^\alpha_\beta (\Pi_{lm})^\beta_\gamma & = (-l^\alpha n_\beta + m^\alpha \bar{m}_\beta)(-l^\beta n_\gamma + m^\beta \bar{m}_\gamma) = \\
& = -l^\alpha n_\gamma + m^\alpha \bar{m}_\gamma = (\Pi_{lm})^\alpha_\gamma,
\end{aligned}
\end{equation*}
where we used the self- and cross-normalization conditions.  
\item $\Pi_{lm}$ \textbf{leaves $l^\mu$ and $m^\mu$ invariant:} the terms $-l^\alpha n_\beta$ and $+m^\alpha \bar{m}_\beta$ guarantee that when the projector acts on $l^\mu$ or $m^\mu$ returns 
$$ (\Pi_{lm})^\alpha_\beta l^\beta = -l^\alpha n_\beta l^\beta + m^\alpha \underbrace{\bar{m}_\beta l^\beta}_{\mathclap{=0}} = + l^\alpha$$
and 
$$ (\Pi_{lm})^\alpha_\beta m^\beta = -l^\alpha \underbrace{n_\beta m^\beta}_{\mathclap{=0}} + m^\alpha \bar{m}_\beta m^\beta = + m^\alpha.$$
Thus the projector preserves vectors that lie exclusively on the $l-m$ plane. 
\item $\Pi_{lm}$ \textbf{annihilates $n^\mu$ and $\bar{m}^\mu$:} similarly, 
$$ (\Pi_{lm})^\alpha_\beta n^\beta = -l^\alpha \underbrace{n_\beta n^\beta}_{\mathclap{=0}} + m^\alpha \underbrace{\bar{m}_\beta n^\beta}_{\mathclap{=0}} = 0$$
and 
$$ (\Pi_{lm})^\alpha_\beta \bar{m}^\beta = -l^\alpha \underbrace{n_\beta \bar{m}^\beta}_{\mathclap{=0}} + m^\alpha \underbrace{\bar{m}_\beta \bar{m}^\beta}_{\mathclap{=0}} = 0,$$
thus the projector `kills' every component of a vector that is orthogonal to the $l-m$ plane. 
\end{itemize}
This prescription can be generalized to yield a projector onto any plane spanned by any two vectors belonging to the tetrad: for instance, $(\Pi_{ln})^{\alpha}_\beta = -l^\alpha n_\beta - n^\alpha l_\beta$ and $(\Pi_{m\bar{m}})^{\alpha}_\beta = m^\alpha \bar{m}_\beta + \bar{m}^\alpha m_\beta$ are projectors onto the $l-n$ and $m-\bar{m}$ planes respectively. In light of the above, notice that the identity map - given by Eq. \eqref{eq:NPFormalismIdentityDecomposition} - can be seen as a sum over appropriate projectors
\begin{equation}
\begin{aligned}
\delta^{\alpha}_\beta &= - l^\alpha n_\beta - n^\alpha l_\beta + m^\alpha \bar{m}_\beta + \bar{m}^\alpha m_\beta \\
& = (\Pi_{ln})^\alpha_\beta + (\Pi_{m\bar{m}})^\alpha_\beta = (\Pi_{lm})^\alpha_\beta + (\Pi_{n\bar{m}})^\alpha_\beta \\
& = (\Pi_{l\bar{m}})^\alpha_\beta + (\Pi_{nm})^\alpha_\beta
\end{aligned}
\end{equation}
meaning that the set of projectors form a complete set of orthogonal projections. Knowing that $l^\alpha n_\alpha = -1$ and $m^\alpha \bar{m}_\alpha = +1$, the above resolution of the identity could be rewritten as 
\begin{equation}
\begin{aligned}
\delta^{\alpha}_\beta &= - l^\alpha n_\beta - n^\alpha l_\beta + m^\alpha \bar{m}_\beta + \bar{m}^\alpha m_\beta \\
& = \frac{l^\alpha n_\beta}{l^\gamma n_\gamma} + \frac{n^\alpha l_\beta}{n^\gamma l_\gamma} + \frac{m^\alpha \bar{m}_\beta}{m^\gamma \bar{m}_\gamma} + \frac{\bar{m}^\alpha m_\beta}{\bar{m}^\gamma m_\gamma}.
\end{aligned}
\end{equation}
where we have effectively terms of the type outer-/inner- cross-products between the null vectors.

\section{The Irrational Newman-Penrose Formalism}
\label{app:irrationalNPFrames}

In Section \ref{subsec:Hyperbolic_Coxeter_Elements} we said that the eigenbasis of the Coxeter element forms such an \textit{irrational NP frame} in Minkowski spacetime $V \simeq \MM^d$ in terms of the real bilinear form $\langle {\bf v}, {\bf w} \rangle = v^\mu \eta_{\mu\nu} w^\nu$. Here, we would like to show this a bit more explicitly.  

\subsection{The Two Possible Quadratic Forms}

When passing to the complex eigenbasis of the Coxeter element, there are essentially two quadratic forms one can use to calculate norms with. 
\begin{itemize}
    \item \textbf{The real bilinear form:} the first one is associated with the vector space $V$ of dimension $d$ we started from\footnote{Here, $V$ is assumed to be Lorentzian but the argument applies unchanged also to the Euclidean case; simply replace the indefinite form $\eta$ with the definite form $\delta$ in the inner product.}, and is the (symmetric, bilinear non-degenerate and indefinite) inner product $\langle \cdot , \cdot \rangle_{\eta} : V \times V \rightarrow \mathbb{R}$, which takes two vectors $\mathbf{x},\mathbf{y} \in V$ and returns the real number $\langle\mathbf{x},\mathbf{y} \rangle = \mathbf{x}^T \eta \mathbf{y}$. Here, we refer to this as $\langle {\bf x}, {\bf y} \rangle_\eta$
    \item \textbf{The complex sesquilinear form:} the second one is formally defined in the complexification $V_{\mathbb{C}} = V \otimes \mathbb{C}$ of the original vector space $V$, where the complex eigenbasis lives, which is of real dimension $\dim_{\mathbb{R}}(V_\mathbb{C}) = 2 \dim_\mathbb{R}(V) = 2d$. Let this (\textit{sesquilinear}, indefinite) quadratic form be denoted as $\langle \cdot , \cdot \rangle_\eta^\mathbb{C} : V_\mathbb{C} \times V_\mathbb{C} \rightarrow \mathbb{C}$ (or simply $\langle \cdot , \cdot \rangle_\mathbb{C}$ if no ambiguities arise) and be defined such that for any two complex vectors $\mathbf{x,y} \in V_\mathbb{C}$, their inner product is the complex number $\langle \mathbf{x},\mathbf{y}\rangle_\mathbb{C} = \mathbf{x}^\dagger \eta \mathbf{y}$.
\end{itemize}
Generally, the two forms will lead to different norms for a complex vector $\mathbf{v}$, whereas the norms agree if $\mathbf{v}$ is real. In particular, the two norms agree for the real pair of Coxeter eigenvectors $\mathbf{v}_0^\pm$, for any linear combination $\alpha_j \mathbf{v}_j^+ + \alpha_j^* \mathbf{v}_j^-$ with $\alpha_j \in \mathbb{C}$ of any pair of complex-conjugate eigenvectors $\mathbf{v}_j^\pm$, and for the rational eigenvector $\mathbf{v}_{-1}$.

Given the inner product $\langle \cdot , \cdot \rangle_\eta^\mathbb{C}$ and any transformation $K$ acting on $V_\mathbb{C}$, we can define the notion of the \textit{adjoint} transformation - denoted as $K^\star$ - as follows
\begin{equation}
\langle \mathbf{x} , K \mathbf{y}\rangle_\eta^\mathbb{C} = \langle K^\star \mathbf{x},\mathbf{y} \rangle_\eta^\mathbb{C}, \quad \forall \mathbf{x,y} \in V_\mathbb{C}.
\label{eq:AdjointTransformation}
\end{equation}
A map $K$ is said to be \textit{self-adjoint} if $K^\star = K$. Note that we could have used the indefinite bilinear form $\langle \cdot, \cdot \rangle_\eta$ instead (or the positive-definite one $\langle \cdot ,\cdot \rangle_\delta$), but since the sesquilinear form is more general, we use this for the definition; to pass on to the bilinear form, replace daggers with transposes and $\eta$ with $\delta$ to further pass on to the Euclidean case. Now, given the explicit form of the inner product, a relationship between $K^\star$ and $K$ (or better its conjugate-transpose) can be derived as follows: 
\begin{equation}
\begin{aligned}
& \langle \mathbf{x} , K \mathbf{y}\rangle_\eta^\mathbb{C} = \mathbf{x}^\dagger \eta K \mathbf{y}
= (K^\star \mathbf{x})^\dagger \eta \mathbf{y} = 
\langle K^\star \mathbf{x},\mathbf{y} \rangle_\eta^\mathbb{C}\\
& \Rightarrow \mathbf{x}^\dagger \eta K \mathbf{y}
= \mathbf{x}^\dagger (K^\star)^\dagger \eta \mathbf{y} \Rightarrow \eta K \eta^{-1} = (K^\star)^\dagger\\
& \Rightarrow \boxed{K^\star = \eta^{-1} K^\dagger \eta}
\end{aligned}
\label{eq:AdjointTransAndConjugateTranspose}
\end{equation}
If $K$ is further taken to be a metric-preserving - orthogonal $K \in O(V)$ - transformation, 
\begin{equation}
\begin{aligned}
& \langle K\mathbf{x},K\mathbf{y}\rangle_\eta^\mathbb{C} = \mathbf{x}^\dagger K^\dagger \eta K \mathbf{y} = \mathbf{x}^\dagger \eta \mathbf{y} =\langle \mathbf{x},\mathbf{y}\rangle_\eta^\mathbb{C}\\
& \Rightarrow \boxed{\eta = K^\dagger \eta K}
\end{aligned}
\label{eq:MetricPreservingTransformation}
\end{equation}
Combing Eqs. \eqref{eq:AdjointTransAndConjugateTranspose}, \eqref{eq:MetricPreservingTransformation}, we conclude that $K^\star = \eta^{-1} \eta K^{-1} = K^{-1}$ for any orthogonal transformation $K$. On the other hand, for $K$ to be self-adjoint with respect to $\langle \cdot , \cdot \rangle_\eta^\mathbb{C}$, we need $\eta K = K^\dagger \eta$.

\subsection{The Coxeter Eigenbasis Satisfies the NP-Properties}

We are now ready to show the following lemma about the Coxeter eigenbasis:
\begin{lemma}[Nullness of the Coxeter Eigenbasis]
All the eigenvectors of a Coxeter element $C$ (for a finite Coxeter system) are null with respect to the inner product $\langle \cdot , \cdot \rangle_\eta$ in $V$, unless their associated eigenvalue squares to $+1$. In terms of the inner product $\langle \cdot , \cdot \rangle_\eta^\mathbb{C}$ in the complexification $V_\mathbb{C}$, eigenvectors whose eigenvalues lie on $\mathbb{S}^1$ have non-zero norm, while those with eigenvalues on the positive real line $\mathbb{R}^+$ are null. 
\label{lemma:NormCoxeterEigenvectors}
\end{lemma}
\begin{proof} 
Let $C \mathbf{v}_{\lambda} = \lambda  \mathbf{v}_{\lambda}$ be the eigenvalue equation for the Coxeter element $C$ of some Coxeter group with a finite number of fundamental reflections, thus $\lambda \in \mathrm{spec}(C) = \mathbb{S}^1 \cup \mathbb{R}^+$ (see Section \ref{subsec:Coxeter_Elements}). Furthermore, let $\langle \cdot , \cdot \rangle_\eta$ and $\langle \cdot, \cdot \rangle_\eta^\mathbb{C}$ be the bilinear and  sesquilinear forms in $V$ and $V_\mathbb{C}$ respectively. We can compute the inner products $\langle C \mathbf{v}_\lambda,C \mathbf{v}_\lambda \rangle_\eta$ and $\langle C \mathbf{v}_\lambda,C \mathbf{v}_\lambda \rangle_\eta^\mathbb{C}$ in two different ways to obtain:
\begin{equation}
\begin{aligned}
& \langle C \mathbf{v}_\lambda,C \mathbf{v}_\lambda \rangle_\eta = \left\{ 
\begin{aligned}
& \lambda^2 \langle \mathbf{v}_\lambda,\mathbf{v}_\lambda \rangle_\eta \\
& \langle C^\star C \mathbf{v}_\lambda,\mathbf{v}_\lambda \rangle_\eta = \langle \mathbf{v}_\lambda,\mathbf{v}_\lambda \rangle_\eta \\
\end{aligned}
\right. \\
& \Rightarrow \langle \mathbf{v}_\lambda,\mathbf{v}_\lambda \rangle_\eta (\lambda^2 - 1) = 0,
\end{aligned}
\label{eq:NormCoxeterEigenvecBilinear}
\end{equation}
and similarly
\begin{equation}
\begin{aligned}
& \langle C \mathbf{v}_\lambda,C \mathbf{v}_\lambda \rangle_\eta^\mathbb{C} = \left\{ 
\begin{aligned}
& \lambda \lambda^* \langle \mathbf{v}_\lambda,\mathbf{v}_\lambda \rangle_\eta^\mathbb{C} \\
& \langle C^\star C \mathbf{v}_\lambda,\mathbf{v}_\lambda \rangle_\eta^\mathbb{C} = \langle \mathbf{v}_\lambda,\mathbf{v}_\lambda \rangle_\eta^\mathbb{C} 
\end{aligned}
\right.\\
& \Rightarrow \langle \mathbf{v}_\lambda,\mathbf{v}_\lambda \rangle_\eta^\mathbb{C} \big(||\lambda||^2 - 1\big) = 0.
\end{aligned}
\label{eq:NormCoxeterEigenvecSesquilinear}
\end{equation}
For the first calculation, we used the eigenvalue equation and the properties of the inner products to remove the two Coxeter elements $C$ from inside the inner product and pick up either a factor of $\lambda^2$ or a factor of $||\lambda||^2 = \lambda \lambda^*$. For the second calculation, we moved one $C$ from the second to the first slot by using the adjoint Coxeter transformation; then, we noticed that $C$ is indeed an orthogonal transformation that preserves $\eta$, thus $C^\star = C^{-1}$ and $C^\star C = \mathbf{1}$. 

As a result, from Eq. \eqref{eq:NormCoxeterEigenvecBilinear}, we infer that either $\langle \mathbf{v}_\lambda,\mathbf{v}_\lambda \rangle_\eta = 0$ or $\lambda^2 = 1$, which can only happen if $\lambda$ is $\pm 1$. This is the case, for example, for the eigenvector $\mathbf{v}_{-1}$ of a hyperbolic Coxeter element when $d$ is odd (see Section \ref{subsec:Hyperbolic_Coxeter_Elements}). For a finite Coxeter group, the eigenvalues $\lambda$ are always primitive $h$-roots of unity and satisfy $\zeta_h^h = 1$, thus we would need $h=2$, which can only happen for $A_1 \simeq SU(2)$ or $I_2(2) \simeq D_4$. From Eq. \eqref{eq:NormCoxeterEigenvecSesquilinear}, we conclude that either $\langle \mathbf{v}_\lambda,\mathbf{v}_\lambda \rangle_\eta^\mathbb{C} = 0$ or $ \lambda \lambda^* = 1.$, which can only happen when $\lambda \in \mathbb{S}^1$. This means that all the complex-conjugate pairs of eigenvectors for a hyperbolic Coxeter element will be of non-zero norm (including $\mathbf{v}_{-1}$), while the real pair of Coxeter eigenvectors with eigenvectors strictly inside and outside $\mathbb{S}^1$ will be null in terms of $\langle \cdot , \cdot \rangle_\eta^{\mathbb{C}}$. For finite Coxeter groups, we only have primitive $h$-roots of unity as eigenvalues so no eigenvector will be null in $V_C$. This concludes the proof.
\end{proof}

So Lemma \ref{lemma:NormCoxeterEigenvectors} tells us that all irrational Coxeter eigenvectors are null with respect to $\eta$ in $\MM^d$. This is one of the conditions mentioned in the previous Section, the self-normalization. The vanishing of any contraction across different pairs follows from a trivial modification of the proof of Lemma \ref{lemma:NormCoxeterEigenvectors}. To see this, take two different eigenvectors ${\bf v}_{\lambda_{1,2}}$ of $C$ and consider their inner products as shown below: 
\begin{equation}
    \langle C {\bf v}_{\lambda_1},C {\bf v}_{\lambda_2} \rangle_\eta =
    \left\{
    \begin{aligned}
        & \lambda_1 \lambda_2 \langle C {\bf v}_{\lambda_1}, C {\bf v}_{\lambda_2} \rangle \\
        & \langle C^\star C {\bf v}_{\lambda_1}, {\bf v}_{\lambda_2} \rangle_\eta = \langle {\bf v}_{\lambda_1}, {\bf v}_{\lambda_2}\rangle_\eta 
    \end{aligned}
    \right.
    \Rightarrow \langle {\bf v}_{\lambda_1}, 2 {\bf v}_{\lambda_2} \rangle_\eta (\lambda_1 \lambda_2 - 1) = 0,
\end{equation}
which implies that whenever $\lambda_1 \lambda_2 \neq 1$, any two Coxeter eigenvectors ${\bf v}_{\lambda_{1,2}}$ will have zero cross-product. \textit{Within the same pairs} of complex conjugate ${\bf v}_j^\pm$ or real ${\bf v}_0^\pm$ eigenvectors, we either have $\lambda_2 = \lambda_1^\ast$ or $\lambda_2 = \lambda_1^{-1}$ respectively, thus $\lambda_1 \lambda_2$ is always $+1$ and the inner products $\langle {\bf v}_0^+,{\bf v}_0^- \rangle_\eta$ and $\langle {\bf v}_j^+, {\bf v}_j^- \rangle_\eta$ are non-vanishing. \textit{Across different pairs}, on the contrary, the condition $\lambda_1 \lambda_2 \neq 1$ is always met and any cross product $ \langle { \bf v}_{\lambda_1},{ \bf v}_{\lambda_2} \rangle_\eta$ will vanish identically.

Using the two quadratic forms $\langle \cdot , \cdot \rangle_\eta$ and $\langle \cdot , \cdot \rangle_\eta^\CC$, we can build two Gram matrices of inner products, the real one $G_{\eta} = \langle {\bf v}_i, {\bf v}_j\rangle_\eta$ and the complex one $G_{\eta}^\CC = \langle {\bf v}_i, {\bf v}_j \rangle_\eta^\CC$ with $i,j=1,\dots 2k + (d \mathrm{mod}2)$ (the last index is $2k+1$ only if $d$ is odd; see Section \ref{subsec:Hyperbolic_Coxeter_Elements}). Since the Coxeter eigenbasis satisfies the self-normalization and contraction conditions, $G_\eta$ reads:
\begin{equation}
  \setlength{\arraycolsep}{2pt}
  \setlength{\delimitershortfall}{0pt}
  \newcommand*{\myfbox}[1]{%
    \fcolorbox{red}{red!20!white}{$#1$}%
  }
  \newcommand{\Oplus}{\ensuremath{\vcenter{\hbox{\scalebox{3.5}{$\oplus$}}}}}
     G_{\eta} = 
    \begin{pmatrix}\,
        \myfbox{\begin{matrix}
            0 & \langle \mathbf{v}_0^+ , \mathbf{v}_0^- \rangle_\eta 
            \\[1ex]
            \langle \mathbf{v}_0^- , \mathbf{v}_0^+ \rangle_\eta & 0 \\
        \end{matrix}} & 0 & 0 \\
        0 & \Oplus_{j=1}^{k-1}\,\,\myfbox{\begin{matrix}
            0 & \langle \mathbf{v}_{j}^+ , \mathbf{v}_{j}^- \rangle_\eta \\[1ex]
            \langle \mathbf{v}_j^- , \mathbf{v}_{j}^+ \rangle_\eta& 0\\
        \end{matrix}} & 0 \\[+5ex]
         0 & 0 & \myfbox{\langle \mathbf{v}_{-1},\mathbf{v}_{-1}\rangle_\eta}
    \,\,
    \end{pmatrix}
    \label{eq:GramMatrixCoxeterEigenbasisBilinear}
\end{equation}
The spectrum of $G_\eta$ gives the signature of $V$; we find 
\begin{equation}
    \mathrm{spec}(G_\eta) = \bigg\{ \pm \langle {\bf v}_0^+, {\bf v}_0^- \rangle_\eta, \big\{ \pm \langle {\bf v}_j^+,{\bf v}_j^-\rangle_\eta \big\}_{j={1,\dots,k-1}} 
    \bigg\} \quad \big( \cup \{\langle {\bf v}_{-1}, {\bf v}_{-1} \rangle_\eta\} \big).
    \label{eq:SpectrumOfBilinearGramMatrix}
\end{equation}
Therefore, in $\MM^d$ of signature $(d-1,1)$, the Gram matrix $G_\eta$ implies a split-signature $(d/2,d/2 + d\mathrm{mod}2)$, since we chose to describe the real vector space $V$ with complex null eigenvectors and insisted on using the real bilinear form. 

With respect to the complex sesquilinear form, though, the Gram matrix $G_\eta^\CC$ reads:
    \begin{equation}
  \setlength{\arraycolsep}{2pt}
  \setlength{\delimitershortfall}{0pt}
  \newcommand*{\myfbox}[1]{%
    \fcolorbox{red}{red!20!white}{$#1$}%
  }
  \newcommand{\Oplus}{\ensuremath{\vcenter{\hbox{\scalebox{3.5}{$\oplus$}}}}}
     G_{\eta}^\CC = 
    \begin{pmatrix}\,
        \myfbox{\begin{matrix}
            0 & \langle \mathbf{v}_0^+ , \mathbf{v}_0^- \rangle_\eta
            \\[1ex]
            \langle \mathbf{v}_0^- , \mathbf{v}_0^+ \rangle_\eta & 0 \\
        \end{matrix}} & 0 & 0 \\
        0 & \Oplus_{j=1}^{k-1}\,\,\myfbox{\begin{matrix}
            \langle \mathbf{v}_{j}^+ , \mathbf{v}_{j}^+\rangle_\eta^\CC & 0 \\[1ex]
            0& \langle \mathbf{v}_j^- , \mathbf{v}_{j}^- \rangle_\eta^\CC\\
        \end{matrix}} & 0 \\[+5ex]
         0 & 0 & \myfbox{\langle \mathbf{v}_{-1},\mathbf{v}_{-1}\rangle_\eta}
    \,\,
    \end{pmatrix}
    \label{eq:GramMatrixCoxeterEigenbasisSesquilinear}
\end{equation}
with eigenvalues
\begin{equation}
    \mathrm{spec}(G_\eta^\CC) = \bigg\{ \pm \langle {\bf v}_0^+, {\bf v}_0^- \rangle_\eta, \big\{ \langle {\bf v}_j^+,{\bf v}_j^+\rangle_\eta^\CC, \langle {\bf v}_j^-,{\bf v}_j^-\rangle_\eta^\CC \big\}_{j={1,\dots,k-1} }
    \bigg\} \quad \big( \cup \{\langle {\bf v}_{-1}, {\bf v}_{-1} \rangle_\eta\} \big).
    \label{eq:SpectrumOfSesquilinearGramMatrix}
\end{equation}
Note that the first and last blocks in $G_{\eta}$ and $G_{\eta}^{\CC}$ corresponding to the real eigenvectors are the same in the two matrices. Notice also that 
$$ \langle {\bf v}_j^+, {\bf v}_j^+ \rangle_\eta^\CC = ({\bf v}_j^+)^\dagger \eta {\bf v}_j^+ = ({\bf v}_j^-)^T \eta {\bf v}_j^+ = \langle {\bf v}_j^- , {\bf v}_j^+ \rangle_\eta = \langle {\bf v}_j^+ , {\bf v}_j^- \rangle_\eta,
$$
and similarly for $\langle {\bf v}_j^- , {\bf v}_j^- \rangle_\eta^\CC$, so the entries of $G_\eta$ and $G_{\eta}^\CC$ for the complex eigenvectors are the same but crucially they arrange themselves either in a 2-by-2 or in a 1-by-1 block diagonal form respectively. The signature of $V$ according to $G_{\eta}^\CC$ must agree with that of $V$ in the Cartesian basis, that is, it must be $(d-1,1)$. Since there is only one negative sign in the first block, we conclude that $\langle {\bf v}_j^+ , {\bf v}_j^+ \rangle_\eta^\CC = \langle {\bf v}_j^-,{\bf v}_j^+ \rangle_\eta$ must all be positive and $\langle {\bf v}_0^+ , {\bf v}_0^- \rangle_\eta $ must be negative. 

We can now normalize them to $+1$ and $-1$ respectively, for instance by dividing each ${\bf v}_0^\pm$ with $\sqrt{-\langle {\bf v}_0^+,{\bf v}_0^-\rangle_\eta}$ and each ${\bf v}_j^\pm$ with $\sqrt{+\langle {\bf v}_j^+, {\bf v}_j^- \rangle_\eta}$, yielding the cross-normalization condition of the NP frame. In this normalized Coxeter eigenbasis, the two Gram matrices become (we put a hat to denote that they are now ``normalized''):
\begin{equation}
  \setlength{\arraycolsep}{2pt}
  \setlength{\delimitershortfall}{0pt}
  \newcommand*{\myfbox}[1]{%
    \fcolorbox{red}{red!20!white}{$#1$}%
  }
  \newcommand{\Oplus}{\ensuremath{\vcenter{\hbox{\scalebox{3.5}{$\oplus$}}}}}
     \hat{G}_{\eta} = 
    \begin{pmatrix}\,
        \myfbox{\begin{matrix}
            0 & -1 \\
            -1 & 0 \\
        \end{matrix}} & 0 & 0 \\
        0 & \bigoplus_{j=1}^{k-1}\myfbox{\mathbb{J}_j} & 0 \\[+1ex]
         0 & 0 & \myfbox{1}
    \,\,
    \end{pmatrix}, 
    \qquad
    \text{ with }
    \mathbb{J}_j = \begin{pmatrix}
        0 & 1 \\
        1 & 0
    \end{pmatrix} \forall j
\label{eq:GramMatrixNormalizedCoxeterEigenbasisBilinear}
\end{equation}
and 
\begin{equation}
  \setlength{\arraycolsep}{2pt}
  \setlength{\delimitershortfall}{0pt}
  \newcommand*{\myfbox}[1]{%
    \fcolorbox{red}{red!20!white}{$#1$}%
  }
  \newcommand{\Oplus}{\ensuremath{\vcenter{\hbox{\scalebox{3.5}{$\oplus$}}}}}
     \hat{G}_{\eta}^\CC = 
    \begin{pmatrix}\,
        \myfbox{\begin{matrix}
            0 & -1\\
            -1 & 0 \\
        \end{matrix}} & 0 & 0 \\
        0 & \bigoplus_{j=1}^{k-1}\myfbox{\mathbb{I}_j} & 0 \\[+1ex]
         0 & 0 & \myfbox{1}
    \,\,
    \end{pmatrix}, 
    \qquad
    \text{ with }
    \mathbb{I}_j = \begin{pmatrix}
        1 & 0 \\
        0 & 1
    \end{pmatrix} \forall j
\label{eq:GramMatrixNormalizedCoxeterEigenbasisSesquilinear}
\end{equation}
respectively

\chapter{The N-V Test in More Detail}
\label{appendix:NVTestDetails}

This Appendix contains supplementary material for the N-V test of Section \ref{sec:NVCorrespondence}.

\section{Causal Diamonds in Minkowski Spacetime}
\label{app:CausalDiamondsInVariousDimensions}

Crucial to the N-V test is sampling causal diamonds $\cdiamond$ from a point set living in Minkowski spacetime $\MM^d$. Below, we describe how to parametrize $\cdiamond$ in different dimensions and how to control their shape under generic boosts in $\MM^d$. Typical references when it comes to the geometry of causal diamonds are \cite{Gibbons_2007,Berthiere_2015}.

\subsection{Lorentz Boosts and Lightlike Coordinates in $(1+1)$D}

In $(1+1)D$, the Lorentz group $SO(1,1)$ contains only a single generator, a boost generator in the $(t,x)$-plane. A Lorentz boost $L(\varpi)$ by a hyperbolic angle $\varpi$ (also called the \textit{rapidity}) in $(t,x)$ coordinates ($c=1$) is given by 
\begin{equation}
\mathbb{R}^{1,1} \ni \mathbf{x} = 
    \begin{pmatrix}
        t \\
        x
    \end{pmatrix} \longmapsto \mathbf{x}' = L(\varpi) \mathbf{x}, \quad \text{ with } \quad 
    L(\varpi) =     
    \begin{pmatrix}
        \cosh{\varpi} & - \sinh{\varpi}\\
        -\sinh{\varpi} & \cosh{\varpi} 
    \end{pmatrix} 
\label{eq:LorentzTransformation1+1D}
\end{equation}
where $\det L = \cosh^2{\varpi} - \sinh^2{\varpi} = +1$. When dealing with causal diamonds, the natural choice of basis adapted to the lightcones is the \textbf{lightlike (or null) coordinates} $(\rm u,v)$ defined as 
\begin{equation}
    \left\{\begin{aligned}
    & {\rm u} = \kappa_{\rm uv} (t+x) \\
    & {\rm v} = \kappa_{\rm uv} (t-x)
    \end{aligned} \right.
     \quad \text{ or } \quad
     {\bf x}_{\rm null} = 
     \begin{pmatrix}
         {\rm u} \\
         {\rm v} 
     \end{pmatrix} = B_{\rm uv} {\bf x}
     \quad \text{ with } \quad 
     B_{\rm uv} =
     \kappa_{\rm uv} 
     \begin{pmatrix}
         1 & 1\\
         1 & -1
     \end{pmatrix}.
     \label{eq:NullCoordinates1Plus1DDef}
\end{equation}
Here, $B_{\rm uv}$ is the change-of-basis matrix for some choice of $\kappa_{\rm uv}\in \RR$ with $\det B_{\rm uv} = -2\kappa_{\rm uv}^2$. 
In the $\rm (u,v)$ basis, the Lorentz boost becomes simpler:
\begin{equation}
   {\bf x}_{\rm null} \mapsto {\bf x}'_{\rm null} = B_{\rm uv}^{-1} L(\varpi) B_{\rm uv}^{-1}    {\bf x}_{\rm null}  = 
   \frac{1}{2\kappa_{\rm uv}^2}
   \begin{pmatrix}
       e^{-\varpi} & 0 \\
       0 & e^{+\varpi}
   \end{pmatrix}
   {\bf x}_{\rm null}  
   \label{eq:LorentzTransformationNullBasis}
\end{equation}
The Minkowski metric $\eta_{\mu \nu} = {\rm diag}(-1,+1)$ in $(t,x)$-coordinates now becomes 
\begin{equation}
    \eta \mapsto \eta_{\rm null} = (B_{\rm uv}^{-1})^T \eta B_{\rm uv}^{-1} = -\frac{1}{2 \kappa_{\rm uv}^2} 
    \begin{pmatrix}
        0 & 1 \\
        1 & 0
    \end{pmatrix}
    \label{eq:NullMinkowskiMetric1Plus1D}
\end{equation}
in the $(\rm u,v)$-coordinates, with $\det \eta_{\rm null} = -1/4\kappa_{\rm uv}^4$. We note that for a generic coordinate transformation (including the Lorentz transformation), a metric (a bilinear form) always transforms with a \textit{congruence transformation} (i.e.~with the transpose, like in Eq.~\eqref{eq:NullMinkowskiMetric1Plus1D}, and not with the inverse, as in Eq.~\eqref{eq:LorentzTransformationNullBasis}). Such transformations do not preserve the determinant, as can be seen from Eq.~\eqref{eq:NullMinkowskiMetric1Plus1D}: while $\det \eta = -1$, after the transformation we get $\det \eta_{\rm null} = -1/(\det B_{\rm uv})^2$. A matrix $M$, on the other hand, transforms with a \textit{similarity transformation} under a change of basis: $M \mapsto B^{-1} M B$.   

In $(1+1)$D, a Lorentz-boosted causal diamond of Lorentzian volume $V_{}$ is specified completely by its centre $(t_c, x_c)$ and boost parameter $\varpi$. In the $(\rm u,v)$-plane, the centre is at $(\rm u_c,v_c)$ and $\cdiamond$ looks like a rectangle $\ABsquare{ppBlueAccentL60}{0}$. Fixing its volume $V$, we get
\begin{equation}
    V = \int_{\cdiamond} \sqrt{-\det \eta_{\rm null}}\dd^2\mathbf{x} = \frac{1}{2\kappa_{\rm uv}^2}\int_{\cdiamond} {\rm du}{\rm dv} = \frac{ab}{2 \kappa_{\rm uv}^2},
    \label{eq:VolumeCausalDiamond1Plus1DNullCoords}
    \end{equation}
where $a,b$ are the diamond's null extents. Under a Lorentz boost \eqref{eq:LorentzTransformationNullBasis}, each null extent picks up a boost-factor $e^{\pm \varpi}$; as a function of $\varpi$, $a$ and $b$ become 
\begin{equation}
    a(\varpi,V) = \sqrt{2\kappa_{\rm uv}^2 V}e^{\varpi} \quad ({\rm u}\text{-direction})\qquad b(\varpi,V) = \sqrt{2\kappa_{\rm uv}^2 V} e^{-\varpi} \quad ({\rm v}\text{-direction})
    \label{eq:NullExtentsCDiamond1+1D}
\end{equation}
Their product is fixed for any boost: $a(\varpi,V)b(\varpi,V) = 2\kappa_{\rm uv}^2 V$. As $\varpi > 0$ increases, the diamond $\cdiamond$ stretches in ${\rm u}$-direction and contracts in the ${\rm v}$-direction (and vice versa for $\varpi < 0$). In any case, $V$ is a Lorentz-invariant quantity. After a boost by some $\varpi$, the four tips of the diamond are at null coordinates $$
({\rm u}_c \pm \tfrac{1}{2}a(\varpi,V), {\rm v}_c \pm \tfrac{1}{2}b(\varpi,V)), \text{ where } {\rm u}_c = \kappa_{\rm uv}(t_c + x_c), {\rm v}_c = \kappa_{\rm uv}(t_c - x_c)$$ are the coordinates of its centre in the null plane.

\subsection{Parametrizing Boosts in $(2+1)$D}

\begin{figure}[p]
\centering
    \begin{tabular}{cc}
        \subfloat[Unboosted $\cdiamond$. 
        ]{\includegraphics[width = 0.4 \linewidth]{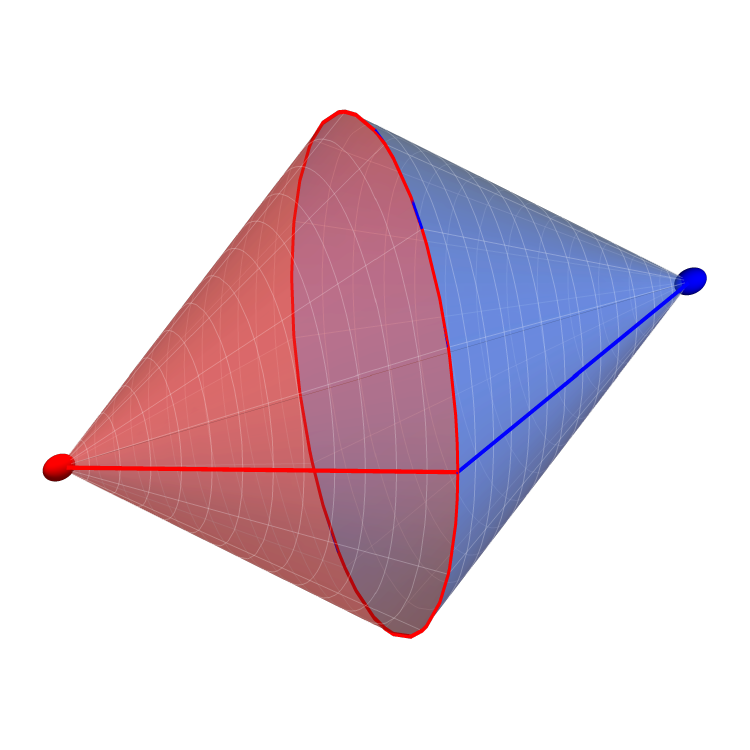}}
        \subfloat[2D cross sections of the $\cdiamond$ at $t=0.65T$ and $t=0$. 
        \label{fig:HistogramSampling1000Diamonds}]{ 
        \raisebox{9mm}{
            \includegraphics[width=0.6\linewidth]{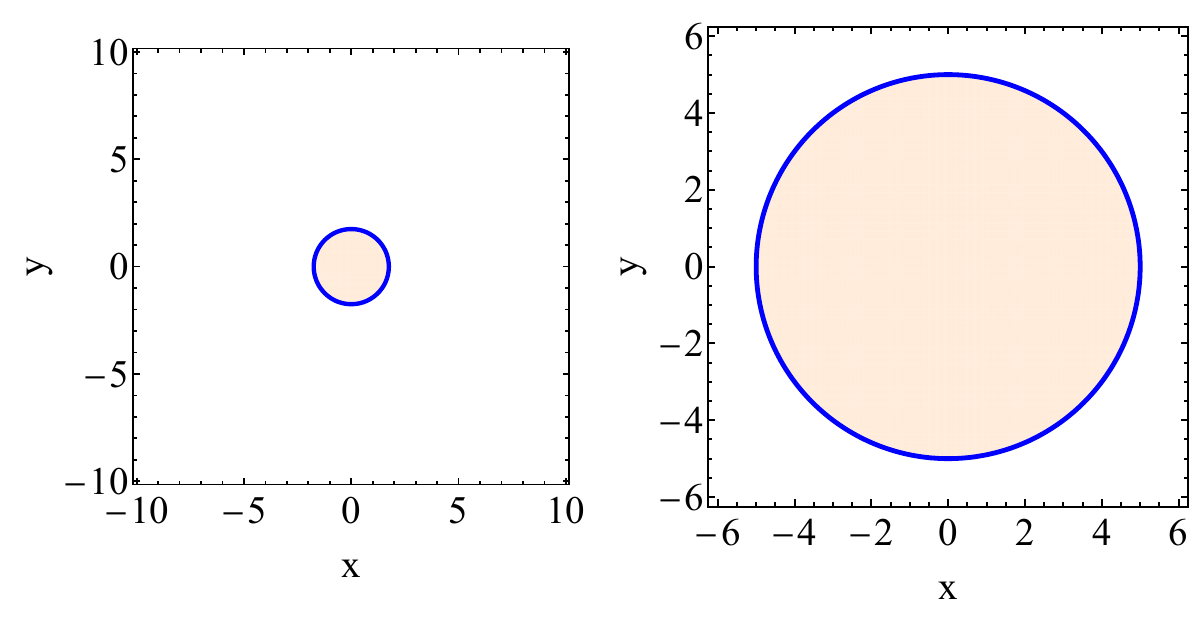}
        }}
    \end{tabular}
    \vspace{+2ex}
    \begin{tabular}{cc}
        \subfloat[Boosted $\cdiamond$ ($\varpi = 1, \varphi=\pi/4$).
        ]{\includegraphics[width = 0.4 \linewidth]{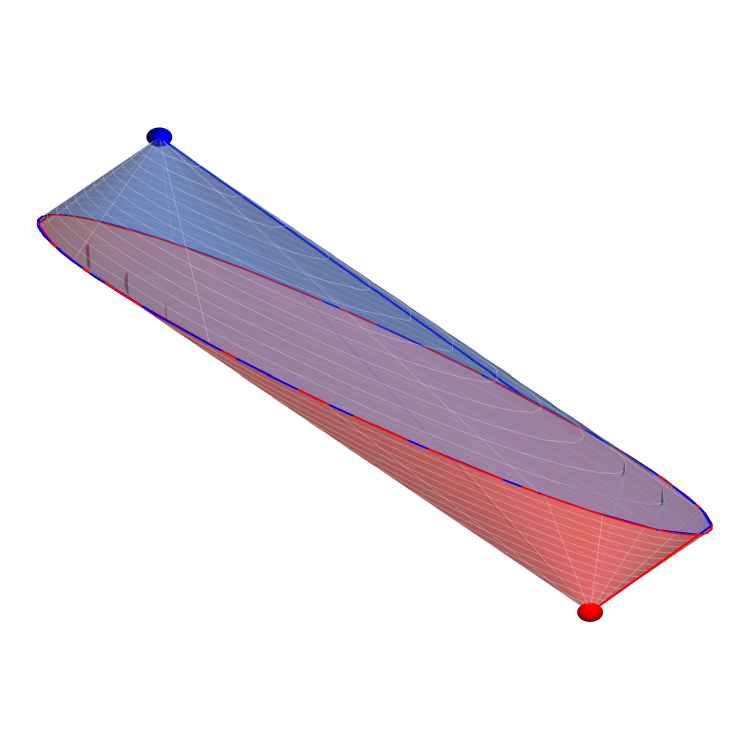}}
        \subfloat[2D cross sections of the $\cdiamond$ at $t=0.65T$ and $t=0$.]{ 
        \raisebox{9mm}{
            \includegraphics[width=0.6\linewidth]{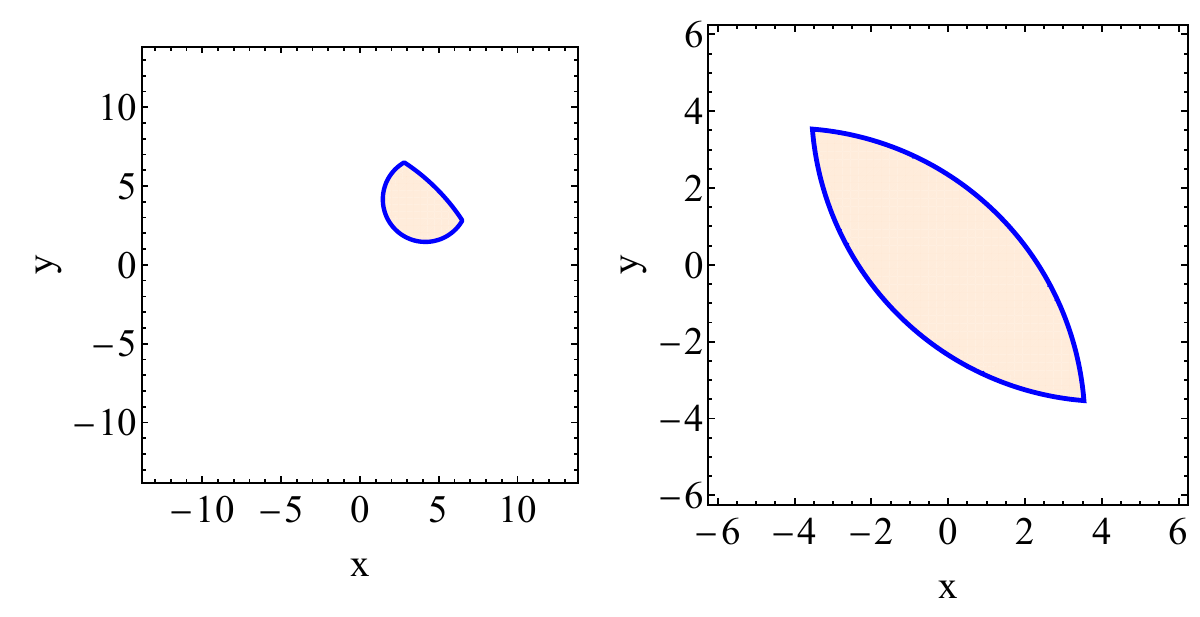}
        }}
    \end{tabular}
    \caption[Boosted and unboosted causal diamond in $(2+1)$D.]{Boosted and unboosted causal diamond of volume $V_\cdiamond = 5$ in $(2+1)$D together with two of its $2D$ cross-sections. For the unboosted diamond, its past and future tips (in blue and red respectively) are in the same rest frame and all its cross-sections are circles. For a boosted diamond, the tips gain a spatial component and do not lie directly above one another in the temporal direction; the cross-sections become distorted.}
\label{fig:BoostedUnboostedDiamond1Plus2D}
\end{figure}

Unfortunately, a description of $\cdiamond$ in null coordinates becomes more complicated in higher dimensions, which is why we stay in some global Minkowski coordinates. In $(2+1)$D, the geometry of a causal diamond is specified by three numbers:  \textit{half-height} $T$, rapidity $\varpi$ and azimuthal angle $\varphi$. The first is related to its Lorentz-invariant volume:
$$V_\cdiamond = \frac{2\pi}{3}T_\cdiamond^3 \quad \Longleftrightarrow \quad T(V_\cdiamond) = \left(\frac{3V_\cdiamond}{2\pi}\right)^{1/3}$$
and $T_\cdiamond$ is the temporal half-extent of $\cdiamond$ in its rest frame. The last two completely specify a generic boost in $\MM^3$: contrary to the previous case in $(1+1)$ dimensions, now one needs to specify not only the rapidity $\varpi$, but also a direction along the circle $\Sph^1$ in order to characterize a boost. Since any combination of boosts can be described by the composition of a boost and a rotation (this is known as a \textit{Wigner or Thomas} rotation \cite{Gourgoulhon_2013}), we are able to parametrize all possible boosts this way. Letting ${\bf x}_c = (t_c,\Vec{x}_c)$ be the centre of a boosted $\cdiamond$, its past and future tips ${\bf p_\cdiamond,q_\cdiamond}$ are:
$${\bf p_\cdiamond} = {\bf x}_c -T_\cdiamond(\cosh \varpi, \sinh{\varpi} \cdot {\hat{\bf n}}) ={\bf x}_c - T_\cdiamond \bigl(\cosh\varpi, \sinh\varpi\cos\varphi, \sinh\varpi\sin\varphi\bigr)$$
$${\bf q_\cdiamond} = {\bf x}_c + T_\cdiamond(\cosh \varpi, \sinh{\varpi} \cdot {\hat{\bf n}}) ={\bf x}_c + T_\cdiamond\bigl(\cosh\varpi, \sinh\varpi \cos\varphi, \sinh\varpi\sin\varphi\bigr)$$
If $\varpi = 0$, we have an unboosted diamond whose tips lie directly above/below its centre along the time direction. If $\varpi > 0$, the tips get tilted towards the spatial direction $\hat{{\bf n}} = (\cos\varphi, \sin\varphi)$. In Fig.~\ref{fig:BoostedUnboostedDiamond1Plus2D} we show some examples of a boosted $\cdiamond$ in $\MM^3$.

\subsection{Parametrizing Boosts in $(3+1)$D}

Adding one more spatial direction means that in $(3+1)$D a generic boost needs the rapidity $\varpi$ together with a unit direction ${\bf \hat{n}} = (\cos \varphi \sin \theta, \sin \varphi \sin \theta, \cos \theta) $ along the sphere $\Sph^2$ (compared to a circle $\Sph^1$), which is parametrised by the azimuthal angle $\varphi$ and a polar angle $\theta$. The (rest-frame) volume of a causal diamond $\cdiamond$ now becomes
$$
V_\cdiamond = \frac{2\pi}{3}T_\cdiamond^4 \Leftrightarrow T(V_\cdiamond) =  \left(\frac{3V}{2\pi}\right)^{1/4}
$$
where $T_\cdiamond$ is again its temporal half-extent, and its past and future tips are
$$
{\bf p}_\cdiamond = {\bf x}_c -T_\cdiamond(\cosh \varpi, \sinh{\varpi} \cdot {\hat{\bf n}})
$$
$$
{\bf q}_\cdiamond = {\bf x}_c + T_\cdiamond(\cosh \varpi, \sinh{\varpi} \cdot {\hat{\bf n}})
$$

\section{Sampling Causal Diamonds in the N-V Test}
\label{appendix:SamplingNVDiamonds}

\subsection{$(1+1)$ Dimensions}

To start the N-V test, we first fix our sampling volume levels $V_{\rm sample}$; we pick $$V_{\rm sample} = \{ 1,5,10,20,30,40,50,60,70,80,90,100\}$$ for most practical tests to match \cite{Saravani_2014}. Then, we pick some maximum number of samples $M_{\rm samples}$ (usually equal to 1000) and a maximum boost rapidity $\varpi_{\max}$. To make sure that we sample the same physical volumes across all patterns of potentially different densities, we scale $V_{\rm sample}$ by the density of each pattern; in the end, we sample $V_\cdiamond = V_{\rm sample}/\rho_{\rm pattern}$ off of a pattern with density $\rho_{\rm pattern}$. The volume $V_\cdiamond$ always refers to physical volumes calculated in the coordinates of the pattern $P$ from which we sample $\cdiamond$. For patterns with unit density, $V_{\rm sample}$ and $V_\cdiamond$ are the same.

Having decided on the point structures to be tested, we generate a large enough patch for each one of them. Let its size be $L_{\rm patch}$; in $(1+1)D$, most patches are rectangular with $t$ and $x$ extents equal to $L_{\rm patch}$. In practice, the \textbf{size of the patch constraints the size of the largest boost we can perform for a given volume} in $V_\cdiamond$. For each trial, the rapidity $\varpi$ is drawn uniformly from the interval $[-\varpi_{\max}, + \varpi_{\max}]$. For a given rapidity and patch-size, we calculate a containment margin equal to $$ \textrm{margin} = \frac{1}{4}\big(a(\varpi,V_\cdiamond)+b(\varpi,V_\cdiamond)\big)$$ in order for the sampled diamond to fit the patch. The trial is rejected if $\textrm{margin} > L_{\rm patch}/2$; otherwise, its centre $(\rm u_c,v_c)$ is drawn uniformly within the interval $[\textrm{margin},L_{\rm patch} - \textrm{margin}]^2$ in $(t,x)$. 

\begin{figure}[t!]
   \centering
    \begin{tabular}{cc}
        \subfloat[Sampling a single $\cdiamond$ from a Poisson sprinkling. 
\label{fig:SamplingSingleCausalDiamondPoisson}
        ]{\includegraphics[scale=0.40]{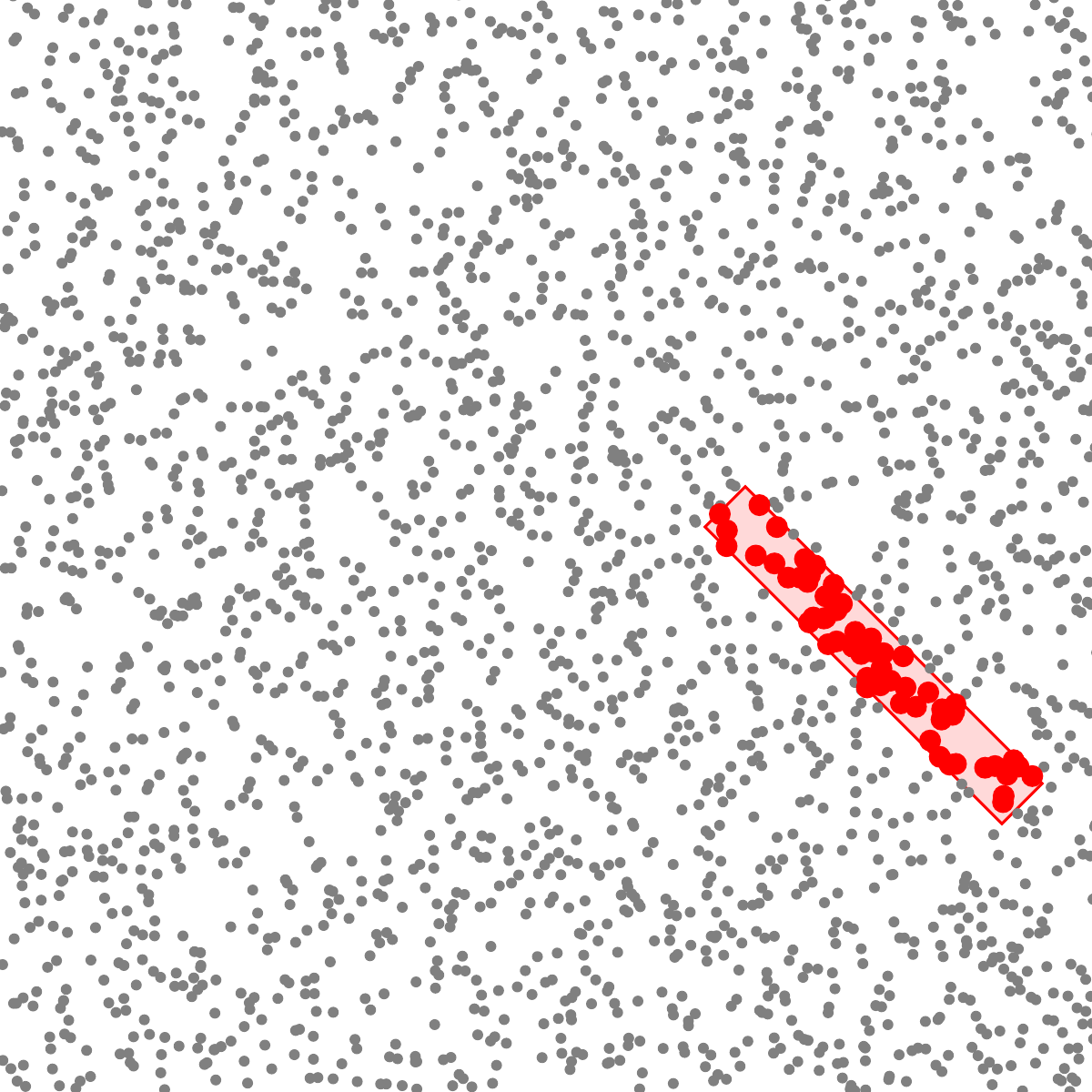}}
        \subfloat[Statistics from sampling 1000 such $\cdiamond$.
        \label{fig:HistogramSampling1000Diamonds}]{ 
        \raisebox{9mm}{
            \includegraphics[scale=0.40]{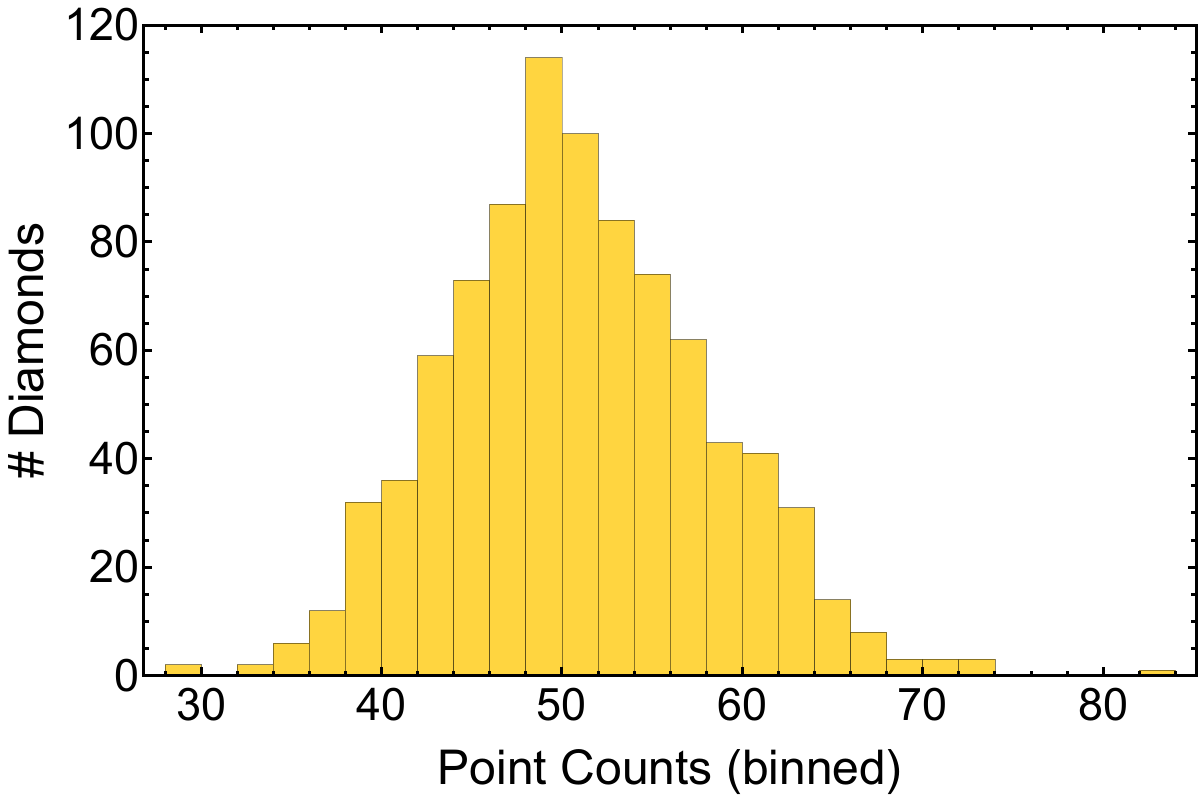}
        }}
    \end{tabular}
    \caption[An example of sampling a causal diamond from a Poisson sprinkling in $(1+1)$ dimensions.]{On the left, we see an example of sampling a single causal diamond from a Poisson sprinkling in $(1+1)$ dimensions. The diamond's volume is $V_\cdiamond = 50$ and is boosted by $\varpi = -1$. On the right, we have the statistics arising from sampling 1000 such diamonds from a $L_{\rm patch} \times L_{\rm patch}$ Poisson sprinkling ($L_{\rm patch} = 400$) with $\varpi \in [-5,5]$. Out of the 1000 proposed sampled, 890 are accepted; the mean is $\langle N_{\cdiamond} \rangle = 50.3944$ with $\sigma_{N_\cdiamond} = 7.18708$ -- to be compared with the expected average $\langle N_\cdiamond \rangle_{\rm Poisson} = V_\cdiamond = 50$ and standard deviation $(\sigma_{N_\cdiamond})_{\rm Poisson} = 5\sqrt{2} \approx 7.07107$.}
\label{fig:CausalDiamondSampleFromPoisson1Plus1D}
\end{figure}

A point $(\rm u,v)$ is inside the causal diamond $\cdiamond$ if and only if
$$ 
| {\rm u} - {\rm u}_c | \leq \tfrac{1}{2}a(\varpi,V_\cdiamond) \quad \textit{ and } \quad
| {\rm v} - {\rm v}_c | \leq \tfrac{1}{2}b(\varpi,V_\cdiamond)
$$
Calculating this for all points in a point pattern $P$, we get the number count $N_\cdiamond$ for a single diamond $\cdiamond$ from
\begin{equation}
N_{\cdiamond} = \sum_{\rm ({\rm u,v}) \in P} \Theta\big(\tfrac{1}{2}a(\varpi,V_\cdiamond) - |{\rm u - u_c}| \big) \cdot \Theta\big(\tfrac{1}{2}b(\varpi,V_\cdiamond) - |{\rm v - v_c}| \big),
\label{eq:CountDiamond1Plus1D}
\end{equation}
where $\Theta(x-x_0)$ is Heaviside's step function, giving $1$ if $x-x_0 > 0$ and zero otherwise. We then repeat this process for all $M_{\rm sample}$ volumes, collect the statistics and obtain a probability distribution, from which we calculate the average number count $\langle N_\cdiamond \rangle$ for each volume in $V_\cdiamond$ and the corresponding standard deviation $\sigma_{N_\cdiamond}$. An example of this process is given in Fig.~\ref{fig:CausalDiamondSampleFromPoisson1Plus1D}; this process is similar in higher dimensions as well. We also keep track of the boosts performed at each trial and calculate the average rapidity $\langle \varpi \rangle$ at each volume level. 

\subsection{Tested Patches and Data Tables for the N-V Test in $(1+1)D$}
\label{subsec:TestedPatches1Plus1D}

Below, we give the tables for the N-V test in $(1+1)$D that led to Figs.~\ref{fig:NVTest1Plus1DMean} and \ref{fig:NVTest1Plus1DStandardDeviation} in Section \ref{subsec:NVTestIn1+1D}, and explain the properties of the patches used.

For the Poisson sprinkling and the two lattices (Fibonacci and integer), we used a rectangular $(t,x)$ patch with extent $L_{\rm patch} = 400$. These patches contain around 160000 points each, and we used null coordinates with $\kappa_{\rm uv} = 1$. The maximum boost was set to $\varpi_{\max} = 5$. The sampling procedure was identical to what was described in the previous subsection, where $V = V_{\rm sample} = \rho_{\rm pattern} V_{\cdiamond}$. The results for the Poisson sprinkling, the Fibonacci lattice and the integer lattice are given in Tables \ref{tab:NVTest1Plus1DDataTablePoisson}, \ref{tab:NVTest1Plus1DDataTableFibonacci} and \ref{tab:NVTest1Plus1DDataTableInteger} respectively.

\begin{table}[]
    \centering
\begin{tabular}{c|c|c|c|c|c}
$V_{\rm sample}$ & $V_\cdiamond = V_{\rm sample}/\rho_{\rm Poisson}$ & $\langle N_\cdiamond \rangle$ & $\sigma_{N_\cdiamond}$ & $\langle \varpi \rangle $ & $M_{\rm samples}$ \\
\hline \hline
1 & 1 & 0.98 & 0.994278 & 2.51305 & 1000 \\
5 & 5 & 4.988 & 2.22729 & 2.5179 & 1000 \\
10 & 10 & 9.98 & 3.2308 & 2.46611 & 1000 \\
20 & 20 & 19.9979 & 4.52019 & 2.36139 & 973 \\
30 & 30 & 30.4234 & 5.46622 & 2.24658 & 933 \\
40 & 40 & 39.8719 & 6.37252 & 2.25944 & 890 \\
50 & 50 & 50.1377 & 6.74055 & 2.13532 & 893 \\
60 & 60 & 60.1525 & 7.99744 & 2.17812 & 846 \\
70 & 70 & 69.7099 & 8.8189 & 2.09655 & 855 \\
80 & 80 & 80.0282 & 9.08909 & 2.10185 & 816 \\
90 & 90 & 90.4083 & 9.44492 & 2.03452 & 791 \\
100 & 100 & 100.404 & 10.1374 & 2.02724 & 822 \\
\hline
\end{tabular}

    \caption{N-V test in $(1+1)D$: Poisson sprinkling ($\rho_{\rm Poisson} = 1$)}
    \label{tab:NVTest1Plus1DDataTablePoisson}
\end{table}

\begin{table}[]
    \centering
\begin{tabular}{c|c|c|c|c|c}
$V_{\rm sample}$ & $V_\cdiamond = V_{\rm sample}/\rho_{\rm Fib}$ & $\langle N_\cdiamond \rangle$ & $\sigma_{N_\cdiamond}$ & $\langle \varpi \rangle $ & $M_{\rm samples}$ \\
\hline \hline
1 & 1.11803 & 1.014 & 0.517757 & 2.48319 & 1000 \\
5 & 5.5901 & 5.007 & 0.609362 & 2.50462 & 1000 \\
10 & 11.1803 & 10.016 & 0.655877 & 2.53723 & 1000 \\
20 & 22.3607 & 20.0021 & 0.691791 & 2.40358 & 958 \\
30 & 33.541 & 30.0241 & 0.704757 & 2.26405 & 914 \\
40 & 44.7214 & 39.9866 & 0.700215 & 2.33639 & 894 \\
50 & 55.9017 & 50.0378 & 0.697105 & 2.10841 & 873 \\
60 & 67.082 & 60.0071 & 0.727573 & 2.11901 & 851 \\
70 & 78.2624 & 69.9938 & 0.711859 & 2.05739 & 812 \\
80 & 89.4427 & 80.0267 & 0.738963 & 2.13119 & 824 \\
90 & 100.623 & 90.0214 & 0.760497 & 1.94223 & 794 \\
100 & 111.803 & 100.026 & 0.752594 & 2.02798 & 784 \\
\hline
\end{tabular}

    \caption{N-V test in $(1+1)D$: Fibonacci lattice ($\rho_{\rm Fib} = \sqrt{5}/2$)}
    \label{tab:NVTest1Plus1DDataTableFibonacci}
\end{table}

\begin{table}[]
    \centering
\begin{tabular}{c|c|c|c|c|c}
$V_{\rm sample}$ & $V_\cdiamond = V_{\rm sample}/\rho_{\rm Int}$ & $\langle N_\cdiamond \rangle$ & $\sigma_{N_\cdiamond}$ & $\langle \varpi \rangle $ & $M_{\rm samples}$ \\
\hline \hline
1 & 1 & 0.725 & 2.15778 & 2.48888 & 1000 \\
5 & 5 & 5.482 & 15.4223 & 2.50881 & 1000 \\
10 & 10 & 10.643 & 26.8151 & 2.47342 & 1000 \\
20 & 20 & 20.3946 & 38.7233 & 2.42874 & 968 \\
30 & 30 & 30.5739 & 42.9512 & 2.24491 & 941 \\
40 & 40 & 39.5759 & 47.0941 & 2.237 & 903 \\
50 & 50 & 54.6884 & 58.9363 & 2.23478 & 876 \\
60 & 60 & 61.4373 & 62.2396 & 2.24539 & 853 \\
70 & 70 & 69.8318 & 63.2673 & 2.13437 & 850 \\
80 & 80 & 80.5588 & 59.308 & 2.01838 & 816 \\
90 & 90 & 89.8407 & 59.3684 & 2.01302 & 816 \\
100 & 100 & 100.11 & 62.7761 & 2.00836 & 816 \\
\hline
\end{tabular}

    \caption{N-V test in $(1+1)D$: integer lattice ($\rho_{\rm Int} = 1$)}
    \label{tab:NVTest1Plus1DDataTableInteger}
\end{table}

For the $\mathrm{I}_{3,1}$ spacetime quasicrystal, we used a causal-diamond patch of null extent $L_{\rm sQC \, patch} = 700$ (in the physical Coxeter eigenbasis, with coordinates $\alpha_0^\pm$). In the $\alpha_{0}^\pm$-coordinates, the induced metric is given by Eq.~\eqref{eq:GramMatrixI13CoxeterEigenbasis}; comparing this to Eq.~\ref{eq:NullMinkowskiMetric1Plus1D}, we find that $\kappa_{\rm uv}^2 = 1/2G_0^{+-}$ in this case, where $G_0^{+-} = (17+5\sqrt{17})/2$ is the metric's scaling factor. The diamond's null extents \eqref{eq:NullExtentsCDiamond1+1D} read 
$$ a(\varpi,V_\cdiamond) = \sqrt{\frac{V_\cdiamond}{G_0^{+-}}} e^{+\varpi} \quad \text{ and } \quad 
b(\varpi,V_\cdiamond) = \sqrt{\frac{V_\cdiamond}{G_0^{+-}}}e^{-\varpi}
$$ with $V_\cdiamond = V_{\rm sample}/\rho_{\rm sQC}^{\mathrm{I}_{3,1}} V_\cdiamond$ (see Section \ref{sec:ExamplesSpacetimeQuasicrystals}). This patch contains around 50 million points and has volume $V_{\rm patch} = G_0^{+-} L_{\rm sQC \, patch}^2$ (using Eq.~\eqref{eq:VolumeCausalDiamond1Plus1DNullCoords} with $a=b=L_{\rm sQC \, patch}$). 

For this structure, we sample entirely in the $(\alpha_0^+,\alpha_0^-)$ null plane; Eq.~\eqref{eq:CountDiamond1Plus1D} remains true with the appropriate coordinate changes. As this is a new structure, we made sure to modify the sampling process so that we do not reject any samples. We use an \textit{adaptive boost bound}: for diamond volume $V_{\cdiamond}$ and patch length $L_{\rm sQC \, patch}$, the maximum boost that keeps the diamond inside the patch is (setting $a = b = L_{\rm sQC \, patch}$ in the above expressions)
$$ \varpi_{\max} (V_\cdiamond) = \ln \left( \frac{L_{\rm sQC \, patch}}{\sqrt{V_{\cdiamond}/G_0^{+-}}}\right)
$$
For each $V_{\cdiamond}$, then, we find $\min\big(\varpi_{\max}=5,\varpi_{\max}(V_{\cdiamond})\big)$ and boosts are drawn uniformly from a symmetric interval around 0 whose end points are given by this maximum boost. The centre null coordinates $(\alpha_0^+, \alpha_0^-)_c$ are drawn uniformly from the interval $$[\tfrac{1}{2}a(\varpi,V_\cdiamond), L_{\rm sQC \, patch} - \tfrac{1}{2}a(\varpi,V_\cdiamond)] \times [\tfrac{1}{2}b(\varpi,V_\cdiamond), L_{\rm sQC \, patch} - \tfrac{1}{2}b(\varpi,V_\cdiamond)].$$ The results are given in Table \ref{tab:NVTest1Plus1DDataTableI31sQC}.

\begin{table}[]
    \centering
\begin{tabular}{c|c|c|c|c|c}
$V_{\rm sample}$ & $V_\cdiamond = V_{\rm sample}/ \rho_{\rm sQC}^{\mathrm{I}_{3,1}}$ & $\langle N_\cdiamond \rangle$ & $\sigma_{N_\cdiamond}$ & $\langle \varpi \rangle $ & $M_{\rm samples}$ \\
\hline \hline
1 & 0.0880392 & 1.0135 & 0.662221 & 3.99006 & 2000  \\
5 & 0.440196 & 5. & 1.03997 & 4.01224 & 2000  \\
10 & 0.880392 & 9.983 & 1.2492 & 3.97476 & 2000  \\
20 & 1.76078 & 20.085 & 1.4659 & 3.91383 & 2000  \\
30 & 2.64118 & 30.0565 & 1.62931 & 3.78801 & 2000  \\
40 & 3.52157 & 39.9985 & 1.70996 & 3.67756 & 2000  \\
50 & 4.40196 & 49.988 & 1.817 & 3.60551 & 2000  \\
60 & 5.28235  & 60.0175 & 1.88174 & 3.57768 & 2000  \\
70 & 6.16275 & 70.0185 & 1.94963 & 3.57072 & 2000  \\
80 & 7.04314 & 80.003 & 2.01222 & 3.5198 & 2000  \\
90 & 7.92353 & 90.0055 & 2.0417 & 3.46372 & 2000  \\
100 & 8.80392 & 100.031 & 2.12057 & 3.46676 & 2000  \\
\hline
\end{tabular}

    \caption{N-V test in $(1+1)D$: sQC ($\rho_{\rm sQC}^{\mathrm{I}_{3,1}} \approx 11.3585$).}
    \label{tab:NVTest1Plus1DDataTableI31sQC}
\end{table}

\subsection{The Bimodal Model for the Boost Catastrophe in $(1+1)D$}

We saw in Fig.~\ref{fig:BoostCatastopheIntegerLattice} that, if we boost the integer lattice $\mathrm{I}_{1,1}$ beyond a certain threshold $\varpi_{\rm crit}$, then the variance grows without limit. Here, we prove the equation used in that figure to approximate the variance $\sigma_{N_\cdiamond}$ in that regime. 

For a very thin diamond of volume $V_\cdiamond$ with $b(\varpi,V_\cdiamond)$ and $\varpi > \varpi_{\rm crit}$, we essentially have $b<\delta {\rm v}_{\min} = 1$ (for $\mathrm{I}_{1,1}$). This means that the diamond's ${\rm v}$-window $[{\rm v}_c - b(\varpi,V_{\cdiamond})/2,{\rm v}_c + b(\varpi,V_{\cdiamond})/2]$ contains at most one integer ${\rm v}_0$. This happens if the window intersects some null-row. There are exactly two outcomes: 
\begin{itemize}
    \item \textbf{Hit:} exactly one integer ${\rm v}_0$ falls in the window's ${\rm v}$-extent. At that $v_0$, the allowed ${\rm u}$-values form a spacing-2 sequence, that is, in a given ${\rm v}$-row, the ${\rm u}$’s are equally-spaced, two integers away from one another. As a result, the window intersects the ${\rm v}_0$ row the number of points that fall within its ${\rm u}$-range of width $a(\varpi,V_\cdiamond)$ is $N_\cdiamond =  a(\varpi,V_\cdiamond)/2$. The factor of $1/2$ in the hit-count comes from the parity structure of $\mathrm{I}_{1,1}$: at the captured ${\rm v}_0$, only one parity class of ${\rm u}$-values is present. This happens with probability $P(N_\cdiamond = \alpha(\varpi,V)/2) = $ $b(\varpi,V)$.
    \item \textbf{Miss:} no integer ${\rm v_0}$ falls within the window's ${\rm v}$-extent and the diamond contains no points at all, $N_\cdiamond = 0$. This happens with probability $P(N_\cdiamond=0) = 1-b(\varpi,V_\cdiamond)$  
\end{itemize}
This gives a bimodal distribution $N_\cdiamond \in [0, a(\varpi,V)/2]$. The mean is
$$\langle N_\cdiamond \rangle = \sum N_\cdiamond \cdot P(N_\cdiamond) = \frac{a(\varpi,V_\cdiamond)}{2} \cdot b(\varpi,V_\cdiamond) = \frac{\sqrt{2V_\cdiamond}e^\varpi}{2} \cdot \sqrt{2V_\cdiamond}e^{-\varpi} = V_\cdiamond$$
and is unaffected by the boost catastrophe — the two outcomes average correctly and the structure satisfy the N-V correspondence. The variance, on the other hand, is heavily affected: we obtain
$$\mathrm{Var}(N_\cdiamond) = \sigma_{N_\cdiamond}^2 = \sum (N_\cdiamond - \langle N_\cdiamond \rangle)^2 P(N_\cdiamond) = \left(\frac{a(\varpi,V)}{2}\right)^2 b(\varpi,V)(1-b(\varpi,V)).$$
For $b \ll 1$ (deep in the catastrophic regime), we can drop the term $1-b(\varpi,V)$ to obtain:
$$\mathrm{Var}(N_\cdiamond) \approx \frac{a(\varpi,V)^2}{4}b(\varpi,V) = \frac{2Ve^{2\varpi}}{4} \cdot \sqrt{2V_\cdiamond} e^{-\eta} = \frac{(2V_\cdiamond)^{3/2}}{4}e^{\varpi},$$
so
$$\boxed{\sigma_{N_\cdiamond}(\varpi) \approx \frac{(2V_\cdiamond)^{3/4}}{2}e^{\varpi/2}},$$
which is the equation we stated in the main text. The $e^{\varpi/2}$ growth comes from the competition between two effects: the hit count ($a(\varpi,V)/2$) grows as $e^\varpi$, while the hit probability $b(\varpi,V)$ shrinks as $e^{-\varpi}$, and $\sigma_{N_\cdiamond} \sim (a(\varpi,V)/2)\sqrt{b(\varpi,V)} \sim e^\varpi \cdot e^{-\varpi/2} = e^{\varpi/2}$.

\subsection{$(2+1)$ Dimensions}

The sampling procedure in $(2+1)$D is very similar to the one in $(1+1)$D. For each volume $V_\cdiamond$, we sample uniformly a rapidity $\varpi$ in some interval $[\varpi_{\min}, \varpi_{\max}]$ (usually taken to be $[0,5]$) and an angle $\varphi \in [0,2\pi]$; we restrict to only positive $\varpi$ and allow $\varphi$ to run in the entire $\Sph^1$, which allows us to capture the negative boosts as well.  

From these parameters, we calculate the tips ${\bf p_\cdiamond,q_\cdiamond}$ of $\cdiamond = \Cint{{\bf p}}{\bf q}$. A point ${\bf x} = (t,x,y)$ is inside a diamond $\cdiamond$ if and only if it is inside the future lightcone of its past tip \textit{and} inside the past lightcone of its future tip. Mathematically, we demand that (see Eq.~\eqref{eq:CausalInterval})
$$ \left\{
\begin{aligned}
    & {\bf x} \in \textrm{Fut}({\bf p_\cdiamond}) \Rightarrow \Delta t_p \ge 0 \quad \text{ and} \quad  \Delta t_p^2 \ge s_p^2 \\
    &  {\bf x} \in \textrm{Past}({\bf q_\cdiamond}) \Rightarrow \Delta t_q \ge 0 \quad \text{ and} \quad \Delta t_q^2 \ge s_q^2
\end{aligned}
\right.
$$
where $\Delta t_p = t - t_p$ and $\Delta t_q = t_q - t$ are the temporal differences and $s_p^2 = (x-x_p)^2+(y-y_p)^2$ and $s_q^2 = (x-x_q)^2+(y-y_q)^2$ are the spatial ones. The number of points from some pattern $P$ inside a given diamond $\cdiamond$ is
\begin{equation}
N_\cdiamond = \sum_{(t,x,y) \in P} \Theta(\Delta t_p) \cdot \Theta(\Delta t_p^2 - s_p^2) \cdot \Theta(\Delta t_q) \cdot \Theta(\Delta t_q^2 - s_q^2)
\label{eq:CountDiamond1Plus2D}
\end{equation}
Here, we do not use pre-generated patches: since all structures are lattices apart from the sprinkling, we know analytically the points and we can enumerate which ones belong to a given $\cdiamond$ once we know the tips of a boosted diamond. This allows us to test all boosts up to our choice for $\varpi_{\max}$ without having to care about finite-size effects and their restrictions on $\varpi_{\max}$ for each $V_\cdiamond$. Finding the points in the interior of a causal diamond amounts to solving the above inequalities over the integers in \texttt{Mathematica}. 

\subsection{Tested Patches and Data Tables for the N-V Test in $(2+1)$D}

Here, we show the data tables that led to Figs.~\ref{fig:1Plus2DNVTestComparisonPlots} and \ref{fig:1Plus2DNVTestStandardDeviationComparisonPlots}, and discuss the relevant patches. 

For the Poisson sprinkling, we actually used a patch of dimensions $[0,75] \times [0,75] \times [0,75]$ in $(t,x,y)$ coordinates; so $L_{\rm patch} = 75$. For a given volume $V_\cdiamond$ and boost $(\varpi,\varphi)$, a $\cdiamond$ is contained entirely in a box of half-extent $T \cosh \varpi$; we use this as a margin, and we sample the diamond's centre uniformly from the interval $[T\cosh{\varpi},L_{\rm patch} - T\cosh\varpi]^3$, ensuring that the diamond will fit in the patch and that we do not reject any samples. We also use a \textit{adaptive boost bound:} for a volume level $V_\cdiamond$, the maximum boost we can perform is given as
$$ \varpi_{\max}(V_\cdiamond) = \cosh^{-1}\bigg(\frac{L_{\rm patch}}{2T(V_\cdiamond)}\bigg)
$$
The results for this run are given in Tab.~\ref{tab:NVTest1Plus2DDataTablePoisson}.

For the rest point structures, we do not generate any patches but rather enumerate analytically the points inside some $\cdiamond$. In Tables \ref{tab:NVTest1Plus2DDataTableInteger} and \ref{tab:NVTest1Plus2DDataTableHexagonal}, we show the results for the integer and spatially hexagonal lattice respectively. For the Fibonacci 3D lattice, we have three different datasets: in Tabs.~\ref{tab:NVTest1Plus2DDataTableFibonacciAnyBoost}, \ref{tab:NVTest1Plus2DDataTableFibonacciTXBoost} and \ref{tab:NVTest1Plus2DDataTableFibonacciTYBoost} we see the results of the N-V test when the boosts are (i) \textit{generic} (any $\varpi$ and $\varphi$), (ii) along the $(t,x)$ plane (any $\varpi$ but $\varphi = 0$), and (iii) along the $(t,y)$-plane (any $\varpi$ but $\varphi = \pi/2$) respectively.

\begin{table}[]
    \centering
\begin{tabular}{c|c|c|c|c|c}
$V_{\rm sample}$ & $V_\cdiamond = V_{\rm sample}/\rho_{\rm Poisson}$ & $\langle N_\cdiamond \rangle$ & $\sigma_{N_\cdiamond}$ & $\langle \varpi \rangle $ & $M_{\rm samples}$ \\
\hline \hline
1 & 1 & 1.045 & 1.02079 & 2.27575 & 1000  \\
5 & 5 & 5.104 & 2.22129 & 1.98422 & 1000 \\
10 & 10 & 9.98 & 3.14125 & 1.88598 & 1000  \\
20 & 20 & 20.018 & 4.56777 & 1.7507 & 1000  \\
30 & 30 & 30.005 & 5.54344 & 1.70097 & 1000  \\
40 & 40 & 39.93 & 6.23233 & 1.70105 & 1000  \\
50 & 50 & 50.24 & 6.97259 & 1.64699 & 1000 \\
60 & 60 & 60.261 & 7.70612 & 1.52742 & 1000 \\
70 & 70 & 70.089 & 8.60006 & 1.56145 & 1000  \\
80 & 80 & 79.876 & 9.05133 & 1.57772 & 1000  \\
90 & 90 & 89.595 & 9.64116 & 1.45806 & 1000  \\
100 & 100 & 99.336 & 10.2663 & 1.58328 & 1000 \\
\hline
\end{tabular}
    \caption{N-V test in $(2+1)D$: Poisson sprinkling ($\rho_{\rm Poisson} = 1$)}
    \label{tab:NVTest1Plus2DDataTablePoisson}
\end{table}

\begin{table}[]
    \centering
\begin{tabular}{c|c|c|c|c|c}
$V_{\rm sample}$ & $V_\cdiamond = V_{\rm sample}/\rho_{\rm Int}$ & $\langle N_\cdiamond \rangle$ & $\sigma_{N_\cdiamond}$ & $\langle \varpi \rangle $ & $M_{\rm samples}$ \\
\hline \hline
1 & 1 & 0.965 & 1.15979 & 2.5177 & 1000 \\
5 & 5 & 4.977 & 4.51185 & 2.60841 & 1000 \\
10 & 10 & 9.646 & 5.99138 & 2.46513 & 1000 \\
20 & 20 & 19.911 & 11.0956 & 2.54518 & 1000 \\
30 & 30 & 30.818 & 25.9734 & 2.59047 & 1000 \\
40 & 40 & 39.412 & 16.1791 & 2.57081 & 1000 \\
50 & 50 & 50.453 & 27.7533 & 2.49096 & 1000 \\
60 & 60 & 60.434 & 33.4833 & 2.58903 & 1000 \\
70 & 70 & 71.293 & 39.1393 & 2.50677 & 1000 \\
80 & 80 & 79.819 & 29.1574 & 2.52409 & 1000 \\
90 & 90 & 90.173 & 37.2398 & 2.51372 & 1000 \\
100 & 100 & 100.598 & 43.6304 & 2.47702 & 1000 \\
\hline
\end{tabular}
    \caption{N-V test in $(2+1)D$: integer lattice ($\rho_{\rm Int} = 1$)}
    \label{tab:NVTest1Plus2DDataTableInteger}
\end{table}

\begin{table}[]
    \centering
\begin{tabular}{c|c|c|c|c|c}
$V_{\rm sample}$ & $V_\cdiamond = V_{\rm sample}/\rho_{\rm Hex}$ & $\langle N_\cdiamond \rangle$ & $\sigma_{N_\cdiamond}$ & $\langle \varpi \rangle $ & $M_{\rm samples}$ \\
\hline \hline
1 & 1 & 0.965 & 0.747217 & 2.45692 & 1000 \\
5 & 5 & 5.079 & 3.19671 & 2.4976 & 1000 \\
10 & 10 & 9.768 & 3.12409 & 2.60492 & 1000 \\
20 & 20 & 20.093 & 6.3788 & 2.50606 & 1000 \\
30 & 30 & 30.076 & 6.78441 & 2.48175 & 1000 \\
40 & 40 & 40.391 & 10.6118 & 2.50612 & 1000 \\
50 & 50 & 49.571 & 8.63107 & 2.50399 & 1000 \\
60 & 60 & 59.731 & 9.0101 & 2.46032 & 1000 \\
70 & 70 & 69.934 & 12.2519 & 2.43476 & 1000 \\
80 & 80 & 79.71 & 12.438 & 2.54192 & 1000 \\
90 & 90 & 89.694 & 14.2987 & 2.48098 & 1000 \\
100 & 100 & 100.542 & 17.7806 & 2.45298 & 1000 \\
\hline
\end{tabular}
    \caption{N-V test in $(2+1)D$: hexagonal lattice ($\rho_{\rm Hex} = 1$)}
    \label{tab:NVTest1Plus2DDataTableHexagonal}
\end{table}

\begin{table}[]
    \centering
\begin{tabular}{c|c|c|c|c|c}
$V_{\rm sample}$ & $V_\cdiamond = V_{\rm sample}/\rho_{\rm Fib}$ & $\langle N_\cdiamond \rangle$ & $\sigma_{N_\cdiamond}$ & $\langle \varpi \rangle $ & $M_{\rm samples}$ \\
\hline \hline
1 & 1.11803 & 1.061 & 1.27864 & 2.44845 & 1000 \\
5 & 5.59017 & 5.002 & 3.63637 & 2.47971 & 1000 \\
10 & 11.1803 & 10.016 & 5.31451 & 2.51572 & 1000 \\
20 & 22.3607 & 19.959 & 8.06948 & 2.47242 & 1000 \\
30 & 33.541 & 30.286 & 12.5071 & 2.54616 & 1000 \\
40 & 44.7214 & 39.627 & 16.2237 & 2.41429 & 1000 \\
50 & 55.9017 & 49.195 & 16.8717 & 2.53936 & 1000 \\
60 & 67.082 & 60.87 & 19.9169 & 2.51264 & 1000 \\
70 & 78.2624 & 71.539 & 24.8563 & 2.60207 & 1000 \\
80 & 89.4427 & 78.856 & 16.6356 & 2.50735 & 1000 \\
90 & 100.623 & 89.464 & 24.2882 & 2.50537 & 1000 \\
100 & 111.803 & 100.597 & 32.6206 & 2.51088 & 1000 \\
\hline
\end{tabular}
    \caption{N-V test in $(2+1)D$: Fibonacci lattice ($\rho_{\rm Fib} = \sqrt{5}/2$), any boost.}
    \label{tab:NVTest1Plus2DDataTableFibonacciAnyBoost}
\end{table}

\begin{table}[]
    \centering
\begin{tabular}{c|c|c|c|c|c}
$V_{\rm sample}$ & $V_\cdiamond = V_{\rm sample}/\rho_{\rm Fib}$  & $\langle N_\cdiamond \rangle$ & $\sigma_{N_\cdiamond}$ & $\langle \varpi \rangle $ & $M_{\rm samples}$ \\
\hline \hline
1 & 1.11803 & 0.993 & 0.646041 & 2.468 & 1000 \\
5 & 5.59017 & 4.983 & 0.922801 & 2.51202 & 1000 \\
10 & 11.1803 & 9.952 & 1.1675 & 2.59428 & 1000 \\
20  & 22.3607 & 20.003 & 1.45158 & 2.4847 & 1000 \\
30  & 33.541 & 30. & 1.50142 & 2.46573 & 1000 \\
40  & 44.7214 & 39.914 & 1.75433 & 2.43494 & 1000 \\
50  &55.9017 & 49.916 & 1.82655 & 2.53837 & 1000 \\
60  &  67.082 & 60.126 & 1.98899 & 2.5341 & 1000 \\
70  & 78.2624 & 69.975 & 2.2551 & 2.4259 & 1000 \\
80 & 89.4427 & 79.841 & 2.19831 & 2.43318 & 1000 \\
90  & 100.623 & 90.071 & 2.35659 & 2.41644 & 1000 \\
100  & 111.803 & 100.013 & 2.5926 & 2.55402 & 1000 \\
\hline
\end{tabular}
    \caption{N-V test in $(2+1)D$: Fibonacci lattice ($\rho_{\rm Fib} = \sqrt{5}/2$), TX boost.}
    \label{tab:NVTest1Plus2DDataTableFibonacciTXBoost}
\end{table}

\begin{table}[]
    \centering
\begin{tabular}{c|c|c|c|c|c}
$V_{\rm sample}$ & $V_\cdiamond = V_{\rm sample}/\rho_{\rm Fib}$  & $\langle N_\cdiamond \rangle$ & $\sigma_{N_\cdiamond}$ & $\langle \varpi \rangle $ & $M_{\rm samples}$ \\
\hline \hline
1 & 1.11803 & 0.983 & 0.647403 & 2.46653 & 1000 \\
5 & 5.59017 & 5.018 & 0.945817 & 2.3968 & 1000 \\
10 & 11.1803 & 9.957 & 1.19188 & 2.50372 & 1000 \\
20 & 22.3607 & 20.005 & 1.40176 & 2.52662 & 1000 \\
30 & 33.541 & 30.032 & 1.53144 & 2.52091 & 1000 \\
40 & 44.7214 & 39.886 & 1.73725 & 2.50087 & 1000 \\
50 & 55.9017  & 49.924 & 1.80873 & 2.49391 & 1000 \\
60 & 67.082 & 60.11 & 1.99245 & 2.46668 & 1000 \\
70 & 78.2624 & 69.99 & 2.26694 & 2.46807 & 1000 \\
80 & 89.4427 & 80.133 & 2.2424 & 2.48918 & 1000 \\
90 & 100.623 & 90.069 & 2.39332 & 2.43656 & 1000 \\
100 & 111.803 & 100.018 & 2.50239 & 2.52428 & 1000 \\
\hline
\end{tabular}
    \caption{N-V test in $(2+1)D$: Fibonacci lattice ($\rho_{\rm Fib} = \sqrt{5}/2$), TY boost.}
    \label{tab:NVTest1Plus2DDataTableFibonacciTYBoost}
\end{table}

\subsection{$(3+1)$ Dimensions}

The sampling procedure in $(3+1)$D is very similar to the one in $(2+1)$D. For each volume $V_\cdiamond$, we sample uniformly a rapidity $\varpi$ in some interval $[\varpi_{\min}, \varpi_{\max}]$ and
a unit direction ${\bf \hat{n}} \in \Sph^2$ (we draw each component from a normal Gaussian distribution and normalize to make the resulting vector of unit length). We let the angles vary throughout their full range on $\Sph^2$, so we only consider positive rapidities $\varpi$. Deciding where to place a $\cdiamond$ depends on whether or not we use a pre-generated patch or now; we discuss this in the next section. Once we know the centre ${\bf x}_c$ of a diamond, though, we can find its tips ${\bf p}_\cdiamond,{\bf q}_\cdiamond$ and check whether or not a point belongs to the interior of $\cdiamond$ by exactly the same condition as in the $(2+1)$D test, now translated in 4D coordinates $(t,x,y,z)$. The only difference is that $s_p^2 = (x-x_p)^2 + (y-y_p)^2 + (z-z_p)^2$ and $s_q^2 = (x-x_q)^2 + (y-y_q)^2 + (z-z_q)^2$; which these trivial modifications, Eq.~\eqref{eq:CountDiamond1Plus2D} gives us the number of points $N_\cdiamond$ inside a sampled diamond $\cdiamond$.

\subsection{Tested Patches and Data Tables for the N-V Test in $(3+1)$D}
\label{subsec:TestedPatched1Plus3D}

Here, we show the data tables that led to Figs.~\ref{fig:NVTest1Plus3DMean} and \ref{fig:NVTest1Plus3DStandardDeviation} and discuss the relevant patches.

\begin{table}[]
    \centering
\begin{tabular}{c|c|c|c|c|c}
$V_{\rm sample}$ & $V_\cdiamond = V_{\rm sample}/\rho_{\rm Poisson}$  & $\langle N_\cdiamond \rangle$ & $\sigma_{N_\cdiamond}$ & $\langle \varpi \rangle $ & $M_{\rm samples}$ \\
\hline \hline
1 & 1 & 1.01 & 1.00644 & 2.54202 & 1000 \\
5 & 5 & 5.121 & 2.28902 & 2.4564 & 1000 \\
10 & 10 & 9.927 & 3.14795 & 2.46237 & 1000 \\
20 & 20 & 19.816 & 4.27228 & 2.54494 & 1000 \\
30 & 30 & 30.22 & 5.54819 & 2.46016 & 1000 \\
40 & 40 & 40.338 & 6.42706 & 2.4825 & 1000 \\
50 & 50 & 49.927 & 7.16987 & 2.59398 & 1000 \\
60 & 60 & 60.003 & 7.56414 & 2.48187 & 1000 \\
70 & 70 & 70.073 & 8.502 & 2.49255 & 1000 \\
80 & 80 & 80.403 & 8.8611 & 2.43021 & 1000 \\
90 & 90 & 90.013 & 9.72941 & 2.46886 & 1000 \\
100 & 100 & 99.831 & 10.3843 & 2.46956 & 1000 \\
\hline
\end{tabular}
    \caption{N-V test in $(3+1)D$: Poisson sprinkling ($\rho_{\rm Poisson} = 1$).}
    \label{tab:NVTest1Plus3DDataTablePoisson}
\end{table}

\begin{table}[]
    \centering
\begin{tabular}{c|c|c|c|c|c}
$V_{\rm sample}$ & $V_\cdiamond = V_{\rm sample}/\rho_{\rm Int}$  & $\langle N_\cdiamond \rangle$ & $\sigma_{N_\cdiamond}$ & $\langle \varpi \rangle $ & $M_{\rm samples}$ \\
\hline \hline
1 & 1 & 1.007 & 0.929416 & 2.52378 & 1000 \\
5 & 5 & 4.951 & 2.6324 & 2.52973 & 1000 \\
10 & 10 & 10.042 & 4.02174 & 2.54322 & 1000 \\
20 & 20 & 19.85 & 7.56655 & 2.49271 & 1000 \\
30 & 30 & 30.257 & 9.06604 & 2.48891 & 1000 \\
40 & 40 & 40.221 & 17.0982 & 2.46428 & 1000 \\
50 & 50 & 49.979 & 11.5253 & 2.4379 & 1000 \\
60 & 60 & 60.036 & 11.3537 & 2.49943 & 1000 \\
70 & 70 & 69.621 & 14.6767 & 2.53047 & 1000 \\
80 & 80 & 79.81 & 16.5752 & 2.52031 & 1000 \\
90 & 90 & 89.639 & 15.5471 & 2.39631 & 1000 \\
100 & 100 & 99.15 & 22.3733 & 2.56779 & 1000 \\
\hline
\end{tabular}
    \caption{N-V test in $(3+1)D$: integer lattice ($\rho_{\rm Int} = 1$).}
    \label{tab:NVTest1Plus3DDataTableInteger}
\end{table}

For the Poisson sprinkling and the integer lattice, we use no pre-generated patch and sampling goes exactly as described so far. In Tables \ref{tab:NVTest1Plus3DDataTablePoisson} and \ref{tab:NVTest1Plus3DDataTableInteger} we give the results of an N-V test for these two structures respectively.

In $(3+1)$D, we test two different spacetime quasicrystals, one from the self-dual integer lattice $\mathrm{I}_{7,1}$ (see Section \ref{sec:ExamplesSpacetimeQuasicrystals}) and the icosahedral one from the root-lattice $HE_7^+$ (see Section \ref{sec:ExamplesNewSpacetimeQuasicrystals}). For both of these structures we use a pre-generated patch but with slightly different setups, as we now explain. 

For the $\mathrm{I}_{7,1}$ quasicrystal, we use a \textit{tubular patch}: on the Lorentzian plane $V_0$, we have a causal diamond of null extent (in the physical coordinates $\alpha_0^\pm$) equal to $L_{\rm sQC \, I_{7,1}} = 30$ and on the Euclidean plane $V_1$ a disk of radius $R_{\rm phys \, disk} = 15$. The patch has volume equal to 
$$ V_{\rm sQC \, I_{7,1}} = (L_{\rm sQC \, I_{7,1}}^2) (2\pi R_{\rm phys \, disk}^2) $$ and it contains 50 229 647 points; its density is given by Eq.~\eqref{eq:DensitysQCI17}, or simply by dividing the number of points by the volume given above. Here, our null coordinates $(\alpha_0^+,\alpha_0^-)$ are in terms of the normalized Coxeter eigenbasis, and can be mapped to Eq.~\eqref{eq:NullCoordinates1Plus1DDef} by choosing $\kappa_{\rm uv} = 1/\sqrt{2}$, which is why the volume of the causal diamond on $V_0$ is simply $L_{\rm sQC \, I_{7,1}}^2$ according to Eq.~\eqref{eq:VolumeCausalDiamond1Plus1DNullCoords}. 

When sampling $\varpi$, we must consider finite-size effects as the boosted diamond must fit inside the finite patch. Two independent geometric constraints are computed from $\hat{{\bf n}}$ and $T_{\cdiamond}$ before drawing $\varpi$ from $[\varpi_{\min},\varpi_{\max}]$ (usually, $\varpi_{\min} = 0$):
\begin{itemize}
    \item \textbf{Null-patch constraint:} requiring that the maximum null extents of the diamond lie within $[-L_{\rm sQC \, I_{7,1}}/2,L_{\rm sQC \, I_{7,1}}/2]$ in both null coordinates, we get the maximum boost on the $V_0$ plane. Letting $|\hat{n}_x|$ be the absolute value of the $x$-coordinate of $\hat{\bf n}$ and $l_\cdiamond = (L_{\rm sQC \, I_{7,1}}/\sqrt{2})/T_\cdiamond$, we get 
    $$ \varpi_{\max}^{\rm null}(\cdiamond) = \ln \left( \frac{l_\cdiamond + \sqrt{l_\cdiamond^2 - 1 + |\hat{n}_x|^2}}{1+|\hat{n}_x|} \right).
    $$
    Here, $l_\cdiamond$ is the ratio of the null half-width of the patch ($L_{\rm sQC \, I_{7,1}}/\sqrt{2}$ (since each of the null coordinates of the patch has half-width $L_{\rm sQC \, I_{7,1}}/2$, this gives a null diagonal of $L_{\rm sQC \, I_{7,1}}/\sqrt{2}$) to the diamond's half-height $T_\cdiamond$. It is a dimensionless measure of how many diamond-lengths fit across the patch. If $\hat{n}_x = 1$ (boost entirely in the TX plane), then $\varpi_{\max}^{V_0} = \ln (l_\cdiamond)$, which is the same result we found in the $(1+1)$D case.
    \item \textbf{Transverse-disk constraint:} when $|\hat{n}_x|<1$ and the boost `sees' the other spatial directions as well, our boosted diamond must also fit within the disk. The spatial cross-section of $\cdiamond$ in the $(y,z)$-plane has maximum radius 
    $$ r_{\cdiamond}^{\max} =  T_\cdiamond\sqrt{\cosh^2\varpi - \sinh^2\varpi |\hat{n}_x|^2}
    ,$$
    which must not exceed $R_{\rm phys \, disk}$; for $|\hat{n}_x|=1$, we find the uniform extent $T_\cdiamond \cosh \varpi$ we mentioned in the $(2+1)$D case. This gives one more constraint:
    $$ \varpi_{\max}^{\rm disk}(\cdiamond) = \mathrm{arcsinh}\left(\sqrt{\frac{(R_{\rm phys \, disk}/T_\cdiamond)^2 - 1}{1 - |\hat{n}_x|^2}}\right)$$
\end{itemize}
Using these two constraints, we find that $\varpi_{\max}(\cdiamond) = \min \left\{ \varpi_{\max}^{\rm null}(\cdiamond),\varpi_{\max}^{\rm disk}(\cdiamond) \right\}$

To find a valid centre to place a diamond on the patch, we first find its null extents $a_0^+,b_0^-$ in the physical coordinates $\alpha_0^\pm$ 
$$a_{0}^+ = \frac{T_\cdiamond(\cosh\varpi + \sinh\varpi \hat{n}_x)}{\sqrt{2}}, \qquad b_{0}^- = \frac{T_\cdiamond(\cosh\varpi - \sinh\varpi \hat{n}_x)}{\sqrt{2}}$$
To guarantee the diamond stays within the patch in both null directions, the null centre coordinates are sampled independently in the safe intervals:
$$\alpha_{0,c}^+ \in \left[-\tfrac{1}{2}L_{\rm sQC \, I_{7,1}} + a_0^+, \tfrac{1}{2}L_{\rm sQC \, I_{7,1}} - a_0^+\right]$$
$$\alpha_{0,c}^- \in \left[-\tfrac{1}{2}L_{\rm sQC \, I_{7,1}} + b_0^-, \tfrac{1}{2}L_{\rm sQC \, I_{7,1}} - b_0^-\right],$$
which are converted back to $(t_c,x_c)$ coordinates. For the transverse $(y,z)$-centre, the maximum transverse extent of the diamond is $r_{\cdiamond}^{\max}$, leaving a safe radius $R_{\rm phys \, disk} -r_{\cdiamond}^{\max}$. The transverse centre is drawn area-uniformly: $r_c \in [0,R_{\rm phys \, disk} - r_{\cdiamond}^{\max}]$ and $(y_c,z_c) = (\cos\theta,\sin\theta)$ with $\theta$ sampled uniformly from $[0,2\pi]$.

\begin{table}[]
    \centering
\begin{tabular}{c|c|c|c|c|c}
$V_{\rm sample}$ & $V_\cdiamond = V_{\rm sample}/\rho_{\rm sQC \, I_{7,1}}$  & $\langle N_\cdiamond \rangle$ & $\sigma_{N_\cdiamond}$ & $\langle \varpi \rangle $ & $M_{\rm samples}$ \\
\hline \hline
1 & 0.0253306 & 1.035 & 0.876135 & 2.16286 & 200 \\
5 & 0.126653 & 4.825 & 1.94765 & 1.93573 & 200 \\
10 & 0.253306 & 9.915 & 2.41182 & 1.99568 & 200 \\
20 & 0.506611 & 19.645 & 3.22038 & 1.88324 & 200 \\
30 & 0.759917 & 29.775 & 4.35883 & 1.84863 & 200 \\
40 & 1.01322 & 39.775 & 4.60111 & 1.82417 & 200 \\
50 & 1.26653 & 50.215 & 5.31791 & 1.684 & 200 \\
60 & 1.51983 & 60.515 & 5.8713 & 1.70034 & 200 \\
70 & 1.77314 & 70.33 & 5.7144 & 1.6254 & 200 \\
80 & 2.02644 & 79.155 & 6.69516 & 1.62926 & 200 \\
90 & 2.27975 & 89.815 & 6.84287 & 1.72698 & 200 \\
100 & 2.53306 & 100.78 & 6.91961 & 1.60063 & 200 \\
\hline
\end{tabular}
    \caption{N-V test in $(3+1)D$: $\mathrm{I}_{7,1}$ sQC ($\rho_{\rm sQC \, I_{7,1}} \approx 39.4784 $), any boost.}
    \label{tab:NVTest1Plus3DDataTablesQCI71AnyBoost}
\end{table}

\begin{table}[]
    \centering
\begin{tabular}{c|c|c|c|c|c}
$V_{\rm sample}$ & $V_\cdiamond = V_{\rm sample}/\rho_{\rm sQC \, I_{7,1}}$  & $\langle N_\cdiamond \rangle$ & $\sigma_{N_\cdiamond}$ & $\langle \varpi \rangle $ & $M_{\rm samples}$ \\
\hline \hline
1 & 0.0253306 & 1.02 & 0.814196 & 2.23735 & 200 \\
5 & 0.126653 & 4.81 & 1.47812 & 1.96776 & 200 \\
10 & 0.253306 & 10.09 & 1.94701 & 1.77721 & 200 \\
20 & 0.506611 & 20.085 & 2.79586 & 1.78302 & 200 \\
30 & 0.759917 & 30.135 & 3.18709 & 1.7995 & 200 \\
40 & 1.01322 & 40.245 & 3.25236 & 1.45939 & 200 \\
50 &1.26653  & 50.18 & 3.4984 & 1.55549 & 200 \\
60 & 1.51983 & 59.82 & 4.16224 & 1.58671 & 200 \\
70 & 1.77314 & 69.695 & 3.93311 & 1.56873 & 200 \\
80 & 2.02644 & 79.7 & 4.94187 & 1.5655 & 200 \\
90 & 2.27975 & 90.14 & 4.5391 & 1.57516 & 200 \\
100 & 2.53306  & 100.085 & 4.87304 & 1.45386 & 200 \\
\hline
\end{tabular}
    \caption{N-V test in $(3+1)D$: $\mathrm{I}_{7,1}$ sQC ($\rho_{\rm sQC \, I_{7,1}} \approx 39.4784 $), TX boost.}
    \label{tab:NVTest1Plus3DDataTablesQCI71TXBoost}
\end{table}

\begin{table}[]
    \centering
\begin{tabular}{c|c|c|c|c|c}
$V_{\rm sample}$ & $V_\cdiamond = V_{\rm sample}/\rho_{\rm sQC \, I_{7,1}}$  & $\langle N_\cdiamond \rangle$ & $\sigma_{N_\cdiamond}$ & $\langle \varpi \rangle $ & $M_{\rm samples}$ \\
\hline \hline
1 & 0.0253306 & 0.83 & 0.863032 & 2.33323 & 200 \\
5 & 0.126653 & 5.175 & 1.94248 & 2.17459 & 200 \\
10 & 0.253306 & 10.215 & 2.67104 & 2.0709 & 200 \\
20 & 0.506611 & 19.875 & 3.88133 & 2.00536 & 200 \\
30 & 0.759917 & 30.37 & 4.16169 & 2.07798 & 200 \\
40 & 1.01322 & 39.96 & 5.12246 & 2.0225 & 200 \\
50 & 1.26653 & 50.38 & 5.05702 & 1.95567 & 200 \\
60 & 1.51983 & 60.385 & 6.52146 & 1.99421 & 200 \\
70 & 1.77314 & 69.785 & 7.28382 & 1.9944 & 200 \\
80 & 2.02644 & 80.23 & 5.91881 & 1.79493 & 200 \\
90 & 2.27975  & 90.555 & 6.91128 & 1.90763 & 200 \\
100 & 2.53306 & 100.62 & 7.03299 & 1.86155 & 200 \\
\hline
\end{tabular}
    \caption{N-V test in $(3+1)D$: $\mathrm{I}_{7,1}$ sQC ($\rho_{\rm sQC \, I_{7,1}} \approx 39.4784 $), TYZ boost.}
\label{tab:NVTest1Plus3DDataTablesQCI71TYZBoost}
\end{table}

In Tables \ref{tab:NVTest1Plus3DDataTablesQCI71AnyBoost}, \ref{tab:NVTest1Plus3DDataTablesQCI71TXBoost} and \ref{tab:NVTest1Plus3DDataTablesQCI71TYZBoost} we give the results for the N-V test for (1) any generic boost (any $\varpi$, $\hat{\bf n} \in \Sph^2$), (ii) boost along the TX plane (any $\varpi$, $\hat{\bf n} = (1,0,0)$), and (iii) boost along the TYZ plane (any $\varpi$, $\hat{\bf n} = (0, \cos \theta, \sin \theta)$). Note that here we have reduced the number of samples from 1000 to just 200. This is because these runs are computationally expensive: while the run with $M_{\rm samples} = 200$ takes approximately 16-20 minutes on the author's laptop, a run with $M_{\rm samples} = 1000$ will take around 1 hour and 20 minutes on the same machine keeping the rest of the parameters the same. To be able to report the preliminary results of this thesis, we did reduce the samples; a more advanced analysis will follow for the subsequent publication.

For the $HE_7^+$ quasicrystal, we use a \textit{causal diamond patch}: its half-extent is $T_{\rm sQC \, HF_4} = L_{\rm sQC \, HF_4}/2 = 15$ (where $L_{\rm sQC \, HF_4}$ is the null extent of the causal diamond in the $V_0$ plane) and its volume is 
$$ V_{\rm sQC \, HF_4} = \frac{\pi}{24} L_{\rm sQC \, HF_4}^4. 
$$
The density is given by Eq.~\eqref{eq:sQCDensityHF4} and the patch contains 630 993 points. 

Again, we have to take into account finite-size effects that restrict the maximum $\varpi$ for a given volume $V_\cdiamond$. 
Here, we need to make sure that a diamond with half-height $T_\cdiamond$, centre ${\bf x}_c = (t_c,\Vec{x}_c)$ and boost $(\varpi, \hat{\bf n})$ fits entirely within the background diamond patch. 
For some given $T_\cdiamond$, the most generous constraint comes from when $\cdiamond$ is centred exactly at the patch's centre; this gives a global ceiling $$\varpi_{\max}(\cdiamond) = \ln \left( \frac{T_{\rm sQC \, HF_4}}{T_\cdiamond}\right),$$ 
which is, in fact, the maximum rapidity achievable for any centre position (because for any other centre $\varpi$ will be smaller). For a fixed boost $(\varpi, \hat{\bf n})$, we then find a valid centre window to sample ${\bf x}_c$ from, similarly to the case before.

In Tables \ref{tab:NVTest1Plus3DDataTablesQCHF4AnyBoost}, \ref{tab:NVTest1Plus3DDataTablesQCHF4TXBoost} and \ref{tab:NVTest1Plus3DDataTablesQCHF4TYZBoost} we give the results for the N-V test for (1) any generic boost (any $\varpi$, $\hat{\bf n} \in \Sph^2$), (ii) boost along the TX plane (any $\varpi$, $\hat{\bf n} = (1,0,0)$), and (iii) boost along the TYZ plane (any $\varpi$, $\hat{\bf n} = (0, \cos \theta, \sin \theta)$). Note that here we have restated the number of samples back to 1000. This is because the patch we use is way smaller than the patch for the $I_{7,1}$ sQC, so 1000 samples are still within reach (in fact, each run takes less than 10 minutes on the author's laptop). However, it is very computationally expensive to generate a larger patch for the $HF_4$ quasicrystal: the current patch was generated in both the author's laptop and in Perimeter's supercomputer and it took approximately 2.5 hours (so no advantage from the supercomputer). We have only tried one version of the algorithm -- based on \texttt{C} and ran through \texttt{Mathematica} -- but for the future publication we will attempt to run our parallelized algorithm 
on the supercomputer, which is expected to lead to some computational advantage. 

Note that for both quasicrystals, we first sample a valid boost $\varpi$ and then find appropriate centre for the boosted $\cdiamond$. This effectively means that the higher the boost, the more likely it is that the sampled diamond is very close to the centre of the patch, so we introduce a bias there that is unavoidable. This does not affect the N-V test; after all, the density is everywhere the same. However, it becomes important for the hyperuniformity tests of Section \ref{subsec:SpacetimeHyperuniformity}.

\begin{table}[]
    \centering
\begin{tabular}{c|c|c|c|c|c}
$V_{\rm sample}$ & $V_\cdiamond = V_{\rm sample}/\rho_{\rm sQC \, HF_4}$  & $\langle N_\cdiamond \rangle$ & $\sigma_{N_\cdiamond}$ & $\langle \varpi \rangle $ & $M_{\rm samples}$ \\
\hline \hline
1 & 0.168035 & 1.004 & 0.827447 & 1.71252 & 1000 \\
5 & 0.840174 & 5.108 & 1.72549 & 1.52374 & 1000 \\
10 & 1.68035 & 10.097 & 2.39193 & 1.38522 & 1000 \\
20 & 3.3607 & 20.109 & 3.02064 & 1.25041 & 1000 \\
30 & 5.04104 & 30.026 & 3.82125 & 1.2478 & 1000 \\
40 & 6.72139 & 39.965 & 4.2282 & 1.17042 & 1000 \\
50 & 8.40174 & 50.001 & 4.84649 & 1.18285 & 1000 \\
60 & 10.0821 & 59.824 & 5.21538 & 1.15706 & 1000 \\
70 & 11.7624 & 69.679 & 5.49729 & 1.11836 & 1000 \\
80 & 13.4428 & 80.067 & 5.73615 & 1.10774 & 1000 \\
90 & 15.1231 & 90.009 & 5.69468 & 1.11611 & 1000 \\
100 & 16.8035 & 100.082 & 6.29579 & 1.11072 & 1000 \\
\hline
\end{tabular}
    \caption{N-V test in $(3+1)D$: $HE_7^+$ sQC ($\rho_{\rm sQC \, HF_4} \approx 5.951 $), any boost.}
\label{tab:NVTest1Plus3DDataTablesQCHF4AnyBoost}
\end{table}

\begin{table}[]
    \centering
\begin{tabular}{c|c|c|c|c|c}
$V_{\rm sample}$ & $V_\cdiamond = V_{\rm sample}/\rho_{\rm sQC \, HF_4}$  & $\langle N_\cdiamond \rangle$ & $\sigma_{N_\cdiamond}$ & $\langle \varpi \rangle $ & $M_{\rm samples}$ \\
\hline \hline
1 &0.168035 & 0.998 & 0.841846 & 1.64014 & 1000 \\
5 & 0.840174 & 5.003 & 1.70467 & 1.43966 & 1000 \\
10 & 1.68035 & 10.1 & 2.20428 & 1.37631 & 1000 \\
20 & 3.3607 & 19.906 & 2.73727 & 1.30627 & 1000 \\
30 & 5.04104 & 30.025 & 3.49007 & 1.23768 & 1000 \\
40 & 6.72139 & 40.219 & 4.12092 & 1.1864 & 1000 \\
50 & 8.40174 & 50.067 & 4.40112 & 1.1825 & 1000 \\
60 & 10.0821 & 59.845 & 4.90561 & 1.14215 & 1000 \\
70 & 11.7624 & 69.947 & 4.91084 & 1.15165 & 1000 \\
80 & 13.4428 & 80.012 & 5.09213 & 1.13976 & 1000 \\
90 & 15.1231 & 90.016 & 5.11448 & 1.10713 & 1000 \\
100 & 16.8035 & 99.896 & 5.40374 & 1.08423 & 1000 \\
\hline
\end{tabular}
    \caption{N-V test in $(3+1)D$: $HE_7^+$ spacetime quasicrystal ($\rho_{\rm sQC \, HF_4} \approx 5.951 $), TX boost.}
\label{tab:NVTest1Plus3DDataTablesQCHF4TXBoost}
\end{table}

\begin{table}[]
    \centering
\begin{tabular}{c|c|c|c|c|c}
$V_{\rm sample}$ & $V_\cdiamond = V_{\rm sample}/\rho_{\rm sQC \, HF_4}$  & $\langle N_\cdiamond \rangle$ & $\sigma_{N_\cdiamond}$ & $\langle \varpi \rangle $ & $M_{\rm samples}$ \\
\hline \hline
1& 0.168035  & 1.004 & 0.89038 & 1.66267 & 1000 \\
5 & 0.840174 & 4.958 & 1.66884 & 1.45349 & 1000 \\
10 & 1.68035 & 9.817 & 2.2932 & 1.35432 & 1000 \\
20 & 3.3607 & 20.137 & 3.11672 & 1.29035 & 1000 \\
30 & 5.04104 & 30.078 & 3.84769 & 1.30157 & 1000 \\
40 & 6.72139 & 40.047 & 4.41591 & 1.22678 & 1000 \\
50 & 8.40174 & 50.097 & 4.7821 & 1.14853 & 1000 \\
60 & 10.0821 & 60.059 & 5.35446 & 1.13717 & 1000 \\
70 & 11.7624 & 70.063 & 5.43899 & 1.10486 & 1000 \\
80 & 13.4428 & 79.914 & 5.90978 & 1.13134 & 1000 \\
90 & 15.1231 & 89.907 & 6.01403 & 1.10648 & 1000 \\
100 & 16.8035 & 100.327 & 6.2094 & 1.10101 & 1000 \\
\hline
\end{tabular}
    \caption{N-V test in $(3+1)D$: $HE_7^+$ spacetime quasicrystal ($\rho_{\rm sQC \, HF_4} \approx 5.951 $), TYZ boost.}
\label{tab:NVTest1Plus3DDataTablesQCHF4TYZBoost}
\end{table}

\subsection{N-V Test as a Hyperuniformity Test: Tables}

We used the same patch described in \ref{subsec:TestedPatches1Plus1D} with the difference that now we the adaptive sampling mentioned in Section \ref{subsec:NVTestIsHyperuniformityTest}; the results are shown in Table \ref{tab:NVTestForHyperuniformitysQCI31}.

\begin{table}[]
    \centering
\footnotesize 
\begin{tabular}{c|c|c|c|c|c|c}
$V_{\rm sample}$ & $V_\cdiamond = V_{\rm sample}/\rho_{\rm sQC \, I_{3,1}}$  & $\langle N_\cdiamond \rangle$ & $\sigma_{N_\cdiamond}$ & $\langle \varpi \rangle $ & $M_{\rm samples}$ & $M_{\max}(\cdiamond)$\\
\hline \hline
1 & 0.0880392 & 1.0135 & 0.662221 & 3.99006 & 2000 & 104657913 \\
5 & 0.440196 & 5 & 1.03997 & 4.01224 & 2000 & 20926524 \\
10 & 0.880392 & 9.983 & 1.2492 & 3.97476 & 2000 & 10461367 \\
20 & 1.76078 & 20.085 & 1.4659 & 3.91383 & 2000 & 5229344 \\
30 & 2.64118 & 30.0565 & 1.62931 & 3.78801 & 2000 & 3485544 \\
40 & 3.52157 & 39.9985 & 1.70996 & 3.67756 & 2000 & 2613724 \\
50 & 4.40196 & 49.988 & 1.817 & 3.60551 & 2000 & 2090674 \\
60 & 5.28235 & 60.0175 & 1.88174 & 3.57768 & 2000 & 1741998 \\
70 & 6.16275 & 70.0185 & 1.94963 & 3.57072 & 2000 & 1492960 \\
80 & 7.04314 & 80.003 & 2.01222 & 3.5198 & 2000 & 1306192 \\
90 & 7.92353 & 90.0055 & 2.0417 & 3.46372 & 2000 & 1160937 \\
100 & 8.80392 & 100.031 & 2.12057 & 3.46676 & 2000 & 1044738 \\
200 & 17.6078 & 199.994 & 2.55357 & 3.23065 & 2000 & 521945 \\
500 & 44.0196 & 500.127 & 3.2202 & 3.10144 & 2000 & 208442 \\
750 & 66.0294 & 749.964 & 3.46225 & 2.94496 & 2000 & 138824 \\
1000 & 88.0392 & 1000.08 & 3.84495 & 2.88529 & 2000 & 104032 \\
2000 & 176.078 & 1999.91 & 4.70017 & 2.69961 & 2000 & 51882 \\
4000 & 352.157 & 3999.98 & 5.65796 & 2.55924 & 2000 & 25847 \\
6000 & 528.235 & 5999.98 & 6.47613 & 2.38608 & 2000 & 17183 \\
8000 & 704.314 & 7999.88 & 6.95313 & 2.35942 & 2000 & 12857 \\
10000 & 880.392 & 10000 & 7.73173 & 2.31458 & 2000 & 10264 \\
12500 & 1100.49 & 12499.9 & 8.12833 & 2.2277 & 2000 & 8192 \\
15000 & 1320.59 & 15000.2 & 8.85425 & 2.219 & 2000 & 6812 \\
20000 & 1760.78 & 20000.1 & 9.32285 & 2.14241 & 2000 & 5090 \\
25000 & 2200.98 & 25000 & 9.84283 & 2.10992 & 2000 & 4058 \\
50000 & 4401.96 & 49999.8 & 11.6897 & 1.95727 & 2000 & 2003 \\
75000 & 6602.94 & 74999.8 & 12.573 & 1.84499 & 1321 & 1321 \\
100000 & 8803.92 & 99999.2 & 13.4507 & 1.7367 & 983 & 983 \\
150000 & 13205.9 & 150000 & 14.5449 & 1.6165 & 646 & 646 \\
200000 & 17607.8 & 200000 & 15.4017 & 1.61698 & 478 & 478 \\
250000 & 22009.8 & 250000 & 15.9881 & 1.41128 & 378 & 378 \\
300000 & 26411.8 & 300000 & 16.3647 & 1.42554 & 312 & 312 \\
350000 & 30813.7 & 350000 & 17.8015 & 1.47939 & 265 & 265 \\
400000 & 35215.7 & 399999 & 17.9761 & 1.42048 & 230 & 230 \\
450000 & 39617.7 & 449999 & 16.9116 & 1.4054 & 203 & 203 \\
500000 & 44019.6 & 499999 & 18.2741 & 1.37765 & 181 & 181 \\
600000 & 52823.5 & 599999 & 17.1112 & 1.34518 & 149 & 149 \\
700000 & 61627.5 & 699997 & 19.6473 & 1.33721 & 126 & 126 \\
800000 & 70431.4 & 799999 & 19.8383 & 1.12659 & 108 & 108 \\
900000 & 79235.3 & 899995 & 20.6388 & 1.31706 & 95 & 95 \\
1000000 & 88039.2 & 1000000 & 20.8671 & 1.19851 & 85 & 85 \\
\hline
\end{tabular}
    \caption{N-V test in $(1+1)D$ for testing spacetime hyperuniformity of the ${\rm I}_{3,1}$ spacetime quasicrystal ($\rho_{\rm sQC \, I_{3,1}} \approx 11.85$).}
\label{tab:NVTestForHyperuniformitysQCI31}
\end{table}

\subsection{Can We Sample in a Lorentz-Invariant Way?}

We should mention that the sampling of causal diamonds performed here could be improved, and one improvement would be to sample somehow with a Lorentz-invariant measure. Of course, since the Lorentz group is non-compact, such a measure is difficult to find, which is why we decided to sample boosts uniformly in an interval. Perhaps what we could do is to use the metric and sample tips of causal diamonds based on that (i.e.~fixing a volume and a tip and then using the metric to find the other tip). We leave all this for future work, but, to the best of our knowledge, our setup here seems to match the setup in \cite{Saravani_2014}, if not being better. 

\section{Showing the Boost Catastrophe of Lattices}
\label{app:PlotsBoostCatastropheLattices}

\subsection{$(2+1)$ Dimensions}

The setup for the following plots is similar to the standard N-V test with the different that now we consider a boost range $\varpi \in [0,5]$ and we split it into bins of width $\delta \varpi = 0.05$ and sample $M = 1000$ $\cdiamond$ in each bin. From the collected statistics, we plot $\sigma_{N_\cdiamond}$ versus $| \varpi |$ to get a boost-stratified N-V test. The structures we test are the same lattices appearing in the N-V test in $(2+1)$D in Section \ref{subsec:NVTest2Plus1D}.  

\begin{figure}[h!]
    \centering
    \includegraphics[width=0.75\linewidth]{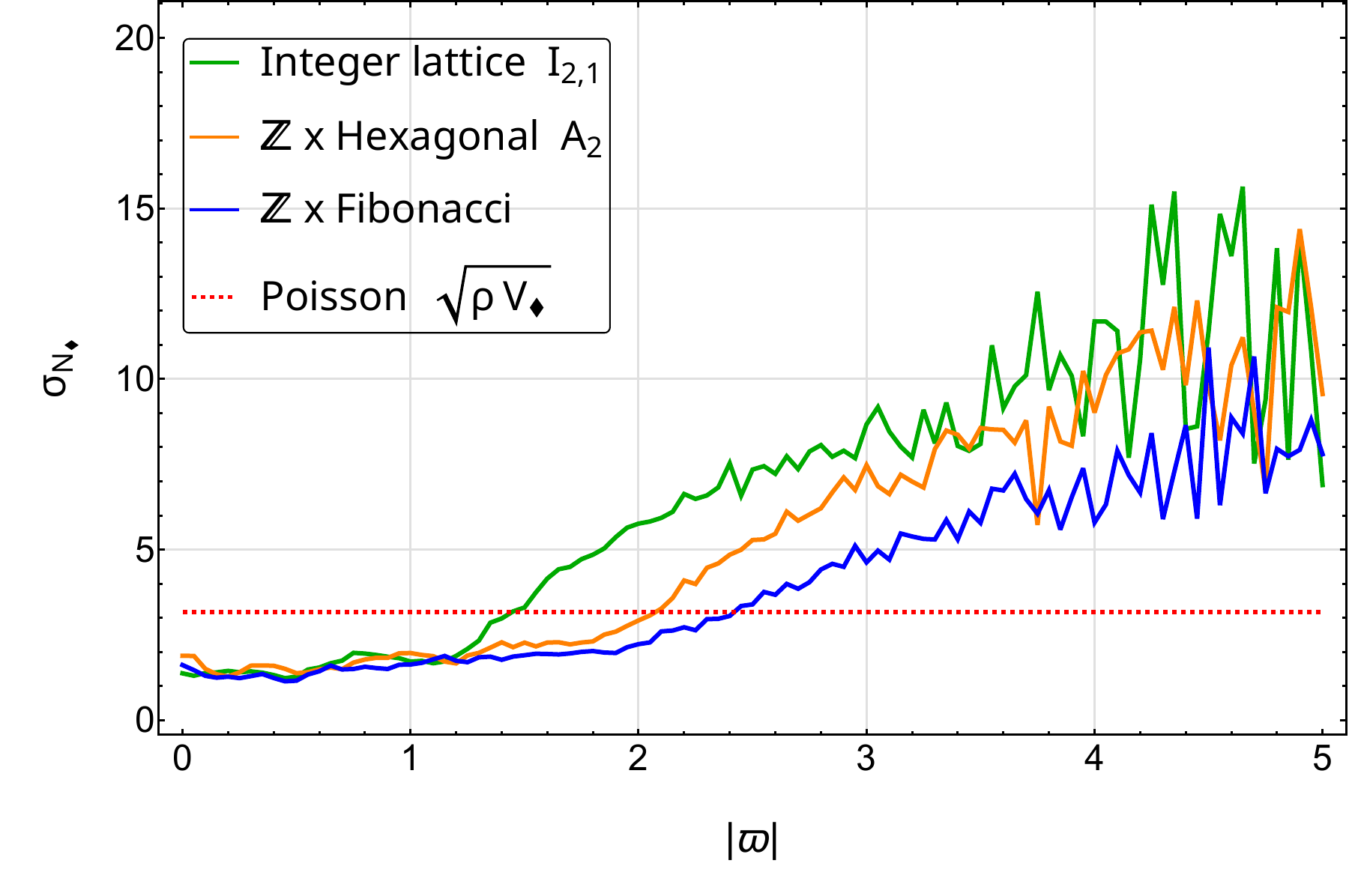}
    \caption{Boost catastrophe of lattices in $(2+1)$D for $V_{\rm sample} = 10.$}
\label{fig:BoostCatastropheLattices1Plus2D_Vol10}
\end{figure}

\begin{figure}[h!]
    \centering
    \includegraphics[width=0.75\linewidth]{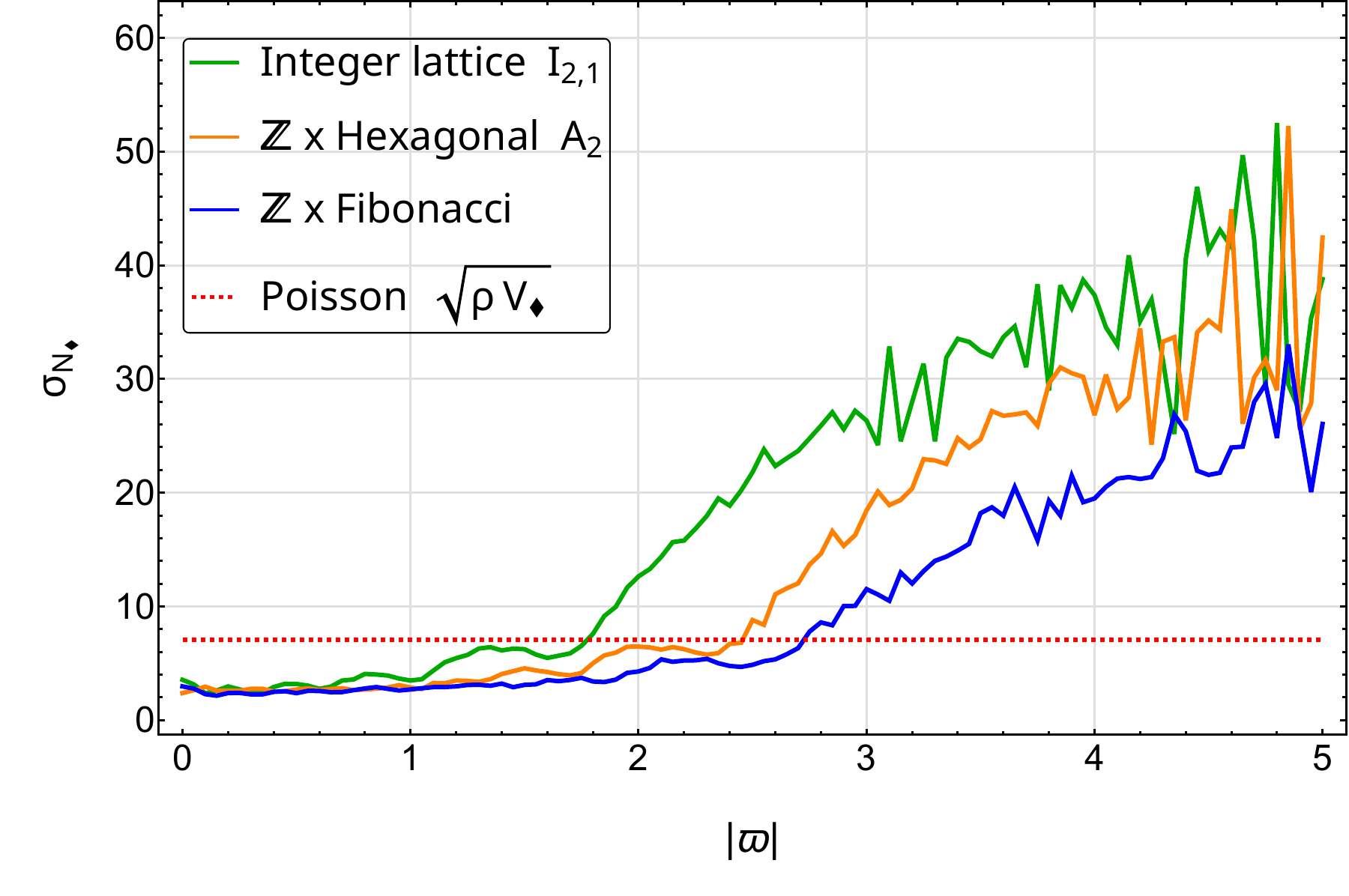}
    \caption{Boost catastrophe of lattices in $(2+1)$D for $V_{\rm sample} = 50.$}
\label{fig:BoostCatastropheLattices1Plus2D_Vol50}
\end{figure}

\begin{figure}[h!]
    \centering
    \includegraphics[width=0.75\linewidth]{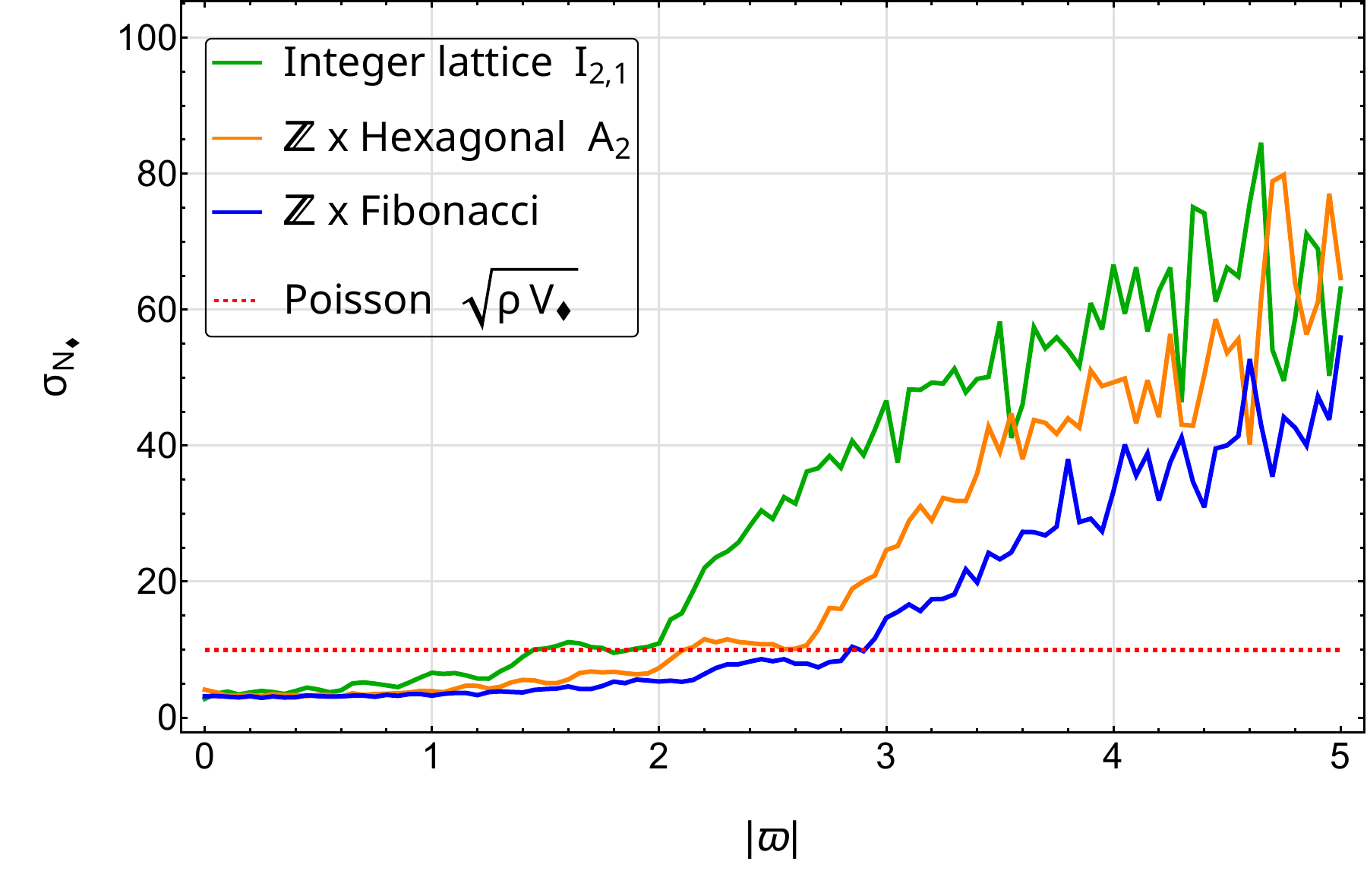}
    \caption{Boost catastrophe of lattices in $(2+1)$D for $V_{\rm sample} = 100.$}
\label{fig:BoostCatastropheLattices1Plus2D_Vol100}
\end{figure}

To investigate the behaviour of the Fibonacci lattice a bit more, we perform two more test: (1) we fix a boost to some high value $\varpi = 3.5$ and we only vary the angle $\varphi$ in the interval $[0,\pi/2]$ with step $\delta \varphi = 1.5^\circ$, (2) we fix two angles $\varpi = 0$ (the TX plane that has boost symmetry) and $\varphi  =60^\circ$ (that has no symmetry) and bin the boosts as before. The results are shown in the following figures. 

\begin{figure}[t!]
   \centering
    \begin{tabular}{cc}
        \subfloat[Angular test
\label{fig:AngularTestFib1Plus2D}
        ]{\includegraphics[width=0.49\textwidth]{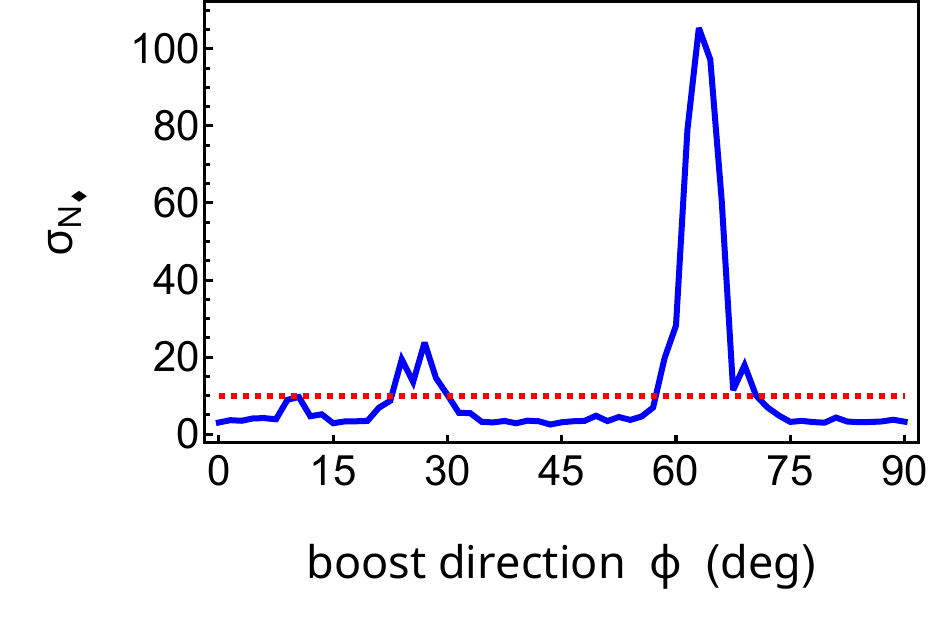}}
        \subfloat[Directional test
        \label{fig:DirectionalTestFib1Plus2D}]{ 
            \includegraphics[width=0.49\textwidth]{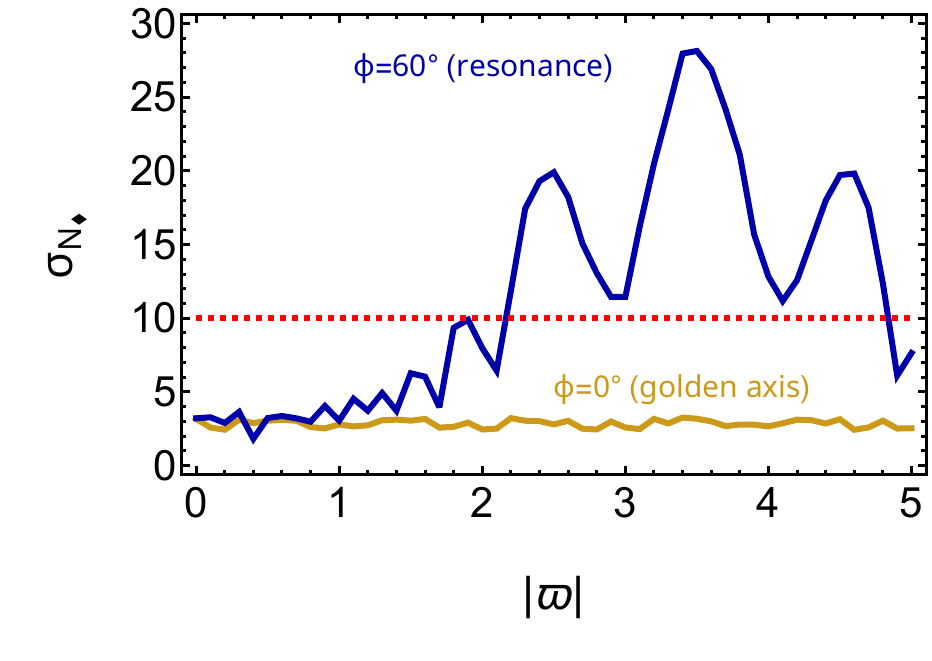}
        }
    \end{tabular}
    \caption[Angular and directional N-V test for the Fibonacci lattice in $(2+1)D$.]{Angular and directional N-V test for the Fibonacci lattice in $(2+1)D$. On the left, we see that the Fibonacci lattice is sub-Poissonian for $\varphi = 0, \pi/2$ (the planes in which it has discrete Lorentz boost symmetry), and it has the maximum super-Poissonian behaviour for around $\varphi \sim 60^\circ$. On the right, we see that the golden direction $\varphi = 0$ remains sub-Poissonian throughout the test, but the direction $\varphi = 60^\circ$ varies wildy in the tested range.}
\label{fig:DirectionalAngularTestFibonacci1Plus2D}
\end{figure}

\subsection{$(3+1)$ Dimensions}

\begin{figure}[t!]
   \centering
    \begin{tabular}{cc}
        \subfloat[$V_\cdiamond = 10$ 
\label{fig:BoostCatastropheIntegerLatticeV100}
        ]{\includegraphics[width=0.5\linewidth]{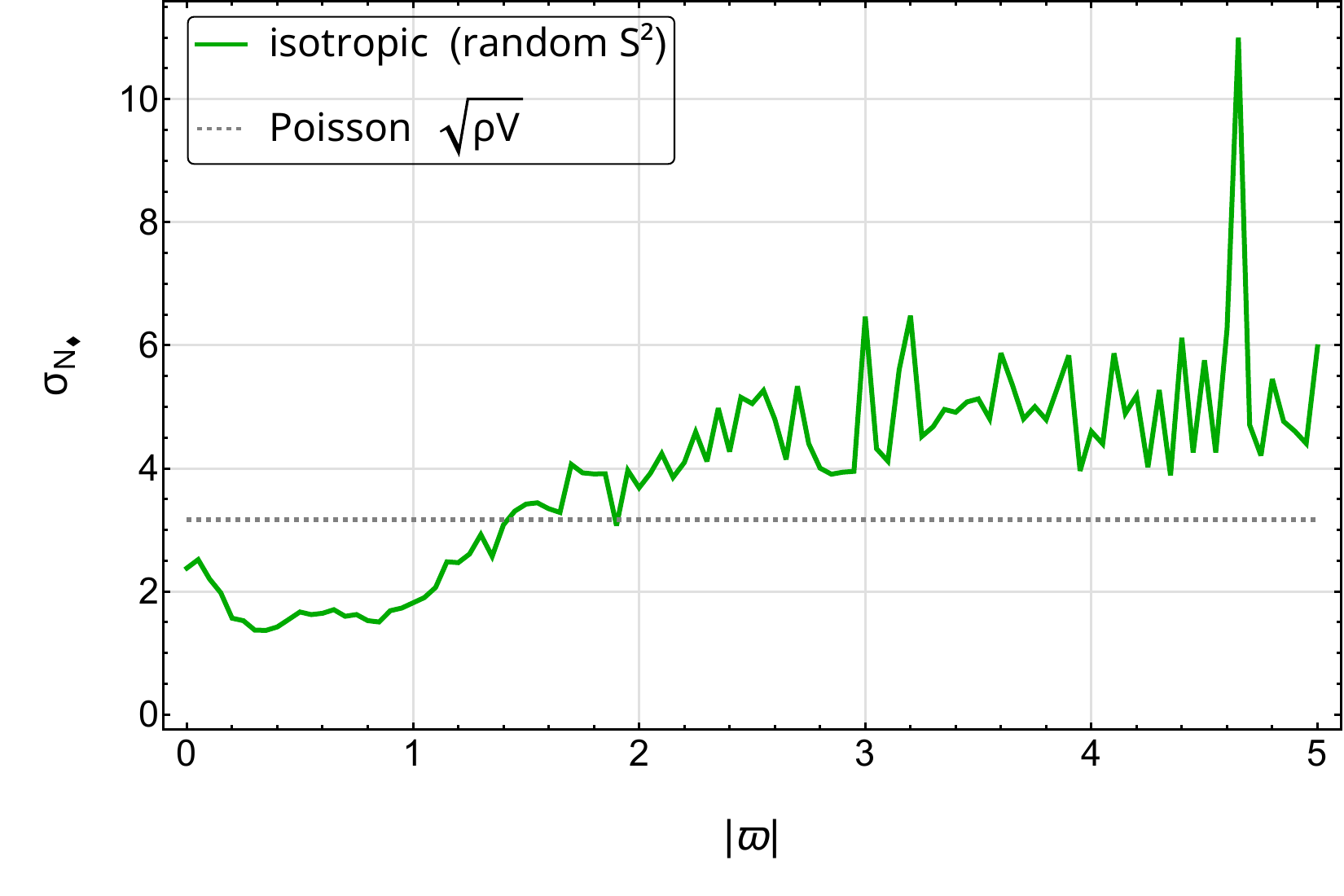}}
        \subfloat[$V_\cdiamond=100$
        ]{ \includegraphics[width=0.5\linewidth]{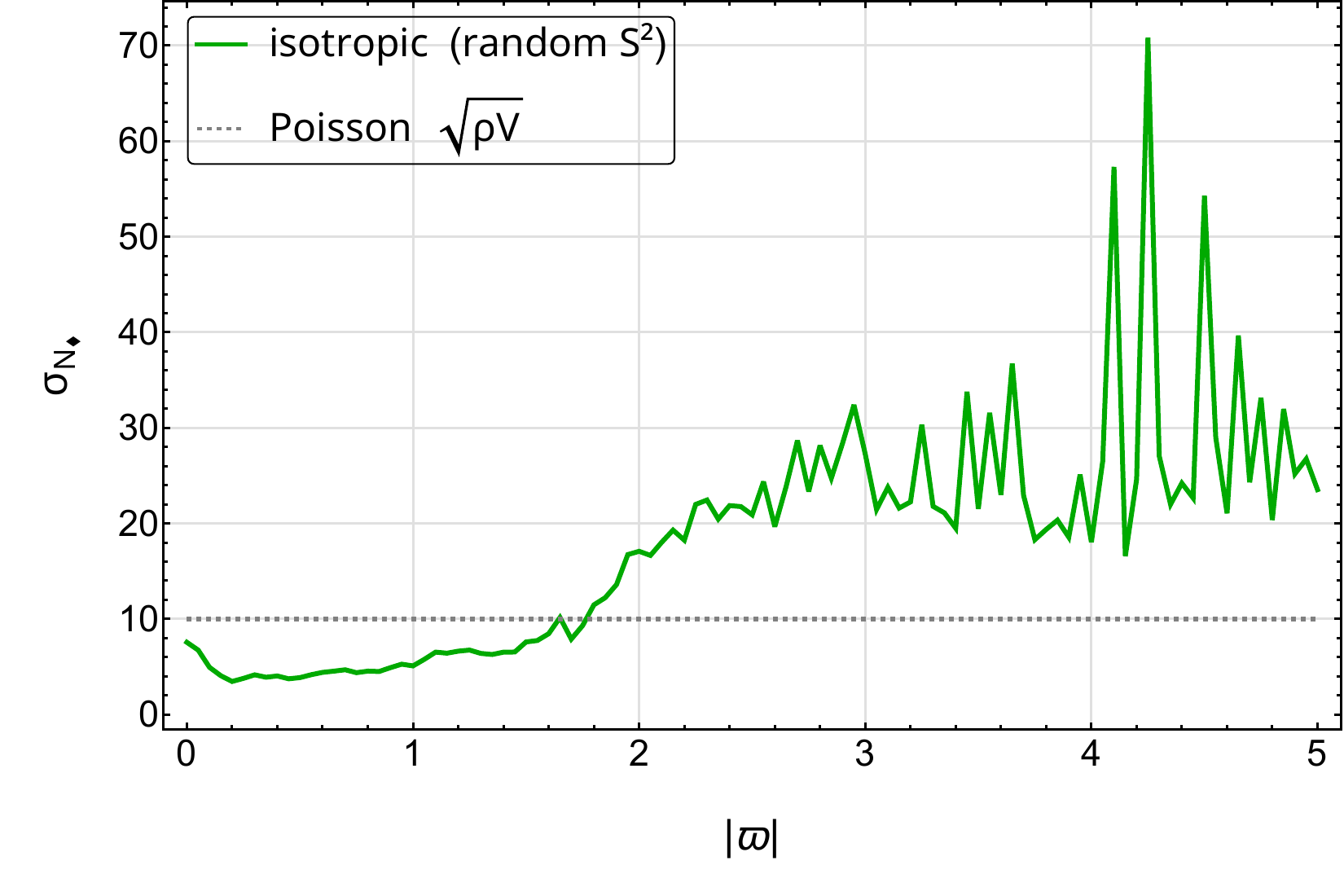}
        }
    \end{tabular}
    \caption{The boost catastrophe of the integer lattice in $(3+1)$D.}
\label{fig:BoostCatastrophe1Plus3DInteger}
\end{figure}


\end{document}